# Ice Giants
## Pre-Decadal Survey Mission Study Report

**Science Definition Team Chairs** | Mark Hofstdater (JPL), Amy Simon (GSFC)

**Study Manager** | Kim Reh (JPL)    **Study Lead** | John Elliott (JPL)

**NASA Point of Contact** | Curt Niebur

**ESA Point of Contact** | Luigi Colangeli

JPL D-100520

National Aeronautics and Space Administration

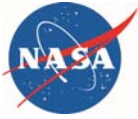

National Aeronautics and
Space Administration

**Jet Propulsion Laboratory**
California Institute of Technology
Pasadena, California

# ICE GIANTS PRE-DECADAL STUDY FINAL REPORT

**Solar System Exploration Directorate**
**Jet Propulsion Laboratory**
**for**
**Planetary Science Division**
**Science Mission Directorate**
**NASA**

**June 2017**

**JPL D-100520**

# ACKNOWLEDGMENTS

While there was a large number of contributors to this study, we would like to acknowledge those who provided study leadership and key inputs to this report:

*Science Definition Team:*

Mark Hofstadter (JPL), Co-Chair

Amy Simon (GSFC), Co-Chair

Sushil Atreya (University of Michigan)

Donald Banfield (Cornell)

Jonathan Fortney (UCSC)

Alexander Hayes (Cornell)

Matthew Hedman (University of Idaho)

George Hospodarsky (University of Iowa)

Adam Masters (Imperial College)

Kathleen Mandt (SwRI)

Mark Showalter (SETI Institute)

Krista Soderlund (University of Texas)

Diego Turrini (INAF-IAPS/UDA)

Elizabeth Turtle (APL)

*Mission Study Team:*

John Elliott (JPL), Study Lead

Kim Reh (JPL), Study Manager

Parul Agrawal (ARC)

Theresa Anderson (JPL)

David Atkinson (JPL)

Nitin Arora (JPL)

Chester Borden (JPL)

Martin Brennan (JPL)

Jim Cutts (JPL)

Helen Hwang (ARC)

Minh Le (JPL)

Young Lee (JPL)

Anastassios Petropoulos (JPL)

Sarag Saikia (Purdue)

Tom Spilker (SSSE)

William Smythe (JPL)

The Aerospace Corporation

The JPL Foundry, especially the members of the A-Team and Team X

Ames Research Center

Purdue University

In addition, we'd like to acknowledge the contributions of Liz Barrios De La Torre for graphic design, Samantha Ozyildirim for document services, and Shawn Brooks for assistance with the ring-hazard analysis.

The research was carried out at the Jet Propulsion Laboratory, California Institute of Technology, under a contract with the National Aeronautics and Space Administration.

## Disclaimer

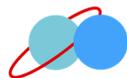

# TABLE OF CONTENTS

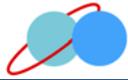

## Appendices

## List of Tables

## List of Figures

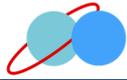

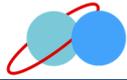

# 1 EXECUTIVE SUMMARY

## 1.1 Introduction to NASA's Ice Giants Pre-Decadal Study

The Ice Giants Study was commissioned by NASA to take a fresh look at science priorities and concepts for missions to the Uranus and Neptune systems in preparation for the third Planetary Science Decadal Survey. This study was led by a Science Definition Team (SDT) and Jet Propulsion Laboratory (JPL) with participation from Langley Research Center, Ames Research Center, The Aerospace Corporation, and Purdue University. The SDT was appointed by NASA and co-chaired by Mark Hofstadter (JPL) and Amy Simon (Goddard Space Flight Center).

The study team assessed and prioritized science objectives taking into account advances since the last Decadal Survey, current and emerging technologies, mission implementation techniques and celestial mechanics. This study examined a wide range of mission architectures, flight elements, and instruments. Six of the prioritized concepts were studied via JPL's Team X process and resulting cost estimates were subjected to independent assessment by The Aerospace Corporation. Results presented herein show that high-value flagship-class mission concepts to either Uranus or Neptune are achievable within ground rule budgetary constraints.

## 1.2 Ground Rules

NASA's Planetary Science Division announced their intent to conduct an ice giants study at Outer Planet Assessment Group (OPAG) in August of 2015. This study was initiated in November following the establishment of ground rules (Section 2.2); key rules are listed here.

- Establish a Science Definition Team
- Address both Uranus and Neptune systems
- Determine pros/cons of using a common spacecraft design for missions to both planets
- Identify missions across a range of price points, with a full life cycle cost not to exceed $2B ($FY15)
- Independent cost estimate and reconciliation with study team estimate
- Identify model payload for accommodation assessment for each candidate mission
- Constrain missions to fit on a commercial launch vehicle
- Identify benefits/cost savings if Space Launch Services (SLS) were available
- Launch dates from 2024 to 2037 (focus on the next decadal period)
- Evaluate use of emerging technologies; distinguish mission specific vs. broad applicability
- Identify clean interface roles for potential international partnerships

Selection of the full SDT was completed in December 2015 and mission concept studies were performed at JPL through CY2016.

## 1.3 Approach to Conduct of the Study

This study was initiated with the establishment of the SDT. Concurrent with this, the design team began a series of interplanetary trajectory and orbit evaluations (Appendix A). These activities fed into a JPL "A-Team" facilitated tradespace exploration workshop in which many conceptual mission architectures were developed, evaluated and ranked from a science value perspective. This resulted in a set of prioritized mission concepts for further investigation. The A-Team sessions included all members of the SDT as well as experts in the development of mission architectures, trajectories, flight systems and technologies.

Output of the A-Team workshop (Appendix B) was then studied by a technical design team to further mature the mission architectures and notional flight elements for each concept. The design team continually iterated with the SDT to ensure their intended science value was maintained/enhanced as the designs progressed.

Once a prioritized set of mission concepts had been identified, JPL's Team X was used to develop a detailed point design and cost estimate for each.

Upon completion of the Team X studies, The Aerospace Corporation performed an Independent Cost Estimate (ICE) for four of the most attractive concepts. Members of The Aerospace Corporation and Ice Giants Study teams met to review and reconcile ICE results. This led to minor revisions and Aerospace's completion of their independent assessment; final results are included in Appendix E of this report.

Finally, the study team assessed the balance between science prioritization, cost, and risk to establish study recommendations.

## 1.4 Ice Giant Science

Exploration of at least one ice giant system is critical to advance our understanding of the Solar System, exoplanetary systems, and to advance our understanding of planetary formation and evolution. Three key points highlight the importance of sending a mission to our ice giants, Uranus and Neptune.

First, they represent a class of planet that is not well understood, and which is fundamentally different from the gas giants (Jupiter and Saturn) and the terrestrial planets. Ice giants are, by mass, about 65% water and other so-called "ices," such as methane and ammonia. In spite of the "ice" name, these species are thought to exist primarily in a massive, super-critical liquid water ocean. No current model for their interior structure is consistent with all observations.

A second key factor in their importance is that ice giants are extremely common in our galaxy. They are much more abundant than gas giants such as Jupiter, and the majority of planets discovered so far appear to be ice giants. Exploration of our local ice giants would allow us to better characterize exoplanets.

The final point to emphasize about ice giants is that they challenge our understanding of planetary formation, evolution and physics. For example, models suggest they have a narrow time window for formation: their rock/ice cores must become large enough to gravitationally trap hydrogen and helium gas just as the solar nebula is being dissipated by the early Sun. Forming earlier would cause them to trap large amounts of gas and become like Jupiter, and forming later would not allow them to trap the gas they have (perhaps 10% of their total mass). But if their formation requires such special timing, why are they so common? Examples of other observations that are challenging to explain are the energy balance of their atmospheres and their complex magnetic fields. For these reasons and others, ice giant exploration is a priority for the near future.

We have identified 12 priority science objectives for ice giant exploration. They are consistent with the Planetary Science Decadal Survey as reported in the *Vision & Voyages* document (released in 2011), but advances since then have changed their prioritization. The two most important objectives relate to the formation, structure, and evolution of ice giants (**Figure 1-1**):

- Constrain the structure and characteristics of the planet's interior, including layering, locations of convective and stable regions, and internal dynamics.
- Determine the planet's bulk composition, including abundances and isotopes of heavy elements, He and heavier noble gases.

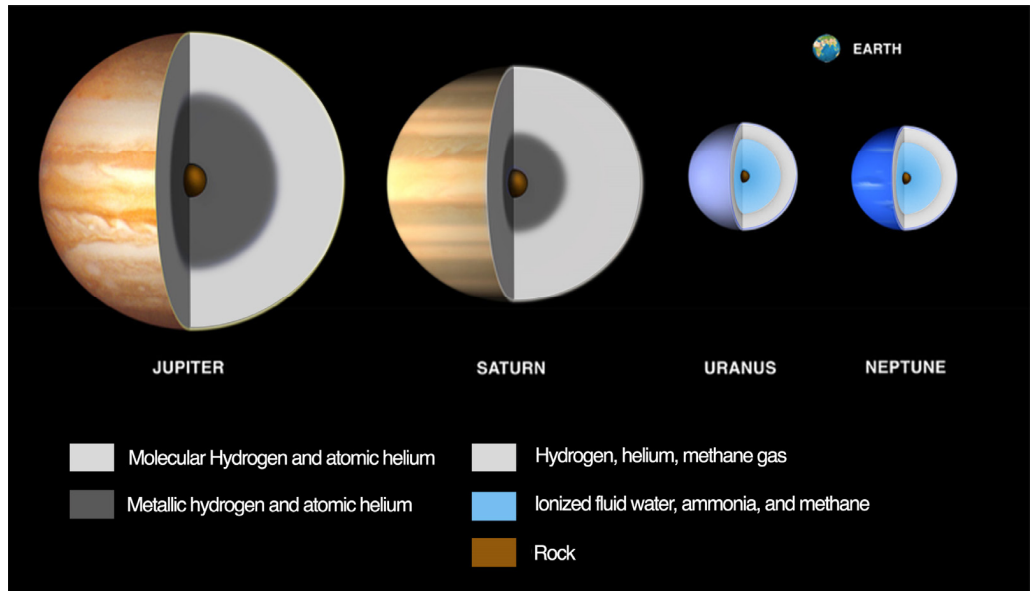

**Figure 1-1.** Illustration of compositional differences among the giant planets and their relative sizes. Earth is shown for comparison. Jupiter and Saturn are primarily made of hydrogen and helium, the terrestrial planets are almost pure rock, while Uranus and Neptune are thought to be largely supercritical liquid water.

The remaining ten objectives, which are of equal importance, touch upon all aspects of the ice giant system:

- Improve knowledge of the planetary dynamo
- Determine the planet's atmospheric heat balance
- Measure the planet's tropospheric 3-D flow (zonal, meridional, vertical) including winds, waves, storms and their lifecycles, and deep convective activity
- Characterize the structures and temporal changes in the rings
- Obtain a complete inventory of small moons, including embedded source bodies in dusty rings and moons that could sculpt and shepherd dense rings
- Determine the surface composition of rings and moons, including organics; search for variations among moons, past and current modification, and evidence of long-term mass exchange / volatile transport
- Map the shape and surface geology of major and minor satellites
- Determine the density, mass distribution, and internal structure of major satellites and, where possible, small inner satellites and irregular satellites
- Determine the composition, density, structure, source, spatial and temporal variability, and dynamics of Triton's atmosphere
- Investigate solar wind-magnetosphere-ionosphere interactions and constrain plasma transport in the magnetosphere

Our study emphasizes that key science questions exist for all aspects of the ice giant system: the planetary interior, atmosphere, rings, satellites, and magnetosphere.

We also find that Uranus and Neptune are equally compelling as a scientific target. The value of sending a spacecraft to Uranus is comparable to sending the same spacecraft to Neptune (though costs may differ). This does not mean the two planets are equivalent, however. Each planet has something important to teach us that the other cannot. A prime example of this relates to their satellites. Triton, the major satellite in the Neptune system, is a captured Kuiper Belt Object (KBO). Voyager's flyby of Triton in 1989 led to the discovery of active geysers and young terrains. It is of great interest to study this object in more detail, particularly now that we have information on another KBO, Pluto, for comparison. The capture of Triton, however, is thought to have ejected or collisionally destroyed all native major satellites of Neptune. Therefore, to study the composition of native ice-giant satellites one must go to Uranus, where Miranda and Ariel are also seen to have relatively young terrains. It is also worth noting that both ice giants host potential ocean-world satellites in their systems. This study finds that exploration of both ice giants is highly desirable if programmatically feasible.

To address all science objectives, an orbiter and an atmospheric probe would be required at one of the ice giants. A probe is the only way to measure heavy noble gases, isotopic ratios, and the bulk abundance of certain species. An orbiter is required to give us vantage points and enough time in the system to understand variable phenomena (e.g., magnetospheric responses to the varying solar wind or weather variations), to allow us to encounter several moons, and to observe all components of the system under varying geometries. Having an orbiter also opens up the possibility of serendipitous discovery and follow-up, which the Cassini mission at Saturn has demonstrated as incredibly valuable (e.g., Enceladus' plumes, or Titan's seas and lakes).

The two critical instruments for an atmospheric probe to carry would be a mass spectrometer and an atmospheric structure package (measuring temperature, pressure, and density). The most important instruments on the orbiter would be an atmospheric seismology instrument such as a Doppler Imager (providing novel measurements of interior structure), a camera, and a magnetometer. There are a range of additional instruments that should be added to maximize the science return, limited primarily by cost constraints. We find a 50 kg orbiter payload is a minimum package to consider for a Flagship mission concept. That payload (with a probe) would achieve the most important two science objectives, and partially address several others. A 150 kg orbiter payload would be needed to achieve all SDT priority science objectives. (For comparison, the Cassini orbiter at Saturn has a science payload of approximately 270 kg.) We also find that a payload near 90 kg, while not achieving all objectives, does allow significant advances in all areas. An important finding of our study is that there is a near linear relationship between mission cost and science return; there is no "plateau" of limited return. (See **Figure 1-2** at the end of Section 1.6.)

## 1.5 Survey of Mission Architecture Trade Space

This study team performed a broad and comprehensive survey of feasible mission concept architectures as detailed in Appendix A. This included an overview of mission designs, which were then mapped to notional flight system architectures to generate conceptual mission options. Launch vehicle options were evaluated, as were a variety of potential propulsion implementations. Tens of thousands of trajectories using various propulsion options, with up to four planetary flybys were investigated. The impact of using different launch vehicles (including SLS) on flight time, delivered mass, propellant throughput, and mission architecture were studied. Details of atmospheric probe coast, entry, and spacecraft orbit insertion at either of the two planets were evaluated. Finally, a procedure for computing dual spacecraft trajectories, capable of delivering individual spacecraft to both planets on a single launch, was developed and exercised.

The study finds that launches to an ice giant are possible any year within the study timeframe, but there are significant variations in performance and available science targets. The availability of Jupiter gravity assist maximizes delivered mass to an ice giant resulting in preferential launch windows for Uranus missions in the 2030–2034 timeframe and a corresponding window of 2029–2030 for Neptune. In these favorable periods chemical trajectories could deliver ample mass for the Uranus missions studied in an 11-year flight time, using a launch performance capability similar to the Atlas V 551. Neptune trajectories utilizing solar electric propulsion (SEP) can deliver a similar mass to Neptune orbit in 13 years using launch performance capability similar to the Delta IVH. There are no all-chemical trajectories to Neptune, even using a Delta IVH, that yield a mission duration less than 15 years, a design target chosen to be consistent with Radioisotope Power System (RPS) design life and mission reliability. Significant science can be done during gravity assists at a gas giant, particularly if a Doppler Imager-type instrument is on board. If a Saturn flyby is preferred over the Jupiter gravity assist, only trajectories to Uranus are available in the time period studied, and launch must occur before mid-2028.

The use of SEP for inner solar system thrusting has the potential to significantly reduce flight times to Uranus and/or increase delivered mass. A variety of trajectories to Uranus and Neptune were evaluated considering a range of SEP power levels, assuming inclusion of an additional SEP flight element (referred to as a SEP stage). The SEP stage would carry solar arrays and ion thrusters and would be used in the inner solar system as far out as 6 AU, at which point solar power is insufficient to provide additional thrusting and the SEP stage would be jettisoned. SEP-enhanced mission concept designs also see a slight preference in launch dates corresponding to availability of Jupiter gravity assists, but well-performing trajectories are possible in any year of the period studied.

Early A-Team studies suggested low-mass SEP stages are possible which would provide significant performance enhancements at both Uranus and Neptune. The more detailed Team X design suggested much higher masses, negating the usefulness of SEP to Uranus. It may be valuable to perform a detailed assessment of an optimized SEP stage design for outer planet missions to confirm the optimal uses of SEP.

There are no trajectories that allow a single spacecraft to encounter both Uranus and Neptune. A single SLS launch vehicle could, however, launch two spacecraft, one to each ice giant.

## 1.6   Assessment of Costs

Based on SDT recommendations (Section 3.5), point design maturation and higher fidelity cost assessments were carried out for four key mission architectures. These architectures were chosen to span the mission concept parameter space in such a way as to allow reliable interpolation of their costs to all other missions considered. The four architectures are:

- Neptune orbiter with ~50-kg payload and atmospheric probe
- Uranus flyby spacecraft with ~50-kg payload and atmospheric probe
- Uranus orbiter with ~50-kg payload and atmospheric probe
- Uranus orbiter with ~150-kg payload but without a probe

Initial architecture assessments included a SEP stage to provide mission flexibility, decreased mission duration, and/or increased mass delivered to the target body. As stated earlier, detailed point designs indicated that the SEP stage would result in significantly increased flight system mass and cost for the Uranus missions, although SEP remains attractive for the Neptune orbiter mission. To avoid these detrimental impacts, chemically propelled ballistic architectures were developed for the Uranus orbiter mission concepts.

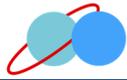

Team X results for these four architectures are summarized in **Table 1-1**.

Team X cost estimates were generated for each mission option using JPL Institutional Cost Models that are based on historical results from missions that have flown as well as other factors. The cost model algorithms, developed by JPL doing organizations, represent "most likely" estimates (i.e., 50% confidence level without reserves). These models have been shown to estimate total mission lifecycle cost within a range of -10% to +20% relative to historical actuals. The model produces cost estimates to Work Breakdown Structure (WBS) Level 3 for the flight system and instruments. NASA's Radioisotope Power System Program Office provided estimated costs for enhanced multi-mission radioisotope thermoelectric generators (eMMRTGs) and radioisotope heater units (RHUs) and the additional launch services provider costs for support of RPS-powered missions.

Following completion of the Team X studies, detailed reports were provided to The Aerospace Corporation as input for their ICE. The Aerospace ICE provides a probabilistic estimate using a combination of models and analogies. Multiple estimates are used for all cost elements to bolster confidence in results and to feed into cost risk analysis. Both cost models and analogous project costs are used to tie the estimated cost to historical actuals. Actual cost of analogies are adjusted based on functional relationships found in traditional cost estimating relationships. Aerospace cost estimates (including reserves) are made to the 70% confidence level.

**Table 1-1.** Mission concept analysis summary.

|  | 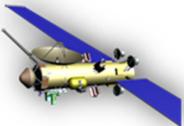 | 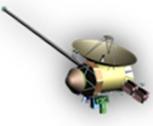 | 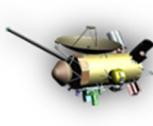 | 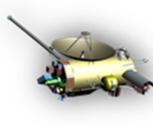 |
|---|---|---|---|---|
| **Case Description** | Neptune Orbiter with probe and ~50 kg science payload. Includes SEP stage for inner solar system thrusting. | Uranus Flyby with probe and ~50 kg science payload. Chemical only mission. | Uranus Orbiter with probe and ~50 kg science payload. Chemical only mission. | Uranus Orbiter without a probe, but with ~150 kg science payload. Chemical only mission. |
| **Science** | Highest priority plus additional system science (rings, sats, magnetospheres) | Highest priority science (interior structure and composition) | Highest priority plus additional system science (rings, sats, magnetospheres) | All remote sensing objectives |
| **Payload** | 3 instruments* + atmospheric probe | 3 instruments* + atmospheric probe | 3 instruments* + atmospheric probe | 15 instruments** |
| **Payload Mass MEV (kg)** | 45 | 45 | 45 | 170 |
| **Launch Mass (kg)** | 7365 | 1524 | 4345 | 4717 |
| **Launch Year** | 2030 | 2030 | 2031 | 2031 |
| **Flight Time (yr)** | 13 | 10 | 12 | 12 |
| **Time in Orbit (yr)** | 2 | Flyby | 3 | 3 |
| **Total Mission Length (yr)** | 15 | 10 | 15 | 15 |
| **RPS use/ EOM Power** | 4 eMMRTGs/ 376W | 4 eMMRTGs/ 425W | 4 eMMRTGs/ 376W | 5 eMMRTGs/ 470W |
| **LV** | Delta IVH + 25 kW SEP | Atlas V 541 | Atlas V 551 | Atlas V 551 |
| **Prop System** | Dual Mode/NEXT EP | Monopropellant | Dual Mode | Dual Mode |

*includes Narrow Angle Camera, Doppler Imager, Magnetometer  **includes Narrow Angle Camera, Doppler Imager, Magnetometer, Vis-NIR Mapping Spec., Mid-IR Spec., UV Imaging Spec., Plasma Suite, Thermal IR, Energetic Neutral Atoms, Dust Detector, Langmuir Probe, Microwave Sounder, Wide Angle Camera

Team X and independent cost estimates from Aerospace align well within model uncertainty as shown in **Table 1-2**. Differences in end-to-end cost are typically about $300M. Aerospace costs are higher due to higher operations and flight system cost estimates, and their higher confidence level (70% vs. JPL's 50%). Both cost models indicate that scientifically compelling mission options fall within the $2B ground rule, though the preferred mission concepts are above that target. NASA and European Space Agency (ESA) collaboration can enable the preferred options while keeping NASA costs below the $2B constraint.

Key conclusions from cost comparisons in **Table 1-2**:

- For similar science return, a Neptune mission costs about $300M more than a Uranus mission, driven primarily by the cost of the SEP stage needed for Neptune. (Compare the Neptune orbiter with probe and 50 kg payload in **Table 1-1** against the Uranus orbiter with probe and 50 kg payload.)

- A Uranus orbiter with a 50 kg science payload and an atmospheric probe fits within the $2B cost target. The SDT considers such a mission a science floor, but recommends a larger orbiter payload.

- A Uranus orbiter with a 150 kg science payload and an atmospheric probe is estimated to cost between $2.3B (JPL) and $2.6B (Aerospace). (For this estimate, we add the $300M cost of a probe—not shown in the above table—to the mission in the far right column.)

**Table 1-2.** Mission concept cost summary.

| Case Description | Neptune Orbiter with probe and ~50 kg science payload. Includes SEP stage for inner solar system thrusting. | Uranus Flyby with probe and ~50 kg science payload. Chemical only mission. | Uranus Orbiter with probe and ~50 kg science payload. Chemical only mission. | Uranus Orbiter without a probe, but with ~150 kg science payload. Chemical only mission. |
|---|---|---|---|---|
| **Team X Cost Estimates ($M, FY15)** | | | | |
| **Total Mission Cost\*** | 1971 | 1493 | 1700 | 1985 |
| **Phase A-D Cost (incl. Reserves)** | 1637 | 1293 | 1406 | 1418 |
| **Phase E Cost (incl. Reserves)** | 334 | 200 | 295 | 568 |
| **Aerospace ICE ($M, FY15)** | | | | |
| **Total Mission Cost\*** | 2280 | 1643 | 1993 | 2321 |
| **Phase A-D Cost (incl. Reserves)** | 1880 | 1396 | 1559 | 1709 |
| **Phase E Cost (incl. Reserves)** | 400 | 247 | 433 | 612 |

*Includes cost of eMMRTGs, NEPA/LA, and standard minimal operations, LV cost not included*

In **Figure 1-2**, we plot the relative science value (Section 3.4.3) against JPL-estimated cost for a subset of the mission architectures considered. All spacecraft in this chart carry the 50 kg payload (Section 3.3.2). An orbiter with probe to Uranus costs $1.7B, and two-planet missions (with both spacecraft carrying identical small payloads, see below and Section 4.9) start at $2.5B. In the figure, the center of each descriptive phrase indicates the cost for that mission concept. This is intentionally vague because, to maximize the number of concepts on this chart, concepts are plotted together which have differing degrees of fidelity and used slightly different model or instrument assumptions. The relative costs and trends are meaningful, however. See **Table 1-2** for our most accurate cost estimates on four key mission concepts, Section 4.9.8 for dual-planet costs, and Section 6.2 for a cost summary. Looking at the trend in **Figure 1-2**, the relative science score of a mission is almost linear with cost, highlighting that we are not in a regime of diminishing returns. This means the "science per dollar" value is constant across the parameter space, and increasing the investment in the mission provides a correspondingly larger increase in science return.

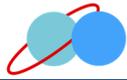

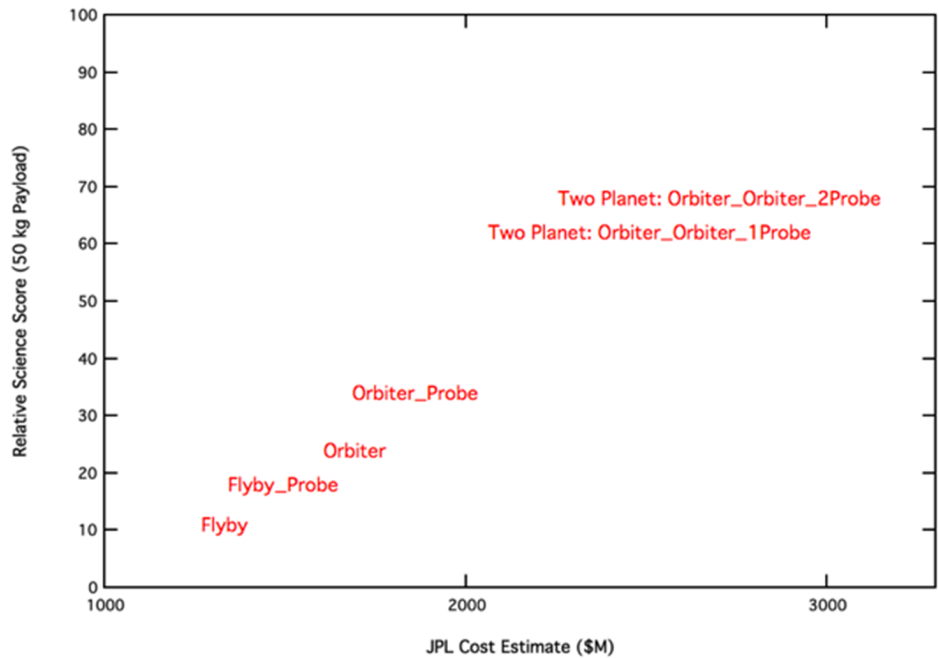

**Figure 1-2.** Relative science return of mission concepts versus cost. This is a subset of architectures considered in the study. All spacecraft carry the small (50 kg) payload. Single-planet missions would target Uranus. Costs are from JPL estimates; Aerospace costs are $100M to $300M higher. See text for a discussion of cost details.

Regarding the architectures that could fly to both Uranus and Neptune (Section 4.9), the study finds that two spacecraft and an SLS launch vehicle would be required (if the ground-rule requiring only one launch vehicle is relaxed, two smaller launch vehicles could be used). Cost estimates for these two-spacecraft, two-planet mission concepts are less accurate, but indicate that flying our smallest orbiter plus a probe to each planet would cost $3.2±0.5B, with the lower end of that range representing the cost for two identical spacecraft built at the same time. If a two-planet, two-spacecraft option is pursued, a ~$4B investment would be scientifically far superior, allowing fully instrumented spacecraft.

## 1.7 Assessment of Technology

In preparation for the Ice Giants Study, we examined the status of a number of technologies with the potential for enhancing the science, reducing the cost or reducing mission duration. A guiding philosophy adopted for the study however, was to develop missions with existing technology, only introducing new technologies where their application would enable or significantly enhance a given mission concept. Only two new technologies, both of which are currently under development, were deemed enabling for the mission described: a potential eMMRTG for the spacecraft and Heatshield for Extreme Entry Environment Technology (HEEET) for the entry probe. The eMMRTG (**Figure 1-3**) would provide a significant improvement in specific power over the existing MMRTG technology at beginning of life (BOL) but, more importantly, a much larger gain at end of life, which is critical, given the duration of an ice giant mission. HEEET would be enabling for the entry conditions of probes at both Uranus and Neptune. Other currently available heat shield materials such as phenolic impregnated carbon ablator (PICA) put severe constraints on probe entry conditions which could not be met by most architectures. The carbon phenolic used on the Galileo Jupiter probe is no longer available.

A number of new technologies were not found to be necessary to the mission concepts described but if they were available they could have an impact on the performance and/or cost of the mission: 1) Aerocapture technology could enable trip times to be shortened, delivered mass to be increased or both. 2) Cryogenic propulsion could have similar but not as pronounced effects. 3) Advanced RPS technologies, with even better specific power than the proposed eMMRTG, such as a segmented modular RTG, could enable more mass or power for instruments or both. 4) Optical communications could dramatically increase the data return from an outer planet mission and 5) Advanced mission operations technologies could drive down cost and permit more adaptive missions operations than are envisaged in the missions reported here. Further details on these advanced technologies and their mission applicability are included in Appendix D.

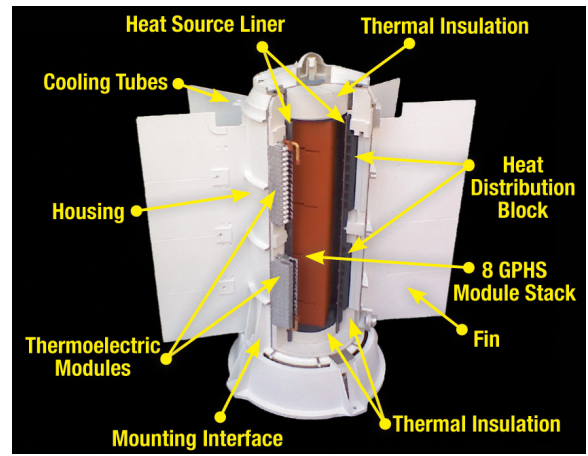

**Figure 1-3.** Proposed eMMRTG configuration.

## 1.8 Recommendations

We reaffirm the scientific importance and high priority of implementing an ice giant mission, as recommended in the *Vision and Voyages* report from NASA's second Planetary Science Decadal Survey. To ensure that the most productive mission is flown, we recommend the following:

- An orbiter with probe be flown to one of the ice giants
- The orbiter carry a payload between 90 and 150 kg
- The probe carry at minimum a mass spectrometer and atmospheric pressure, temperature, and density sensors
- The development of eMMRTGs and HEEET be completed as planned
- Two-planet, two-spacecraft mission options be explored further
- Investment in ground-based research, both theoretical and observational, to better constrain the ring-crossing hazard and conditions in the upper atmosphere (both of which are important for optimizing the orbit insertion trajectory)
- Mature the theory and techniques of atmospheric seismology
- International collaborations be leveraged to maximize the science return while minimizing the cost to each partner
- A joint NASA/ESA study be executed that uses refined ground-rules to better match the programmatic requirements each agency expects for a collaborative mission

The study validated that NASA could likely implement a mission to the ice giants for under $2B (FY15) that would achieve a worthy set of science objectives. Opportunities exist to achieve all priority science objectives for less than $3B. A partnership with another space agency has the potential to significantly increase the science return while limiting the cost to each partner. Given the development time scale of outer solar system missions, the time of the best launch opportunities, and—for Uranus missions—the desire to arrive at the optimal season, now is the time to begin formulating the next mission to the ice giants.

Note that the information in this report is predecisional, for planning and discussion purposes only.

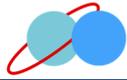

## 2    INTRODUCTION

The Ice Giants Pre-Decadal Study was requested by NASA HQ to provide inputs to the next Decadal Survey. There have been advances in science and technology since the 2013–2022 Decadal Survey (*Vision and Voyages*) that warrant a reexamination of missions to Uranus and Neptune. Additionally, the passage of time since the earlier Decadal Survey necessitates a reevaluation of interplanetary trajectories for the next decade. NASA PSD announced their intent to conduct an Ice Giants Study at OPAG in August 2015. The study was initiated in November following the establishment of ground rules. Selection of the full SDT was completed in December 2015. Mission concept studies were performed at JPL through 2016 with the participation of Ames Research Center (ARC) and Purdue University. Independent cost assessment was provided by The Aerospace Corporation. This report presents the results of this effort.

### 2.1    Study Goal and Objectives

The goal of the Ice Giants Study was to "Assess science priorities and affordable mission concepts & options for exploration of the Ice Giant planets, Uranus and Neptune in preparation for the next Decadal Survey." The objective was stated as, "Evaluate alternative architectures to determine the most compelling science mission(s) that can be feasibly performed within a price limit of $2B ($FY15)" with specific direction to:

- Identify mission concepts that can address science priorities based on what has been learned since the 2013–2022 Decadal Survey
- Identify potential concepts across a spectrum of price points
- Identify enabling/enhancing technologies
- Assess capabilities afforded by SLS

### 2.2    Study Guidelines, Assumptions, and Ground Rules

To begin the study a set of ground rules was developed and iterated with NASA to ensure a well-understood set of assumptions for concept evaluation.

- Establish a Science Definition Team (SDT)
  - Mission science objectives shall be guided by the SDT
  - Science objectives to be established based on 2013–2022 Decadal Survey revised with recent developments in science and technology
- Study to address both Uranus and Neptune systems
- Determine pros/cons of using one spacecraft design for both missions (possibility of joint development of two copies)
- Identify mission concepts across a range of price points, with a cost not to exceed $2B ($FY15) per mission
  - All costs in $FY15
  - Assume Class B (per NASA Procedural Requirements [NPR] 8705.4), Category I (per NPR 7120.5) mission
  - Exclude launch vehicle (LV) from cost
  - Include cost of RPS (cost according to New Frontiers [NF] 4 guidelines as stated in Community Announcement and draft Announcement of Opportunity [AO])
  - Include cost of National Environmental Policy Act (NEPA) / launch approval (LA)

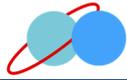

- Include operations (full life cycle mission cost)
- Include Deep Space Network (DSN) as separate line item
- Include minimum 30% reserves (A–D), 15% (E–F)
  - Reserves excluded on RPS and LV
  - Provide risk-based rationale for reserves
- Exclude cost of technology development from the mission cost estimate; provide a separate cost estimate for technology development
- Assume no foreign contributions to reduce costs but identify areas where such contributions would be beneficial to NASA in terms of cost and interfaces

- Independent cost estimate and reconciliation with study team
- Power profile and mass of RPS units will be provided by the RPS Program Office. Limit RPS to MMRTG, eMMRTG; RHUs for heating are allowed.
- Identify model payload for accommodation assessment for each candidate mission. SDT should also identify other candidate instruments not in the accommodated model payload that can address the science objectives.
- Identify clean interface roles for potential international partnerships
- Constrain missions to fit on a commercial LV
  - Also identify benefits/cost savings if SLS were available (e.g., time, trajectory, etc.)
- Launch dates from 2024 to 2037 (focus on the next decadal period)
  - Identify any potential game-changing opportunities through the 2040s
- Evaluate use of realistic emerging enabling technologies; distinguish mission specific vs. broad applicability
- Studies shall include an assessment of mission and implementation risk
- Studies should use margins per institutional guidelines in all areas. The margin is defined as:

$$\text{Margin} = \text{Max Possible Resource Value} - \text{Proposed Resource Value}$$
$$\text{Margin (\%)} = \text{Margin} \times 100 \,/\, \text{Max Possible Resource Value}$$

- DSN capability. A new architecture is expected for the DSN for the mission lifetime used for these studies. Detailed plans for this new architecture are not yet complete, and firm schedule and budget commitments are not in place. As a result, DSN performance expectations for this timeframe are not well defined and could be over optimistic. Studies should confine themselves to the broad assumptions below.
  - Study teams should assume that Ka band (32 GHz; 500 MHz bandwidth) is available for downlink from the spacecraft. Tests with New Horizons show using both right hand circular and left hand circular polarization in the downlink can increase data return by almost a factor of 2.
  - Studies should assume that the existing 34 m antenna sub-network is available. Spacecraft telecom systems on each mission must be able to communicate the required mission datasets through these 34 m systems now in place. Arrays of 34 m stations may be considered.
  - A 70 m equivalent capability can be assumed only for critical events and safing. The impact on data return of using 70 m capability shall be determined.
  - Studies should assume the DSN ground system can handle a throughput of 100 Mbits/sec.

- Planetary protection requirements are documented in NPR document 8020.12C; details are provided in specification sheets appended to that NPR. Assume Planetary Protection Categories I and II. These categories require documentation only, and should refer to NPR 8020.12 for details. Missions targeting such objects, but that will encounter Mars or another object of greater concern as part of the orbital trajectory, must meet the more stringent requirements. For additional discussion and clarification of planetary protection requirements, including sample return, please contact the NASA Planetary Protection Officer (cassie.conley@nasa.gov).
- Deliverables shall consist of a final written report, in addition to presentations for status and final reviews

## 2.3    Study Approach

The study began with the establishment of the SDT. Concurrent with this, the JPL design team began a series of trajectory evaluations to establish an initial tradespace, and complete study ground rules were developed and negotiated with NASA HQ. These three activities fed into a JPL "A-Team" facilitated workshop in which mission architectures were evaluated and ranked to develop a set of high value mission concepts for further investigation. The A-Team sessions integrated representation from all members of the SDT, as well as experts in mission architecture, trajectory and flight system design.

Output of the A-Team workshop (Appendix B) was then studied by a technical design team to develop mission architectures and notional flight elements of each concept, resulting in a number of notional designs. Iteration of concepts between the design team and the SDT continued through this process to ensure science value was maintained/enhanced as the designs progressed.

Once a prioritized set of mission concepts had been fleshed out, JPL's Team X was used to provide a detailed point design and cost for each concept.

Upon completion of the Team X study reports, the Aerospace Corporation was contracted to perform an Independent Cost Estimate (ICE) for four of the most attractive concepts. These ICE results were then reconciled with the team and their results included in this report (Appendix E).

The study team reviewed the cost results in light of science prioritization to develop final recommendations.

### 2.3.1    A-Team

To support its wide range of mission activities, JPL maintains an organizational capability specifically dedicated to concept development activities, providing services to NASA, as well as a variety of other government, academic, and industry users. Specialized teams and tools allow exploration of a diverse range of trade spaces, providing capabilities that can take an idea all the way from a sketch on a napkin to a detailed mission architecture, space vehicle and ground system design.

An important early step in the execution of this study was the engagement of the A-Team at JPL to facilitate exploration of the wide range of potential Ice Giants science objectives and practical mission architectures. The A-Team is a standing group at JPL formed with the purpose of addressing the need for rapid, effective early tradespace exploration and mission concept development as a precursor to the traditional "point design" development stage. The A-Team approach integrates a small team of broad-thinking, multidisciplinary experts to work with the client team to generate and explore a wide range of possibilities driven by the mission objectives.

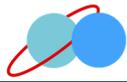

Group brainstorming and trade space analyses are conducted at a higher level of assessment across multiple mission architectures to enable rapid assessment of a set of diverse, innovative concepts. By rapidly examining a large number of varied science and implementation options early, the A-Team approach avoids a typical design team's natural tendency to drive to a baseline architecture prematurely and seeks to avoid being constrained early on by an architecture that does not adequately address the mission objectives or has unacceptable cost or risk.

The Ice Giants SDT met with the A-Team in March 2016. The study consisted of 3 full-day sessions.

In the first session, findings of the last decadal survey were summarized highlighting changes in the state of knowledge that had occurred since it was completed. Members of the SDT then gave presentations describing ice giant science that could be addressed by a dedicated mission. This was followed by a report on results of an ice giant preparation study, held the week prior to the science study to provide a first look at cost, possible trajectories, power options, possible architectures, etc. Presentations on radio science and Doppler Imaging (for giant-planet seismology) were also given.

The second session began with the SDT prioritizing the science investigations in their working science traceability matrix (STM). The full team then discussed and identified nominal small (30 kg), medium (100 kg) and large (200 kg) payloads on flyby or orbiting space craft that could address the prioritized science.

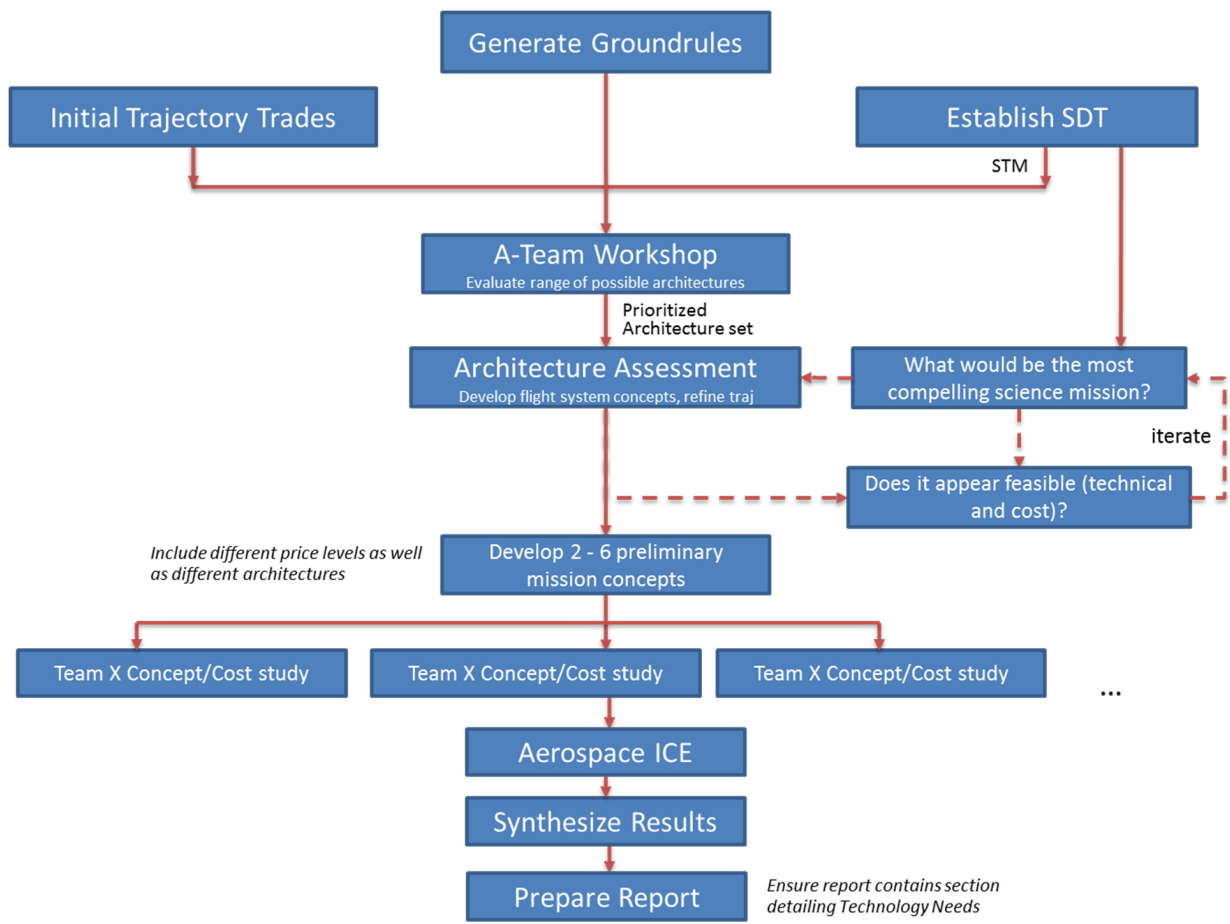

**Figure 2-1.** Ice Giants Study logic diagram.

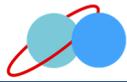

In the third session, a Science Value Matrix was developed, which assessed how well various architectures were able to address the prioritized science. The top ranking architectures were then assigned rough costs based on the A-Team Concept Maturity Level (CML) 2 cost tool. This enabled an evaluation of the science value achieved per unit cost ($) for the top architectures. Actions needed going forward were then discussed and identified. Section 3.4 provides details of the A-Team study and its results.

### 2.3.2 Team X

Once the SDT and technical team had identified the top candidates for point design development, JPL's Team X was used to flesh out mission and flight system concepts, as well as estimate cost. Team X, developed by JPL in the mid-1990s to explore new, more cost-effective ways of doing spacecraft and mission development, is composed of about 20 discipline experts, drawn from the ranks of the organizations responsible for each area at JPL, operating in a collaborative manner and using tools designed and approved by these same organizations. Studies are conducted in a dedicated facility with networked workstations sharing data on a common server. The team's work can range from simple mass or cost estimation of existing designs to more complex optimization trades (e.g., trading trajectory vs. communications data rate vs. telecom energy requirements vs. battery mass vs. science return), to extended architecture studies (is a rover the most cost effective way to gather samples or would a hopper or group of low-cost landers produce better science?).

For the Ice Giants Study, three representative mission architectures were initially presented for Team X development; a Uranus orbiter with a 50-kg science payload and a probe, a Neptune orbiter, also with a 50-kg science payload and a probe, and a Uranus orbiter mission without a probe, but with a 150-kg science payload. Prior to these three studies, a separate design session had been held to develop a point design and cost for the Uranus and Neptune atmospheric probes.

All three of these initial architectures included a solar electric propulsion (SEP) stage for use in the inner solar system as a means to shorten trip time and increase mass delivered to the target body. In the course of the first two studies, it was found that the mass of the SEP stage was likely to be considerably higher than had been initially assumed by the design team when developing the mission architectures, to the point that the SEP stage was actually harming the mission performance and significantly increasing overall mission cost. As a result of this finding, the team elected to perform two additional Team X studies, looking at options for chemical-only missions to Uranus. The Neptune orbiter/probe option was determined to still benefit from the SEP stage and so this option was not revisited.

Finally, in an attempt to try to bound a lower cost mission that would still be scientifically compelling, Team X was tasked with the study of a mission consisting of a Uranus flyby spacecraft with atmospheric probe. Section 4 provides details of the Team-X studies and their results.

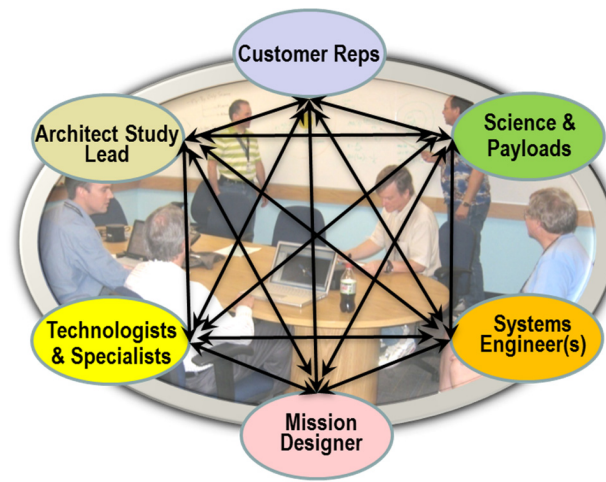

**Figure 2-2.** Small architecture team working in a concurrent setting enables efficient interactions and member contributions across the entire architectural trade space.

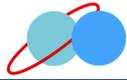

### 2.3.3 Mission/Trajectory Design

Over the past few decades, since at least the 1960s, there have been many studies which have investigated mission design options for exploration of Uranus and Neptune. These range from pure chemical trajectories to electric propulsion trajectories, both with and without gravity assists, with various underlying assumptions on the mission architecture.

In the present study, a large set of trajectories for a mission to the ice giants launching between 2024 and 2037, with a variety of launch vehicles, propulsion options, and gravity assist paths of up to four planetary flybys, were identified and documented. Four major classes of trajectory were studied: 1) ballistic (chemical) trajectories, 2) SEP trajectories, 3) radioisotope electric propulsion (REP) trajectories, and 4) dual-spacecraft, single-launch trajectories capable of delivering two spacecraft, one to each planet on a single launch. For all these classes, key trajectory trades based on launch C3 (launch energy in $km^2/s^2$), flight time, delivered mass, approach velocity, choice of gravity-assist bodies, spacecraft power, number of EP thrusters and $\Delta V$ requirements were studied. The effect of powered flybys and deep space maneuvers (DSMs) between flybys were also quantified. In most cases, a leveraging DSM was found to benefit mission performance by reducing flight time or launch C3, and/or by improving the phasing. Additional trades for the dual-spacecraft trajectories involved looking for hybrid trajectory pairs that would separate either directly after launch or after one of the common flybys. The role of mission architecture in determining the mission design was also examined, for example, dropping a probe from a flyby spacecraft is constrained in flyby velocity by the permissible acceleration loads that the probe can withstand.

The general solution approach consisted of a broad search followed by local pruning and optimization. An impulsive, patched-conic–based search algorithm (called STAR), capable of adding optimal powered flybys and impulsive leveraging/deep space maneuvers between flybys, was used for the initial broad search. For EP trajectories, the magnitude of the DSM was selected to reflect the maximum thrust capability of the spacecraft. Mission goals and constraints were then used to further prune the solution space. A subset of the remaining large set of trajectories was then refined using JPL's Mission Analysis Low-Thrust Optimizer (MALTO) software, a preliminary trajectory design tool. The dual-spacecraft trajectories were identified by examining the large set of single spacecraft trajectories and adding constraints in the optimization step. Analytical and heuristic relationships were used to identify promising dual-spacecraft trajectory pairs. A new tool developed in Julia, a new high-performance computing language, was used as the main driver layer for controlling execution and flow of data between STAR and MALTO.

### 2.3.4 Entry Probe Analysis Method

Probe atmospheric entry is one of the most critical pieces of a science mission. The selection of entry trajectories as well as thermal protection system (TPS) materials depend on mission elements such as communications, launch window, launch vehicle, payload allocation, hyperbolic access velocity etc. For this study, the entry trajectories and the aerothermal environments for the probe missions to Uranus and Neptune were calculated using TRAJ (Allen et al. 1998, 2005), a NASA Ames-developed 3-DoF (depth of field) atmospheric entry simulation. TRAJ determines the entry trajectories based on the entry vectors generated from the interplanetary trajectory models. TRAJ also calculates the convective and radiative heating at the stagnation point using engineering correlations. These heating correlations utilize equilibrium gas thermodynamics based on the Gordon and McBride program, CEA (Chemical Equilibrium with Applications) (Gordon and McBride 1994; McBride and Gordon 1996). This methodology is consistent with other early-phase studies and helps to quickly explore the parameter space. Higher-fidelity calculations, such as computational fluid dynamics (CFD) simulations to calculate the convective heating environment, lead to lower uncertainties but are computationally more expensive and thus were not used in this study.

Based on the heating and pressure environments of the Uranus and Neptune trajectories, the selected forebody TPS material was HEEET (Heatshield for Extreme Entry Environments Technology). HEEET is a 3-D woven, dual-layered material. The high-density top recession layer (RL) is designed to manage the recession during entry while the bottom lower density insulation layer (IL) accommodates the heat load during entry. The HEEET layers were sized to the aerothermal environments generated by TRAJ using FIAT (Chen and Milos 1999; Milos and Chen 2013), a 1-D ablation and thermal analysis simulation developed at NASA Ames. The thermal response model for HEEET was developed based on material property testing, thermochemical calculations and arcjet testing data of the material. As a comparison, sizing for full density carbon phenolic (CP) is also presented in Appendix A.

### 2.3.5 Instrumentation

The mission concepts identified for study include three proposed instrument complements; probe instruments (33 kg), a low-mass (~50 kg) orbiter payload, and a high-mass (~150 kg) orbiter payload. Outer planet flagship missions have previously carried somewhat more massive payloads (Voyager: 105 kg, 10 instruments; Galileo orbiter: 240 kg, 13 instruments; Cassini orbiter: 270 kg, 12 instruments). However, it was found for this mission that the 150 kg orbiter payload allocation was sufficient to fulfill all primary science objectives of the Ice Giant missions (see Section 3.3), which can be attributed to advances in instrument miniaturization in the intervening years.

There are payload accommodation challenges associated with these architectures. The requisite long cruise times, exceeding 10 years, may require selective or full redundancy for the instruments (instruments are typically single string). This could place a lien on mass and volume. Instrument lifetime issues are mitigated by the fact that orbital operation durations are no more than four years for any option.

The available power is limited for the options studied, requiring power cycling of the instruments. Power cycling is a consumable (especially for solder joints) and can affect the lifetime of the instruments.

The available telemetry is limited because of the large range to Earth, making spacecraft storage, data selection, and data compression an important attribute of the instruments. Spacecraft data storage is required to buffer high rate observations—especially the Doppler Imager seismology experiment, feature tracks, and high rate fields and particles samples. The avionics design used for the options detailed in Section 4 include sufficient memory to buffer several orbits. The need for data selection implies that arbitration for telemetry bandwidth between science disciplines will be complex. It is anticipated that all data compression can be accommodated in the spacecraft avionics, though many of the exemplar instruments include data selection and compression within the instrument.

Remote sensing instruments have to accommodate the low irradiance (and the related low temperature emissions) associated with the large heliocentric distances through the usual trades of increasing aperture, increasing integration time, or decreasing spectral (or spatial) resolution, The relatively slow encounter velocities (5–6 km/sec compared to ~8 km/sec for prior Jupiter and Saturn tours) slightly ameliorate this.

Use of a 3-axis stabilized spacecraft creates challenges for particles & fields instruments—a spinning platform would greatly improve the sampling strategies. The current assumption for the options studied for this report is that the orbiter would be spun at a low rate for these measurements in the same manner as Cassini.

### 2.3.6 Spacecraft System Concepts

Spacecraft system concepts and performance estimates were developed at increasing levels of fidelity throughout the study. For early trajectory development mass estimates were used for notional flight elements based on a range of past studies including orbiters, probes, and SEP stages. Notional master equipment lists (MELs) were developed based on these past concepts to facilitate trades.

These flight element concepts were used to inform the A-Team study and develop initial cost elements with the caveat that although they were likely optimistic (based on the tendency to see mass growth as design maturity increases) they should produce cost estimates that would be fairly accurate in relative terms.

Once prioritized mission concepts were settled on by the SDT, the Study Team prepared introductory material to provide to Team X for their studies. This included descriptions of architecture for each mission option, and analogous designs, based both on flight projects and recent concept development studies. Team X then took these inputs and worked with the Study Team to develop point designs for the various elements for each mission option. The probe design was developed in a single Team X study as a common element that could be used at either Uranus or Neptune. Aerothermal analysis and entry system TPS design were supported by ARC personnel outside the Team X environment, with their output being fed back into Team X for the final report.

Orbiter and SEP stage designs were developed by Team X in a collaborative engineering environment and iterated with the Study Team both during and after the Team X sessions.

### 2.3.7 Technology

The ground rule for the study was to evaluate the use of realistic emerging enabling technologies and to distinguish mission specific vs. broad applicability. This was approached by first conducting an A-Team assessment of technologies that might have an application to a mission to the outer solar system even before the science team was convened to develop specific requirements.

Only two new technologies were deemed enabling for the mission described: a proposed eMMRTG for the spacecraft and HEEET for the entry probe. The eMMRTG would provide a significant improvement in specific power over the existing MMRTG technology at beginning of life (BOL) but, more importantly, a much larger gain at end of life, which is critical, given the duration of an ice giant mission. MMRTG output after 17 years is estimated to be 61 W versus 94 W for the eMMRTG. HEEET is enabling for the entry conditions of probes at both Uranus and Neptune. Other currently available heat shield materials such as phenolic impregnated carbon ablator (PICA) are not adequate for the environments experienced in this mission and carbon phenolic used on the Galileo Jupiter probe is no longer available.

A number of new technologies were not found to be necessary to the mission concepts described but if they were available they could have an impact on the performance and/or cost of the mission: 1) aerocapture technology could enable trip times to be shortened, delivered mass to be increased or both; 2) cryogenic propulsion could have similar but not as pronounced effects;

3) advanced RPS technologies, with even better specific power than eMMRTG, such as the segmented modular RTG could enable more mass or power for instruments or both; 4) optical communications could revolutionize the data return from an outer planet mission; and 5) advanced mission operations technologies could drive down cost and permit more adaptive missions operations than are envisaged in the missions reported here. Further details on these advanced technologies and their mission applicability are included in Appendix D.

### 2.3.8  Cost

A-Team cost estimates were generated for each mission option considered using a parametric cost model that was run by a costing subject matter expert. The model, developed specifically for early concept development, uses a small number of inputs that can be defined at early CMLs. It is meant to establish the feasibility of a mission within a cost range. The model uses wraps, cost estimating relationships (CERs), and rules of thumb based on project actuals as well as hundreds of Team X studies to estimate costs. Total project costs generated by this model have been found to be consistent with actuals within a range of ±30%.

Team X cost estimates were generated for each mission option using JPL Institutional Cost Models (ICMs). The ICM algorithms have been provided by the JPL doing organizations and represent the "most likely" estimates (i.e., 50% confidence level without reserves). The ICMs have been shown to estimate total mission lifecycle cost within a range of -10% to +20% of historical actuals.

The ICMs were run in Team X by subject matter experts and output costs to Level 3 of the Work Breakdown Structure (WBS) for the flight system and instruments. Each ICM estimates the levels of workforce, procurements, services, and subcontracts that would be needed to complete the mission assuming a JPL in-house development. A MEL was generated during the Team X study and was used for estimating the cost of all flight system components and instruments. The number of flight units, flight spares, prototypes, engineering models, and testbed units were specified by each subject matter expert and were included in the costs estimated for each subsystem. JPL dual-string reference bus electronics heritage was assumed. JPL-approved labor and burden rates were applied to the resource estimates generated by the models. Instrument costs were estimated using NASA Instrument Cost Model (NICM) Version VII in system mode. For multi-element missions, each of the subsystems have bookkept all of their subsystem management and system engineering costs with the primary element (i.e., the orbiter or flyby spacecraft). NASA's RPS Program Office provided costs for eMMRTGs and RHUs and the additional Launch Services Provider costs for support of nuclear missions.

Both A-Team and Team X costs represent total lifecycle costs (phases A through F including reserves) without launch vehicle. All costs were generated in FY15$ and were costed as Mission Risk Class B per the study guidelines. A "typical" flagship mission development schedule was assumed for each option (20-month phase A, 16-month phase B, 47-month phase C/D). Costs are reported using the JPL Standard WBS, which is consistent with the NASA standard except that NASA includes Mission Design under WBS 2.0 rather than JPL's WBS 12.0. It is assumed that WBS 11.0, Education and Public Outreach, is not required. No costs were included in project estimates for any needed new technology development. Phases A–D reserves were set at 30% of the development costs without LV, DSN tracking, and costs provided by the RPS Program Office. Phase E–F reserves were set at 15% of the Mission Operations and Data Analysis (MO&DA) costs without tracking. This is consistent with JPL guidelines for missions at this level of design maturity.

# 3   SCIENCE

Since the brief Voyager 2 flybys of Uranus (1986) and Neptune (1989), the ice giant systems have intrigued and tantalized.  They represent a distinct class of planet, fundamentally different from the better explored gas giants, Jupiter and Saturn (**Table 3-1 and Figure 3-1**). The gas giants are composed mostly of hydrogen and helium (more than 80% by mass).  Their gas envelopes are thought to extend all the way to their relatively small rock/ice cores, with molecular $H_2$ beginning a transition to ionized, metallic hydrogen at mega-bar pressures (Guillot 2005; Lissauer and Stevenson 2007).  While Uranus and Neptune also possess hydrogen and helium envelopes, the envelopes are much smaller, accounting for less than 20% of the planets' masses and never making the transition to metallic hydrogen (Guillot 2005; c.f. Nellis 2015).  The bulk composition of these planets is dominated by much heavier elements.  Based on cosmic abundances, oxygen, carbon, nitrogen, and sulfur are the likely candidates.  Since these species are thought to have been incorporated into proto-planets primarily as ices—either as solids themselves or as gas trapped in water-ice (e.g., Hersant et al. 2004)—the term "ice giants" has been adopted.  Today, however, there is probably very little ice in Uranus and Neptune, a supercritical fluid being the preferred phase of $H_2O$ at depth.

**Table 3-1.**  Approximate composition by mass of the known planet types.

| Planet Type | Gas | Ice | Rock | Total Mass ($M_{Earth}$) |
|---|---|---|---|---|
| Terrestrial (Earth) | ~0% | ~0% | ~100% | 1 |
| Gas Giant (Jupiter/Saturn) | ~85% | ~10% | ~5% | 300/100 |
| Ice Giant (Uranus/Neptune) | ~10% | ~65% | ~25% | 15 |

Not surprising for planets with such different composition, the ice giants and their systems are a natural laboratory to challenge our understanding of fundamental processes.  The compositions of Uranus and Neptune are the expression of a formation environment different from that of the gas giants and result in vastly different interior structures.  That interior structure in turn generates a magnetic field and magnetosphere whose external appearance is unlike those found in the gas giant or even terrestrial planets. Both ice giants have complex magnetic fields whose dipole terms are centered 1/3 of a planetary radius away from the center of mass, and tilted by up to 60° from the rotation axis. The lack of alignment between the rotation and magnetic poles creates unique and variable orientations to the solar wind, particularly on Uranus whose rotation axis is almost in the plane of its orbit. The tilt of Uranus is one of several intriguing differences between Uranus and Neptune. Another major distinction is that, unlike all the other the giant planets, Uranus emits almost no internal heat. Why that is and how it affects both the interior and atmospheric circulation are areas of active research.

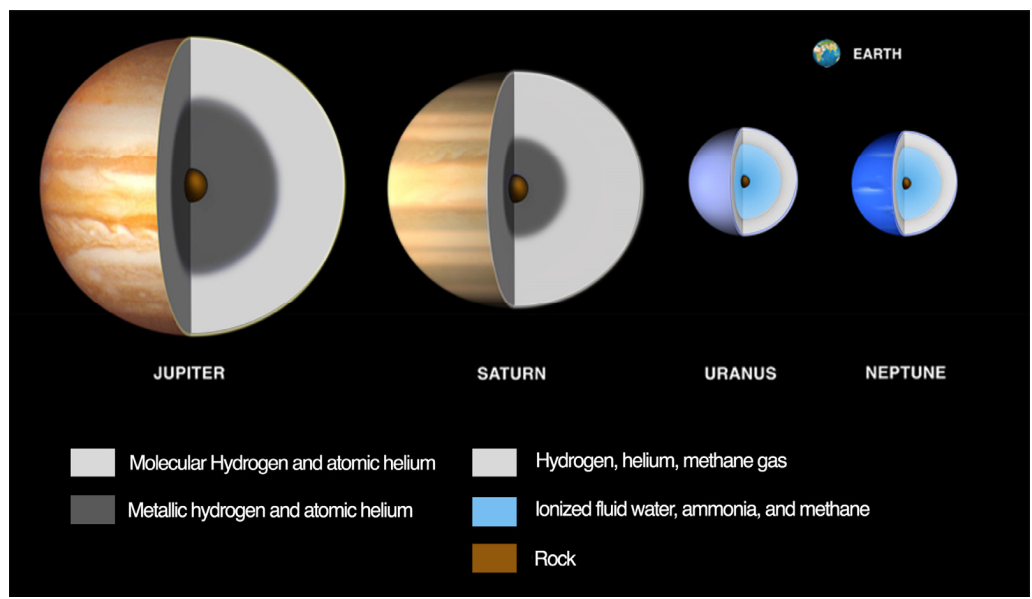

**Figure 3-1.**  Illustration of compositional differences among the giant planets and their relative sizes.  Earth is shown for comparison.  Jupiter and Saturn are primarily made of hydrogen and helium, the terrestrial planets are almost pure rock, while Uranus and Neptune are thought to be largely supercritical liquid water.

The rings and satellites of the ice giants also differ markedly from those of the gas giants and from each other. Uranus hosts several mid-sized moons whose surface ices are different from those of Jupiter and Saturn's satellites, as would be expected given the colder temperatures in the zone of the ice giants. These mid-sized moons show features indicative of endogenic activity, such as Miranda's patchwork geologic morphology, the flow-like features on the floors of Ariel's graben, and Umbriel's bright polar feature. Even one of the smaller Uranian satellites, Mab, is associated with a mysterious ring, though the relationship is not understood (Showalter & Lissauer 2006; de Pater et al. 2006; Kumar et al. 2015). Uranus' classical rings are narrow and dense, quite different from the broad expanse of Saturn's or the tenuous ones at Jupiter. Also of interest, the Uranian rings are gravitationally entwined with a densely packed system of smaller moons that interact chaotically on short time scales (Duncan & Lissauer 1997; French & Showalter 2012; French et al. 2015). Neptune's satellite system is dominated by the captured Kuiper Belt object (KBO), Triton. That capture is believed to have ejected or destroyed any larger, native moons of Neptune (e.g., Masters et al. 2014 and references therein), leaving only a family of small native moons today. But Triton itself is of great interest, having an atmosphere, active geysers, and unusual geology reminiscent of its sister KBO, Pluto. And finally, Neptune's rings display their own unique features, dominated by large clumps that evolve on decadal time-scales (Porco et al. 1995; de Pater et al. 2005).

Studying the ice giants is of great scientific interest for anyone seeking to understand our solar system's formation, evolution, and current state. The smaller amounts of hydrogen and helium in these worlds is often attributed to the slower accretion rates at larger distances from the Sun. However, recent models of solar system formation suggest that Uranus and Neptune may have undergone substantial radial migration during the early parts of the solar system's history, complicating efforts to understand the conditions under which the ice giants formed (e.g., Turrini et al. 2014 and references therein). Furthermore, Uranus' extreme obliquity and Neptune's capture of Triton suggest that both systems experienced dramatic and/or violent events in their early history, which perhaps reflect drastic changes in the structure of the early outer solar system. Information about these early events could potentially be encoded in the structure and composition of the planets and their moons/rings (e.g., Turrini et al. 2014 and references therein).

The importance of understanding Uranus and Neptune extends beyond our solar system. In 2004, the first ice giant candidate was reported around another star (Butler et al. 2004). Today, the vast majority of the more than 3,000 planets we know of in our galaxy (see http://exoplanetarchive.ipac.caltech.edu) appear to be ice giants as opposed to gas giant or terrestrial planets. Correcting for selection effects, which favor larger planets, there appear to be about 9× more ice-giant-sized planets than gas giants in our galaxy (Borucki et al. 2011, Fressin et al. 2013). There is some uncertainty in the detailed statistics because we do not know enough about ice giants to reliably say, based only on radius or mass and radius, whether a planet's composition is more like Jupiter or Uranus. But for unknown reasons, planets the size of ice giants are clearly more common than gas giants in planetary systems other than our own (**Figure 3-2**).

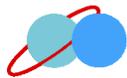

Given the ice giants' importance to our understanding the formation of our solar system and the processes that have shaped it, as well as how common they are in exoplanetary systems, the *Vision and Voyages Planetary Decadal Survey* (NRC 2011; hereafter referred to as V&V) made these poorly studied worlds a priority. An ice giant mission was identified as the next priority for a Flagship-class mission after Mars sample return and a dedicated Europa mission. The science team of the current study reaffirms the importance of a Flagship mission to an ice giant. The science goals and objectives we have defined are consistent with those in V&V, with updates in some areas due to advances in technology or in our understanding. (Advances since V&V are highlighted in Sections 3.1 and 5.0.)

## 3.1    Science Objectives

Studies of the ice giant systems encompass all disciplines of planetary science, with much cross-disciplinary overlap, particularly when looking at system-wide interactions. Although each discipline (interiors, atmospheres, magnetospheres, classical satellites, small satellites and rings) was first considered individually, broad themes quickly emerged. The SDT compiled 12 main science objectives (**Table 3-2**), which answered more than 50 science questions (Appendix F). That this list is by no means all-encompassing underscores the great breadth of science that could be achieved at either of these planets. Each of the 12 objectives is discussed in subsections below, along with measurements that could meet that objective. Payload elements are discussed in Section 3.3.

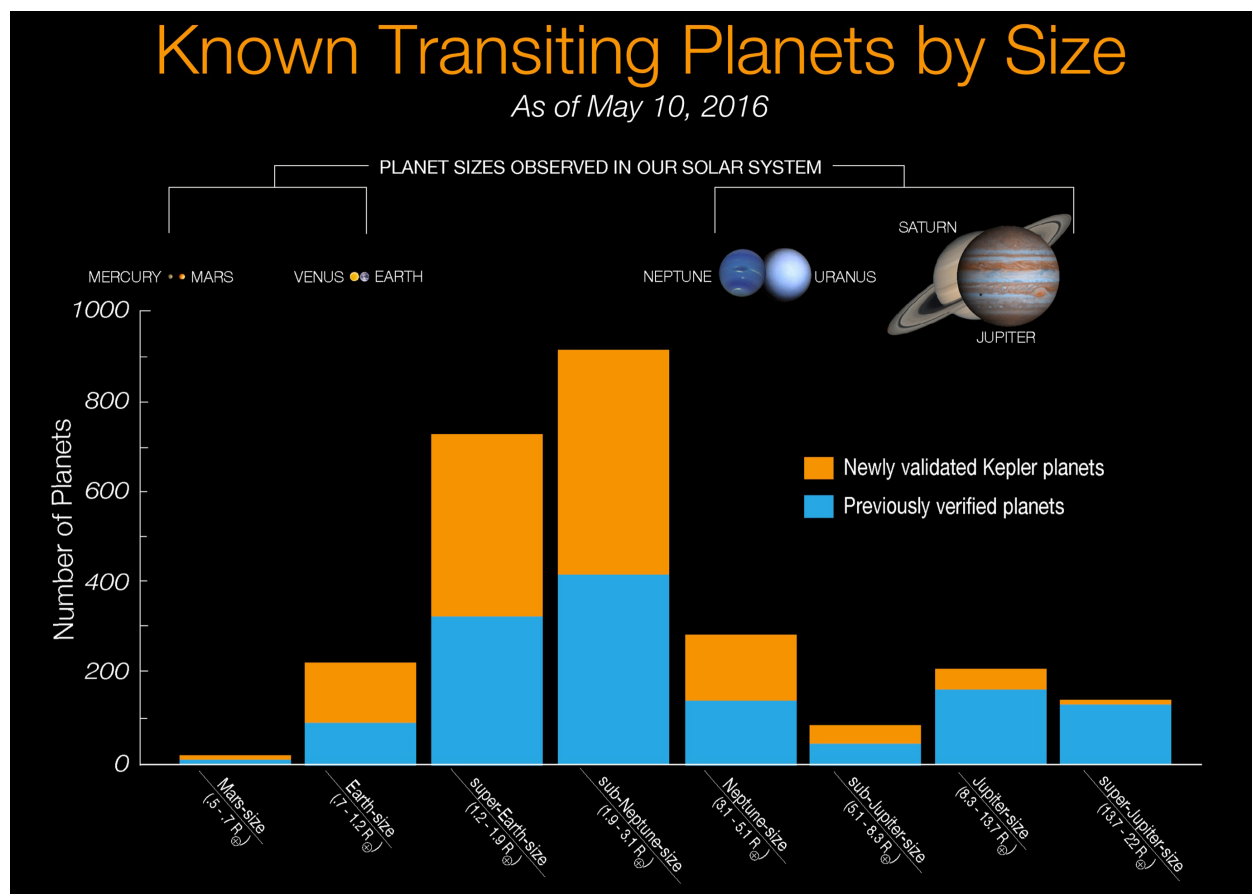

**Figure 3-2.** Number of confirmed exoplanets of known radius, sorted by size. When selection effects are accounted for, which favors detection of the largest planets, ice giants (labeled sub-Neptune- and Neptune-size in the diagram) are estimated to be about 9× more common than gas giants in our galaxy (Borucki et al. 2011). Credit: NASA Ames/W. Stenzel

The most important science investigations are ones that address the fundamental questions "What is an ice giant?" and "How do they form?" We therefore consider the objectives of determining the interior structure of an ice giant (Section 3.1.1) and the bulk composition (including noble gases and key isotopic ratios, Section 3.1.2) as the highest-priority science to be done. The SDT does not prioritize among our other 10 objectives, and they are listed in no particular order. We emphasize the wealth of important work to be done in each area and know that a "Eureka!" discovery may lie anywhere in the ice giant system.

### 3.1.1 Determine the Interior Structure

Uranus and Neptune are known as "ice giants," which suggests an interior structure that is dominated by water. However, while this is presumed to be true, there is very little direct evidence to support this assertion. It is commonly thought that both planets have three layers: an outer H/He envelope, below that a thick fluid ionic water layer (or water mixed with other ices like ammonia and methane), and at the center a small rocky core. But available gravity field data from the Voyager 2 flybys actually show that both planets are not as centrally condensed as this simple picture, and there is no strong motivation, beyond computational convenience, for a 3-layer structure (Helled et al. 2010).

Both planets have bulk densities a bit less than that of compressed water, necessitating a low-density component (H/He), which we can readily see in the visible atmosphere. Below this H/He dominated layer, though, little is known. Three-layer structural models for both planets can only fit the gravitational field constraints with a water-dominated interior that includes a small admixture of ammonia, methane, or H/He, to lower the envelope density, and a very small rocky core (Nettelman et al. 2013) (**Figure 3-1**). However, the high interior ice:rock ratios implied (~10:1) are far larger than expectations based on solar system abundances or the ice:rock ratios of any of the large icy moons. Furthermore, mixtures of high pressure H/He mixed with rock can mimic the density of water, leading to significant compositional degeneracies. Acceptable models could have an arbitrary number of layers beyond three.

Models that fit the gravity data can fail to reproduce the measured intrinsic fluxes from the planets, particularly for Uranus. Fully adiabatic 3-layer models for Neptune cool to Neptune's observed flux in 4.5 Gyr, but this in itself is not a strong confirmation of an adiabatic interior, as other unmodeled energetically important processes (such as sedimentation of He or carbon) could also be at play. However, the low flux from the interior of Uranus necessitates a structure quite different from a simple 3-layer model. Recent work (Nettelmann et al. 2016) achieves good fits using an ad hoc statically stable layer at the H/He—water layer interface. This layer stops convection and also leads to a much hotter interior, leading to a much lower density water envelope and a lower (more "reasonable") water:rock ratio. But this is just one possible interior profile.

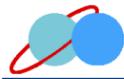

**Table 3-2.** Science Traceability Matrix.

| SDT Science Objective (First two are highest priority, others all of equal priority) | Scientific Measurement Requirements | | Instrument | Instrument Requirements | Mission Requirements | Comment |
|---|---|---|---|---|---|---|
| | Physical Parameters | Observables | | | | |
| 1. Constrain the structure and characteristics of the planet's interior, including layering, locations of convective and stable regions, internal dynamics | Planetary oscillations | Top-of-atmosphere radial velocities, temperatures, or geoid | Doppler Imager or similar seismology instrument | Detect velocities of a few cm/s with a spatial resolution of ~100 pixels across the disk of the planet. | Allow imaging on approach for tens-of-days with a 1- to 2-minute cadence (to achieve micro-Hz frequency resolution). | Oscillations may also be detectable through studies of the rings. |
| | Magnetic field geometry and time variability | Magnetic field strength and orientation | Magnetometer | | Magnetic cleanliness of spacecraft | |
| | Gravity moments, J2-J6 (not a driver at Uranus) | Perturbations to s/c orbit (also useful to make astrometric observations of rings and satellites) | USO | | Close periapse passess | Gravity is still of interest at Uranus, but of lesser importance for interior structure than at Neptune. If Doppler tracking is a problem (dangerous to fly close to planet), can investigate astrometry further. |
| 2. Determine the planet's bulk composition, including abundances and isotopes of heavy elements, He and heavier noble gases | Atmospheric composition | CH₄, noble gases (He, Ne, Ar, Kr, Xe), and isotopic abundances (C, Ne, Ar, Kr, Xe) at two tropospheric pressure levels such as 1 bar and 10 bars, N, S, and O isotopes to 20 bars. NH₃, H₂S, H₂O below their respective cloud bases. | Probe with mass spec (and TLS if pressures >10 bar) | Probe to 10 bars for noble gases, CH₄ and most isotopic ratios. Probe to 20 bar for N, S, and O isotopes. Probe to > several kilobars for NH₃, H₂S, and H₂O. All measurements ±10% | Probe relay. Instrument survival and performance in the extreme p,T environment, especially if deployed to >10 bar. | Probe to 10 bars will not give well-mixed H₂S, H₂O, S and O isotopes, and NH₃ only marginally. Probe to 100 bars could give NH₃ and H₂S, but not H₂O. Probe to 10's-100's kilobars is required to confidently get H₂O, especially if an ionic ocean is present, as predicted by models and lab work. |
| 3. Improve knowledge of the planetary dynamo | In situ magnetic field direction and magnitude | Magnetic field direction and magnitude | 3-axis Magnetometers on boom | 0.1 to 20,000 nT, 1 second cadence | Multiple close orbits; longitude and latitude coverage for degree and order at least 4, preferably 15 | |
| | Remote sensing of magnetic field footprint. | UV and IR emission from auroral and satellite footprints | IR, UV spectral imager(s) | 1600-1800 Ang imaging and 3.4–4 micron imaging. | | Should not drive the UV or IR instrument, but should be considered regarding instrument capability and operations |
| | | Auroral radio emission | Radio Receiver, at least 2 axis electric antenna | 10 kHz to 1 MHz, direction finding ability | Multiple close orbits, good longitude and latitude coverage | |
| 4. Determine the planet's atmospheric heat balance | He/H₂ abundance | He/H₂ abundance | Probe mass spectrometer | He/H₂ ± 5% | Measure at P ≥ 1 bar | Solar uncertain by 2%; protosolar somewhat greater, hence ±5%. |
| | Net thermal emission | Broadband thermal IR emission | Thermal IR bolometer | 5–900 cm⁻¹, accuracy 1% | Full phase angle coverage | Range based on ~1% of peak for Neptune, accuracy based on 0.1% of peak, See Li et al, 2010, Fig. 1 as reference |
| | Bond albedo | Visible wavelength bond albedo | Photometer, or suitably calibrated imager or spectrometer | 0.3–1.6 µm, accuracy 1% | Full phase angle coverage | Range based on ~10% of peak flux of Sun, and similar to Voyager IRIS shortwave radiometer bandpass. Accuracy assumed to reduce error bars to 1% of total. May be too stringent. |
| 5. Measure planet's tropospheric 3-D flow (zonal, meridional, vertical) including winds, waves, storms and their lifecycles, and deep convective activity | Vertical, zonal, and meridional profiles of wind speed | Repeat maps of cloud tracers at multiple wavelengths including methane bands | Visible/Near-IR imager, 8 channels | 30km/pixel spatial resolution, 3 emission angles (nadir,near limb, intermediate), 425 ± 25nm, 500 ± 25nm, 619 ± 5nm, 653 ± 25nm, 727 ± 5nm, 750 ± 5nm, 890 ± 10nm, 925 ± 5nm, repeat one rotation later, Absolute I/F's to 5%, SNR:50 for all but SNR:100 for 750nm & 653nm for winds | Global mapping and feature tracking. Each region viewed at 3 emission angles on 2 consecutive rotations. | |
| | | In situ wind speeds to 15 bars (100 bar goal) | Probe USO for Doppler tracking | Velocity to 20 m/s | Probe relay to 15 bars (100 bar goal) | |
| | Distribution of condensible and disequilibrium species. | Hydrogen ortho/para ratio at P ≥ 3 bar | Mid-IR spectrometer | ~1,000 km horizontal spatial resolution, mixing ratio ±20% | Global mapping. | Cloud particle size and density distribution was proposed to be added to this objective, but that is a separate, lower-ranked science objective. |
| | | 3-D distribution of hydrocarbons, GeH₄, AsH₃, PH₃, in the stratosphere | Near to mid-IR spectrometer | ~1,000 km horizontal spatial resolution, mixing ratio ±20% | Global mapping. | Probe can give vertical distribution ONLY in one location |
| | | 3-D distribution of CO, PH₃ in stratosphere | Millimeter/submm spectrometer | ~1,000 km horizontal spatial resolution, mixing ratio ±20% | Global mapping | |
| | | 3-D distribution of CH4 in upper troposphere | Visible/Near-IR spectrometer | ~1,000 km horizontal spatial resolution, mixing ratio ±20% | Global mapping | |
| | | 3-D distribution of CO, PH₃, NH₃, H₂S, in upper troposphere | Millimeter spectrometer, centimeter radiometer | ~1,000 km horizontal spatial resolution, mixing ratio ±20% | Global mapping | |
| | | 3-D distribution of NH₃, H₂S, H₂O in deep troposphere | Centimeter to meter-wavelength radiometer | ~1,000 km horizontal spatial resolution, mixing ratio ±20% | Global mapping | |

| SDT Science Objective (First two are highest priority, others all of equal priority) | Scientific Measurement Requirements | | Instrument | Instrument Requirements | Mission Requirements | Comment |
|---|---|---|---|---|---|---|
| | Physical Parameters | Observables | | | | |
| 6. Characterize the structures and temporal changes in the rings | Fine-scale radial structure and particle size constraints in selected rings | Ring optical depth in the visible | Imager or photometer (observe stellar occultations) | Imaging with spatial resolution <100 m | Observe rings at multiple longitudes. Repeat > 8 times for spatial variations and ring shapes | |
| | | Ring optical depth at radio wavelengths | Transmitter (preferably at two wavelengths) | High-gain antenna diffraction pattern on rings < 100 m | Spacecraft pass behind the rings as seen from Earth or other flight element at multiple longitudes. Repeat > 3 times for spatial variations and ring shapes | Uplink radio occultations may be useful. |
| | Spreading rates and other temporal variations | Visible reflectivity | Imager | Spatial resolution < 100 m | Measure selected rings multiple times over months to years | Constrain evolution of longitudinal structures over at least 10 keplerian spreading times. Constrain radial evolution over 10 viscous spreading times |
| | Dusty ring structure and particle properties | Visible and near-IR reflectivity | Imager | High rejection of off-axis scattered light | Shadow passages for very high phase angle observations. | Search for dust throughout system, from above planet to beyond major satellites. Characterize rings at phase angles >160 degrees |
| 7. Obtain a complete inventory of small moons, including embedded source bodies in dusty rings and moons that could sculpt and shepherd dense rings. | Location and orbital parameters of small moons, constrain their size | Sunlight reflected from small moons | Imager (framing camera) | Identify all bodies >0.5 km in radius within 500,000 km of the planet's center. Repeated observations for orbit determinations. | Time and data volume to search orbits within 500,000 km of planet center | |
| 8. Determine surface composition of rings and moons, including organics; search for variations among moons, past and current modification, and evidence of long-term mass exchange / volatile transport | Surface composition and microphysical properties of rings and moons | Surface composition and phase properties | Visible/Near-IR spectral imager | Wavelength: visible to ~5 μm; pixel scale ≤500 m globally, ≤100 m locally/regionally observations at different phase angles [rings and satellites] | Multiple flybys for global mapping of multiple satellites | |
| | Geologic context on moons for compositional variations | Surface morphology | Visible wavelength imager | Global mapping ≤1 km pixel scale local high-resolution imaging ≤50 m pixel scale, panchromatic as a minimum, stereo views | Multiple flybys for global mapping of multiple satellites, stereo views | |
| | Search for plumes and identify spatial and temporal variability in activity and dynamics | Scattered sunlight from plumes | Narrow-angle visible imager | ≤ 10km/pixel for all satellites, ≤1 km for Triton | High phase angle and terminator imaging, repeated to constrain variability and dynamics | Narrow-angle visible camera can identify plumes with distant observations; NUV sensitivity useful, but panchromatic can be sufficient (and has more total photons for better SNR) |
| | Constrain the relative abundances of major constituents (e.g., water ice, methane ice, amorphous carbon) on the rings and satellite surfaces | Vis/NIR spectral features | Visible/Near-IR spectral imager | Wavelength range and spectral resolution similar to Cassini VIMS (256 channels between 0.3 and 5 microns). Spatial resolution sufficient to assess compositional trends at level of 1/10th body, and geological context at 1/100th body on all major moon and select small satellites | Flybys of all major moons and select small satellites and distances sufficient to meet spatial resolution requirements (see instrument requirements) | |
| | Thermal properties of rings and satellite surfaces | Map thermal IR emission as a function of local time-of-day | Thermal mapper (in conjunction with albedo maps from vis/near-IR imaging) | Resolution of a few km | Multiple satellite flybys, inclined orbits for rings | |
| | Pickup ions in the magnetosphere that originated as neutrals sputtered from satellite surfaces | Measure the composition of pickup ions in the magnetosphere | Plasma and energetic particle detectors | ~1 eV to 1 MeV | Multiple, close (~50 km) satellite flybys. Requires supporting measurements from a 3-axis magnetometer on a boom, measuring 0.1 to 20,000 nT, 0.05 second cadence | This is a secondary way to achieve this objective |

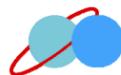

| SDT Science Objective (First two are highest priority, others all of equal priority) | Scientific Measurement Requirements | | Instrument | Instrument Requirements | Mission Requirements | Comment |
|---|---|---|---|---|---|---|
| | Physical Parameters | Observables | | | | |
| 9. Map the shape and surface geology of major and minor satellites | Global-scale surface and structural mapping of major satellites and regional mapping of topography and slopes | Visible reflectivity | Imager | ~80% surface coverage <0.5-km pixel scale. Monochromatic sufficient. | Multiple satellite flybys, stereo views. | Map the shape and surface geology of major and minor satellites to: understand past and current surface modification processes; constrain crater distributions and relative surface ages; determine topography; provide evidence of subsurface structure; correlate with gravity; search for sources of endogenic heating. NAC can accomplish with observations on approach/departure. Note that a radar sounder for sub-surface structures was considered for this science objective, but was considered secondary due to accommodation issues. |
| | Local high-resolution surface mapping, topography, slopes | Visible reflectivity | Narrow-angle visible camera | <100-m pixel scale. Monochromatic sufficient. | Multiple satellite flybys, stereo views | |
| | Spectral mapping of geologic units | Visible, IR, and UV reflectivity | Spectral imager at visible, IR, UV wavelengths | Visible to ~5 μm, pixel scale 0.5 km; UV pixel scale ≤500 m globally, ≤100 m locally. | Multiple satellite flybys, observations at different phase angles | |
| | Distribution of mass | Gravity field | USO | Measure J2 | Requires multiple satellite flybys, s/c tracking *during* flybys, ~50 km altitude. | Would like to measure k2. |
| | Thermal properties of surface | Thermal emission, thermal inertia | Thermal mapper | Spatial resolution ~3 km. | Multiple satellite flybys. Also need albedo maps from vis/near-IR imaging | Considered secondary because of concerns about ability to accommodate a thermal mapper. |
| 10. Determine the density, mass distribution, internal structure of major satellites and, where possible, small inner satellites and irregular satellites | Mass, and mass distribution | Gravitational moments | USO for Doppler tracking | | Multiple, close (~50 km) satellite flybys, tracking during flybys | This objective includes detection of internal oceans and bulk composition. Of the irregular satellites, Neptune's moon Nereid is of particular interest. |
| | Mass distribution and rheology | Libration, Cassini state, pole position, orbital motion, tides | Narrow-angle visible camera | 100 m - 1 km pixel scale imaging | Global imaging at 100 to 1000 m pixel scale, repeated at different orbital true anomalies | |
| | Orbit determination | Satellite positions | Narrow-angle visible camera | n/a | frequent distant observations | |
| | Density | Global shape | Narrow-angle visible camera Radio science | 100 m - 1 km pixel scale imaging | Global imaging at 100 to 1000 m pixel scale occultations | |
| | Presence of a conducting layer or dynamo | induced/intrinsic magnetic fields, direction and magnitude | 3-axis magnetometers on boom | 0.1 to 20,000 nT, 1-s second cadence | Multiple, close (<50 km) satellite flybys | |
| 11. Determine the composition, density, structure, source, spatial and temporal variability, and dynamics of Triton's atmosphere | Vertical profiles of aerosols, gas and aerosol scattering properties; velocity of atmospheric features | Vis/Near-IR and UV spectral reflectivity and phase function | Spectral imager at visible, IR, and UV wavelengths | Global coverage at km scales, ~100 m resolution in select regions. | Global context and repeat imaging over months to years (minimum would be two observations, Ideally want to be able to look regularly (every ~1–2 months) over a period of at least 1–2 years) to constrain variability and dynamics; high-phase angle imaging opportunities | |
| | Atmospheric composition | Neutral and charged particle characterization | Mass spectrometer | mass range 1 - 64 amu, mass resolution >1,000 M/dM | Multiple Close Flybys, good latitude and LT coverage | |
| | | | Plasma and particle detectors | ~1 eV to 1MeV | | |
| | | | Langmuir probe | sweep ~20 seconds | | |
| | | | Radio and plasma wave instrument | 10's Hz to 100's kHz, 30 second cadence (required cadence may depend on flyby velocity) | | |

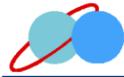

| SDT Science Objective (First two are highest priority, others all of equal priority) | Scientific Measurement Requirements | | Instrument | Instrument Requirements | Mission Requirements | Comment |
|---|---|---|---|---|---|---|
| | Physical Parameters | Observables | | | | |
| 12. Investigate solar wind-magnetosphere-ionosphere interactions and constrain plasma transport in the magnetosphere | Measure ion and electron energy spectra vs time | Measure ion and electron energy spectra vs time | Plasma and energetic particle detectors | ~1 eV to 1MeV | Multiple Orbits, dayside apogees >25 planetary radii, nightside apogees >50 radii, good LT and Latitude coverage. 54 days upstream (of planetary bowshock) continuous measurements. | |
| | Measure magnetic field direction and magnitude | Measure magnetic field direction and magnitude | 3-axis magnetometer on boom | 0.1 to 20,000 nT, 0.05 second cadence | Multiple Orbits, dayside apogees >25 planetary radii, nightside apogees >50 radii, good LT and Latitude coverage. 54 days upstream (of planetary bowshock) continuous measurements | |
| | Measure electric and magnetic wave fields to determine wave-particle interaction, electron density, boundaries, magnetospheric activity | Measure electric and magnetic wave fields | Plasma wave receiver, 2-axis electric, 3-axis search coil. | few Hz to 100's kHz, 30 second cadence, burst mode ablility | Multiple Orbits, dayside apogees >25 planetary radii, nightside apogees >50 radii, good LT and Latitude coverage. 54 days upstream (of planetary bowshock) continuous measurements. | |
| | Measure ion composition to separate magnetospheric and solar wind plasma | Measure ion composition | Plasma and energetic particle detectors | ~1ev to ~300 keV | Multiple Orbits, dayside apogees >25 planetary radii, nightside apogees >50 radii, good LT and Latitude coverage | |
| | Measure ENAs | Measure ENAs | ENA imaging instrument (can also measure energetic ion and electrons with the same instrument) | 10 to 150 keV/nucleon | Orbits with large apogee distances, good LT and Latitude coverage | |

The fluid ionic "watery" sea is likely the region that generates each planet's internal dynamo. However, we have not yet reached a sufficient level of understanding of either the interior structure or the dynamo mechanism for creating the planetary fields to allow one to place firm constraints on the other.

A mission to the ice giants should address whether Uranus and Neptune are fully convective planets and related questions. What parts of these planets are convective and what are the resulting flow characteristics in the deep interior? Where within the planets are their dynamos generated? What is the bulk composition of the planets, in particular the ratios of H/He:water:rock? Are there distinct compositional boundaries in the interior? Is the change from a predominantly H/He atmosphere to heavier elements in the deep interior a gradual one? To what degree is the outer H/He envelope enhanced in heavy elements compared to solar abundances?

Probing the gravity fields of Uranus or Neptune to determine J2, J4, and J6 to a higher precision than currently available can give important information on aspects of the structure of each planet, particularly for Neptune. The relatively smaller error bars on J2 and J4 for Uranus already well-constrain the heavy element enrichment of the inner and outer envelope of Uranus, within the framework of a 3-layer model. However, large uncertainties for Neptune still remain (Nettelmann et al. 2013). Another interesting gravity field result is that the available J4 data suggest that the dynamics of the visible atmosphere are confined to a thin weather layer no more than about 1,000 km deep on both planets (Kaspi et al. 2013), though deeper extents cannot be ruled out (Zhang et al. 2015; Kaspi et al. 2016). Refined measurements of J4 and J6 could further constrain this limit.

While the gravity field provides overall constraints on interior structure, it cannot be used to understand layering or deviations from an adiabatic structure. Observing planetary oscillations, akin to helioseismology, holds promise for fundamentally new information that would lead to real breakthroughs in understanding boundaries, layering, and regions stable to convection within the interior. A Doppler imager (described in Section 3.3.1) is one way of measuring such oscillations. An entry probe into the atmosphere, carrying a mass spectrometer and temperature, pressure, density sensors, will also help us to understand if He sedimentation has occurred and shed light on the overall super-solar enrichment of carbon (in methane). Measurements of the strength, morphology, and time variability of the magnetic fields will help to constrain regions of the interior that must be able to sustain dynamo action as well as the convective flows responsible for its generation (see also Section 3.1.11).

### 3.1.2 Determine the Bulk Composition, including Noble Gases and Key Isotopic Ratios

The composition and associated isotopic abundances of the bulk atmosphere are key to the question of how the giant planets formed and evolved. Furthermore, composition can provide insight into possible migration of the giant planets. Of the two main models of giant-planet formation—core accretion and gravitational instability—the former is generally favored. The best evidence of the core accretion model comes from the observed enrichment of the heavy elements in Jupiter (mass greater than $^4$He), presence of first solids (millimeter-size chondrules and calcium aluminum inclusions) at the very beginning of the solar system, and greater frequency of exoplanets around metal-rich stars (Atreya et al. 2017, and references therein).

In the core accretion model, non-gravitational collisions between small grains of dust, metals, refractory material, ices, and trapped volatiles lead to larger particles that grow to tens to hundreds of kilometer sized planetesimals and eventually form an embryo, the core. When the core is large enough, i.e., 10–15 Earth masses, it is able to gravitationally capture the most volatile of the gases, hydrogen, helium, and neon, from the surrounding protoplanetary nebula, thereby completing the formation of the giant planet. Volatiles trapped in the core are released during accretionary heating and form the atmosphere of the giant planet together with $H_2$, He and Ne. There are a number of variations to this scenario, in particular growth by pebbles, but the basic theme of core accretion for giant planets (formation of a large solid core first, followed by gravitational capture of nebular gas) is preserved (e.g., Levison et al. 2015 and references therein).

This scenario of core accretion is generally a slow process, taking up to several million years to form Jupiter and Saturn and up to tens of millions of years to form Uranus and Neptune. However, the solar nebula from which all planets formed dissipated in less than 5 million years, which is not long enough to form the ice giants, unless they underwent the majority of their formation close to where the gas giants formed and then migrated out to their present orbits. Recent work (e.g., Bitsch et al. 2015) based on the pebble accretion scenario, however, indicates that it is possible to form the cores of all four giant planets within 3 million years (so within the expected lifetime of the solar nebula) farther away from the Sun than their present orbits and then migrate the planets inward due to disc-planet interactions. Heavy element abundances will be important to understand possible migration of these planets.

Due to their large mass and relatively low exospheric temperatures, the giant planets are not expected to have suffered significant loss of volatiles from their atmospheres by escape since the time of their formation. Thus, their *bulk* atmospheric compositions should reflect the *proto*solar composition. In the Sun, the most abundant elements after H and He are O, C, Ne, N, Si, Mg, S, Fe, P, etc. In the giant planets, Si, Mg, Fe and other solids would be sequestered in the core, whereas O, C, Ne, N, S, P, and the noble gases are expected to be present in their atmospheres. Since these heavy elements made up the bulk of the original core, their abundances and isotopic ratios in the *well-mixed* atmosphere reflect the history of the formation and evolution of the giant planets (**Figure 3-3**). In the atmospheres of the giant planets, the key heavy elements, O, C, N, S and P are present predominantly in $H_2O$, $CH_4$, $NH_3$, $H_2S$ and $PH_3$, respectively, and the noble gases, Ne, Ar, Kr and Xe are present in their elemental form. $H_2O$, $NH_3$, $H_2S$ are condensable in the tropospheres of all giant planets, whereas $CH_4$, and possibly $PH_3$, can also condense in the atmospheres of Uranus and Neptune due to their much colder temperatures compared to Jupiter and Saturn. The well-mixed abundance of the condensable gases would thus be found below their respective condensation levels, but the noble gases would be well-mixed everywhere, with the exception of Ne, which is miscible in liquid helium.

At levels in Jupiter and Saturn where helium raindrops are expected to form at pressures of several megabars, Ne would be removed, and it was indeed found to be depleted relative to solar abundances in the upper troposphere of Jupiter by the Galileo entry probe. Measurements of Ne have not yet been made at Saturn.) The possibility of helium rain in smaller Uranus and Neptune is questionable, but not entirely ruled out, so Ne may or may not be depleted in the upper troposphere. In any event, determination of its abundance is important to understand both the planetary formation and the interior processes. Along with Ne, a precise measurement of $He/H_2$ in the troposphere is essential to understand the interior processes and the planetary heat balance.

For the most part, the well-mixed atmospheres of the giant planets are too deep to probe by remote sensing, and the noble gases, despite being well-mixed in the stratosphere, cannot be measured by remote sensing either. The Galileo probe was thus designed to determine the bulk composition, and hence the heavy element abundances, of Jupiter's atmosphere. The probe succeeded in carrying out measurements to 22 bars, well below the expected condensation level of the deepest cloud, water, which was calculated to be at a level between 5–10 bar for O/H between 1–10× solar on the basis of equilibrium cloud condensation models (ECCM; Atreya et al. 1999). The Galileo probe successfully measured the well-mixed abundances of all but one of the key heavy elements (oxygen) and found them to be enriched by a factor of 4±2 × solar (Atreya et al. 2017). Ne was severely depleted and He was slightly sub-solar. The only heavy element whose well-mixed abundance was not reached even at 22 bars was water (Wong et al. 2004), due to the Galileo probe having entered a very dry region of subsiding air. As water was presumably the original carrier of the heavy elements that formed Jupiter, it is crucial to measure its well-mixed abundance, i.e., the O/H ratio. Thus, the microwave radiometer on the Juno spacecraft is designed to map the distribution of water to several hundred bars in Jupiter's atmosphere. Nonetheless, the Galileo probe data are a good demonstration of the role of dynamics in the troposphere of Jupiter, showing that the well-mixed abundances of the condensable gases may be found only at pressure levels much, much deeper than their respective condensation levels. From limited data on ammonia this seems to hold for Saturn also, and could well be the case at Uranus and Neptune.

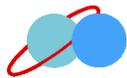

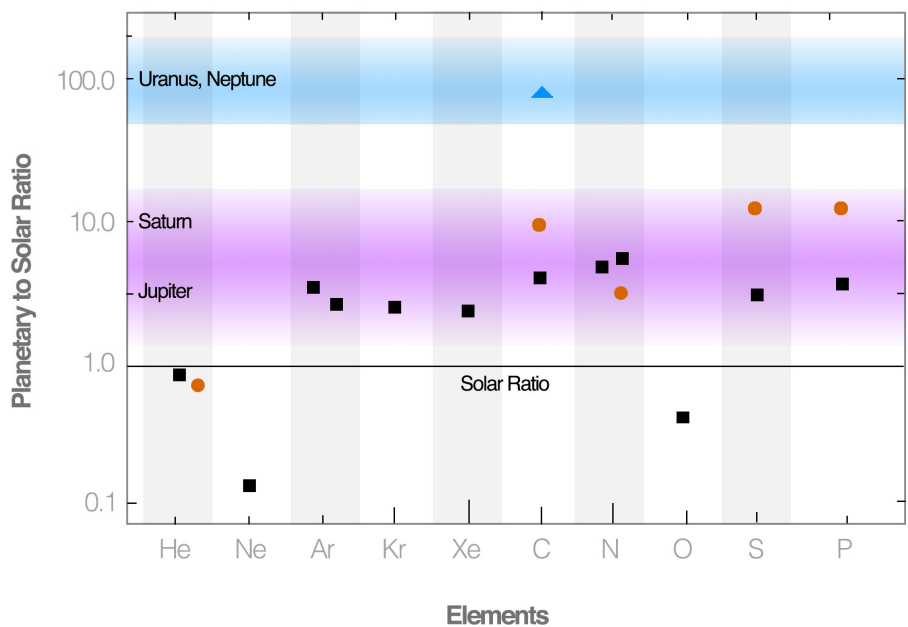

**Figure 3-3.** The abundance of key species in the giant planets, relative to solar abundances (based on Atreya et al. 2017). Jupiter values are black squares, Saturn abundances are brown circles, and the ice giants are the blue triangle. Planetary formation models predict certain ratios for each planet. The colored horizontal bands show schematically the values predicted by various models. Measurements on Jupiter and Saturn are consistent with some but not all of these models (evolutionary processes and/or dynamics are believed to create the sub-solar abundances of He, Ne, and O). Only carbon, in the form of methane, has been reliably measured on Uranus and Neptune, limiting our ability to constrain models of their formation.

For Uranus and Neptune, currently only remote sensing constraints on composition are available. The only heavy element identified in their atmospheres is carbon from $CH_4$, and that too has large uncertainty. The C/H ratio is found to be 80±20 × solar, or greater, in both Uranus (Sromovsky et al. 2011; E. Karkoschka and K. Baines, personal communication, 2015) and Neptune (Karkoschka and Tomasko 2011). The increasing C/H ratio from 4 × solar at Jupiter to 8 × solar at Saturn to ≥80 × solar at Uranus and Neptune is what the core accretion model predicts. However, it would be premature to draw conclusions about this or any other formation scenario in the absence of data on the remaining suite of heavy elements in the ice giants. Measuring those elements requires an entry probe.

The ECCM calculations for Uranus and Neptune predict cloud bases of $CH_4$, $NH_3$, $NH_4SH$ and $H_2O$ (ice), and $H_2O$-$NH_3$ solution to be present, respectively, at 0.75, 10, 30, 53 and 88 bars for uniform 1× solar ratios for all elements, and at approximately 1, 13, 44, 39, and 495 bars for 80× solar ratios (Atreya and Wong 2004, 2005, but revised using current solar elemental abundances from Asplund et al. 2009; intermolecular force corrections at high pressures are from Atreya and Wong 2004). Considering the effects of dynamics, as seen in the Galileo probe entry data at Jupiter, the well-mixed levels of these condensable volatiles could lie much, much deeper, perhaps at 2–10× the cloud-base levels from the ECCM. For the 80× solar O/H case, well-mixed water would be at least at the several kilobar level. In fact, it might be even deeper than that because of the possible presence of an ionic ocean at the tens to hundreds of kilobar level (Atreya and Wong 2005).

Molecular dynamics calculations predict a superionic phase of water at temperatures above 2000 K and pressures above 30 GPa (Goldman et al. 2005). Such a phase change would not only result in the depletion of water at ≥300 kilobar, it may deplete ammonia as well due to the solubility of $NH_3$ in $H_2O$. A likely composition of such an ionic ocean is $H_3O^+ \cdot NH_4^+ \cdot OH^-$, together with free electrons in the plasma. The dynamo that drives the internal magnetic field of Uranus and Neptune may be the result of such an ionic ocean at Uranus and Neptune (Ness et al. 1986; Ness et al. 1989). The large depletion of $NH_3$ observed in the troposphere of these planets by the VLA (Mark Hofstadter, personal comm., 2016; de Pater et al. 1991; Gulkis et al. 1978) could also be a result of the removal of ammonia in the deep ionic ocean.

Considering that the well-mixed regions of $NH_3$ and $H_2O$ may lie at tens to hundreds of kilobar, levels to which current technology does not permit measurements, it's unlikely that unassailable determinations are possible for O/H, N/H, and possibly also S/H ($H_2S$ is taken up in the $NH_4SH$ cloud as well as possibly the deeper ionic ocean). This limitation would not be a showstopper, however, for constraining formation, evolution, and migration models of Uranus and Neptune, provided that robust measurements are carried out of the abundances of the noble gases, He, Ne, Ar, Kr, Xe, their isotope ratios, and certain stable gas isotopes, in particular $^{13}C/^{12}C$ and D/H. An entry probe capable of making mass spectrometer measurements down to only 10 bars would be able to accomplish all of that, and possibly also yield $^{15}N/^{14}N$ and $^{34}S/^{32}S$, depending on the actual degree of depletion of $NH_3$ and $H_2S$ at deeper levels. A comparative analysis of these data at Uranus and Neptune with those for Jupiter, and possibly Saturn in the future, and the solar abundances would provide excellent insight into the formation and evolution of the outer solar system in particular and extrasolar giant planets, in general.

### 3.1.3 Characterize the Planetary Dynamo

The intrinsic magnetic fields of Uranus and Neptune were discovered by Voyager 2 in 1986 and 1989, respectively (Ness et al. 1986, 1989). The spacecraft passed within 4.2 Uranian radii ($R_U$=25,600 km) and 1.2 Neptunian radii ($R_N$=24,765 km), allowing large-scale field components to be mapped (e.g., Connerney et al. 1987, 1991; Holme & Bloxham 1996). To first approximation, Uranus' (and Neptune's) field can be characterized by a dipole of moment 0.23 gauss $R_U^3$ (0.14 gauss $R_N^3$) that is offset by 0.31 $R_U$ (0.55 $R_N$) from the planet center and tilted by 59° (47°) away from the rotation axis. In a spherical harmonic representation, the magnetic fields are characterized by quadrupole and octupole components that are comparable to or greater than the dipole, yielding mean surface field strengths of approximately 0.3 G (30 µT). These in situ magnetic field measurements are complemented by auroral observations, which provide additional high-latitude constraints on the internal field (e.g., Sandel et al. 1990; Herbert 2009; Lamy et al. 2012). At present, spherical harmonics greater than degree four are below the limits of spatial resolution, and there is no information about any secular variations.

These multipolar, non-axisymmetric magnetic fields were a surprise upon their discovery, and it is still not understood why these bodies generate remarkably different fields compared to all other planets in our solar system (**Figure 3-4**), whose intrinsic fields are dipole-dominated and nearly aligned with their rotation axes (Earth, Jupiter, Saturn, Ganymede, and Mercury to a lesser extent), extinct (Mars, the Moon), or unknown (Venus). Because of their unique characteristics, the ice giants serve as excellent laboratories for determining the fundamental dynamical and chemical processes responsible for generating all planetary magnetic fields. For the magnetic and plasma environments of Uranus and Neptune, knowledge of their internal magnetic field characteristics and temporal evolution is also essential for correctly interpreting magnetospheric configurations and interactions with rings and moons (see also Section 3.1.11 on magnetospheres as well as Sections 3.1.7, 3.1.8, and 3.1.9 on satellite composition, geology, and interiors).

Magnetic fields originate in the electrically conducting fluid regions of planetary bodies and are likely driven by convectively driven dynamo action, which converts kinetic energy into magnetic energy (e.g., Stevenson 2003; Jones 2011). As a result, an understanding of the dynamo processes that control the magnetic field strength, morphology, and temporal evolution of ice giant planets is critically dependent on their poorly constrained interior structures (Section 3.1.1), bulk compositions (Section 3.1.2), heat balance (Section 3.1.6), and dynamics (Section

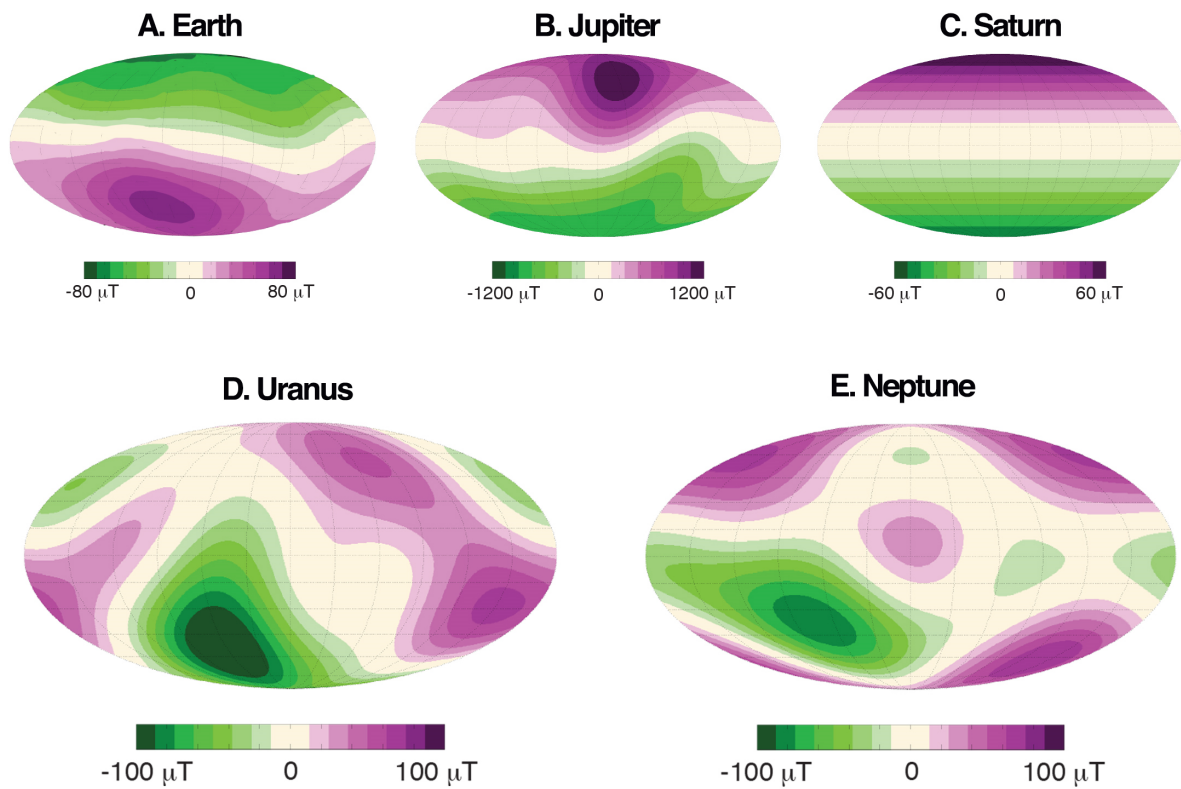

**Figure 3-4.** Radial magnetic fields measured on (A) Earth, (B) Jupiter, and (C) Saturn are contrasted against those measured on (D) Uranus and (E) Neptune. The colors represent field intensity where purple (green) indicates outward (inward) directed fields. The measurements on Uranus and Neptune have the lowest spatial resolution (to spherical harmonic degree 3), so all planets are shown with that resolution. This comparison illustrates the ice giants' unique magnetic field morphologies. Credit: Adapted from Schubert and Soderlund (2011)

3.1.11). It is typically assumed that the magnetic fields are generated in the planets' "watery" oceans (e.g., Ness et al. 1986, 1989; Ruzmaikin and Starchenko 1991; Nellis 2015). This mixture undergoes progressive dissociation and ionization with depth due to increasing pressures and temperatures, resulting in electrical conductivities large enough to support dynamos in the ice giants at depths below approximately 0.8 planetary radii (e.g., Redmer et al. 2011; Lee & Scandolo 2011; Knudson et al. 2012). This relatively shallow dynamo region is consistent with the prominence of the higher-order spherical harmonics of the planets' magnetic fields. Moreover, if the transition between the watery ocean and the overlying atmosphere is continuous, these two regions may be dynamically coupled, linking the dynamo to atmospheric dynamics and its thermal emissions (Soderlund et al. 2013). Layers of thermal and/or compositional stratification that could lead to double diffusive convection may also exist within the deep interiors, which would further complicate the interior structure, dynamical, and thermal evolution of the ice giants (e.g., Podolak et al. 1991; Hubbard et al. 1995; Redmer et al. 2011; Leconte & Chabrier 2012; Chau et al. 2011; Nettelmann et al. 2013, 2016).

A variety of competing numerical dynamo models have been developed to explain the ice giants' magnetic fields (e.g., Stanley and Bloxham 2004, 2006; Gomez-Perez and Heimpel 2007; Soderlund et al. 2013). For example, Stanley and Bloxham (2004, 2006) find multipolar dynamos when moderately strong convection occurs in a thin spherical shell overlying a stably stratified fluid layer. Gomez-Perez and Heimpel (2007), in contrast, find that dynamo models tend to become multipolar with relatively low electrical conductivity and strong zonal flows. Soderlund et al. (2013) and King & Aurnou (2013) both argue that convection in the deep interiors of Uranus and Neptune are weakly constrained by rotation, leading to a fundamentally different convective planform compared to the rotationally constrained dynamo action of the Earth, Jupiter, and Saturn and explaining the associated dichotomy in magnetic field characteristics. The importance of inertia (i.e., the dynamo fluid's momentum) for the transition from dipolar to multipolar magnetic field generation has also been shown in geodynamo and gas giant dynamo models (e.g., Sreenivasan and Jones 2006; Christensen & Aubert 2006; Soderlund et al. 2012; Gastine et al. 2012; Oruba & Dormy 2014). Further modeling progress in combination with additional constraints on the magnetic fields and the internal structures and dynamics derived from a second mission to Uranus and/or Neptune would aid development of more realistic models that capture the relevant physics and yield better predictions about the evolution of planetary magnetic fields. By understanding the dynamos of our solar system, we would be able to predict the magnetic field strengths and morphologies of exoplanetary dynamos with more confidence as well (e.g., Tian & Stanley 2013).

A mission to Uranus and/or Neptune could answer key questions that characterize the intrinsic magnetic fields and constrain the dynamo processes responsible for their generation:

- What is the configuration of the ice giants' intrinsic magnetic fields? Has secular variation occurred since the Voyager 2 observations? What is the rotation rate of the bulk interior and how does it compare to the radio rotation rate?
- What is the internal density distribution? How do composition and temperature vary with radius? What regions of the planet are unstable to convection versus stably stratified? What are the characteristics of zonal winds, meridional circulations, and turbulent convective flows in the deep interior?

New magnetic field measurements close to the planet at a variety of latitudes and longitudes would allow characterization of the ice giants' higher degree magnetic field structure. This field determination could be further improved through imaging of auroral and satellite footprints that provide additional high-latitude constraints (e.g., Herbert, 2009). The internal magnetic field may have undergone temporal change since the Voyager 2 epoch so new observations, even from a flyby, would provide constraints on secular variation and potentially identify changes in the locations of flux patches that are indicative of zonal and/or meridional winds in the deep interior. To help place the zonal winds in context of the deep interior requires better determination of the planets' rotation rates. The depth of strong zonal wind penetration as well as the radial density distribution within these planets may be established through measurements of the gravity field. As above, observations of planetary oscillations are expected to further constrain interior flows and to identify layers of convection versus stable stratification, while deep remote sensing would allow inference of deep meridional circulations and determination of key elemental abundances, such as water, in the deep atmosphere. The emitted thermal flux and Bond albedos, critical for understand the planets' energy balances and internal heat fluxes, are also required. The probe instruments would supplement these wind, composition, and temperature measurements.

### 3.1.4  Measure the Atmospheric Heat Balance

The Voyager 2 flybys of Uranus and Neptune provided our only measurements of the energy balance of these atmospheres. These data sets not only led to important advances in our understanding of the planets, but also to several enduring mysteries. The Bond albedo (which is a measure of reflectivity over all wavelengths, weighted by the solar spectrum, over all phase angles) was determined for each planet, with relatively large error bars of 16% and 23% for Uranus and Neptune, respectively (Pearl et al. 1990; Pearl and Conrath 1991). Measurement of the total emitted power in the infrared also had relatively large error bars of 2 and 5% for Uranus and Neptune, respectively. Comparing the visible-wavelength energy absorbed by the atmosphere with the emitted infrared energy leads to a determination of the intrinsic power being provided from the interior. Here a dramatic dichotomy was seen, with Neptune's intrinsic flux at least 10 times larger than Uranus, and Uranus' value so low that the large error bars allowed only an upper limit to be determined, with zero being a possible value (**Figure 3-5**). We have no knowledge if the energy balance of any of the giant planets is time variable.

The internal heat fluxes of Jupiter, Saturn, and Neptune can be understood in terms of secular cooling from an initial hot state, and evolutionary models of these planets with fully convective interiors reasonably reproduce the observed thermal fluxes at 4.5 Gyr. Two kinds of explanations have been put forth to explain the low heat flux from Uranus. The first connects to the planet's interior (see also Section 3.1.1). It has been suggested that a dramatically different interior structure for Uranus compared to Neptune, perhaps due to deep-seated composition gradients within Uranus, could suppress convection in the interior, leading to very low fluxes transported to the visible atmosphere (Hubbard et al. 1991, Nettelmann et al. 2016). Alternatively, Gierasch and Conrath (1987) suggest that the low heat flux at Uranus could be due to an atmospheric effect, rather than related to the deep interior, and could be strongly temporally variable, a side effect of the ortho-para controlled intermittent overturning. Further questions for these two scenarios are given below:

- Deep interior scenario:
  What causes the break in adiabatic energy transport in the interior?
  How would that manifest itself in the magnetic field and atmospheric motions?
  What other observables can constrain this?

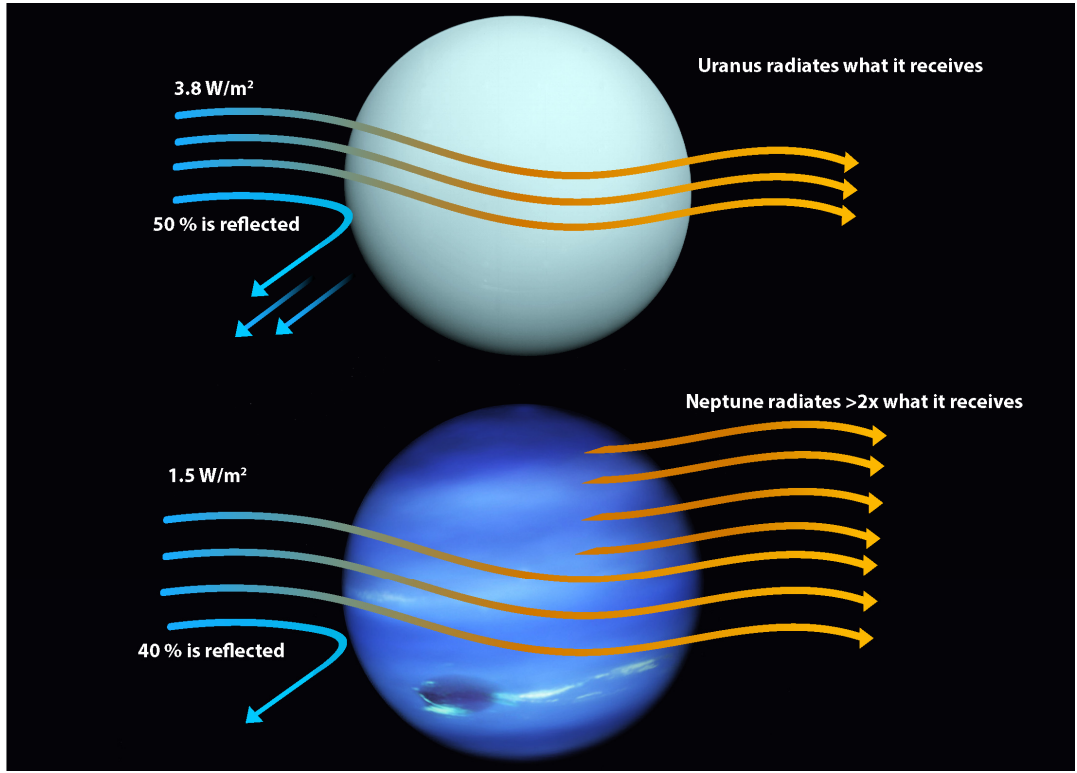

**Figure 3-5.** The atmospheric energy balance of Uranus and Neptune. All giant planets are expected to radiate significantly more energy than they receive from the Sun, due to the release of internal heat from planetary formation. Uranus is the only giant planet that does not.

- Atmospheric scenario:
  How is the atmosphere storing energy and over what timescales?
  What bounds on temporal variability can we place using feasible observations?

For the atmospheric heat balance, key questions include: Is the thermal emission from the ice giants time-variable? If so, what are the time constant and controlling physics? Is the intrinsic flux spatially inhomogeneous? Is there a hemispheric dichotomy? Can the intrinsic flux of Uranus be detected?

To answer these questions, we must directly measure the emitted thermal flux and the Bond albedo. Measurements should be made of the 3-D meridional circulation of the atmosphere via remote sensing. As described in Section 3.1.1, planetary oscillations can be used to understand detailed internal structure, which affects the transport of heat from the interior. Additionally, the mission should determine the helium mixing-ratio to understand internal convection via an atmospheric entry probe. The probe is also needed to measure the ortho-para hydrogen fraction to determine how much stored energy is in the weather layer. In conjunction, the probe should measure temperature and pressure profiles, as well as make direct measurements of the location and particle properties in cloud layers.

### 3.1.5  Characterize Atmospheric Dynamics

Uranus and Neptune have very different obliquities and internal heat fluxes, parameters expected to have major impacts on their atmospheric behavior. Neptune has moderate obliquity, not too different from Earth's, making for familiar seasonal behavior, and a relatively large internal heat flux (compared to its solar flux). In contrast, Uranus has high obliquity, yielding extreme seasonal variations, and a very small internal heat flux. Nevertheless, their atmospheric dynamics are superficially quite similar with broad, slightly retrograde, equatorial jets, and prograde jets poleward of the mid-latitudes on both planets (Smith et al. 1986; Limaye and Sromovsky 1991; Hammel et al. 2005). Through analysis of gravity data from Voyager 2, these zonal winds are predicted to be limited to depths of 1,000 km below the visible surface (Kaspi et al. 2013); however, deeper extents cannot be ruled out given the lack of consensus on gravity inversion techniques for giant planets (e.g., Zhang et al. 2015, Kaspi et al. 2016). Because of their long orbital periods, we have only observed a portion of a year for each planet since the Voyager flybys and the availability of good telescopic resolution (e.g., the Hubble Space Telescope and large ground-based telescopes with adaptive optics). Nevertheless, Uranus has exhibited dramatic seasonal changes, going from a near featureless atmosphere during the Voyager encounter to one with significant bright storms (**Figure 3-6**) and a polar hood that switched hemispheres near the recent equinox (Sromovsky et al. 2012). Neptune's disk-averaged brightness has changed over many decades, suggesting seasonal changes are slowly occurring there too (Lockwood et al. 1991; Hammel and Lockwood 2007).

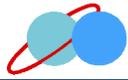

During the Voyager encounter, Neptune had multiple storms that lasted throughout the several months when Voyager was close enough to image them. However, these storms were gone and replaced by others when Hubble observed Neptune a few years later (Hammel et al. 1995). This same evolving set of storms has continued since then, suggesting a very dynamic atmosphere. Uranus had barely discernible discrete storms at the Voyager encounter, but now shows storms with high-contrast and varying lifetimes, differing length-scales and amounts of contrast in both polar regions, and wave features in the tropics (**Figure 3-6** and Sromovsky et al. 2012b, 2014, 2015; de Pater et al. 2015). Spectroscopic studies reveal that the storms on both planets are a mix of deeper clouds (or lack thereof) and high methane clouds that could be orographically related to deeper circulations of the storms (Sromovsky et al. 2012a; Irwin et al. 2016a, 2016b). On both ice giants, the preferred latitudinal locations of the storms seems less controlled by the meridional profile of the zonal winds when compared to the highly structured flows on Saturn, and especially Jupiter. In addition to the large-scale storm and wave structures, there is also suggestion that mesoscale waves may be a significant source of energy to the upper atmosphere (Uckert et al. 2014).

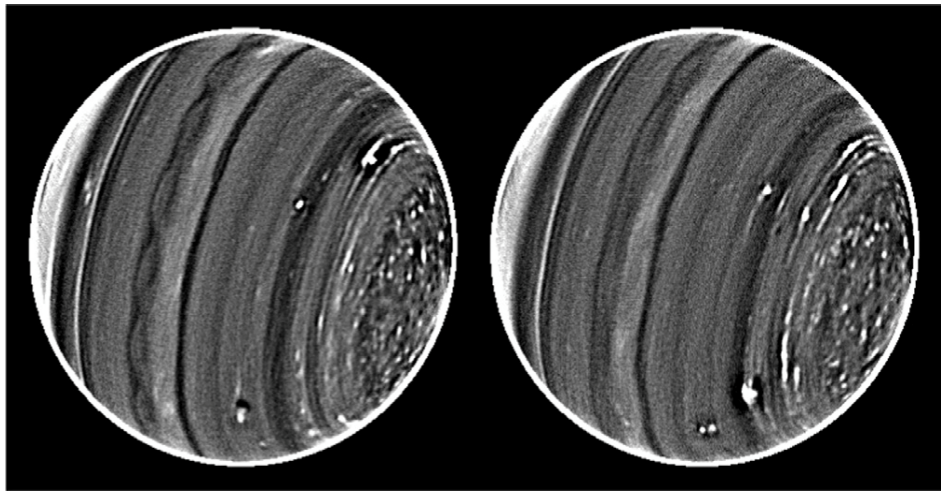

**Figure 3-6.** The appearance of Uranus on two dates in July 2012, from Sromovsky et al. (2015). This complex and time-varying atmosphere, imaged near equinox, is quite different from the featureless "billiard ball" seen by Voyager 2, which encountered Uranus near solstice.

Combining the detailed studies of the background and storm-cloud structures, the zonal wind profiles from feature tracking, and deeper probing using radio-wavelength sounding, pictures of the meridional circulation on both planets is beginning to emerge. Neptune may have a single layer of meridional cells, with a thermally direct circulation at low latitudes and a single indirect circulation poleward of that (de Pater et al. 2014). Uranus may have stacked layers of meridional cells (**Figure 3-7**), with an indirect circulation at the top near the equator and at depth (Hofstadter and Butler 2003; Sromovsky et al. 2014). For both ice giant planets, dynamo simulations predict large circulation cells in the deep interior with upwelling near the equator; depending on the thickness of the convecting layer (see Section 3.1.1), polar meridional cells may also be present (Soderlund et al. 2013). These poorly characterized circulations, which carry the internal heat from some level deep in each planet, along with the absorbed solar radiation, up to a level where they can be radiated to space, are the main driver of all of the atmospheric motions. Complicating the circulations is the fact that significant heat can also be transported not just in the sensible heat flux, but also in latent heat effects from condensation and in the conversion of hydrogen from its ortho- to para-state and vice versa. As the seasons change and the solar input distribution is modified, the dynamical response of each atmosphere will change. The details of this complex interplay of dynamics, condensation, and hydrogen spin-state conversion are only beginning to be known, with many questions remaining before we can claim to understand what controls the visible face of the ice giants, and by analogy a vast number of exoplanets.

An ice giant mission can provide significant advancements in our understanding of these processes on Uranus and Neptune. The key questions are:

- What is the 3-D circulation on each planet?
- How does the circulation vary with season?
- Are differences in the two planet's circulation due to differences in internal heat loss?
- How does obliquity effect the circulation?

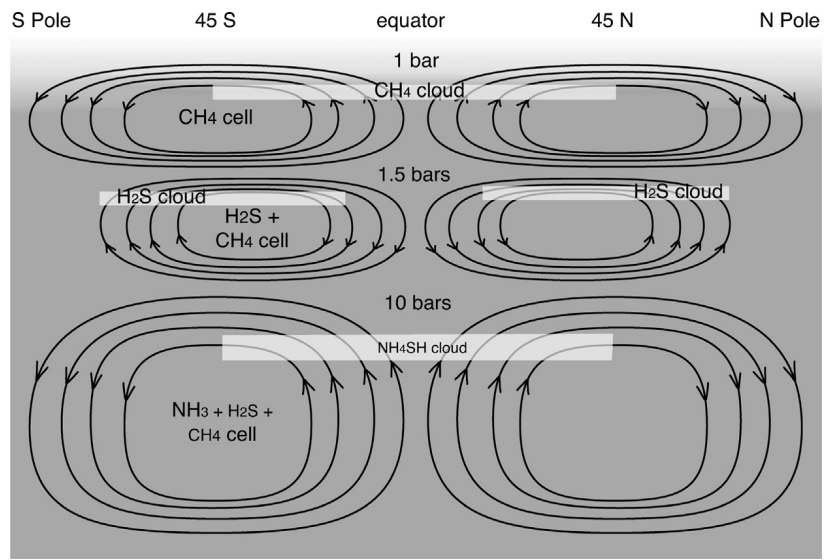

**Figure 3-7.** A possible tropospheric circulation for Uranus, from Sromovsky et al. (2014). As discussed in that paper, this circulation explains some, but not all observations. The meridional circulation is likely what distributes the seasonally varying solar input and the internal heat from below to yield a nearly homogeneous emission to space. Presumably this circulation controls not only the heat fluxes, but also the zonal winds, the cloudy (and clear) latitude bands, and the behavior of discrete storms. Deducing and confirming this structure for the ice giants and allowing extrapolation to the circulation on extrasolar planets, is a key goal of the future study of the ice giants.

- Why do large storms appear different on Uranus and Neptune?
- How do these varied atmospheres place the other known atmospheres (including exoplanets) in context?

The 3-D circulation of the atmospheres are best observed at high spatial resolution and with short and intermediate time-scale temporal resolution. While seasonal changes won't be directly observed from a single mission to the ice giants, comparing the close-up view from Voyager with those some 40–50 years later will highlight a significant seasonal change for both planets. Ground-based data can fill in some seasonal gaps for Uranus, given its unique tilt, but feature scales are necessarily limited by telescopic resolution and atmospheric seeing. Comparing detailed observations of Uranus and Neptune, we can try to understand the significance of differences in obliquity and internal heat flux on a giant planet's (whether ice- or gas-) circulation. The close-up view of either planet will help better constrain the detailed behavior and the controlling effects of large storms, as well as directly allow observations of mesoscale waves that may be important drivers for the upper atmosphere. An entry probe into either planet would directly reveal the stability structure of the atmosphere as well as the exact locations of cloud layers, and the profile of hydrogen para-fraction. These in situ measurements would be key to understanding the meridional circulation on either planet, and such observations can also serve as a touchstone to better understand the rich variety likely in exoplanets.

A range of instruments can contribute to our understanding of atmospheric dynamics. On the orbiter, a narrow angle camera (NAC) and/or wide angle camera (WAC) would not only be useful to track the zonal winds and winds in storms, but using filters in and near absorption lines could reveal the 3-D structure of those winds and features. Similarly, a VIS/NIR (VISible Near InfraRed) mapping spectrometer could yield the 3-D structure of the clouds and storms, possibly even their winds, at deeper levels than are available at visible wavelengths. A thermal IR (or mid-IR) spectrometer would help to understand the connection between the winds and temperatures, the ortho-para disequilibrium and its spatial distribution, and the total heat flux (when combined with an imager to determine the visible albedo) and its spatial distribution. A Doppler imager could be used to directly measure upwellings in storms, complementary to divergence measurements from a camera. A microwave sounder would be useful to further constrain the 3-D meridional circulation by sounding the deeper aspects of the temperature and condensation structure of the planet, both in and out of discrete storms.

Probe instruments would also be key to fully investigating the 3-D structure of the atmospheric circulation. A mass spectrometer or tunable laser spectrometer could yield measures of disequilibrium species as an indicator of the strength of the mixing at deeper levels in the atmosphere. An ultra-stable oscillator (USO) would provide a measure of the horizontal wind profile by Doppler tracking the probe's carrier. A temperature, pressure, and density package could directly measure the atmospheric temperature and any vertical winds, providing the vertical sensible heat fluxes along the probe trajectory. An ortho-para sensor could add the vertical heat flux carried in hydrogen spin-state conversion, as well as its use as a passive tracer of fluid motions to help diagnose the 3-D circulation. Finally, a nephelometer could help measure the vertical heat fluxes carried by latent heat. A nephelometer would also serve to provide an in situ reference for the remote sensing from visible-wavelength cameras and VIS/NIR mapping spectrometers' radiative transfer analysis of cloud structure, removing significant ambiguity that necessarily remains in the remote sensing retrievals.

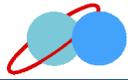

### 3.1.6 Determine the Structure and Dynamics of Rings; Perform an Inventory of Small Satellites

Due to the similarity of measurement approaches and common scientific goals, two objectives are considered together here: determine the structure and dynamics of rings, and perform an inventory of small satellites.

Both Uranus and Neptune are surrounded by complex systems of small moons and faint and/or narrow rings (**Figure 3-8**). The roles and origins of rings and small moons are deeply intertwined. Small unseen moons probably serve as the source bodies for a number of faint dust rings. These moons' gravity fields may also help to confine rings and to impose some of the structure seen within them. Conversely, some moons could have formed via the accretion of dust and other debris outside the Roche Limit (Crida and Charnoz 2012; Tiscareno et al. 2013). Ring-moon systems are snapshots into the dynamical history of the central planet, providing fundamental information about how the system formed and evolved. Small moons and rings can also be used to probe their host-planet's gravity field and, consequently, their internal structure as well (see also Section 3.1.1).

The ring system of Uranus is dominated by nine narrow, dense rings with sharp edges. Two small moons—Ophelia and Cordelia—"shepherd" the outermost dense ring (ε) and sculpt the γ and δ rings via orbital resonances (Porco & Goldreich 1987). The remaining ring boundaries remain unexplained. Because dissipative collisions among ring particles should cause the material to spread radially on time scales of just a few thousand years, it has long been speculated that additional moons, too small to be detected by Voyager 2's cameras, are probably

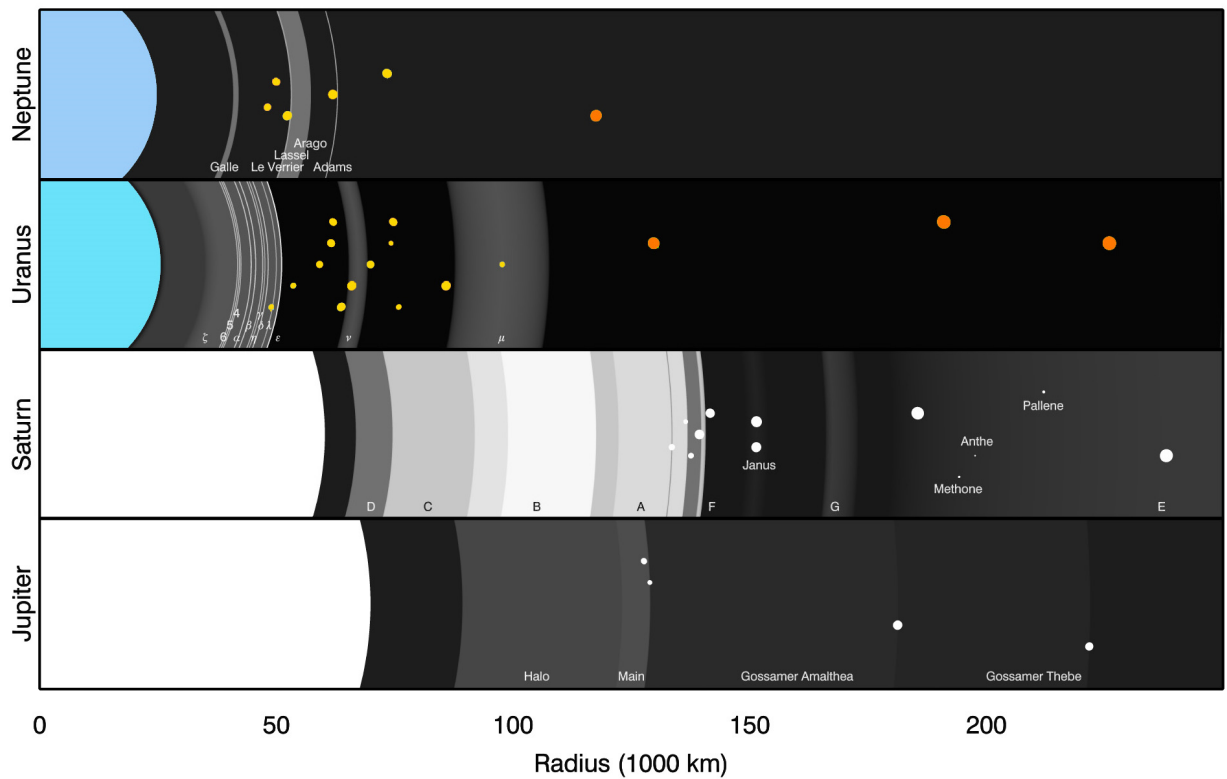

**Figure 3-8.** Diagram comparing the ring-moon systems of the giant planets. The extents of the various rings are shown to scale, with greyscale levels indicating their relative optical depths. The locations of various moons are indicated by dots. These dots do not indicate the size of the moons relative to the rings, though the relative sizes of the satellites are approximately correct. Credit: Matthew Hedman

shepherding the other ring edges as well. The two narrow rings of Neptune may be similarly confined. However, Cassini has not detected shepherding moons near many of the similarly sharp edges in the rings of Saturn, indicating that our understanding of ring confinement remains incomplete. The ice-giants' narrow rings are ideal targets for investigating in detail how ring material can be confined.

Voyager occultation data also revealed that Uranus' tightly confined dense rings contain abundant structure on scales smaller than a few kilometers. (Neptune's rings were not observed at such fine detail.) These structures probably arise from various combinations of external perturbations and inter-particle interactions, but the actual origins of many of these structures are still poorly understood. For example, Uranus' δ ring exhibits a variation in its optical depth that is very similar to density waves in Saturn's rings (Horn et al. 1989). Such waves are normally found at resonances with satellites, but there is no resonance with any known moon that could explain this structure. Recently, additional features and the α and β rings resemble moonlet wakes, similar to that generated by Pan in Saturn's rings (Chancia & Hedman 2016). These features could point to undiscovered bodies in the Uranian system, but they could also potentially be generated by resonances with oscillations or asymmetries inside the planet, in the same way that waves generated by Saturn have recently been identified in its rings (Hedman & Nicholson 2013, 2014). High-resolution observations of the narrow, dense rings at Uranus and Neptune should clarify the processes responsible for sculpting these systems.

The ring systems of Uranus and Neptune are also filled with optically-thin bands of dust. Because micron-sized dust particles have limited lifetimes (due to Poynting-Robertson drag, plasma drag, and other non-gravitational forces), they must be continually replenished. We believe that these dust rings emerge from the ejecta generated by micrometeoroid impacts into the surfaces of embedded macroscopic bodies, perhaps meters to a few km in size. In a few cases the source bodies are known: Uranus's ~10-km moon Mab orbits in the middle of the faint μ ring (Showalter & Lissauer 2006), and Voyager imaged a narrow ringlet in the orbit of Galatea at Neptune. However, the source bodies for the vast majority of dusty rings are unknown. The μ ring is especially puzzling because its color is distinctly blue (de Pater et al. 2006), unlike any other known planetary dust ring except Saturn's E ring. In the case of the E ring, the peculiar color derives from the narrow size distribution of ice grains ejected from the plumes of Enceladus. However, Mab is likely too small to be geologically active, so the color of its ring remains a mystery.

Although we once thought that planetary rings were stable for millions of years, recent observations show changes on time scales of decades. The innermost dust ring of Uranus, designated ζ, is in a different location today than it was in the Voyager images from 1986 (de Pater et al. 2007). In 1989, Voyager imaged a set of four arcs orbiting within Neptune's narrow Adams ring. Such arcs should spread in time scales of years, but have persisted for decades. They may be confined by a resonant interaction with the nearby moon Galatea (Porco 1991, Foryta & Sicardy 1996, Namouni & Porco 2002), but the details remain mysterious (de Pater et al. 2005; Renner et al. 2014). Furthermore, recent observations from Hubble indicate that the two leading arcs have vanished and the trailing arcs are slowly fading (Showalter et al. 2013). These observations build on results from the Cassini mission that show many rings to be variable over decadal time scales. We are still struggling to understand some of the rapid changes being observed in planetary rings, and so close inspection of these rings for a protracted time period would be extremely useful.

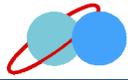

Uranus and Neptune also serve as hosts to a family of small, inner moons. Of the thirteen known inner moons of Uranus, nine (Bianca to Perdita) orbit within a narrow radial span of just 27,000 km. This is the most densely packed set of moons in the solar system, and dynamical simulations show that the system is chaotic, with some moons surviving for time scales of only $10^6$ years (Duncan & Lissauer 1997; French & Showalter 2012; French et al. 2015). We do not know how such an unstable satellite system can have survived for the lifetime of Uranus. The Neptune system has also shown some dynamical quirks. The innermost moon, Naiad, was recently recovered, but at an orbital longitude 88° off its ephemeris (Showalter et al. 2014). In 2013, Neptune's ~10-km moon S/2004 N 1 was detected in Hubble images, orbiting between the much larger moons Larissa and Proteus (Showalter et al. 2013). We do not know how this tiny object survives in a stable orbit between two objects each ~$10^5$ times larger.

New missions to Uranus and Neptune are critical to advancing our understanding of these ring-moon systems. Such a mission would allow the rings to be observed at higher resolution than has been possible since Voyager, revealing new details of the processes that sculpt the ring material. Observations at higher phase angles will allow the rich structure of the dusty rings to be clearly seen, which should provide further information about how these fine particles are generated, lost, and transported throughout the system. A complete survey of the system at low phase angles would obtain a complete inventory of as yet unseen km-sized bodies. Satellite astrometry can also reveal the subtle interactions between the moons and rings, providing our first determinations of the moons' masses.

In summary, the key questions about the rings and small satellites that could be addressed with a new mission include:

- How are the narrow rings confined and maintained?
- What is responsible for the fine-scale structure in the ice giant's denser rings?
- What are the sources, sinks and transport mechanisms for the dusty rings?
- What allows ring structures to change on decadal time scales?
- What is the full inventory of small satellites in these systems?
- How did the small satellites and rings originate?
- How are the small satellites and rings currently coupled to each other dynamically?

Important measurements for studying the rings and small satellites include:

- Images capable of resolving kilometer-scale structures in Uranus' and Neptune's rings. Such observations would also be capable of detecting km-sized satellites and obtaining precision astrometry of all the moons and rings.
- Stellar and radio occultation experiments, which would provide the best opportunity to obtain information about ring structures at sub-kilometer scales.
- Imaging with sufficient signal-to-noise to discern features in the planets' dusty rings (high phase angles being particularly useful in this regard).
- Near-IR spectrometers could not only provide information about surface composition, but also help to constrain the particle sizes of these rings.
- A dust detector could potentially provide detailed information about the composition and spatial distribution of finer particles, but it is not yet clear if any of the relevant rings could be safe enough for a spacecraft to fly through.

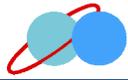

### 3.1.7    Determine the Surface Composition of Rings and Moons

The surface composition of the satellites of Uranus and Neptune is poorly known.  Due to the lack of a near-infrared spectrometer on-board Voyager 2, the compositional information we possess is derived from telescope observations from the ground and from near-Earth orbit.  As a consequence, the correlation between the compositional and the geological information on these planetary objects is resolution-limited and fine-grain details are currently unavailable.  Given the complex geologic histories of these satellites revealed by the images of Voyager 2, this lack of information represents a major issue for our understanding of the formation and the histories of these bodies and the planets they orbit.

The surfaces of the satellites of Uranus are mainly characterized by a mixture of crystalline $H_2O$ ice (see Dalton et al. 2010 and references therein) and a dark, spectrally neutral non-ice constituent responsible for their low albedo (~0.25–0.5).  The composition of this dark material is as yet undetermined, although spectral modeling suggests it may be carbonaceous in nature (see e.g., Clark and Lucey 1984 and Cartwright et al. 2015 and references therein).  The spectral features of Ariel, Umbriel, and Titania also show the presence of $CO_2$ ice (**Figure 3-9** see Grundy et al. 2006; Dalton et al. 2010 and references therein), which recent results suggest could be pure and segregated from other constituents (see Cartwright et al. 2015 and references therein).  The $CO_2$ ice is associated with the trailing hemispheres of the satellites and is present in decreasing concentrations with increasing orbital distance from the planet (Grundy et al. 2006; Cartwright et al. 2015).  While $CO_2$ ice has not yet been directly observed on Oberon (Grundy et al. 2006), it was recently suggested that it could also be present on this most distant Uranian regular satellite (Cartwright et al. 2015).  Given that $CO_2$ ice should be unstable over timescales of the order of $10^6$ years, the presence of this ice is likely the result of continuous emplacement, the source of which has been suggested to be the bombardment of charged particles driven by Uranus' magnetosphere (see Grundy et al. 2006; Dalton et al. 2010).  In the case of Miranda, the possible presence of ammonia hydrate has also been reported, but both the presence of the spectral band and its interpretation are yet to be confirmed (see Dalton et al. 2010 and references therein).  Confirmation of the presence of ammonia would be of great importance due to its potential to act as anti-freeze, depressing the freezing temperatures of potential liquids in the satellite interiors.

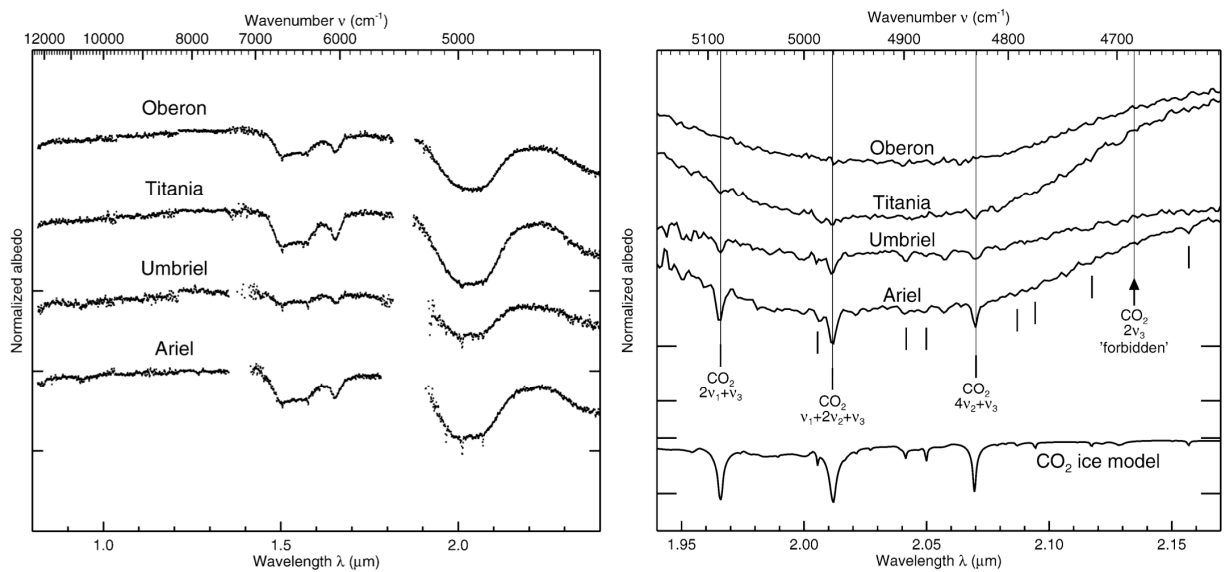

**Figure 3-9.**  Spectra of Oberon, Titania, Umbriel, and Ariel from IRTF/SpeX, covering (left) the 0.8–2.4 μm spectral range and (right) the 2-μm region, where narrow $CO_2$-ice absorption features superimposed on the broader $H_2O$-ice absorption band indicate the presence of $CO_2$ ice on Ariel, Umbriel, and Titania; no $CO_2$-ice absorption is seen on Oberon.  Longitudinal mapping showed $CO_2$ ice to be most abundant on Ariel, Umbriel, and Titania's trailing hemispheres.  Mapping of $H_2O$-ice absorptions showed deeper $H_2O$ bands on the leading hemispheres of Ariel, Umbriel, and Titania, with the opposite pattern being observed on Oberon.  After Grundy et al. (2010) Figs 1 and 3.

Spatially resolved infrared spectroscopy of satellite surfaces will constrain the composition of specific geomorphologic features, such as the dark material (organic-rich?) that fills the floors of major impact craters on Oberon and the bright deposits surrounding Wunda and in the central peaks of Vuver on Umbriel.  Albedo and morphologic variability amongst the multitude of potentially cryovolcanic deposits observed throughout the Uranian system have been attributed to differences in cryomagma composition and the conditions at the time of eruption (Croft & Soderblom 1991).  Infrared spectroscopy can attempt to confirm this hypothesis and, combined with high-resolution images and topography, start to reconstruct how and when the observed deposits were emplaced.  For example, spectra and topography of the viscous flows observed on Ariel and Titania may help discern differences between early and late-stage resurfacing events within the Uranian System.  Infrared spectroscopy would also verify tentative detections made by Earth-based observations, such as the discovery of ammonia hydrate on Miranda by Bauer et al. (2002) and $CO_2$ ice, potentially the result of outgassing, delivery during impacts, and/or radiolysis by magnetospheric charged particle bombardment of dark material from the rings, on Ariel, Umbriel, and Oberon (Grundy et al. 2006; Cartwright et al. 2015).

Spectra of Triton display absorption bands of $N_2$, $CH_4$, $C_2H_6$, CO, (the last three in solution in $N_2$), $CO_2$, and $H_2O$ ices (see Dalton et al. 2010 and references therein).  Detection of the HCN ice band has also been reported (see Dalton et al. 2010 and references therein), suggesting the presence of more complex materials of astrobiological interest.  Triton's surface undergoes seasonal cycles of sublimation and re-condensation that are responsible for the generation of the tenuous $N_2$ and CO atmosphere of the satellite (see also Section 3.1.10).  Images taken by Voyager 2 revealed active plumes rising above the surface of Triton (**Figure 3-12**), which indicate that the surface of the satellite is active, although whether due to endogenic geologic activity or the result of insolation is not certain (e.g., Hansen and Kirk 2015).  At present Triton is not tidally heated (see Schubert et al. 2010 and references therein), but as discussed later (see Section 3.1.9) Triton and some of the Uranian satellites may still harbor subsurface oceans that formed earlier in their histories.  The relatively high mean density of Triton (of the order of 2 $g \cdot cm^{-3}$, comparable to that of dwarf planet Ceres) reveals that its bulk composition is dominated in large part by rocks and metals.  Such non-ice components should be mostly confined to the interior of the satellite if it is differentiated, but they could contribute to its surface composition through exchange processes with the subsurface (see Sohl et al. 2010, Tobie et al. 2010 and references therein).  If any ice giant satellite possesses a subsurface ocean of liquid water (see Section 3.1.9), the satellite's surface composition may provide essential clues as to the habitability of the ocean.

Alongside their major satellites, both Uranus and Neptune possess rings, small inner satellites and families of outer irregular satellites (which could possibly prove to be more abundant than those of Jupiter and Saturn, Jewitt and Sheppard 2005) that are poorly compositionally characterized, and spectral data obtained using the Keck telescope do not show evidence for the ices seen in the larger satellites (de Kleer et al. 2013).  The composition of the rings and the innermost small satellites would provide important constraints on the material from which the satellite systems of these planets originally formed.  The irregular satellites (of which Triton is the largest sample in the solar system) are captured objects extracted from the trans-neptunian region and injected into the outer solar system by the gravitational perturbations of Neptune (see Jewitt and Haghighipour 2007, Nesvorny et al. 2007, Turrini et al. 2009, Mosqueira et al. 2010), analogous to the contemporary population of the Centaurs. Collisional evolution of the irregular satellites results in the production of dust that, under the influence of non-gravitational forces, can migrate inward toward the planet and contaminate the surfaces of the major satellites (Tamayo et al. 2013), as exemplified by the transport of dust from Phoebe to Iapetus in Saturn's system (Tosi et al. 2010; Tamayo et al. 2011). The compositions of the irregular satellites would therefore provide essential information to discriminate between contaminants modifying the surfaces of the major satellites and indigenous material. Compositionally characterizing the irregular satellites, moreover, would provide details on the flux of ice-rich bodies that crossed the solar system throughout its history (see Turrini et al. 2014 and references therein), while at the same time offering the possibility to explore more easily accessible members of the trans-neptunian population.

Based on what we know of the history of the solar system and of the processes governing planetary formation, it is likely the two ice giants did not form near their current orbits as previously thought. Some scenarios place their formation region nearer to the Sun (Tsiganis et al. 2005; Levison et al. 2011; Walsh et al. 2011) while others predict they formed farther away (Bitsch et al. 2015). All scenarios agree, however, that they reached their present positions as a result of one or more migration events. Scenarios that predict the ice giants formed nearer to the Sun generally propose that the formation region is between the present orbits of Saturn and Uranus (Tsiganis et al. 2005; Levison et al. 2011; Walsh et al. 2011), but the number and the details of subsequent migration events are currently a matter of debate. The surface composition of the regular satellites of the ice giants, being a reflection of the material composing the building blocks from which these bodies were assembled, provides fundamental information to understand where the ice giants formed and to constrain how and when they migrated. The composition of the irregular satellites provides a parallel constraint on the orbital regions from which the ice giants were able to extract material through capture and, therefore, on their dynamical evolution (see Turrini et al. 2014 and references therein). Finally, the surface composition of the major satellites could reveal details on the evolution of their systems. The high obliquity of Uranus is thought to be the result of one or two giant impacts in the first half billion years of the solar system (see Arridge et al. 2014, Turrini et al. 2014 and references therein): if the resulting change of obliquity proceeded more rapidly than the time required for the satellite system to follow it, the satellite system might have been destabilized and collisionally destroyed (see Coradini et al. 2010 and references therein). In this case, the present satellites of Uranus would be a second-generation population, a fact that could be reflected in their bulk, and potentially surface, composition. In the case of Neptune, it is thought that the capture of Triton could have been accompanied by the collisional removal of the original satellite system (the original system being expected to have about the same mass as the Uranian satellite system), providing the necessary removal of Triton's angular momentum (Cuk and Gladman 2005). Depending on the rate of resurfacing activity, residual material from these ancient collisions could still contaminate the surface of Triton.

Understanding the diversity of the satellites helps us to address:
- Where did the ice giants and the satellites form?
- What were the building blocks of the satellites (and the planetary cores)?
- How far did the ice giants migrate?
- Can the satellites constrain whether Uranus and Neptune exchanged positions during their dynamical evolution?

- Are the satellites of Uranus primordial or did they form after the change in the tilt of the spin axis?
- Where did the irregular satellites originate?
- How much does dust produced by the collisional evolution of the satellites contaminate the regular satellites?

A key goal is to compositionally characterize the rings and the satellites (both regular and irregular) of the ice giants using a VIS-NIR mapping spectrometer. This information can be complemented by color imaging at visible wavelengths, which also provides higher-resolution context information on the morphology and the geologic structures on the satellite surfaces. Another important piece of information is in situ measurement of the isotopic abundances and ratios (e.g., D/H) of the satellites that can be provided by a mass spectrometer during the passage through their exospheres or, in the case of Triton, through the topmost layers of its atmosphere.

### 3.1.8    Determine the Shape and Surface Geology of Satellites

When Voyager 2 flew by Uranus in 1986, its trajectory was nearly 90° from the orbital plane of the major satellites. While the geometry permitted ~300-m resolution images of Miranda (Croft & Soderblom 1991), it resulted in significantly lower resolution images (1 – 12+ km/line-pair) of the remaining major satellites (**Figure 3-10**). During its Neptune flyby in 1989, Voyager 2 came within 40,000 km of Triton and obtained images with resolutions of up to 750 m at closest approach (Croft et al. 1995). At the times of both ice giant encounters, only the southern hemispheres of the satellites were illuminated, limiting spatial imaging coverage to ≤50% for the Uranian satellites and ~40% for Triton. Despite these constraints, Voyager revealed that the surfaces of the major Uranian satellites and Triton exhibit extreme geologic diversity, including signs of endogenic resurfacing through multiple episodes of cryovolcanic and tectonic activity (e.g., Croft and Soderblom 1991; Schenk 1991; Plescia 1987; Strom 1987; Smith et al. 1986; Smith et al. 1989; Croft et al. 1995). Triton even shows evidence of currently active geologic processes, including plumes of $N_2$ gas and dust erupting from the southern polar cap (Soderblom et al. 1990; see 3.1.10). Thought to be a captured KBO (see Section 3.1.10), Triton is the only large moon in the solar system that did not form around its host planet. Triton provides a unique opportunity to study an icy dwarf planet from the Kuiper Belt that has been subject to processes (e.g., tidal, radiolytic, and collisional) typically reserved for indigenous icy satellites (Masters et al. 2014). In the Saturn system, the Voyager flybys showed that Enceladus and Titan were interesting bodies, but the subsequent results from the Cassini orbiter dwarfed expectations of just how complex their histories are and the levels of ongoing geologic activity (e.g., Hsu et al. 2015; Spencer et al. 2006; Porco et al. 2006; Hayes 2016; Stofan et al. 2007). An orbital mission to Uranus and/or Neptune can be expected to do the same for the ice giant satellites.

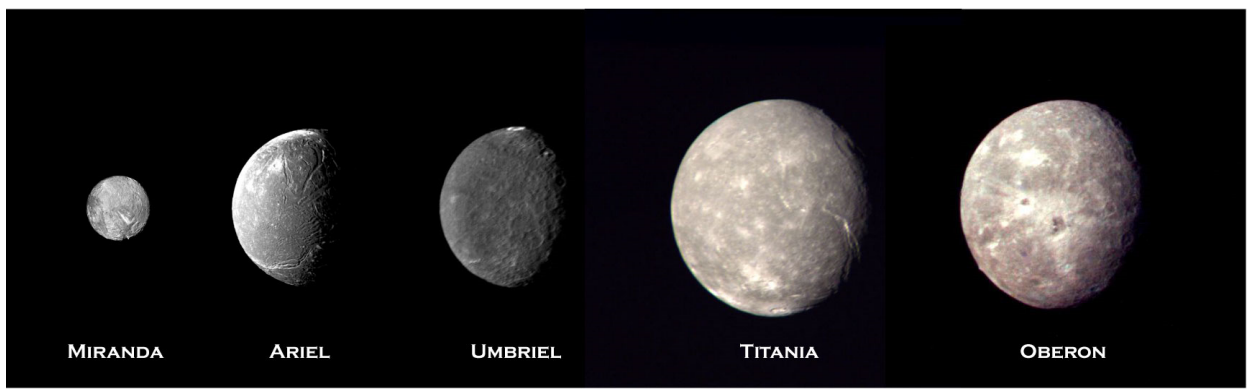

**Figure 3-10.** Voyager 2 images of the 5 major satellites of Uranus. Satellites are shown with their correct relative sizes. The inner three (left) are black and white images, the outer two are false color. Credit: NASA/JPL

Uranus boasts the most densely packed system of satellites in the solar system, including 13 low-mass inner moons, five large moons, and dozens of irregular satellites thought to represent captured Centaurs, trans-neptunian objects, and comets (Delsanti & Jewitt 2006). The low-mass inner moons may be in a state of dynamical transition, with collisions leading to destruction and re-accretion of the satellites on ~$10^6$ yr timescales (French & Showalter 2012). Similar to the ~12 low-mass inner satellites of Saturn, these innermost satellites were originally predicted to be primitive bodies with uniformly cratered surfaces (Smith et al. 1986). Cassini has shown, however, that the innermost satellites of Saturn instead show a wide range of morphologic and compositional diversity that correlate with several dynamical niches associated with the rings and larger satellites (Thomas et al 2013). Their origins are strongly linked to the rings (Charnoz et al. 2010). It stands to reason that similar correlations will be found in the Uranian system, e.g., the rapid changes in the orbit of the moon Mab and its interaction with the μ-ring (Kumar et al 2015).

Neptune has 14 known satellites, at least eight of which have diameters greater than 40 km. The largest, Triton, has a diameter of 2,707 km and is between the sizes of Europa (3,122 km) and Titania (1,577 km). The next two largest Neptunian moons, Proteus (418 km), and Nereid (340 km), are similar in size to Mimas (396 km). Unlike Mimas, which is the smallest object known to be nearly rounded in shape by self-gravity (Dermott and Thomas, 1988), Proteus is a transitional object that lies between irregular and spherical shapes, with a global figure that is relaxed but surface features that remain unrelaxed (Croft 1992). The shape of Nereid is unknown. Modeling suggests that, following Triton's capture, Neptune's irregular satellites were captured as a result of planet-planet encounters during the late heavy bombardment (Nesvorny et al. 2007). Neptune's inner satellites are driven to crossing orbits by perturbations from Triton that result in high-velocity catastrophic collisions (Goldreich et al. 1989), suggesting that the current inner satellites are either shards left over from this process or a newer generation of satellites that have accreted from the rings and debris disk (Crida and Charnoz 2012).

The five largest moons of Uranus, Miranda, Ariel, Umbriel, Titania, and Oberon (**Figure 3-10**) are comparable in size and orbital configuration to the medium-sized moons of Saturn, but are characterized by larger mean densities (~1,500 kg/m³ vs. ~1000 kg/m³), subjected seasonally to pole-centered solar insolation patterns, and are considerably lower in albedo than their Saturnian counterparts, with the exception of Phoebe and the dark hemisphere of Iapetus. There is a notable lack of current orbital resonances between the Uranian satellites, unlike the situations at Jupiter and Saturn, although endogenic activity could have been driven by resonances in the past (see below). Within the Uranian system itself, Miranda, Ariel, and Umbriel orbit within the magnetosphere while Oberon orbits outside the magnetosphere, leading to differences in the way space weathering affects the surface (Arridge et al. 2014). Titania spends some time inside the magnetosphere and some time outside of it.

All of the major Uranian satellites appear to show evidence for various degrees of cryovolcanic and tectonic resurfacing (Croft & Soderblom 1991). In particular, the two innermost satellites, Miranda and Ariel, have undergone extensive endogenic activity. Intense tidal heating during sporadic passages through orbital resonances is expected to have induced internal melting, potentially triggering the geologic activity responsible for the observed cryovolcanic morphologies and leading to the late resurfacing of Ariel (Tittemore & Wisdom 1988, Tittemore & Wisdom 1989, Tittemore & Wisdom 1990). Ariel (578 km in radius), the brightest of the Uranian satellites, shows a paucity of large craters (Plescia 1987, Strom 1987) and is characterized by cratered plains cross-cut by a series of tectonic scarps, canyons, and ridges. Intriguingly, Ariel's tectonic activity appears to have been accompanied by extensive emplacement of viscous flows (Croft & Soderblom 1991, Jankowski & Squyres 1988, Schenk 1991, Smith et al. 1986, Stevenson & Lunine 1986), suggestive of extrusive cryovolcanism that emplaced a series of smooth units embaying and partially burying craters, surrounding nunataks of cratered plains materials, and filling graben floors with convex deposits with bounding and often medial troughs. Umbriel (radius of 584 km) is similar to Ariel in size, density (~1,500 kg/m$^3$), and semi-major axis (10.5 R$_u$ vs. 7.5 R$_u$), but is more heavily cratered and exhibits a dark, low-contrast, surface with two bright deposits associated with the craters Vuver and Wunda (Croft & Soderblom 1991). The two outermost satellites, Titania and Oberon, have radii of 789 km and 761 km, respectively, and may be large enough to support subsurface oceans generated during past melting events (Hussmann et al. 2006). Titania, the largest of the Uranian satellites, is deficient in craters >100 km as compared to Oberon and exhibits a system of ridges similar to some of Europa's ridged linea that may represent compressional tectonics (Culha et al. 2014). Along with the other Uranian satellites, Titania exhibits a strong wavelength-dependent opposition effect suggesting an open, porous regolith texture (Croft & Soderblom 1991). Oberon, the outermost satellite, is heavily cratered and exhibits multiple generations of scarps and canyons. The youngest units are a dark terrain that may represent cryovolcanic flooding of crater floors and tectonically controlled lows. Miranda (radius ~240 km), similar in size to Saturn's Mimas and Enceladus, is perhaps the most geologically bizarre object in the solar system. Miranda exhibits striking geological structures, including ridges and grooves that may be the result of internal differentiation processes (Janes & Melosh 1988) or the surface expression of large-scale upwelling (Pappalardo et al. 1997). For a detailed review of the post-Voyager view of Uranian satellite geology, see Croft and Soderblom (1991).

The areas of Triton's surface that have been observed are young. Crater counts indicate an average surface age of several tens to hundreds of millions of years (Stern and McKinnon 2000), with some locations possibly as young as a few million years (McKinnon and Kirk 2007). This differs from Pluto, where a wider variety of crater densities indicate terrain ages ranging from a few million years to a few billion years (Stern et al. 2015). Investigating the similarities and differences between Triton and Pluto may shed light on some of the effects of Triton's capture by Neptune. The ~40% of Triton's surface imaged by Voyager showed a large polar cap of nitrogen ice and two dominant geologic terrain types; cantaloupe terrain and smooth undulating plains (Croft et al. 1995). Cantaloupe terrain (**Figure 3-12**) is characterized by regularly spaced cavi, which are quasi-circular shallow depressions tens of kilometers across (Croft et al. 1995). The cavi are cross-cut by ridges and troughs morphologically similar to, but significantly less numerous than, Europa's lineae (Prockter et al. 2005). Cantaloupe terrain may be formed by instability and overturn (diapirism) of Triton's crust (Schenk and Jackson, 1993). Triton's smooth plains are inset by a variety of morphologic features including terraces and depressions filled with smooth material that may represent extrusive cryovolcanic flows (Croft et al. 1995). The smooth plains onlap, or encroach upon, the cantaloupe terrain suggesting that the cantaloupe terrain is older.

While the Voyager flybys revealed Uranus to have a rich and diverse satellite system and Triton to have been subject to a complex history of (ongoing) surface processes, a detailed understanding of their geologic histories and tectonic evolution is prevented by the low resolution and incomplete coverage of imaging. While the highest resolution images of Miranda approach 300 m, the best views of the remaining Uranian satellites were acquired at 1–12 km/line-pair. Many of the most diagnostic geologic units on Miranda, such as the bright albedo chevron (Smith et al 1986), would only fill a few pixels in the best images of the larger satellites. At Neptune, the best images of Triton have a resolution of ~750 m, but the majority of the observed surface was imaged at km-scale resolutions (Croft et al. 1995). Cassini's experience at Enceladus emphasizes the need for high-resolution imaging and a variety of types of instrumentation to resolve fundamental ambiguities in interpreting geologic structures, processes, and history. High-resolution decameter-scale (or better) images of the Uranian and Neptunian satellites will reveal the nature of and interaction between observed morphologic features, illuminate never-before-seen terrain in the northern hemispheres of the satellites, and permit generation of more complete crater distributions to be used in surface age estimates. Comparative planetology, both between the Uranian satellites themselves as well as between the satellite systems of Uranus, Jupiter, Saturn, and Neptune and with Pluto and other KBOs, is an essential tool in the study of solar system formation and evolution. Miranda, in particular, provides a unique and compelling opportunity to understand how small moons can become so active (Castillo-Rogez & Lunine 2012). Ariel's evidence for extrusive viscous cryovolcanism may be unique in the solar system, and of the Uranian satellites Ariel seems to offer the highest potential for geologically recent endogenic activity (Castillo-Rogez & Turtle 2012). Comparative planetology between Triton and other KBOs, including both captured irregular satellites and dwarf planets such as Pluto, will help with understanding the origin and evolution of outer solar system objects in general.

The key instruments required to study the geology of the ice giant satellites include:

- Both narrow- and wide-angle multi-spectral cameras that can provide decameter to hectometer regional-scale mapping and meter-scale images of targeted geomorphologic features. Such cameras would also provide topographic information through stereo photogrammetry, photoclinometry, and limb profile images as well as photometric information from images acquired at low and high phase.
- A near- to mid-infrared hyperspectral imager is essential to understanding the composition and provenance of satellite surface features
- A magnetometer to search for induced magnetic fields, indicative of conductive subsurface oceans
- Radio science to determine the gravitational potential field and tidal Love numbers of the satellites, as well as search for any tenuous atmospheres

### 3.1.9 Determine the Internal Structure of the Major Satellites

Much of what is understood about the interiors of the large satellites of Uranus and Neptune, beyond their bulk densities, has necessarily been the result of modeling. The five largest moons of Uranus—Miranda, Ariel, Umbriel, Titania, Oberon—are similar in size to Enceladus, Mimas, Dione, Rhea, Tethys, and Iapetus at Saturn and are somewhat higher in density on average. As discussed above, larger Triton is thought to be a captured KBO. **Figure 3-11** illustrates the relative sizes, bulk densities, and levels of available tidal heating and observed geologic activity of icy satellites at Jupiter, Saturn, Uranus, and Neptune.

The histories of these satellites hold important clues to the evolution of their planetary systems, for example, whether the Uranian satellites pre- or post-date the event that gave Uranus its high obliquity, and whether Triton swept up some of Neptune's original satellites in the process of being captured. Tidal interactions are also key to satellite evolution; as in other

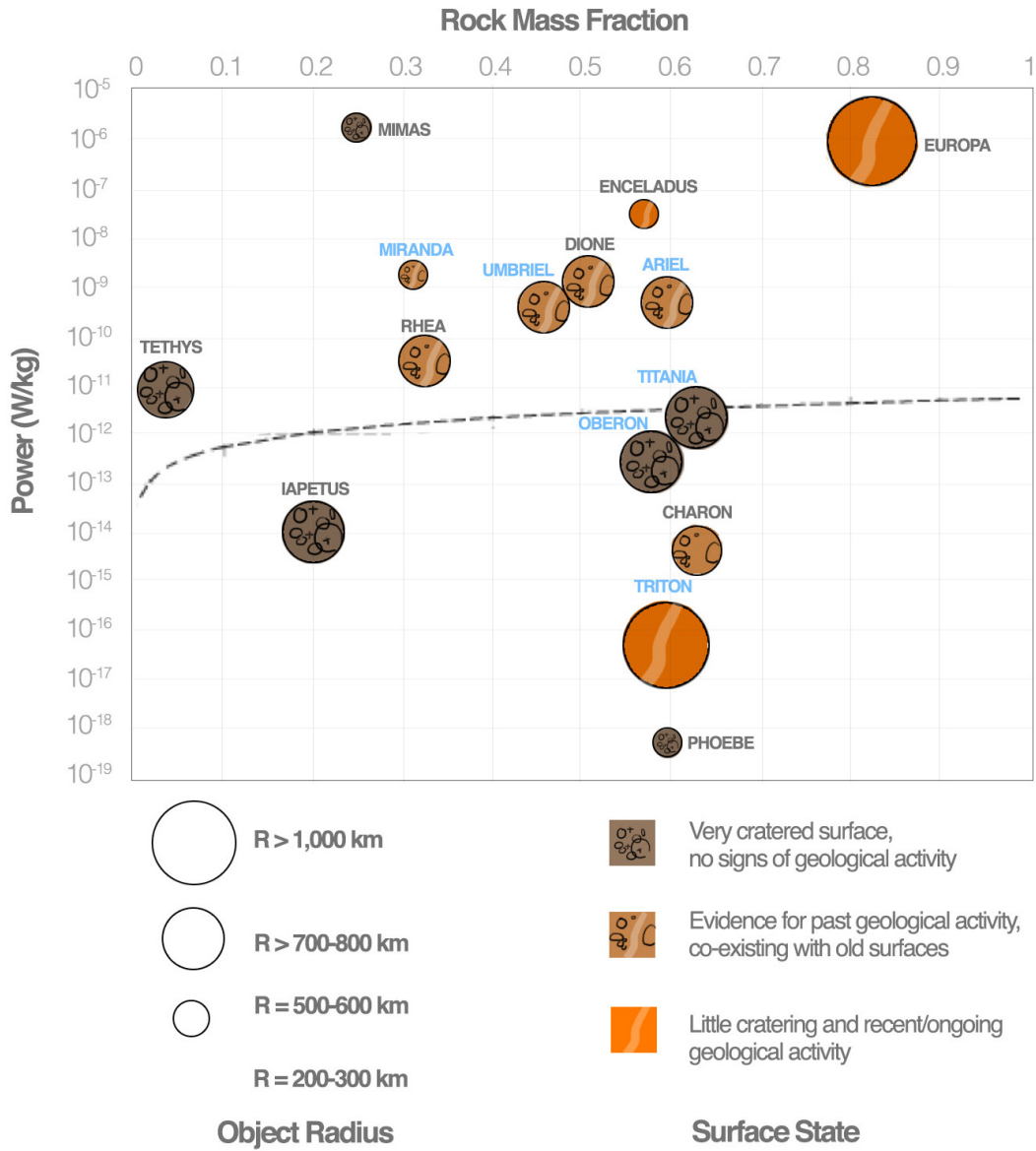

**Figure 3-11.** Icy-satellite rock mass fractions, sizes, and available tidal heating power, with schematic representation of observed surface geology, which can reflect the potential for endogenic activity and the (past or current) presence of a subsurface ocean. Only a sampling of satellites is shown, with ice giant satellites labeled in blue. The dashed line indicates current-day heating due to natural decay of radioisotopes in the rock. Credit: Based on Castillo-Rogez and Lunine 2012, with updated information on Charon from New Horizons data (Castillo-Rogez personal communication)

satellite systems, heating as satellites pass through resonances could have induced internal melting and possibly endogenic geologic activity (Tittemore and Wisdom 1988, 1989, 1990; Tittemore 1990). The largest of the ice-giant satellites—Titania, Oberon, and Triton—might still be able to support subsurface oceans if they had formed in the past (Hussmann et al. 2006; Sohl et al. 2010; Gaeman et al. 2012). More recent modeling (Castillo-Rogez and Turtle 2012) suggests that Ariel, too, could have undergone melting in its past, consistent with its dramatic evidence for endogenic activity including possible extrusive cryovolcanic flows (see also Section 3.1.8). Heating Miranda sufficiently to result in significant melting is difficult, yet the satellite's tortured surface hints at a comparably complex interior structure (e.g., Janes and Melosh 1988; Pappalardo et al. 1997; Marzari et al. 1998).

Of particular interest is understanding how common subsurface oceans are during the evolution of giant planet satellites. Identifying the conditions under which oceans have formed (or were prevented from forming) and where they could still be present is essential to determining how far the habitable zone extends in our own solar system, as well as in exoplanetary systems. Exploring the Uranian and Neptunian satellites to better determine their internal structures will provide important additional data points regarding the prevalence of ocean worlds.

An ice giant mission with multiple satellite flybys would not only reveal fundamental information about their surface geology and composition (discussed in more detail in Sections 3.1.7 and 3.1.8), but also provide key constraints on internal structure to address high-priority science questions regarding the satellite interiors and their histories:

- What were the conditions during satellite formation?
- How are satellite thermal and orbital evolution and internal structure related?
- How have endogenic processes influenced their interiors?
- Are there subsurface oceans and what are their characteristics and histories?
- How do satellites interact with the magnetospheres of their parent planets (and do they generate their own)?
- What energy sources have been available to sustain life?

Measurements to meet this objective include:

- Refined mass values and constraints on internal mass distribution via gravitational moments during satellite flybys;
- Detection of induced (and/or intrinsic) magnetic fields and measurement of their direction and magnitude, which are diagnostic of the presence of a conducting layer (or dynamo);
- Higher resolution imaging to constrain satellite global shape and refine density values; and
- Repeated distant imaging of satellites at different true anomalies to constrain libration, Cassini state, pole position, orbital motion, tides, and orbital parameters.

In addition, laboratory work to better understand material behavior and rheological parameters under appropriate conditions is crucial to improving models of satellite evolution.

### 3.1.10 Characterize Triton's Atmosphere

Triton is the largest moon of Neptune and the only large moon in the solar system with a retrograde orbit. The fact that Triton's orbit is retrograde and highly inclined suggests that it is a captured moon presumed to have originated in the Kuiper Belt (McKinnon 1984; Goldreich et al. 1989; McKinnon & Leith, 1995; Agnor and Hamilton 2006). This means that Triton is likely to have formed in the protosolar nebula (PSN) and should have a composition similar to Pluto and to objects in the Kuiper Belt (KBOs). Triton's bulk density of 2.07 $g/cm^3$ is larger than the bulk densities of Pluto (Nimmo et al. 2016) and the icy moons of Uranus and Saturn, suggesting that Triton has a larger rock/ice ratio (Lissauer et al. 1995), which is potentially indicative of a loss of volatiles as a result of Triton's capture. Triton's surface is made up primarily (~55%) of $N_2$ ice, with water and $CO_2$ ices making up the remaining ~45%. Trace amounts of $CH_4$ and CO have also been detected on its surface as well as in the atmosphere (Lellouch et al. 2010).

The current state, origin, and evolution of Triton are of great interest, but challenging to understand due to its unique formation history and complex seasonal variations. In order to be captured, Triton must have slowed significantly due to a collision (Goldreich et al. 1989) or have been part of a binary system that dissociated when Neptune drew close (Agnor & Hamilton 2006). Triton's initial orbit would have been highly eccentric and would have gradually become circular. In the process, Triton might have collided with many of Neptune's original small moons. A significant portion of its mass (>20%) could have accumulated from such collisions (Ćuk & Gladman 2005), thus changing the bulk composition from a primordial value similar to KBOs to a combination of KBOs and the small moons of Neptune. Tidal heating during the decay of Triton's orbit would have led to differentiation of the interior and could have significantly impacted the volatile evolution of Triton.

In the present era, Triton's orbit (with which Triton's equator is almost exactly aligned) is tilted 23° from the plane of Neptune's equator. Neptune's equator is tilted 29.6° from Neptune's orbital plane. This geometry leads to extreme changes in season for Triton, where the subsolar point reaches high latitudes during northern and southern summers. Migration of volatiles across the surface due to seasonal variation in insolation would have a major impact on atmospheric dynamics and evolution (e.g., Trafton 1984). Observations comparing Triton's surface frost temperature during the Voyager flyby in 1989 with ground-based measurements in 1997 suggest that the atmospheric pressure doubled in ten years due to seasonal variations (Elliot et al. 1998). One of the most exciting discoveries at Triton was finding at least four active plumes (**Figure 3-12**) that introduce several hundred kg/s of $N_2$ from the surface into the atmosphere (Soderblom et al. 1990). A study of surface aeolian features suggested seasonal control of plume activity related to the subsolar latitude (Hansen et al. 1990). The energy source for these plumes has been attributed to solar-powered sublimation of $N_2$ ice (Kirk et al. 1990), although the lack of active plumes observed on Pluto and the existence of water plumes from Saturn's moon Enceladus (Dougherty et al. 2006) have provided reasons to reexamine this hypothesis (Hansen and Kirk 2015).

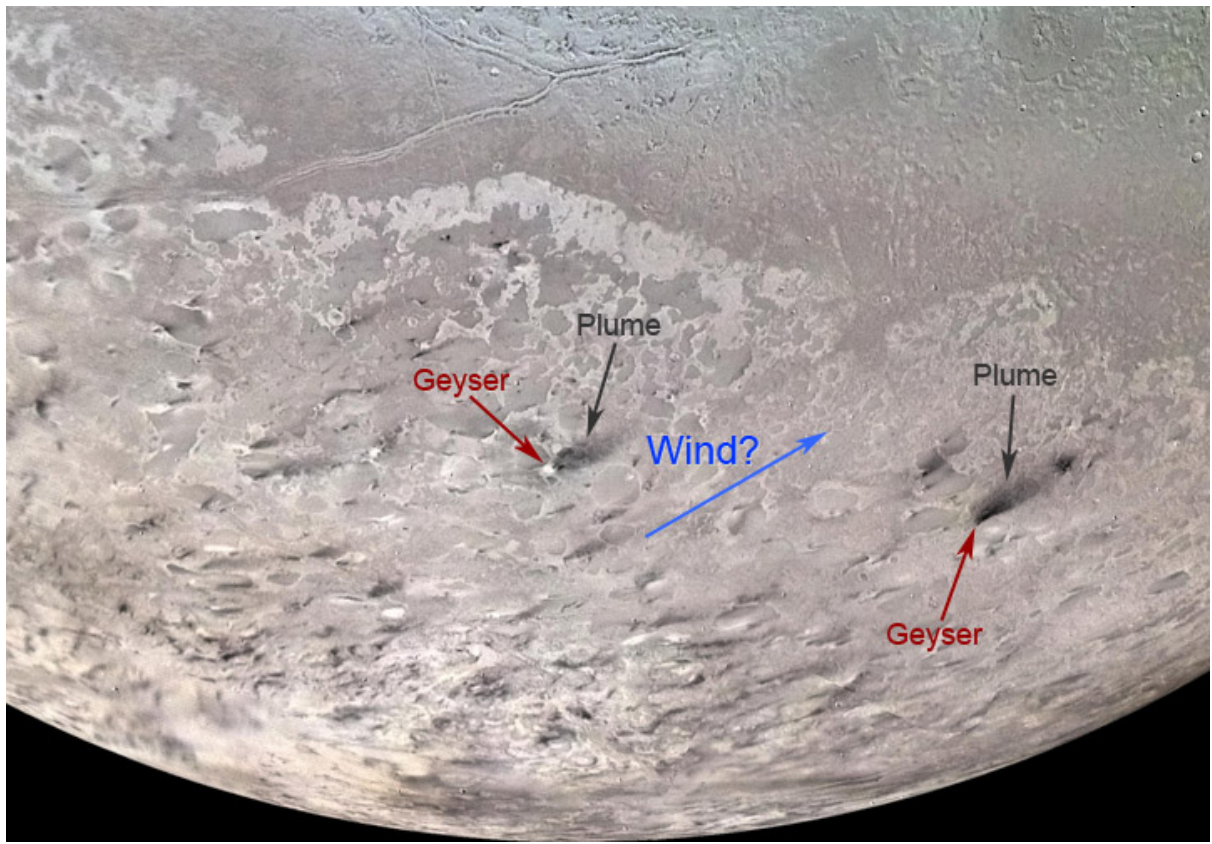

**Figure 3-12.** Mosaic of Triton's south polar region taken by *Voyager 2*, whose limited exploration of Triton revealed a complex interplay of geologic processes and a surprisingly young surface. Even more intriguingly, dark streaks were observed to originate and extend ~100 km downwind from narrow, vertical columns of material ~8-km high, thought to be the result of geyser-like eruptions (Soderblom et al. 1990), the driven energy for which remains the subject of on-going debate (e.g., Hansen and Kirk 2015). Triton's cantaloupe terrain (Section 3.1.8) is seen near the top of the image. Credit: NASA/JPL/USGS

Triton's atmosphere is composed primarily of $N_2$ with trace amounts of CO and $CH_4$ (Tyler et al. 1989; Broadfoot et al. 1989; Strobel et al. 1990; Elliot et al. 1998; Lellouch et al. 2010). Photochemical modeling suggests the presence of hydrocarbons and nitriles, as well as loss of N, H, C and O by escape processes (Krasnopolsky & Cruikshank 1995). Beyond this, the composition, dynamics and chemistry of Triton's atmosphere are poorly understood. What little is known suggests that Triton's atmosphere is very similar to Pluto's, presenting an opportunity for comparative planetology studies as a follow up to the New Horizons mission.

Significant efforts have been made to model the migration of volatiles across the surface of Triton, the dynamics driving the plumes, and the chemistry influencing the composition of the atmosphere. However, observations for constraining these studies are very limited. Therefore, the highest priority questions about Triton's atmosphere that could be addressed by a future mission are:

- What is the composition of Triton's atmosphere?
- How does Triton's atmosphere vary spatially and temporally?
- What is the origin of Triton's volatiles and what can they tell us about the origin of Triton and of Kuiper Belt Objects?

The measurement objectives that would address these questions include:

- Measure the composition of Triton's atmosphere, including $N_2$, $CH_4$, CO, hydrocarbons, nitriles and stable isotope ratios D/H, $^{14}N/^{15}N$, $^{12}C/^{13}C$, $^{16}O/^{17}O$ and $^{16}O/^{18}O$
- Measure the vertical profile of aerosols, aerosol scattering properties, and velocities of atmospheric features

Observations of Triton's atmosphere during a future mission are best done with a combination of in situ and remote sensing techniques. In situ measurements of atmospheric composition could be made using a similar approach to what has been done at Titan with Cassini. These observations have provided unprecedented understanding of the composition (e.g., Magee et al. 2009; Cui et al. 2009), chemistry (e.g., Vuitton et al. 2006; 2007; De la Haye et al. 2007; Mandt et al. 2012a), dynamics (e.g., Yelle et al. 2008; Bell et al. 2010a, 2010b, 2011, 2014), and origin and evolution (e.g., Mandt et al. 2009, 2012b, 2014) of Titan's atmosphere. However, in situ observations are limited spatially and temporally, and some constituents are difficult to measure due to instrument effects (e.g., hydrogen cyanide; Magee et al. 2009). Remote sensing of Titan's atmosphere in the ultraviolet (UV) has provided observations with greater spatial coverage than can be obtained with a mass spectrometer (Stevens et al. 2015) and has provided constraints on constituents that are difficult to measure in situ due to instrument effects (Koskinen et al. 2011; Kammer et al. 2013). The best observations to date of Triton's atmosphere were obtained by Voyager through remote sensing in the ultraviolet (Broadfoot et al. 1989), and UV observations have proven valuable for understanding the atmosphere of Pluto (Gladstone et al. 2016). Therefore, the best approach for addressing the measurement objectives for Triton's atmosphere is a combination of in situ and remote sensing observations.

### 3.1.11 Characterize the Magnetosphere

The magnetized plasma environments of the ice giants are highly irregular and very different from all other magnetospheres in the solar system (Bagenal 1986, 1992; Arridge 2015), though they may be common among exoplanets. At both Uranus and Neptune, the angle between the planetary magnetic dipole axis and the planetary rotation axis is larger than that of any of the other magnetized planets (~59° and ~47° respectively) (Ness et al. 1986, 1989; Connerney et al. 1987, 1991; Holme and Bloxhom 1996; Herbert 2009), which leads to dramatic magnetospheric dynamics in both cases (**Figures 3-13** and **3-14**). However, differing planetary obliquities and the presence of Triton in the Neptunian system create important differences between the two systems.

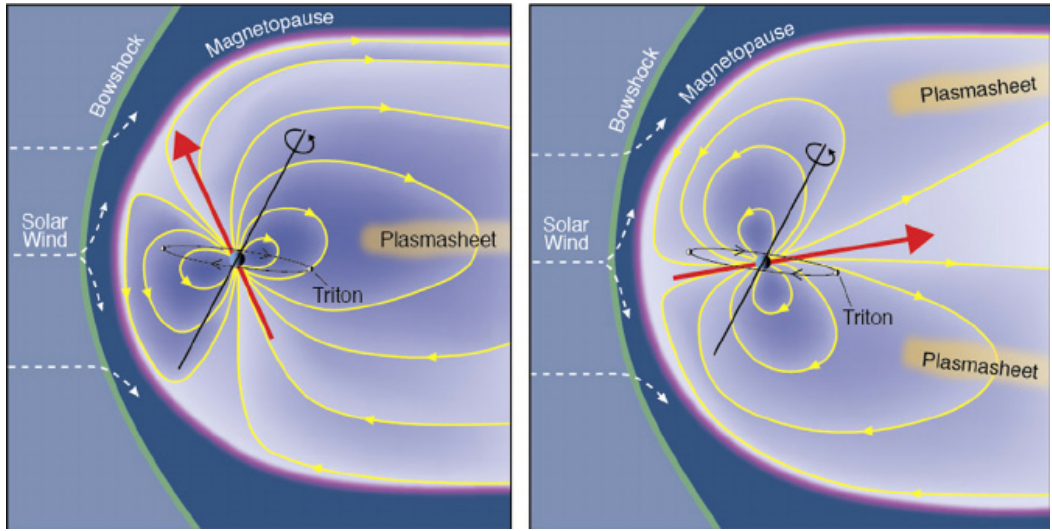

**Figure 3-13.** Diurnal variation of ice giant magnetospheric geometry with respect to the solar wind; schematic of the changes over half a planetary rotation. The case of Neptune during the Voyager 2 encounter of 1989 is shown. The Sun is to the left in both panels, the planet's rotation axis is shown in black, the planetary magnetic dipole axis is shown in red, and magnetic field lines are shown in yellow. This orientation produces a complex interaction between the solar wind and the magnetosphere as the planet rotates. It changes from a more "closed" configuration (left), where the planetary field lines block the solar wind, to a more "open" configuration (right) where the planetary field can connect to the solar wind. The fields of the terrestrial and gas giant planets are always in the "closed" configuration. Uranus' magnetosphere will have a similar geometry in the late 2030s when Uranus is between solstice and equinox. Credit: Steve Bartlett and Fran Baegenal

Present understanding of how these magnetospheres work is very limited, based only on the single-point in situ measurements made during the Voyager 2 flyby (Stone and Miner 1986, 1989) and results of numerical modeling (e.g., Cao and Paty 2015; Mejnertsen et al. 2016). Progress in this area is not only essential for understanding energy flow through the space environments surrounding the ice giants and for improving our understanding of fundamental space plasma processes, but is also essential for efforts to reveal the origin of the planetary magnetic dynamos and interior structure (see Sections 3.1.1 and 3.1.3), for determining auroral energy input to the atmosphere (a possible explanation for the unusually high upper-atmospheric temperatures seen on all giant planets), for constraining the extent of space weathering of planetary moon surfaces (see Section 3.1.7), and for understanding the possible magnetospheres of exoplanets.

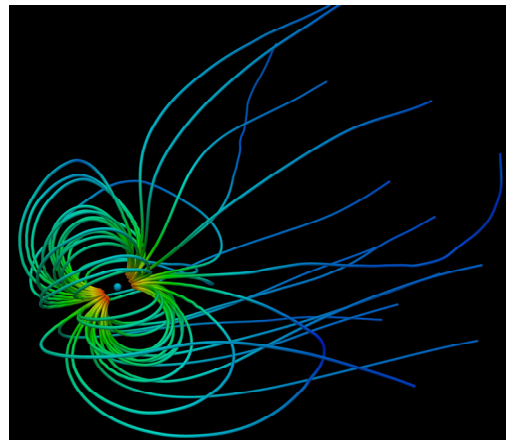

**Figure 3-14.** Recent numerical modeling of Neptune's magnetosphere. The solar wind is incident from the bottom left. Magnetic field lines are colored by the local field strength. Note the complex twisting of field lines down the magnetotail to the right. Credit: Lars Mejnertsen, Imperial College London

The ice giants are exposed to a more tenuous solar wind than any of the planetary magnetospheres found closer to the Sun. This provides a number of unique opportunities to explore the physics of fundamental processes (e.g., magnetic reconnection) in parameter regimes that are inaccessible closer to the Sun (Cowley 2013; Masters 2014, 2015). Despite the low-density and weak embedded magnetic field of the solar wind at both ice giants it has been proposed that energy flow through the magnetosphere may be strongly driven by changes in the solar wind (Vasyliunas 1986; Selesnick and Richardson 1986; Mauk et al. 1987, 1991; Selesnick, 1988, 1990; Schulz et al. 1995; Cowley 2013). Establishing whether this is the case would be a result with broad significance across magnetospheric science.

As introduced above, due to their highly irregular planetary magnetic fields and the large angles between their magnetic dipole and rotation axes, the ice giant magnetospheres are subject to dramatic diurnal and seasonally dependent reconfigurations (e.g., a twisted topology of each planet's magnetic tail: Tóth et al. 2004; Arridge 2015; Cao and Paty 2015; Mejnertsen et al. 2016). However, the nature of the asymmetries present in these magnetospheres and exactly how these systems globally reconfigure on a timescale of hours is essentially unknown. Understanding the dynamic nature of the ice giant magnetospheres is crucial for revealing how mass and energy are transported through each system. Determining the sources and sinks of plasma and dust is a priority, as well as the extent of particle trapping and acceleration to form radiation belts (Mauk et al. 1987, 1991; Cheng et al. 1991; Kempf et al. 2005). The variable interaction with the solar wind may be relevant for this problem (e.g., Vasyliunas 1986), and characterizing the intense plasma wave (e.g., whistler wave) activity in these systems is necessary (e.g., Kurth and Gurnett 1991).

The in situ "snapshot" provided by Voyager 2, ground-based observations, and space-based observations by the Hubble Space Telescope (HST) have provided unambiguous evidence for the existence of auroral emissions at a range of wavelengths at both the ice giants (e.g., Zarka et al. 1995; Ballester et al. 1998; Bhardwaj and Gladstone 2000; Herbert 2009; Lamy et al. 2012). However, detailed information about the auroras is lacking. Note also that the search for exoplanetary auroral radio emissions is ongoing and that a significant fraction of exoplanets detected to date are of similar size to Uranus and Neptune. Observations to date suggest that at the ice giants, both the auroral processes and the auroral response to the solar wind could be radically different from those at other planets (e.g., Lamy et al. 2012; Cowley 2013). How the solar wind, magnetosphere, and ionosphere are coupled together is a major open issue concerning both Uranus and Neptune (e.g., Arridge 2015).

The interaction between planetary satellites, rings, and the surrounding magnetosphere is highly relevant for determining the mass budget of the space environment and also the surface environments of the satellites themselves. The importance of each Uranian and Neptunian satellite as a source of neutral particles and magnetospheric plasma is undetermined. Triton in particular may play a significant role in mass loading Neptune's magnetosphere due to the presence of a nitrogen-rich atmosphere (Broadfoot et al. 1989; Tyler et al. 1989; Richardson et al. 1991). Furthermore, in the case of Triton and also other satellites at both planets, the highly dynamic background magnetosphere produces a class of satellite-magnetosphere interaction that is unique in the Solar System (e.g., Strobel et al. 1990). The combination of large temporal variability in the background magnetic field and a conducting layer at any satellite (e.g., a subsurface ocean, see also Section 3.1.5) could lead to a detectable induced magnetic field that provides information about the conductor (cf., Jupiter's Galilean moons Europa and Ganymede). Sampling the satellite-magnetosphere interaction also has potential to reveal dynamic satellite atmospheres, like that of Saturn's moon Enceladus (Dougherty et al. 2006), which also may be related to subsurface oceans.

Since the publication of the Planetary Science Decadal Survey in 2011, a number of HST campaigns have taken place (Lamy et al. 2012) and a number of studies have considered the complex solar wind-magnetosphere interaction and magnetotail physics at both the ice giants (Cowley 2013; Masters, 2014, 2015; Arridge 2015). In addition, modern supercomputer resources have been applied to the problem of magnetospheric dynamics at Uranus and Neptune (Cao and Paty 2015; Mejnertsen et al. 2016). These recent results are valuable input to future mission design, demonstrating the importance of near-planet solar wind measurements and the need for broad spatial coverage of in situ measurements.

Key questions concerning the magnetospheres of the ice giant planets include, but are not limited to:

- How does the coupled solar wind-magnetosphere-ionosphere system work?
- How does the magnetosphere reconfigure on a timescale of hours and what are the implications?
- What are the sources and sinks of plasma in the system, and how is plasma transported and accelerated?
- How do the planetary satellites interact with the surrounding magnetosphere?

Relevant instruments for ice giant magnetospheric science are all those that measure electric and magnetic fields, detect neutral and charged particles (including dust), and instruments that provide remote observations of the planetary auroral emissions. Of these instruments, a magnetometer, a radio and plasma wave receiver, and a particle instrument that covers low-energy to energetic particles and has energetic neutral atom imaging capabilities are identified as the key instruments. The combination of these three instruments is essential for answering the above key questions; having only one or two of these instruments is insufficient. Furthermore, near-planet solar wind measurements (e.g., during cruise) are essential (see first key question above), and broad spatial coverage in situ measurement opportunities are needed in order to answer all key magnetospheric science questions (e.g., Mejnertsen et al. 2016).

## 3.2 Uranus-Neptune Comparative Planetology

When planning a mission in a cost-constrained environment, the obvious question is "Which ice giant system is more important to explore?" And from a purely scientific point of view, the related question is "Can we fundamentally advance our knowledge of planetary systems by only studying one ice giant?" Regarding the first question, the SDT concludes that Uranus and Neptune are of equal importance, and does not rank one as more important the other. At the highest level, each planet is similar in size and gross composition, and is a potential archetype for the ice giant class of planets. Each planet has a system of rings and satellites with unusual features, evolving dynamically on decadal time scales. Each has a magnetic field with complexity unseen in the terrestrial and gas-giant planets, and a magnetosphere that has unique diurnally and seasonally varying coupling to the solar wind. Both planetary systems can teach us about violent processes that are likely common in planetary formation; Uranus experienced a giant impact creating its 98° obliquity, and Neptune captured Triton with catastrophic consequences for its original satellite system. (See Section 3.1 for discussion and references on all the above points.)

To say that the two planets are equally important is not to say that they are equivalent. The brief Voyager flyby of each, along with Earth-based observations and modeling, have allowed us to see some fundamental ways in which the Uranus and Neptune systems differ. They each have things to teach us that the other cannot, and we answer the second question posed at the start of this section with "Yes, our understanding of planets and planetary systems will be fundamentally advanced by visiting either ice giant." As an example, consider Triton. An orbiter at Neptune offers the opportunity to explore this captured KBO in detail, and leverages the knowledge we are gaining from the recent New Horizons flyby of Pluto, to better understand KBOs. But in the process of capturing Triton, Neptune appears to have lost its larger, native satellites, and perhaps fundamentally disrupted what a "normal" ice giant ring/moon system looks like (see Coradini et al. 2010, Mosqueira et al. 2010, Turrini et al. 2014 and references therein). To explore a native ice giant satellite system, and study how it formed and evolved, one must go to Uranus. Another example involves the internal structures of Uranus and Neptune, which may be different (e.g., Nettelmann et al. 2013). Whether the differing interior structures are the cause or the result of Uranus releasing minimal heat from its interior is an area of active research. Regarding the release of internal heat, the fact that Neptune releases 10x more than Uranus does suggests Uranus' atmospheric energy balance may be dominated by sunlight, while Neptune's may be dominated by heating from the interior, at least at the times of their respective Voyage flybys (see discussion in Section 3.1.4).

The SDT concludes that a mission to either ice giant system can fundamentally advance our understanding of ice giants and processes at work in planetary systems. We affirm the conclusion in the V&V Decadal Survey that an ice giant Flagship-class mission should be flown as a logical and scientifically compelling next step in planetary exploration. We also emphasize that, while the planets are equally and individually compelling, the highest science return will come from an exploration of both systems. Not only does each planet provide information the other cannot, but by comparing the two we see how planets react to differing physical inputs (e.g., sunlight vs. internal heat), and we get a better idea of what properties might be common among ice giant exoplanets.

## 3.3 Model Instruments and Payload

To achieve a realistic ice giant mission design, cost estimate, and assessment of science value, it is necessary to know what instruments are included in the payload and what constraints they place upon the spacecraft. Constraints include the mass, volume, power, and data volume needed by the instruments, as well as requirements on pointing and viewing geometries. In the detailed science discussions above (Section 3.1) reference was already made to the types of instruments that could address each objective, and our Science Traceability Matrix (STM, **Table 3-2**) includes a summary of key measurement objectives and instrument requirements. Here we provide some of the background and details.

The SDT has deliberately chosen to be somewhat generic in its instrument descriptions. This is intended to avoid the appearance of an endorsement of specific instruments at this early stage of the mission life cycle. We also wish to emphasize that the payload packages we discuss, while representing our best judgment of a compelling science package, are intended as models for assessing the science potential of a given mission architecture and for estimating the constraints likely to be placed on the spacecraft and mission design. Our payload selections serve as a useful starting point for future ice giant mission studies, but are not intended to be the final word.

### 3.3.1 Instrumentation

**Table 3-3** lists all instrument types identified in the science discussions as useful to achieving important science objectives. For each instrument, the table identifies objectives for which that instrument is critical, and objectives for which it plays a supporting role. Only our "top 12" science objectives, listed in the STM, are included in the table. The next few paragraphs provide factors the SDT considered for each instrument which are not obvious from the table by itself, or other background information. There is no significance to the order in which instruments are listed.

**Table 3-3.** Instruments identified as addressing science objectives.

| Instrument | Science Objectives for which the Instrument Makes' | |
|---|---|---|
| | Primary Contributions | Secondary Contributions |
| Mass spectrometer (on main s/c) | 11 | |
| Vis/NIR imaging spectrometer | 8, 5, 11 | |
| Thermal-IR bolometer/Vis photometer | 4 | |
| Narrow-angle camera (NAC) | 6, 9, 7 | 10 |
| Wide-angle camera (WAC) | | 10, 6, 9, 7 |
| Radio and plasma waves | 12 | |
| Plasma low-energy particles | 12 | |
| Plasma high-energy particles | 12 | |
| Langmuir probe | 12 | |
| Energetic neutral atoms (ENA) | 12 | |
| Magnetometer (on boom) | 12, 3 | 1 |
| Doppler imager | 1 | |
| UV imaging spectrometer | 12 | 8, 11 |
| Dust detector | | 6 |
| Radar | § | |
| USO (for radio science) | 10, 11 | 1 |
| Microwave sounder | 5 | |
| Mid-IR imaging spectrometer | 8, 5, 11 | |
| Small satellites/CubeSats | Many, see text | Many, see text |
| Lander(s) | Many, see text | Many, see text |
| Probe: Mass spectrometer | 2 | |
| Probe: ASI (density, pressure, temperature) | | 4, 5 |
| Probe: Nephelometer | | 5 |
| Probe: Ortho-para $H_2$ instrument | | 4, 5 |
| Probe: Net-flux radiometer | | 4, 5 |
| Probe: Helium abundance | | 2 |

' The 12 most important science objectives are numbered in the order they are presented in the STM (Table 3-2):
1: Planet Interior
2: Planet bulk composition
3. Planetary dynamo
4: Atmospheric heat balance
5: Planet's tropospheric 3-D flow
6. Structure and variability of rings
7. Inventory of small moons
8: Surface composition of rings and moons
9. Shape and geology of satellites
10: Satellite density, internal structure
11: Triton's atmosphere
12: Solar wind-magnetosphere-ionosphere interactions
§ Instrument discussed by the SDT but it does not directly address one of the 12 priority objectives.

*Mass Spectrometer:* A mass spectrometer could be carried by an atmospheric probe or on board a flyby/orbiting spacecraft, so it appears in two places.

*Vis/NIR Imaging Spectrometer:* See STM for performance requirements.

*Thermal-IR bolometer/Vis Photometer:* This is envisioned as an atmospheric energy balance instrument. The bolometer/photometer combination was chosen as the simplest instrument to achieve that science objective. We note that imaging or spectroscopic capabilities could be added to increase the science return of the instrument; alternatively, a WAC and mid-IR spectrometer could be enhanced to achieve similar functionality. These alternatives were considered when selecting model payloads (Section 3.2.2).

*Narrow Angle Camera (NAC):* The SDT discussed whether both wide-angle and narrow-angle cameras were needed. While useful to have both, only the NAC was considered critical to reach our science requirements for spatial resolution on the satellites and precision tracking of atmospheric dynamics.

*Wide Angle Camera (WAC):* The SDT discussed whether both wide-angle and narrow-angle cameras were needed. While useful to have both, only the NAC was considered critical.

*Radio and Plasma Waves:* See STM for performance requirements. We later refer to the "Plasma Suite" of instruments, which collectively refers to the following 3 instruments: radio and plasma waves; plasma low-energy particles; plasma high-energy particles.

*Plasma Low-Energy Particles:* See STM for performance requirements. We later refer to the "Plasma Suite" of instruments, which collectively refers to the following 3 instruments: radio and plasma waves; plasma low-energy particles; plasma high-energy particles.

*Plasma High-Energy Particles:* See STM for performance requirements. We later refer to the "Plasma Suite" of instruments, which collectively refers to the following 3 instruments: radio and plasma waves; plasma low-energy particles; plasma high-energy particles.

*Langmuir Probe:* See STM for performance requirements.

*Energetic Neutral Atoms (ENA):* See STM for performance requirements.

*Magnetometer (with boom):* See STM for performance requirements. A 10-meter boom is believed necessary for the desired sensitivity based on Cassini (11 m). Actual boom length is driven by the combination of spacecraft electromagnetic cleanliness requirements and the maximum spacecraft magnetic field that can be tolerated at the end of the boom.

*Doppler Imager:* The Doppler imager is representative of a new class of instruments that have the potential to probe giant-planet interiors in a fundamentally new way (Gaulme et al. 2015 and references therein). Based on the same principles used in helioseismology, it looks for planetary-scale oscillations whose temporal and spatial frequencies are diagnostic of interior structure. A Doppler imager measures the velocity of atmospheric aerosols that are entrained in the oscillations, but this instrument could be considered a placeholder for ones that measure adiabatic heating and cooling associated with oscillations, or changes in the shape of an atmospheric surface of constant pressure. Though envisioned as an interior-structure instrument, these measurements on local scales can be diagnostic of smaller-scale weather patterns and motions. In Section 3.3.2, we discuss some of the risks associated with the instrument. Recent work with Cassini observations of Saturn's rings suggests that planetary oscillations have excited detectable waves in the rings themselves (Hedman & Nicholson 2013; 2014). It may be possible to use the rings of Uranus and Neptune in a similar fashion.

*UV Imaging Spectrometer:* See STM for performance requirements.

*Dust Detector:* While useful, it was not considered necessary for any of our top-priority objectives.

*Radar:* While useful, it was not considered necessary for any of our top-priority objectives.

*Ultra-Stable Oscillator (USO):* The USO, used in radio-science experiments, would be a primary instrument for studying the mass and radial density distribution of satellites and the atmosphere of Triton (both "top 12" objectives). The USO also aids in precise measurements of a planet's gravitational field, which traditionally has been an important way to study the interior structure of the giant planets. We list the USO has a "secondary" instrument for interior studies, however, because the SDT determined that a seismology instrument like the Doppler imager has the potential to provide revolutionary advances in our understanding of planetary interiors, while additional gravity measurements, particularly at Uranus, would be more incremental in nature and would not be likely to distinguish among currently competing models for the interiors. The other secondary science objectives of the USO are to improve measurements of the planet's atmosphere and rings during occultations.

*Microwave Sounder:* A microwave instrument is one of the few remote sensing techniques that can probe the deep troposphere (pressures > 10 bars). Current designs for deep sounding (similar to the Juno mission) do require relatively large antennas.

*Mid-IR Imaging Spectrometer:* See STM for performance requirements.

*Additional Flight Elements (Small satellites, CubeSats, Landers, Atmospheric Probes):* These were discussed in the context of instruments, as they take resources that would otherwise be devoted to a traditional instrument. For CubeSats, some of the uses discussed include: releasing on the order of 10 magnetometer-equipped CubeSats from a flyby or orbiting spacecraft to create a 3-D snapshot of the planetary magnetic field; sending many precision-trackable CubeSats between the rings and atmosphere (where the risk of ring-particle collisions may be high) to map the gravity field; and releasing a CubeSat before approach to act as a "pathfinder", arriving ahead of the main spacecraft to measure conditions necessary for safe operation of the main spacecraft (e.g., upper atmospheric properties for aerocapture). For small satellites, we discussed releasing one or more to make close flybys of the natural satellites.

Regarding landers, we recognize that they could achieve a wide range of satellite science objectives. In early discussions, however, the SDT decided not to explore detailed lander designs. The main reason for this is that the SDT feels it is unlikely a mission could accommodate both a lander and an atmospheric probe, and the probe is the higher-priority science. We also noted the difficulty in designing a lander and picking a target for it, given how little we know of the ice giant satellites.

Turning finally to probes, we discussed both extremely shallow probes (to pressures <100 mbar), and very deep ones (wondering if we could reach 100's of kbars in order to probe a possible ionic ocean). After informal discussions with engineers, it was decided that extremely deep probes—if even possible—would likely consume all available resources and be the sole instrument of the mission. The SDT felt the science return from such a probe did not warrant the expense at this stage of our study of ice giant systems. Similarly, it was felt that extremely shallow probes (to micro or perhaps 1-mbar pressures), while of scientific interest, are not a high priority at this time. Finally, the resources needed to get a probe to ~100 mbar are not much less than that which is required to get a probe to several bars, so a deeper probe is preferred. Therefore, we chose to consider a probe which collects data in the 0.1 to 10 bar pressure region of the atmosphere.

Of the six probe instruments listed in **Table 3-3**, five have already been flown on probes, such as Galileo and Huygens. A new instrument we have considered is a device to measure the relative amounts of ortho- and para-hydrogen in the atmosphere (Banfield et al. 2005). Ortho- and para-hydrogen describe the quantum mechanical state of the molecule (whether the two nuclear spins are aligned or anti-aligned, respectively). This ratio is temperature-sensitive, but the time needed to equilibrate with ambient conditions is long, meaning a measurement of the ratio tells us about the temperature-history of the air parcel. This can shed light on vertical convection in the atmosphere. Knowing the ortho-para ratio is also important because the heat capacity of the two forms of $H_2$ is different, and the ratio affects the atmospheric temperature profile.

Of the probe instruments listed, the mass spectrometer is by far the highest priority. It is the only instrument that can meaningfully address our goal of determining the bulk composition of the planet, particularly the noble gases and key isotopic ratios. The second priority on-board the probe is the ASI instrument, providing fundamental atmospheric structure information which is useful to all other atmospheric studies. The remaining instruments are all lower priority than the mass spectrometer and ASI. We do not see strong arguments for one being more important than the other, though we did choose an $H_2$ ortho-para instrument for our model payload (discussed further in Section 3.3.2).

For future reference, we note that the instruments required to address our highest-priority science (the interior structure and bulk composition of the planets) are the Doppler Imager and the probe mass spectrometer. The magnetometer and USO provide supporting information for those objectives. We also note that the instrument that addresses the widest range of science objectives is the NAC, followed closely by the imaging spectrometers (Vis/NIR, UV, and mid-IR), the magnetometer, and the USO.

Having identified the most important types of instruments, we also need estimates of the mass and power needs of each. Wherever possible we based our estimates on existing instruments that have flown, though we often modified those numbers to account for additional capabilities we desired or for the different conditions expected at the ice giants. Details are provided in **Table 3-4**.

**Table 3-4.** Assumed instrument mass and power.

| Instrument | Mass (kg) | Power (W) | Flight-Qualified Analog | Other Analogs Considered | Comments |
|---|---|---|---|---|---|
| Mass spectrometer (on main s/c) | 16 | 19 | DFMS/Rosetta | INMS/Cassini | The less capable INMS (10 kg) is a fallback. |
| Vis/NIR imaging spectrometer | 16.5 | 8.8 | OVIRS/OSIRIS-REx | Ralph-LEISA/NH | Add imaging capability and longer IR wavelengths to OVIRS. |
| Thermal-IR bolometer/Vis photometer | 12 | 25 | DIVINER/LRO | E-THEMIS/Europa | The mass and power levels chosen are well above what a minimal bolometer/photometer instrument would need. DIVINER is a close analog: eliminate its scan capability, add imaging. Desire longer wavelengths than E-THEMIS. |
| Narrow-angle camera (NAC) | 12 | 16 | LORRI/NH | EIS/Europa | Desire filters and pushbroom capability, making EIS (Turtle et al. 2016, LPSC) the better analog. |
| Wide-angle camera (WAC) | 4 | 10 | MDIS-WAC/MESSENGER | | Overestimates mass and power: MDIS has extra components. |
| Radio and plasma waves | 5.6 | 2.7 | LPW/MAVEN | Waves/Juno | Includes 2 kg for a second antenna. Prefer the more capable Waves (12.7 kg, 8W). |

| Instrument | Mass (kg) | Power (W) | Flight-Qualified Analog | Other Analogs Considered | Comments |
|---|---|---|---|---|---|
| Plasma low-energy particles | 3.3 | 2.3 | SWAP/NH | JADE/Juno | Prefer the more capable JADE (17.5 kg. 10.6W). |
| Plasma high-energy particles | 1.5 | 2.5 | PEPPSI/NH | JEDI/Juno or JENA/JUICE | Prefer the capabilities of JEDI (6.4 kg, 4.7W). |
| Langmuir probe | 1 | 0.1 | RPWS-LP/Cassini | LPW/MAVEN | Used RPWS-LP as analog. MAVEN is 1 kg, 2.7W, but that includes an EUV instrument. |
| Energetic neutral atoms (ENA) | 6.9 | 3 | INCA/Cassini | HENA/IMAGE | Used INCA. HENA/IMAGE is 12.9 kg, 9.9W. |
| Magnetometer (on boom) | 14 | 10 | MAG/Cassini | MAG/ Juno or MAG/ JUICE | Includes 10-m boom at 5 kg. 9 kg for the magnetometers seems high, expect it could be reduced by 3 to 5 kg. For Team-X, used Galileo magnetometer at 4.7 kg (no boom) and 8 W. |
| Doppler imager | 20 | 20 | None | ECHOES/ JUICE | Since Uranus is fainter than Jupiter and the instrument is low maturity, increased mass by 6kg over ECHOES proposal. |
| UV imaging spectrometer | 7 | 10 | Alice/NH or UVS/Juno | UVS/Europa | Used UVS/Europa as the analog, as may want more capability than Alice (4 kg) and less than UVS/Juno (21 kg). |
| Dust detector | 5 | 6.5 | SDC/NH | CDA/Cassini or SUDA/Europa | Used SDC/NH values. CDA is 17 kg and 1–17W, SUDA is 4 kg. |
| Radar | 21 | 50 | MARSIS/MarsExp | SHARAD/MRO or REASON/ Europa | Used MARSIS. SHARAD is 17 kg, 25W. REASON is 32 kg, 55W. |
| USO (for radio science) | 2 | 2 | Many | | JUICE USO is 2 kg and 6 W. |
| Microwave sounder | 42 | 33 | MWR/Juno | | |
| Mid-IR imaging spectrometer | 6.3 | 10.8 | OTES/OSIRIS-REx | Ralph/NH | Used OTES for its longer wavelengths. Ralph is 10.3 kg, 6.3W. |
| Probe: Mass spectrometer | 17.4 | 66 | Proprietary proposal | Galileo, Huygens | Peak power is listed. Average 50W. Galileo Probe Mass Spec was 13.2 kg, 25 W. Huygens probe GCMS was 17.3 kg, 41 W avg. |
| Probe: ASI | 2.5 | 3.5 | Galileo (subset of instruments) | | Team-X used 2.5 kg, 10 W average power, 7 W standby. Includes wind speed and direction, temperature, pressure, humidity, acceleration. |
| Probe: Nephelometer | 2.3 | 4.6 | Galileo Probe | | Team-X report says 4.4 kg, 11 W avg 5 W standby. |
| Probe: Ortho-para $H_2$ | 1 | 3.5 | None | | Peak power is listed, average ~1W. Mass does not include boom to place it outside the boundary layer. Based on Banfield et al. 2005. Team-X used 4 W average, 1 W standby. |
| Probe: Net-flux radiometer | 2.3 | 4.6 | Galileo Probe | | |
| Probe: Helium abundance | 1.4 | 0.9 | Galileo Probe | | |

### 3.3.2   Model Payloads

Model payloads were assembled for the purpose of understanding mass, power, and data volume demands placed upon the spacecraft, as well as understanding operational scenarios (e.g., when must instruments be turned on as opposed to downlinking data to Earth, or the need to duty-cycle instruments). Model payloads are also necessary to rank the science return from a mission architecture. As mentioned at the start of Section 3.2, we have attempted to select payloads that optimize the science. The mission that ultimately flies, however, may have different ground-rules than this study or may benefit from new technology or new information about the ice-giant systems, any of which should trigger a re-thinking of the optimal payload.

During our A-Team study (the A-Team methodology is described in Section 2.3.1), the SDT was asked to select model payloads in three mass classes (not-so-surprisingly referred to as small, medium and large), for missions with and without an atmospheric probe. We also considered whether the primary spacecraft was a flyby or orbiter, and whether the target was Uranus or Neptune. That led to consideration of 24 different payloads. It quickly became apparent in our discussions, however, that the payload we desired to have on the main spacecraft was the same regardless of whether or not it was a flyby or obiter, or whether or not a probe was carried. With two exceptions (discussed later), the payloads were also the same whether we flew to Uranus or Neptune. This led to the welcome simplification of considering just three model payloads, distinguished by their total mass. The mass targets changed over time, as we gained experience with the capabilities of launch vehicles and the masses of the individual instruments, but ultimately we settled on approximately 50, 90, and 150 kg for our science payloads, not including any probe (which is the highest-priority additional flight element).

### Small (50 kg) Science Payload on the Main Spacecraft

The SDT feels a 50-kg payload represents the science floor for a Flagship mission. A payload of that size is needed to touch upon science questions across multiple elements of an ice-giant system (interior, atmosphere, rings, satellites, magnetosphere), and to have some capacity for serendipitous discovery in this relatively unexplored corner of our solar system. The model payload chosen for this mass-level is

- Doppler Imager (analog for any atmospheric seismology instrument)
- NAC
- Magnetometer

The Doppler Imager was selected because of its potential to dramatically advance our understanding of giant planet interiors, one of the two highest-priority science objectives we identified. Furthermore, it probes the interior on approach to the target planet, therefore giving it the same diagnostic abilities on a flyby mission as on an orbiter. That allows it to perform ground-breaking science observations during any gravity assists at Jupiter or Saturn, potentially providing a head-to-head comparison of a gas- and ice giant planet. Another argument in favor of a Doppler Imager-type instrument is that the traditional way to address the high-priority interior structure question, with gravity data, may be problematic given that a) The ring impact hazard may prevent the main spacecraft from flying close enough to measure higher-order gravitational moments, and b) It is not clear that additional gravity data can distinguish between currently competing models of Uranus' interior, though it may do so for Neptune. A final argument in favor of putting a Doppler imager on our model payload is that we considered it a good "stress test" of mission architectures, because it generates a large data volume (operational details are discussed in Section 4).

There are several significant risks associated with a Doppler imager-type instrument, however, which must be assessed before selecting it for any actual ice-giant flight opportunity. The one easily addressed is the TRL level (currently 6) which—while a common level for a proposal—is the lowest for any instrument considered for the main spacecraft. More problematic is that while oscillations have likely been detected on Jupiter (Gaulme et al. 2011) and Saturn (Hedman and Nicholson 2013), we do not know if the oscillation amplitudes on Uranus or Neptune will be detectable, and their excitation mechanism is not well-enough understood to even make an accurate prediction from what we see on the gas giants. Furthermore, unlike Jupiter or Saturn, even with the largest telescopes available it does not seem possible to determine whether the oscillations exist from the Earth or from near-Earth spacecraft (Neil Murphy, personal comm., Gaulme et al. 2015). We are aware of at least one effort (Matt Hedman, personal comm.) to look for evidence of planetary oscillations in Voyager ring data, which could at least partially mitigate this risk.

Overall, the SDT feels the potential benefits outweigh the risks, and an atmospheric seismology instrument is included on our 50-kg model payload. Because it is a relatively large, power- and data volume-hungry instrument, however, a Doppler Imager can easily be replaced on the payload without taxing the spacecraft design. If necessary, our team would replace it with a Vis/NIR imaging spectrometer, or a mid-IR imaging spectrometer. Either of those instruments would actually increase the number of our key science questions addressed by the payload, but the payload then might not address our highest-priority investigations. This question of weighing the quantity vs. quality of science objectives will be discussed further in Section 3.4.

### *Medium (90 kg) Science Payload on the Main Spacecraft*

A 90-kg payload is able to at least touch upon all 12 of our primary science objectives, and is therefore the threshold we would like to see considered for an ice giant Flagship mission. The model payload chosen for this mass-level includes the three instruments of the 50-kg case, and adds:

- Vis/NIR imaging spectrometer
- Thermal IR/visible instrument
- Plasma suite (Radio, low-energy, and high-energy plasma instruments)
- Mid-IR spectrometer for Uranus, UV spectrometer for Neptune

The thermal-IR instrument envisioned here is not a bare-bones energy-balance instrument as discussed in Section 3.3.1 (an IR-bolometer paired with a visible-wavelength photometer), but is instead a more capable instrument as described in **Table 3-4**, with imaging capabilities and several broad-band filters.

Both the mid-IR spectrometer and the UV spectrometer are good candidates to round out the 90-kg payload, whichever planet is targeted. Both will contribute to atmospheric studies and to constraining the surface composition of rings and satellites. The UV spectrometer will be good for auroral studies and the thinnest or highest-altitude atmospheres, while the mid-IR targets thicker/deeper atmospheres. The SDT does not have a strong preference for one or the other, but suggests using the mid-IR spectrometer at Uranus, and the UV spectrometer at Neptune because the relatively thin atmosphere of Triton is likely a good UV target.

### *Large (150 kg) Science Payload on the Main Spacecraft*

The 150-kg payload is able to make significant progress on all 12 of our primary science objectives, and is well equipped to make the type of unplanned discoveries that made the Cassini mission so productive at Saturn. That makes this payload class the one that we hope to baseline in an ice giant Flagship mission. The model payload chosen for this mass-level includes all the instruments of the 90-kg case, and adds

- WAC
- USO
- ENA
- Dust detector
- Langmuir probe
- Microwave sounder (Uranus), mass spectrometer (Neptune)

With this large a payload, we get close to being able to fly every instrument identified by our team as key for addressing our 12 primary science objectives. The radar was not selected, as its primary science (mapping near-surface geologic structures on satellites), while potentially important, was not picked as one of our primary science objectives. Sub-spacecraft (e.g., CubeSats or SmallSats) were also not selected, the SDT instead choosing to invest the mass allocation in more traditional instruments. (We discuss CubeSats and SmallSats further in Section 5, Technology.) We also note that flying a Juno-analog microwave instrument would put us a bit over the target mass (159 vs. 150 kg), but one could descope the longest-wavelength for a significant mass reduction. A final comment on this payload is that when flying to Neptune, a mass spectrometer for the main spacecraft is preferred over the microwave sounder because of a desire to make in-situ measurements of Triton's atmosphere; the composition, noble gas abundances, and isotopic ratios being important factors to help understand the formation and history of Triton and the Kuiper Belt.

### Probe Payload

For purposes of this study, the SDT selected the following payload for the atmospheric probe:

- Mass spectrometer
- ASI (temperature, pressure, density)
- Nephelometer
- $H_2$ ortho-para sensor

Only the mass spectrometer and ASI are considered "must have" instruments. The mass spectrometer comes first because it is the only instrument that can address one of our two highest-priority science objectives: to determine the bulk composition of the planet including noble gases and isotopic ratios. ASI is important because it provides fundamental information that helps interpret virtually all other atmospheric measurements (including the mass spectrometer). If a probe can accommodate more than those two instruments, any or all of the remaining 4 listed in **Table 3-4** is a viable candidate. The SDT had a minor preference for the nephelometer and ortho-para sensor. There is no special reason the SDT chose to carry a total of 4 instruments on the probe as opposed to 3 or 5. Four seems like an intermediate number, which is more than a minimal probe but still small enough to require some science prioritization and compromises.

### 3.4 Architecture Prioritization

This section first discusses how we prioritized the science potential of a range possible mission architectures, and presents results from our scoring system (Section 3.4.1). In Section 3.4.2, we discuss limitations of our scoring, and the one major area in which the SDT did not reach a consensus (involving the relative merit of dual-planet missions vs. enhanced single-planet missions). In Section 3.4.3, we summarize our conclusions regarding architecture prioritization.

### 3.4.1    Prioritization Approach and Raw Results

At the start of our mission study, during the first A-Team session, we had a brainstorming session to come up with as many possible mission architectures, which we thought had scientific merit, with minimal consideration of cost and technical feasibility.  They involved various combinations of flyby and orbiter spacecraft, large and small sub-satellites, landers and probes, and targeting one or both ice giants.  **Figure 3-15a** summarizes in cartoon form the architectures considered.  We initially distinguished between a simple Uranus or Neptune orbiter, which did not attempt a satellite tour (referred to as "Orbiter" in the figure), and an orbiter which also performs an extended satellite tour ("Orbiter with Moon Tour").  In later work, we assume any orbiter would also perform a moon tour, and drop that distinction.  We also use the term "Dual Orbiter" to refer to a mission that sends two orbiters to the same planet, and "Two Spacecraft, Two Planets" to refer to a mission that sends one spacecraft to Uranus and one to Neptune.  (The two spacecraft, two planets architecture has previously been explored in the framework of ESA's call for the scientific themes of its L2 and L3 missions.  See Turrini et al. 2013, 2014.)

**Figure 3-15b** highlights conceptual multiple-body missions at and beyond the ice giants, but it should be kept in mind that important science can be done on the flight out to the ice giants. Chemical trajectories in particular (Section 4.1) typically incorporate gravity assists at terrestrial planets and Jupiter or Saturn.  A flyby of a gas giant offers an important opportunity for comparative planetology with an ice giant.  This is particularly true for the Doppler Imager

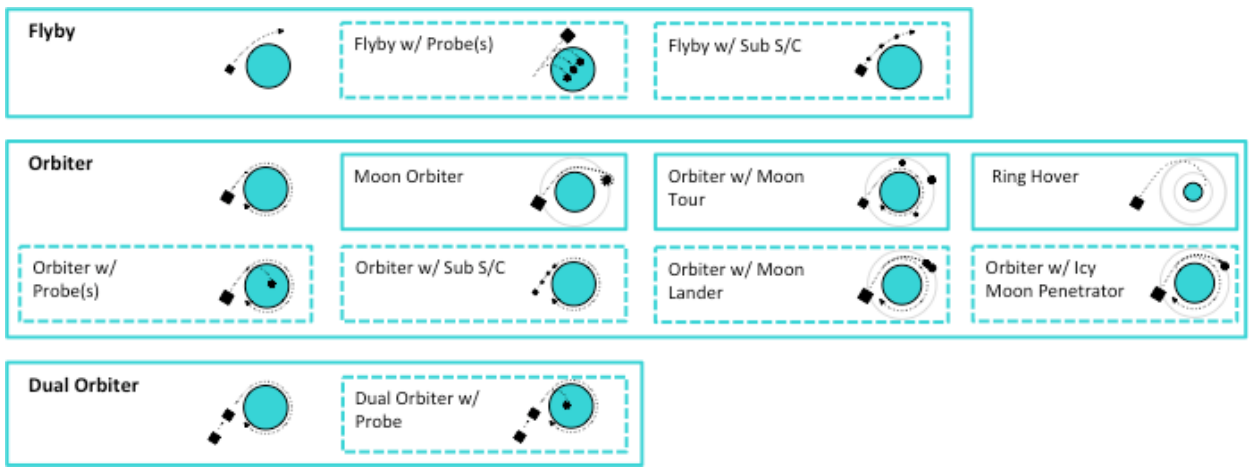

**Figure 3-15a.**  Cartoon summary of conceptual mission architectures considered at the A-Team level, targeting either Uranus or Neptune.  For Neptune missions, the Moon Orbiter, Moon Lander, and Moon Penetrator options would target Triton.  Figure 3-14b considers missions targeting multiple bodies.

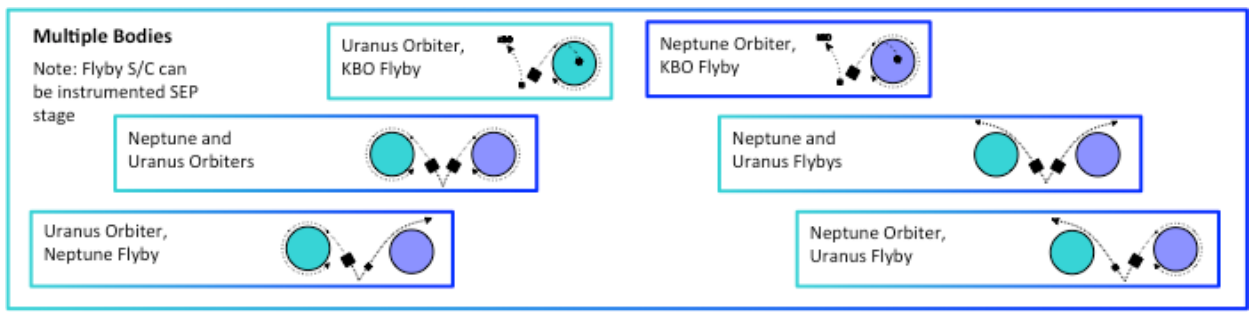

**Figure 3-15b.**  Cartoon summary of conceptual mission architectures considered at the A-Team level, targeting multiple bodies. We do not present here options that flyby Jupiter, Saturn, or inner-solar system objects on their way out to the ice giants, but it should be kept in mind that such encounters do take place in most chemical trajectories, enabling additional science.

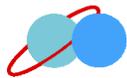

instrument discussed in Section 3.3, which can perform detailed interior-structure measurements during a Jupiter or Saturn flyby. Because these opportunistic science benefits are launch-year and launch-vehicle dependent, we did not consider them when ranking the science value of architectures or payloads. This allows for a clearer comparison of the basic architectures.

The SDT first ranked the 32 mission architectures summarized in **Figures 3-15a and -15b** assuming an unlimited payload. This allowed us to assess the scientific potential of each architecture. We also ranked the architectures assuming each of our 3 payloads described in Section 3.3. The relative ordering of architectures did not change as the payload was altered, though the relative distances between them did. For example, the lowest ranked mission is always a flyby of a single ice giant, and the highest a mission that flies an orbiter and probe to each ice giant (one of our two-spacecraft, two-planet scenarios). The science value of those two missions is much further apart with a large payload, however, than it is with a small one. This makes sense, as a small payload is less capable of leveraging the opportunities that arise from orbiting as opposed to flying by.

The ranking of architectures is done in the following way. For each science objective, an architecture is given a score from 0 to 5. Zero means the objective is not addressed at all and a score of 5 means the objective is fully met (or is met as well as is possible). Intermediate scores rank partial achievement of objectives, for example a score of 1 indicates a minimal amount of relevant information is collected, while a score of 3 indicates significant advances will be made. Scores are agreed upon by consensus among the SDT. The sum of the science scores for an architecture is then its "raw science value." The SDT then reviewed the scores and made some adjustments (described later), to more accurately reflect our consensus assessment of the relative science values. **Table 3-5** shows the raw scores given for the 32 architectures and 3 payload masses, before normalization.

For the scoring of two spacecraft, two planet architectures, the SDT decided to simply add the scores of each element as if it were flown alone. For example, the score of a two-planet mission that places an orbiter at Uranus and does a flyby of Neptune is given a score equal to the score of a Uranus orbiter by itself plus the score of a Neptune flyby by itself. While some science information may be redundant between the two planets (which would make the sum of scores an over-estimation of science value), this is balanced by the increased value in being able to perform comparative planetology studying the significant ways in which the planets differ as well as how they are similar (see Section 3.2).

In our science ranking of architectures, we dropped explicit reference to flybys of KBOs from the list of multi-target missions. The SDT does not want KBO science to influence the architecture chosen for our first dedicated ice giant mission. We would encourage any flyby spacecraft, however, to then target a KBO. We also removed the Ring Hover architecture from our ranking because higher-priority science will be done by missions that do not focus so strongly on the rings. We certainly encourage that ring science be part of any mission flown, and would welcome, for example, a mission that could incorporate aspects of a ring hover as part of its moon tour.

The SDT also chooses to remove architectures with landers, penetrators, and sub-spacecraft (e.g., CubeSats) from the rankings. Landers and penetrators are removed because they require resources similar to that needed by a probe, and the SDT finds a probe is of higher priority. Similarly, architectures with sub-spacecraft are not ranked because the SDT finds it more valuable to allocate mass resources to carried instruments (i.e., the payloads identified in Section 3.3.2).

It should also be noted that at the time we did the ranking, we used 10 science objectives instead of the 12 in our final STM. The "Satellite Inventory" and "Planetary Dynamo" objectives were not explicitly scored. The SDT chose not to go back and re-score, as it was felt interior studies and satellite studies were already given enough weight in the existing score; each of those studies has 3 objectives contributing to the raw score. This highlights one of the limitations of our scoring approach. Listing related science objectives separately (e.g., "Bulk composition," and "Isotopic Ratios," and "Noble gases" vs. just using "Bulk composition" in an inclusive sense) can weigh the score toward a particular discipline. Limitations of our scoring are discussed further in the next section. Due to those limitations, it is important to not over interpret details of the raw scores. After discussing our handling of the limitations identified, we will summarize our consensus ranking of architectures in Section 3.4.3 and **Figure 3-16**.

### 3.4.2  Limitations of Our Ranking System

No scoring system will be perfect, so the SDT reviewed the raw scores, looking for unintended biases. One such bias arose from having a science objective specific to Triton. We already described in Section 3.2 why Uranus and Neptune are equally valuable scientifically, but having an explicit Triton objective gave Neptune missions the potential of reaching higher scores. We could have corrected this bias by adding an explicit objective related to native ice giant satellites, or pointing out that the energy balance objective is likely to be more significant for Uranus. But instead it seemed simpler just to drop the Triton objective from our raw score. (In fact, a careful look at **Table 3-5** will show the Triton objective listed separately and not included in the raw score.) This has the intended consequence of giving missions to Uranus and Neptune identical science scores.

Another aspect of our scoring system to highlight is that it does not give any extra weight to the two science objectives we identify as most important (interior structure and bulk composition). As a result of this, an architecture addressing two lower-priority science objectives is scored higher than an architecture addressing one high-priority objective. We discussed applying a multiplicative factor of 2 or 3 to the score assigned to a high-priority objective, but did not reach consensus on that. Some on the SDT felt it was appropriate to consider breadth of science more important than any single goal, while others felt the highest-priority goals should carry extra weight. While the SDT did not converge on a universal way to score quantity vs. quality of science objectives, we were able to review each architecture and identify the ones that were in dispute and which were not.

The SDT agreed that the relative ranking of all single-planet missions was "correctly" scored. For example, a flyby with probe (addressing our two highest-ranked science objectives) is scored lower than an orbiter without a probe (addressing one high-ranked and many secondary objectives). The SDT also agreed that all two-planet missions that included at least one probe and one orbiter, should be ranked much higher than any single planet mission. The only missions on which we failed to reach a consensus ranking were ones involving flybys of both planets, and two-planet missions not having any probe. Since the two-planet missions that do not have a consensus ranking are not among the missions the SDT recommends, we decided it is unnecessary to rank those 6 (out of 32) architectures.

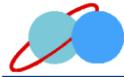

**Table 3-5.** Raw science scores assigned to the 32 conceptual architectures considered during A-Team studies. The rows highlighted in yellow are the raw scores for the three payloads considered. These scores are not our final ranking, see text for details.

| Scale 0 = Does not address / 1 = Poor / 3 = Average / 5 = Great | Uranus Flyby | Flyby + Probe | Flyby + Sub-S/C | Orbiter | Moon Orbiter | Orbiter + Moon Tour + Probe | Orbiter + Sub-S/C | Orbiter + Moon Tour | Dual Orbiter | Dual Orbiter + Probe | Neptune Flyby | Flyby + Probe | Flyby + Sub-S/C | Orbiter | Orbiter + Moon Tour + Probe | Triton Orbiter | Orbiter + Sub-S/C | Orbiter + Moon Tour | Dual Orbiter | Dual Orbiter + Probe | Uranus Orbiter + Neptune Flyby | Uranus Orbiter + Neptune Flyby + Probe | Uranus Orbiter + Neptune Flyby + 2 Probes | Uranus Orbiter + Neptune Orbiter | Uranus Orbiter + Neptune Orbiter + Probe | Uranus Orbiter + Neptune Orbiter + 2 Probes | Neptune Orbiter + Uranus Flyby | Neptune Orbiter + Uranus Flyby + Probe | Neptune Orbiter + Uranus Flyby + 2 Probes | Uranus and Neptune Flybys | Uranus and Neptune Flybys + Probe | Uranus and Neptune Flybys + 2Probes |
|---|---|---|---|---|---|---|---|---|---|---|---|---|---|---|---|---|---|---|---|---|---|---|---|---|---|---|---|---|---|---|---|---|
| Constrain Interior | 4 | 4 | 4 | 5 | 5 | 5 | 5 | 5 | 5 | 5 | 4 | 4 | 4 | 5 | 5 | 5 | 5 | 5 | 5 | 5 | 9 | 9 | 9 | 10 | 10 | 10 | 9 | 9 | 9 | 8 | 8 | 8 |
| Bulk Composition | 0 | 5 | 0 | 2 | 2 | 5 | 2 | 2 | 2 | 5 | 0 | 5 | 0 | 2 | 5 | 2 | 2 | 2 | 2 | 5 | 2 | 5 | 10 | 4 | 7 | 10 | 2 | 5 | 10 | 0 | 5 | 10 |
| Heat Balance | 2 | 3 | 2 | 4 | 4 | 5 | 4 | 4 | 4 | 5 | 2 | 3 | 2 | 4 | 5 | 4 | 4 | 4 | 4 | 5 | 6 | 7 | 8 | 8 | 9 | 10 | 6 | 7 | 8 | 4 | 5 | 6 |
| Tropospheric 3D Flow | 2 | 3 | 3 | 4 | 4 | 5 | 4 | 4 | 4 | 5 | 2 | 3 | 3 | 4 | 5 | 4 | 4 | 4 | 4 | 5 | 6 | 7 | 8 | 8 | 9 | 10 | 6 | 7 | 8 | 4 | 5 | 6 |
| Ring Structure | 1 | 1 | 3 | 4 | 4 | 4 | 5 | 4 | 5 | 5 | 1 | 1 | 3 | 4 | 4 | 4 | 5 | 4 | 5 | 5 | 5 | 5 | 5 | 8 | 8 | 8 | 5 | 5 | 5 | 2 | 2 | 2 |
| Satellite Surface Composition | 3 | 3 | 3 | 4 | 5 | 5 | 4 | 5 | 5 | 5 | 3 | 3 | 3 | 4 | 5 | 5 | 4 | 5 | 5 | 5 | 8 | 8 | 8 | 10 | 10 | 10 | 8 | 8 | 8 | 6 | 6 | 6 |
| Satellite Surface Geology | 3 | 3 | 3 | 4 | 5 | 5 | 4 | 5 | 5 | 5 | 3 | 3 | 3 | 4 | 5 | 5 | 4 | 5 | 5 | 5 | 8 | 8 | 8 | 10 | 10 | 10 | 8 | 8 | 8 | 6 | 6 | 6 |
| Satellite Bulk Density | 1 | 1 | 2 | 3 | 5 | 5 | 4 | 5 | 5 | 5 | 1 | 1 | 2 | 3 | 5 | 5 | 4 | 5 | 5 | 5 | 6 | 6 | 6 | 10 | 10 | 10 | 6 | 6 | 6 | 2 | 2 | 2 |
| Planetary Magnetosphere | 1 | 1 | 2 | 4 | 4 | 4 | 4 | 4 | 5 | 5 | 1 | 1 | 2 | 4 | 4 | 4 | 4 | 4 | 5 | 5 | 5 | 5 | 5 | 8 | 8 | 8 | 5 | 5 | 5 | 2 | 2 | 2 |
| **Science Score, 150 kg payload** | 17 | 24 | 22 | 34 | 38 | 43 | 36 | 38 | 40 | 45 | 17 | 24 | 22 | 34 | 43 | 38 | 36 | 38 | 40 | 45 | 55 | 60 | 67 | 76 | 81 | 86 | 55 | 60 | 67 | 34 | 41 | 48 |
| **Science Score, 90 kg payload** | 17 | 24 | 22 | 32 | 36 | 43 | 34 | 36 | 38 | 45 | 17 | 24 | 22 | 32 | 43 | 36 | 34 | 36 | 38 | 45 | 53 | 60 | 67 | 72 | 79 | 86 | 53 | 60 | 67 | 34 | 41 | 48 |
| **Science Score, 50 kg payload** | 11 | 18 | 16 | 24 | 28 | 34 | 26 | 32 | 30 | 36 | 11 | 18 | 16 | 24 | 34 | 28 | 26 | 28 | 30 | 36 | 43 | 45 | 52 | 60 | 62 | 68 | 39 | 45 | 52 | 22 | 29 | 36 |
| Extra Investigations | | | | | | | | | | | | | | | | | | | | | | | | | | | | | | | | |
| Triton's Atmosphere | 0 | 0 | 0 | 0 | 0 | 0 | 0 | 0 | 0 | 0 | 2 | 2 | 3 | 4 | 4 | 5 | 4 | 5 | 5 | 5 | 2 | 2 | 2 | 4 | 4 | 4 | 4 | 4 | 4 | 2 | 2 | 2 |
| U+N Comparative Planetology & Exoplanets | 0 | 0 | 0 | 0 | 0 | 0 | 0 | 0 | 0 | 0 | 0 | 0 | 0 | 0 | 0 | 0 | 0 | 0 | 0 | 0 | 1 | 1 | 1 | 1 | 1 | 1 | 1 | 1 | 1 | 1 | 1 | 1 |

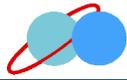

### 3.4.3    Ranking Results

The consensus ranking of architectures is presented graphically in **Figure 3-16**. Consistent with the discussion in previous sections, for a given architecture, the science value of flying to Uranus is the same as flying to Neptune, so we do not distinguish between the two targets. The numeric values plotted in the figure assume our 150 kg payload for the flyby or orbiter spacecraft. Other payload sizes have the same relative ranking (e.g., a flyby is always less valuable than an orbiter), but the smaller payloads have systematically lower values, and slightly smaller differences between the lowest and highest ranked architectures. This is consistent with expectations; a smaller payload not only does less science, it also cannot take full advantage of the opportunities afforded by getting into orbit, making the difference between flyby and orbiting spacecraft smaller.

The SDT recommends an orbiter with probe as the next ice giant Flagship mission, with the orbiter having a payload mass of at least 90 kg, preferably 150 kg. This is consistent with the recommendation in V&V, and achieves both the highest priority science (interior structure and composition), and is able to achieve some science in all other disciplines (atmosphere, rings, satellites and magnetosphere) by virtue of the time it spends in the system. Maximizing time within an ice giant system also enables serendipitous discoveries and follow-up which the Cassini mission at Saturn has demonstrated to be extremely valuable (e.g., the plumes and sub-surface ocean at Enceladus).

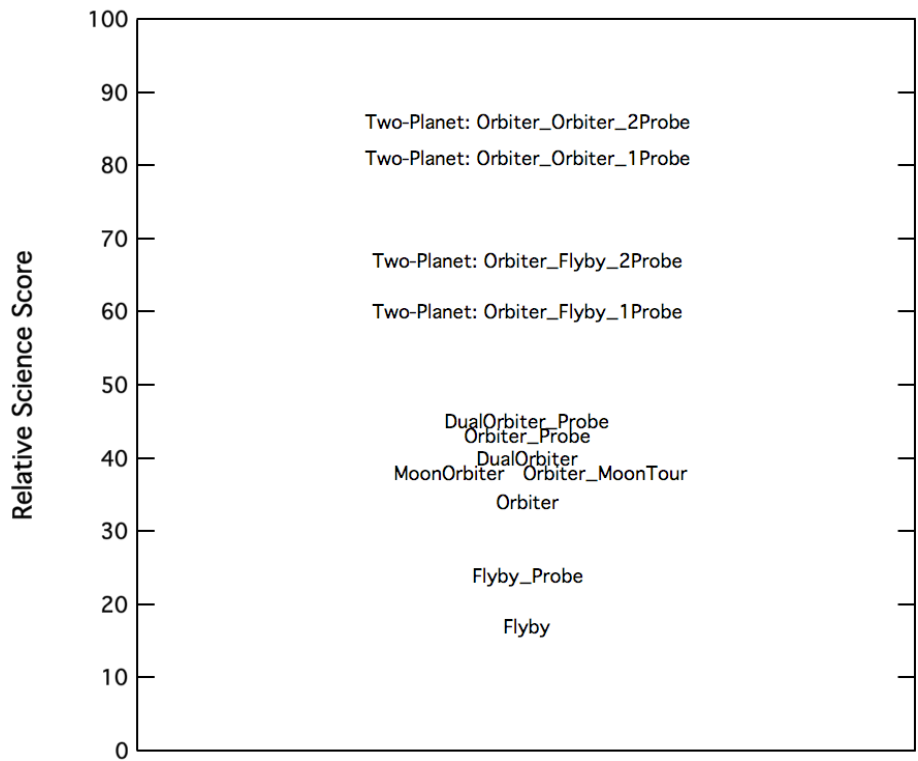

**Figure 3-16.** Graphical representation of the relative science value of various conceptual mission architectures. The values shown are for the 150 kg payload option. Other payload masses would not change the ordering of architectures, but would reduce the separation between them (see text).

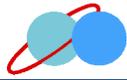

For reasons discussed in Section 3.4.2, the SDT did not reach consensus on how to rank all dual-planet missions against single planet ones. While there are details upon which there is no consensus, it is still clear that a mission that includes an orbiter at one planet, a spacecraft (whether flyby or orbiter) at the other planet, and a single probe at either planet is far more valuable than any mission flown to a single planet.

Key conclusions from **Figure 3-16** are

- Flybys of a single planet have the lowest science return.
- A significant increase in science value is achieved by getting into orbit at one ice giant.
- Adding a probe to an orbiter is another significant increase in science return, and it is the recommended ice giant Flagship mission.
- The missions with by far the highest science return are ones that incorporate an orbiter at one ice giant, an orbiter or flyby at the other, and at least one probe.

Mission costs will be discussed in Section 4 and summarized in Section 6.2, and there are a myriad of details and caveats associated with them. None-the-less, it is useful at this point to consider a plot of science value versus cost for some of the architectures considered. This is presented in **Figure 3-17**. To maximize the number of architectures plotted, we started with A-Team results (see Section 2.3.1). A-Team cost estimates, however, are not as accurate as the ones derived in Team-X (Section 4). Comparing the studies, we found A-Team costs to average about 20% below those from Team-X, so we multiplied A-Team costs by 1.2 for this chart. We also only plot architectures for which the SDT reached consensus on their science value (Section 3.4.2). Finally, to clarify relationships, we only plot missions with a 150 kg payload on the spacecraft, and we do not include Neptune-only missions (their costs are slightly higher and science values identical to the Uranus missions plotted). In such a chart, it is useful to look for "cliffs," where there is a sharp rise in science value but little increase in cost, and "plateaus," where increasing cost is not returning much additional science.  Important things to note in **Figure 3-16** are:

- The single planet orbiter and probe mission concept represents the cheapest mission on a science plateau.
- Sending two orbiters to the same planet adds significant cost but little science (creating a plateau).  If we invest in building two spacecraft, it is far better to target both ice giants (moving up the science value cliff).
- The one-planet two-orbiter missions are unusual cases.  Ignoring them, there is a fairly linear relation between cost and science return.  Thus, the increase in science from two-planet missions is commensurate with the increased cost.

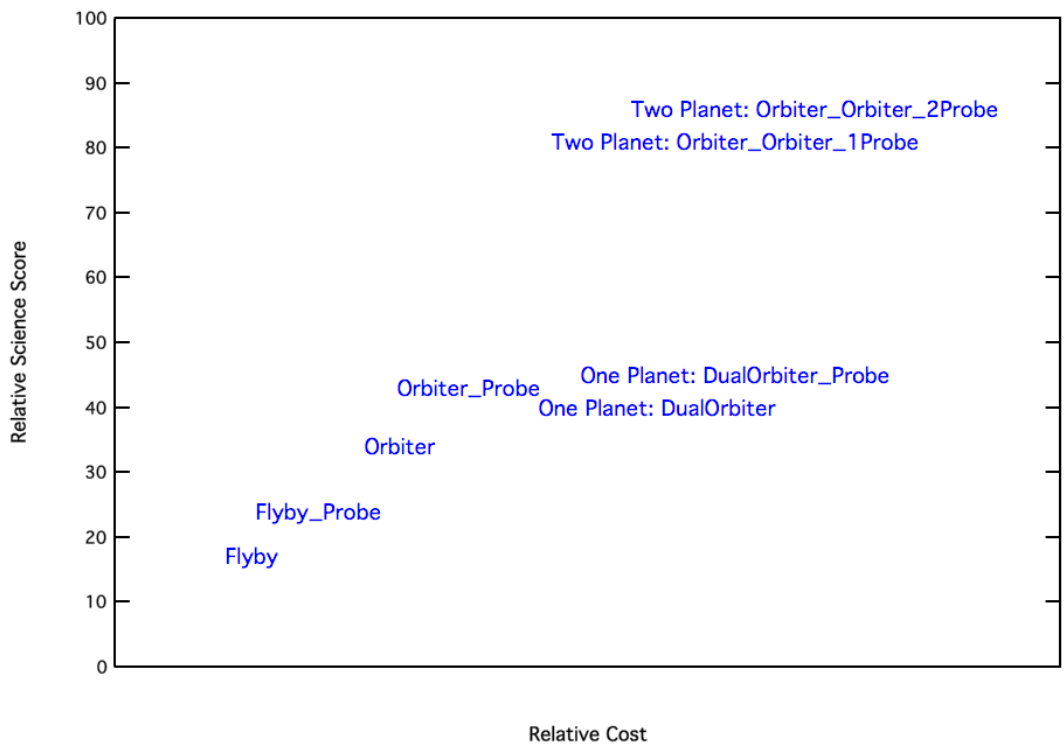

**Figure 3-17.** Science value vs. relative cost for a subset of conceptual architectures considered. See the text for a discussion.

## 3.5    Recommendations for Detailed Point Designs

At the conclusion of the A-Team study, the SDT identified 11 architectures for which it would like to see refined designs with higher-fidelity cost estimates.  The list includes all high-value architectures, as well as lower-value missions which (according to the preliminary A-Team cost estimates) cover the full targeted cost range of $1B to $2B.  By intent, the requested studies sample a wide range of architectures.  The initial list is:

- Uranus orbiter with probe
- Neptune orbiter with probe
- Triton orbiter
- Two-planet, two-orbiter mission
- Two-planet, two-orbiter mission with one probe at one planet
- Neptune orbiter
- Uranus orbiter
- Neptune flyby
- Neptune flyby with probe
- Two-plant, two-flyby mission with a probe at one planet
- Uranus orbiter and Neptune flyby

Based on cost and schedule considerations, it was felt that no more than 6 Team-X (point design) studies could be conducted.  The SDT leadership, in consultation with the engineering and full science teams and with study management, trimmed the above list to four architectures.  The shorter list is intended to bracket parameter space, allowing better cost estimates and guidance on accommodation issues for all architectures, but not specific point designs for all architectures.  If time and funding allowed (which it did not), results from these four point designs would be used to select the most useful additional architectures for study.  The four architectures selected are:

- Uranus orbiter with a 50 kg science payload and atmospheric probe,
- Uranus orbiter with a 150 kg science payload (no probe),
- Neptune orbiter with a 50 kg science payload and atmospheric probe, and
- Uranus flyby with a 50 kg science payload plus atmospheric probe.

These four missions are expected to bracket the target cost range, they bracket the range of recommended payloads, they allow a direct comparison of similar Uranus and Neptune missions, they include both orbiter and flyby missions, and they include missions with and without probes. The next section describes the detailed point designs carried out.

# 4    MISSION IMPLEMENTATION OPTIONS

## 4.1    Architecture (Trajectory) Trade Space

Evaluation of implementation options for the ice giant mission began with a comprehensive assessment of feasible mission architectures. The team began with an overview of possible mission designs, which were then mapped to notional flight system architectures to generate possible mission options. Launch vehicle options were evaluated for the missions, as were a variety of potential propulsion implementations. Trajectories using various propulsion options, with up to four planetary flybys were investigated. The impact of using different launch vehicles (including the SLS), on flight time, delivered mass, propellant throughput and mission architecture were studied. Details of the probe coast, entry and spacecraft orbit insertion at either of the two planets were also evaluated. Finally, a procedure for computing dual spacecraft trajectories capable of delivering individual spacecraft to both planets on a single launch, was developed. The results show that exceptional launch opportunities for a variety of multi-element mission architectures exist and these have been documented. For the mission options described below, trajectories were chosen that are representative of a family of possible solutions. This should ensure that the designs developed will be equally applicable for a variety of launch opportunities. Details on the wide range of trajectory analyses performed for the study are contained in Appendix A.

## 4.2    Recommended Architectures for Team X Design/Costing

Based on recommendations from the Science Definition Team (SDT; Section 3.5), four basic mission architectures were recommended for point design analysis for the study:

1.  Uranus orbiter with ~50-kg payload and atmospheric probe (with SEP stage)
2.  Uranus orbiter with ~150-kg payload without a probe (with SEP stage)
3.  Neptune orbiter with ~50-kg payload and atmospheric probe (with SEP stage)
4.  Uranus flyby spacecraft with ~50-kg payload and atmospheric probe

Initial architecture assessment by the systems and mission design team recommended the inclusion of a SEP stage in the mission architecture to provide flexibility to decrease mission time and/or increase mass delivered to the target body. This recommendation was based on earlier study results; however, upon completing the point designs for these missions, it was determined that the increased flight system mass incurred by the inclusion of the SEP stage was actually having a detrimental impact on the two Uranus orbiter missions, although it was felt to remain applicable to the Neptune orbiter. For this reason, two additional mission architecture point designs were studied for the Uranus orbiter missions that were able to complete the missions with solely chemical propulsion, avoiding the extra mass and cost of the SEP stage. These two options were:

5.  Uranus orbiter with ~50-kg payload and atmospheric probe (chem. only)
6.  Uranus orbiter with ~150-kg payload without a probe (chem. only)

The following sections describe the point design studies carried out by Team X for each of these options. Details on the Team X results can be found in Appendix C.

## 4.3    Mission Option 1: Uranus Orbiter and SEP Stage with 50-Kg Payload and Atmospheric Probe

### 4.3.1    Overview

The first option brought to Team X for analysis and costing was the architecture consisting of a Uranus orbiter with 50 kg of science payload, in addition to an atmospheric probe. The orbiter would make use of a SEP stage in the inner solar system to provide additional delta-v. The use of an appropriately sized SEP stage was considered advantageous to the architecture as a way to shorten trip time to the target planet, and/or increase delivered mass capability.

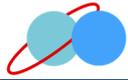

The mission design chosen for this concept would launch in July of 2030, and would execute a ~11 year trajectory using Venus-Venus-Earth (VVE) gravity assists. The atmospheric probe would be released 60 days prior to entry and the orbiter would then perform a divert maneuver to optimize geometry for probe relay and Uranus orbit injection (UOI). Probe entry into Uranus' atmosphere would be followed by ~1-hour descent to a pressure level of 10 bar, with data relay to the orbiter during this time. Immediately after the probe mission is complete, the orbiter would slew to its burn attitude and prepare for UOI, which would take place ~2 hours after probe entry.

The UOI burn puts the orbiter into a 150-day orbit around Uranus, which is reduced during the subsequent 4-year orbital phase to approximately 50 days, and which includes multiple satellite flybys.

### 4.3.2    Science

This mission represents the minimum the science team felt an ice giant Flagship mission should accomplish. The instrumentation, discussed below and in Section 3.3.2, allows the highest-priority science to be accomplished; measurement of the bulk composition including noble gases and isotopic ratios (primarily achieved by the atmospheric probe), and determination of interior structure (primarily achieved by the Doppler imager, but supported by the magnetometer). The combination of carrying a camera system and remaining in orbit within the uranian system allows other priority goals to be achieved and still more to be partially addressed. The orbiter mission also enables the study of time-varying features within the system, and opens the possibility of serendipitous discovery and follow-up.

### 4.3.3    Instrumentation

Instruments comprising the "50 kg" orbiter payload for this option include:
- Narrow angle camera (based on the Europa Imaging System [EIS])
- Doppler imager (based on Echoes from Juice)
- Magnetometer, based on that flown on the Galileo probe

In addition, the mission includes an atmospheric probe, equipped with its own science payload:
- Gas Chromatograph Mass Spectrometer (GCMS; based on GSFC heritage designs)
- Atmospheric Structure Instrument (ASI; based on Galileo and Huygens instruments)
- Nephelometer (based on Galileo probe design)
- Ortho/Para Hydrogen Experiment (OPH)

Payload characteristics were derived using existing flight instruments to ensure realism of instrument parameters. There are two exceptions—the OPH for the probe and the Doppler imager, both of which have not flown. A sound velocity-based OPH instrument has been described but has not been implemented in a space flight mission. At least two versions of a Doppler imager have been designed and proposed for flight—one based on an interferometer and one based on magneto-optical filters. The instruments have undergone bench top testing but may experience significant changes in mass, power, volume and cost when implemented for flight. For this reason, the OPH and the Doppler imager properties require higher levels of contingency. Characteristics of the probe instruments are summarized in **Table 4-1** and the orbiter instruments are summarized in **Table 4-2**.

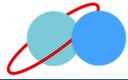

**Table 4-1.** Probe instruments.

| Instrument | Deployments | Nominal Data Volume (mbits) | Accommodation | Comments |
|---|---|---|---|---|
| ASI | none | 5.0 | P: 8-cm pitot tube to manifold T:  sensor outside hull | |
| Ortho/Para | See nephelometer | 1.0 | Share nephelometer arm | |
| MassSpec | Inlet Cover | 10 | | |
| Nephelometer | Optical Arm | 4.0 | Deployable arm | |

**Table 4-2.** Orbiter payload instruments.

| Instrument | Deployments | Cooler Deg. C | Bus Rate kbps | FOV Degrees | Comments |
|---|---|---|---|---|---|
| Doppler Imager | Cover | -35 | 10 | 0.4, 7 μrad IFOV | Sample 2/minute Higher bus data rate if images windowed in CDS |
| NAC | Cover | -35 | 1280 | 2.3×1.2 10 μrad IFOV | |
| MAG | Boom | none | 128 | 4π | 3 axes, inboard & outboard, 10 meter boom |
| Radio Science | None | none | NA | See telecom | 2 band, 2 way coherent tracking |

#### 4.3.3.1   Probe Instrument Accommodation

The probe instrument accommodation is based largely on the Galileo probe.  The probe capabilities are very similar with respect to high entry g loads (up to ~200 g), descent duration (~1 hour), and available data return (a few megabits).  Payload power is expected to be similar—though some savings may be realized through the advances in electronics over the 50 years that will have passed since Galileo.

The ASI pressure and temperature measurements, the nephelometer, the ortho-para hydrogen experiment and the mass spectrometer all require penetration through the probe hull.  These penetrations extend beyond the boundary layer of the probe to about 8 centimeters.  The ASI pressure measurements use a pitot tube that connects to a manifold feeding pressure sensors having different pressure ranges.  The ASI temperature sensors extend beyond the hull (in the Galileo version, the sensor is a coil of platinum wire).  The nephelometer has a deployed optical arm (deployed after egress from the entry system) that reflects scattered light into the inboard sensors.  This arm may also be used for the OPH instrument.  The mass spectrometer has an inlet near the apex of the probe and exits at the minimum pressure point inside the probe (for Galileo, this provided a 6 mbar pressure gradient to the ambient flow system).  The probe instrument locations, and penetrations, are shown in **Figure 4-1**.

Initial probe measurements are made by the accelerometers to measure turbulence on entry.  A goal is to measure the accelerations at 20 to 50 khz, but the data volume would exceed the total available return from the probe. Data selection may include reporting the changes in acceleration that exceed certain tolerances in a similar manner to Galileo.

Instrument mechanical deployments occur after deployment of the probe chute and measurements begin at about 0.1 bar. The mass spectrometer has critical measurements at the top and bottom of the descent, and acquires selected measurements at points in between. The remaining instruments take measurements at regular intervals (typically 2 seconds) during descent. In general, the sensor data is highly commutated within the instruments to fit within the available data return.

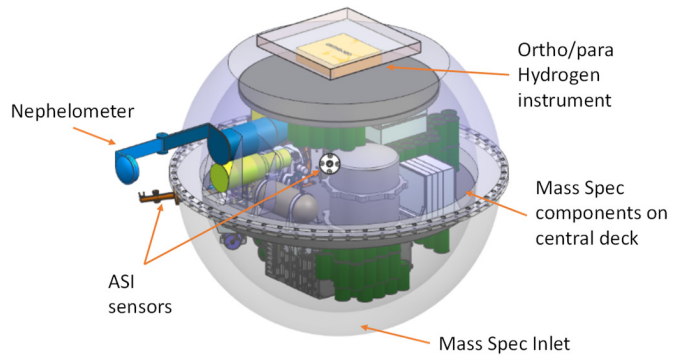

**Figure 4-1.** Probe instrument accommodation.

### 4.3.3.2 Orbiter Instrument Accommodation

The spacecraft includes a Doppler imager, a narrow angle camera, a magnetometer, and radio science. **Table 4-1** provides the nominal instrument parameters and **Figure 4-1** shows their locations on the spacecraft.

The Doppler imager and the narrow angle camera are mounted on a fixed platform. The position of the Doppler imager field-of-view (FOV) is fixed with respect to the (fixed) telecom antenna. The angle is set to allow telecom to operate while the Doppler image measurements are occurring on approach (-85 d to -30 days)—which is required to return the relatively high data volume generated by the Doppler imaging experiment. This reduces the ability to make major changes to the approach trajectory once the spacecraft design is finalized.

Both the Doppler imager and the NAC focal plane array are passively cooled to -35 centigrade, requiring clear fields of view for their radiators. The instrument's bore sights are coaligned to minimize spacecraft motion during approach observations. No source of scattered light should be within 12 degrees of the NAC bore sight.

The Doppler imager has a fast steering mirror to stabilize the image on approach, with a range of motion under 5 degrees. The model NAC has a gimbal to access multiple targets with fixed spacecraft pointing. The range of motion for the NAC gimbal is sufficient to avoid the need for articulating the high gain antenna.

The magnetometer is located on a deployable 10-m boom. There are both inboard and outboard sensors to compensate for fields generated by the spacecraft.

Radio science requires continuous tracking before and after closest approach and around apoapse, utilizing two bands (X and Ka). It was deemed possible to achieve the science goals without an ultra-stable oscillator, so it is not included in this configuration.

### 4.3.3.3 Instrument Descriptions

#### Doppler Imager

The Doppler imager measures atmospheric motions by sensing the Doppler shift of solar absorption lines in sunlight reflected off atmospheric particles. The model Doppler imager utilizes magneto-optical filters. The Doppler imager returns full disk images of approximately 100×100 pixels at the end of the approach sequence (UOI-30 days). The image size at beginning of the approach sequence (UOI-85 days) may have fewer pixels (~35×35) or may also be 100×100 (using pixel averaging for the closer images). In the latter case, the IFOV would be approximately 7 μrad, in the former about 18 μrad. The Doppler shift measurements require narrow bandwidths, so irradiance and signal-to-noise ratio (SNR) are key issues that must be assessed to verify the applicability of the technique to Uranus and Neptune.

#### Magnetometer

The magnetometer includes both vector sensors and a scalar sensor.

#### NAC

The NAC has a large format focal plane array (4k×2k) and six or more filters. The model NAC includes a gimbal to allow imaging offset from the nominal platform L-vector. The NAC can acquire images in both framing and pushbroom modes. The pushbroom mode supports time-delay integration (TDI) for image acquisition at low light levels.

### Radio Science

The radio science experiment supports dual band tracking and requires 2-way coherent tracking. It does not include a USO.

#### 4.3.3.4    Payload Utilization

A description of the observation strategies appears in the Concept of Operations (Section 4.3.5). The following brief description summarizes the instrument utilization.

The Doppler imager begins its campaign at UOI-85 days and is subdivided into 5 campaigns of 11 days each. Imaging is continuous at a cadence of two per minute. The experiment is complete at UOI -30 days.

The imager obtains rotation movies on approach and performs feature tracking within 15 days of approach.

For orbital operations, the Doppler imager will take only 20 images over a 16-hour period each orbit to measure cloud particle velocities in small-scale weather features. The NAC instrument will be used to track motions of atmospheric features and to image the illuminated portions of the rings and of satellite surfaces at global and local (postage stamp) scales. The magnetometer will be on continuously, with higher rate modes near periapse during encounters with satellite. Radio science requires 2-way coherent tracking through periapse and at apoapse.

Spacecraft data storage is sufficient to store all observations acquired for several (~8) orbits.

### 4.3.4    Mission Design

The mission design objective for the first mission option was to enable a Uranus orbiter and a Uranus atmospheric probe. The orbiter was sized for a ~50-kg payload and the probe was designed to take science measurements at up to 10 bar. The high-level mission design guidelines for this option were:

1. Launch between 2024–2037; preferred launch date between 2029–2031
2. Total mission lifetime <15 years (including Uranus Science phase)
3. Launch on an existing commercial launch vehicle
4. Avoid Uranus rings during orbit insertion
5. Design probe coast, entry and descent trajectory with feasible orbiter—probe telecom geometry
6. Uranian moon tour with two flybys each of the five major moons

As this was the first architecture evaluated by Team X for the current Ice Giants Study, the orbiter dry mass was not known during the initial stages of the mission concept design. To allow for flexibility in delivered mass, a SEP-based mission architecture was selected. A purely chemical propulsion based architecture (without SEP stage) was also studied later (see mission option 5). The orbit insertion was carried out using a chemical (bipropellant) maneuver. Given that Uranus is almost 20 AU away from the Sun, reaching the planet in a short amount of time results in high approach velocity, leading to a large orbit insertion ΔV. **Figure 4-2** depicts the SEP baseline mission architecture.

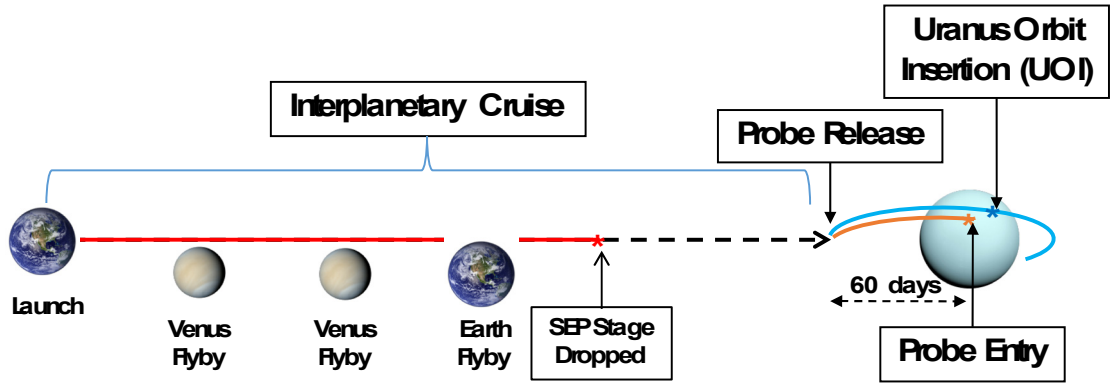

**Figure 4-2.** Mission Concept Option 1: SEP mission design overview.

### 4.3.4.1 Launch and Interplanetary Cruise

The baseline interplanetary trajectory relies on a 25-kW SEP stage powered by 2 NEXT ion engines, to propel the spacecraft while in the inner solar system. The spacecraft is assumed to launch on a Delta-IV Heavy. The spacecraft then performs two Venus flybys followed by one Earth flyby to gain momentum. The SEP stage thrusting in the inner solar system makes up for the relatively low launch energy while enjoying high propellant efficiency. After thrusting, the SEP stage is jettisoned at ~6 AU, as the solar flux beyond this point is insufficient to power the ion engines. The SEP stage dry mass is significant and dropping it before UOI results in significant propellant savings on the orbiter.

**Table 4-3** lists two trajectories: the first one (in blue) is the in-session interplanetary trajectory, which was used during the Team X session for initial design; the second trajectory (in black) represents further refinements made to the in-session trajectory after the probe mass, SEP stage dry mass, and orbiter dry mass became known. **Note that the Team X in-session and the refined trajectory both deliver the same or more mass into Uranus orbit than the allocated amount in the MEL**. Also, note that the refined trajectory uses ~400 kg less Xenon propellant. After the Team X session, structural refinements were made to the SEP stage, which resulted in significant mass savings, allowing this propellant reduction. The refined SEP stage was still sized to carry the larger Xenon load (~1,037 kg), and could be refined further to take advantage of the ~400 kg reduction in Xenon.

The quantity "*Arrival Mass*" refers to mass of the spacecraft **before** probe release (assumed Uranus probe mass ~321 kg) but after SEP stage separation. The probe is released ~60 days before UOI. Orbit insertion ΔV for the baseline is calculated for a notional 140-day capture orbit with periapsis at 1.05 Uranus radii. **Figures 4-3** and **4-4** show the two trajectories listed in **Table 4-3**. The red arrows on these two figures indicate SEP thrusting phases of the mission.

**Table 4-3.** Option 1 Team X in-session (blue) and refined baseline (black) mission trajectory.

| Flyby Sequence | Launch Vehicle | Launch Date | Launch C3 (km²/s²) | IP TOF (yrs.) | Xenon Mass (kg) | SEP Stage Dry Mass (kg) | Arrival Mass (kg) | Orbit Insertion ΔV (km/s) | Mass in Orbit (kg) |
|---|---|---|---|---|---|---|---|---|---|
| Earth-VVE-Uranus | Delta-IV Heavy | 9/12/2030 | 15.3 | 11.0 | 1037 | 1779.3 | 5028.1 | 2.3 | 2305.7 |
| Earth-VVE-Uranus | Delta-IV Heavy | 9/26/2030 | 26.5 | 11.0 | 603 | 1486.1 | 4319.1 | 2.2 | 1988.3 |

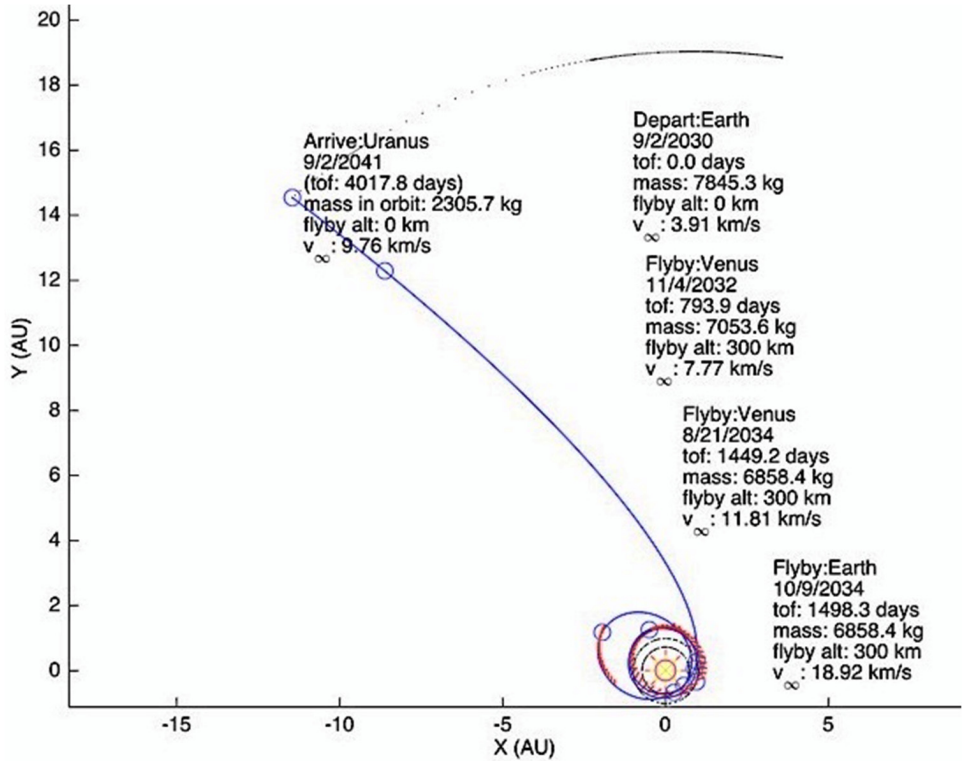

**Figure 4-3.** Option 1 Team X in-session trajectory.

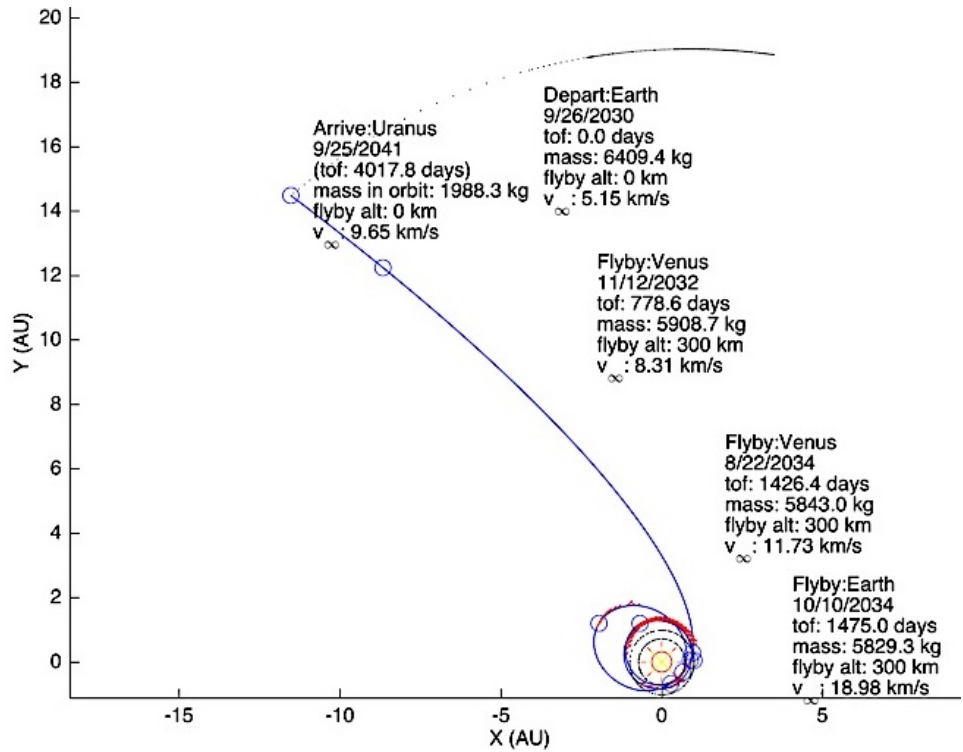

**Figure 4-4.** Option 1 refined baseline trajectory.

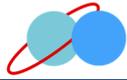

#### 4.3.4.2   Probe Coast, Entry and Descent + Uranus Orbit Insertion

The baseline mission trajectory assumes the probe is released 60 days before atmospheric entry. A probe targeting maneuver (PTM) is performed by the orbiter prior to probe release, followed by an Orbiter Divert and Periapsis Targeting Maneuver (ODPTM) to target the UOI periapsis. The probe enters Uranus atmosphere at an Entry Flight Path Angle (EFPA) of -30 degrees. A steep EFPA is chosen to alleviate orbiter-probe relay geometry issue and to reduce total accumulated heat load on the probe. Details on some of the probe entry parameters are given in **Table 4-4**.

The probe sequence is shown in **Figure 4-6**. The probe descent to 10 bars lasts for ~1 hr.

The orbiter performs the UOI ΔV of ~2.26 km/s at a periapsis altitude of ~1.05 Uranus radii, which occurs two hours after probe entry (one hour after completion of the probe relay). Post UOI, the orbiter enters a 140-day orbit around Uranus. An orbit insertion altitude relatively close to the atmosphere is chosen to mitigate potential ring particle impingement issues. The low UOI altitude also helps in reducing the UOI ΔV magnitude. **Figure 4-5** highlights UOI and probe entry geometry.

#### *Probe Link Geometry Considerations*

Following are some of the findings from the UOI analysis and probe-orbiter telecommunications geometry optimization exercise. A hyperbolic probe entry (with orbiter relay) at Uranus must trade the following design variables:

1. UOI ΔV magnitude
2. Probe g-load tolerance
3. Probe–orbiter relay telecommunications requirements (aspect angle and range)

UOI ΔV is sensitive to the orbiter periapsis altitude (see Appendix A). Higher orbiter periapsis provides better relay line-of-sight and longer persistence (lower angular rate relative to probe), but higher UOI ΔV. The baseline design results in a UOI ΔV of ~2.2 km/s at 1.08 Uranus radii (Ru), but 2.86 km/s at 1.75 Ru. Shallow EFPA reduces probe g-load, but presents challenging relay geometry and increases cumulative heat load.

Another factor to consider is the time between probe entry and UOI. Currently, there are two hours allocated between probe entry and UOI, which is a critical event. It may be operationally challenging to sequence both the probe relay and UOI on the orbiter within this time window. Increasing the separation will make the geometry more challenging for telecommunications. Probe-orbiter geometry also needs to deal with issues like uncertainties regarding the Uranus atmosphere and potential signal attenuation.

A steeper EFPA (-30 deg) was hence chosen to alleviate the telecommunications geometry and UOI ΔV issue, while still having acceptable g loads on the probe (<200 g). Detailed navigation and telecommunications analyses are recommended for a follow up study.

**Table 4-4.** Option 1 Probe entry parameters.

| Parameter | Value |
|---|---|
| Interface Altitude | 1000 km |
| Entry Velocity | 23.1 km/s |
| Entry Flight Path Angle | -30 degrees |
| Max g load | 165 g |
| Stagnation Pressure | 9 atm. |
| Cumulative Heatload | 41114 J/cm² |

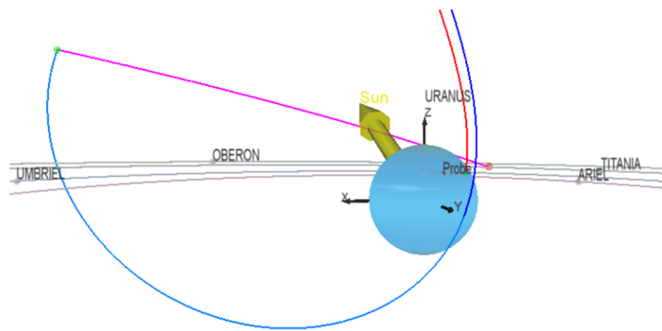

**Figure 4-5.** Option 1 UOI and probe entry.

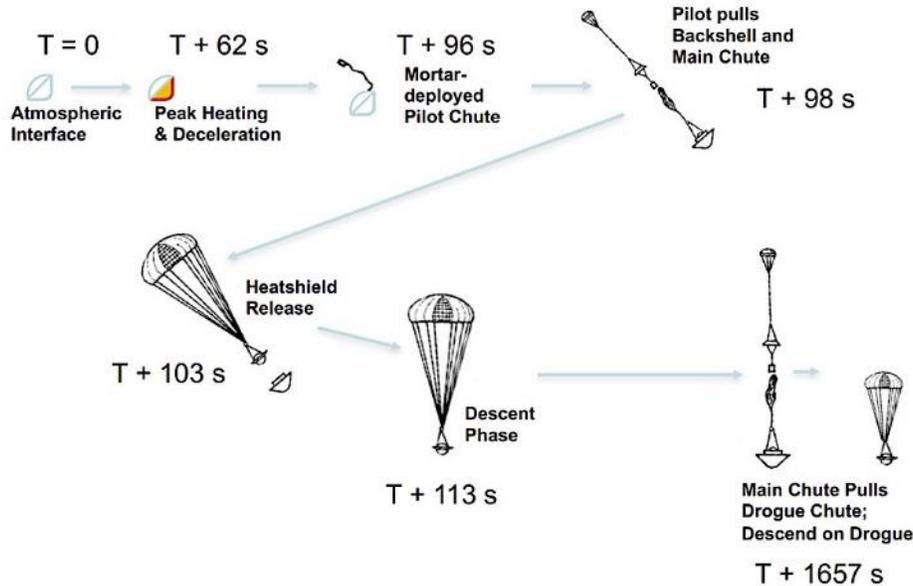

**Figure 4-6.** Probe entry sequence (timing represents Uranus example).

### 4.3.4.3    Uranus Tour Phase

Noting that the Uranus obliquity is 97.7 degrees, this has consequences for a uranian satellite tour. To achieve relatively low flyby speeds (< 6 km/s) at the five major uranian moons, the orbiter needs to change its orbital plane (with respect to Uranus ring plane) after orbit insertion.

For the baseline trajectory, the orbiter goes into ~69 degree inclined orbit (with respect to Uranus equatorial plane) around Uranus after orbit insertion. For Option 1, Team X assumed a 4-year tour at Uranus with at least two flybys of the five major moons (Miranda, Umbriel, Ariel, Titania, and Oberon). Subsequent tour design efforts (presented in Option 5) revealed that a basic uranian moon tour that achieves all the science objectives could be accomplished in 2 to 2.5 years after UOI. The flyby velocities at the moons are expected to be <5 km/s. These flyby velocities are attainable due to the Apoapsis Twist and Targeting Maneuver (ATTM; see **Figure 4-5**). Further details on the tour design can be found in Option 5.

#### *Mission Delta-V Summary*

**Table 4-5** shows the summary ΔV table for this mission option. The chemical ΔV is broken down into monoprop and biprop burns, based on ΔV magnitude and criticality of the maneuver. The main chemical ΔV driver for a mission to Uranus is the orbit insertion due to relatively high approach $V_\infty$. Following UOI, the orbiter coasts towards its capture orbit apoapsis where it performs an ATTM. The ATTM is designed to achieve an efficient orbital plane change and target one of the major moons of Uranus (Titania).

### *4.3.5    Concept of Operations/GDS*

After a long and relatively dormant 11-year cruise characteristic of outer planet missions, a densely packed 4-year tour studying Uranus and its major satellites is conducted. The highest priority in the mission operations design is to maintain a safe and fully functional spacecraft throughout each phase of the mission lifetime. The mission operations concept for Option 1 (and similarly for the following options) draws upon many successful current and past missions (including Galileo, Cassini, and Rosetta). One of the most ubiquitous themes found in such missions' "Lessons

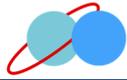

**Table 4-5.** Mission option 1 ΔV summary for refined baseline trajectory.

| | Biprop (m/s) | Monoprop (m/s) | Comments |
|---|---|---|---|
| Interplanetary TCMs | 0 | 25 | Most of the TCMs are immediately after launch and after SEP stage separation |
| PTM | 0 | 5 | Probe targeting maneuver, 60 days before UOI |
| ODPTM | 20 | 0 | 60 days before UOI |
| UOI | 2260 | 0 | <1 hr. burn using two 890N engines |
| UOI-CU (2%) | 45 | 0 | |
| ATTM | 300 | 0 | Estimated from Option 5 tour design |
| ATTM-CU (2%) | 0 | 6 | |
| Tour Deterministic | 45 | 40 | Multiple maneuvers |
| Tour Margin | 0 | 20 | For extended mission and contingency |
| Tour Statistical | 0 | 30 | |
| De-orbit and Disposal | 10 | 0 | Impact planet |
| **Total** | **2670** | **126** | |

Learned" reports is the need for early consideration of science operations when designing flight and ground systems. Each of the ice giants mission options is designed to produce optimal science return through tailored mission architectures alongside streamlined operations, leveraging modern numerical analysis tools to simplify ground and flight operations. Utilizing modern automation tools in mission planning and science planning (similar to ASPEN as used for the Rosetta mission) not only simplifies the operations processes, but also provides a thorough analysis of potential science opportunities. These numerical tools are driven by detailed science observation requirements provided by the science teams, therefore rapid iteration with the science designers will hone activity schedules and allow for a more adaptable operations process. A more detailed discussion of automated numerical tools and other advanced mission operations systems can be found in Appendix D.

Shortly after launch, the mission begins its checkout processes and deployment of all flight systems. During the first four weeks of cruise, continuous DSN coverage is required for thorough characterization of all flight systems and accommodating variable commanding schedules typical of early checkout operations. When checkout is complete, the post-launch phase configures the spacecraft to low thrust SEP navigation and the DSN coverage is reduced to only 1 pass per week. Throughout interplanetary cruise, each critical mission event (TCM, gravity assist, etc.) is expected to have daily DSN coverage for commanding and tracking for 2 weeks approaching the event, especially for the nuclear safety maneuvers prior to any Earth gravity assists. Continuous coverage is required during the days surrounding the gravity assist or maneuvering. The SEP stage is jettisoned after about 6.5 years around the 6AU distance, where the solar illumination is insufficient to support thrusting. Though the Team X study assumes minimal operations work throughout most of the interplanetary cruise, additional checkouts and characterizations of flight systems and instruments could be performed during the gravity assist encounters of Venus and/or Earth.

The long ballistic cruise to Uranus after the SEP low-thrust phase will have a further reduction in coverage using a 15-minute beacon 3 out of 4 weeks. Within 6–12 months of Uranus Arrival, ground operations support ramps up (including mission/science planning and execution staff). Starting at 9 months until Entry/UOI, DSN coverage increases to daily tracks in preparation for the Approach Science phase and Probe release. At 85 days to Entry/UOI, the Approach Science phase begins. Requirements for this phase are driven by 55 days of near-continuous Doppler mapping observations of Uranus. The first 22 days have a lower Doppler observing cadence of 1 image every 30 seconds, reserving the 33 days closer to the planet for high data rates of 1 image every 2 seconds as the planet fills the Doppler field of view. The science plan for Mission Option 1 is detailed in **Table 4-6**, showing number/cadence of images for optical instruments and data rate levels of continuously observing MAG instrument. During the Approach phase, Doppler and NAC instruments perform uninterrupted observing while the HGA is continuously Earth-pointed. The Earth-pointed HGA provides flexibility in DSN scheduling to ensure maximum science data return during this period.

**Table 4-6.** Mission Option 1 instrument science plan for major mission phases: Approach, Satellite Flyby, Apoapse, and Periapse Science.

| Ice Giants Mission #1 Instrument Science Plan | Approach Obs Info | Satellite Flyby Obs Info | Apoapse Obs Info | Periapse Obs Info |
|---|---|---|---|---|
| NAC | Mov: 200im - 1clr | 6Gb (1/2 Map) | 10-rot/2hr - 4clr | 5-10 F.trax - 1clr |
| Doppler Imager | 22dy/30s,33dy/2s | 100 images | 3 images/rot | 20 images/rot |
| Magnetometer | Low | Very High | Nominal | High |

The baseline DSN plan is to allocate a single 34-meter Beam WaveGuide antenna (34BWG) station at 24/7 continuous tracking and downlink support for approach science. The power system is able to accommodate up to 20 hours straight Ka-band downlink with continuous approach science. Alongside the Doppler science, the MAG instrument will be continuously measuring the solar wind and distant fields of Uranus at a low rate. The NAC will create a color mosaic of the planetary system, as well as an approach movie (200 single-color images) towards the end of the approach phase. The approach science will collect a total of 53 Gb of compressed data (184 Gb uncompressed). Using modern compression techniques, a compression ratio of 3.5X is used for optical instruments and 2.0X for particle and fields instruments with minimal loses. It will be assumed that all recorded and transmitted data discussed in the Concept of Operations sections are addressing uncompressed data values and rates, while any discussion of compressed data values will be explicitly defined. The data budget averaged per day on approach is balanced at about 0.96 Gb/day of compressed data recorded and 1.10 Gb/day 34BWG downlink capability (13% margin). The detailed data outline for each instrument is shown in **Table 4-7**. Any overflow approach science data not returned is stored on high capacity 1024 Gb solid-state data recorders and downlinked alongside on-orbit science after UOI. At 60 days prior to Entry/UOI, the approach science pauses to release the probe. The orbiter executes a small divert maneuver to adjust the final approach trajectory for probe relay and UOI. After these maneuvers are complete and verified, the approach science finishes out at 30 days to Entry/UOI. Throughout the approach phase, periodic optical navigation observations are performed with the NAC to reduce the orbital uncertainty of the Uranus system satellites.

**Table 4-7.** Mission Option 1 instrument science data outline for major mission phases: Approach, Apoapse, Periapse, and Satellite Flyby Science.

| Ice Giants Mission #1 Instrument Science Data | Min Data Rate (Kbps) | Nom Data Rate (Kbps) | Max Data Rate (Kbps) | Approach Data (Gb) | Satellite Flyby Data (Gb) | Orbit Data (Gb) | Apoapse Data (Gb) | Periapse Data (Gb) |
|---|---|---|---|---|---|---|---|---|
| NAC | - | 134Mb/im/clr | - | 32.16 | 6.00 | 49.58 | 45.56 | 4.02 |
| Doppler Imager | - | 0.1Mb/im | - | 148.90 | 0.01 | 0.03 | 0.02 | 0.01 |
| Magnetometer | 0.50 | 1.00 | 3–12 | 2.38 | 0.30 | 4.84 | 4.06 | 0.78 |
| Optical Compression: 3.5X Fields & Particles Compression: 2.0X | | | Science Data Totals: | 183.4 Gb | 6.3 Gb | 54.4 Gb | | |
| | | | Uncompressed: | 183.4 gb | Uncompressed: | 60.8 gb | | |
| | | | Compressed: | 52.9 gb | Compressed: | 18.5 gb | | |
| | | | Approach Data Rate: | 0.96 Gb/day | Orbit Data Rate: | 0.37 Gb/day | | |
| | | | Approach Data Return: 1.10 Gb/day 1X 34BWG (15 Kbps) | | Orbit Data Return: 1X 34BWG (15 Kbps) | 0.37 Gb/day | | |

The final checkouts for the probe are completed during the last 30 days before Entry, while the orbiter is prepared for UOI. The probe begins its entry and continuous data relay with the orbiter at 3 hours prior to UOI. The probe is expected to continue transmission of increasingly high value science for up to 1 hour from entry until loss of signal. The total data volume collected from the probe and relayed to the spacecraft is on the order of 2–4 Mb. Though the probe data is of very high importance, its size is negligible relative to the approach and orbital data volumes; therefore, it is not identified on the total mission data volume tables/charts. After the probe science is completed, the spacecraft is reoriented to its UOI maneuver attitude one hour before UOI. The large UOI burn will require 1-hour to complete, inserting the spacecraft into its initial 180-day orbit about Uranus.

During the first weeks in orbit, the instrument suite is configured/calibrated for orbiter science and the final Operational Readiness Tests (ORTs) are performed. Performing regular optical navigation activities is vital to understanding the Uranus system dynamics and reducing their orbital uncertainties to operational levels. Additionally, periodic extended 10–12 hour navigation DSN passes will be implemented to provide about 4–6 hours of two-way coherent tracking for orbit determination, driven by the long Earth-Uranus OWLT (one-way light time) up to 2.9 hours. Structuring the orbital science sequences will be tied to the orbital trim maneuver (OTM) schedule and satellite flybys, from which a reference trajectory is maintained. Similar to the other missions' system tours, trajectory uncertainties are allowed to deviate from the reference trajectory outside these OTMs and satellite flybys. To maintain the science plans for pointing-sensitive observations, these uncertainties are countered by strategically updating onboard vectors with the latest orbit determination models. A precise 3-axis stabilized reaction wheel attitude control scheme is required, especially for geometrically sensitive observations such as stellar occultations, radio science activities, etc. During satellite flybys and certain periapse activities where faster turning rates may be required, the thruster-based reaction control system (RCS) may be used to accommodate science collection. The baseline optical instrument configuration is to co-align the NAC and Doppler boresites for common targeting on a static mounting. The Doppler instrument has an internal fast steering mirror for small adjustments to its pointing. The NAC can also decouple its pointing with a 2-axis gimbal, allowing additional attitude flexibility for simultaneous independent targeting.

The Orbital Science phase can be divided into 3 categories of observation activities that are repeated throughout: apoapse science, periapse science, and satellite flyby science. The generalized orbital science activities are based on a typical 50-day orbit, assuming at most a single flyby per orbit. Apoapse science is denoted outside of 20 Uranus radii ($R_U$), reserving the activities inside of 20 $R_U$ as periapse science. The apoapse science comprises primarily NAC mapping observations of Uranus and the ring system, imaging 10 rotations of the planet at a 2-hour cadence with full color. Throughout apoapse nominal rate MAG data of the outer fields will be recorded, while the Doppler imager only performs intermittent targeted observations of Uranus atmospheric features (~3 images per Uranus rotation). The outer ranges of the periapse science (~20–10 $R_U$) provide excellent opportunities for NAC observations of the Uranian satellites and Rings, while the inner ranges (<10 $R_U$) through periapse are reserved for high rate (3 Kbps) MAG investigations of the inner magnetic fields, as well as NAC and Doppler feature tracking on the planet or in the rings. The NAC is allocated 5–10 feature tracks, assuming 3 images (single color) for each tracking observation. The Doppler imager will perform its high-resolution feature tracks with an average cadence of about 20 images per planet rotation.

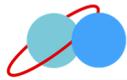

Throughout the orbital mission, the spacecraft will execute at least two flybys of each of the five major satellites. During a typical flyby, the NAC and MAG instruments will be the primary instruments observing the satellite and its interactions with Uranus's magnetic field. The NAC is able to map half of the satellite surface during each flyby with 5 Gb (uncompressed) of mosaic images along with 1Gb (uncompressed) of high resolution and stereo coverage. The MAG instrument will increase its collection data rate to 6–9 Kbps within ±6 hours of closest approach and up to 12 Kbps during the hour of closest approach. Each satellite flyby may also implement a small set of exploratory Doppler observations (~100 images) of the satellite on approach and/or departure. The repeated gravity assist flybys will torque the orbiter's inclination more and more in plane with the satellites (from the original 65 deg inclination), while also reducing the period down to 50 days. These final orbital conditions yield longer and better quality flyby science.

Within the course of each orbit about 19 Gb of compressed science data (61 Gb uncompressed) will be collected, including 2 Gb of satellite flyby compressed science data. In order to efficiently return all of the science data each orbit, the baseline DSN plan is for daily 8-hour passes using only a single 34BWG station. Each pass downlinks 367 Mb of data (assuming 15% overhead) at a flat rate of 15 kbps. With at least one extended 10–12 hour navigation tracking pass every week, the orbital science has a downlinked data margin of about 6%. A graph of science data recorded compared to the DSN capacity for the Approach science and Orbit science phases is shown in **Figure 4-7**. The Approach science plot displays the continuous Doppler data collection, then ramping up towards the end with the NAC movie recording. The Orbit science plot begins with apoapse data, collecting all of the NAC images of the planet, with continuous MAG measurements throughout apoapse. Then the 3-day spike of high rate periapse data is recorded, followed by densely packed satellite flyby science for about a day.

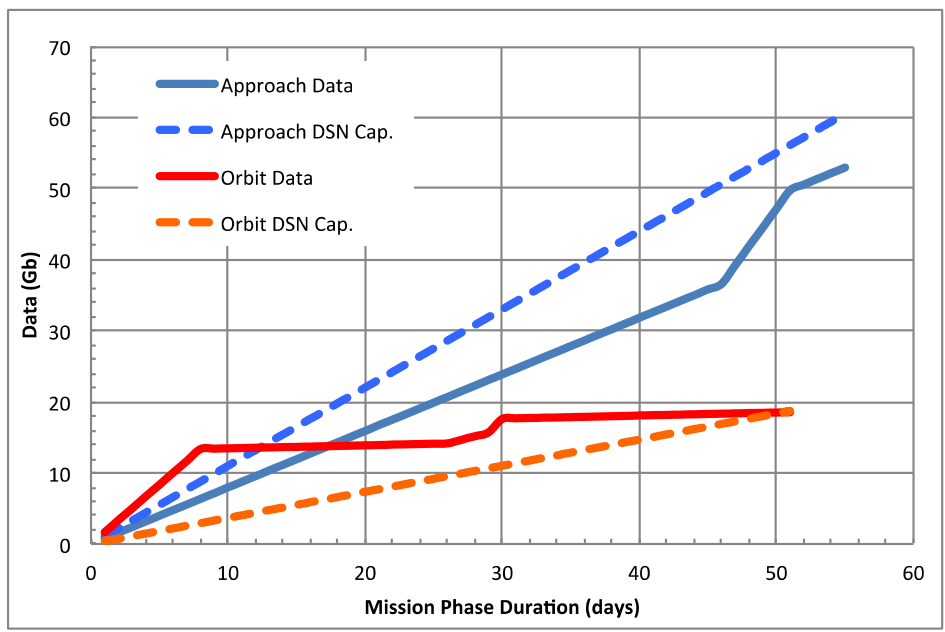

**Figure 4-7.** Mission Option 1 science cumulative data volume recorded compared to DSN downlink capacity for Approach science and Orbit science (starting at apoapse for a typical ~50 day orbit).

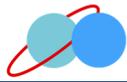

A noteworthy modification to the DSN plan would be to adopt a Cassini-like 5-level "wedding-cake" optimized downlink rate configuration, which would increase the total downlinked data volume by up to 50%. This configuration also allows for longer 9-hour pass duration, further improving the total data return. Although the flat rate downlink method is slightly more robust to late changes in DSN support, the modern mission planning tools developed can aid in rapidly reconfiguring the rate structure to accommodate such changes.

After the Uranus mission is complete, the baseline plan for decommissioning and disposal of the spacecraft is to reduce its trajectory's periapse close to the planet for valuable magnetic fields data collection followed by entering into Uranus's atmosphere.

### 4.3.6    Flight System/Probe Design

The basic architecture brought to Team X for this study consisted of three flight elements; an orbiter, a SEP stage, and an atmospheric probe. The orbiter and SEP stage designs were completed by Team X during the study. The Probe design had been developed in an earlier Team X study and was adopted for this option as-designed.

#### 4.3.6.1    Probe Design

The SDT had developed a requirement for an atmospheric probe that could be inserted into Uranus or Neptune and take data down to a pressure of 10 bar. Four instruments were specified for the probe as described in Section 3.3. Entry conditions were developed by the trajectory design team for the initial Team X probe design session, and modified during each Team X mission study to meet the specific requirements of the missions.

Given the similarities in atmospheric conditions and measurements desired, a single probe design was developed that could be used at either planet. Entry geometry for designing telecom links used Uranus as the worst case, and Neptune entry conditions were used as the worst case for entry and TPS design.

The probe design drew on heritage from the Galileo and Pioneer Venus probes, using current technologies and instrument designs. The probe is spin-stabilized during its coast to the planet and is powered by primary batteries. Survival heating during the 60-day coast period is provided by radioisotope heater units (RHUs). The overall configuration is illustrated in **Figure 4-8**.

The probe descent module is a truncated sphere, approximately 73 cm in diameter. The descent module is vented, allowing an equalization of pressure inside the probe with the external atmosphere during its descent. Apertures in the probe sides provide instrument access to the atmosphere. Telecom uses a flat patch antenna on the top of the probe to maintain a communications link with the orbiter during the ~1 hr science mission.

The probe entry system consists of a $45^0$ sphere-cone heat shield scaled from Galileo to 1.2 m in diameter, and a spherical backshell with a radius of curvature originating at the vehicle center of gravity (CG), as shown in **Figure 4-8**. The heat shield thermal protection system (TPS) uses HEEET material and the backshell TPS is C-PICA. Mass of the probe at entry is estimated to be 308 kg. The probe entry scenario is shown in **Figure 4-6**.

Further discussion of the entry and descent characteristics for Uranus and Neptune point designs can be found in Appendix A.

Probe subsystems were maintained as simple as possible for the descent module (**Figure 4-9**).

The probe is spin stabilized and deployed from the orbiter on a ballistic trajectory ~60 days prior to entry. It requires no propulsion or GNC subsystems. Command and data handling (C&DH) assumes JPL's Sphynx smallsat computer system (**Figure 4-10**). The Sphynx is a Leon 3 processor-based system currently under development and scheduled to fly in 2018 on several smallsat missions. It provides all the processing power and interfaces needed to operate the probe and instruments and route data to the telecom system during the science mission. The Sphynx C&DH is packaged together with another JPL smallsat development, the IRIS radio, operating in the UHF-band. Though designed for deep space applications, IRIS and Sphynx share a cubesat form factor, allowing efficient packaging in a single stack for each of two redundant strings on the upper equipment deck which aids accommodation in the small volume of the probe. The telecom subsystem also includes redundant 20W solid state power amplifiers (SSPAs) to support the data link to the orbiter. A 25-cm square patch antenna is located on the top of the descent module.

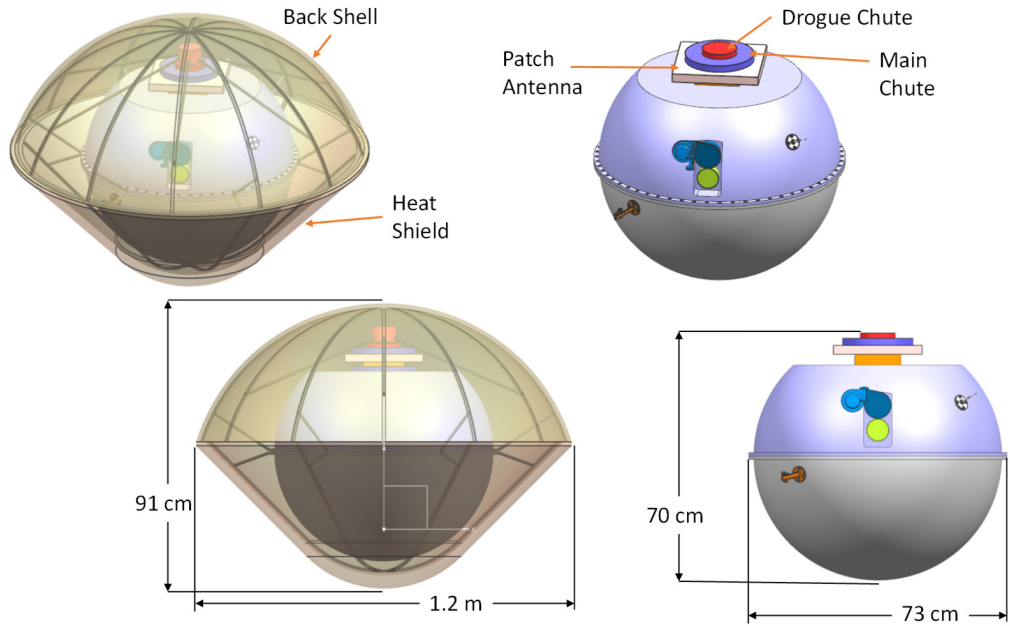

**Figure 4-8.** Probe concept configuration.

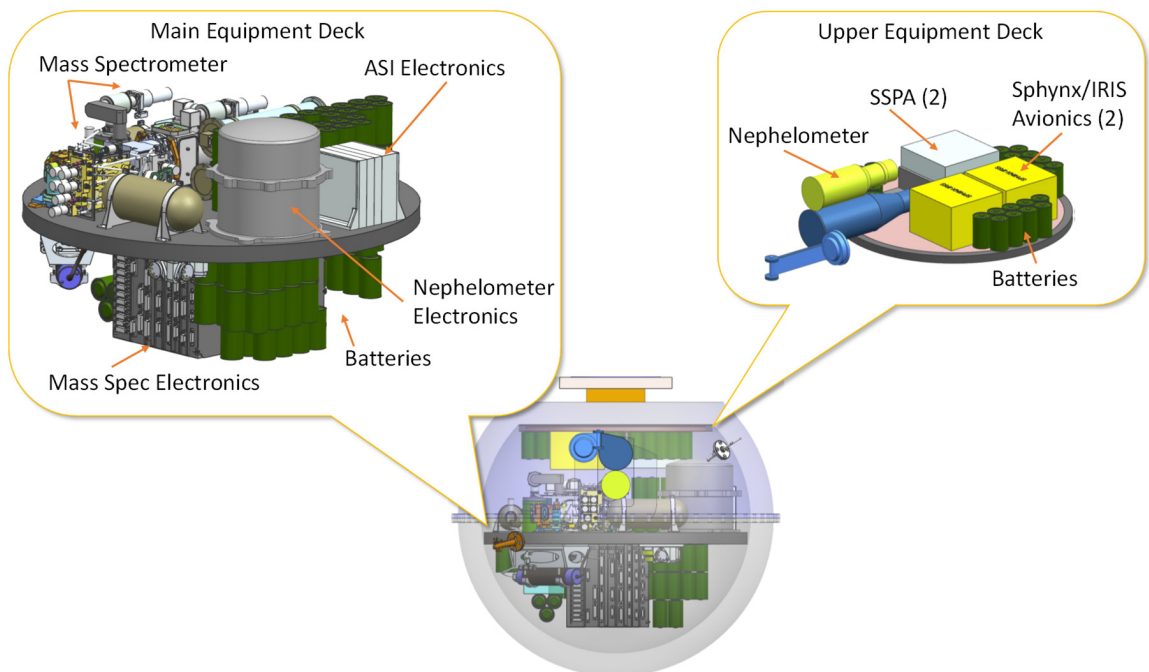

**Figure 4-9.** Probe concept equipment layout.

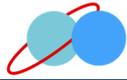

Power is provided by 19 nine-cell strings of Li-ion primary batteries providing a capacity of 32 Ahr after a 13-year cruise. Power electronics are provided in two block redundant strings, each comprising a power switching/Event Timer Module (ETM) slice and a power distribution/pyro slice. The power electronics is based on JPL heritage designs and uses the CPCI 3U form factor. Survival and checkout power is provided from the orbiter during cruise through the power switching slice. During the coast phase following deployment from the orbiter, probe subsystems are turned off. RHUs provide survival heating and the ETM counts down the time to probe activation prior to entry.

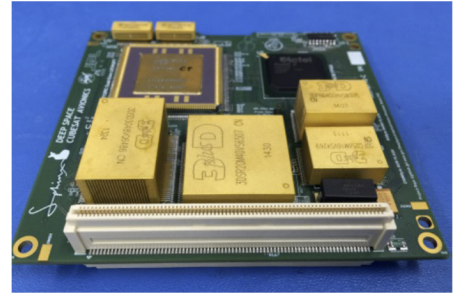

**Figure 4-10.** Sphynx computer.

Thermal control design begins with a thermal insulation shield consisting of a honeycomb insulated shell with aerogel in the core, enclosing the internals of the descent module. The shell is designed to reduce the heat leak from internal components to the cold (as low as 50 K) hydrogen atmosphere during descent. This enclosure forms the outer surface of the Descent Module. Coast phase survival temperatures are maintained through the incorporation of 19 RHUs distributed through the Descent Module. Two passive heat switches are assumed to regulate the RHU heat to a 0.2 m$^2$ radiator located on the outside of shell looking at the interior surface of the aeroshell (spinning aeroshell temperature at Neptune is estimated at 55 K). This provides autonomous, passive regulation of the temperature to 20°C for probe internals. There is a warm-up period before the release of the aeroshell to condition the mass spec instrument to its operating temperature to be ready for data taking at 0.1 bar atmosphere. The warm-up heater is turned off by software control.

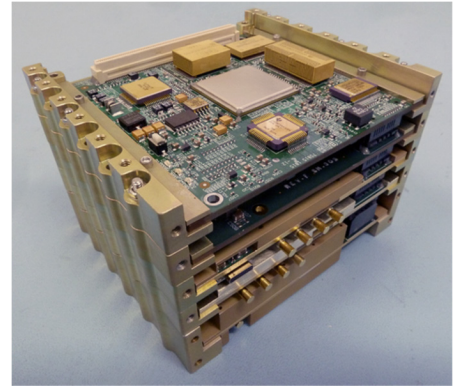

**Figure 4-11.** IRIS UHF radio.

Mass-Equipment Lists (MELs) for the Uranus and Neptune probes are given in **Tables 4-8** and **4-9**. The MEL for the Descent Module is identical for the two probes. This MEL shows a current best estimate (cbe) mass for the Descent Module of 121.3 kg. The Maximum Expected Value (MEV) mass, including heritage-based contingency, is 154.2 kg. Finally, per JPL Design Principles, the fully margined mass is estimated at 174 kg.

While the Descent Module design was assumed to be identical for Uranus and Neptune, there are slight differences in the Entry System mass, resulting from the differing entry conditions at the

**Table 4-8.** Uranus Probe concept MEL.

| Uranus Probe | CBE Mass (kg) | Contingency (%) | Total Mass (kg) | Heritage/Comments |
|---|---|---|---|---|
| Instruments | 25.3 | 29% | 32.5 | GCMS, ASI, Neph, OP |
| C&DH | 0.6 | 17% | 0.7 | Sphynx dual string |
| Power | 20.1 | 26% | 25.4 | Batteries |
| Telecom | 6.2 | 26% | 7.8 | IRIS UHF |
| Structures | 61.3 | 30% | 79.7 | Honeycomb Shell |
| Thermal | 8.0 | 3% | 8.2 | |
| **Descent Module Total** | **121.5** | **27%** | **154.3** | |
| System Margin | | | 19.4 | |
| **Dry Mass Total** | | **43%** | **173.7** | |
| Entry System | 102.8 | 43% | 147.0 | Aeroshell + parachutes |
| **Probe Entry Mass Total** | | | **320.7** | |

…

**Table 4-9.** Neptune Probe MEL.

| Neptune Probe | CBE Mass (kg) | Contingency (%) | Total Mass (kg) | Heritage/Comments |
|---|---|---|---|---|
| Instruments | 25.3 | 29% | 32.5 | |
| C&DH | 0.6 | 17% | 0.7 | Sphynx dual string |
| Power | 20.1 | 26% | 25.4 | Batteries |
| Telecom | 6.2 | 26% | 7.8 | IRIS UHF |
| Structures | 61.3 | 30% | 79.7 | Honeycomb Shell |
| Thermal | 8.0 | 3% | 8.2 | |
| **Descent Module Total** | **121.5** | **27%** | **154.3** | |
| System Margin | | | 19.4 | |
| **Dry Mass Total** | | **43%** | **173.7** | |
| Entry System | 103.4 | 43% | 147.9 | Aeroshell + parachutes |
| **Probe Entry Mass Total** | | | **321.5** | |

two targets. Entry system mass (MEV) varies by about 1 kg between the two planets, as a result of differing TPS requirements.

### 4.3.6.2 Orbiter Design

Design of the Uranus orbiter was driven largely by the high delta V requirements of the mission, as well as the integrated design challenges of power and thermal control for such a long duration mission at large distances from the Sun. For this option, the orbiter also incorporates a SEP stage for use in the inner solar system. The SEP stage uses a 2+1 (two active + one spare) arrangement of NEXT ion engines to provide a delta V of ~5,600 m/s. ROSA solar arrays provide ~30 kW of power at Beginning of Mission (BOM) at 1 AU. The SEP stage is jettisoned about 6.5 years after launch at a solar range of ~6 AU.

The orbiter configuration is shown in **Figure 4-12**. It is a 3-axis stabilized spacecraft with a baseline power system that includes four enhanced Multi-Mission Radioisotope Generators (eMMRTGs).

The flight system includes a fixed 3-m high gain antenna (HGA) using Ka band for science data downlink. The planning payload of three instruments plus radio science is accommodated with the two imaging instruments located near the upper end of the spacecraft and the magnetometer on a deployable boom. Accommodation for the atmospheric probe is provided on the upper deck. Four eMMRTGs are mounted on the lower end of the spacecraft, providing a total of about 376 W at end of mission (EOM, 15 years after launch). The baseline power subsystem also incorporates a 10 A-hr Li-ion battery sized to provide energy balance throughout the science phase of the mission.

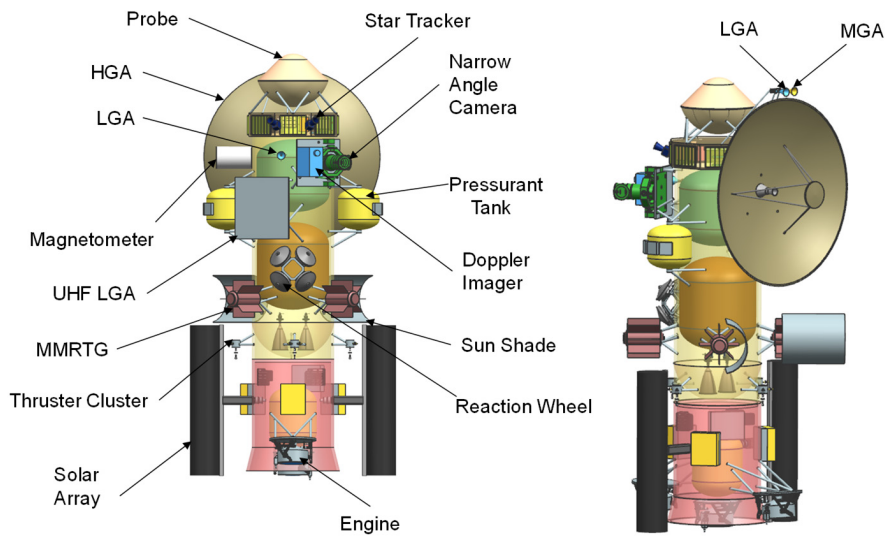

**Figure 4-12.** Option 1 Flight System in launch configuration.

The C&DH subsystem is built on JPL's "reference bus" platform, using a design derived from recent missions including MSL and SMAP. The data processing and handling architecture includes a dual string RAD750 computer that is capable of performing all science and engineering functions. Memory includes 1,024 Gb of storage, providing more than 100% margin over orbital requirements.

Spacecraft attitude is controlled primarily with reaction wheels during science operations. Small 1 N monopropellant RCS thrusters are used to reduce post-launch separation rates, as well as to provide attitude control during cruise. These thrusters will also be used for desaturation of the reaction wheels.

The propulsion system has a dual-mode architecture, which includes two 890 N bipropellant main engines in addition to the eight 1 N RCS thrusters and four 20 N thrusters to provide attitude control during main engine burns. Total $\Delta$V capability of the orbiter is ~2,600 m/s.

Waste heat from the eMMRTGs is exploited for thermal control to the maximum extent practical to reduce the use of electrical power for heaters. RHUs and variable RHUs are employed to further reduce electrical requirements.

The flight system also incorporates a SEP stage for efficient $\Delta$V augmentation during the first part of the cruise trajectory. The SEP stage was developed as a simple, bolt-on augmentation built around and incorporating the function of a Launch Vehicle Adapter (LVA). The basic LVA structure is used to support two 15 kW (BOM, at 1 AU) roll-out solar array (ROSA) wings, as well as three NEXT ion thrusters, power processing units (PPUs), Xenon tanks, and electronics necessary to the control and operation of this self-contained stage. Interfaces with the launch vehicle and orbiter have been maintained as simple as possible to allow the flexibility to operate with or without the SEP stage without significant changes to orbiter configuration.

The integrated flight system has a total wet launch mass of 6,886 kg, and comprises a 1,686 kg dry orbiter with 2,354 kg of bipropellants, a 1,486 kg dry SEP stage with 1,040 kg of Xenon propellant for its NEXT-based ion thrusters, and a 321 kg atmospheric probe. See **Table 4-10** for the flight system mass summary.

The atmospheric probe is slated to be released 60 days prior to UOI. The orbiter propellant mass is sized to accommodate this release scenario.

Subsystem engineers evaluated the maturity of their designs and applied appropriate contingency at the component level. Then, system level mass (222 kg on the orbiter and 183 kg on the SEP stage) was added in order to achieve a 30% margin as required by JPL Design Principles for a concept at this stage of development. The eMMRTG mass was provided as a not-to-exceed allocation and no additional contingency or system margin is held against this line item.

### 4.3.7    New Technology

A guiding philosophy of the mission architectures selected was that new technologies should be used only where their inclusion would make a significantly enhancing impact either by increasing mission return or reducing cost. In the case of Option 1, no new technologies were required to execute the mission. The SEP stage may be considered an advanced engineering development in that a separable stage of this type has not previously been flown, however all components assumed in its design represent current state-of-the-art technologies. Interestingly, a finding from this particular option as the SEP stage was developed by Team X was that the overall mass of the stage was negating its benefit and it was found that non-SEP stage mission designs could achieve the mission objectives at a significantly lower cost, as shown in Option 5. This conclusion was not evident based on initial assumptions for SEP stage mass, and it is a recommendation of the study that further work to perform detailed design of a specialized SEP stage be completed to more fully investigate the value of this architecture option.

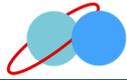

**Table 4-10.** Option 1 MEL.

| Orbiter | CBE Mass (kg) | Contingency (%) | Total Mass (kg) | Heritage/Comments |
|---|---|---|---|---|
| Instruments | 36.7 | 23% | 45.2 | |
| C&DH | 21.6 | 10% | 23.8 | |
| Power | 216.6 | 2% | 220.8 | No contingency on eMMRTGs |
| Telecom | 59.4 | 16% | 68.9 | |
| Structures | 462.7 | 30% | 601.5 | |
| Harness | 86.3 | 30% | 112.3 | |
| Thermal | 112.7 | 23% | 138.9 | |
| Propulsion | 173.3 | 5% | 182.7 | |
| GN&C | 63.5 | 10% | 69.8 | |
| **Orbiter Total** | **1232.8** | **19%** | **1463.9** | |
| System Margin | | | 221.7 | No margin on RPS |
| **Dry Mass Total** | | **43%** | **1685.5** | |
| | | | | |
| **Propellant** | | | **2354.0** | |
| | | | | |
| **Wet Mass Total** | | | **4039.5** | |
| SEP Stage | CBE Mass (kg) | Contingency (%) | Total Mass (kg) | Heritage/Comments |
| C&DH | 1.6 | 0.1 | 1.7 | MREU |
| Power | 263.9 | 0.3 | 340.9 | 30 kW ROSA |
| Mechanical | 444.0 | 0.3 | 577.2 | |
| Thermal | 78.7 | 0.0 | 78.7 | |
| Propulsion | 245.0 | 0.2 | 297.8 | |
| GN&C | 6.0 | 0.1 | 6.4 | gimbal drives, sun sensors |
| **SEP Stage Total** | **1039.2** | **25%** | **1302.7** | |
| System Margin | | | 183.4 | |
| **Dry Mass Total** | | **43%** | **1486.1** | |
| | | | | |
| **Propellant** | | | **1040.0** | |
| Xenon | | | 1040.0 | |
| **Wet Mass Total** | | | **2526.1** | |
| Mission System | CBE Mass (kg) | Contingency (%) | Total Mass (kg) | Heritage/Comments |
| Probe | | | 320.7 | |
| Orbiter | | | 4039.5 | |
| SEP Stage | | | 2526.1 | |
| **Launch Mass Total** | | | **6886.4** | |
| Injected Mass Cap. | | | 10120.0 | Delta IVH |
| **Remaining LV Cap.** | | | **3233.6** | |

### 4.3.8    Cost

Team X estimated the full mission cost of Option 1 to be $1.93B ($FY15) as shown in **Table 4-11**. This cost includes 30% reserves for Phases A–D and 15% in Phase E. Note that per the groundrules, no reserves were carried on eMMRTG and DSN costs.

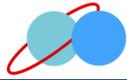

**Table 4-11.** Option 1 Team X Cost Summary.

| WBS Elements | NRE | RE | 1st Unit |
|---|---|---|---|
| **Project Cost (no Launch Vehicle)** | **$1351.0 M** | **$594.4 M** | **$1945.4 M** |
| **Development Cost (Phases A - D)** | **$1004.5 M** | **$594.3 M** | **$1598.8 M** |
| 01.0 Project Management | $47.3 M | | $47.3 M |
| 02.0 Project Systems Engineering | $23.7 M | $0.5 M | $24.3 M |
| 03.0 Mission Assurance | $52.9 M | $0.0 M | $52.9 M |
| 04.0 Science | $24.8 M | | $24.8 M |
| 05.0 Payload System | $80.2 M | $48.3 M | $128.5 M |
| 5.01 Payload Management | $7.8 M | | $7.8 M |
| 5.02 Payload Engineering | $5.8 M | | $5.8 M |
| Orbiter Instruments | $33.5 M | $24.3 M | $57.8 M |
| Narrow Angle Camera | $11.6 M | $8.4 M | $20.0 M |
| Doppler Imager | $17.4 M | $12.6 M | $30.0 M |
| Magnetometer | $4.5 M | $3.3 M | $7.8 M |
| Probe Instruments | $33.1 M | $24.0 M | $57.1 M |
| Mass Spectrometer | $22.9 M | $16.6 M | $39.6 M |
| Atmospheric Structure Investigation (ASI) | $3.4 M | $2.5 M | $5.9 M |
| Nephelometer | $5.3 M | $3.8 M | $9.1 M |
| Ortho-para H2 meas. Expt. | $1.5 M | $1.1 M | $2.6 M |
| 06.0 Flight System | $496.1 M | $386.7 M | $882.8 M |
| 6.01 Flight System Management | $5.0 M | | $5.0 M |
| 6.02 Flight System Systems Engineering | $51.1 M | | $51.1 M |
| Orbiter | $297.2 M | $236.6 M | $533.8 M |
| SEP Stage | $50.6 M | $106.1 M | $156.7 M |
| Probe | $26.8 M | $18.1 M | $44.9 M |
| Entry System | $57.1 M | $24.4 M | $81.5 M |
| Ames/Langley EDL Engineering/Testing | $3.8 M | $0.0 M | $3.8 M |
| 6.14 Spacecraft Testbeds | $4.5 M | $1.5 M | $6.0 M |
| 07.0 Mission Operations Preparation | $27.0 M | | $27.0 M |
| 09.0 Ground Data Systems | $22.1 M | | $22.1 M |
| 10.0 ATLO | $21.1 M | $21.7 M | $42.8 M |
| 11.0 Education and Public Outreach | $0.0 M | $0.0 M | $0.0 M |
| 12.0 Mission and Navigation Design | $30.9 M | | $30.9 M |
| Development Reserves | $178.3 M | $137.1 M | $315.5 M |
| **Operations Cost (Phases E - F)** | **$313.5 M** | **$0.1 M** | **$313.6 M** |
| 01.0 Project Management | $27.1 M | | $27.1 M |
| 02.0 Project Systems Engineering | $0.0 M | $0.1 M | $0.1 M |
| 03.0 Mission Assurance | $3.6 M | $0.0 M | $3.6 M |
| 04.0 Science | $69.2 M | | $69.2 M |
| 07.0 Mission Operations | $149.5 M | | $149.5 M |
| 09.0 Ground Data Systems | $28.8 M | | $28.8 M |
| 11.0 Education and Public Outreach | $0.0 M | $0.0 M | $0.0 M |
| 12.0 Mission and Navigation Design | $0.0 M | | $0.0 M |
| Operations Reserves | $35.3 M | $0.0 M | $35.4 M |
| **8.0 Launch Vehicle** | **$33.0 M** | | **$33.0 M** |
| Launch Vehicle and Processing | $0.0 M | | $0.0 M |
| Nuclear Payload Support | $33.0 M | | $33.0 M |

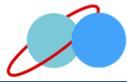

## 4.4    Mission Option 2: Uranus Orbiter and SEP Stage with ~150-Kg Payload and No Probe

### 4.4.1    Overview

The second mission option assessed by Team X was an architecture comprising a Uranus orbiter with an expanded 150 kg science payload, without an atmospheric probe. As with Option 1, the orbiter would make use of a SEP stage in the inner solar system to provide additional delta-V. The SEP stage for this option uses a 2+1 (two active + one spare) arrangement of NEXT ion engines to provide a delta V of ~5,600 m/s. ROSA solar arrays provide ~30 kW of power at BOM at 1 AU.

The mission design for this concept is essentially the same as Option 1, launching in July of 2030, and executing a ~11-year trajectory using VVE gravity assists. The SEP stage is jettisoned about 6.5 years after launch at a solar range of ~5 AU. Upon arrival, the orbiter performs a 2,260 m/s UOI burn, putting it into an initial 150-day orbit around Uranus, which is dropped during the subsequent 4-year orbital phase to approximately 50 days, and which includes multiple satellite flybys.

### 4.4.2    Science

This architecture was selected for detailed study to explore the effects of having a large orbiter science payload. A probe was not included both because A-Team work suggested it would not fit within our cost guidelines and to allow us to explore the "no probe" region of parameter space. This mission achieves all priority science goals except the one related to bulk composition including noble gases and isotopic ratios. That goal requires an atmospheric probe and is one of our two highest-priority goals. Because of the probe's importance, the SDT would prefer not to fly this mission, though we find it does achieve significant science. The SDT also notes that this architecture would be highly desired should a separate mission execute a probe mission.

### 4.4.3    Instrumentation

Instruments comprising the "150 kg" orbiter payload for this option include the three instruments from the 50 kg payload of Option 1:

- Narrow angle camera
- Doppler imager
- Magnetometer

These are augmented by an additional 12 instruments:

- Vis.-near IR mapping spectrometer
- Mid-IR spectrometer
- UV imaging spectrometer
- Radio waves
- Low energy plasma
- High energy plasma
- Thermal IR
- Energetic and neutral atoms
- Dust detector
- Langmuir probe
- Microwave sounder
- Wide angle camera

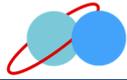

Features of the individual instruments are summarized in **Table 4-12**. An illustration of their location on the orbiter is shown in **Figures 4-1** and **4-12**.

**Table 4-12.** Option 2 Orbiter 150-kg payload instruments.

| Instrument | Deployments | Cooler Temp | Bus Data Rate (kbps) | FOV | Comments |
|---|---|---|---|---|---|
| Doppler imager | Cover | -35°C | 10 | 0.4, 7 µrad IFOV | Principle use on approach |
| NAC | Cover | -35°C | 1280 | 2.3×1.2 10 µrad IFOV | |
| WAC | Cover | -35°C passive | 1,680 | 10.5×10.5 degrees, 175 µrad IFOV | |
| Thermal infrared | Cover | none | 90 | 67 mrad 6.7×3.4 mrad ifov, crosstrack | 9 bands, 21 pix/band |
| VNIR | Cover | 105 K passive | 625 | 4×4 mrad fov | Spot spectrometer 0.4–4.3 microns, R=300 |
| MidIR spectrometer | Cover | None | 72 | 8 mrad circular | 5–50 microns |
| Ultraviolet spectrometer imager | Cover | None | 180 | 6 degrees long, IFOV~ 0.6 degrees | 520–1870å |
| MAG | Boom | None | 128 | 4π | 3 axes, inboard and outboard, 10 m boom |
| Plasma wave spectrometer | 2 antennae | None | 1500 | | 7.1 meter whip antennae |
| Low energy plasma | Cover | None | 50 | 276×10 degrees | 35 ev to 7.5 kev |
| High energy plasma | Cover | None | 50 | 160×12 degrees | 25 Kev to 1MEV |
| Energetic neutral atoms | Cover | None | 350 | 90×120 degrees | |
| Dust | None | None | 100 | ~2 pi | 12 flat plate collectors |
| Langmuir probe | 0.9 m boom | None | 10 | ~4pi | 5 cm sphere |
| Microwave sounder | None | None | 10 | 20 to 150 degrees | 6 bands, 5 flat antennae, 1 horn |
| Radio science | None | None | NA | See Telecom | 2 band, 2 way coherent tracking |

#### 4.4.3.1   Instrument Descriptions

##### *Narrow Angle Camera (NAC)*
See Section 4.3.3.3.

##### *Wide Angle Camera (WAC)*
The wide angle camera has a 1k×1k focal plane and 12 filters. Its IFOV is approximately 20 times that of the NAC camera resulting in an FOV of about 10×10 degrees. The WAC is utilized for tracking atmospheric features at closest approach, large scale images of the rings and global imaging of the satellites. No sources of scattered light may be with 15 degrees of center of boresite.

##### *Doppler Imager (DI)*
See Section 4.3.3.3.

##### *Visible-Near IR Mapping Spectrometer (VNIRMS)*
The visible-NIR mapper is a point spectrometer utilizing a wedge filter over the range 0.4 to 4.3 microns and an HgCdTe (mercury cadmium telluride) focal plane array operated at 105 K.

### Mid-IR Spectrometer (MIS)

The mid-IR spectrometer is a FTIR point spectrometer over the range 5.0 to 50 microns. The detector is uncooled.

### Thermal Imager (TI)

The thermal imager is an imaging thermal spectrometer with 9 bands spanning the range 0.35 to 400 microns. Each band has 21 pixels. The detector is an array of microthermopiles and is not cooled.

### Ultraviolet Imaging Spectrometer (UIS)

The ultraviolet instrument has a passband of 520–1870 microns. Its boresight is co-aligned with imaging.

### Magnetometer (MAG)

See Section 4.3.3.3.

### Radio Waves (RW)

The radio waves experiment has two 7.1-meter whip antennas that are deployed after launch. The radio waves experiment has a high rate mode for operations near flux tubes and rapidly changing magnetosphere, and several lower rates for monitoring the magnetosphere throughout the orbit.

### Low Energy Plasma (LEP)

The low energy plasma requires a clear FOV. The preferred orientation is in the ram direction while the spacecraft is in 3-axis stabilized mode.

### High Energy Plasma (HEP)

The high energy plasma requires a clear FOV. The preferred orientation is in the ram direction while the spacecraft is in 3-axis stabilized mode.

### Energetic and Neutral Atoms (ENA)

The Energetic and Neutral Atoms experiment bore sight is aligned with NAC to support approach imaging of the magnetosphere.

### Dust Detector (DD)

The dust detector has 12 flat plate collectors, placed on the instrument platform. Each detector observes 2Pi sky.

### Langmuir Probe (LM)

The Langmuir probe has a 5 centimeter titanium ball supported by a 0.8 meter boom.

### Microwave Sounder (MWS)

The microwave sounder is a radiometer receiver with six bands having 6 corresponding antennae. The highest frequency antenna is a feed horn. The other frequencies use flat plate antennas co-located on one side of the spacecraft. The largest antenna is 1.7×1.7 meters. The antenna are co-aligned with the NAC bore sight to enable simultaneous imaging of the atmosphere.

### Radio Science

See Section 4.3.3.3.

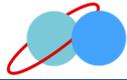

#### 4.4.3.2    Payload Utilization

The more complex 150 kg payload can, to simplify discussion of operations, be binned into remote sensing instruments (RS suite) (NAC, WAC, DI, VNIRMS, MIS, TI, UIS, MWS), fields and particle instruments (F&P suite) (RW, MAG, LEP, HEP, ENA) and other (RS, Dust). Generally, the F&P instruments and dust are on continuously in a low data rate mode. Higher rate observations occur at closest approach to the primary and satellites, and in critical parts of the magnetosphere. The remote sensing instruments obtain global views (movies) of the primary at apoapse, feature tracks and high-spatial resolution observations near periapse, and mapping and high spatial resolution observations of the satellites near closest approach.

The approach sequence for the instruments also included in the 50 kg payload remains the same as for Option 1. Additional activities include imaging of the magnetosphere by ENA and global coverage of the primary by all remote sensing instruments. The dust instrument remains on at all times.

With 5 RTGs in the baseline power subsystem there would be sufficient power to operate all instruments, and telemetry, at the same time. However, telemetry bandwidth is sufficiently small that high data volume observations must be carefully allocated. Thus the Doppler imager is nominally used only on approach, only occasional use of the full capability of the high-rate instruments (especially cameras, spectrometers, and ENA and PW) is possible, and the F&P instruments have modes that support low data volume monitoring of long temporal duration.

The RS suite has their boresights co-aligned to enable simultaneous observations. However, the fields of view and modes of operation of the instruments are quite different, so allocation and design of observing sequences could become quite complex. This could be ameliorated with modifications to instrument design, and perhaps addition of articulation to select instruments or the RS suite during spacecraft design.

### 4.4.4    Mission Design

The mission design objective for the second mission option was to enable a Uranus orbiter with an enhanced science payload (~150 kg). The high-level mission design guidelines for this options were:

1. Launch between 2024–2037, with a preferred launch date between 2029–2031
2. Total mission lifetime <15 years (including Uranus Science phase)
3. Launch on an existing commercial Launch Vehicle
4. Avoid Uranus rings during orbit insertion
5. Uranian moon tour with two flybys each of the major five moons

As with the first Team X study, it was assumed that an SEP-enabled architecture would be needed for this mission. A purely chemical propulsion based architecture (without SEP stage) was also studied later (see Mission Option 6). The SEP stage was augmented with a biprop stage on the spacecraft which was primarily used for orbit insertion and the uranian tour (same as Mission Option 1). Given that Uranus is almost ~20 AU away from the Sun, reaching the planet in a reasonable amount of time results in high approach velocity, leading to large orbit insertion ΔV. Hence, dropping the SEP stage before orbit insertion saves a significant amount of propellant. **Figure 4-13** depicts the SEP baseline mission architecture, which can be divided into three mission phases:

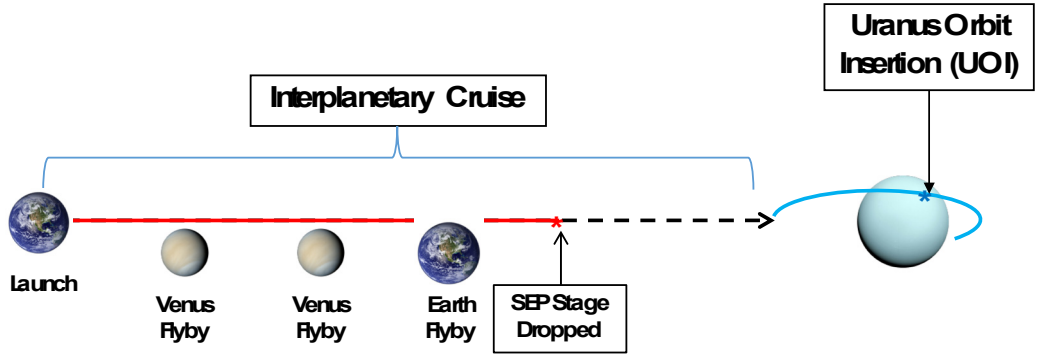

**Figure 4-13.** Mission Concept Option 2: SEP mission design architecture.

### 4.4.4.1    Launch and Interplanetary Cruise

Similar to Option 1, the baseline interplanetary trajectory relies on a 25-kW SEP stage powered by 2 NEXT ion engines, to propel the spacecraft within the inner solar system. The spacecraft launches on a Delta-IV Heavy, one of NASA's highest performing launch vehicles. The spacecraft then performs two Venus flybys followed by one Earth flyby to gain momentum in the inner solar system. The SEP stage thrusting in the inner solar system makes up for the relatively low launch energy while enjoying high propellant efficiency. After thrusting in the inner solar system the SEP stage is dropped at ~6 AU, as the solar flux beyond this point is insufficient to power the ion engines. The SEP stage dry mass is significant and dropping it before UOI results in significant propellant savings on the orbiter.

**Table 4-13** lists two trajectories. The first one (in blue) is the in-session interplanetary trajectory which was used during the Team X session for initial design. The second trajectory (in black) represents further post-session refinements made to the in-session trajectory after the Probe mass, SEP stage dry mass and orbiter dry mass became known. **Note that the Team X in-session and the refined trajectory both deliver the same or more mass into Uranus orbit than the allocated amount in the MEL.** Also, note that the refined trajectory uses ~300 kg less Xenon propellant. After the Team X session, structural refinements were made to the SEP stage which resulted in significant mass savings, allowing this propellant reduction. The refined SEP stage was still sized to carry the larger Xenon load (~1,034 kg), and could be refined further to reflect the ~300 kg reduction in Xenon. A future study may expect further SEP stage dry mass reductions which could be translated into reductions in mission cruise time or increased payload mass.

**Table 4-13.** Option 2 Team X (blue) and refined baseline (black) mission trajectory

| Flyby Sequence | Launch Vehicle | Launch Date | Launch C3 (km²/s²) | IP TOF (yrs.) | Xenon Mass (kg) | SEP-Stage Dry Mass (kg) | Arrival Mass (kg) | Orbit Insertion ΔV (km/s) | Mass in Orbit (kg) |
|---|---|---|---|---|---|---|---|---|---|
| Earth-VVE-Uranus | Delta-IV Heavy | 9/03/2030 | 15.4 | 11.0 | 1034 | 1700.0 | 5099.8 | 2.3 | 2498.9 |
| Earth-VVE-Uranus | Delta-IV Heavy | 9/23/2030 | 21.9 | 11.0 | 741 | 1484.9 | 4753.5 | 2.3 | 2361.6 |

The quantity "*Arrival Mass*" refers to mass of the spacecraft before UOI and after SEP stage separation. Orbit insertion ΔV for the baseline is calculated for a notional 140-day capture orbit with periapsis at 1.05 Uranus Radii. **Figures 4-14** and **4-15** show the two trajectories listed in **Table 4-13**.

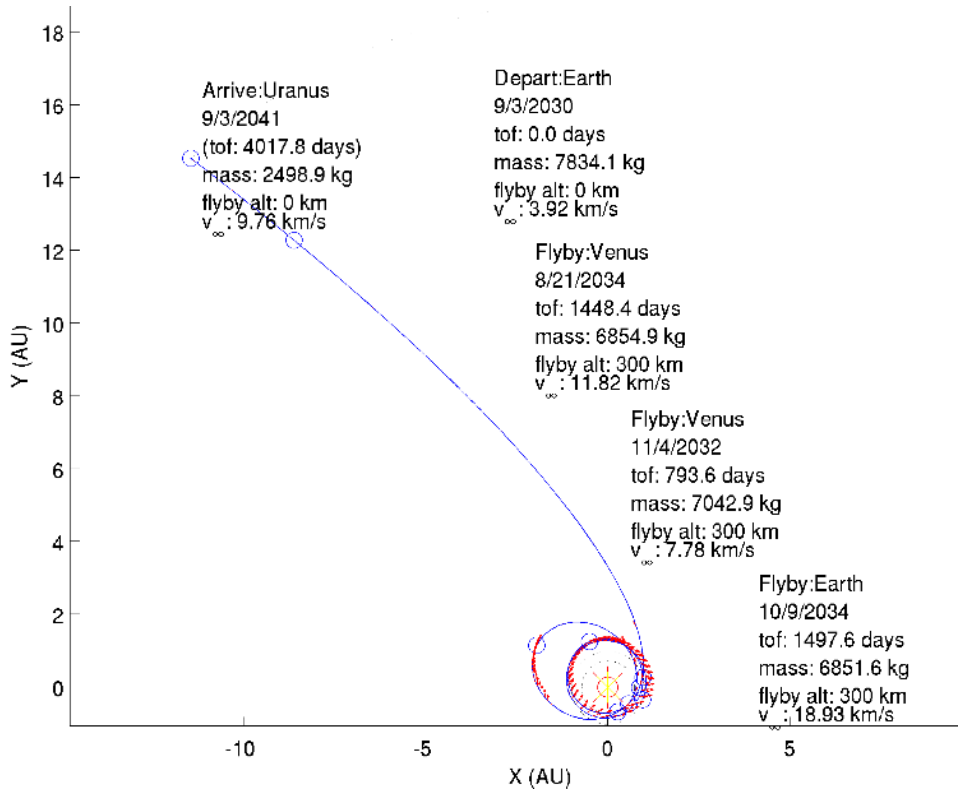

**Figure 4-14.** Option 2 Team X in-session trajectory.

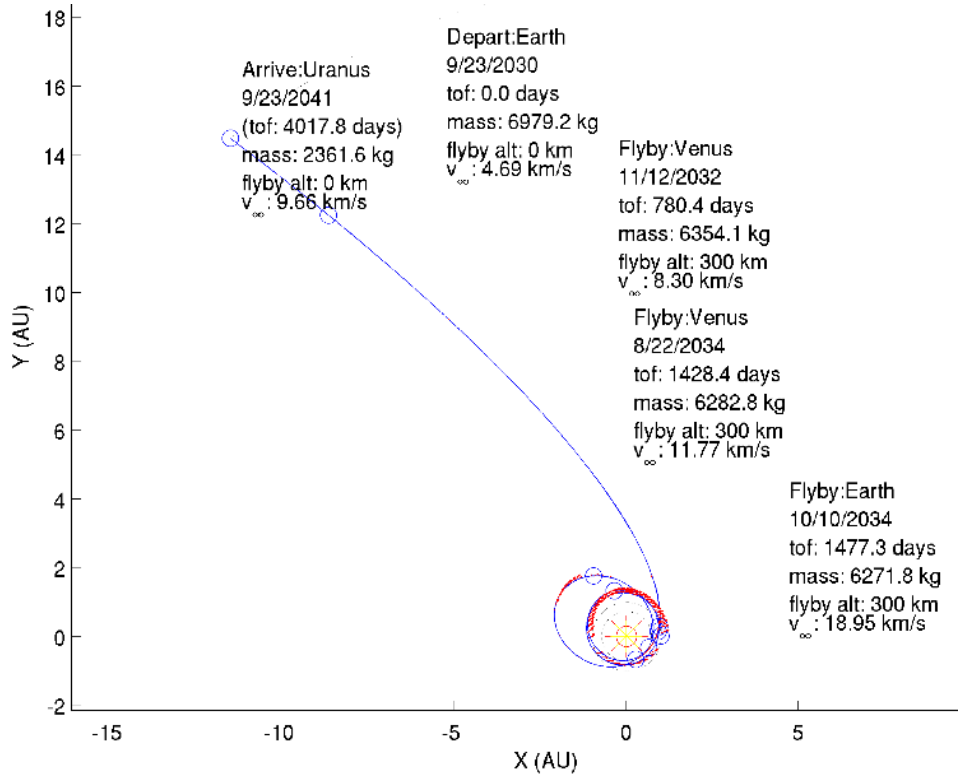

**Figure 4-15.** Option 2 refined baseline trajectory.

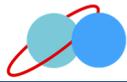

### 4.4.4.2   Uranus Orbit Insertion

The in-session baseline mission trajectory performs a UOI periapsis targeting maneuver (UPTM) 10 days before the target UOI. The orbiter performs a UOI ΔV of ~2.26 km/s at a periapsis altitude of ~1.05 Uranus radii. The UOI burn lasts for ~1 hr. using the two 890N engines on the spacecraft. Post UOI the orbiter enters a ~140-day orbit around Uranus. An orbit insertion altitude relatively close to the atmosphere is chosen to mitigate potential ring particle impingement issues. The low

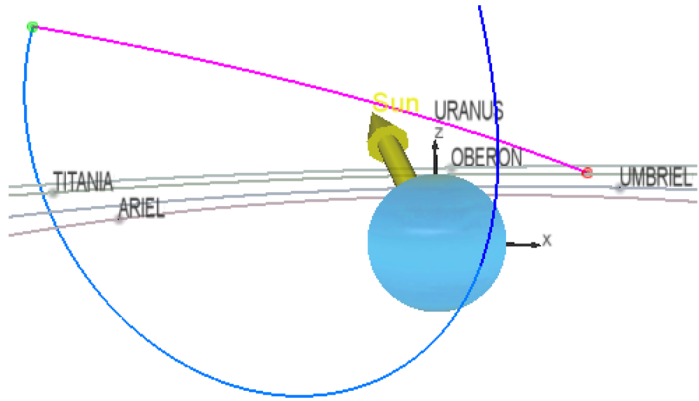

**Figure 4-16.** Option 2 UOI and ATTM.

UOI altitude also helps in reducing the UOI ΔV magnitude. **Figure 4-16** shows the UOI, ATTM, which results in targeting Titania.

### 4.4.4.3   Uranus Tour Phase

Noting that the Uranus obliquity is 97.7 degrees, this has consequences for a uranian satellite tour. To achieve relatively low flyby speeds (<6 km/s) at the five major uranian moons, the orbiter needs to change its orbital plane (wrt. Uranus ring plane) after orbit insertion.

For the baseline trajectory, the orbiter goes into ~69 degree inclined orbit (wrt. Uranus equatorial plane) around Uranus after orbit insertion. For Option 1, Team X assumed a 4-year tour at Uranus with at least two flybys of the five major moons (Miranda, Umbriel, Ariel, Titania, and Oberon). Subsequent tour design efforts (presented in Option 5) revealed that a basic Uranian moon tour that achieves all the science objectives could be accomplished in 2 to 2.5 years after UOI. The flyby velocities at the moons are expected to be <5 km/s. Low flyby velocities are attainable due to the ATTM (see **Figure 4-16**). Further details on the tour design can be found in Option 5.

### *Mission Delta-V Summary*

**Table 4-14** shows the summary ΔV table for this mission option. The chemical ΔV is broken down into monoprop and bi-prop burns, based on ΔV magnitude and criticality of the maneuver. The main chemical ΔV driver for a mission to Uranus is the orbit insertion due to relatively high approach $V_\infty$. Following UOI, the orbiter coasts towards its capture orbit apoapsis where it

**Table 4-14.** Mission Option 2 ΔV summary.

| | Biprop (m/s) | Monoprop (m/s) | Comments |
|---|---|---|---|
| Interplanetary TCMs | 0 | 25 | Most of the TCMs are immediately after launch and after SEP stage separation |
| UPTM | 0 | 10 | 60 days before UOI |
| UOI | 2260 | 0 | <1 hr. burn using two 890N engines |
| UOI-CU (2%) | 45 | 0 | |
| ATTM | 300 | 0 | Estimated from Option 5 tour design |
| ATTM-CU (2%) | 0 | 6 | |
| Tour Deterministic | 45 | 40 | Multiple maneuvers |
| Tour Margin | 0 | 20 | For extended mission and contingency |
| Tour Statistical | 0 | 30 | |
| De-orbit and Disposal | 10 | 0 | Crash into planet |
| **Total** | **2650** | **131** | |

performs an ATTM. ATTM is designed to achieve an efficient orbital plane change and targeting one of the major moons of Uranus (Titania). Tour ΔV is based on the tour design work done for mission option 5, scaled to take into account difference in post UOI conditions for this option.

### 4.4.5 Concept of Operations/GDS

The science goals of Option 2 are met by delivering a spacecraft with a large suite of scientific instruments into orbit about the Uranus to tour and investigate the system. This mission's operations concept differs from the previous option in dispensing with operations associated with deployment and relay for the atmospheric probe, and accommodation of observations from an expanded instrument suite.

Launch and SEP cruise operations are the same as those detailed for Option 1, as are the post-SEP cruise activities until the beginning of the Approach science phase at 85 days before UOI. The 55-day Approach Science phase is similar to that of Option 1, being driven by the Doppler science activities. The NAC will create a 4-color mosaic of the planetary system, as well as an approach movie (200 4-color images) towards the end of the approach phase. Intermittent observations with other instruments are also planned during the Approach Science phase. The additional activities include a small set of 10 Vis/NIR Spectrometer images, 2 rotations of the planet (34.4 hours) with the Mid-IR Spectrometer, periodic low rate UV Spectrometer observations at about 6 hours per day, and a small set of low rate ENA measurements. The fields & Particles Suite and Dust Detector are allocated continuous low rate operation throughout the Approach science phase. The science plan for Mission Option 2 is detailed in **Table 4-15**, showing number/cadence of images for optical instruments and data rate levels of continuously observing Fields & Particles instruments.

**Table 4-15.** Mission Option 2 instrument science plan for major mission phases: Approach, Apoapse, Periapse, and Satellite Flyby Science.

| Ice Giants Mission #2 Instrument Science Plan | Approach Obs Info | Satellite Flyby Obs Info | Orbit Obs Info | Apoapse Obs Info | Periapse Obs Info |
|---|---|---|---|---|---|
| WAC | - | 1Gb | 1Gb | - | - |
| NAC | Mov: 200 im - 4clr | 6Gb (1/2 Map) | - | 10-rot/2hr - 4clr | 5-10 F.Trax - 1clr |
| Doppler Imager | 22dy/30s, 33dy/2s | 100 im | - | 3 im/rot | 20im/rot |
| Vis/NIR Map. Spec. | 10 im | 20 im | 20 im | - | - |
| Mid-IR Spectrometer | 2-rot (34.4hr) | 6hr | - | 2-rot (34.4hr) | 6hr |
| UV Imag Spec | Low 6hr/day | High 8hr | - | Nom 12hr/day | High |
| Thermal IR Imager | Nom <50RU | High 3hr | Nom Full Map | - | - |
| Fields & Particles Suite: | Low | Very High | - | Nom | High |
| Radio & Plasma Waves | - | - | - | - | - |
| Low-Energy Plasma | - | - | - | - | - |
| High-Energy Plasma | - | - | - | - | - |
| Magnetometer | - | - | - | - | - |
| Langmuir probe | - | - | - | - | - |
| Dust Detector | Nom | High 24hr | - | Nom | High |
| ENA | Low 10 Obs | High 3 Obs | Nom 10 Obs | - | - |
| Microwave Sounder | - | 6hr | - | - | 1-rot (17hr) |

The baseline DSN plan is to allocate a single 34BWG station at 24/7 continuous tracking and downlink support for approach science. The baseline power system of 5 eMMRTGs is able to accommodate 24 hours/day of continuous Ka-band downlink alongside approach science, if necessary. The approach science will collect a total of 77 Gb of compressed data (267 Gb uncompressed). Using modern compression techniques, a compression ratio of 3.5X is used for optical instruments and 2.0X for particle and fields instruments with minimal loses. The data budget averaged per day on approach is balanced at about 1.4 Gb/day of compressed data recorded and 2.2 Gb/day 34BWG downlink capability (37% margin). The detailed data outline for each instrument is shown in **Table 4-16**. Considering the full instrument suite and anticipating a large amount of on-orbit data, the approach data return rate of 30 kbps is achieved by using a high power communications system (70W radiated, double that of Option 1). A similar high data rate could be achieved by utilizing an array of 2X 34BWG stations, although additional benefits from an extra eMMRTG drove the decision to use the high power communications system. With such a high data rate and downlink margin, the mission will be much more flexible in handling unexpected DSN support issues or conflicts. Any overflow approach science data not returned, however unlikely, is safely stored on high capacity 1024 Gb solid-state data recorders and downlinked alongside on-orbit science after UOI. The approach science finishes out at 30 days before UOI. Throughout the approach phase, periodic optical navigation observations are performed with the NAC to reduce the uncertainty of the Uranus system satellites.

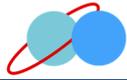

**Table 4-16.** Mission Option 2 instrument science data outline for major mission phases: Approach, Apoapse, Periapse, and Satellite Flyby Science.

| Ice Giants Mission #2 Instrument Science Data | Min Data Rate (Kbps) | Nom Data Rate (Kbps) | Max Data Rate (Kbps) | Approach Data (Gb) | Satellite Flyby Data (Gb) | Orbit Data (Gb) | Apoapse Data (Gb) | Periapse Data (Gb) |
|---|---|---|---|---|---|---|---|---|
| WAC | - | 134 Mb/im/clr | - | - | 1.00 | 1.00 | - | - |
| NAC | - | 134 Mb/im/clr | - | 112.56 | 6.00 | 49.58 | 45.56 | 4.02 |
| Doppler Imager | - | 0.1 Mb/im | - | 148.90 | 0.01 | 0.03 | 0.02 | 0.01 |
| Vis/NIR Map. Spec. | - | 100 Mb/im | 200 Mb/im | 1.00 | 2.00 | 4.00 | - | - |
| Mid-IR Spectrometer | - | 4.00 | - | 0.50 | 0.09 | 0.58 | 0.49 | 0.09 |
| UV Imag Spec | 0.27 | 1.36 | 5.50 | 0.32 | 0.16 | 4.19 | 2.76 | 1.43 |
| Thermal IR Imager | - | 2.27 | 20.00 | 0.26 | 0.22 | 0.02 | | |
| Fields & Particles Suite: | 0.50 | 1.00 | 10 - 100 | 2.38 | 0.79 | 6.65 | 4.06 | 2.59 |
| Radio & Plasma Waves | - | - | - | - | - | - | - | - |
| Low-Energy Plasma | - | - | - | - | - | - | - | - |
| High-Energy Plasma | - | - | - | - | - | - | - | - |
| Magnetometer | - | - | - | - | - | - | - | - |
| Langmuir Probe | - | - | - | - | - | - | - | - |
| Dust Detector | 0.05 | 0.52 | 4.19 | 0.24 | 0.36 | 3.35 | 2.26 | 1.09 |
| ENA | 50 Mb/Obs | 80 Mb/Obs | 150 Mb/Obs | 0.50 | 0.45 | 0.80 | - | - |
| Microwave Sounder | - | 0.72 | - | - | 0.02 | 0.04 | - | - |

| | | |
|---|---|---|
| Science Data Totals: | 266.6 Gb | 11.1 Gb | 70.2 Gb |
| Uncompressed: | 266.6 gb | Uncompressed: | 81.3 Gb |
| Compressed: | 76.9 gb | Compressed: | 25.9 Gb |
| Approach Data Rate: | 1.40 Gb/day | Orbit Data Rate: | 0.52 Gb/day |
| Approach Data Return: 1X 34BWG (30 Kbps) | 2.20 Gb/day | Orbit Data Return: 1X 34BWG (30 Kbps) | 0.73 Gb/day |

Optical Compression: 3.5 X
Fields & Particles Compression: 2.0 X

none
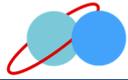

Final instrument systems checkouts are completed during the 30 days prior to UOI. If able to accommodate during the final days/hours before UOI, the thermal IR imager will take observations of the planet when the spacecraft is within ~50 R$_U$. Upon reaching periapse, the large UOI burn will require 1 hour to complete, inserting the spacecraft into its initial 120-day orbit about Uranus.

Initial orbital activities are also similar to those for Option 1. The instrument suite is configured/calibrated for orbiter science and the final ORTs are performed.

As with Option 1, the Orbital Science phase is divided into 3 categories of observation activities: apoapse science and periapse science (divided at 20 R$_U$), as well as satellite flyby science. The major objective of the apoapse science is NAC mapping observations of Uranus and the ring system, imaging 10 rotations of the planet at a 2-hour cadence with full 4-color images. The WAC will capture strategic full system images/mosaics at far distances and high resolution tracks up close to the planet, totaling to about 1 Gb each orbit. Alongside the NAC/WAC, other optical instruments (Vis/NIR, Mid-IR, & UV Spectrometers, and thermal IR imager) are also active during the apoapse period.

The Vis/NIR mapping spectrometer is allocated a total of 20 spectral images each orbit for mapping the planet and ring system. The Mid-IR spectrometer will perform planetary atmospheric observations for up to 2 rotations of the planet during apoapse. The UV imaging spectrometer will observe the planetary atmosphere at a cadence of 12 hours a day during apoapse, capturing auroral movies and limb scans. The thermal IR imager is able to collect enough images for between a half and full global map of the planet during each orbit, with higher resolution mapping near periapse. Throughout the apoapse period, the Fields & Particles Suite and Dust Detector instruments take measurements of the outer Uranus system environment at nominal rates. The ENA instrument will collect up to 10 observations at nominal rates throughout each orbit to get a full picture of the neutral torus around the system. The Doppler imager only performs intermittent targeted observations of Uranus planetary features during the apoapse science phase at about 3 images per planet rotation.

The outer ranges of the periapse science (~20–10 R$_U$) provide excellent opportunities for most optical instruments (NAC, Vis/NIR, Mid-IR, & UV Imaging Spectrometers, and thermal IR imager) to conduct observations of the Uranian satellites and rings. The mid-IR spectrometer will conduct up to 6 hours of observing during periapse. The UV imaging spectrometer will switch to its high rates throughout the periapse period. There will also be opportunities within the periapse science range to perform radio science measurements of the planet and its major satellites. The inner ranges (<10 R$_U$) through periapse are reserved primarily for the Fields & Particles Suite instruments to investigate the inner magnetic fields and particle environment. Throughout the periapse period, the Field & Particles Suite and the Dust Detector instruments will be switched to high rates (10 and 4.19 Kbps respectively) to capture the rapid changes in the environment. The proximity also offers good opportunities for WAC, NAC, Doppler, and other optical instrument feature tracking on the planet or in the rings. The NAC is allocated 5–10 feature tracks, assuming 3 images (single color) for each tracking observation. The Doppler imager will perform its high-resolution feature tracks with an average cadence of about 20 images per planet rotation. The Microwave Sounder will probe the deep atmosphere of the planet for as much of 1-rotation as possible surrounding periapse.

During a typical satellite flyby, nearly all instruments (not the Doppler imager) are used at high rates to observe the satellite's features and interactions with Uranus's fields/particles environment. It is assumed that the NAC is able to map half of the satellite surface during each flyby with 5 Gb of mosaic images along with 1 Gb of high resolution and stereo coverage. The WAC will also capture strategic observations of the satellite throughout closest approach with up to 1 Gb provisioned. The Vis/IR mapping spectrometer is allocated 20 spectral images during the satellite flyby. The Mid-IR spectrometer will observe within ±3 hours of closest approach, while the UV imaging spectrometer will switch to high rate observations within ±4 hours of closest approach. The thermal IR imager will observe for the 3 hours surrounding the closest approach to capture a half global map of the satellite. The Fields & Particles Suite instruments will increase their collection data rate to 10–20 Kbps within ±6 hours of closest approach and up to 100 Kbps during the hour of closest approach. The Dust Detector is switched to high within ±12 hours of the satellite flyby. The flyby will also include up to 3 ENA observation periods. Each satellite flyby may also implement a small set of exploratory Doppler observations (~100 images) of the satellite on approach and/or departure. The microwave sounder is also employed for about 6 hours during closest approach. Throughout the tour, repeated gravity assists will torque the orbiter's inclination more and more in plane with the satellites (from the original 48 deg inclination), while also reducing the period down to 50 days. These final orbital conditions yield longer and better quality flyby science. Within the course of each orbit about 26 Gb of compressed science data (82 Gb uncompressed) will be collected, including 3.5 Gb of compressed satellite flyby science data. In order to efficiently return all of the science data each orbit, the baseline DSN plan is for daily 8 hour passes using only a single 34BWG station at 30kbps. Each pass is capable of downlinking 734Mb of data (assuming 15% overhead) at a flat rate of 30kbps. The data budget averaged per day is 0.52 Gb/day of compressed orbital science recorded, compared to the 0.73 Gb/day downlink capability. As a result, there is a healthy 29% data margin to aid in absorbing any disruptions in DSN support and allow for additional discretionary science activities.

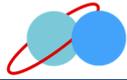

A graph of science data recorded compared to the DSN capacity for the Approach science and Orbit science phases is shown in **Figure 4-17**. The Approach science plot displays the continuous Doppler data collection and Fields & Particles background activities, then ramping up towards the end with the NAC movie recording. The Orbit science plot begins with apoapse data, collecting all of the NAC images of the planet, with continuous Fields & Particles measurements throughout apoapse. Then the 3-day spike of high-rate periapse data is recorded, followed by densely packed satellite flyby science for about a day.

The decommissioning and disposal plan for the mission is also the same as for Option 1.

### 4.4.6 Flight System Design

Design of the Option 2 Uranus flight system is similar to that of Option 1, with major changes limited to those associated with elimination of the probe and accommodation of a greatly enhanced instrument suite. This in turn led to the decision to add one eMMRTG to the baseline power subsystem, and increase telecom RF power to 70 W. As with Option 1, the orbiter also incorporates a SEP stage for use in the inner solar system.

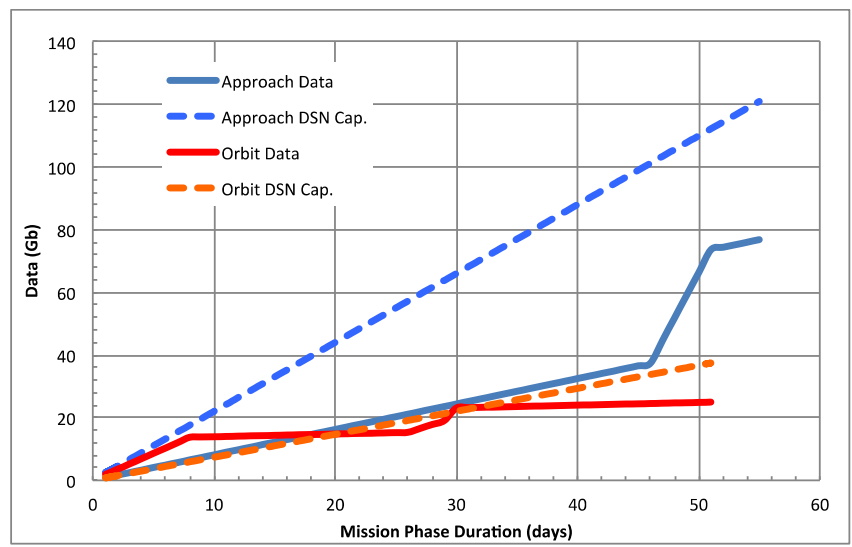

**Figure 4-17.** Mission Option 2 science data volume recorded compared to DSN downlink capacity for Approach science and Orbit science (starting at apoapse for a typical ~50-day orbit).

The orbiter configuration is shown in **Figure 4-18**. It is a 3-axis stabilized spacecraft with a baseline power subsystem that includes five eMMRTGs. The flight system includes a fixed 3-m high-gain antenna (HGA) using Ka band for science data downlink. The planning payload of 15 instruments plus radio science is accommodated with an instrument deck on the upper end of the spacecraft (**Figure 4-19**) and the magnetometer on a deployable boom. The Langmuir Probe and MW Sounder instruments are mounted on the sides of the orbiter. Five eMMRTGs are mounted on the lower end of the spacecraft, providing a total of about 470 W at end of mission (EOM, 15 years after launch). The eMMRTGs are mounted in two stacks of two units each plus one single unit to minimize the number of launch vehicle fairing doors required. The fifth eMMRTG was added to this option to provide extra power for telecom. In this design the telecom subsystem uses a 70 W RF TWTA rather than the 35 W TWTA used in Option 1. The extra RF power is needed to ensure full data return from the greatly expanded instrument suite. The power subsystem also incorporates a 10 A-hr Li-ion battery sized to provide energy balance throughout the science phase of the mission.

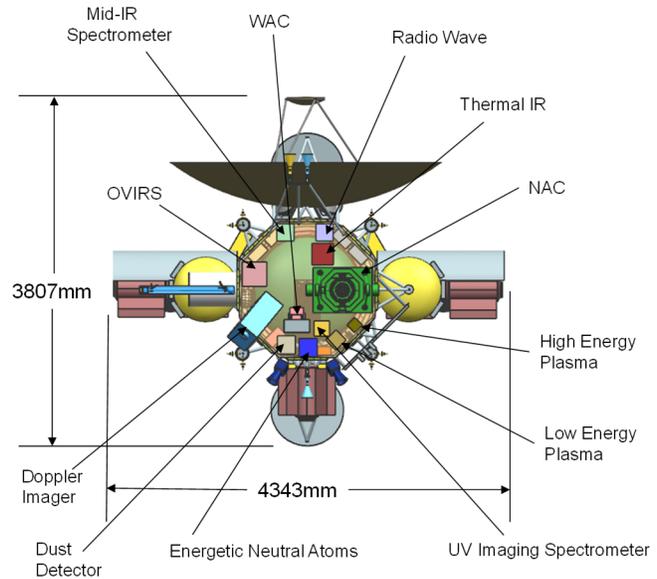

**Figure 4-18.** Option 2 Conceptual Instrument accommodation.

The remaining orbiter subsystems and SEP stage are identical to the design for Option 1.

The integrated flight system has a total wet launch mass of 6203 kg, and comprises a 1608 kg dry orbiter with 2526 kg of bi-propellants, a 785 kg dry SEP stage with 451 kg of Xenon propellant for its NEXT-based ion thrusters. See **Table 4-17** for the flight system mass summary.

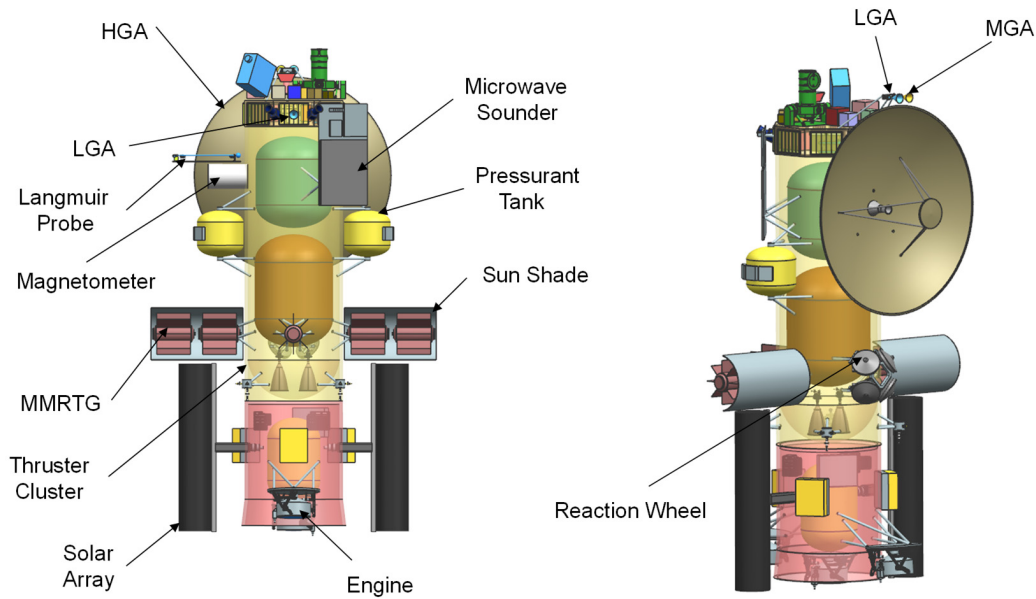

**Figure 4-19.** Option 2 Flight System in launch configuration.

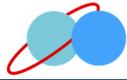

**Table 4-17.** Option 2 MEL.

| Orbiter | CBE Mass (kg) | Contingency (%) | Total Mass (kg) | Heritage/Comments |
|---|---|---|---|---|
| Instruments | 144.8 | 17% | 169.5 | 15 instruments |
| C&DH | 27.6 | 18% | 32.4 | JPL Reference Bus |
| Power | 265.3 | 2% | 269.7 | No contingency on eMMRTGs |
| Telecom | 55.2 | 15% | 63.4 | 70 W Ka band, 3m HGA |
| Structures | 516.4 | 30% | 671.3 | |
| Harness | 103.8 | 30% | 135.0 | |
| Thermal | 121.1 | 23% | 148.5 | |
| Propulsion | 188.8 | 5% | 198.9 | Dual Mode |
| GN&C | 63.5 | 10% | 69.8 | |
| **Orbiter Total** | **1486.5** | **18%** | **1758.6** | |
| System Margin | | | 270.3 | |
| **Dry Mass Total** | | **43%** | **2028.9** | |
| | | | | |
| **Propellant** | | | **2770.0** | |
| | | | | |
| **Wet Mass Total** | | | **4798.9** | |
| SEP Stage | CBE Mass (kg) | Contingency (%) | Total Mass (kg) | Heritage/Comments |
| C&DH | 1.6 | 0.1 | 1.7 | MREU |
| Power | 265.5 | 0.3 | 342.9 | 30 kW ROSA |
| Mechanical | 441.6 | 0.3 | 574.1 | |
| Thermal | 78.7 | 0.0 | 78.7 | |
| Propulsion | 245.0 | 0.2 | 297.8 | |
| GN&C | 6.0 | 0.1 | 6.4 | gimbal drives, sun sensors |
| **SEP Stage Total** | **1038.4** | **25%** | **1301.6** | |
| System Margin | | | 183.3 | |
| **Dry Mass Total** | | **43%** | **1484.9** | |
| | | | | |
| **Propellant** | | | **1040.0** | |
| Xenon | | | 1040.0 | |
| **Wet Mass Total** | | | **2524.9** | |
| Mission System | CBE Mass (kg) | Contingency (%) | Total Mass (kg) | Heritage/Comments |
| Orbiter | | | 4798.9 | |
| SEP Stage | | | 2524.9 | |
| **Launch Mass Total** | | | **7323.9** | |
| Injected Mass Cap. | | | 10120.0 | Delta IVH |
| **Remaining LV Cap.** | | | **2796.1** | |

## 4.4.7 New Technology

As with Option 1, no new technology was required for the flight system, although engineering developments associated with the SEP stage would be required.

## 4.4.8 Cost

Team X estimated the full mission cost of Option 2 to be $2.24B ($FY15) as shown in **Table 4-18**. This cost includes 30% reserves for Phases A–D and 15% in Phase E. Note that per the ground rules, no reserves were carried on eMMRTG and DSN costs.

**Table 4-18.** Option 2 Team X cost summary.

| WBS Elements | NRE | RE | 1st Unit |
|---|---|---|---|
| **Project Cost (no Launch Vehicle)** | **$1628.3 M** | **$630.8 M** | **$2259.1 M** |
| **Development Cost (Phases A - D)** | **$1000.0 M** | **$630.8 M** | **$1630.7 M** |
| 01.0 Project Management | $47.3 M | | $47.3 M |
| 02.0 Project Systems Engineering | $24.9 M | $0.8 M | $25.7 M |
| 03.0 Mission Assurance | $54.1 M | $0.0 M | $54.1 M |
| 04.0 Science | $66.2 M | | $66.2 M |
| 05.0 Payload System | $147.9 M | $86.3 M | $234.1 M |
| 5.01 Payload Management | $15.8 M | | $15.8 M |
| 5.02 Payload Engineering | $12.9 M | | $12.9 M |
| Orbiter Instruments | $119.1 M | $86.3 M | $205.4 M |
| Narrow Angle Camera | $11.6 M | $8.4 M | $20.0 M |
| Doppler Imager | $17.4 M | $12.6 M | $30.0 M |
| Magnetometer | $4.5 M | $3.3 M | $7.8 M |
| Vis-Near IR Mapping Spectrometer | $9.7 M | $7.0 M | $16.7 M |
| Mid-IR Spectrometer | $7.1 M | $5.2 M | $12.3 M |
| UV Imaging Spectrometer | $5.8 M | $4.2 M | $10.0 M |
| Radio Waves | $3.4 M | $2.4 M | $5.8 M |
| Low Energy Plasma | $2.5 M | $1.8 M | $4.2 M |
| High Energy Plasma | $2.2 M | $1.6 M | $3.8 M |
| Thermal IR | $14.7 M | $10.6 M | $25.3 M |
| Energetic Neutral Atoms | $4.4 M | $3.2 M | $7.7 M |
| Dust Detector | $5.7 M | $4.1 M | $9.8 M |
| Langmuir Probe | $1.1 M | $0.8 M | $1.9 M |
| Microwave Sounder | $23.3 M | $16.9 M | $40.2 M |
| WAC | $5.7 M | $4.1 M | $9.8 M |
| 06.0 Flight System | $391.3 M | $380.5 M | $771.8 M |
| 6.01 Flight System Management | $5.0 M | | $5.0 M |
| 6.02 Flight System Systems Engineering | $35.9 M | | $35.9 M |
| 6.03 Product Assurance (included in 3.0) | | | $0.0 M |
| Orbiter | $295.2 M | $272.6 M | $567.9 M |
| SEP Stage | $50.6 M | $106.3 M | $156.9 M |
| 6.14 Spacecraft Testbeds | $4.6 M | $1.5 M | $6.1 M |
| 07.0 Mission Operations Preparation | $32.2 M | | $32.2 M |
| 09.0 Ground Data Systems | $28.7 M | | $28.7 M |
| 10.0 ATLO | $16.2 M | $17.7 M | $33.9 M |
| 11.0 Education and Public Outreach | $0.0 M | $0.0 M | $0.0 M |
| 12.0 Mission and Navigation Design | $20.8 M | | $20.8 M |
| Development Reserves | $170.4 M | $145.6 M | $315.9 M |
| **Operations Cost (Phases E - F)** | **$595.3 M** | **$0.1 M** | **$595.4 M** |
| 01.0 Project Management | $27.1 M | | $27.1 M |
| 02.0 Project Systems Engineering | $0.0 M | $0.1 M | $0.1 M |
| 03.0 Mission Assurance | $3.6 M | $0.0 M | $3.6 M |
| 04.0 Science | $243.7 M | | $243.7 M |
| 07.0 Mission Operations | $208.3 M | | $208.3 M |
| 09.0 Ground Data Systems | $45.2 M | | $45.2 M |
| 11.0 Education and Public Outreach | $0.0 M | $0.0 M | $0.0 M |
| 12.0 Mission and Navigation Design | $0.0 M | | $0.0 M |
| Operations Reserves | $67.4 M | $0.0 M | $67.5 M |
| **8.0 Launch Vehicle** | **$33.0 M** | | **$33.0 M** |
| Launch Vehicle and Processing | $0.0 M | | $0.0 M |
| Nuclear Payload Support | $33.0 M | | $33.0 M |

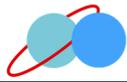

## 4.5    Mission Option 3: Neptune Orbiter and SEP Stage with Probe and 50-Kg Payload

### 4.5.1    Overview

The third option recommended for study by the SDT consists of a Neptune orbiter mission with the same 50 kg payload as Option 1. This mission option also delivers an atmospheric probe, the design and science payload of which is identical to the Uranus probe as described in Section 4.3.6.1. A SEP stage is included in the mission architecture, as a way of trimming transit time to Neptune to a reasonable duration. The SEP stage for this option uses a 3+1 (three active + one spare) arrangement of NEXT ion engines to provide an inner solar system delta V of ~3,000 m/s. One extra thruster is used in this design compared to Options 1 and 2 to provide enhanced thrust in the early stages of the mission. ROSA solar arrays provide ~30 kW of power at BOM at 1 AU.

The mission design for this concept consists of a launch in July of 2030, and executing a ~13-year trajectory using Earth and Jupiter gravity assists. The SEP stage is jettisoned about 6.5 years after launch at a solar range of ~5 AU. Upon arrival the orbiter performs a 2,800 m/s NOI burn, putting it into an initial 180 day orbit around Neptune, which is dropped during the subsequent 2-year orbital phase to approximately 50 days, and which includes multiple satellite flybys.

### 4.5.2    Science

Scientifically, this mission is equivalent to the similar Uranus orbiter with probe considered under Option 1, and it achieves the minimum the science team expects for an ice giant Flagship mission. The instrumentation, discussed below and in Section 3.3.2, allows the highest-priority science to be accomplished; measurement of the bulk composition including noble gases and isotopic ratios (primarily achieved by the atmospheric probe), and determination of interior structure (primarily achieved by the Doppler imager, but supported by the magnetometer). The combination of carrying a camera system and remaining in orbit within the neptunian system allows other priority goals to be achieved and still more to be partially addressed. The orbiter mission also enables the study of time-varying features within the system, and opens the possibility of serendipitous discovery and follow-up.

### 4.5.3    Instrumentation

Instrumentation for Option 3 comprises the ~50-kg payload suite of three instruments plus radio science as described for Option 1 (Section 4.3.3).

The Doppler imager begins its campaign at NOI – 85 days and is subdivided into 5 campaigns of 11 days each. Imaging is continuous at a cadence of two per minute. The experiment is complete at NOI – 30 days.

The imager obtains rotation movies on approach and performs feature tracking within 15 days of approach.

For orbital operations, the Doppler imager will take only 20 images over a 16-hour period each orbit to measure cloud particle velocities in small-scale weather features. The NAC instrument will be used to track motions of atmospheric features and to image the illuminated portions of the rings and of satellite surfaces at global and local (postage stamp) scales. The magnetometer will be on continuously, with higher rate modes near periapse during encounters with satellite. Radio science requires 2-way coherent tracking through periapse and at apoapse.

Spacecraft data storage is sufficient to store all observations acquired for several (~8) orbits.

### 4.5.4    Mission Design

The mission design objective for this third Team X mission option was to enable a flagship class Neptune orbiter (with ~50 kg payload) and a Neptune atmospheric probe. The probe was designed to survive at up to 10 bar. The mission design guidelines for this options were:

1. Launch between 2024–2037, with a preferred launch date between 2029–2031
2. Total mission lifetime <15 years (including Neptune Science phase)
3. Launch on an existing commercial launch vehicle
4. Avoid Neptune rings during orbit insertion
5. Design Probe coast, entry and descent trajectory with feasible orbiter–probe telecom geometry
6. Design Neptune orbit insertion keeping in mind both Neptune and Triton science
7. Neptune moon tour with at least 10 flybys of Triton at different latitudes and longitudes

Neptune is far from Earth. It can be further than Pluto, depending upon where Pluto's position is, in its orbit. Early back of the envelope estimates concluded that a flagship class orbiter to Neptune would require >1,000 kg dry mass. This mass, coupled with mass of a Neptune atmospheric probe led to the conclusion that an SEP-based flight architecture (capable of delivering large mass into Neptune orbit) is most suited for this mission option. A purely chemical propulsion based architecture was not deemed feasible for a Neptune mission for a 13-year interplanetary cruise flight time. **Figure 4-20** depicts the SEP baseline mission architecture, which can be divided into three mission phases:

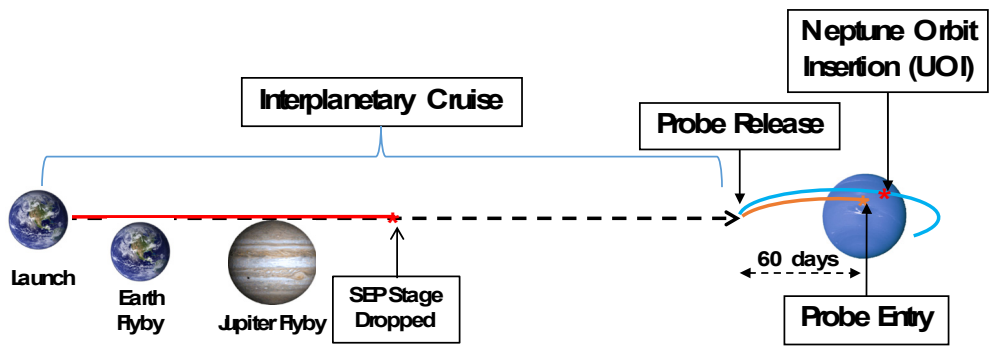

**Figure 4-20.** Mission Concept Option 3: SEP mission design architecture.

### 4.5.4.1    Launch and Interplanetary Cruise

The baseline interplanetary trajectory relies on a 25-kW SEP stage powered by 3 NEXT Ion Engines to propel the spacecraft within the inner solar system. The extra ion thruster, compared to Uranus SEP stage, allows for extra thrust in the inner solar system necessary to satisfy the interplanetary flight time and delivered mass constraints. Post-launch the spacecraft uses the SEP stage to gain momentum and perform an Earth flyby, which is followed by a Jupiter flyby, leading into a long coast phase before Neptune arrival. The SEP stage makes up for the relatively low launch energy while enjoying high propellant efficiency. Similar to the Uranus missions discussed previously, after thrusting in the inner solar system the SEP stage is dropped at ~6 AU, as the solar flux beyond this point is insufficient to power the ion engines. The SEP stage dry mass is significant and dropping it before Neptune Orbit Insertion (NOI) results is significant propellant savings on the orbiter.

**Table 4-19** lists two trajectories. The first one (in blue) is the Team X in-session interplanetary trajectory, which was used during the Team X session for initial design. The second trajectory (in black) represents further post-session refinements made to the in-session trajectory after the SEP stage dry mass and orbiter dry mass became known. **Note that the Team X in-session and the refined trajectory both deliver the same or more mass into Neptune orbit than the allocated amount in the MEL**. Also, note that the refined trajectory uses ~200 kg less Xenon propellant. After the Team X session structural refinements were made to the SEP stage which resulted in significant mass savings, allowing this propellant reduction. The refined SEP stage was still sized to carry the larger Xenon load, and could be refined further to resize the stage to reflect the ~400 kg reduction in Xenon. A future study may expect further SEP stage dry mass reductions, which could be translated into reductions in mission cruise time or increased payload mass.

**Table 4-19.** Option 3 Team X in-session (blue) and refined baseline (black) mission trajectory.

| Flyby Sequence | Launch Vehicle | Launch Date | IP TOF (years) | Xenon Mass (kg) | Assumed SEP-Stage Dry Mass (kg) | Arrival Mass (kg) | NOI ΔV (km/s) | Mass into Orbit (kg) |
|---|---|---|---|---|---|---|---|---|
| Earth-EJ-Neptune | Delta-IV Heavy | 04/28/2030 | 13 | 582.2 | 1817.5 | 4975.3 | 2.8 | 1937.0 |
| Earth-EJ-Neptune | Delta-IV Heavy | 05/04/2030 | 13 | 436.2 | 1603.9 | 5032.7 | 2.7 | 2012.2 |

The quantity "*Arrival Mass*" refers to mass of the spacecraft **before** probe drop-off (Neptune probe mass ~321 kg), which is ~60 days before NOI. Orbit insertion ΔV is calculated for a 252-day capture orbit with periapsis at 1.05 Neptune radii for the refined trajectory. The mass into orbit includes the propellant required for TCMs and Neptune tour maintenance. **Figures 4-21** and **4-22** plots these two trajectory options.

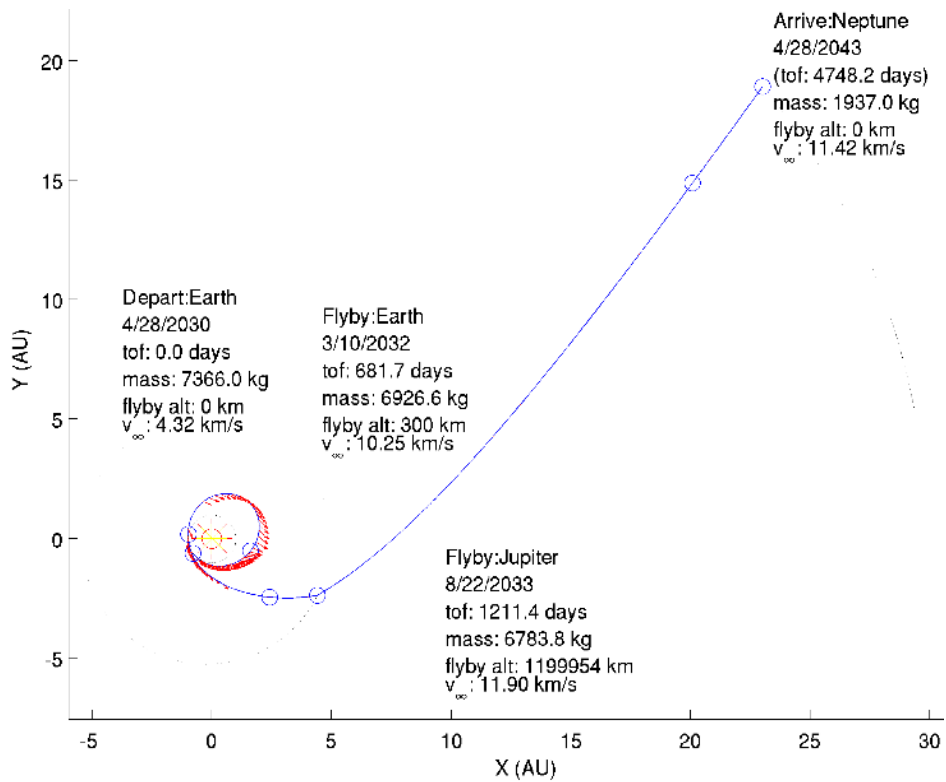

**Figure 4-21.** Option 3 Team X in-session trajectory.

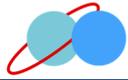

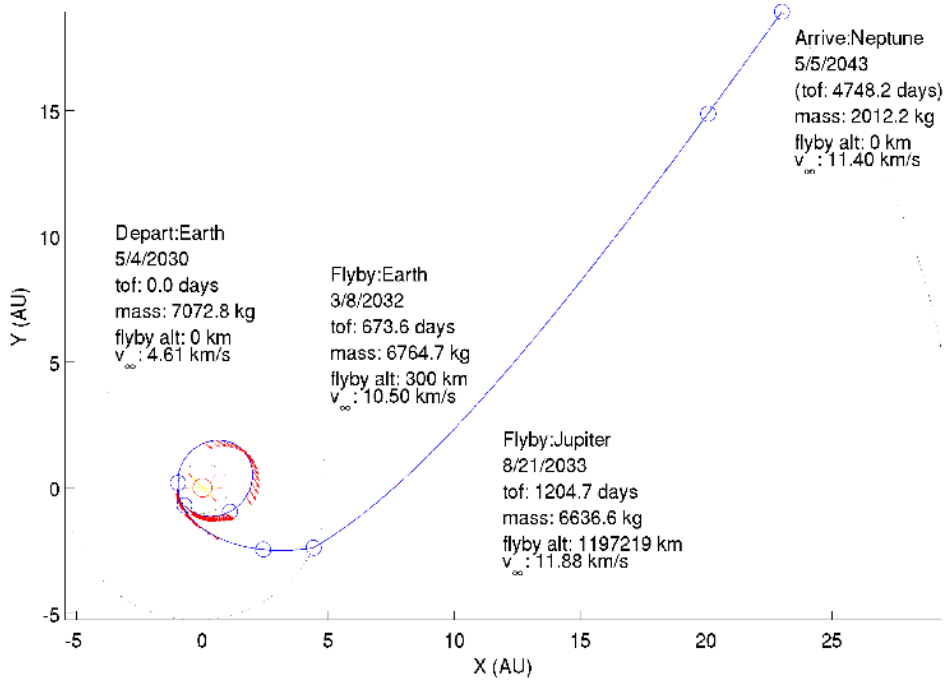

**Figure 4-22.** Option 3 refined baseline trajectory.

### 4.5.4.2 Probe Coast, Entry and Descent + Neptune Orbit Insertion

The baseline mission trajectory releases the probe ~60 days before Neptune atmospheric entry. A PTM is performed by the orbiter prior to probe release, followed by an Orbiter Divert and Periapsis Targeting Maneuver (ODPTM) to target the required NOI conditions. The NOI burn is designed to last ~1.2 hrs and places the orbiter in a retrograde orbit around Neptune. A retrograde orbit insertion is required to achieve low flyby velocities at Triton, a major science driver. Triton is in a retrograde orbit around Neptune. Hence, to achieve low flyby velocities at Triton, the orbiter also need to be in a retrograde orbit around Neptune. A retrograde entry does increase the 'g' load and maximum heat load on the probe but is found to be within acceptable limits.

The probe enters Neptune's atmosphere (retrograde entry) at an EFPA of -20 degrees. The relatively shallow EFPA angle negatively impacts the probe-orbiter telecommunication geometry but is required to reduce probe g loads and heat loads. Due to the relatively low data rate requirements, an EPFA of -20 was found to suffice. Note that based on the analysis done by Ames (see Appendix A) the probe will encounter conditions during entry that are within the capability of the TPS material but beyond the currently available TPS test facilities. The probe descent to 10 bar lasts for ~1 hr, of which the first ~30 mins represent the entry sequence (similar to that shown in Figure 4-32).. Details on the probe entry are given in **Table 4-20**.

The orbiter performs a large NOI ΔV (~2.7 km/s) at an altitude of ~1.05 Neptune Radii and enters in a 252-day retrograde orbit around Neptune. As explained in Section 4.1.6, a relatively close to atmosphere orbit insertion altitude is chosen to mitigate potential ring crossing issues. The low NOI altitude also helps in lowering the NOI ΔV. **Figure 4-23** show the NOI location, orbiter and the probe approach and descent trajectory (in red and gray).

**Table 4-20.** Option 3 probe entry parameters.

| Parameter | Value |
|---|---|
| Interface Altitude | 1000 km |
| Entry Velocity | 22.5 km/s |
| Entry Flight Path Angle | -20 |
| Max G load | 208 G |
| Stagnation Pressure | 11.5 atm. |
| Cumulative Heatload | 109,671 J/cm$^2$ |

**Risks and Concerns**

Following are some of the findings from Neptune Orbit Insertion analysis and probe–orbiter telecomm geometry optimization. A hyperbolic probe entry (with orbiter relay) at Neptune must trade the following design considerations:

1. Neptune Orbit Insertion ΔV

2. Probe g-load and heat load tolerance

3. Relay telecomm. requirements

NOI ΔV is sensitive to the orbiter periapsis altitude (see Appendix A). Higher orbiter periapsis provides better relay line-of-sight and longer persistence (lower angular rate relative to probe), but higher NOI ΔV. Shallow EFPA reduces probe g-load, but presents challenging telecomm geometry and more TPS mass on the probe due to higher accumulated heat loads. Relay antenna must point zenith since the probe rotational phase during EDL cannot be easily predicted. One potential solution that was not explored is to baseline an omnidirectional antenna, or have multiple antennae on the probe.

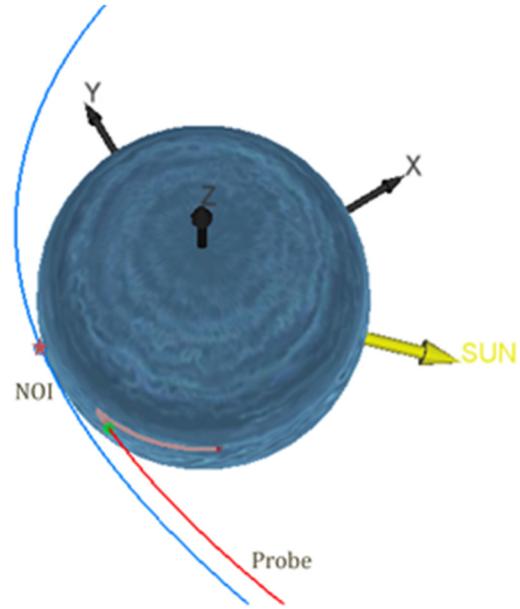

**Figure 4-23.** Option 3 NOI and probe entry.

Another factor to consider is the time between Probe Entry and NOI. Currently, there are two hours allocated between probe entry and NOI, a critical event. It may be operationally challenging to sequence both the probe relay and NOI on the orbiter within this time window. Increasing the separation will make the geometry more challenging for telecomm. Probe-orbiter geometry also needs to deal with issues like uncertainties regarding the Neptune atmosphere and potential signal attenuation. A potential solution would be to perform the NOI burn post periapsis at the cost of increased orbit insertion ΔV.

### 4.5.4.3    Neptune Tour Phase

The orbiter inserts into a ~128 degree inclined, 252 days, retrograde orbit (wrt. Neptune equatorial plane) around Neptune. Approximately, 105 days after orbit insertion the spacecraft performs a Periapsis Raise and Triton Targeting Maneuver (PRTTM). The spacecraft encounters Triton ~235 days after PRTTM and this marks the beginning of the Triton tour. Team X in-session design for Option 3 assumed a 2-year tour at Neptune with multiple flybys of Triton at high velocities. Subsequent tour design efforts revealed that a low hyperbolic flyby velocity of ~3.6 km/s at Neptune is possible within the ΔV budget.

Multiple tours of Triton and Neptune were investigated during the study. A tour with 32 Triton flybys was chosen as the baseline. **Figure 4-24** shows the flyby map for Triton for the baseline tour. A detailed tour design description is given in **Table 4-21**. As noted, a fairly complete coverage of Triton at different longitudes and latitudes is possible. Given that Triton's orbit period is ~6 days, it is possible to encounter Triton after every 6 days near the end of the mission. Even though this opens up exciting scientific possibilities of getting Triton passes every week, it leads to a challenging spacecraft navigation campaign.

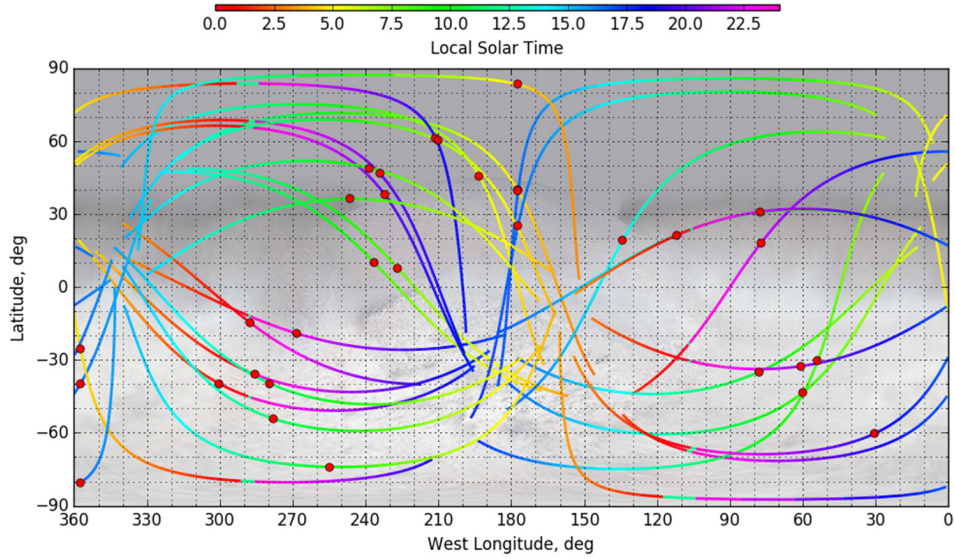

**Figure 4-24.** Option 3 Triton baseline flyby tour 2, tour duration ~2.0 years, 32 Triton flybys, Triton V∞ = 3.6 km/s.

**Table 4-21.** Option 3 Triton baseline flyby tour, shaded rows show nightime flybys.

| Flyby No. | Body | TOF (days) | Date (ET) | S/C Rev | Body Rev | Orbit Period (days) | Flyby Alt (km) | LST | Lat. (deg) | W Lon. (deg) | V-inf (km/s) | V (speed) (km/s) |
|---|---|---|---|---|---|---|---|---|---|---|---|---|
| 1 | Triton | 35.3 | 2044-02-01 22:21:22.40 | 1 | 6 | 35.26 | 100 | 22:41:25 | -19.1 | 268.4 | 3.60 | 3.86 |
| 2 | Triton | 17.6 | 2044-02-19 13:29:21.50 | 1 | 3 | 17.63 | 100 | 23:50:47 | -35.9 | 285.6 | 3.60 | 3.86 |
| 3 | Triton | 11.8 | 2044-03-02 07:34:40.90 | 1 | 2 | 11.75 | 100 | 00:49:45 | -39.7 | 300.3 | 3.60 | 3.86 |
| 4 | Triton | 17.6 | 2044-03-19 22:42:40.00 | 1 | 3 | 17.63 | 100 | 08:48:57 | -43.5 | 60.0 | 3.60 | 3.86 |
| 5 | Triton | 14.7 | 2044-04-03 15:19:18.16 | 1.1 | 2.5 | 13.11 | 1667 | 23:58:45 | -14.8 | 287.4 | 3.60 | 3.73 |
| 6 | Triton | 11.8 | 2044-04-15 09:24:37.55 | 1 | 2 | 11.75 | 100 | 05:41:59 | 45.5 | 193.2 | 3.60 | 3.86 |
| 7 | Triton | 23.5 | 2044-05-08 21:35:16.34 | 3 | 4 | 7.84 | 100 | 08:35:00 | 10.0 | 236.4 | 3.60 | 3.86 |
| 8 | Triton | 5.9 | 2044-05-14 18:37:56.03 | 1 | 1 | 5.88 | 100 | 07:57:01 | 7.8 | 226.8 | 3.60 | 3.86 |
| 9 | Triton | 5.9 | 2044-05-20 15:40:35.73 | 1 | 1 | 5.88 | 100 | 16:39:18 | -39.8 | 357.3 | 3.60 | 3.86 |
| 10 | Triton | 5.9 | 2044-05-26 12:43:15.43 | 1 | 1 | 5.88 | 100 | 04:39:13 | 39.8 | 177.3 | 3.60 | 3.86 |
| 11 | Triton | 14.7 | 2044-06-10 05:19:55.76 | 1.8 | 2.5 | 8.27 | 199 | 21:59:00 | 18.1 | 77.2 | 3.60 | 3.85 |
| 12 | Triton | 11.8 | 2044-06-21 23:25:15.15 | 1 | 2 | 11.75 | 100 | 08:25:54 | -30.4 | 53.9 | 3.60 | 3.86 |
| 13 | Triton | 11.8 | 2044-07-03 17:30:34.53 | 1 | 2 | 11.75 | 100 | 04:39:50 | -25.3 | 357.3 | 3.60 | 3.86 |
| 14 | Triton | 11.8 | 2044-07-15 11:35:53.92 | 1 | 2 | 11.75 | 100 | 16:39:41 | 25.3 | 177.3 | 3.60 | 3.86 |
| 15 | Triton | 17.6 | 2044-08-02 02:43:53.00 | 1 | 3 | 17.63 | 100 | 13:47:17 | 19.3 | 134.1 | 3.60 | 3.86 |
| 16 | Triton | 17.6 | 2044-08-19 17:51:52.07 | 1 | 3 | 17.63 | 100 | 16:40:00 | 39.7 | 177.2 | 3.60 | 3.86 |
| 17 | Triton | 11.8 | 2044-08-31 11:57:11.46 | 1 | 2 | 11.75 | 100 | 20:20:09 | 38.0 | 232.2 | 3.60 | 3.86 |
| 18 | Triton | 17.6 | 2044-09-18 03:05:10.53 | 2 | 3 | 8.82 | 100 | 20:28:14 | 46.6 | 234.1 | 3.60 | 3.86 |
| 19 | Triton | 23.5 | 2044-10-11 15:15:49.28 | 3 | 4 | 7.84 | 100 | 18:56:34 | 61.2 | 211.1 | 3.60 | 3.86 |
| 20 | Triton | 11.8 | 2044-10-23 09:21:08.66 | 1 | 2 | 11.75 | 100 | 12:20:11 | 21.4 | 112.0 | 3.60 | 3.86 |
| 21 | Triton | 20.6 | 2044-11-12 23:00:26.47 | 1.1 | 3.5 | 19.09 | 217 | 10:05:03 | -35.0 | 78.1 | 3.60 | 3.85 |
| 22 | Triton | 11.8 | 2044-11-24 17:05:45.84 | 1 | 2 | 11.75 | 100 | 11:30:26 | -39.7 | 279.4 | 3.60 | 3.86 |
| 23 | Triton | 17.6 | 2044-12-12 08:13:44.90 | 2 | 3 | 8.82 | 100 | 11:24:40 | -54.2 | 277.9 | 3.60 | 3.86 |
| 24 | Triton | 17.6 | 2044-12-29 23:21:43.96 | 2 | 3 | 8.82 | 100 | 16:43:40 | -80.6 | 357.6 | 3.60 | 3.86 |
| 25 | Triton | 23.5 | 2045-01-22 11:32:22.69 | 3 | 4 | 7.84 | 100 | 09:52:07 | -74.0 | 254.6 | 3.60 | 3.86 |
| 26 | Triton | 5.9 | 2045-01-28 08:35:02.38 | 1 | 1 | 5.88 | 100 | 09:20:11 | 36.2 | 246.6 | 3.60 | 3.86 |
| 27 | Triton | 5.9 | 2045-02-03 05:37:42.06 | 1 | 1 | 5.88 | 100 | 04:44:05 | 83.4 | 177.5 | 3.60 | 3.86 |
| 28 | Triton | 23.5 | 2045-02-26 17:48:20.79 | 3 | 4 | 7.84 | 100 | 20:58:03 | -32.6 | 61.0 | 3.60 | 3.86 |
| 29 | Triton | 11.8 | 2045-03-10 11:53:40.15 | 1 | 2 | 11.75 | 100 | 22:05:05 | 30.9 | 77.7 | 3.60 | 3.86 |
| 30 | Triton | 17.6 | 2045-03-28 03:01:39.20 | 2 | 3 | 8.82 | 100 | 08:48:41 | 48.9 | 238.5 | 3.60 | 3.86 |
| 31 | Triton | 23.5 | 2045-04-20 15:12:17.92 | 3 | 4 | 7.84 | 100 | 06:56:02 | 60.3 | 210.3 | 3.60 | 3.86 |
| 32 | Triton | 17.6 | 2045-05-08 06:20:16.95 | 2 | 3 | 8.82 | 100 | 18:56:50 | -60.3 | 30.4 | 3.60 | 3.86 |

## Mission Delta-V Summary

**Table 4-22** shows the summary ΔV table for this mission options. The chemical ΔV is broken down into monoprop and biprop. The main ΔV driver for any missions to Neptune is the orbit insertion maneuver due to relatively high approach $V_\infty$ (~11.4 km/s). Following NOI, the orbiter coasts towards its capture orbit apoapsis. Approximately, 20 days before apoapsis the orbiter performs a PRTTM. P RTTM is designed to achieve a low $V_\infty$ at Triton (3.6 km/s).

**Table 4-22.** Mission Option 3 ΔV summary.

| | Biprop (m/s) | Monoprop (m/s) | Comments |
|---|---|---|---|
| Interplanetary TCMs | 0 | 20 | Most of the TCMs are just after launch and after SEP orbit separation |
| PTM | 0 | 5 | Probe targeting maneuver, ~60 days before NOI |
| ODPTM | 20 | 0 | ~60 days before NOI |
| NOI | 2700 | 0 | <1.2 hr. burn on two 890 N engines |
| SOI-CU (2%) | 54 | 0 | |
| PTTM | 226 | 0 | |
| PRM-CU (2%) | 0 | 5 | |
| Tour Deterministic | 45 | 20 | Multiple maneuvers |
| Tour Margin | 0 | 20 | For future tour design work |
| Tour Statistical | 0 | 30 | |
| De-orbit and Disposal | 10 | 0 | Impact Neptune |
| **Total** | **3055** | **100** | |

### 4.5.5    Concept of Operations/GDS

Option 3 presents a concept of operations very similar to Option 1, where both have the same instrument suite, planetary probe, and SEP stage, but with different target planets. The flight time to Neptune is about two years longer in comparison with the Uranus mission, resulting in a Neptune system tour of two years, rather than four for Uranus.

The lengthy and challenging mission trajectory is enabled by using a SEP stage, allowing a 13-year flight time to Neptune. Operations and tracking post-launch and during the SEP cruise portion of the trajectory are identical to those described for Option 1.

The long ballistic cruise to Neptune after the SEP low thrust phase will have a reduced coverage using a 15-minute beacon 3 out of 4 weeks. Within 6–12 months of Neptune arrival, the ground and flight operations are identical to those in Option 1, including Approach science and probe activities. One particular difference is the communications link, where a reasonable downlink data rate (15 kbps, similar to the Uranus mission options) can be achieved at Neptune using the same telecom subsystem as Option 1 by arraying 3X 34BWG stations. The spacecraft is able to accommodate up to 20 hours straight of Ka-band downlink with continuous approach science. The approach science data outline is identical to that of Option 1, including the associated tables and graphs.

After the large NOI burn, the spacecraft is inserted into its initial 180-day orbit about Neptune. The on-orbit checkouts, ORTs, and instrument configuration is the similar to Option 1. The navigation and science pointing methodology is also the same.

As with the Uranus options, the Neptune Orbital Science phase can be divided into 3 categories of observation activities: apoapse science, periapse science, and satellite flyby science. The Neptune apoapse and periapse science activities are also identical to Option 1, except with Neptune, its Rings, and Triton. A tour about the Neptune system will have frequent flybys of Triton in order to map its surface and features. During the tour, the repeated gravity assists of Triton will torque the orbiter's inclination more and more in plane with Triton's orbit (from the original 9.1 deg inclination), while also reducing the period down to 50 days. These final orbital conditions yield longer and better quality Triton flyby science.

Assuming the same data rate of 15 kbps is achieved by arraying 3X 34BWG, the orbit science data scheme is identical to Option 1, including the associated tables and graphs. The spacecraft decommissioning and disposal into Neptune is also similar to the disposal described in Option 1.

### 4.5.6 Flight System Design

The Option 3 Neptune orbiter flight system is very similar to the Option 1 design for Uranus (**Figure 4-25**). Modifications to the Option 1 design have been limited to changes required to accommodate the higher delta-V requirements of the Neptune mission. The SEP stage adds one extra NEXT thruster over the Option 1 design to exploit the opportunity for higher thrust in the early stages of the mission.

The instrument complement on the orbiter and probe is unchanged from Option 1. As mentioned in Section 4.3.6.1 the probe descent module design is identical for Neptune and Uranus, however the entry system is slightly heavier for the Neptune mission, reflecting harsher entry conditions relative to Uranus (**Table 4-9**).

The integrated flight system has a total wet launch mass of 7364 kg, and comprises a 1756 kg dry orbiter with 3088 kg of bi-propellants, a 1604 kg dry SEP stage with 595 kg of Xenon propellant for its NEXT-based ion thrusters, and a 322 kg atmospheric probe. See **Table 4-23** for the flight system mass summary.

The atmospheric probe is slated to be released 60 days prior to NOI. The orbiter propellant mass is sized to accommodate this release scenario.

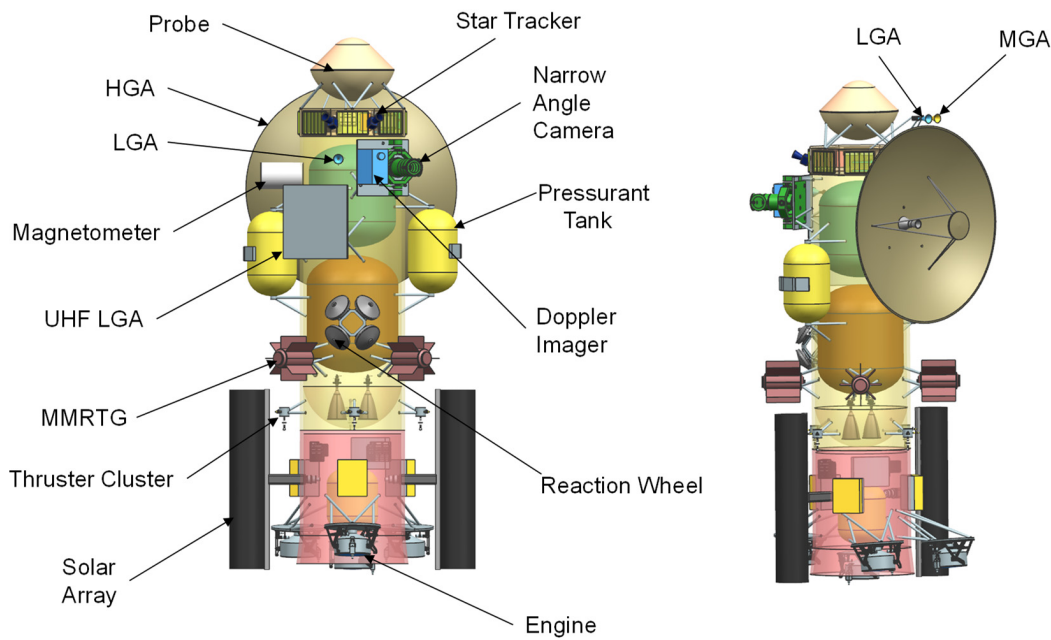

**Figure 4-25.** Option 3 launch configuration.

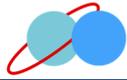

**Table 4-23.** Option 3 MEL.

| Orbiter | CBE Mass (kg) | Contingency (%) | Total Mass (kg) | Heritage/Comments |
|---|---|---|---|---|
| Instruments | 36.7 | 23% | 45.2 | 3 instruments |
| C&DH | 21.6 | 10% | 23.8 | JPL Reference Bus |
| Power | 217.6 | 2% | 222.2 | No contingency on eMMRTGs |
| Telecom | 59.4 | 16% | 68.9 | 35 W TWTA, 3 m HGA |
| Structures | 490.4 | 30% | 637.6 | |
| Harness | 87.8 | 0.3 | 114.2 | |
| Thermal | 118.1 | 0.2 | 145.9 | |
| Propulsion | 186.8 | 0.1 | 196.9 | Dual Mode |
| GN&C | 63.5 | 0.1 | 69.8 | |
| | | | | |
| **Orbiter Total** | **1282.1** | **19%** | **1524.5** | |
| System Margin | | | 231.5 | |
| **Dry Mass Total** | | **43%** | **1756.0** | |
| **Propellant** | | | **3088.0** | |
| **Wet Mass Total** | | | **4844.0** | |
| **SEP Stage** | **CBE Mass (kg)** | **Contingency (%)** | **Total Mass (kg)** | **Heritage/Comments** |
| C&DH | 1.6 | 0.1 | 1.7 | MREU |
| Power | 265.5 | 0.3 | 342.9 | 30 kW ROSA |
| Mechanical | 478.7 | 0.3 | 622.3 | |
| Thermal | 78.9 | 0.0 | 78.9 | |
| Propulsion | 289.9 | 0.2 | 358.1 | |
| GN&C | 7.0 | 0.1 | 7.5 | gimbal drives, sun sensors |
| **SEP Stage Total** | **1121.6** | **26%** | **1411.4** | |
| System Margin | | | 192.5 | |
| **Dry Mass Total** | | **43%** | **1603.9** | |
| | | | | |
| **Propellant** | | | **595.0** | |
| Xenon | | | 595.0 | |
| **Wet Mass Total** | | | **2198.9** | |
| **Mission System** | **CBE Mass (kg)** | **Contingency (%)** | **Total Mass (kg)** | **Heritage/Comments** |
| Probe | | | 321.5 | |
| Orbiter | | | 4844.0 | |
| SEP Stage | | | 2198.9 | |
| **Launch Mass Total** | | | **7364.4** | |
| Injected Mass Cap. | | | 7575.0 | Delta IVH to C3 of 18.66 |
| **Remaining LV Cap.** | | | **210.6** | |

### 4.5.7 New Technology

As with the previous options no new technology was required for the flight system, although engineering developments associated with the SEP stage would be required.

### 4.5.8 Cost

Team X estimated the full mission cost of Option 3 to be $1.97B ($FY15) as shown in **Table 4-24**. This cost includes 30% reserves for Phases A–D and 15% in Phase E. Note that per the groundrules, no reserves were carried on eMMRTG and DSN costs.

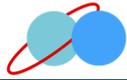

**Table 4-24.** Option 3 Team X Cost Summary.

| WBS Elements | NRE | RE | 1st Unit |
|---|---|---|---|
| **Project Cost (no Launch Vehicle)** | **$1365.9 M** | **$606.0 M** | **$1971.9 M** |
| **Development Cost (Phases A - D)** | **$1000.6 M** | **$605.9 M** | **$1606.6 M** |
| 01.0 Project Management | $47.3 M | | $47.3 M |
| 02.0 Project Systems Engineering | $23.7 M | $0.5 M | $24.3 M |
| 03.0 Mission Assurance | $53.1 M | $0.0 M | $53.1 M |
| 04.0 Science | $24.8 M | | $24.8 M |
| 05.0 Payload System | $80.2 M | $48.3 M | $128.5 M |
| 06.0 Flight System | $495.4 M | $395.6 M | $891.1 M |
| 6.01 Flight System Management | $5.0 M | | $5.0 M |
| 6.02 Flight System Systems Engineering | $51.1 M | | $51.1 M |
| 6.03 Product Assurance (included in 3.0) | | | $0.0 M |
| Orbiter | $295.4 M | $236.6 M | $532.0 M |
| SEP Stage | $51.5 M | $114.9 M | $166.4 M |
| Probe | $27.2 M | $18.2 M | $45.4 M |
| Entry System | $57.1 M | $24.4 M | $81.5 M |
| Ames/Langley EDL Engineering/Testing | $3.8 M | $0.0 M | $3.8 M |
| 6.14 Spacecraft Testbeds | $4.5 M | $1.5 M | $6.0 M |
| 07.0 Mission Operations Preparation | $26.6 M | | $26.6 M |
| 09.0 Ground Data Systems | $22.1 M | | $22.1 M |
| 10.0 ATLO | $21.1 M | $21.7 M | $42.8 M |
| 11.0 Education and Public Outreach | $0.0 M | $0.0 M | $0.0 M |
| 12.0 Mission and Navigation Design | $28.8 M | | $28.8 M |
| Development Reserves | $177.4 M | $139.8 M | $317.3 M |
| **Operations Cost (Phases E - F)** | **$332.3 M** | **$0.1 M** | **$332.3 M** |
| 01.0 Project Management | $27.1 M | | $27.1 M |
| 02.0 Project Systems Engineering | $0.0 M | $0.1 M | $0.1 M |
| 03.0 Mission Assurance | $3.6 M | $0.0 M | $3.6 M |
| 04.0 Science | $69.2 M | | $69.2 M |
| 07.0 Mission Operations | $171.3 M | | $171.3 M |
| 09.0 Ground Data Systems | $28.2 M | | $28.2 M |
| 11.0 Education and Public Outreach | $0.0 M | $0.0 M | $0.0 M |
| 12.0 Mission and Navigation Design | $0.0 M | | $0.0 M |
| Operations Reserves | $32.8 M | $0.0 M | $32.8 M |
| **8.0 Launch Vehicle** | **$33.0 M** | | **$33.0 M** |
| **Launch Vehicle and Processing** | **$0.0 M** | | **$0.0 M** |
| **Nuclear Payload Support** | **$33.0 M** | | **$33.0 M** |

Following completion of the Team X study, the Aerospace Corporation performed an independent cost estimate (ICE) for this option, using the Team X design as documented in the Team X study report. Aerospace estimated this mission to be about $2.28B ($FY15), placing it above the target cap. Differences were largely related to a differing assessment of the cost of the orbiter SEP stage, and a more conservative estimate of the cost of operations in the lengthy cruise phase of the mission.

**Table 4-25.** Aerospace ICE Option 3 Cost Estimate.

| FY15 $M | JPL Estimate | Aerospace ICE | Difference FY15 $M | Difference % |
|---|---|---|---|---|
| Phase A | $ - | $ 24.7 | $ 24.7 | |
| Subtotal | $ - | $ 24.7 | $ 24.7 | |
| Phase B/C/D | | | | |
| Mission PM/SE/MA | $ 124.5 | $ 158.0 | $ 33.5 | 26.9% |
| Payload[1] | $ 128.5 | $ 124.6 | $ (3.9) | -3.0% |
| Flight System[2] | $ 932.2 | $ 1,012.4 | $ 80.2 | 8.6% |
| Pre-Launch GDS/MOS | $ 102.3 | $ 120.7 | $ 18.4 | 18.0% |
| *Launch Vehicle* | *$ 33.0* | *$ 33.0* | *$ -* | *0.0%* |
| Reserves | $ 316.6 | $ 405.0 | $ 88.4 | 27.9% |
| Subtotal | $ 1,637.1 | $ 1,853.7 | $ 216.6 | 13.2% |
| Phase E/F | | | | |
| MO&DA - Science | $ 301.3 | $ 347.8 | $ 46.5 | 15.4% |
| Reserves | $ 33.0 | $ 52.8 | $ 19.8 | 59.9% |
| Subtotal | $ 334.3 | $ 400.5 | $ 66.2 | 19.8% |
| **Total** | **$ 1,971.4** | **$ 2,279.0** | **$ 307.6** | **15.6%** |

Note: *Italics* indicates project values; treated as pass throughs
1. Includes orbiter and probe instruments, as applicable
2. Includes probe, entry system, orbiter/flyby bus, SEP as applicable

## 4.6 Mission Option 4: Uranus Flyby with Probe and 50-Kg Payload

### 4.6.1 Overview

For the fourth option recommended for study by the SDT the team modeled a mission consisting of a Uranus flyby-only spacecraft with an atmospheric probe. Although not an orbiter, the same "50 kg" instrument suite was determined to be appropriate to the flyby vehicle to perform Uranus science on approach, transit and departure from the system. This option was considered to be a low cost alternative to the more complex orbiter missions that would still provide a rich science data return.

### 4.6.2 Science

As was discussed in Section 3.4.3, the science team does not recommend a single-planet flyby mission for an ice giant Flagship. While our two highest-priority science goals can be achieved (related to bulk composition and interior structure), the breadth of science enabled by getting into orbit causes any flyby to be ranked significantly lower than any orbiter mission (see **Figure 3-15**). This mission was nonetheless selected for detailed study because its expected low cost helps complete our attempts to bracket the entire mission parameter space. This mission design also informs our two-planet, two-spacecraft mission options, particularly a highly-ranked one in which one spacecraft is an orbiter and the other is a flyby.

### 4.6.3 Instrumentation

Instrumentation for Option 4 is identical to that for the probe and Orbiter used in Option 1.

### 4.6.4 Mission Design

The mission design objective for the fourth mission option was to enable a Uranus flyby (with ~50 kg payload) and an atmospheric probe. The high-level mission design guidelines for this options were:

1. Launch between 2024–2037, with a preferred launch date between 2029–2031

2. Total mission lifetime < 12 years (including Uranus Science phase)
3. Launch on an existing commercial Launch Vehicle
4. Uranian flyby with an atm. probe.
5. Design Probe coast, entry and descent trajectory with feasible Orbiter – Probe telecom geometry.

A purely chemical propulsion based architecture was selected as a baseline for this option. **Figure 4-26** depicts the baseline mission architecture, which can be divided into two mission phases:

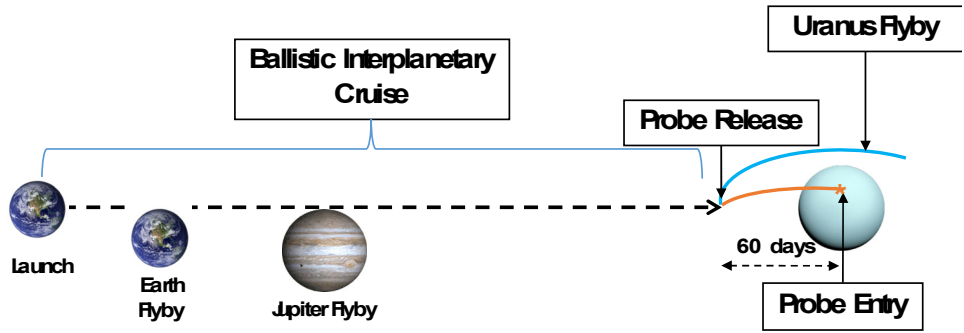

**Figure 4-26.** Mission Option 4: mission design architecture.

### 4.6.4.1 Launch and Interplanetary Cruise

Given that we are not required to enter Uranus orbit, the baseline interplanetary trajectory is a purely chemical trajectory requiring 10 years to reach Uranus. The spacecraft launches on an Atlas V (541) commercial launch vehicle. After launch, the spacecraft performs a Deep Space Maneuver (DSM, ~240 m/s) leading into an Earth flyby, to gain momentum in the inner solar-system, followed by a Jupiter flyby before encountering Uranus. **Table 4-26** describes the baseline interplanetary trajectory.

The quantity "*Arrival Mass*" refers to mass of the spacecraft before Uranus encounter but after Probe separation. The spacecraft does a flyby of Uranus at a periapsis of 1.65 Uranus Radii. **Figure 4-27** show the baseline trajectory listed in **Table 4-26**.

### 4.6.4.2 Uranus Flyby and Probe Entry

The baseline mission trajectory performs a Probe Targeting Maneuver (PTM) 60 days before the target Uranus periapsis. The orbiter performs a Uranus flyby at a periapsis altitude of 1.05 Uranus radii. The high UOI altitude also helps to avoid Uranus rings and allows for a favorable Orbiter-Probe relay geometry. **Table 4-27** shows the probe entry parameters. The probe descent to 10 bar lasts for ~1 hr., of which the first ~30 mins represent the entry sequence, as shown in **Figure 4-29**.

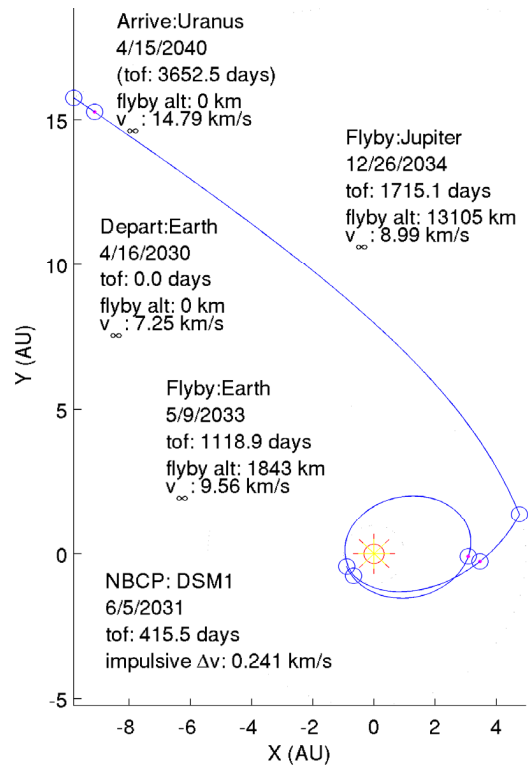

**Figure 4-27.** Option 4 Team X in-session trajectory.

**Table 4-26.** Option 4 baseline mission trajectory.

| Flyby Sequence | Launch Vehicle | Launch Date | Launch C3 (km²/s²) | IP TOF (yrs.) | DSM (m/s) | Arrival Mass (kg) |
|---|---|---|---|---|---|---|
| Earth-EJ-Uranus | Atlas V (541) | 4/14/2030 | 52.6 | 10 | 240 | 1345 |

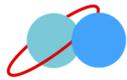

**Table 4-27.** Option 4 probe entry parameters.

| Parameter | Value |
|---|---|
| Interface Altitude | 1000 km |
| Entry Velocity | 23.1 km/s |
| Entry Flight Path Angle | -30 degrees |

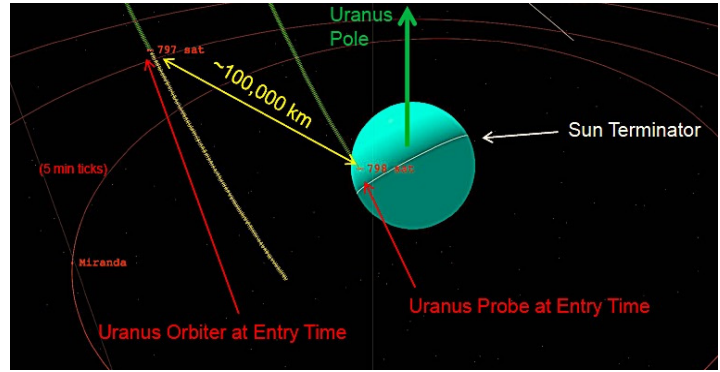

**Figure 4-28.** Option 4 UOI and ATTM.

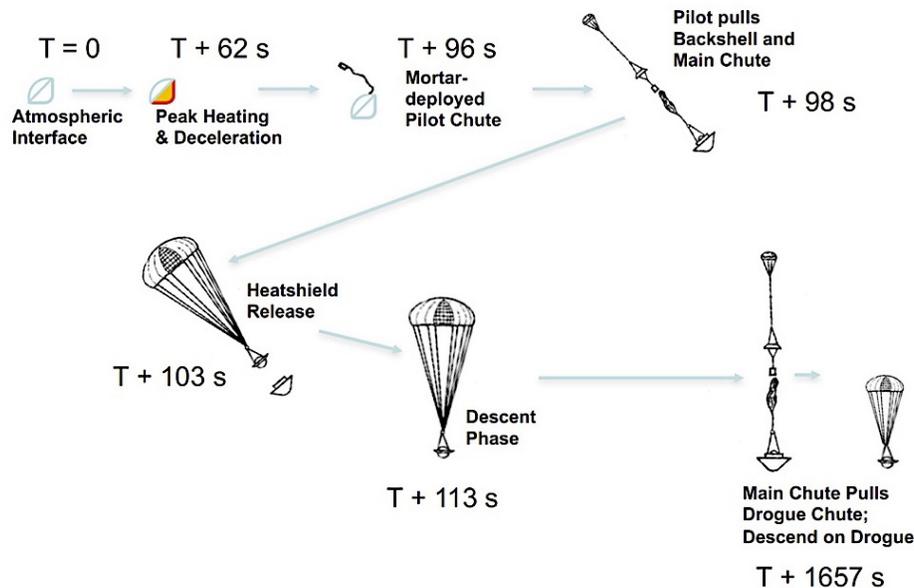

**Figure 4-29.** Probe entry sequence (timing represents Uranus example).

### Mission Delta-V Summary

**Table 4-28** shows the summary ΔV table for this mission options. Compared to other mission options, the ΔV requirement is quite low due to lack of Uranus orbit insertion.

**Table 4-28.** Mission Option 4 ΔV summary for refined baseline trajectory

| | Monoprop (m/s) | Comments |
|---|---|---|
| Interplanetary TCMs | 40 | |
| DSM | 240 | Only maneuver on the main monoprop |
| UPTM | 10 | 60 days before UOI |
| Orbiter Divert | 20 | |
| **Total** | **310** | |

### 4.6.5 Concept of Operations/GDS

The goals of Option 4 are met by delivering a spacecraft with a core set of scientific instruments in minimal flight time for a close flyby encounter of the Uranus system and retrieving direct atmospheric measurements of the planet with a probe. One of the most difficult challenges for any mission to Uranus is the vast distance that must be trekked in order to reach the outer planet. A flyby mission is able to minimize flight time and spacecraft mass by removing the need for the massive quantity of orbit insertion propellant. The reduced flight time and mission cost are traded for a limited window of time for unique/valuable scientific observations. Compared to the similar orbiter mission Option 1, the flyby option reduces the flight time by 2 years and does not make use of a SEP stage. The spacecraft mission operations up until orbit insertion are nearly identical between the flyby option and the all-chemical propulsion orbiter mission Option 5.

Shortly after launch, the mission begins its checkout processes and deployment of all flight systems. During the first four weeks of cruise, continuous DSN coverage is required for thorough characterization of all flight systems and accommodating variable commanding schedules typical of early checkout operations. When checkout is complete, the post-launch phase configures the spacecraft to a stable dormant state with minimal operation and the DSN coverage is reduced to only 1 pass per week. Throughout interplanetary cruise, each critical mission event (TCM, DSM, gravity assist, etc.) is expected to have daily DSN coverage for commanding and tracking for 2 weeks approaching the event, especially for the nuclear safety maneuvers prior to any Earth gravity assists. Continuous coverage is required during the days surrounding the gravity assist or maneuvering. Though the Team X study assumes minimal operations work throughout most of the interplanetary cruise, additional checkouts and characterizations of flight systems and instruments could be performed during the gravity assist encounters of Earth and/or Jupiter.

The long ballistic cruise to Uranus after the Jupiter gravity assist will have a further reduction in coverage using a 15-minute beacon 3 out of 4 weeks. Within a year of Uranus Arrival, ground operations support ramps up (including mission/science planning and execution staff). Starting at 6 months until Entry and closest approach, DSN coverage increases to daily tracks in preparation for the Approach Science phase and Probe release. The final checkouts and Operational Readiness Tests (ORTs) are completed, while the spacecraft is prepared for the Uranus system encounter.

The Approach Science phase begins at 85 days to UOI, with very similar activities to Option 1. The 55-day approach Doppler and NAC science is identical to Option 1. The MAG instrument's continuous observations in the background during approach is set to the nominal 1kbps data rate (compared to minimum rate in Option 1). The increased MAG science plan is to ensure a full data set from this single planetary encounter. The approach science plan details are shown in **Table 4-29**. Using the 4 eMMRTG baseline power system and 15kbps communications system, the spacecraft is able to accommodate up to 20 hours straight of Ka-band downlink with continuous approach science. The approach science will collect a total of 54 Gb of compressed data (186 Gb uncompressed). The data budget averaged per day on approach is balanced at about 0.98 Gb/day of compressed data recorded and 1.10 Gb/day 34BWG downlink capability (11% margin). The detailed data outline for each instrument is shown in **Table 4-30**. Any overflow approach science data not returned is stored on high capacity 1024 Gb solid-state data recorders and downlinked alongside Uranus flyby science after encounter. The optical navigation activities are periodically performed by the NAC instrument throughout approach, similar to other mission options.

**Table 4-29.** Mission Option 4 instrument science plan for major mission phases: Approach, Apoapse, Periapse, and Satellite Flyby Science.

| Ice Giants Mission #4 Instrument Science Plan | Aproach Obs Info | Satellite Flyby Obs Info | Pre/Post Periapse Obs Info | Periapse Obs Info |
|---|---|---|---|---|
| NAC | Mov: 200im - 1clr | 6Gb (1/2 Map) | 15-rot/1hr - 4clr | 5-10 F.Trax - 4clr |
| Doppler Imager | 22dy/30s,33dy/2s | 100 im | 3 im/rot | 100 im/rot |
| Magnetometer | Nom | Very High | High | Very High |

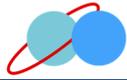

**Table 4-30.** Mission Option 4 instrument science data outline for major mission phases: Approach, Apoapse, Periapse, and Satellite Flyby Science.

| Ice Giants Mission #4 Instrument Science Data | Min Data Rate (Kbps) | Nom Data Rate (Kbps) | Max Data rate (Kbps) | Approach Data (Gb) | Satellite Flyby Data (Gb) | System Encounter Data (Gb) | Pre/Post Periapse Data (Gb) | Periapse Data (Gb) |
|---|---|---|---|---|---|---|---|---|
| NAC | - | 134Mb/im/clr | - | 32.16 | 6.00 | 152.76 | 136.68 | 16.08 |
| Doppler Imager | - | 0.1Mb/im | - | 148.90 | 0.01 | 0.07 | 0.03 | 0.04 |
| Magnetometer | 0.50 | 1.00 | 3 - 12 | 4.75 | 0.06 | 18.66 | 15.55 | 3.11 |

| | | | |
|---|---|---|---|
| Optical Compression: 3.5 X | Science Data Totals: | 185.8 Gb | 6.1 Gb | 171.5 Gb |
| Fields & Particles Compression: 2.0 X | Uncompressed: | 185.8 gb | Uncompressed: | 177.6 Gb |
| | Compressed: | 54.1 gb | Compressed: | 54.7 Gb |
| | Approach Data Rate: | 0.98 Gb/day | Encounter Data Rate: | 0.91 Gb/day |
| | Approach Data Return: 1X 34BWG (15 Kbps) | 1.10 Gb/day | Data Return Rate: 1X 34BWG (15 Kbps) | 0.37 Gb/day |
| | | | Data Return Duration: | 149 day |

After the approach science phase is completed, the flyby encounter activities begins at 30 days until closest approach. The MAG instrument will be collecting high rate (3 kbps) magnetic field measurements of the Uranian system throughout the encounter. The NAC will focus on mapping the system with repeated mosaics of Uranus and its rings at the beginning of the encounter period, covering 15 rotations of the planet and at a faster cadence compared to other mission options. Next, as the optical instruments' resolution improves, the NAC and Doppler imager will target specific features on the planet and in the rings (with full color imaging). During the week prior to closest approach, the NAC is able to resolve better images of the satellites. In the final days, the spacecraft will be busy capturing high-resolution images of all the major Uranian satellites, as well as continuing its mosaics mapping the planet, rings and feature tracks. The NAC is allocated 5–10 feature tracks, assuming 3 images (full 4-color) for each tracking observation. The Doppler imager will perform its high-resolution feature tracks with an average cadence of about 100 images per planet rotation. Within 8 hours of closest approach, the spacecraft will focus on at least one flyby of a major satellite, providing very high resolution mapping of its surface features. Some of the most important magnetic field observations are obtained during the Uranus closest approach and the satellite flyby, revealing any interactions within the system (such as caused by a sub-surface ocean), therefore the MAG data collection is switched to the max 12 Kbps rates. The outbound sequence of science activities reflects that of the inbound, observing the Uranus satellites, the planet, and its rings under high phase angle lighting conditions, which could highlight the small-particle environment in the system.

During the encounter, 55 Gb of compressed science data (178 Gb uncompressed) will be collected, including 1.8 Gb of satellite flyby compressed data. In order to efficiently return all of the science data collected during the encounter, the baseline DSN plan is for daily 8-hour passes for 6 months after the encounter using only a single 34M BWG station. Each pass provides about 367 Mb of science data downlinked (assuming 15% overhead). At this rate, 5 months of daily passes are required until all of the encounter data is returned, providing a month of margin to absorb disruptions in DSN support and any other issues.

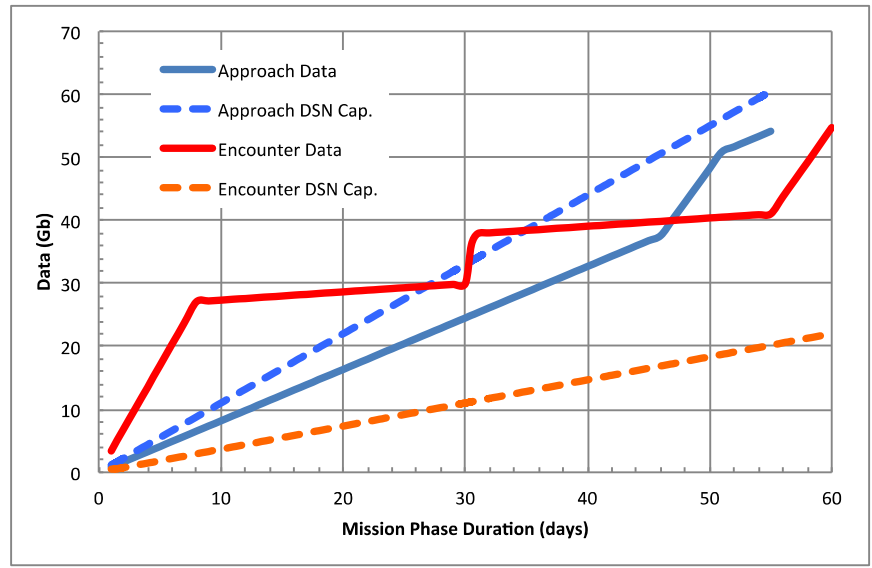

**Figure 4-30.** Mission Option 4 science data volume recorded compared to DSN downlink capacity for Approach science and Encounter science.

A graph of science data recorded compared to the DSN capacity for the Approach science and Encounter science phases is shown in **Figure 4-30**. The Approach science plot displays the continuous Doppler data collection and Fields & Particles background activities, then ramping up towards the end with the NAC movie recording. The Encounter science plot begins with pre-periapse data, collecting 2/3 of the NAC images of the planet (10-rotations), with continuous Fields & Particles measurements throughout apoapse. Then the short spike of high rate periapse data is recorded for about 16 hours, followed by densely packed 2-hour satellite flyby science. The post-periapse data collection reflects the pre-periapse activities, where the final NAC images capture the Uranus system from the unique high phase angle perspective for 5-rotations of the planet. As the graph suggests, the recorded encounter data will far outpace the DSN downlink capability upon exiting the planetary system, yet after 5 months of daily passes all of the data is returned.

With the spacecraft exiting the solar system at a hyperbolic escape velocity, there is little concern for planetary protection, easing the process of decommissioning the spacecraft. Although, an extended mission in the Kuiper belt could be considered after successful completion of the Uranus flyby mission.

### 4.6.6    Flight System/Probe Design

Being a flyby-only mission, the flight system design for Option 4 differs in some ways from the earlier options. It is still a three-axis stabilized flight system, but its much lower delta V requirement allows downsizing and simplification of the propulsion system by going to a fully monopropellant implementation. With a single, much smaller propellant tank, the flight system size and mass is significantly reduced, as illustrated in **Figure 4-31**.

Apart from propulsion and structures, the remainder of the subsystem designs remain substantially unchanged from the earlier options. In this case the four eMMRTGs are mounted in two stacks of two to allow them to fit without interfering with the 3 m HGA and UHF LGA. This also helps minimize the number of payload fairing access doors required.

The instrument complement and Uranus probe design are unchanged from Option 1 and accommodation on the spacecraft bus was likewise kept the same.

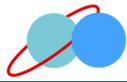

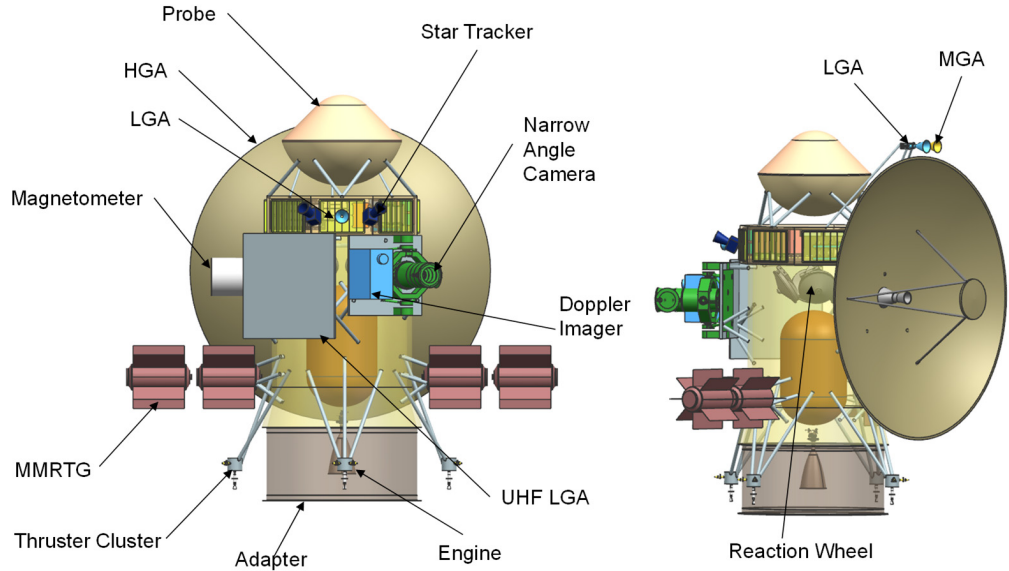

**Figure 4-31.** Option 4 launch configuration.

The integrated flight system has a total wet launch mass of 1525 kg, and comprises a 1024 kg dry flyby spacecraft with 180 kg of monopropellant hydrazine and a 321 kg atmospheric probe. See **Table 4-31** for the flight system mass summary.

The atmospheric probe is slated to be released 60 days prior to UOI. The orbiter propellant mass is sized to accommodate this release scenario.

**Table 4-31.** Option 4 MEL.

| Orbiter | CBE Mass (kg) | Contingency (%) | Total Mass (kg) | Heritage/Comments |
|---|---|---|---|---|
| Instruments | 36.7 | 0.2 | 45.2 | 3 instruments |
| C&DH | 21.6 | 0.1 | 23.8 | JPL Reference Bus |
| Power | 214.9 | 0.0 | 218.7 | No contingency on eMMRTGs |
| Telecom | 59.4 | 0.2 | 68.9 | 35 W TWTA, 3 m HGA |
| Structures | 246.6 | 29% | 317.6 | |
| Harness | 70.3 | 0.3 | 91.4 | |
| Thermal | 39.5 | 0.2 | 46.9 | |
| Propulsion | 31.5 | 0.0 | 32.5 | Monoprop |
| GN&C | 49.5 | 0.1 | 54.4 | |
| **Orbiter Total** | **770.1** | **17%** | **899.5** | |
| System Margin | | | 124.3 | |
| **Dry Mass Total** | | **43%** | **1023.8** | |
| | | | | |
| **Propellant** | | | **180.0** | |
| | | | | |
| **Wet Mass Total** | | | **1203.8** | |
| Mission System | CBE Mass (kg) | Contingency (%) | Total Mass (kg) | Heritage/Comments |
| Probe | | | 320.7 | |
| Orbiter | | | 1203.8 | |
| **Launch Mass Total** | | | **1524.5** | |
| Injected Mass Cap. | | | 1775.0 | Atlas 541 to C3 of 52.6 |
| **Remaining LV Cap.** | | | **250.5** | |

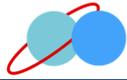

### 4.6.7 New Technology

The flyby spacecraft for Option 4 relies on flight proven subsystem and component designs. Given the relatively simple nature of this flight system, no new technologies should be required.

### 4.6.8 Cost

Team X estimated the full mission cost of Option 4 to be $1.49B ($FY15) as shown in **Table 4-32**. This cost includes 30% reserves for Phases A–D and 15% in Phase E. Note that per the groundrules, no reserves were carried on eMMRTG and DSN costs.

**Table 4-32.** Option 4 Team X Cost Summary.

| WBS Elements | NRE | RE | 1st Unit |
|---|---|---|---|
| Project Cost (no Launch Vehicle) | $1064.6 M | $428.2 M | $1492.8 M |
| **Development Cost (Phases A - D)** | **$831.8 M** | **$428.1 M** | **$1259.9 M** |
| 01.0 Project Management | $47.3 M | | $47.3 M |
| 02.0 Project Systems Engineering | $23.7 M | $0.5 M | $24.2 M |
| 03.0 Mission Assurance | $42.3 M | $0.0 M | $42.3 M |
| 04.0 Science | $24.8 M | | $24.8 M |
| 05.0 Payload System | $80.2 M | $48.3 M | $128.5 M |
| 06.0 Flight System | $383.2 M | $259.8 M | $643.0 M |
| 6.01 Flight System Management | $5.0 M | | $5.0 M |
| 6.02 Flight System Systems Engineering | $49.1 M | | $49.1 M |
| 6.03 Product Assurance (included in 3.0) | | | $0.0 M |
| Orbiter | $237.0 M | $215.8 M | $452.7 M |
| SEP Stage | $0.0 M | $0.0 M | $0.0 M |
| Probe | $26.8 M | $18.1 M | $44.9 M |
| Entry System | $57.1 M | $24.4 M | $81.5 M |
| Ames/Langley EDL Engineering/Testing | $3.8 M | $0.0 M | $3.8 M |
| 6.14 Spacecraft Testbeds | $4.5 M | $1.5 M | $6.0 M |
| 07.0 Mission Operations Preparation | $26.6 M | | $26.6 M |
| 09.0 Ground Data Systems | $24.0 M | | $24.0 M |
| 10.0 ATLO | $21.7 M | $20.8 M | $42.5 M |
| 11.0 Education and Public Outreach | $0.0 M | $0.0 M | $0.0 M |
| 12.0 Mission and Navigation Design | $11.5 M | | $11.5 M |
| Development Reserves | $146.3 M | $98.8 M | $245.1 M |
| **Operations Cost (Phases E - F)** | **$199.8 M** | **$0.1 M** | **$199.9 M** |
| 01.0 Project Management | $19.3 M | | $19.3 M |
| 02.0 Project Systems Engineering | $0.0 M | $0.1 M | $0.1 M |
| 03.0 Mission Assurance | $2.5 M | $0.0 M | $2.5 M |
| 04.0 Science | $46.4 M | | $46.4 M |
| 07.0 Mission Operations | $83.6 M | | $83.6 M |
| 09.0 Ground Data Systems | $23.5 M | | $23.5 M |
| 11.0 Education and Public Outreach | $0.0 M | $0.0 M | $0.0 M |
| 12.0 Mission and Navigation Design | $0.0 M | | $0.0 M |
| Operations Reserves | $24.5 M | $0.0 M | $24.6 M |
| **8.0 Launch Vehicle** | **$33.0 M** | | **$33.0 M** |
| Launch Vehicle and Processing | $0.0 M | | $0.0 M |
| Nuclear Payload Support | $33.0 M | | $33.0 M |

Following completion of the Team X study, the Aerospace Corporation performed an independent cost estimate (ICE) for this option, using the Team X design as documented in the Team X study report. Aerospace estimated this mission to be about $1.64B ($FY15), somewhat above the JPL estimate, but still well below the target cost. Differences in this case resulted from a higher modeled spacecraft cost, as well as a higher estimate on operations cost related to the lengthy duration of inner solar system cruise.

**Table 4-33.** Aerospace ICE Option 4 Cost Estimate.

| FY15 $M | JPL Estimate | Aerospace ICE | Difference FY15 $M | Difference % |
|---|---|---|---|---|
| Phase A | $ - | $ 19.5 | $ 19.5 | |
| Subtotal | $ - | $ 19.5 | $ 19.5 | |
| Phase C/D | | | | |
| Mission PM/SE/MA | $ 113.8 | $ 121.8 | $ 8.0 | 7.0% |
| Payload[1] | $ 128.5 | $ 124.6 | $ (3.9) | -3.1% |
| Flight System[2] | $ 685.7 | $ 714.1 | $ 28.4 | 4.1% |
| Pre-Launch GDS/MOS | $ 86.9 | $ 89.0 | $ 2.1 | 2.5% |
| *Launch Vehicle* | *$ 33.0* | *$ 33.0* | *$ -* | *0.0%* |
| Reserves | $ 245.1 | $ 294.3 | $ 49.2 | 20.1% |
| Subtotal | $ 1,293.0 | $ 1,376.7 | $ 83.7 | 6.5% |
| Phase E/F | | | | |
| MO&DA - Science | $ 175.3 | $ 204.4 | $ 29.1 | 16.6% |
| Reserves | $ 24.5 | $ 42.6 | $ 18.1 | 73.9% |
| Subtotal | $ 199.8 | $ 247.0 | $ 47.2 | 23.6% |
| **Total** | **$ 1,492.8** | **$ 1,643.2** | **$ 150.4** | **10.1%** |

Note: *Italics* indicates project values; treated as pass throughs
1. Includes orbiter and probe instruments, as applicable
2. Includes probe, entry system, orbiter/flyby bus, SEP as applicable

## 4.7    Mission Option 5: Uranus Orbiter with Probe and 50-Kg Payload (no SEP)

### 4.7.1    Overview

Additional trajectory work spurred by the unexpectedly high mass of the SEP stages developed for Options 1 and 2 led to a reassessment of the use of SEP for these missions. Incorporation of the SEP stage was expected to decrease flight time and keep these missions on a smaller launch vehicle, however the dry mass estimates developed by Team X drove the launch mass to a level that, while feasible, pushed the missions to the highest performance LV available while trimming only about a year off the cruise duration. Given the significant contribution the SEP stage was adding to the total mission cost (~$200M, including reserves) it was determined that the optimal mission architecture for both the Uranus orbiter with probe and the Uranus orbiter with a 150 kg payload and no probe would be all-chemical, foregoing the option of a SEP stage. Team X then added Option 5 and Option 6 to develop point designs and costs for these architectures. As expected, these changes lowered mission cost at the expense of ~one year less science orbit time at Uranus as a result of the slightly longer transfer duration, which still allowed all science objectives to be achieved.

### 4.7.2    Science

Scientifically, this mission is identical to Option 1, the Uranus orbiter with probe using SEP in the inner solar system. That description is repeated here.

This mission represents the minimum the science team felt an ice giant Flagship mission should accomplish. The instrumentation, discussed below and in Section 3.3.2, allows the highest-priority science to be accomplished; measurement of the bulk composition including noble gases and isotopic ratios (primarily achieved by the atmospheric probe), and determination of interior structure (primarily achieved by the Doppler imager, but supported by the magnetometer). The combination of carrying a camera system and remaining in orbit within the uranian system allows other priority goals to be achieved and still more to be partially addressed. The orbiter mission also enables the study of time-varying features within the system, and opens the possibility of serendipitous discovery and follow-up.

### 4.7.3 Instrumentation

Instrumentation for Option 5 is identical to that described for Option 1 (Section 4.3.3).

### 4.7.4 Mission Design

The mission design objective for this Team X mission option was to enable a Uranus orbiter (with ~50 kg payload) and a Uranus atmospheric probe. The high-level mission design guidelines for this options were:

1. Launch between 2024–2037, with a preferred launch date between 2029–2031
2. Fly a purely chemical interplanetary trajectory (no SEP stage)
3. Total mission lifetime < 15 years (including Science phase)
4. Launch on existing commercial Launch Vehicle
5. Avoid Uranus rings during orbit insertion
6. Design Probe coast, entry and descent trajectory with feasible Orbiter – Probe telecom geometry relay
7. Uranus moon tour with two flybys of each of the major 5 moons

Given that this was one of the later Team X studies, the orbiter dry mass was known during the initial stages of the mission design. This allowed for an optimized, ballistic mission trajectory. **Figure 4-32** depicts the baseline mission architecture, which can be divided into three mission phases:

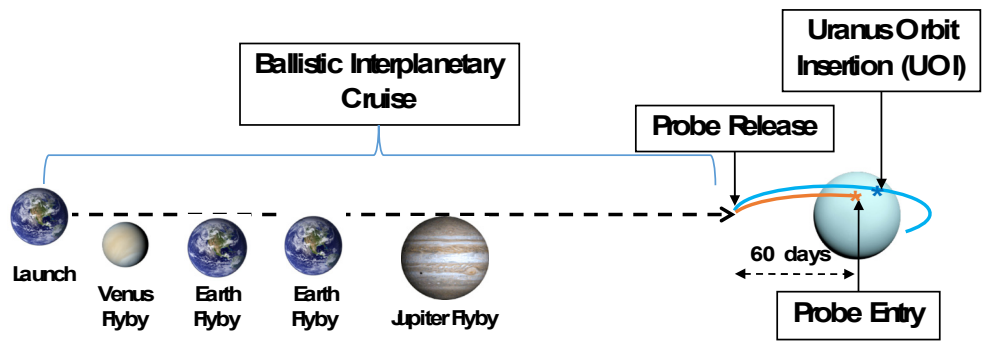

**Figure 4-32.** Mission Concept Option 5: mission design architecture.

### 4.7.4.1 Launch and Interplanetary Cruise

The baseline interplanetary trajectory relies on one Venus and two Earth flybys in the inner solar system. After the last Earth flyby, the spacecraft is boosted out to Jupiter. Post Jupiter flyby the spacecraft coasts to Uranus, taking one year extra to arrive at Uranus as compared to Mission options relying on SEP. Before the Jupiter flyby the spacecraft performs two deep space maneuvers. The first deep space maneuver (DSM) is performed between the two Earth flybys to increase the spacecraft's flyby velocity with respect to Earth. The second DSM is performed after the third Earth flyby and is required for achieving correct flyby conditions at Jupiter. **Table 4-34** lists the Team X in-session interplanetary trajectory.

**Table 4-34.** Option 5 Team X baseline mission trajectory

| Flyby Sequence | Launch Vehicle | Launch Date | Launch C3 (km²/s²) | IP TOF (yrs.) | Deep Space Maneuvers (m/s) | Arrival Mass (kg) | Orbit Insertion ΔV (km/s) | Mass in Orbit (kg) |
|---|---|---|---|---|---|---|---|---|
| **Earth-VEEJ-Uranus** | Atlas-V 541 | 05/25/2031 | 11.9 | 12.0 | 565 | 3582.5 | 1.7 | 1913.4 |

**Figure 4-33**, shows the top view trajectory plot with relevant details. The quantity "*Arrival Mass*" refers to mass of the spacecraft **before** probe drop-off (Uranus probe mass ~321 kg), which is ~60 days before UOI.

If a more powerful launch vehicle were used, then the interplanetary trip time (for a similar useful mass in orbit) could be cut by up to 1.5 years using a commercially available launch vehicle (Delta-IV Heavy) or up to 4 years for a SLS-1B. This assumes an insertion ΔV <4.5 km/s.

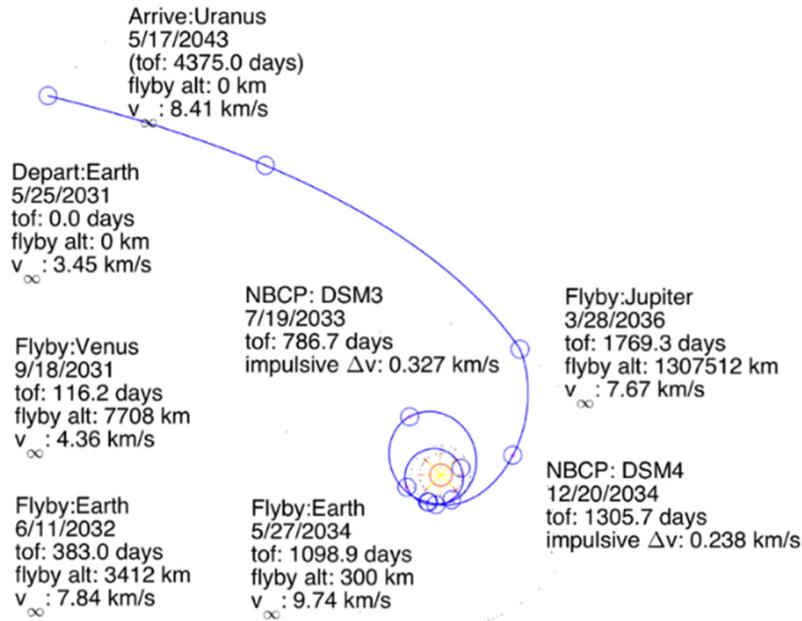

**Figure 4-33.** Option 5 Team X in-session trajectory.

### 4.7.4.2 Probe Coast, Entry and Descent + Uranus Orbit Insertion

The baseline mission trajectory releases the probe 60 days before entry. A PTM is performed by the orbiter prior to probe release, followed by an ODPTM on the orbiter to target UOI. The probe enters Uranus' atmosphere at an entry flight path angle of -30 degrees. A steep angle is chosen to alleviate orbiter-probe relay geometry issues (explained later in this section). The probe descent lasts for ~1 hr, of which the first ~30 mins represent the entry sequence, as shown in **Figure 4-35**. Details on the probe entry are given in **Table 4-35**.

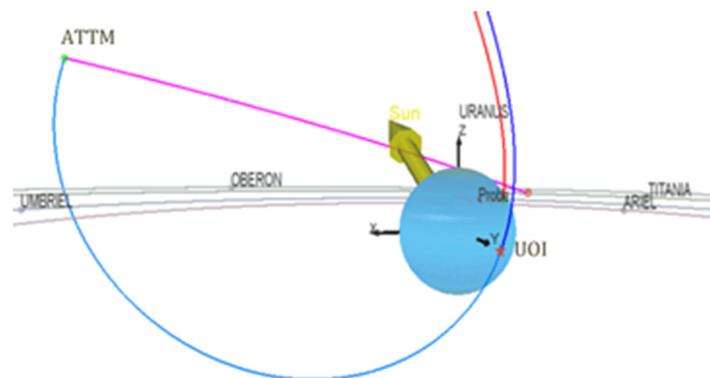

**Figure 4-34.** Option 5 UOI and probe entry.

The orbiter performs the UOI burn at an altitude of ~1.05 Uranus radii and enters in a ~142 days orbit around Uranus. An orbit insertion altitude relatively close to the atmosphere is chosen to mitigate potential ring particle impingement issues. The low UOI altitude also helps in reducing the UOI ΔV magnitude. **Figure 4-35** highlights UOI and probe entry geometry.

**Table 4-35.** Option 5 probe entry.

| Parameter | Value |
|---|---|
| Interface Altitude | 1000 km |
| Entry Velocity | 22.5 km/s |
| Entry Flight Path Angle | -30 |
| Max G load | 165 G |
| Stagnation Pressure | 9 atm |
| Cumulative Heatload | 41114 J/cm² |

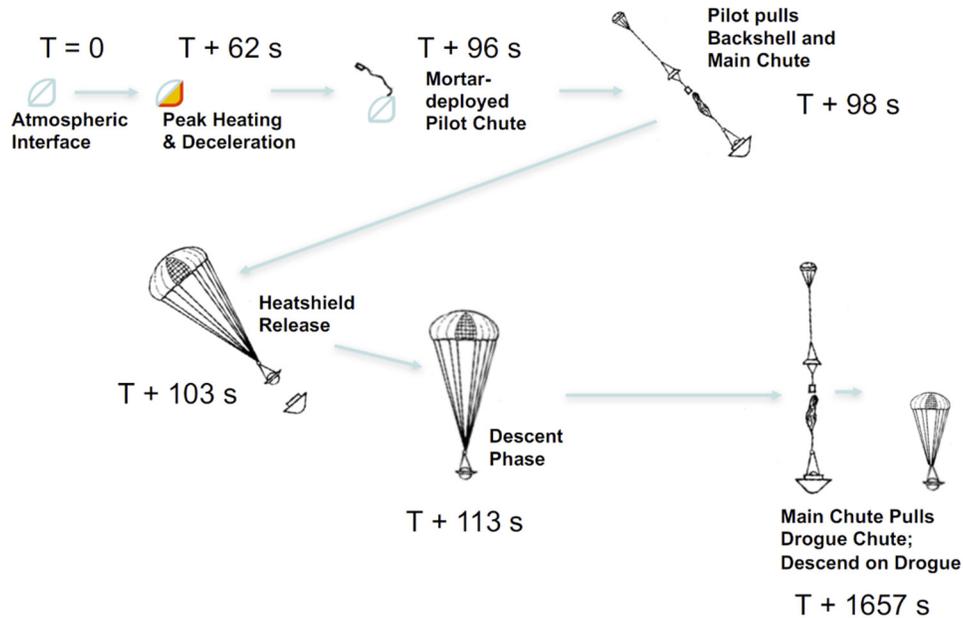

**Figure 4-35.** Probe entry sequence (timing represents Uranus example).

### Risks and Concerns

Following are some of the findings from Uranus Orbit Insertion analysis and Probe-Orbiter Telecomm. geometry optimization. A hyperbolic probe entry (with orbiter relay) at Uranus must trade the following design considerations:

1. Uranus Orbit Insertion (UOI) ΔV

2. Probe g-load tolerance

3. Relay telecomm. requirements

UOI ΔV is sensitive to the orbiter periapsis altitude (see Section 4.3.6). Higher orbiter periapsis provides better relay line-of-sight and longer persistence (lower angular rate relative to probe), but higher UOI DV. The Team X in-session design results in a UOI DV of ~2.27 km/s at 1.08 Uranus radii (Ru), but 2.86 km/s at 1.75 Ru. Shallow EFPA reduces probe g-load, but presents challenging telecomm. geometry. The probe relay antenna must point zenith since the probe rotational phase during EDL cannot be easily predicted. One option is to baseline an omnidirectional antenna, or have multiple antennae.

Another factor to consider is the time between Probe Entry and UOI. Currently, there are 2 hours allocated between probe entry and UOI, a critical event. It may be operationally challenging to sequence both the probe relay and UOI on the orbiter within this time window. Increasing the separation will make the geometry more challenging for telecomm. Probe-orbiter geometry also needs to deal with issues like uncertainties regarding the Uranus atmosphere and potential signal attenuation.

A steeper EFPA (-30 deg) was hence chosen to alleviate the telecom geometry and UOI ΔV issue, while still having acceptable g loads on the probe (<200 g). Detail navigation, telecomm analysis is recommended in a follow up study.

### 4.7.4.3 Uranus Tour Phase:

The orbiter inserts into a ~73 degree (or ~107 degree, depending on the assumed Uranus Pole direction) inclined orbit around Uranus. For UOI orbit to be inclined at ~73 degrees, the resulting Uranus pole direction results in the uranian moons inclined at ~0 degrees (except for the inner most moon Miranda). Post UOI, the spacecraft coasts for ~75 days towards its apoapsis where it performs an ATTM. The ATTM is designed to achieve an efficient orbital plane change (by twisting about the line joining the center of the planet and apoapsis) along with targeting one of the big moons of Uranus (Titania).

Post ATTM, the spacecraft acquires an orbital inclination of ~13 degrees and encounters Titania after ~67 days with a hyperbolic flyby velocity of 4.5 km/s. During the ~2 years Uranus moon tour, the spacecraft does 2 flybys of Titania, 3 flybys of Oberon, 3 flybys of Umbriel, 3 flybys of Miranda and 3 flybys of Ariel. **Table 4-36** summarizes the Uranus moon tour. As you can see, most of the flybys are on the sunlit side of the moons. Note that this tour represents one possible tour option. More day sides flybys of Titania and Oberon can be added with future tour design work. **Figure 4-36** show different views of the uranian moon tour.

**Table 4-36.** Option 5 Uranus Moon Tour (inclination is wrt. Uranus ring plane, orbit period is reported post flyby).

| Encounter | Julian Date | Encounter Date (ET) | TOF per leg (days) | Resonances Body | S/C | Flyby Alt. (km) | B-Plane (deg) | V-inf (km/sec) | Inclination (deg) | Orbit Period (days) | Det. DV per leg (m/s) |
|---|---|---|---|---|---|---|---|---|---|---|---|
| T1 | 2467527.8 | 2043-OCT-05 06:46:02 | 104.5 | 12 | 1 | 627 | -2 | 4.5 | 12.2 | 104.5 | 0.5 |
| T2 | 2467632.3 | 2044-JAN-17 18:03:46 | 77.3 | 8.9 | 1 | 200 | -40 | 4.5 | 11.1 | 77.7 | 10.6 |
| O1 | 2467709.6 | 2044-APR-04 01:23:17 | 67.3 | 5 | 1 | 500 | -51 | 4.0 | 10.7 | 67.3 | 0.5 |
| O2 | 2467776.9 | 2044-JUN-10 09:11:07 | 53.8 | 4 | 1 | 200 | -30 | 4.0 | 9.3 | 53.9 | 1.0 |
| O3 | 2467830.7 | 2044-AUG-03 05:34:28 | 49.2 | 3.5 | 1 | 100 | -60 | 4.0 | 5.9 | 47.5 | 42.7 |
| U1 | 2467879.9 | 2044-SEP-21 10:03:06 | 45.6 | 11 | 1 | 500 | 243 | 5.4 | 4.6 | 45.6 | 2.3 |
| U2 | 2467925.5 | 2044-NOV-06 00:05:56 | 41.4 | 10 | 1 | 200 | -221 | 5.4 | 4.0 | 41.5 | 0.5 |
| U3 | 2467966.9 | 2044-DEC-17 10:46:39 | 79.5 | 9.6 | 2 | 100 | 106 | 5.4 | 3.3 | 39.8 | 15.7 |
| M1 | 2468046.4 | 2045-MAR-06 22:37:36 | 39.6 | 28 | 1 | 500 | -181 | 6.0 | 3.0 | 39.6 | 0.0 |
| M2 | 2468086.0 | 2045-APR-15 12:28:05 | 39.6 | 28 | 1 | 500 | -157 | 6.0 | 3.0 | 39.6 | 1.0 |
| M3 | 2468125.6 | 2045-MAY-25 02:20:44 | 78.5 | 27.7 | 2 | 500 | -141 | 6.0 | 3.0 | 39.1 | 10.9 |
| A1 | 2468204.1 | 2045-AUG-11 13:55:03 | 37.8 | 15 | 1 | 200 | 105 | 5.8 | 2.4 | 37.8 | 1.0 |
| A2 | 2468241.9 | 2045-SEP-18 09:16:39 | 37.8 | 15 | 1 | 100 | 90 | 5.9 | 1.9 | 37.8 | 1.0 |
| A3 | 2468279.7 | 2045-OCT-26 04:35:29 | -- | -- | -- | 100 | 166 | 5.8 | -- | -- | -- |

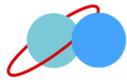

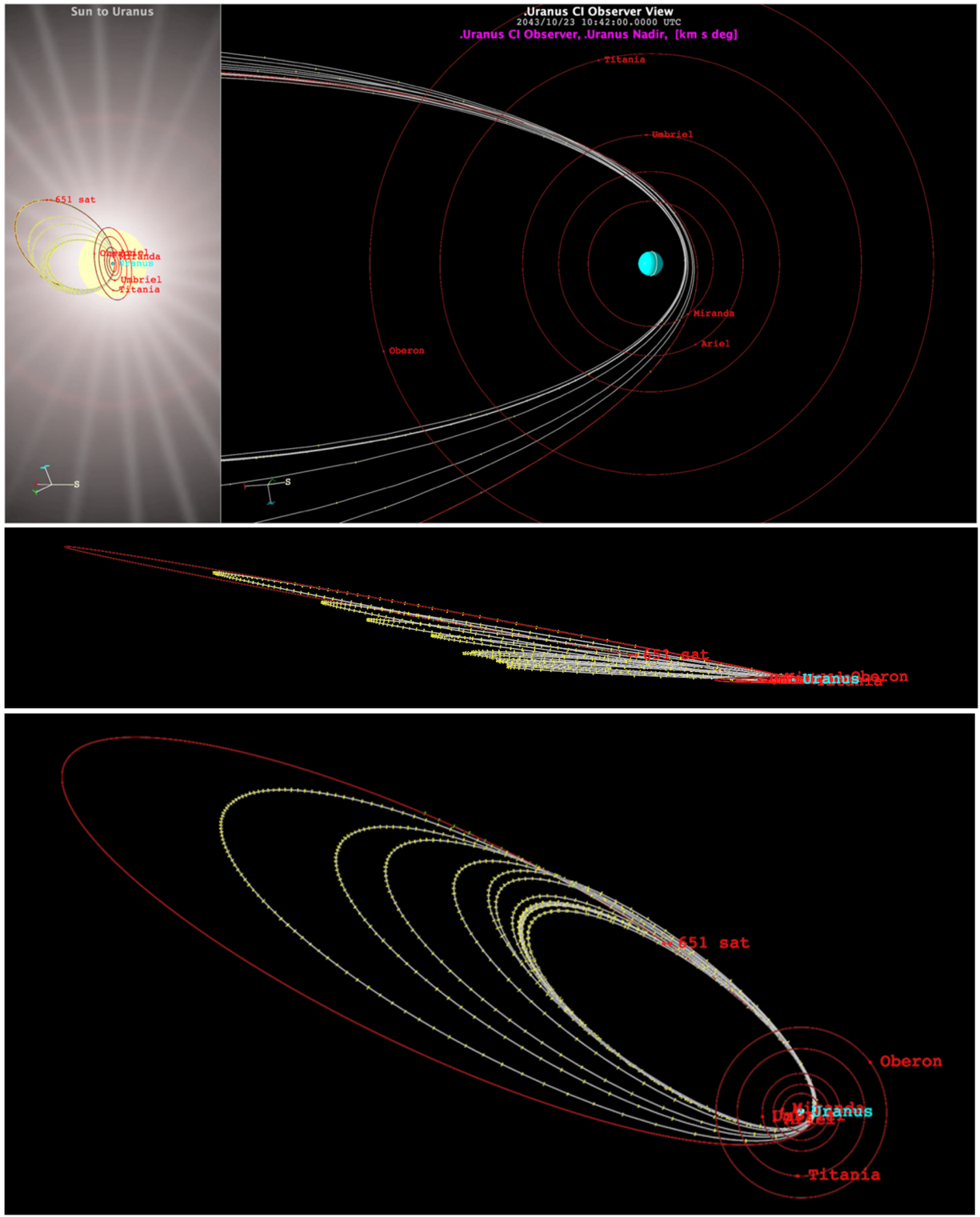

**Figure 4-36.** Option 5 Uranian moon tour views.

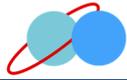

## Mission Delta-V Summary

**Table 4-37** shows the summary ΔV table for this mission options. The chemical ΔV is broken down into monoprop and biprop burns, based on ΔV magnitude and criticality of the maneuver. The main chemical ΔV driver for a mission to Uranus is the orbit insertion due to relatively high approach V∞. Following UOI, the orbiter coasts towards its capture orbit apoapsis where it performs an ATTM. The ATTM is designed to achieve an efficient orbital plane change and targeting one of the major moons of Uranus (Titania). Tour ΔV is based on the tour design work done for this mission option.

**Table 4-37.** Mission Option 5 Delta-V summary.

| | Biprop (m/s) | Monoprop (m/s) | Comments |
|---|---|---|---|
| Interplanetary TCMs | 25 | 25 | Most of the TCMs are just after launch and after SEP orbit separation |
| DSMs | 565 | 10 | Two deep space maneuvers |
| PTM | 0 | 5 | Probe targeting maneuver, ~60 days before UOI |
| ODPTM | 20 | 0 | 60 days before UOI |
| UOI | 1700 | 0 | ~1 hr. burn on two 700 lbf engines |
| UOI-CU (2%) | 34 | 0 | |
| ATTM | 201 | 0 | Apoapsis Twist and Targeting Maneuver |
| ATTM-CU (2%) | 0 | 4 | |
| Tour Deterministic | 45 | 40 | Multiple maneuvers |
| Tour Margin | 0 | 20 | For future tour design work |
| Tour Statistical | 0 | 30 | |
| De-orbit and Disposal | 10 | 0 | Crash into planet |
| **Total** | **2600** | **134** | |

### 4.7.5    Concept of Operations/GDS

Option 5 was explored as a lower cost alternative to Option 1, eliminating the SEP stage and completing the trajectory using only chemical propulsion. The absence of SEP thrusting in the inner solar system simplifies operations during the first years of the mission, however the number of gravity assists in the first part of the cruise will require a typical level of tracking and planning, as described in Option 4.

The long ballistic cruise to Uranus after the Jupiter gravity assist is identical to the operations concept for Option 1, as are the approach phase, probe delivery and science data relay, and Uranus orbital operations. The approach and orbital science plans and data outlines are all also identical to that of Option 1, as well as the decommissioning and disposal plan.

### 4.7.6    Flight System/Probe Design

The flight system design for Option 5 began with the Option 1 design. Only minor changes were required to adapt to the absence of the SEP stage, mostly involved with the elimination of hardware to accommodate SEP stage interface and separation. The flight system configuration is shown in **Figure 4-37**.

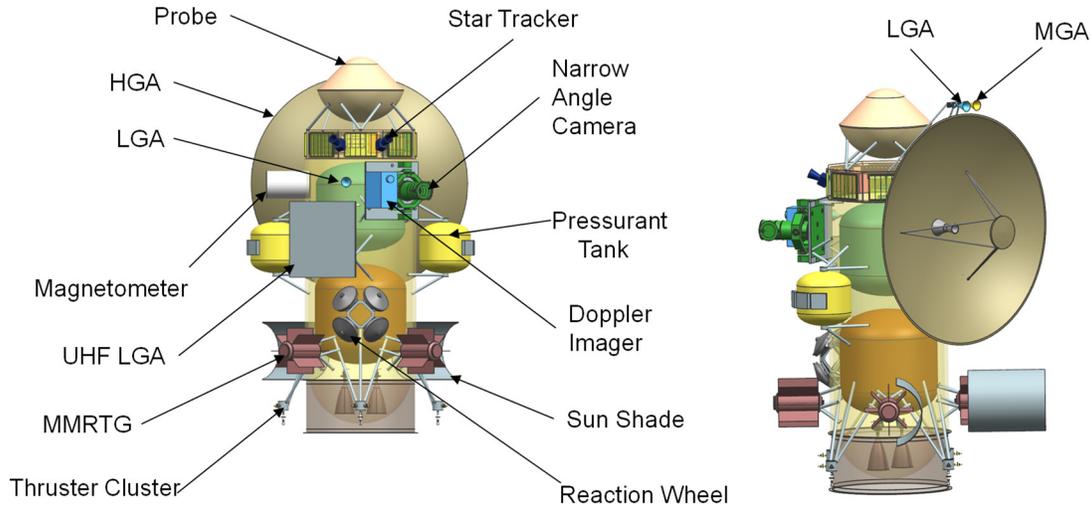

**Figure 4-37.** Option 5 launch configuration.

The instrument complement and Uranus probe design are unchanged from Option 1 and accommodation on the spacecraft bus likewise remains the same.

The integrated flight system has a total wet launch mass of 4,344 kg, and comprises a 1442 kg dry orbiter spacecraft with 2357 kg of bipropellant and a 321 kg atmospheric probe. See **Table 4-38** for the flight system mass summary.

**Table 4-38.** Option 5 MEL.

| Orbiter | CBE Mass (kg) | Contingency (%) | Total Mass (kg) | Heritage/Comments |
|---|---|---|---|---|
| Instruments | 36.7 | 23% | 45.2 | 3 instruments |
| C&DH | 21.6 | 10% | 23.8 | JPL Reference Bus |
| Power | 216.6 | 2% | 220.8 | No contingency on eMMRTGs |
| Telecom | 59.4 | 16% | 68.9 | 35 W TWTA, 3 m HGA |
| Structures | 451.1 | 29% | 581.1 | |
| Harness | 86.3 | 30% | 112.2 | |
| Thermal | 113.1 | 23% | 139.5 | |
| Propulsion | 171.7 | 5% | 181.0 | Dual Mode |
| GN&C | 63.5 | 10% | 69.8 | |
| **Orbiter Total** | **1220.0** | **18%** | **1442.3** | |
| System Margin | | | 224.9 | |
| **Dry Mass Total** | | **43%** | **1667.2** | |
| | | | | |
| **Propellant** | | | **2357.0** | |
| | | | | |
| **Wet Mass Total** | | | **4024.2** | |
| **Mission System** | CBE Mass (kg) | Contingency (%) | Total Mass (kg) | Heritage/Comments |
| Probe | | | 320.7 | |
| Orbiter | | | 4024.2 | |
| **Launch Mass Total** | | | **4344.9** | |
| Injected Mass Cap. | | | 4450.0 | Atlas 541 to C3 of 11.9 |
| **Remaining LV Cap.** | | | **105.1** | |

The atmospheric probe is slated to be released 60 days prior to UOI. The orbiter propellant mass is sized to accommodate this release scenario.

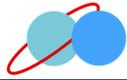

### 4.7.7    New Technology

As with the previous options, no new technologies are required for the flight system. The elimination of the SEP stage also serves to reduce the amount of engineering development required.

### 4.7.8    Cost

Team X estimated the full mission cost of Option 5 to be $1.70B ($FY15) as shown in **Table 4-39**. This cost includes 30% reserves for Phases A–D and 15% in Phase E. Note that per the ground rules, no reserves were carried on eMMRTG and DSN costs.

**Table 4-39.** Option 5 Team X Cost Summary.

| WBS Elements | NRE | RE | 1st Unit |
|---|---|---|---|
| **Project Cost (no Launch Vehicle)** | **$1243.8 M** | **$456.3 M** | **$1700.1 M** |
| **Development Cost (Phases A - D)** | **$916.4 M** | **$456.2 M** | **$1372.6 M** |
| 01.0 Project Management | $47.3 M | | $47.3 M |
| 02.0 Project Systems Engineering | $23.7 M | $0.5 M | $24.2 M |
| 03.0 Mission Assurance | $46.0 M | $0.0 M | $46.0 M |
| 04.0 Science | $24.8 M | | $24.8 M |
| 05.0 Payload System | $80.2 M | $48.3 M | $128.5 M |
| 06.0 Flight System | $445.2 M | $280.5 M | $725.7 M |
| 6.01 Flight System Management | $5.0 M | | $5.0 M |
| 6.02 Flight System Systems Engineering | $51.1 M | | $51.1 M |
| 6.03 Product Assurance (included in 3.0) | | | $0.0 M |
| Orbiter | $297.0 M | $236.5 M | $533.5 M |
| SEP Stage | $0.0 M | $0.0 M | $0.0 M |
| Probe | $26.8 M | $18.1 M | $44.9 M |
| Entry System | $57.1 M | $24.4 M | $81.5 M |
| Ames/Langley EDL Engineering/Testing | $3.8 M | $0.0 M | $3.8 M |
| 6.14 Spacecraft Testbeds | $4.5 M | $1.5 M | $6.0 M |
| 07.0 Mission Operations Preparation | $26.1 M | | $26.1 M |
| 09.0 Ground Data Systems | $22.0 M | | $22.0 M |
| 10.0 ATLO | $21.1 M | $21.7 M | $42.8 M |
| 11.0 Education and Public Outreach | $0.0 M | $0.0 M | $0.0 M |
| 12.0 Mission and Navigation Design | $21.9 M | | $21.9 M |
| Development Reserves | $158.0 M | $105.3 M | $263.3 M |
| **Operations Cost (Phases E - F)** | **$294.4 M** | **$0.1 M** | **$294.5 M** |
| 01.0 Project Management | $27.1 M | | $27.1 M |
| 02.0 Project Systems Engineering | $0.0 M | $0.1 M | $0.1 M |
| 03.0 Mission Assurance | $3.6 M | $0.0 M | $3.6 M |
| 04.0 Science | $69.2 M | | $69.2 M |
| 07.0 Mission Operations | $130.6 M | | $130.6 M |
| 09.0 Ground Data Systems | $30.0 M | | $30.0 M |
| 11.0 Education and Public Outreach | $0.0 M | $0.0 M | $0.0 M |
| 12.0 Mission and Navigation Design | $0.0 M | | $0.0 M |
| Operations Reserves | $33.9 M | $0.0 M | $34.0 M |
| **8.0 Launch Vehicle** | **$33.0 M** | | **$33.0 M** |
| Launch Vehicle and Processing | $0.0 M | | $0.0 M |
| Nuclear Payload Support | $33.0 M | | $33.0 M |

Following completion of the Team X study, the Aerospace Corporation performed an independent cost estimate (ICE) for this option, using the Team X design as documented in the Team X study report. Aerospace estimated this mission to be about $1.99B ($FY15), above the

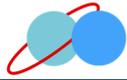

JPL estimate, but still within the target cost. Differences in this case once again resulted from a higher modeled spacecraft cost, as well as a higher estimate on operations cost related to the lengthy duration of inner solar system cruise.

**Table 4-40.** Aerospace ICE Option 5 Cost Estimate.

| FY15 $M | JPL Estimate | Aerospace ICE | Difference FY15 $M | Difference % |
|---|---|---|---|---|
| Phase A | $    - | $   21.8 | $   21.8 | |
| Subtotal | $    - | $   21.8 | $   21.8 | |
| Phase C/D | | | | |
| Mission PM/SE/MA | $  117.5 | $  133.5 | $   16.0 | 13.6% |
| Payload[1] | $  128.6 | $  124.8 | $   (3.8) | -3.0% |
| Flight System[2] | $  768.6 | $  810.0 | $   41.4 | 5.4% |
| Pre-Launch GDS/MOS | $   94.8 | $   99.3 | $    4.5 | 4.7% |
| *Launch Vehicle* | *$   33.0* | *$   33.0* | *$    -* | *0.0%* |
| Reserves | $  263.3 | $  336.9 | $   73.6 | 28.0% |
| Subtotal | $ 1,405.8 | $ 1,537.4 | $  131.6 | 9.4% |
| Phase E/F | | | | |
| MO&DA - Science | $  260.6 | $  361.2 | $  100.6 | 38.6% |
| Reserves | $   34.0 | $   72.1 | | |
| Subtotal | $  294.6 | $  433.3 | $  138.7 | 47.1% |
| **Total** | **$ 1,700.4** | **$ 1,992.6** | **$  292.2** | **17.2%** |

Note: *Italics* indicates project values; treated as pass throughs
1. Includes orbiter and probe instruments, as applicable
2. Includes probe, entry system, orbiter/flyby bus, SEP as applicable

## 4.8    Mission Option 6: Uranus Orbiter with 150-Kg Payload (no Probe or SEP)

### 4.8.1    Overview

As with Option 5, this mission architecture resulted when it was discovered that all-chemical trajectories could be found that enabled meeting the mission requirements of Option 2 without the addition of a SEP stage. In fact, using the trajectory developed for this option the all-chemical architecture was able to trim a year of flight time off the SEP stage architecture, while also lowering the class of launch vehicle from a Delta IVH to an Atlas V.

### 4.8.2    Science

Scientifically, this mission is identical to Option 2, the Uranus orbiter (no probe) using SEP in the inner solar system. That description is repeated here.

This architecture was selected for detailed study to explore the effects of having a large orbiter science payload. A probe was not included both because A-Team work suggested it would not fit within our cost guidelines and to allow us to explore the "no probe" region of parameter space. This mission achieves all priority science goals except the one related to bulk composition including noble gases and isotopic ratios. That goal requires an atmospheric probe and is one of our two highest-priority goals. Because of the probe's importance, the SDT would prefer not to fly this mission, though we find it does achieve significant science. The SDT also notes that this architecture would be highly desired should a separate mission execute a probe mission.

### 4.8.3    Instrumentation

Instrumentation for Option 5 is identical to that described for Option 2 (Section 4.4.3).

### 4.8.4    Mission Design

The mission design objective for this Team X mission option was to enable a Uranus orbiter with a large payload. The high-level mission design guidelines for this options were:

1. Launch between 2024–2037, with a preferred launch date between 2029–2031
2. Fly a purely chemical interplanetary trajectory (no SEP stage)
3. Total mission lifetime <15 years (including Science phase)
4. Launch on existing commercial Launch Vehicle
5. Avoid Uranus rings during orbit insertion
6. Design Probe coast, entry and descent trajectory with feasible orbiter–probe telecomm geometry relay
7. Uranus moon tour with two flybys of each of the major 5 moons

Given that this was one of the later Team X studies, the orbiter dry mass was known during the initial stages of the mission design. This allowed for selection of an optimized, ballistic mission trajectory. **Figure 4-38** depicts the baseline mission architecture, which can be divided into three mission phases:

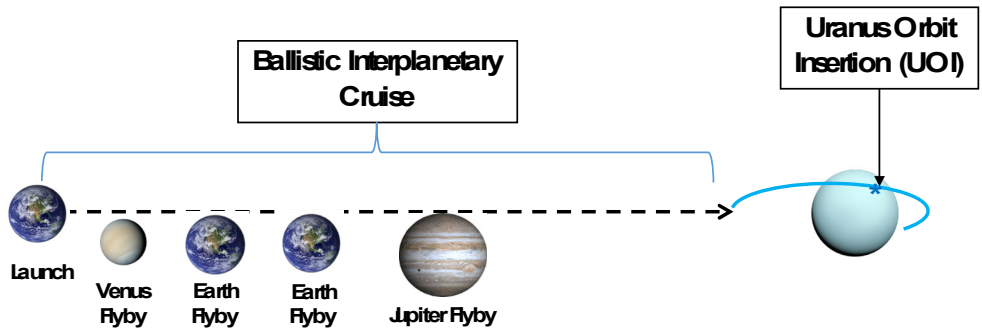

**Figure 4-38.** Mission Option 6: mission design architecture.

### 4.8.4.1   Launch and Interplanetary Cruise

The baseline interplanetary trajectory relies one Venus and two Earth flybys in the inner solar system.  After the last Earth flyby, the spacecraft is boosted up to Jupiter. Post Jupiter flyby the spacecraft coasts to Uranus, taking one year's extra to arrive at Uranus as compared to Mission options relying on SEP.  **Table 4-41** lists Team X in-session interplanetary trajectory.

**Table 4-41.** Option 6 Team X baseline mission trajectory.

| Flyby Sequence | Launch Vehicle | Launch Date | Launch C3 (km²/s²) | IP TOF (yrs.) | Deep Space Maneuvers (m/s) | Arrival Mass (kg) | Orbit Insertion DV (km/s) | Mass in Orbit (kg) |
|---|---|---|---|---|---|---|---|---|
| Earth-VEEJ-Uranus | Atlas-V 551 | 05/25/2031 | 11.9 | 12.0 | 565 | 3903.5 | 1.7 | 2289.9 |

**Figure 4-39** shows the top view trajectory plot with relevant details. The quantity "*Arrival Mass*" refers to mass of the spacecraft **before** UOI.

If a more powerful launch vehicle were used, then the interplanetary trip time (for a similar useful mass in orbit) could be cut by up to 1.5 years using a commercially available launch vehicle (Delta-IV Heavy) or up to 4 years for a SLS-1B.  This assumes an insertion Delta-V <4.5 km/s.

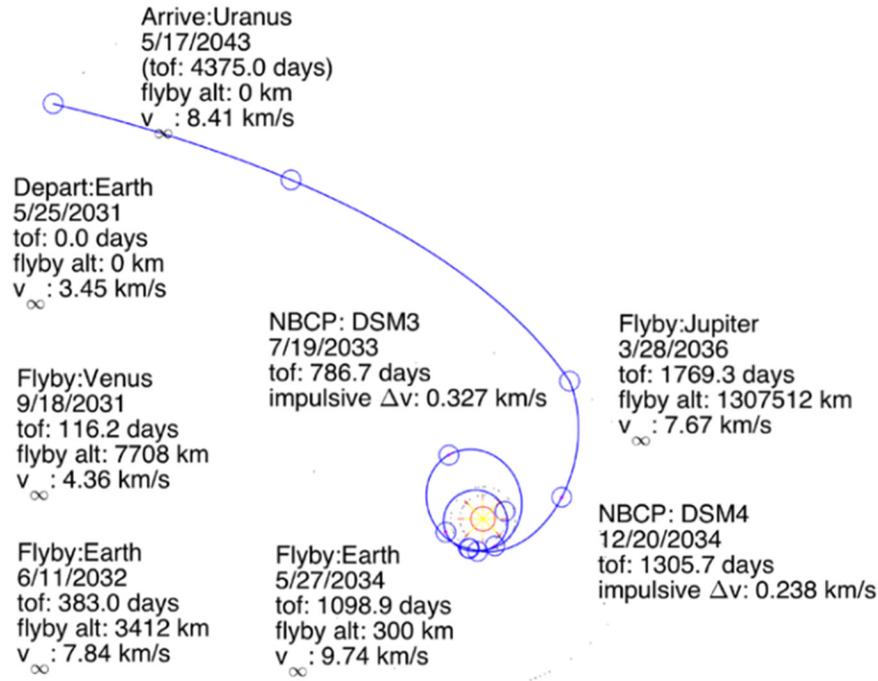

**Figure 4-39.** Option 6 Team X in-session trajectory.

### 4.8.4.2    Probe Coast, Entry and Descent + Uranus Orbit Insertion

The in-session baseline mission trajectory performs a UOI periapsis targeting maneuver (UPTM) 10 days before the target UOI. The orbiter performs a UOI Delta-V of ~1.7 km/s at a periapsis altitude of ~1.05 Uranus radii. The UOI burn lasts for <1 hr. using the two 890N engines on the spacecraft. Post-UOI the orbiter enters a ~142 day orbit around Uranus. An orbit insertion altitude relatively close to the atmosphere is chosen to mitigate potential ring particle impingement issues. The low UOI altitude also helps in reducing the UOI ΔV magnitude. **Figure 4-40** highlights the UOI geometry.

### 4.8.4.3    Uranus Tour Phase:

As with Option 5, the orbiter inserts into a ~107 degree (or ~73 degree, depending on the assumed Uranus Pole direction) inclined orbit around Uranus. For UOI orbit to be inclined at ~107 degrees, the resulting Uranus pole direction means that the uranian moons are inclined at ~180 degrees with respect to the Uranus equatorial plane. Post UOI, the spacecraft coasts for ~75 days towards its apoapsis, at which point it perform an ATTM. The ATTM is designed to achieve an efficient orbital plane change (by twisting about the line joining the center of the planet and apoapsis) along with targeting one of the big moons of Uranus (Titania).

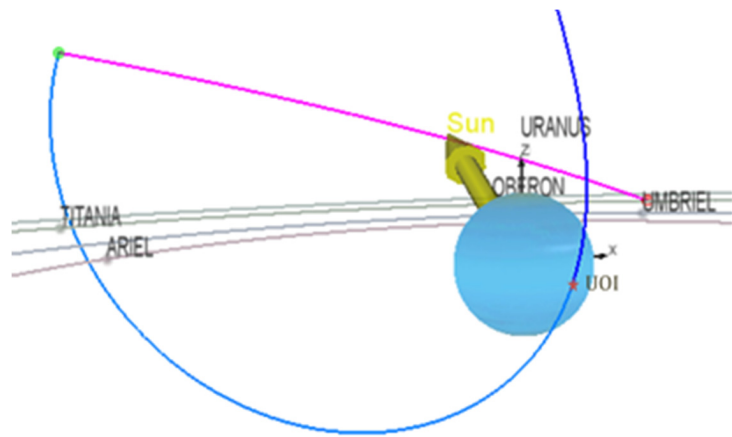

**Figure 4-40.** Option 6 UOI and probe entry.

Post ATTM, the spacecraft changes its orbital inclination to ~167 degrees and encounters Titania after ~67 days with a hyperbolic flyby velocity of 4.5 km/s. During the ~2 years Uranus moon tour, the spacecraft performs 2 flybys of Titania, 3 flybys of Oberon, 3 flybys of Umbriel, 3 flybys of Miranda and 3 flybys of Ariel. **Table 4-42** below summarizes the Uranus moon tour. As can be seen, most of the flybys are on the sunlit side of the moon. Note that this tour represents one possible tour option. More day sides flybys of Titania and Oberon can be added with future tour design work. **Figure 4-41** show different views of the uranian moon tour.

**Table 4-42.** Option 6 Uranus Moon Tour (inclination is wrt. Uranus ring plane, orbit period is reported post flyby).

| Encounter | Julian Date | Encounter Date (ET) | TOF per leg (days) | Resonances Body | Resonances S/C | Flyby Alt. (km) | B-Plane (deg) | V-inf (km/sec) | Inclination (deg) | Orbit Period (days) | Det. DV per leg (m/s) |
|---|---|---|---|---|---|---|---|---|---|---|---|
| T1 | 2467527.8 | 2043-OCT-05 06:46:02 | 104.5 | 12 | 1 | 627 | -2 | 4.5 | 12.2 | 104.5 | 0.5 |
| T2 | 2467632.3 | 2044-JAN-17 18:03:46 | 77.3 | 8.9 | 1 | 200 | -40 | 4.5 | 11.1 | 77.7 | 10.6 |
| O1 | 2467709.6 | 2044-APR-04 01:23:17 | 67.3 | 5 | 1 | 500 | -51 | 4.0 | 10.7 | 67.3 | 0.5 |
| O2 | 2467776.9 | 2044-JUN-10 09:11:07 | 53.8 | 4 | 1 | 200 | -30 | 4.0 | 9.3 | 53.9 | 1.0 |
| O3 | 2467830.7 | 2044-AUG-03 05:34:28 | 49.2 | 3.5 | 1 | 100 | -60 | 4.0 | 5.9 | 47.5 | 42.7 |
| U1 | 2467879.9 | 2044-SEP-21 10:03:06 | 45.6 | 11 | 1 | 500 | 243 | 5.4 | 4.6 | 45.6 | 2.3 |
| U2 | 2467925.5 | 2044-NOV-06 00:05:56 | 41.4 | 10 | 1 | 200 | -221 | 5.4 | 4.0 | 41.5 | 0.5 |
| U3 | 2467966.9 | 2044-DEC-17 10:46:39 | 79.5 | 9.6 | 2 | 100 | 106 | 5.4 | 3.3 | 39.8 | 15.7 |
| M1 | 2468046.4 | 2045-MAR-06 22:37:36 | 39.6 | 28 | 1 | 500 | -181 | 6.0 | 3.0 | 39.6 | 0.0 |
| M2 | 2468086.0 | 2045-APR-15 12:28:05 | 39.6 | 28 | 1 | 500 | -157 | 6.0 | 3.0 | 39.6 | 1.0 |
| M3 | 2468125.6 | 2045-MAY-25 02:20:44 | 78.5 | 27.7 | 2 | 500 | -141 | 6.0 | 3.0 | 39.1 | 10.9 |
| A1 | 2468204.1 | 2045-AUG-11 13:55:03 | 37.8 | 15 | 1 | 200 | 105 | 5.8 | 2.4 | 37.8 | 1.0 |
| A2 | 2468241.9 | 2045-SEP-18 09:16:39 | 37.8 | 15 | 1 | 100 | 90 | 5.9 | 1.9 | 37.8 | 1.0 |
| A3 | 2468279.7 | 2045-OCT-26 04:35:29 | -- | -- | -- | 100 | 166 | 5.8 | -- | -- | -- |

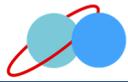

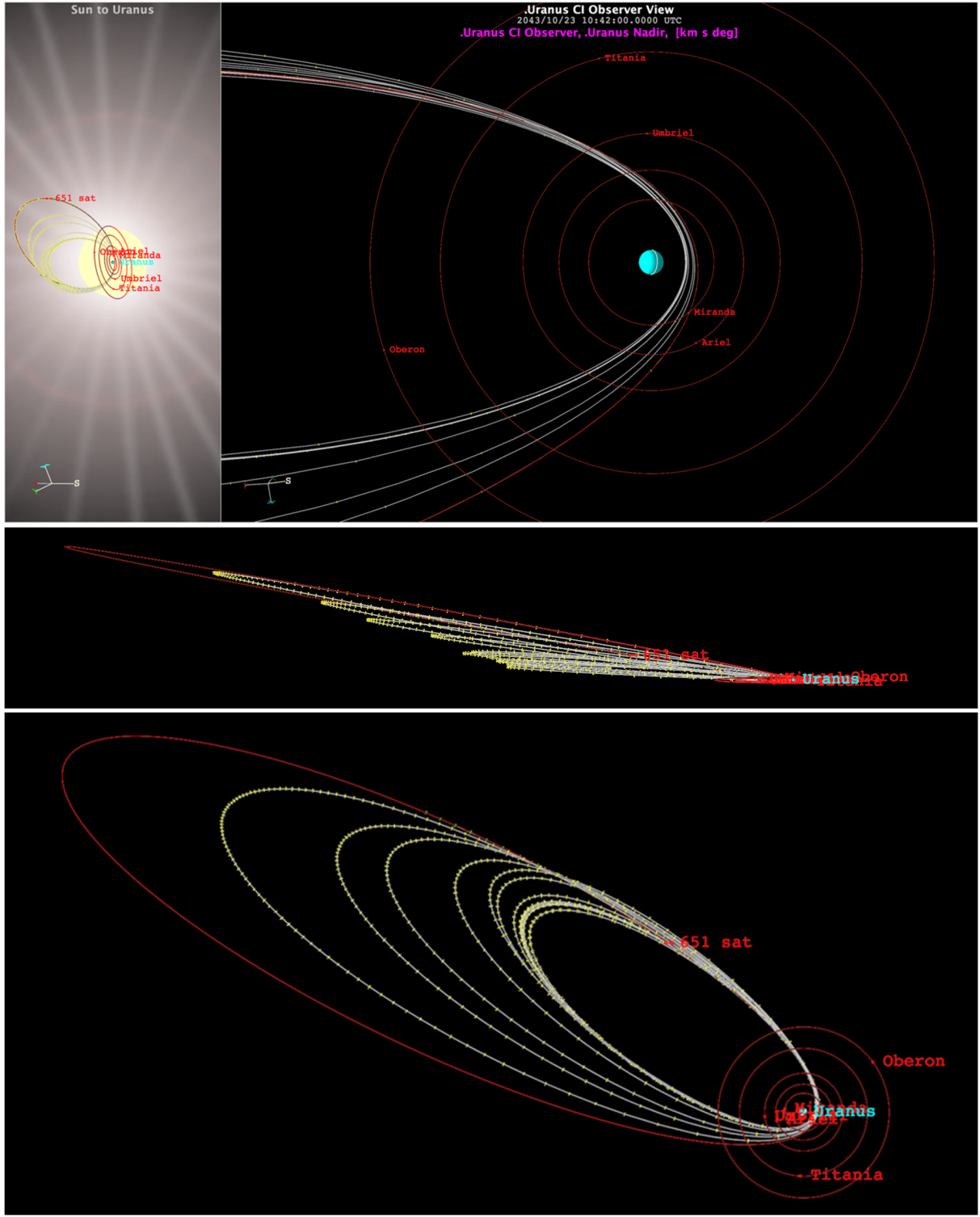

**Figure 4-41.** Option 6 Uranian moon tour views.

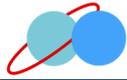

## Mission Delta-V Summary

**Table 4-43** shows the summary ΔV table for this mission options. The chemical ΔV is broken down into monoprop and bi-prop burns, based on ΔV magnitude and criticality of the maneuver. The main chemical ΔV driver for a mission to Uranus is the orbit insertion due to relatively high approach V∞. Following UOI, the orbiter coasts towards its capture orbit apoapsis where it performs an ATTM. The ATTM is designed to achieve an efficient orbital plane change and targeting one of the major moons of Uranus (Titania). Tour ΔV is based on the tour design work done for this mission option.

**Table 4-43.** Mission Option 6 Delta-V summary.

| | Biprop (m/s) | Monoprop (m/s) | Comments |
|---|---|---|---|
| Interplanetary TCMs | 25 | 25 | Most of the TCMs are just after launch and after SEP orbit separation |
| DSMs | 565 | 10 | Two deep space maneuvers |
| UPTM | 5 | 0 | 10 days before UOI |
| UOI | 1700 | 0 | ~1 hr burn on two 700 lbf engines |
| UOI-CU (2%) | 34 | 0 | |
| ATTM | 201 | 0 | Apoapsis Twist and Targeting Maneuver |
| ATTM-CU (2%) | 0 | 4 | |
| Tour Deterministic | 45 | 40 | Multiple maneuvers |
| Tour Margin | 0 | 20 | For future tour design work |
| Tour Statistical | 0 | 30 | |
| De-orbit and Disposal | 10 | 0 | Estimated |
| **Total** | **2585** | **129** | |

### 4.8.5    Concept of Operations/GDS

Option 6 is operationally similar to Option 2, the difference once again being the absence of a SEP stage for inner solar system thrusting. Without the SEP stage, launch and early mission activities are the same as Option 4. Cruise activities, tracking schedules, and mission operations after the Jupiter flyby are identical to option 2, as is the Uranus orbit phase, including the science plans and data outlines. The decommissioning and disposal of the spacecraft is also the same as Option 2.

### 4.8.6    Flight System Design

The flight system for Option 6 was based on that developed for Option 2, removing any accommodation for the SEP stage. Subsystem designs were unchanged, with minor adaptations to propulsion to size for the different propellant load, and mechanical to remove attachment and separation hardware associated with the SEP stage. The orbiter configuration is shown in **Figure 4-42**.

The instrument complement is unchanged from Option 2 and accommodation on the spacecraft bus likewise remains the same.

The integrated flight system has a total wet launch mass of 4718 kg, and comprises a 1995 kg dry orbiter spacecraft with 2723 kg of bipropellant. See **Table 4-44** for the flight system mass summary.

### 4.8.7    New Technology

As with the previous options, no new technologies are required for the flight system. The elimination of the SEP stage also serves to reduce the amount of engineering development required.

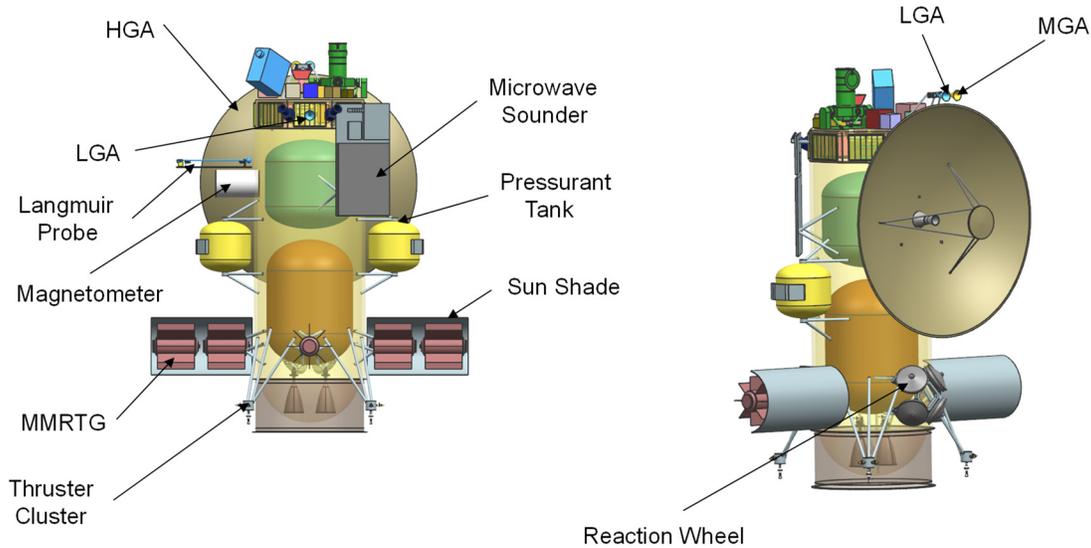

**Figure 4-42.** Option 6 launch configuration.

**Table 4-44.** Option 6 MEL.

| Orbiter | CBE Mass (kg) | Contingency (%) | Total Mass (kg) | Heritage/Comments |
|---|---|---|---|---|
| Instruments | 144.8 | 17% | 169.5 | 3 instruments |
| C&DH | 27.6 | 18% | 32.4 | JPL Reference Bus |
| Power | 265.3 | 2% | 269.7 | No contingency on eMMRTGs |
| Telecom | 55.2 | 15% | 63.4 | 35 W TWTA, 3 m HGA |
| Structures | 501.6 | 29% | 646.3 | |
| Harness | 103.4 | 30% | 134.4 | |
| Thermal | 121.1 | 23% | 148.5 | |
| Propulsion | 179.9 | 5% | 189.7 | Dual Mode |
| GN&C | 63.5 | 10% | 69.8 | |
| **Orbiter Total** | **1462.4** | **18%** | **1723.7** | |
| System Margin | | | 270.8 | |
| **Dry Mass Total** | | **43%** | **1994.5** | |
| **Propellant** | | | **2723.0** | |
| **Wet Mass Total** | | | **4717.5** | |
| Mission System | CBE Mass (kg) | Contingency (%) | Total Mass (kg) | Heritage/Comments |
| Orbiter | | | 4717.5 | |
| **Launch Mass Total** | | | **4717.5** | |
| Injected Mass Cap. | | | 4880.0 | Atlas 551 to C3 of 11.9 |
| **Remaining LV Cap.** | | | **162.5** | |

### 4.8.8    Cost

Team X estimated the full mission cost of Option 6 to be $2.01B ($FY15) as shown in **Table 4-45**. This cost includes 30% reserves for Phases A–D and 15% in Phase E. Note that per the groundrules, no reserves were carried on eMMRTG and DSN costs.

Following completion of the Team X study, the Aerospace Corporation performed an ICE for this option, using the Team X design as documented in the Team X study report. Aerospace estimated this mission to be about $2.32B ($FY15), above the $2B target cost. Differences in this case were driven by higher estimates from Aerospace for the flight system and instruments.

**Table 4-45.** Option 6 Team X Cost Summary.

| WBS Elements | NRE | RE | 1st Unit |
|---|---|---|---|
| **Project Cost (no Launch Vehicle)** | **$1514.0 M** | **$491.1 M** | **$2005.1 M** |
| **Development Cost (Phases A - D)** | $913.6 M | $491.0 M | $1404.6 M |
| 01.0 Project Management | $47.3 M | | $47.3 M |
| 02.0 Project Systems Engineering | $24.8 M | $0.8 M | $25.6 M |
| 03.0 Mission Assurance | $47.3 M | $0.0 M | $47.3 M |
| 04.0 Science | $66.2 M | | $66.2 M |
| 05.0 Payload System | $147.9 M | $86.3 M | $234.1 M |
| 06.0 Flight System | $333.9 M | $273.0 M | $606.9 M |
|     6.01 Flight System Management | $5.0 M | | $5.0 M |
|     6.02 Flight System Systems Engineering | $35.9 M | | $35.9 M |
|     6.03 Product Assurance (included in 3.0) | | | $0.0 M |
|     Orbiter | $288.4 M | $271.5 M | $559.9 M |
|     SEP Stage | $0.0 M | $0.0 M | $0.0 M |
|     6.14 Spacecraft Testbeds | $4.6 M | $1.5 M | $6.1 M |
| 07.0 Mission Operations Preparation | $32.1 M | | $32.1 M |
| 09.0 Ground Data Systems | $28.7 M | | $28.7 M |
| 10.0 ATLO | $16.2 M | $17.7 M | $33.9 M |
| 11.0 Education and Public Outreach | $0.0 M | $0.0 M | $0.0 M |
| 12.0 Mission and Navigation Design | $18.8 M | | $18.8 M |
| Development Reserves | $150.4 M | $113.3 M | $263.7 M |
| **Operations Cost (Phases E - F)** | **$567.4 M** | **$0.1 M** | **$567.5 M** |
| 01.0 Project Management | $27.1 M | | $27.1 M |
| 02.0 Project Systems Engineering | $0.0 M | $0.1 M | $0.1 M |
| 03.0 Mission Assurance | $3.6 M | $0.0 M | $3.6 M |
| 04.0 Science | $243.7 M | | $243.7 M |
| 07.0 Mission Operations | $177.6 M | | $177.6 M |
| 09.0 Ground Data Systems | $47.7 M | | $47.7 M |
| 11.0 Education and Public Outreach | $0.0 M | $0.0 M | $0.0 M |
| 12.0 Mission and Navigation Design | $0.0 M | | $0.0 M |
| Operations Reserves | $67.8 M | $0.0 M | $67.8 M |
| **8.0 Launch Vehicle** | **$33.0 M** | | **$33.0 M** |
| Launch Vehicle and Processing | $0.0 M | | $0.0 M |
| Nuclear Payload Support | $33.0 M | | $33.0 M |

**Table 4-46.** Aerospace ICE Option 6 Cost Estimate.

| FY15 $M | JPL Estimate | Aerospace ICE | Difference FY15 $M | Difference % |
|---|---|---|---|---|
| Phase A | $ - | $ 21.9 | $ 21.9 | |
|   Subtotal | $ - | $ 21.9 | $ 21.9 | |
| Phase C/D | | | | |
|   Mission PM/SE/MA | $ 119.6 | $ 143.0 | $ 23.4 | 19.5% |
|   Payload | $ 218.9 | $ 272.5 | $ 53.6 | 24.5% |
|   Flight System | $ 640.9 | $ 740.4 | $ 99.5 | 15.5% |
|   Pre-Launch GDS/MOS | $ 145.7 | $ 107.5 | $ (38.2) | -26.2% |
|   *Launch Vehicle* | *$ 33.0* | *$ 33.0* | *$ -* | *0.0%* |
|   Reserves | $ 259.1 | $ 390.7 | $ 131.6 | 50.8% |
|   Subtotal | $ 1,417.2 | $ 1,687.0 | $ 269.8 | 19.0% |
| Phase E/F | | | | |
|   MO&DA - Science | $ 499.9 | $ 509.4 | $ 9.5 | 1.9% |
|   Reserves | $ 67.8 | $ 102.8 | | |
|   Subtotal | $ 567.7 | $ 612.2 | $ 44.5 | 7.8% |
| **Total** | **$ 1,984.9** | **$ 2,321.1** | **$ 336.2** | **16.9%** |

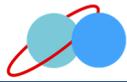

## 4.9    Dual Spacecraft on a Single Launch Vehicle Visiting Both Planets

### 4.9.1    Overview

An option that was ranked very highly by the SDT, but not studied by Team X, was that of an architecture that would include two spacecraft launched on a single launch vehicle. The spacecraft would travel together for some portion of their cruise, then separate to perform individual trajectories to send one to Uranus while the other would target Neptune. (There is no trajectory in the coming decades that allows a single spacecraft to target both planets as Voyager did in the 1980s.) To meet the minimum science requirements for a flagship, as summarized in Section 3.4.3, at least one spacecraft would be an orbiter, and at least one would carry an atmospheric probe. Such a mission would have the highest possible science value, but would cost well above the $2B target. (Note that two-planet mission concepts were not given an independent cost assessment by Aerospace, so all costs in this section are based on JPL estimates.)

Given their high science value, once the Team X studies were complete and design and costing was available for a range of flight elements, a notional set of two-planet missions was evaluated to estimate relative costs. Two costing approaches were used, as described in Section 4.9.8. The first is based on certain assumptions expected to maximize the affordability of this mission architecture, primarily:

- The two spacecraft would be identical and built together
- Probes, if included, would likewise be identical copies
- Instrument suites would be limited to the ~50 kg suite and would be identical for each flight system

Two variants were assessed using this approach, aiming for the lowest-cost missions:

- Orbiter with ~50 kg payload and probe at both Uranus and at Neptune
- Orbiter only with ~50 kg payload (no probe) at Uranus and flyby spacecraft with ~50 kg payload and probe at Neptune

The estimates arrived at by this method should be considered lower-limits. Our second costing approach sets an upper limit by assuming no cost savings in building two spacecraft.

The results suggest that it may be possible to fly two-spacecraft, two-planet options for a cost under $3B, though such missions would fly the smallest payloads (~50 kg) and would not address all 12 of our priority science objectives. Spending $3.5 to $4B would allow us to at least partially address all science objectives at both planets, with ~$4.5B needed to comprehensively address all goals at both planets and maximize the opportunities for comparative planetology.

### 4.9.2    Science

The specific science objectives for each planet do not change if both planets are visited, and the recommended payload sizes and instruments remain the same; all of the discussions in Section 3 apply. Section 3.2 discusses how Uranus and Neptune are equally and individually compelling. It also discusses that the two planets are not equivalent, and points out the highest science return will come from an exploration of both systems. Not only does each planet provide information the other cannot, but by comparing the two we see how similar planets react to differing physical inputs (e.g., at Uranus, sunlight appears to be the dominant energy input to the atmosphere, while at Neptune the atmosphere's energy balance is dominated by internal heat), and we get a better idea of what properties might be common among ice giant exoplanets.

### 4.9.3    Instrumentation

The instruments recommended for each spacecraft remain the same as in the single-planet missions.  We note that, for the larger payloads, the SDT recommended slightly different instrument suits for Uranus than for Neptune.  In the 90 kg payload category, the mid-IR spectrometer was chosen for Uranus and the UV spectrometer for Neptune.  In the 150 kg payload, a microwave sounder was recommended for Uranus and a mass spectrometer replaces it for Neptune.  Both of those instrument swaps were made to optimize the Neptune spacecraft's ability to study Triton's atmosphere.  If the two spacecraft need be identical, we would choose the mid-IR spectrometer in the 90 kg case, and the mass spectrometer in the 150 kg case.  If only a 50 kg science payload is flown, the instrumentation is identical for both planets.

### 4.9.4    Mission Design

Mission design considerations for the two-spacecraft architectures are discussed in Appendix A, Section A.7.

### 4.9.5    Concept of Operations/GDS

The flight operations of the two-spacecraft architectures described above will comprise a combination of the detailed payloads and spacecraft designed for Options 1–6.  The flight systems and cruise operations are dependent on the various interplanetary trajectory options for the two-spacecraft architecture outlined in the Mission Design section.  The trajectories requiring a separate SEP stage for each individual spacecraft would have a flight system and cruise operations similar to the SEP options.  Also, those trajectories that do not require a SEP stage would have flight systems and cruise operations similar to the no-SEP options.  The trajectories with both spacecraft stacked atop a single SEP stage would have similar spacecraft flight systems, while each segment of their cruise operations would be comparable to a similar segment among the SEP and no-SEP options.  Once each spacecraft is on approach to their respective planet, the flight operations will be identical to those detailed in the individual Uranus and Neptune options.  Each spacecraft will complete their approach science, orbit insertion, and orbiter science phases with similar science operations and data collected. If the mission design allows enough mass to include a probe for either or both spacecraft, then these probes will also have similar flight systems and associated operations as the detailed mission options with a probe.  For all two-spacecraft architectures, the decommissioning and disposal of each spacecraft will be the same as the Uranus and Neptune orbiter mission options.

   The ground systems and operations will also be similar to those previously described in the detailed mission options.  The ground systems and operations will remain separate and independent for certain aspects such as DSN support, sequence development, etc.  It is expected that there would be overlap between the science teams of each spacecraft, but each spacecraft's science operations team must be able to function independently since both may be collecting data at the same time.  For most trajectories, the Uranus spacecraft will arrive first, creating the opportunity for "lessons learned" to be applied to Neptune operations.  Other aspects of the ground systems and operations could be combined to provide cooperative support and potentially reduce costs such as software development and tools, data management systems, etc.

### 4.9.6    Flight System Design

For the purposes of this feasibility analysis, orbiter designs were assumed to be identical to each other and were modeled on that for Option 1.  The flyby flight system was assumed to be identical to Option 4.  The SEP stage assumed the design from Option 1.  Modifications necessary to accommodate two spacecraft were accounted for only in the cost analysis, where the cost was boosted by 20% over that for the single orbiter, as a way to account for the likely higher mass of the dual spacecraft stage. No additional design work was performed.

### 4.9.7    New Technology

Given the assumptions of using identical flight systems based on orbiters/flybys similar to those developed in earlier options, it is anticipated that no additional new technologies would be required for the flight system.

### 4.9.8    Cost

To estimate costs for two-spacecraft, two-planet missions, Team X estimates for single-planet missions are used as a starting point. Our first method utilizes the fact that Team X identifies both nonrecurring (NRE) and recurring engineering (RE) costs for all WBS elements. So, for two identical copies of a WBS element, the cost is estimated as 2*RE+NRE. For each of these new options, it is assumed that the orbiter/flyby spacecraft and instruments are identical to those designed in Team X Option 1 (orbiter) or Option 4 (flyby) (Sections 4.3 and 4.6). It is assumed that a single SEP stage is used to carry both flight systems, and the cost of the SEP stage is 20% higher than the Team X estimate because it is more massive. The cost of the probe, probe instruments, and entry system are identical to the Team X cost. Costs are scaled for Science, MOS/GDS, Mission Design, and other elements of the WBS using rules of thumb. For example, the cost during cruise for science and MOS/GDS for a dual orbiter is estimated to be 1.5 times the cost of the science and MOS/GDS during cruise for the single orbiter case. Reserves are set at 30% (without RPS) for development and 15% (without tracking) for operations. It should be noted that these costs are not as reliable as Team X estimates because they have not been vetted by the team. Also, there is no Aerospace cost estimate for comparison.

We can also consider a worst-case situation, where there are no cost savings due to building multiple flight elements or common instruments, software, etc. In this model, the cost of flying a 50 kg orbiting payload and an atmospheric probe to each planet is the sum of the costs of the individual missions. (This approach does not explicitly identify costs associated with, for example, building a larger SEP stage to support two spacecraft, but assumes to zeroth order that such expenses are balanced by savings elsewhere.) Using this approach, costs are significantly higher as shown in **Table 4-47**.

**Table 4-47.** Dual S/C mission cost summary assuming 50 kg science payload and assumptions outlined in text.

| Cost Summary (FY2015 $M) | Orbiter + Probe at both Uranus and Neptune | Uranus Orbiter + Neptune Flyby with Probe |
|---|---|---|
| Project Cost Using Simple Summing of Elements | $3672.0 M | $3465.0 M |
| Project Cost Tracking RE vs NRE | $2707.3 M | $2584.2 M |
| Development Cost | $2135.9 M | $2031.0 M |
| Phase A | $21.4 M | $20.3 M |
| Phase B | $192.2 M | $182.8 M |
| Phase C/D | $1922.3 M | $1827.9 M |
| Operations Cost | $571.4 M | $553.3 M |

The two-spacecraft, two-planet mission that includes a Uranus orbiter (with 50 kg payload) and Neptune flyby (50 kg payload) plus atmospheric probe meets the minimum science objectives we have set. It would cost approximately $3.2B, but that estimate has large variations (±$0.5B) depending on how identical the two spacecraft are. Spending $100 to $200M more would allow a small orbiter to be flown at both planets. Spending $3.8 ±$0.5B would allow the Uranus orbiter to be fully instrumented (150 kg payload) so as to comprehensively address all 12 priority science objectives. Flying orbiters with the largest payload plus probe to each ice giant would cost around $4.4B.

These cost estimates are very crude. If NASA's budget or international partnerships allow consideration of missions costing $3B or more, a more detailed study of dual-planet missions should be performed. That study would be focused both on more accurate cost estimates and on refining the science trades (for example, sending a fully instrumented orbiter to one planet and a minimally-instrumented spacecraft to the other, vs. moderately-instrumented spacecraft to both planets).

## 4.10   Mission Architectures Summary

Features of the six mission architectures studied by Team X are summarized in **Table 4-48**.

**Table 4-48.** Summary of features and characteristics of the six mission architectures for which detailed studies were performed.

| Option | 1** | 2** | 3 | 4 | 5 | 6 |
|---|---|---|---|---|---|---|
| Case Description | Uranus Orbiter with probe and <50 kg science payload. Includes SEP stage for inner solar system thrusting. | Uranus Orbiter without a probe, but with 150 kg science payload. Includes SEP stage for inner solar system thrusting. | Neptune Orbiter with probe and <50 kg science payload. Includes SEP stage for inner solar system thrusting. | Uranus Flyby spacecraft with probe and <50 kg science payload | Uranus Orbiter with probe and <50 kg science payload. Chemical only mission. | Uranus Orbiter without a probe, but with 150 kg science payload. Chemical only mission. |
| Science | Highest priority plus additional system science (rings, sats, magnetospheres) | All remote sensing objectives | Highest priority plus additional system science (rings, sats, magnetospheres) | Highest priority science (interior structure and composition) | Highest priority plus additional system science (rings, sats, magnetospheres) | All remote sensing objectives |
| Team X Cost Estimate* ($k, FY15) | 1927 | 2239 | 1971 | 1493 | 1700 | 1985 |
| Aerospace ICE ($k, FY15) | NA | NA | 2280 | 1643 | 1993 | 2321 |
| Payload | 3 instruments† + atmospheric probe | 15 instruments‡ | 3 instruments† + atmospheric probe | 3 instruments† + atmospheric probe | 3 instruments† + atmospheric probe | 15 instruments‡ |
| Payload Mass MEV (kg) | 45 | 170 | 45 | 45 | 45 | 170 |
| Launch Mass (kg) | 6886 | 7312 | 7365 | 1524 | 4345 | 4717 |
| Launch Year | 2030 | 2030 | 2030 | 2030 | 2031 | 2031 |
| Flight Time (yr) | 11 | 11 | 13 | 10 | 12 | 12 |
| Time in Orbit(yr) | 4 | 4 | 2 | Flyby | 3 | 3 |
| Total Mission Length (yr) | 15 | 15 | 15 | 10 | 15 | 15 |
| RPS use/EOM Power | 4 eMMRTGs/ 376W | 5 eMMRTGs/ 470W | 4 eMMRTGs/ 376W | 4 eMMRTGs/ 425W | 4 eMMRTGs/ 376W | 4 eMMRTGs/ 376W |
| LV | Delta IVH + 25 kW SEP | Delta IVH + 25 kW SEP | Delta IVH + 25 kW SEP | Atlas V 541 | Atlas V 551 | Atlas V 551 |
| Prop System | Dual Mode/NEXT EP | Dual Mode/NEXT EP | Dual Mode/NEXT EP | Monopropellant | Dual Mode | Dual Mode |

*Includes cost of eMMRTGs, NEPA/LA, and standard minimal operations, LV cost not included     ** Determined to be non-optimal. Replaced by Options 5 and 6.
†Includes Narrow Angle Camera, Doppler Imager, Magnetometer     ‡Includes Narrow Angle Camera, Doppler Imager, Magnetometer, Vis-NIR Mapping Spec., Mid-IR Spec., UV Imaging Spec., Plasma Suite, Thermal IR, Energetic Neutral Atoms, Dust Detector, Langmuir Probe, Microwave Sounder, Wide Angle Camera

# 5    TECHNOLOGY

In preparation for the Ice Giants Study and before there was even a clearly defined mission concept, the study team were briefed on the status of a number of technologies with the potential for enhancing the science, reducing the cost or reducing mission duration.  A guiding philosophy adopted for the study however, was to try to develop missions with existing technology, only introducing new technologies where their application would be enabling or significantly enhancing to a given mission concept.  As mission designs and concept architectures were developed, it was found that all the concepts identified by the SDT for detailed study could be implemented without the need for development of advanced technologies.  This section discusses those specific technologies which were considered but not adopted and the benefits that would accrue if they were available for a future missions.  Further details on the technologies described can be found in Appendix D.

## 5.1    Shortening Trip Times to Uranus and Neptune

When opportunities for reaching Uranus or Neptune orbit with shorter flight times were examined, it became clear that the determining factor was not the size and capabilities of the launch vehicle but a means for entering orbit at Uranus and Neptune.  As trip times were shortened, the approach velocity necessarily increased. The mass of a propulsion system using conventional space storable propellants then rose exponentially with the approach velocity.  The three approaches examined involved methods of increasing the specific impulse of the propulsion system which is the denominator in that exponential factor through the use LOX-hydrogen propulsion or radioisotope electric propulsion.  An alternative approach is the use of aerocapture where atmospheric drag is used to rapidly dissipate the approach velocity.  At the current level of understanding there is considerable overlap in the range of applicability of these technologies and it will take considerable further study to determine the appropriate choice for any given mission.

### 5.1.1    Liquid Oxygen – Liquid Hydrogen (LOX LH2) Propulsion

Although LOX-LH2 has an $I_{sp}$ that is only 25% higher than the state of the practice space storable bipropellants, because of the exponential factor it has the potential for at least a factor of two saving in mass in the range beyond 5 km/sec although this saving depends critically on the mass of tankage and other equipment.

In addition to the propulsion system elements required for a typical space storable propulsion system, the use of LOX – $LH_2$ requires some means of storing these cryogenic materials until their point of use without boil off. NASA's Game Changing Development program is currently working on a solar powered space refrigerator which would maintain the temperature so that this propellant system could be used on a human mission (https://spaceflightsystems.grc.nasa.gov/stpo/stmd-gcd/20-w-20-k-cryocooler/).  However, for an ice giant mission, such an active cooling system is unlikely to be necessary and a purely passive cooling approach is adequate.

Not only will ice giant missions spend a great deal of time remote from the heating effects of the sun but the mission profile also makes it possible to shield the propellant tanks from solar radiation and any other sources of heat input such as planetary visible and IR radiation, will be limited, if they occur at all, to brief flybys for gravitational assist.  A combination of a solar radiation shield combined with conductive struts has been shown to work effectively and these technologies have been developed for missions such as Gravity Probe B and space telescopes (Guernsey et al. 2005).

Other opportunities for use of new technology lie with the propellant tanks as these are much larger in physical size because of the low density of liquid hydrogen. The benefits of such technologies were not examined. The availability of LOX LH2 thrusters in the range required still needs to be verified and qualification may be needed.

### 5.1.2    Radioisotope Electric Propulsion (REP)

Electric propulsion offers a factor of 10 improvement in Isp over chemical propellants and is an attractive approach to achieving very large velocity changes. In evaluating the benefits from Isp, it is necessary to consider not only the mass of the propulsion system but also of the baseline power system that provides the electrical energy to drive the propulsion system. For applications at the outer planets, solar power is impractical and radioisotope power has such low specific power that the mass of the power system becomes a major factor in the trade study.

In June 2015, JPL studied the use of REP for achieving orbit at Pluto. That study concluded that a Pluto orbiter was feasible but only for small spacecraft (<500kg) given the levels of radioisotope power that were realistically available. Even under those circumstances, the RPS capabilities assumed for this Ice Giants Study (eMMRTG) would be inadequate to the task. However, performance estimates with proposed small modular RTG (SMTRG) indicated that a Pluto orbiter would be feasible. Reducing the size of the spacecraft below the 500 kg assumed in the Pluto study to 200 kg would have an even larger impact on trip time since the period of time that the vehicle would have to be decelerating in its approach to the target could be cut in half.

In summary, REP systems appear to have the potential for reducing trip time for Neptune missions and probably below that achievable with LOX-LH$_2$.

### 5.1.3    Aerocapture

Aerocapture uses atmospheric drag to slow down a spacecraft so that it can be captured into a planetary orbit. As such, its performance is not characterized by the exponential relationship characteristic of the reaction forces produced by expelling propellant at high velocities. In fact, the dependence of system mass on the desired velocity change was the subject of a study led by Purdue University (Saikia 2016) carried out in parallel with the main Ice Giants Study. The principal findings of that study are as follows:

1. Key parameters important in aerocapture system design are heating rate, total heat load and maximum deceleration. Stagnation pressure may also be a limiting factor on the TPS performance. These constraints place an upper bound on the attainable velocity changes achievable with aerocapture.

2. Because of entry corridor uncertainties, aerocapture systems with large control authority (requiring high L/D ratios) will be needed. These appear to be technically feasible but will increase mass to assure structural integrity at high angles of attack and to provide TPS margins to accommodate the effects of differential ablation.

3. Advanced thermal protection materials such as HEEET will be needed to accommodate the high peak heating rates, stagnation pressures and heat loads required to accommodate the high entry speeds and large Delta Vs for aerocapture applications.

4. Further investigations should be conducted to determine if there are ways of reducing navigation and atmospheric uncertainties (such as pathfinder subsatellites) that could enable aerocapture with the Theoretical Corridor Widths attainable with existing low L/D vehicles.

5. Further work should also be conducted on hybrid aerocapture-chemical approaches where it might be possible to carry out the control functions chemically with minimal recourse to aerodynamic trajectory control.

In summary, aerocapture appears to offer a solution for missions where the Delta V required for orbit insertion exceeds that achievable with conventional bipropellant and an alternative to LOX-LH2. However, heating rates, total heat loads, stagnation pressure and g-loading will set an upper bound to the achievable velocity change. For the High L/D vehicles which are the subject of the current study this appears to be about 9.5 km/sec. For missions requiring DV in excess of 8km/sec REP is the preferred approach.

Unlike REP, there are no restrictions that limit aerocapture to use with small spacecraft. However, small spacecraft typically have smaller ballistics coefficients which reduces heat loading and they may be more tolerant of high g loading. Those factors could make it easier to use aerocapture technology.

## 5.2 Improved Data Return with Optical Communications

Optical communications can offer greatly enhanced data return from deep space relative to state-of-the- art Ka band technology. The Deep Space Optical Communications (DSOC) project is developing four key technologies needed to enable a demonstration of the technology for applications out to a distance of 3AU from the Earth. At this time, these four technologies have reached TRL 4 in NASA's Game Changing Development (GCD) program and are being transitioned to the Technology Demonstration Mission (TDM) program to facilitate infusion in a Discovery mission.

Application of these technology beyond 3AU—in fact far beyond 3AU for ice giant missions—involves some additional challenges which represent one reason the technology was not adopted for the Ice Giant missions. A second reason is the lack of an adequate ground-based infrastructure or firm plans to develop one to fully capitalize on the benefits of optical communications.

The primary problem with operating beyond 3AU is the difficulty of directing the very narrow laser beam carrying downlink optical data to the tracking station on Earth. For distances out to 3AU, this depends on a beacon signal beamed up from Earth with a powerful laser. However, for every doubling of the distance to the spacecraft four times as much beacon power is needed. This presents an environmental hazard and at some point the power absorbed in the atmosphere distorts the waveform so it is no longer even practical. Alternative methods include tracking the Earth in the infrared or referencing nearby stars with star trackers. Research on these alternate approaches are highly desirable.

The high data rates achievable with optical communications using modest amounts of power is attributable in significant part to the ability to direct energy into a very narrow optical beam. Another factor is the use of pulse position modulation which is part of the DSOC capability. A further possibility that can be anticipated in future systems is to code information by the wavelength of the optical beam. Improvements of 1 to 3 orders of magnitude can be anticipated depending on the specifics of the technology.

## 5.3 Higher Specific Power from Radioisotopes

The NASA RPS Program is pursuing technology development for both the thermoelectric (TE) and Stirling power conversion options focused on systems of higher efficiency and improved specific power at end of mission.

### 5.3.1 Advanced Thermoelectric Generators

As a possible follow-on to the proposed eMMRTG which is baselined for ice giants, JPL is leading the development of an advanced thermoelectric generator based on the segmented skutterudite, La3-xTe4, and Zintl couple technology. One potential generator design based on this technology is a Segmented Modular Radioisotope Thermoelectric Generator (SMRTG). The SMRTG was assumed for the Pluto Orbiter study discussed in an earlier section.

The SMRTG would be composed of sections containing General Purpose Heat Source (GPHS) modules. Each section could be one, two, or four GPHS modules high, depending on the final design. The minimum size of the SMRTG would be a single section, while the maximum size of the SMRTG would have 16 GPHS modules. Like the GPHS-RTGs currently in use on Cassini, the SMRTG would operate only in vacuum. An SMRTG would use segmented TEs in a two GPHS module assembly that produces about 43 We BOM. The design is scalable up to 16 GPHS modules in increments of two, producing just under 500 We BOM. The SMRTG would be designed to operate in vacuum at a hot-shoe temperature of 1273K and a cold-shoe of 498K with an overall projected system efficiency of about 8–11%. The operating temperatures and system efficiency would represent major improvements over the eMMRTG, which is projected to achieve about 7% system efficiency at 873K/473K. Some of the key component technologies needed to realize the SMRTG concept include the TE multi-couple module development, lightweight high-temperature MLI, compliant cold-shoes, aerogel encapsulation, sublimation control, and life verification.

### 5.3.2 Advanced Stirling Generators

Glenn Research Center is leading the RPS Stirling technology development that includes research and technology development of hot-end components, cold-end components, and systems/testbeds. Under hot-end components, work has focused on improved thermal insulation materials, Stirling-specific MLI packaging, and heat source backup cooling using variable conductance heat pipes (VCHP). The cold-end components effort has developed advanced NdFeB magnets for higher temperature alternators, high temperature organic adhesives and wire insulation, and titanium-water heat pipes for heat rejection. The systems/testbeds task has demonstrated new fault-tolerant electrical controller architectures for both single- and dual-Stirling convertor systems and pursued mechanical balancers that permit single, unopposed convertors to operate with low vibration.

The High Power Stirling Radioisotope Generator (HPSRG) is a notional group of designs for a higher power generators in the SRG family. The SRG family would use Stirling dynamic power conversion technology to convert thermal energy into electrical with high conversion efficiencies (>25%) and high specific powers. Both distributed heat source systems as well as centralized heat source generators are under consideration. Sharing the heat of multiple GPHS modules may provide the ability to continue operation in the event of a Stirling convertor failure. One conceptual design from the HPSRG design family modifies the basic ASRG layout to increase output power; it can be thought of as a larger ASRG consisting of two dual-opposed Stirling converters, using the thermal output of four (4-GPHS SRG), six (6-GPHS SRG), or eight (8-GPHS SRG) GPHS modules. Though this family of design concept uses technologies developed for the ASRG, the HPSRG remains at a conceptual phase and none of the concepts have been built or tested as a system. This advanced concept has undergone some conceptual study to determine feasibility.

Greater details on these high power RPS concepts can be found in the Nuclear Power Assessment Study Final Report1 and Radioisotope Power Systems Reference Book for Mission Designers and Planners (Version 1.1)2.

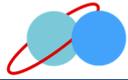

# 6  CONCLUSIONS

We summarize our conclusions in the context of the four specific study objectives (Section 2.1).

## 6.1  Identify Mission Concepts that Can Address Science Priorities Based on What Has Been Learned since the 2013–2022 Decadal Survey

*Science Priorities:* The SDT finds that the Uranus and Neptune systems are equally valuable scientifically, and each is scientifically compelling. We do not prioritize one over the other. We emphasize, however, that the two systems are not equivalent; each planet provides information that the other cannot (see Section 3.2). Exploring both ice giant systems will result in the highest science return.

Our Science Definition Team (SDT) identified 12 priority science objectives that cannot be met with Earth-based measurements (Section 3.1). The two highest priority objectives are

- Constrain the structure and characteristics of the planet's interior, including layering, locations of convective and stable regions, internal dynamics.
- Determine the planet's bulk composition, including abundances and isotopes of heavy elements, He and heavier noble gases.

These two objectives address the most fundamental questions of what ice giants are and how they form, and are likely to have the most impact on larger scientific questions regarding the formation and evolution of our Solar System and attempts to understand exoplanetary systems. We do not prioritize among the remaining 10 objectives, which are listed below. The SDT recognizes that valuable science remains to be done in all aspects of the ice giant systems.

- Improve knowledge of the planetary dynamo.
- Determine the planet's atmospheric heat balance.
- Measure the planet's tropospheric 3-D flow (zonal, meridional, vertical) including winds, waves, storms and their lifecycles, and deep convective activity.
- Characterize the structures and temporal changes in the rings.
- Obtain a complete inventory of small moons, including embedded source bodies in dusty rings and moons that could sculpt and shepherd dense rings.
- Determine surface composition of rings and moons, including organics; search for variations among moons, past and current modification, and evidence of long-term mass exchange / volatile transport.
- Map the shape and surface geology of major and minor satellites.
- Determine the density, mass distribution, internal structure of major satellites and, where possible, small inner satellites and irregular satellites.
- Determine the composition, density, structure, source, spatial and temporal variability, and dynamics of Triton's atmosphere.
- Investigate solar wind-magnetosphere-ionosphere interactions and constrain plasma transport in the magnetosphere.

The above objectives were distilled from an extensive list of initial science questions listed in Appendix F. All our science objectives are consistent with and traceable to the *Vision and Voyages Decadal Survey* (NRC, 2011; referred to as V&V). V&V is discussed later in this section.

*Mission Concepts:* After an extensive discussion of many possible architectures (Sections 3.3 and 3.4), we determine that sending an orbiter with an atmospheric probe to one of the ice giant systems is the minimum we recommend for an Ice Giant Flagship mission. The probe is needed to address one of our two highest priority objectives (composition), and the majority of our science objectives require an orbiter so that enough time is spent in the system to study temporal variability, execute multiple satellite flybys, and to study the entire system from a variety of geometries. Given the value of studying both ice giant systems (Section 3.2), we recommend that an attempt be made to exceed this minimum mission by sending a second spacecraft to the second ice giant. Even a flyby of the second planet would yield significant new science when combined with what is learned from the orbiter, though of course, an orbiter at both planets would be optimal.

Regarding the science payload for the spacecraft (whether flyby or orbiter, see Section 3.3), the highest priority instruments are an atmospheric seismology instrument (e.g., a Doppler Imager), a narrow-angle camera, and a magnetometer. This minimal suite of instruments (totaling less than 50 kg, comparable to the New Horizons spacecraft's payload), along with an atmospheric probe, achieves our two highest priority science objectives and partially addresses several others. Our smallest recommended orbiter payload, however, is near 90 kg. It adds a visible/near-IR imaging spectrometer, a thermal-IR instrument, a plasma suite, and either a mid-IR or a UV spectrometer. This larger package fully addresses both of our highest priority objectives and also makes significant progress in all other objectives. If the orbiter payload is at least 150 kg (the Cassini orbiter's payload is near 270 kg), all 12 of our science objectives can be achieved. This larger payload would include a wide-angle camera, detectors of energetic neutral atoms and dust, a Langmuir probe, a microwave sounder, and potentially a mass spectrometer (primarily for Triton's atmosphere). This largest payload would include all instruments the SDT identified as useful, with the exception of a radar sounder and sub-spacecraft (e.g., CubeSats); those two are not critical to achieving our major objectives.

There are opportunities to launch the recommended spacecraft and payload to either ice giant almost every year in the time-period studied (2024 to 2037), though costs, launch vehicle, and spacecraft design would vary. Optimal launch windows are within a couple years of 2030, and utilize a Jupiter gravity assist. Gravity assists to Uranus using Saturn, which are less efficient than Jupiter assists, are best early in the time period studied. Saturn flybys on the way to Uranus may not be possible beyond 2028. It does not appear to be possible to flyby Saturn on the way to Neptune in the time period studied. Flight times to Neptune are typically 2 years longer than those to Uranus. Interplanetary flight times are 6 to 12 years to Uranus, 8 to 13 years to Neptune, depending on launch year and mission architecture (e.g., SEP vs. chemical); for this study flight times were constrained not to exceed 13 years.

*Changes since V&V:* The basic mission we recommend is similar to that described in V&V, though there are some differences. V&V identified atmospheric dynamics as their top-priority investigation. They listed interior structure and bulk composition (our highest priority objectives) as secondary because they were not confident it was possible to accurately determine them. Technological and scientific advances since V&V have mitigated those concerns, along with opportunities for more efficient trajectories. Specifically, using atmospheric seismology instead of gravity (Sections 3.3.1 and 3.3.2) allows us to study the interior without incurring any ring-plane crossing risk (in fact, interior structure can be determined in a flyby). It should be noted, however, that in Appendix A we discuss that there is likely to be a safe region close to the planet through which the spacecraft can be flown, which allows high-resolution gravity data to be collected. Regarding the ability to measure atmospheric composition, V&V did not have a robust probe design, and the trajectories were marginal for accommodating a probe. Our study incorporates a much more robust and heavier probe which has been designed in detail (Section 4.3.6), and has benefitted both from a more comprehensive exploration of interplanetary trajectories and better launch geometries in the time period considered.

Another difference between V&V and the current study is that we have determined that Uranus and Neptune are equally valuable scientifically, while V&V did not explicitly make that finding. V&V instead focused on a Uranus mission because favorable trajectories were available to it and not to Neptune.

Overall, while there have been scientific and technological advancements since V&V, our conclusions remain quite similar. This is an encouraging sign that we are converging on an appropriate mission concept.

## 6.2 Identify Potential Mission Concepts across a Spectrum of Price Points

We have identified scientifically justifiable missions ranging in cost (all costs in $FY15) from $1.5 billion (a Uranus flyby with probe) to over $4.0 billion (two orbiters, each with 150 kg science payload and probe, one at Uranus and the other at Neptune). Science ranking is discussed in Section 3.4.3, and mission designs are discussed in Section 4 and Appendix A. Our ground rules target a mission costing under $2 billion, and the most compelling mission under that cap would be a Uranus orbiter carrying a ~50 kg science payload and an atmospheric probe ($1.7 to 2.0 billion). Such a mission could achieve the two most important science objectives at the ice giants, and would make limited progress in several others. To achieve all 12 of the key science objectives identified, an orbiter with a 150 kg payload and an atmospheric probe is needed. That mission is estimated to cost between $2.3 and $2.6 billion. An intermediate mission was also studied which would achieve the two most important science objectives, and make significant progress in all others. It would include an atmospheric probe and an orbiter with a 90 kg science payload, and would cost $2 to $2.3 billion. Missions to Uranus are less expensive than missions to Neptune with a similar payload by ~$300M. This is primarily due to Neptune missions requiring a SEP stage and a larger spacecraft to accommodate either additional fuel for orbit insertion or aerocapture technologies.

Given the scientific value of visiting both ice giants (Section 3.2), we explored options for a two-spacecraft mission, sending one to each ice giant. (In the time period studied there are no trajectories that allow a single spacecraft to fly to both planets.) These missions did not receive an independent cost assessment. We find (Section 4.9) that utilizing identical designs, instruments, and simultaneous or sequential construction, two spacecraft with our 50 kg payload and probe could be flown for under $3B. Allowing for design differences, costs would be around $3.5B. Flying the larger, 150 kg payload necessary for achieving all science objectives would likely cost $4B, and could cost as much as $5B if identical designs and concurrent construction are not utilized. It is important to note that the SDT did not reach consensus on the science value of flying some of the least-expensive two-planet missions (see Section 3.4.2). For example, we did not determine whether it was scientifically more valuable to have a fully instrumented orbiter and probe at one ice giant or fly smaller orbiters at each planet without any probes.

There is a nearly linear relationship between cost and science return (Section 3.4.3 and **Figure 3-17**). Foreign partnerships are a viable way to fly a more capable mission while meeting NASA's cost targets.

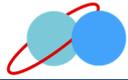

## 6.3    Identify Enabling/Enhancing Technologies

We find that eMMRTG's and HEEET are the only emerging technologies required for an ice giant mission. An eMMRTG is a proposed radioisotope power system based on the current MMRTG design, but with a thermocouple having improved performance overall and – most important for a long duration outer solar system mission – a far slower decay in output over time. HEEET is a new high-performance heat shield material for atmospheric probes which replaces the no-longer-available carbon phenolic heat shield used by the Galileo probe. Both the eMMRTG and HEEET technologies are currently in development. A compelling mission to Uranus or Neptune can be flown without starting any other technology development program. In Section 5, we reviewed several technologies which would, however, be enhancing. Two of the technologies likely to be most useful to the missions studied are summarized below.

*Aerocapture:*    As described in Section 5.1.3, aerocapture has several benefits which are particularly important for missions to Neptune. Much larger payloads and faster trip times are enabled by aerocapture because it eliminates the need for a large chemical burn for capture into orbit. Extremely short trip times to both Uranus (<5 years) and Neptune (<7 years) are possible if aerocapture is used in conjunction with the SLS launch vehicle.

Areas of additional work that would help make aerocapture a reality include

- Development of high L/D aeroshells
- Improved knowledge of the density profile of the uppermost atmospheres of Uranus and Neptune (which potentially eliminates the need for high L/D aeroshells)
- Continued development of HEEET technology

*Optical Communication:*    While all science objectives can be met with the current DSN and arraying of 35-meter antennas, data volume limits do require some restrictions on instrument observing times, particularly the imaging systems. The one to three orders of magnitude improvement in data volume offered by optical communications would be welcome. We note, however, that optical com beyond 3 AU is particularly difficult due to the great distances involved and the need for precise pointing of the downlink laser. Techniques allowing such precise pointing are in need of development.

## 6.4    Assess Capabilities Afforded by SLS

While not enabling of the science we wish to do, the availability of SLS would allow (see Appendix A).

- Reduced flight times and/or increased delivered mass to either ice giant. This allows additional tradeoffs between cost and science return.
- Two-planet, two-spacecraft missions on a single launch vehicle. While there is no scientific penalty to launching two-planet missions on different launch vehicles several years apart, there may be programmatic benefits to utilizing a single SLS launch vehicle.

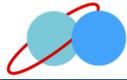

# 7   RECOMMENDATIONS

This report reaffirms the scientific importance of flying an ice giant Flagship mission, as recommended in the Vision and Voyages Planetary Science Decadal Survey. To ensure that the most productive mission is flown, we recommend

- An orbiter with probe be flown to one of the ice giants.
  - If the next Planetary Science Decadal Survey does not prioritize an ice giant Flagship, we recommend allowing for New Frontier options that can achieve a subset of key science objectives.
- The orbiter carry a payload between 90 and 150 kg.
- The probe carry at minimum a mass spectrometer and atmospheric pressure, temperature, and density sensors.
- The development of eMMRTGs and HEEET be completed as planned.
- Two-planet, two-spacecraft mission options be explored.
- International collaborations be leveraged to maximize the science return while minimizing the cost to each partner.
- Investment in ground-based research (theoretical and observational) to allow optimized targeting of periapse altitude for orbit insertion and for collecting gravity data. Specifically,
  - Better constrain the ring-crossing hazard
  - Better constrain the upper atmosphere
- Investment in instruments, research, and observations related to atmospheric seismology. This would
  - Improve the design of instruments
  - Improve our understanding of the forcing of planetary-scale oscillations
  - Improve our ability to interpret oscillation measurements
- An additional mission study be executed using refined programmatic ground rules to better define the mission most likely to fly. The current study was designed to explore a broad range of parameter space, and its limitations include
  - The Team X point designs were selected to allow the best characterization of costs across parameter space instead of accurate costs of the most likely mission.
  - The ground rules used in this study may not represent conditions when the mission flies. In particular, we note that launch vehicle performance may be different than currently assumed, and estimated costs depend on assumptions made about international participation and what components to include or exclude (e.g., launch vehicle).

# 8 REFERENCES

Agnor, C. B., & Hamilton, D. P. (2006). Neptune's capture of its moon Triton in a binary–planet gravitational encounter. Nature, 441(7090), 192-194.

Allen Jr., Gary A., Michael J. Wright, and Peter Gage, "The Trajectory Program (Traj): Reference Manual and User's Guide", NASA TM-2004-212847, 2005.

Allen Jr., Gary A., Peter J. Gage, Ethiraj Venkatapathy, David R. Olynick, and Paul F. Wercinski, "A Web-based analysis system for planetary entry vehicle design", AIAA Paper 98-4826, AIAA/USAF/NASA/ISSMO Symposium on Multidisciplinary Analysis and Optimization, Sept. 2-4, 1998.

Arridge CS, Achilleos N, Agarwal J, Agnor CB, Ambrosi R, et al. 2014. The science case for an orbital mission to Uranus: Exploring the origins and evolution of ice giant planets. Planetary and Space Science 104: 122-40

Arridge, C. S. (2015) Magnetotails of Uranus and Neptune, in Magnetotails in the Solar System (eds A. Keiling, C. M. Jackman and P. A. Delamere), John Wiley & Sons, Inc, Hoboken, NJ. doi: 10.1002/9781118842324.ch7

Asplund, M., Grevesse, N., Sauval, J., et al. (2009). The chemical composition of the Sun. Annu. Rev. Astron. Astrophys, 47, 481–522.

Atreya, S. K., Crida, A., Guillot, T., Lunine, J. I., Madhusudhan, N., Mousis, M. (2017). The Origin and Evolution of Saturn, with Exoplanet Perspective, a chapter in Saturn in the 21st Century (K. Baines, M. Flasar, N. Krupp, and T. Stallard, editors), Cambridge University Press, 2017, in press. Pre-publication pdf of the chapter is available at http://arxiv.org/abs/1606.04510.

Atreya, S. K., Wong, A-S (2004). Clouds of Neptune and Uranus (2004). Proceedings, International Planetary Probe Workshop, NASA Ames, 2004, NASA CP-2004-213456.

Atreya, S. K., Wong, A-S (2005). Coupled Chemistry and Clouds of the Giant Planets – A Case for Multiprobes, in The Outer Planets and their Moons (T. Encrenaz, R. Kallenbach, T. C. Owen, C. Sotin, eds.), pp 121-136, Springer, Berlin-New York- Heidelberg, 2005. Also in Space Sci. Rev., 116, Nos. 1-2, pp 121-136, 2005.

Atreya, S. K., Wong, M. H., Owen, T. C., et al. (1999). A Comparison of the atmospheres of Jupiter and Saturn: deep atmospheric composition, cloud structure, vertical mixing, and origin. Planet. Space Sci., 47, 1243-62.

Bagenal, F. (1986), The double tilt of Uranus, Nature, 321, 809–810.

Bagenal, F. (1992), Giant Planet Magnetospheres, Ann. Rev. Earth Planet. Sci., 20, 289.

Ballester et al. (1998) Proc. Conf. Ultraviolet Astrophysics, beyond the IUE Final Archive, Sevilla, Spain, 21.

Banfield, D., Gierasch, P., and Dissly, R. (2005). "Planetary descent probes: polarization nephelometer and hydrogen ortho_para instruments," in Aerospace, 2005 IEEE Conference, (IEEE, 2005), pp. 1–7.

Bauer, J. M., Roush TL, Geballe TR, Meech KJ, Owen TC, et al. 2002. The near infrared spectrum of Miranda - Evidence of crystalline water ice. Icarus 158: 178-90

Bell, J. M., J. Hunter Waite Jr, Joseph H. Westlake, Stephen W. Bougher, Aaron J. Ridley, Rebecca Perryman and Kathleen Mandt, 2014, Developing a self-consistent description of Titan's upper atmosphere without hydrodynamic escape, J. Geophys. Res., 119, 4956-4972. doi: 10.1002/2014JA019781

Bell, J., Bougher, S.W., Waite, Jr., J.H., Ridley, A.J., Magee, B., Mandt, K., Westlake, J., DeJong, A. D., Bar-Nun, A., Jacovi, R., Toth, G., de la Haye, V., 2010a. Simulating The One-Dimensional Structure of Titan's Upper Atmosphere, Part I: Formulation of the Titan Global Ionosphere-Thermosphere Model and Benchmark Simulations, J. Geophys. Res., 115, E12002.

Bell, J., Bougher, S.W., Waite, Jr., J.H., Ridley, A.J., Magee, B., Mandt, K., Westlake, J., DeJong, A. D., Bar-Nun, A., Jacovi, R., Toth, G., de la Haye, V., 2011. Simulating The One-Dimensional Structure of Titan's Upper Atmosphere, Part III: Mechanisms determining methane escape, J. Geophys. Res., 116, E11002.

Bell, J., Bougher, S.W., Waite, Jr., J.H., Ridley, A.J., Magee, B., Mandt, K., Westlake, J., DeJong, A. D., de la Haye, V., Gell, D., Fletcher, G., Bar-Nun, A., Jacovi, R., Toth, G., 2010b. Simulating The One-Dimensional Structure of Titan's Upper Atmosphere, Part II: Alternative scenarios for methane escape, J. Geophys. Res., 115, E12018.

Bhardwaj, A., and G. R. Gladstone (2000), Auroral emissions of the giant planets, Rev. Geophys., 38, 295.

Bitsch, B., Lambrechts, M., Johansen, A. (2015). The growth of planets by pebble accretion in evolving protoplanetary discs. Astronomy & Astrophysics 582, article id. A112.

Borucki, W.J. and 66 co-authors, 2011. Characteristics of planetary candidates observed by Kepler, II: Analysis of the first four months of data. Ap. J. 736, 19.

Broadfoot, A. L., et al. (1989), Ultraviolet spectrometer observations of Neptune and Triton, Science, 246, 1459.

Broadfoot, A.L., Herbert, F., Holberg, J.B., Hunten, D.M., Kumar, S., Sandel, B.R., Shemansky, D.E., Smith, G.R., Yelle, R.V., Strobel, D.F., Moos, H.W., Donahue, T.M., Atreya, S.K., Bertaux, J.L., Blamont, J.E., McConnell, J.C., Dessler, A.J., Linick, S., Springer, R., 1986. Ultraviolet spectrometer observations of Uranus. Science 233, 74-79.

Butler, R.P., Vogt, S.S., Marcy, G.W., Fischer, D.A., Wright, J.T., Henry, G.W., Laughlin, G., Lissauer, J.J., 2004. A Neptune-mass planet orbiting the nearby M Dwarf GJ 436. Astrophys. J. 617, 580–588.

Cao, X., and C. Paty, "3D Multifluid MHD simulation for Uranus and Neptune: the seasonal variations of their magnetosphere", AGU fall meeting 2015, SM31C-2503.

Cartwright, R. J., Emery, J. P., Rivkin, A. S., Trilling, D. E., & Pinilla-Alonso, N. (2015). Distribution of $CO_2$ ice on the large moons of Uranus and evidence for compositional stratification of their near-surfaces. Icarus, *257*, 428-456.

Castillo-Rogez, J. C., & Lunine, J. I. (2012), Small habitable worlds. *Frontiers of Astrobiology*, 331.

Castillo-Rogez J, Turtle E. 2012. Comparative planetology between the uranian and saturnian satellite systems – focus on Ariel. Presented at American Astronomical Society DPS meeting #44, Reno, NV

Chancia, R. O., and Hedman, M. M. 2016. Are there moonlets near Uranus' alpha and beta rings? Astron. J., in press.

Charnoz S., Salmon J., Crida A. 2010. The recent formation of Saturn's moonlets from viscous spreading of the main rings. Nature 465: 752-4

Chau, R., Hamel, S., & Nellis, W. J. (2011). Chemical processes in the deep interior of Uranus. Nature communications, 2, 203.

Chen, Y.K. and F.S. Milos, "Ablation and Thermal Response Program for Spacecraft Heatshield Analysis," Journal of Spacecraft and Rockets, Vol. 36, No. 3, 1999, pp. 475-483.

Cheng, A. F., et al. (1991), Energetic particles at Uranus, in Uranus, University of Arizona Press.

Christensen, U. R., & Aubert, J. (2006). Scaling properties of convection-driven dynamos in rotating spherical shells and application to planetary magnetic fields. Geophysical Journal International, 166(1), 97-114.

Clark, R. N., Lucey, P. G., (1984). Spectral properties of ice-particulate mixtures and implications for remote sensing. I - Intimate mixtures. Journal of Geophysical Research 89, 6341-6348

Connerney, J. E. P., Acuña, M. H., & Ness, N. F. (1987). The magnetic field of Uranus. Journal of Geophysical Research: Space Physics, 92(A13), 15329-15336.

Connerney, J. E. P., Acuña, M. H., & Ness, N. F. (1991). The magnetic field of Neptune. Journal of Geophysical Research: Space Physics, 96(S01), 19023-19042.

Coradini, A., Magni, G., Turrini, D. (2010). From Gas to Satellitesimals: Disk Formation and Evolution. Space Science Reviews 153, 411-429.

Cowley, S. W. H. (2013), Response of Uranus' auroras to solar wind compression at equinox, J. Geophys. Res. Lett., 118, 2897–2902, doi:10.1002/jgra.50323.

Crida and Charnoz (2012). Formation of regular satellites from ancient massive rings in the Solar System. Science 338, 1196.

Croft (1992). Proteus: Geology, shape, and catastrophic destruction. Icarus 99, issue 2, pp. 402-419.

Croft SK, Soderblom LA. 1991. Geology of the Uranian satellites. In Uranus, ed. JT Bergstralh, ME D., MS Matthews: Univerity of Arizona Press.

Croft et al. (1995), The Geology of Triton, in *Neptune and Triton,* [D. Cruickshank, Ed.], University of Arizona Press.

Cui, J. et al. (2009). Analysis of Titan's neutral upper atmosphere from Cassini Ion Neutral Mass Spectrometer measurements. Icarus, 200(2), 581-615.

Ćuk, M., & Gladman, B. J. (2005). Constraints on the orbital evolution of Triton. The Astrophysical Journal Letters, 626(2), L113.

Culha C, Hayes AG, Manga M, Thomas AM. 2014. Double ridges on Europa accommodate some of the missing surface contraction. Journal of Geophysical Research (Planets) 119: 395-403

Dalton, J. B., et al. (2010). Chemical Composition of Icy Satellite Surfaces. Space Science Reviews 153, 113-154.

de Pater, I., Romani, P.N., Atreya, S. K. (1991). Possible Microwave Absorption by H2S in Uranus and Neptune Atmospheres, Icarus, 91, 220.

de Pater, I., S. Gibbard, E. Chiang, H. B. Hammel, B. Macintosh, F. Marchis, S. Martin, H. G. Roe, and M. Showalter 2005. The dynamic Neptunian ring arcs: Evidence for a gradual disappearance of Liberté and resonant jump of Courage. Icarus 174, 263–272.

de Pater, I., Hammel, H.B., Gibbard, S.G., Showalter, M., 2006. New dust belts of Uranus: one ring, two ring, red ring, blue ring. Science 312 (5), 92–94.

de Pater, I., H. B. Hammel, M. R. Showalter, and M. A. van Dam 2007. The dark side of the rings of Uranus. Science 317, 1888–1890.

de Pater, I., Fletcher, L.N., Luszcz-Cook, S., DeBoer, D., Butler, B., Hammel, H.B., Sitka, M.L., Orton, G., Marcus, P.S. (2014). Neptune's global circulation deduced from multi-wavelength observations. Icarus, 237, 211-238.

de Pater, I., Sromovsky, L.A., Fry, P.M. Hammel, H.B., Baranec, C., Sayanagi, K.M. (2015). Record-breaking storm activity on Uranus in 2014. Icarus, 252, 121-128.

Delsanti A, Jewitt D. 2006. The Solar System Beyond the Planets. In Solar System Update, ed. P Blondel, J Mason, pp. 267-94. Germany.

Dermott and Thomas (1988), The Shape and Internal Structure of Mimas, Icarus 1988, volume 73, issue 1, pp. 25-65.

Dougherty, M. K., K. K. Khurana, F. M. Neubauer, C. T. Russell, J. Saur, J. S. Leisner, and M. Burton (2006), Identification of a dynamic atmosphere at Enceladus with the Cassini magnetometer, Science, 311, 1406–1409.

Duncan, M. J., Lissauer, J. J., 1997. Orbital stability of the Uranian satellite system. Icarus 125, 1-12.

Elliot, J. L. et al. (1998). Global warming on Triton. Nature, 393(6687), 765-767.

Foryta, D. W., and B. Sicardy 1996. The dynamics of the Neptunian Adams ring's arcs. Icarus 123, 129–167.

French, R. G., R. I. Dawson, and M. R. Showalter 2015. Resonances, chaos, and short-term interactions among the inner Uranian satellites. The Astronomical Journal 149, 142–169.

French RS, Showalter MR. 2012. Cupid is doomed: An analysis of the stability of the inner uranian satellites. Icarus 220: 911-21

Fressin F., et al. (2013). The false positive rate of Kepler and the occurrence of planets. The Astrophysical Journal 766, 81-101.

Gaeman, J., Hier-Majumder, S., & Roberts, J. H. (2012). Sustainability of a subsurface ocean within Triton's interior. *Icarus*, *220*(2), 339-347.

Gastine, T., Duarte, L., & Wicht, J. (2012). Dipolar versus multipolar dynamos: the influence of the background density stratification. Astronomy & Astrophysics, 546, A19.

Gaulme P., Mosser B., Schmider F.-X., Guillot T. (2015). Seismology of Giant Planets. In *Extraterrestrial Seismology,* V. Tong and R. Garcia [Editors], Cambridge University Press.

Goldman, N., Fried, E., Kuo, I-F W., Mundy, C. J. (2005). Bonding in the superionic phase of water, Phys. Rev. Lett. 94, 217801.

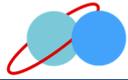

Goldreich, P., Murray, N., Longaretti, P. Y., & Banfield, D. (1989). Neptune's story. Science, 245(4917), 500-504.

Gómez-Pérez, N., & Heimpel, M. (2007). Numerical models of zonal flow dynamos: an application to the ice giants. Geophysical and Astrophysical Fluid Dynamics, 101(5-6), 371-388.

Gordon, Sanford and Bonnie J. McBride, "Computer Program for Calculation of Complex Chemical Equilibrium Compositions and Applications", Part I. Analysis, NASA Reference Publication, 1311, October 1994.

Grundy WM, Young LA, Spencer JR, Johnson RE, Young EF, Buie MW. 2006. Distributions of H2O and CO2 ices on Ariel, Umbriel, Titania, and Oberon from IRTF/SpeX observations. Icarus 184: 543-55

Guillot, T., 2005. The interiors of giant planets: models and outstanding questions. Ann. Rev. Earth Planet. Sci. 33, 493–530.

Gulkis, S., Janssen, M.A., and Olsen, E.T., 1978. Evidence for the depletion of ammonia in the Uranus atmosphere. Icarus 34, 10-19.

Hammel, H.B., and Lockwood, G.W., (2007). Suggestive correlations between the brightness of Neptune, solar variability, and Earth's temperature. Geophys. Res. Lett., 34, L08203, doi:10.1029/2006GL028764.

Hammel, H.B., de Pater, I., Gibbard, S., Lockwood, G.W., and Rages, K., (2005). Uranus in 2003: Zonal winds, banded structure, and discrete features. Icarus 175, 534-545.

Hammel, H.B., Lockwood, G.W., Mills, J.R., and Barnet, C.D., (1995). Hubble Space Telescope Imaging of Neptune's Cloud Structure in 1994. Science, 268, 1740.

Hansen and Kirk (2015). Triton's Plumes — Solar-Driven Like Mars or Endogenic Like Enceladus?, LPSC 46, Abstract #2423.

Hansen, C. J., McEwen, A. S., Ingersoll, A. P., & Terrile, R. J. (1990). Surface and airborne evidence for plumes and winds on Triton. Science, 250(4979), 421-424.

Hayes, A., (2016). The Lakes and Seas of Titan, Ann. Rev. of Earth and Planetary Sci. 44, pp. 57-83.

Hedman, M. M., and P. D. Nicholson 2013. Kronoseismology: Using density waves in Saturn's C ring to probe the planet's interior. Astron. J. 146, 12–27.

Hedman, M. M., and P. D. Nicholson 2014. More Kronoseismology with Saturn's rings. MNRAS 444, 1369-1388.

Helled, R., Anderson, J. D., Podolak, M., & Schubert, G. (2010). Interior models of Uranus and Neptune. Astrophys. J. 726(1), 15.

Herbert, F. (2009), Aurora and magnetic field of Uranus, J. Geophys. Res. Lett., 114, A11206, doi:10.1029/2009JA014394.

Herbert, F. (2009). Aurora and magnetic field of Uranus. Journal of Geophysical Research: Space Physics, 114(A11).

Hersant, F., Gautier, D., Lunine, J.I., 2004. Enrichment in volatiles in the giant planets of the Solar System. Plan. Space Sci. 52, 623–641.

Hofstadter, M.D., Butler, B.J., 2003. Seasonal change in the deep atmosphere of Uranus. Icarus 165, 168-180.

Holme, R., & Bloxham, J. (1996). The magnetic fields of Uranus and Neptune: Methods and models. Journal of Geophysical Research: Planets, 101(E1), 2177-2200.

Holme, R., and J. Bloxham (1996), The magnetic fields of Uranus and Neptune: Methods and models, J. Geophys. Res., 101, 2177–2200, doi:10.1029/95JE03437.

Horn, L. J., P. A. Yanamandra-Fisher, L. W. Esposito, and A. L. Lane 1989. Physical properties of Uranian delta ring from a possible density wave. Icarus 76, 485–492.

Hsu et al. 2015. Ongoing hydrothermal activities within Enceladus, Nature 519, pp. 207-210.

Hubbard, W. B., Podolak, M., & Stevenson, D. J. (1995). The interior of Neptune. Neptune and Triton, 109.

Hussmann H, Sohl F, Spohn T. 2006. Subsurface oceans and deep interiors of medium-sized outer planet satellites and large trans-neptunian objects. Icarus 185: 258-73.

Irwin, P.G.J., Fletcher, L.N., Read, P.L., Tice, D., de Pater, I., Orton, G.S., Teanby, N.A., Davis, G.R. (2016). Spectral analysis of Uranus' 2014 bright storm with VLT/SINFONI. Icarus, 264, 72-89.

Irwin, P.G.J., Fletcher, L.N., Tice, D., Owen, S.J. Orton, G.S., Teanby, N.A., Davis, G.R. (2016). Time variability of Neptune's horizontal and vertical cloud structure revealed by VLT/SINFONI and Gemini/NIFS from 2009 to 2013. Icarus, 271, 418-437.

Janes DM, Melosh HJ. 1988. Sinker Tectonics - an Approach to the Surface of Miranda. Journal of Geophysical Research-Solid Earth and Planets 93: 3127-43

Jankowski DG, Squyres SW. 1988. Solid-State Ice Volcanism on the Satellites of Uranus. Science 241: 1322-5

Jewitt, D., Sheppard, S. (2005). Irregular Satellites in the Context of Planet Formation. Space Science Reviews 116, 441-455.

Jewitt, D., Haghighipour, N. (2007). Irregular Satellites of the Planets: Products of Capture in the Early Solar System. Annual Review of Astronomy & Astrophysics 45, 261-295.

Jones, C. A. (2011). Planetary magnetic fields and fluid dynamos. Annual Review of Fluid Mechanics, 43, 583-614.

Kammer, J. A., Shemansky, D. E., Zhang, X., & Yung, Y. L. (2013). Composition of Titan's upper atmosphere from Cassini UVIS EUV stellar occultations. Planetary and Space Science, 88, 86-92.

Karkoschka, E. and Tomasko, M. G. (2011). The haze and methane distributions on Neptune from HST-STIS spectroscopy. Icarus, 211, 328-340.

Kaspi, Y., Showman, A. P., Hubbard, W. B., Aharonson, O., & Helled, R. (2013). Atmospheric confinement of jet streams on Uranus and Neptune. Nature, 497(7449), 344-347.

Kaspi, Y., Davighi, J. E., Galanti, E., & Hubbard, W. B. (2016). The gravitational signature of internal flows in giant planets: Comparing the thermal wind approach with barotropic potential-surface methods. Icarus, 276, 170-181.

Kempf, S., et al. (2005), High-velocity streams of dust originating from Saturn, Nature, 433, 289.

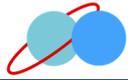

King, E. M., & Aurnou, J. M. (2013). Turbulent convection in liquid metal with and without rotation. Proceedings of the National Academy of Sciences, 110(17), 6688-6693.

Kirk, R. L., L. A. Soderblom, and R. H. Brown (1990). Subsurface energy storage and transport for solar-powered geysers on Triton, *Science*, 250, 424–429.

Knudson, M. D., Desjarlais, M. P., Lemke, R. W., Mattsson, T. R., French, M., Nettelmann, N., & Redmer, R. (2012). Probing the Interiors of the Ice Giants: Shock Compression of Water to 700 GPa and 3.8 g/cm 3. Physical review letters, 108(9), 091102.

Koskinen, T. T., Yelle, R. V., Snowden, D. S., Lavvas, P., Sandel, B. R., Capalbo, F. J., ... & West, R. A. (2011). The mesosphere and lower thermosphere of Titan revealed by Cassini/UVIS stellar occultations. Icarus,216(2), 507-534.

Krasnopolsky, V. A., & Cruikshank, D. P. (1995). Photochemistry of Triton's atmosphere and ionosphere. J. Geophys. Res., 100(E10), 21271-21286.

Kumar K, de Pater I, Showalter MR. 2015. Mab's orbital motion explained. Icarus 254: 102-21

Kurth, W. S., and D. A. Gurnett (1991), Plasma waves in planetary magnetospheres, J. Geophys. Res., 96(S01), 18977–18991, doi:10.1029/91JA01819.

Lamy, L., Prangé, R., Hansen, K. C., Clarke, J. T., Zarka, P., Cecconi, B., ... & Barthélémy, M. (2012). Earth-based detection of Uranus' aurorae. Geophysical Research Letters, 39(7).

Leconte, J., & Chabrier, G. (2012). A new vision of giant planet interiors: Impact of double diffusive convection. Astronomy & Astrophysics, 540, A20.

Lee, M. S., & Scandolo, S. (2011). Mixtures of planetary ices at extreme conditions. Nature communications, 2, 185.

Lellouch, E., De Bergh, C., Sicardy, B., Ferron, S., & Käufl, H. U. (2010). Detection of CO in Triton's atmosphere and the nature of surface-atmosphere interactions. Astronomy & Astrophysics, 512, L8.

Levison, H.F., Kretke, K.A., and Duncan, M.J. (2015). Growing the gas-giant planets by the gradual accumulation of pebbles. Nature 524, 322-324.

Levison, H. F., et al. (2011). Late Orbital Instabilities in the Outer Planets Induced by Interaction with a Self-gravitating Planetesimal Disk. The Astronomical Journal 142, article id. 152.

Limaye, S. S., and Sromovsky, L.A., (1991). Winds of Neptune: Voyager observations of cloud motions. J. Geophys. Res., 96(S01), 18941–18960, doi:10.1029/91JA01701

Lissauer, J.J., and Stevenson, D.J., 2007. Formation of giant planets, in Protostars and Planets V, Reipurth, Jewitt, and Keil, (Eds.). Univ. of Arizona Press.

Lissauer, J. J., Pollack, J. B., Wetherill, G. W., & Stevenson, D. J. (1995). Formation of the Neptune system. In Neptune and Triton (Vol. 1, pp. 37-108). Univ. of Arizona Press Tucson.

Lockwood, G.W., Thompson, D.T., lutz, B.L., and Howell, E.S., (1991). The brightness, albedo, and temporal variability of Neptune. Astrophys. Jour. 368, 287-297.

Magee, B. A., J. H. Waite, Jr., K. E. Mandt, J. Westlake, J. Bell, D. A. Gell, 2009, INMS derived composition of Titan's upper atmosphere: analysis methods and model comparison, Planetary and Space Science, 57, 1895-1916.

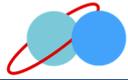

Mandt, K. E. et al., 2012a. Ion densities and composition of Titan's upper atmosphere derived by the Cassini Ion Neutral Mass Spectrometer: analysis methods and model comparison, J. Geophys. Res., 117, E10006. doi:10.1029/2012JE004139.

Mandt, K. E., J. H. Waite, Jr., B. A. Magee, J. Bell, J. Lunine, O. Mousis, D. Cordier, 2009, Isotopic evolution of Titan's main atmospheric constituents, Planetary and Space Science, 57, 1917-1930.

Mandt, K. E., J. H. Waite, Jr., B. Teolis, B. A. Magee, J. Bell, J. Westlake, C. Nixon, O. Mousis, J. Lunine, 2012b, The 12C/13C ratio on Titan from Cassini INMS measurements and implications for the evolution of methane, Astrophys. J., 749, 160.

Mandt, K. E., O. Mousis, J. Lunine & D. Gautier, 2014, Primordial ammonia as the source of Titan's Nitrogen, Astrophys. J., 778, L24.

Marzari et al., Modelling the disruption and reaccumulation of Miranda, Astron & Astrophys, 333, pp. 1082-1091, 1998.

Masters, A. (2014), Magnetic reconnection at Uranus' magnetopause, J. Geophys. Res. Space Physics, 119, 5520–5538, doi:10.1002/2014JA020077.

Masters, A. (2015), Magnetic reconnection at Neptune's magnetopause, J. Geophys. Res. Space Physics, 120, 479–493, doi:10.1002/2014JA020744.

Masters, A., et al. (2014). Neptune and Triton: essential pieces of the solar system puzzle. *Planetary and Space Science*, *104*, 108-121.

Mauk, B. H., E. P. Keath, M. Kane, S. M. Krimigis, A. F. Cheng, M. H. Acuña, T. P. Armstrong, and N. F. Ness (1991), The magnetosphere of Neptune: Hot plasmas and energetic particles, J. Geophys. Res., 96(S01), 19,061–19,084, doi:10.1029/91JA01820.

Mauk, B. H., S. M. Krimigis, E. P. Keath, A. F. Cheng, T. P. Armstrong, L. J. Lanzerotti, G. Gloeckler, and D. C. Hamilton (1987), The hot plasma and radiation environment of the Uranian magnetosphere, J. Geophys. Res., 92, 15,283–15,308, doi:10.1029/JA092iA13p15283.

McBride, Bonnie J. and Sanford Gordon, "Computer Program for Calculation of Complex Chemical Equilibrium Compositions and Applications", Part II. User's Manual and Program Description, NASA Reference Publication, 1311, June 1996.

McKinnon, W. B. & A. C. Leith (1995). Gas drag and the orbital evolution of a captured Triton, Icarus, 118, 392-413.

McKinnon, W.B. (1984). On the origin of Triton and Pluto. Nature 311, 355-358.

Milos, F.S. and Y.K. Chen, "Ablation, Thermal Response, and Chemistry Program for Analysis of Thermal Protection Systems," Journal of Spacecraft and Rockets, Vol. 50, No. 1, 2013, pp. 137-149.

Mejnertsen, L., J. P. Eastwood, J. Chittenden, and A. Masters (2016), Global MHD simulations of Neptune's magnetosphere, J. Geophys. Res. Space Physics, 121, doi:10.1002/2015JA022272.

Mosqueira I, Estrada P., Turrini D. (2010). Planetesimals and Satellitesimals: Formation of the Satellite Systems. Space Science Reviews 153, 431-446.

Namouni, F., and C. C. Porco 2002. The confinement of Neptune's ring arcs by the moon Galatea. Nature 417, 45–47.

National Research Council (NRC), 2011, *Vision and Voyages*. National Academies Press, Washington, DC, DOI: 10.17226/13117.

Nellis, W. J. (2015). The unusual magnetic fields of Uranus and Neptune. Modern Physics Letters B, 29(01), 1430018.

Ness, N. F., Acuña, M. H., Behannon, K. W., Burlaga, L. F., Connerney, J. E., Lepping, R. P., & Neubauer, F. M. (1986). Magnetic fields at Uranus. Science, 233(4759), 85-89.

Ness, N. F., Acuña, M. H., Burlaga, L. F., Connerney, J. E., Lepping, R. P., & Neubauer, F. M. (1989). Magnetic fields at Neptune. Science, 246(4936), 1473-1478.

Nesvorny, D., Vokrouhlicky, D., Morbidelli, A. (2007). Capture of Irregular Satellites during Planetary Encounters. The Astronomical Journal 133, 1962-1976.

Nettelmann, N., Helled, R., Fortney, J. J., & Redmer, R. (2013). New indication for a dichotomy in the interior structure of Uranus and Neptune from the application of modified shape and rotation data. Planetary and Space Science, 77, 143-151.

Nettelmann, N., Wang, K., Fortney, J. J., Hamel, S., Yellamilli, S., Bethkenhagen, M., & Redmer, R. (2016). Uranus evolution models with simple thermal boundary layers. Icarus, 275, 107-116.

Nimmo, F. et al. (2016). Mean radius and shape of Pluto and Charon from New Horizons images. Icarus, in press.

Oruba, L., & Dormy, E. (2014). Transition between viscous dipolar and inertial multipolar dynamos. Geophysical Research Letters, 41(20), 7115-7120.

Pappalardo RT, Reynolds SJ, Greeley R. 1997. Extensional tilt blocks on Miranda: Evidence for an upwelling origin of Arden Corona. Journal of Geophysical Research-Planets 102: 13369-79.

Pearl, J.C. and B.J. Conrath, 1991. The albedo, effective temperature, and energy balance of Neptune, as determined from Voyager data. J. Geophys. Res., 96, 18921-18930.

Pearl, J.C., B.J. Conrath, R.A. Hanel, and J.A. Pirraglia, 1990. The albedo, effective temperature, and energy balance of Uranus, as determined from Voyager IRIS data. Icarus, 84, 12-28.

Plescia J.B., 1987. Geological Terrains and Crater Frequencies on Ariel. Nature 327: 201-4

Podolak, M., Hubbard, W. B., & Stevenson, D. J. (1991). Models of Uranus' interior and magnetic field. In University of Arizona Press; Space Science Series.

Porco, C. C., 1991. An explanation for Neptune's ring arcs. Science 253, 995–1001.

Porco, C. C., and P. Goldreich 1987. Shepherding in the Uranian rings I. Kinematics. Astron. J. 93, 724-729, 778.

Porco et al., (2006). Cassini Observes the Active South Pole of Enceladus, Science 311, pp. 1393-4001.

Porco, C. C., Nicholson, P. D., Cuzzi, J. N., Lissauer, J. J., & Esposito, L. W. (1995). Neptune's ring system. In *Neptune and Triton*, pp. 703-804.

Redmer, R., Mattsson, T. R., Nettelmann, N., & French, M. (2011). The phase diagram of water and the magnetic fields of Uranus and Neptune. Icarus, 211(1), 798-803.

Renner, S., Sicardy, B., Souami, D., Carry, B., and Dumas, C., 2014. Neptune's ring arcs: VLT/NACO near-infrared observations and a model to explain their stability. Astron. & Astrophys. 563, doi:10.1051/0004-6361/201321910.

Richardson, J. D., J. W. Belcher, M. Zhang, and R. L. McNutt Jr. (1991), Low-energy ions near Neptune, J. Geophys. Res., 96, 18,993–19,011, doi:10.1029/91JA01598.

Ruzmaikin, A. A., & Starchenko, S. V. (1991). On the origin of Uranus and Neptune magnetic fields. Icarus, 93(1), 82-87.

Sandel, B. R., Herbert, F., Dessler, A. J., & Hill, T. W. (1990). Aurora and airglow on the night side of Neptune. Geophysical Research Letters, 17(10), 1693-1696.

Sarani, S. (2009). Titan Atmospheric Density Reconstruction Using Cassini Guidance, Navigation, and Control Data. AIAA Guidance, Navigation, and Control Conference 10-13 August 2009, Chicago, IL. DOI: 10.2514/6.2009-5763.

Schenk PM. 1991. Fluid Volcanism on Miranda and Ariel - Flow Morphology and Composition. Journal of Geophysical Research-Solid Earth and Planets 96: 1887-906.

Schenk and Jackson, 1993. Diapirism on Triton: a record of crustal layering and instability. Geology 21, 299-302.

Schubert, G., et al. (2010). Evolution of Icy Satellites. Space Science Reviews 153, 447-484.

Schubert, G., & Soderlund, K. M. (2011). Planetary magnetic fields: Observations and models. Phys. Earth Planet. Int. 187(3), 92-108.

Schulz, M., et al. (1995), Magnetospheric Configuration of Neptune, in Neptune and Triton, ed. D. P. Cruikshank, M. S. Matthews, A. M. Schumann, University of Arizona Press.

Selesnick, R. S. (1988), Magnetospheric convection in the nondipolar magnetic field of Uranus, J. Geophys. Res., 93(A9), 9607–9620, doi:10.1029/JA093iA09p09607.

Selesnick, R. S. (1990), Plasma convection in Neptune's magnetosphere, Geophys. Res. Lett., 17, 1681–1684, doi:10.1029/GL017i010p01681.

Selesnick, R. S., and J. D. Richardson (1986), Plasmasphere formation in arbitrarily oriented magnetospheres, Geophys. Res. Lett., 13, 624–627, doi:10.1029/GL013i007p00624.

Showalter, M. R., I. de Pater, J. J. Lissauer, and R. S. French 2013. The Neptune System Revisited: New Results on Moons and Rings from the Hubble Space Telescope. DPS meeting #45, #206.01.

Showalter, M. R., I. de Pater, J. J. Lissauer, and R. S. French 2013. New Satellite of Neptune: S/2004 N 1. CBET 3586.

Showalter, M. R., and J. J. Lissauer 2006. The second ring-moon system of Uranus: Discovery and dynamics. Science 311, 973–977.

Smith BA, Soderblom LA, Beebe R, Bliss D, Boyce JM, et al. 1986. Voyager-2 in the Uranian System - Imaging Science Results. Science 233: 43-64

Smith, B. A. et al. (1989). Voyager 2 at Neptune: Imaging science results. Science, 246(4936), 1422-1449.

Soderblom, L. A. et al. (1990). Triton's geyser-like plumes: Discovery and basic characterization. Science, 250(4979), 410-415.

Soderlund, K. M., Heimpel, M. H., King, E. M., & Aurnou, J. M. (2013). Turbulent models of ice giant internal dynamics: Dynamos, heat transfer, and zonal flows. Icarus, 224(1), 97-113.

Soderlund, K. M., King, E. M., & Aurnou, J. M. (2012). The influence of magnetic fields in planetary dynamo models. Earth and Planetary Science Letters, 333, 9-20.

Sohl, F., Choukroun, M., Kargel, J., Kimura, J., Pappalardo, R., Vance, S., & Zolotov, M. (2010). Subsurface water oceans on icy satellites: chemical composition and exchange processes. *Space Science Reviews*, *153*(1-4), 485-510.

Spencer et al., (2006). Cassini Encounters Enceladus: Background and the Discovery of a South Polar Hot Spot, Science 311, pp. 1401-1405.

Sreenivasan, B., & Jones, C. A. (2006). The role of inertia in the evolution of spherical dynamos. Geophysical Journal International, 164(2), 467-476.

Sromovsky, L. A., Fry, P. M. and Kim, J. H. (2011). Methane on Uranus: The case for a compact CH4 cloud layer at low latitudes and a severe CH4 depletion at high-latitudes based on re-analysis of Voyager occultation measurements and STIS spectroscopy. Icarus, 215, 292-312.

Sromovsky, L.A., Hammel, H.B., de Pater, I., Fry, P.M., Rages, K.A., Showalter, M.R., Merline, W.J., Tamblyn, P., Neyman, C., Margot, J.-L., Fang, J., Colas, F., Dauvergne, J.-L., Gomez-Forrellad, J.M., Hueso, R., Sanchez-Lavega, A., Stallard, T. (2012). Episodic bright and dark spots on Uranus. Icarus, 220, 6-22.

Sromovsky, L.A., Fry, P.M., Hammel, H.B., de Pater, I., Rages, K.A. (2012). Post-equinox dynamics and polar cloud structure on Uranus. Icarus, 220, 694-712.

Sromovsky, L.A., Karkoschka, E., Fry, P.M., Hammel, H.B., de Pater, I., Rages, K. (2014). Methane depletion in both polar regions of Uranus from HST/STIS and Keck/NIRC2 observations. Icarus, 238, 137-155.

Sromovsky, L.A., de Pater, I., Fry, P.M., Hammel, H.B., Marcus, P. (2015). High S/N Keck and Gemini AO imaging of Uranus during 2012-2014: New cloud patterns increasing activity, and improved wind measurements. Icarus, 258, 192-223.

Stanley, S., & Bloxham, J. (2004). Convective-region geometry as the cause of Uranus' and Neptune's unusual magnetic fields. Nature, 428(6979), 151-153.

Stanley, S., & Bloxham, J. (2006). Numerical dynamo models of Uranus' and Neptune's magnetic fields. Icarus, 184(2), 556-572.

Stern et al. (2015). The Pluto system: Initial results from its exploration by New Horizons. Science 350, Issue 6258.

Stevens, M. H., Evans, J. S., Lumpe, J., Westlake, J. H., Ajello, J. M., Bradley, E. T., & Esposito, L. W. (2015). Molecular nitrogen and methane density retrievals from Cassini UVIS dayglow observations of Titan's upper atmosphere. Icarus, 247, 301-312.

Stevenson DJ, Lunine JI. 1986. Mobilization of Cryogenic Ice in Outer Solar-System Satellites. Nature 323: 46-8

Stevenson, D. J. (2003). Planetary magnetic fields. Earth and planetary science letters, 208(1), 1-11.

Stofan et al., (2007). The lakes of Titan, Nature 445, pp. 61-64.

Stone, E. C., and E. D. Miner (1986), The Voyager 2 encounter with the Uranian system, Science, 233, 39–43.

Stone, E. C., and E. D. Miner (1989), The Voyager 2 encounter with the Neptunian system, Science, 246, 1417–1421.

Strobel, D. F., et al. (1990), Magnetospheric interaction with Triton's ionosphere, Geophys. Res. Lett., 17, 1661.

Strobel, D. F., Simmers, M. E., Herbert, F., & Sandel, B. R. (1990). The photochemistry of methane in the atmosphere of Triton. Geophysical Research Letters, 17(10), 1729-1732.

Strom RG. 1987. The Solar-System Cratering Record - Voyager-2 Results at Uranus and Implications for the Origin of Impacting Objects. Icarus 70: 517-35

Tamayo, D., et al. (2011). Finding the trigger to Iapetus' odd global albedo pattern: Dynamics of dust from Saturn's irregular satellites. Icarus 215, 260-278.

Tamayo, D., Burns, J. A., Hamilton, D. P. (2013). Chaotic dust dynamics and implications for the hemispherical color asymmetries of the Uranian satellites. Icarus 226, 655-662.

Thomas PC, Burns JA, Hedman M, Helfenstein P, Morrison S, et al. 2013. The inner small satellites of Saturn: A variety of worlds. Icarus 226: 999-1019

Tian, B. Y., & Stanley, S. (2013). Interior Structure of Water Planets: Implications for their dynamo source regions. The Astrophysical Journal, 768(2), 156.

Tiscareno, M.S., Hedman, M.M, Burns, J.A., Castillo-Rogez, J. (2013). Compositions and Origins of Outer Planet Systems: Insights from the Roche Critical Density. Astrophys. Journal Letters 765, L28.

Tittemore, W. C. (1990). Tidal heating of Ariel. *Icarus*, *87*(1), 110-139.

Tittemore WC, Wisdom J. 1988. Tidal Evolution of the Uranian Satellites .1. Passage of Ariel and Umbriel through the 5-3 Mean-Motion Commensurability. Icarus 74: 172-230

Tittemore WC, Wisdom J. 1989. Tidal Evolution of the Uranian Satellites .2. An Explanation of the Anomalously High Orbital Inclination of Miranda. Icarus 78: 63-89

Tittemore WC, Wisdom J. 1990. Tidal Evolution of the Uranian Satellites .3. Evolution through the Miranda-Umbriel 3-1, Miranda-Ariel 5-3, and Ariel-Umbriel 2-1 Mean-Motion Commensurabilities. Icarus 85: 394-443

Tobie, G., et al. (2010). Surface, Subsurface and Atmosphere Exchanges on the Satellites of the Outer Solar System. Space Science Reviews 153, 375-410.

Tosi, F., et al. (2010). Probing the origin of the dark material on Iapetus. MNRAS 403, 1113-1130.

Tóth, G., D. Kovács, K. C. Hansen, and T. I. Gombosi (2004), Three-dimensional MHD simulations of the magnetosphere of Uranus, J. Geophys. Res., 109, A11210, doi:10.1029/2004JA010406.

Trafton, L. (1984). Large seasonal variations in Triton's atmosphere. Icarus, 58(2), 312-324.

Tsiganis, K., et al. (2005). Origin of the orbital architecture of the giant planets of the Solar System. Nature 435, 459-461.

Turrini, D., Marzari, F., Tosi, F. (2009). A new perspective on the irregular satellites of Saturn – II: Dynamical and physical origin. MNRAS 392, 455-474.

Turrini, D. et al. (2013). The ODINUS mission concept - The scientific case for a mission to the ice giant planets with twin spacecraft to unveil the history of our Solar System. White Paper submitted in respone to ESA's call for science themes for the L2 and L3 missions, http://sci.esa.int/jump.cfm?oid=52030, pp. 549-569.

Turrini, D. et al. (2014). The comparative exploration of the ice giant planets with twin spacecraft: Unveiling the history of our Solar System. Planetary and Space Science 104, 93-107

Tyler, G. L., et al. (1989), Voyager Radio Science Observations of Neptune and Triton, Science, 246, 1466.

Tyler, G. L., Sweetnam, D. N., Anderson, J. D., Borutzki, S. E., & Campbell, J. K. (1989). Voyager radio science observations of Neptune and Triton. Science, 246(4936), 1466.

Uckert, K., Chanover, N.J., Olkin, C.B., Young, L.A., Hammel, H.B., Miller, C., Bauer, J.M. (2014). An investigation of the temperature variations in Neptune's upper stratosphere including a July 2008 stellar occultation event. Icarus, 232, 22-33.

Vasyliunas, V. M. (1986), The convection-dominated magnetosphere of Uranus, Geophys. Res. Lett., 13, 621–623, doi:10.1029/GL013i007p00621.

Vuitton, V., Yelle, R. V., & Anicich, V. G. (2006). The nitrogen chemistry of Titan's upper atmosphere revealed. The Astrophysical Journal Letters,647(2), L175.

Vuitton, V., Yelle, R. V., & McEwan, M. J. (2007). Ion chemistry and N-containing molecules in Titan's upper atmosphere. Icarus, 191(2), 722-742.

Walsh, K., et al. (2011). A low mass for Mars from Jupiter's early gas-driven migration. Nature 475, 206-209.

Wong, M. H., Mahaffy, P.R., Atreya, S.K., et al. (2004). Updated Galileo probe mass spectrometer measurements of carbon, oxygen, nitrogen, and sulfur on Jupiter. Icarus, 171, 153-170.

Yelle, R. V., Cui, J., & Müller-Wodarg, I. C. F. (2008). Methane escape from Titan's atmosphere. Journal of Geophysical Research: Planets, 113(E10).

Zarka, P., et al. (1995), Radio emission from Neptune, in Neptune and Triton, ed. D. P. Cruikshank, M. S. Matthews, A. M. Schumann, University of Arizona Press.

Zhang, K., Kong, D., & Schubert, G. (2015). Thermal-gravitational wind equation for the wind-induced gravitational signature of giant gaseous planets: mathematical derivation, numerical method, and illustrative solutions. The Astrophysical Journal, 806(2), 270.

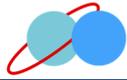

# A    MISSION DESIGN

## A.1    Possible Mission Design Architectures

The comprehensive nature of this study resulted in a vast mission design trade space encompassing a number of mission architectures. **Table A-1** below lists possible mission types along with relevant mission design parameter options that have been considered in this study.

Table A-1.  Possible mission design options.

| Mission Types | Max. Flybys | Primary Propulsion Type | | Power (kW, 1 AU) | Max. # of Engines | | DSM |
|---|---|---|---|---|---|---|---|
| | | Interp. | Orbit Insertion | | EP | Biprop / SC | |
| Chemical orbiter | 4 | Bi-prop | Biprop | - | 2 | 2 | Yes |
| Chemical orbiter + probe | 4 | Bi-prop | Biprop | - | 2 | 2 | Yes |
| Chemical flyby + probe | 4 | Bi-prop | Biprop | - | 2 | 2 | Yes |
| SEP orbiter | 4 | EP | Biprop | 35, 25, 15 | 3, 2 | 2 | No |
| SEP orbiter + probe | 4 | EP | Biprop | 35, 25, 15 | 3, 2 | 2 | No |
| Dual orbiters | 4 | EP/ | Biprop | 75 | 2–6 | 2 | No |
| REP orbiter | 4 | EP | EP/Biprop | 1.65, 2.15 | 1 | 1 | No |

The above listed mission types and mission design options were then mapped to the mission architectures to be studied by Team X with relevant constraints, as listed in **Table A-2**. Dual orbiter and REP based mission concepts (discussed later) were only evaluated from a mission design perspective and did not go on to point designs for Team X. Various limits defined in **Table A-2** were derived from discussions with the science and engineering teams.

Table A-2.  Trajectory broad-search mission parameters and limits.

| # | Mission Options | Payload Mass (kg) | Max. Cruise TOF (yrs.) | Max. OI-DV (km/s) |
|---|---|---|---|---|
| 1 | Uranus orbiter with SEP stage + probe | 50 | 12 | 4.5 |
| 2 | Uranus orbiter with SEP stage | 150 | 12 | 4.5 |
| 3 | Neptune orbiter with SEP stage + probe | 50 | 13 | 4.5 |
| 4 | Uranus flyby + probe | 50 | 12 | - |
| 5 | Uranus orbiter + probe | 50 | 12 | 2.0 |
| 6 | Uranus orbiter | 150 | 12 | 2.0 |

Using the parameters defined in **Table A-2** and the assumptions listed in Section 2.2, a broad interplanetary trajectory search was carried out following the approach discussed in Section 2.3.3.

## A.2    Launch Vehicle Options

**Table A-3** lists performance limits and launch fairing characteristics of the three different launch vehicles considered in this study. It should be noted that the Falcon Heavy could also be considered in this timeframe, however data available to the team at the time of mission analysis indicated its performance for the missions considered would be in the same range as the Delta IV Heavy. Performance improvement to each of the three launch vehicles using an optimal kick stage was also computed.

Table A-3.  Launch vehicle options.

| Launch Vehicle Type | Max. Launch C3 (km²/s²) | | Adapter Type | Fairing Dia. (m) |
|---|---|---|---|---|
| | Nominal | With Optimal Kick Stage | | |
| Atlas V (551) | 60 | 225 | C22 adapter | 5 |
| Delta-IV Heavy | 100 | 225 | 1575-5 payload attach fitting | 5 |
| SLS Block 1-B | 135 | 225 | Notional (assumed to be included in LV performance) | 8.3 |

**Figure A-1** shows the nominal and optimal kick stage-assisted performance for each of the three launch vehicles. Given a kick-stage ISP, Propellant Mass Fraction (PMF) and burn altitude the kick-stage propellant load is optimally chosen to maximize delivered payload mass for a given launch C3. The optimal kick stage enables high C3 launch (>100 km$^2$/s$^2$) and improves launch payload mass capability.

Note that kick-stage doesn't help with low C3 launches. In fact, the kick-stage selection algorithm (developed for this study) chooses zero propellant mass for this case, thereby essentially eliminating the requirement to have a kick-stage.

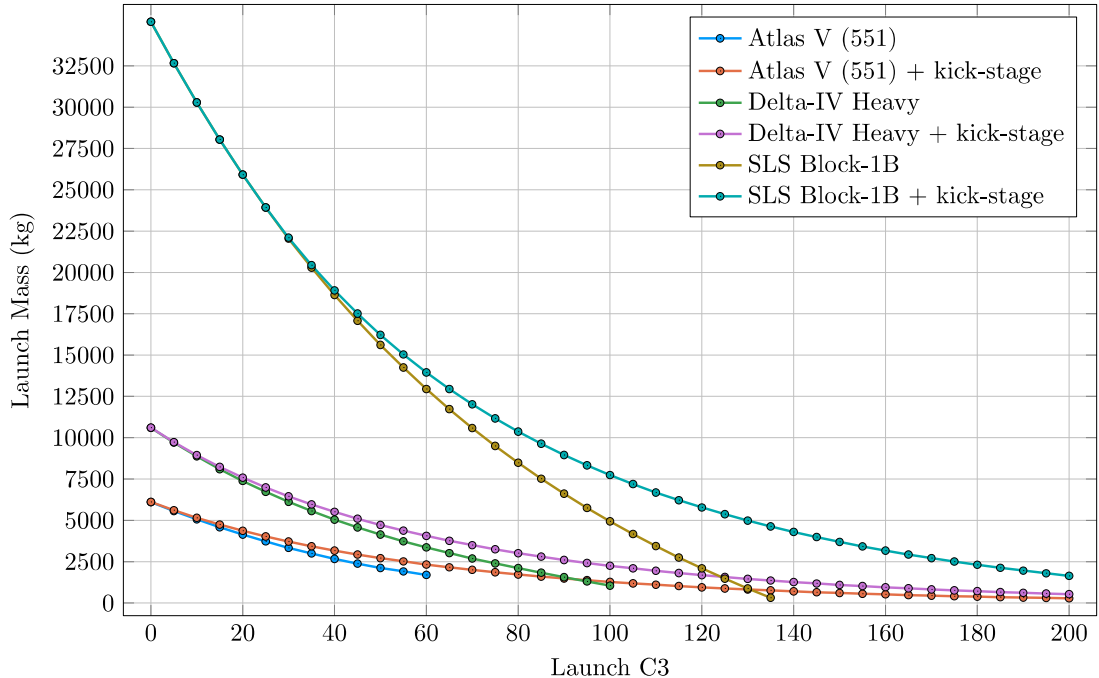

**Figure A-1.** Launch vehicles performance design space.

## A.3 Propulsion System Assumptions

### A.3.1 Chemical Assumptions

All mission options discussed in this study assume that the spacecraft carries a monopropellant system for small burns and attitude control. Furthermore, mission options which go into orbit around Uranus or Neptune also carry a high performance bi-propellant system (with 325 sec ISP). As listed in **Table A-2**, a maximum orbit insertion ΔV constraint of 4.5 km/s was imposed to keep the propulsion subsystem in a reasonable performance range. Please see individual mission options for details on the propulsion system.

### A.3.2 SEP Stage Assumptions

Some mission options listed in **Table A-2** utilize a solar electric propulsion (SEP) stage in the inner solar system to increase delivered mass to the target planet. Trajectories making use of a SEP stage are very sensitive to the stage dry mass; which is assumed to be dropped off before planetary orbit insertion. The SEP stage mass (crudely) depends on number of ion engines, solar array size (or power at 1AU) and propellant throughput.

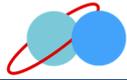

**Table A-4** lists various assumptions made during the SEP trajectory broad search. Some of these assumptions (e.g., SEP stage dry mass), which were based on previous studies, turned out to be somewhat optimistic when assessed by Team X, however they do serve to capture "what-if" scenarios which could be realized with technology advancements in lightweight structures and propellant tanks.

**Table A-4.** SEP stage assumptions.

| Target Planet | # of Ion Engines | Solar Power at 1 AU (kW) | Allowable Propellant Throughput (kg) | SEP Stage Dry Mass (kg) |
|---|---|---|---|---|
| Uranus | 2 | 15, 25 | 1300, 1300 | 600, 700 |
| Uranus | 3 | 35 | 1950 | 1000 |
| Neptune | 2 | 15 | 1300 | 600 |
| Neptune | 3 | 25, 35 | 1950, 1950 | 800, 1000 |

### A.3.3    REP Trajectory Assumptions

The current study also looked at radioisotope electric propulsion (REP) trajectories powered by the radioisotope thermoelectric generators or RTGs. The spacecraft was still assumed to have a bipropellant propulsion system for orbit insertion at the target planet. The REP system was assumed to be part of the core spacecraft and not dropped before orbit insertion. Trajectories were optimized to pick the optimum ratio between orbit insertion propellant usage and xenon usage. Specifically, the mission design trade-space analysis considered advanced RTG (SMRTG) powered trajectories using the assumptions listed in **Table A-5**.

**Table A-5.** REP trajectory assumptions.

| Target Planet | # of XIPS Ion Engines | REP Power (kW) | Allowable Propellant Throughput (kg) | Max. Cruise TOF (yrs.) |
|---|---|---|---|---|
| Uranus | 1+1 | 1.65, 2.15 | 1103, 820 | 12 |
| Neptune | 1+1 | 1.65, 2.15 | 1103, 820 | 13 |

REP mission launched with either the Atlas V (551) or the Delta-IV Heavy launch vehicle were evaluated.

### A.4    Mission Design Trade-Space

In this section, we summarize the results of a broad search of missions to Uranus and Neptune launching between the year 2025 and 2037. The maximum flight time for a mission to Uranus was set to 12 years (see **Table A-2**) and for Neptune was set to 13 years. Along with all-chemical propulsion trajectories, three different SEP power levels + engine configurations were considered. See **Table A-1** and **Table A-4** for different configurations. Two mass metrics are defined for concise presentation of results in this section:

1. **Arrival Mass** (kg) = Launch Mass – Interplanetary Propellant Mass
2. **Useful Inserted Mass** (kg) = Arrival Mass – (1+TMF)Orbit Insertion Propellant Mass

where the interplanetary propellant mass can be either bipropellant or xenon mass consumed during the interplanetary cruise phase of the mission. TMF stands for Tank Mass Fraction of the Bi-propellant tanks which are used for orbit insertion. A constant TMF of 0.1 was assumed for this trade study. The capture orbit assumes an orbit insertion maneuver at a periapsis of 1.05 Planet Radii. This helps in avoiding rings at both Uranus and Neptune and reduces the orbit insertion DV. For consistency, the capture orbit period was fixed to 120 days for this broad search. More details on orbit insertion will be discussed in later sections. The next few sections summarize the trade-space of trajectories to Uranus and Neptune.

## A.4.1    Chemical Trajectories for Uranus

**Figure A-2** highlights the interplanetary trajectory tradespace for a chemical mission to Uranus with up to 4 flybys.  The Atlas V (551) launch vehicle option allows for >4,000 kg arrival mass at Uranus and >1,600 kg of mass into orbit around Uranus with ~11-year interplanetary cruise time.  Using the Delta-IV Heavy launch vehicle results in >2,000 kg mass in Uranus orbit in ~10 years from launch.  Launching on an SLS, we see a dramatic increase in Useful inserted mass at Uranus (>10,000 kg for 11-year flight time).  Using SLS it is also possible to insert >2,000 kg in Uranus orbit in ~8 years from launch.  The limiting factor here is the maximum allowable orbit insertion DV of 4.5 km/s (see **Table A-2**), which in turn limits the arrival velocity at Uranus.

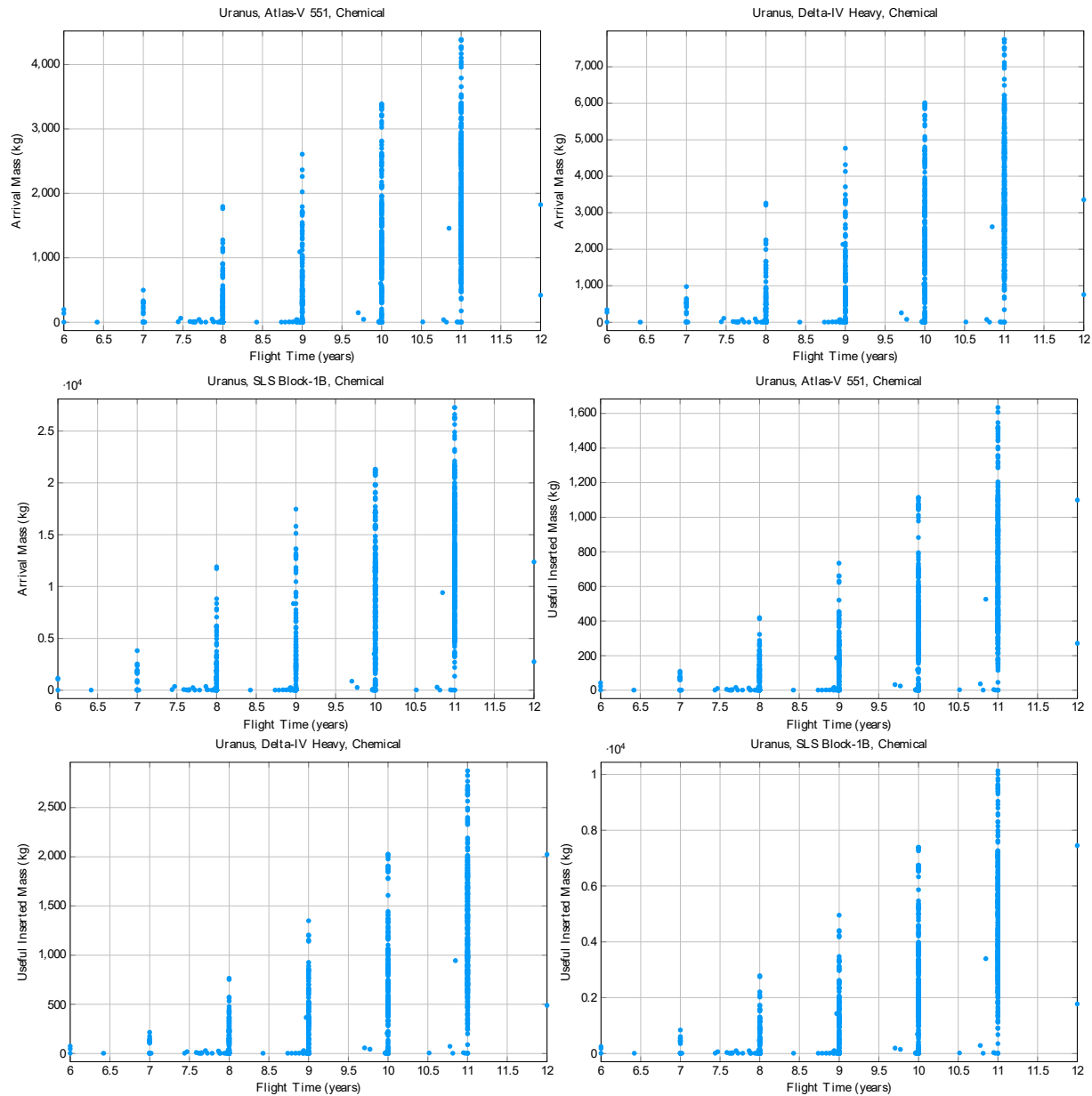

**Figure A-2.**  Chemical trajectory options to Uranus: arrival and useful inserted mass.

**Figure A-3** lists all possible launch opportunities between 2025 and 2037. There is a clear optimal launch period between 2030 and 2034. This corresponds with the availability of a Jupiter gravity assist. High-performing launch opportunities are similar across the three launch vehicle options.

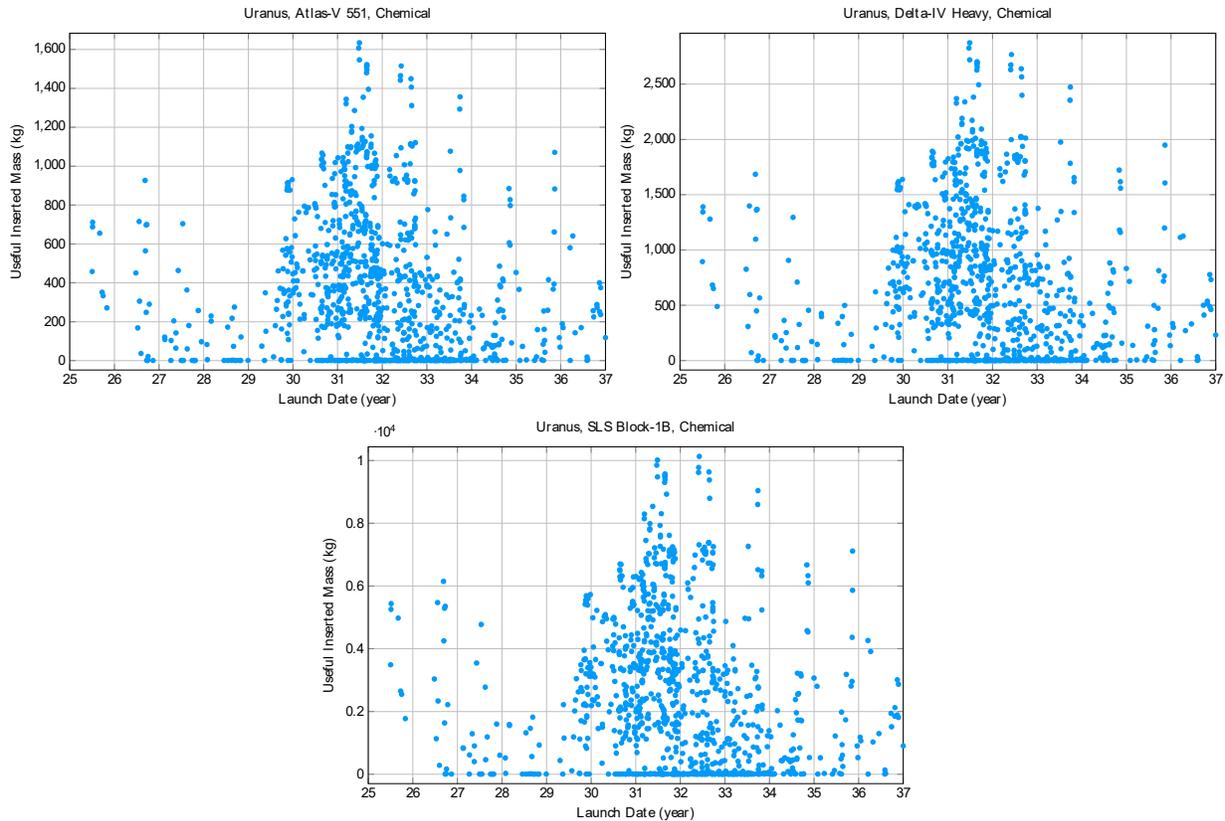

**Figure A-3.** Chemical trajectory options to Uranus: launch opportunities.

## A.4.2 Chemical Trajectories for Neptune

**Figure A-4** highlights the interplanetary trajectory tradespace for a chemical mission to Neptune with up to 4 flybys. Given that Neptune is further out in the solar system, the Atlas V (551) launch vehicle option allows only for >3000 kg arrival mass at Neptune and >800 kg of mass into orbit around Neptune with ~13-year interplanetary cruise time. The low inserted mass is due to high relative velocity at Neptune and larger launch C3 requirements. Using the Delta-IV Heavy launch vehicle results in >1,500 kg mass in Neptune orbit in ~13 years from launch. Launching on an SLS, we again see a dramatic increase in useful inserted mass at Neptune (>5,000 kg for 13-year flight time). Using SLS it is also possible to insert >1,700 kg in Neptune orbit in ~11 years from launch. The limiting factor is again the maximum allowable orbit insertion DV of 4.5 km/s (see **Table A-2**), which in turn limits the arrival velocity at Neptune.

**Figure A-3** lists all possible launch opportunities between 2025 and 2037. There is a clear optimal launch period between 2030 and 2034. This corresponds with the availability of a Jupiter flyby. High-performing launch opportunities are similar across the three launch vehicle options.

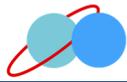

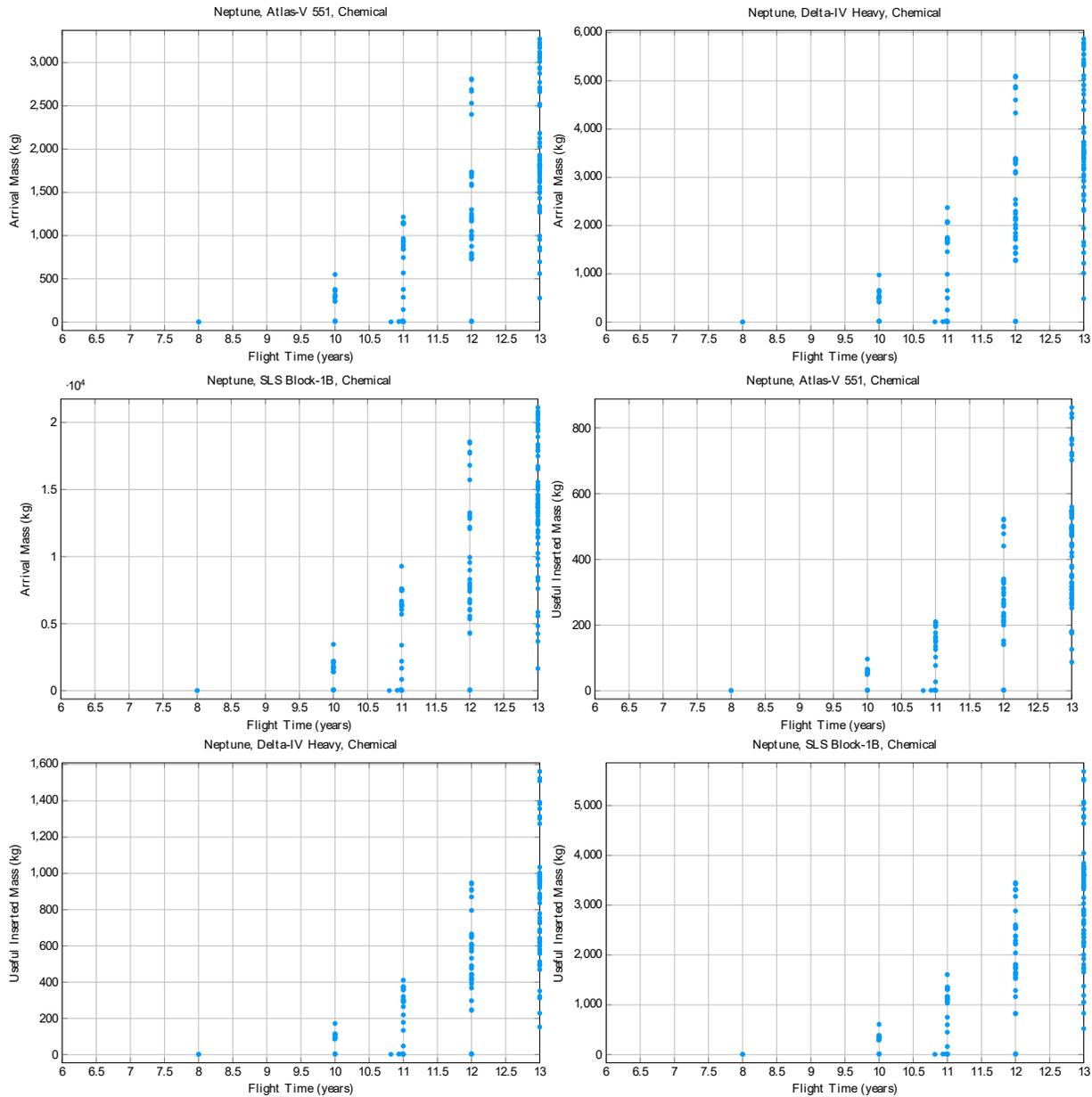

**Figure A-4.** Chemical trajectory options to Neptune: arrival and useful inserted mass.

**Figure A-5** lists all possible launch opportunities between 2025 and 2037. Again, there is a clear, small, optimal launch period between 2029 and 2030. This corresponds with the availability of a Jupiter gravity assist. High-performing launch opportunities perform similarly across the three launch vehicle options.

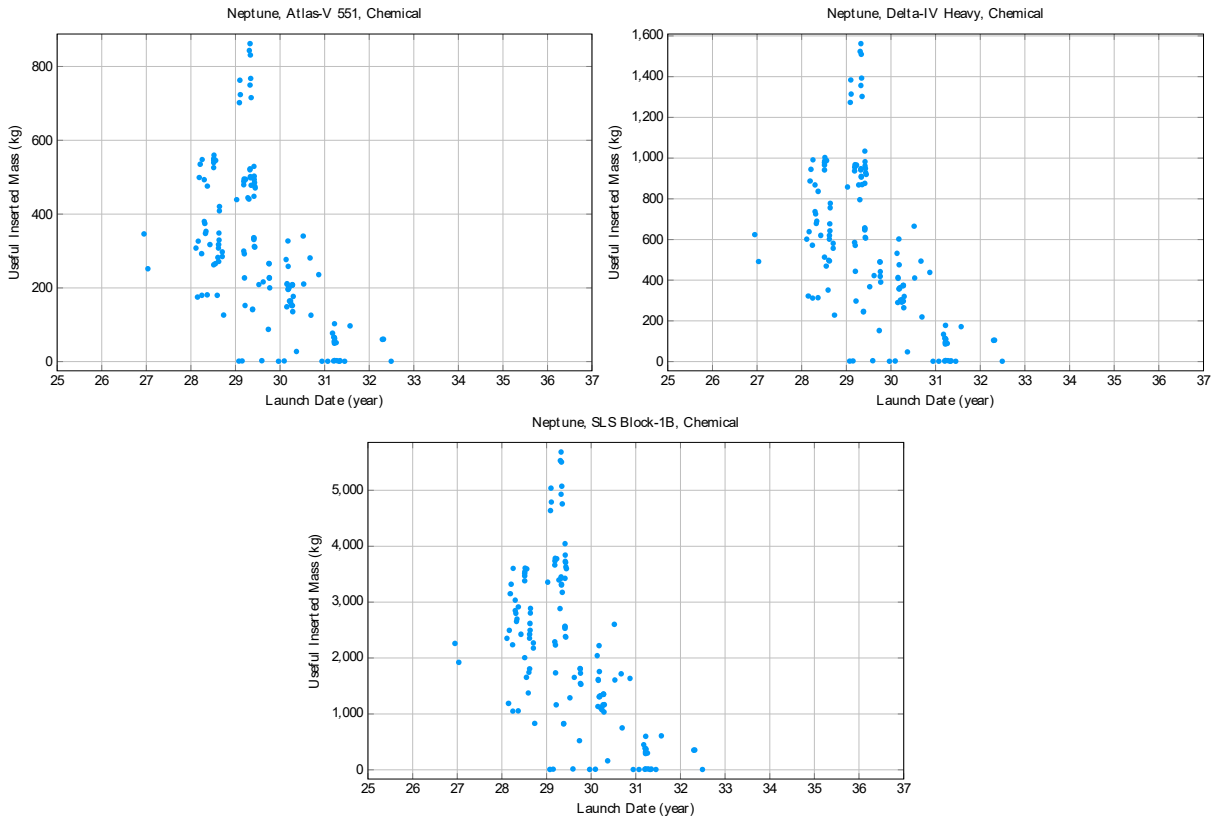

**Figure A-5.** Chemical trajectory options to Neptune: launch opportunities.

### A.4.3 SEP Trajectories for Uranus

**Figure A-6** highlights the interplanetary trajectory tradespace for SEP missions to Uranus with up to 4 flybys, launching on Atlas V 551 and with three different SEP power levels (15 kW, 25 kW and 35 kW). The three SEP power levels result in different SEP stage dry mass and number of engines as shown in **Table A-4**. Using the 15 kW SEP stage, we get ~4,000 kg arrival mass at Uranus in ~9 years from launch. The maximum arrival mass of ~4,900 kg is possible for a 12 years' trajectory. Going to high power levels gives us no benefit in terms for arrival mass when launching on the Atlas V 551 as the trajectory performance is limited by launch C3 and number of SEP engines (maximum thrust). A similar trend is noted for the useful inserted mass metric. A 15 kW SEP stage trajectory allows for >2000 kg mass in to Uranus orbit in ~10 years from launch and this trend remains more or less the same for higher SEP power levels. There appears to be a sweet spot SEP power level between 15 kW and 25 kW and it is recommended that this be further investigated in a follow up study.

Please note that these results are only valid for assumptions listed in **Table A-4**. If the SEP stage is considerably heavier than assumed in **Table A-4** the sweet spot SEP power will shift towards higher power levels of 25 kW. Furthermore, going to a heavier SEP stage will resulted longer interplanetary flight times.

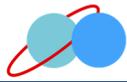

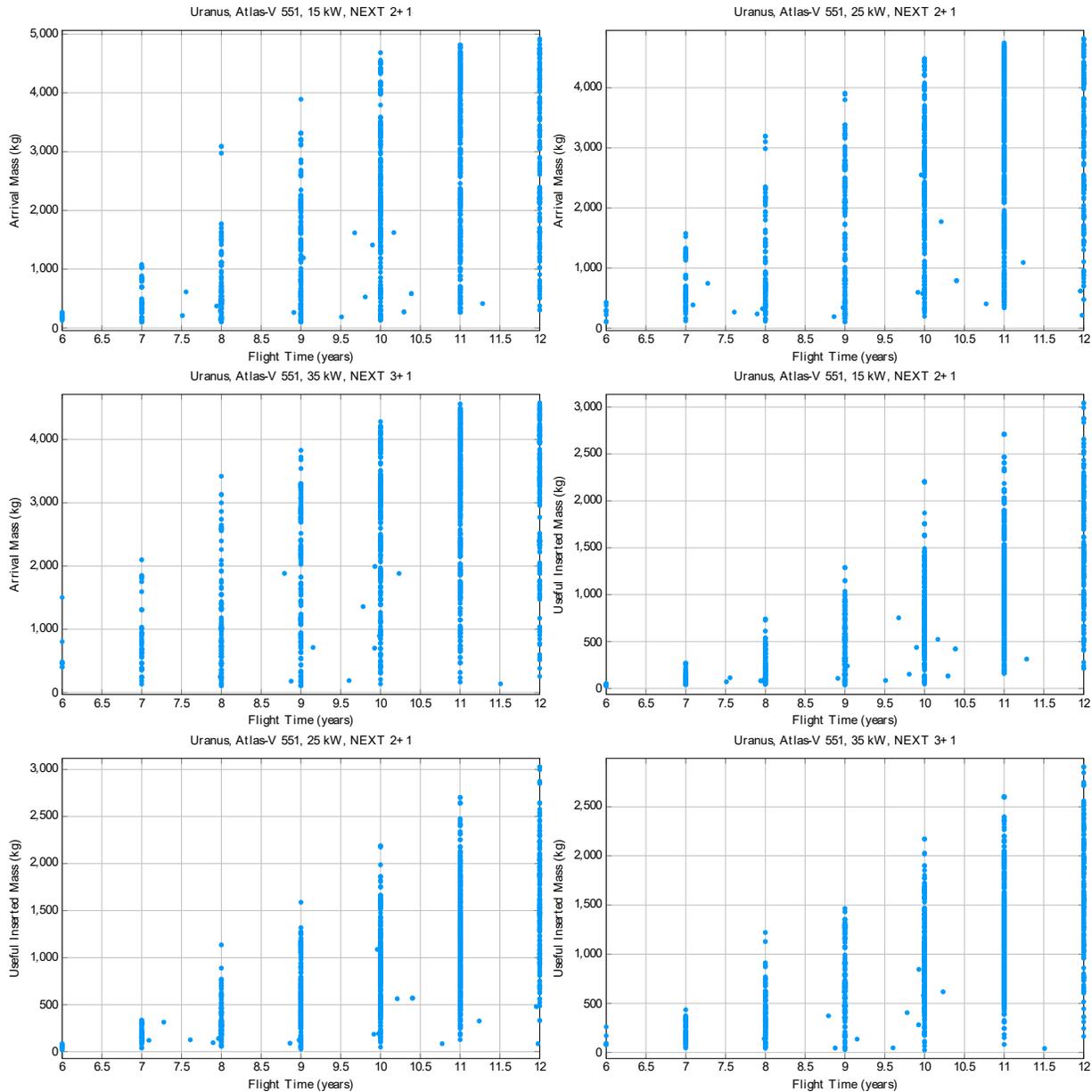

**Figure A-6.** SEP trajectories to Uranus: Atlas V (551) with 15 kW, 25 kW and 35 kW SEP power.

**Figure A-7** highlights the interplanetary trajectory tradespace for SEP missions to Uranus with up to 4 flybys, launching on the Delta-IV Heavy launch vehicle. The performance trends remain similar and with the exception that going to higher power levels results in slight improvement in useful inserted mass. A 15 kW SEP stage trajectory allows for >2,000 kg mass in to Uranus orbit in ~9 years from launch. For higher power levels, we see a steady reduction in flight time for same mass in Uranus orbit. A 15 kW SEP stage can deliver in excess of 3,000 kg in Uranus orbit in 10 years and going to a 35 kW SEP stages increases the inserted mass to almost 4,000 kg.

As noted before, going to a heavier SEP stage results in longer interplanetary flight times, which was the case during the first two Uranus Team-X studies.

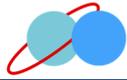

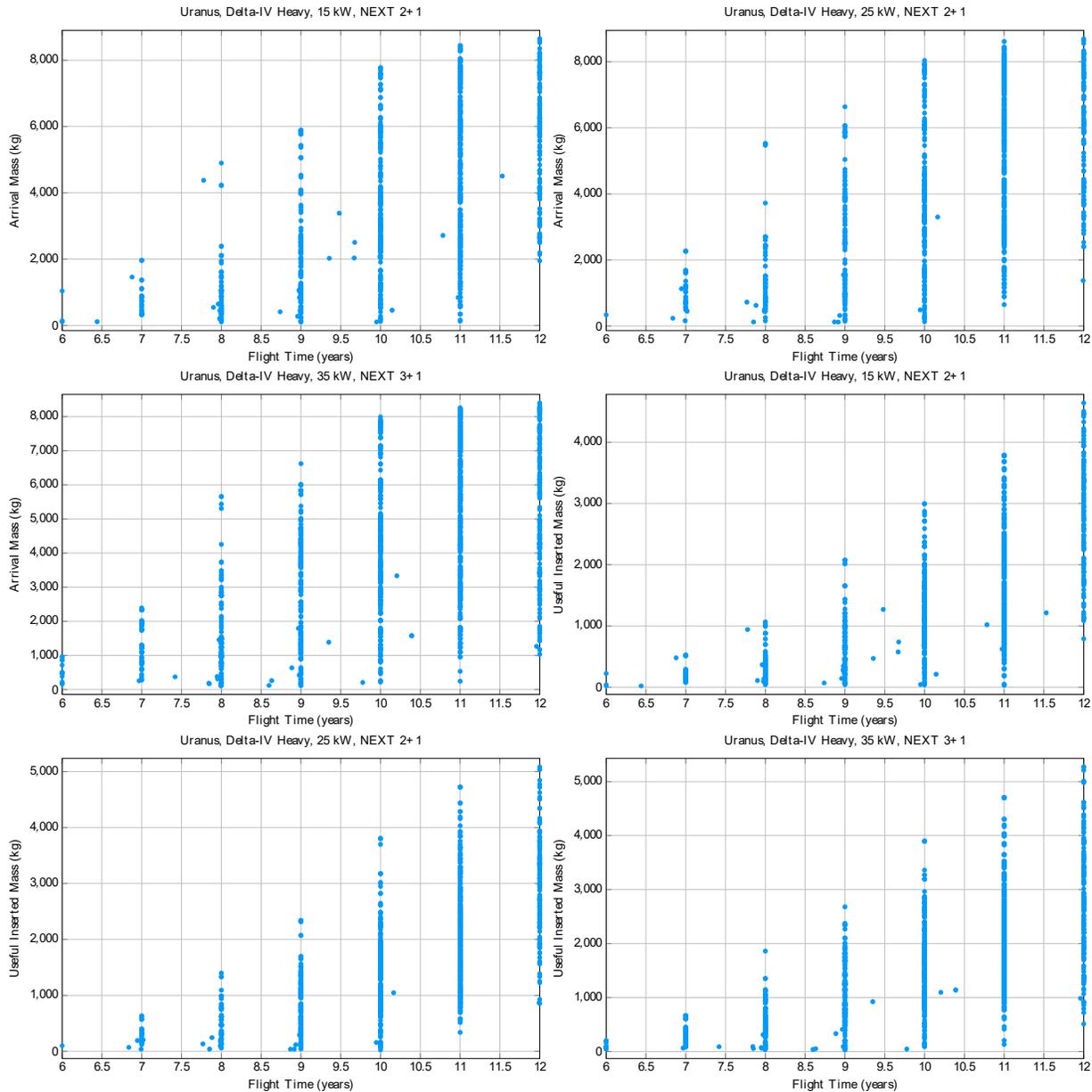

**Figure A-7.** SEP trajectories to Uranus: Delta-IV Heavy with 15 kW, 25 kW and 35 kW SEP power.

**Figure A-8** highlights the interplanetary trajectory tradespace for SEP missions to Uranus with up to 4 flybys, launching on the SLS-Block 1b launch vehicle. SLS provides a dramatic improvement in inserted mass or reduction on flight time at all power levels. The performance trends between the three SEP power levels is more pronounced. Going to higher SEP power levels results in improvement in both arrival mass and useful inserted mass. A 15 kW SEP stage trajectory allows for >2,000 kg mass in to Uranus orbit in ~7 years from launch. For a fixed 2,000 kg useful inserted mass we don't see significant reduction in flight time from going to high power levels. This is due to the fact that we are hitting the orbit insertion DV limit of 4.5 km/s. On the other hand, a 15 kW SEP stage can deliver in excess of 8,000 kg in Uranus orbit in 11 years and going to a 35 kW SEP stages increases the inserted mass to >9,000 kg for the same 11 years' cruise time.

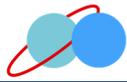

As noted before, going to a heavier SEP stage results in longer interplanetary flight times, which was the case during the first two Uranus Team-X studies.

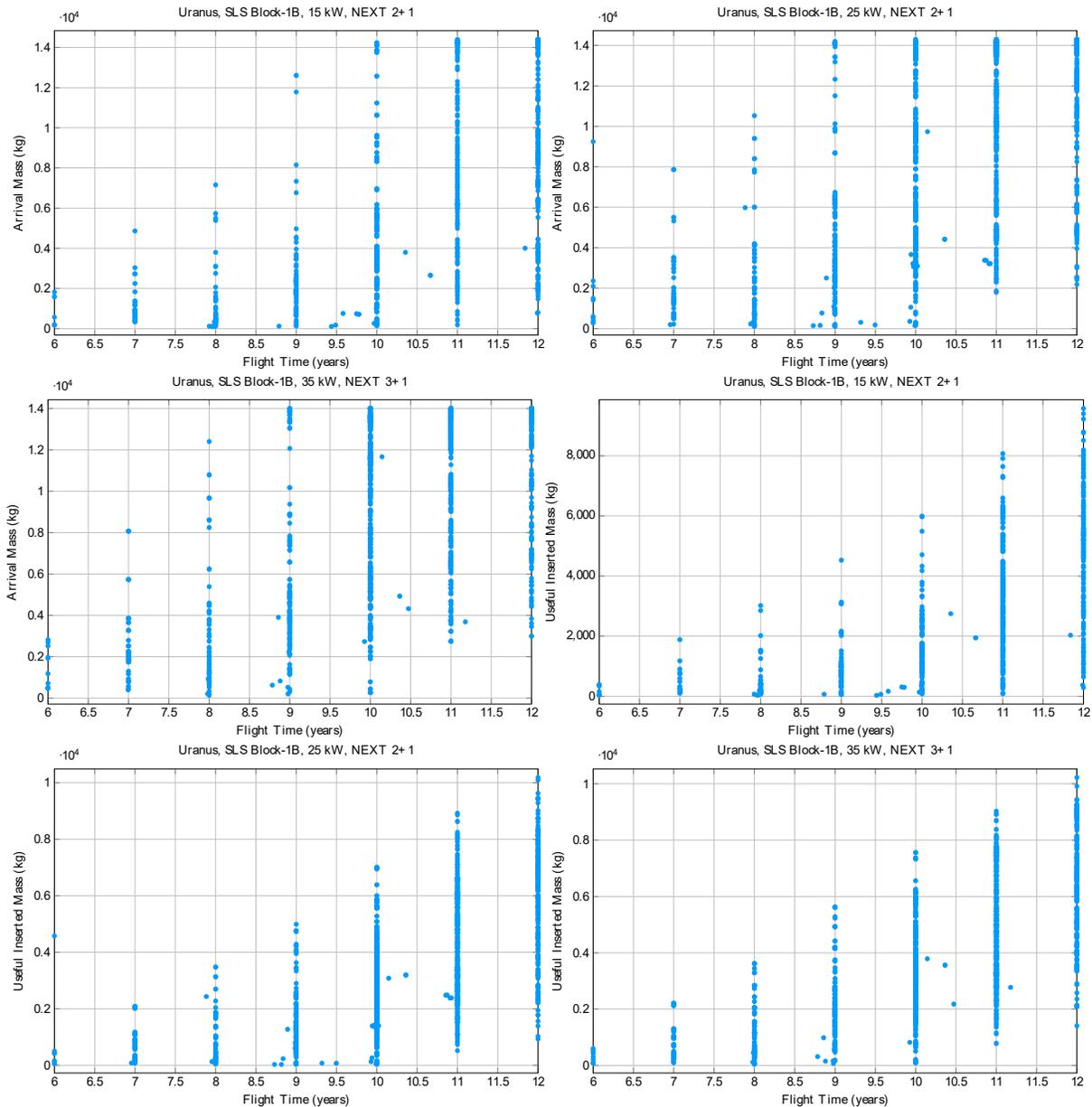

**Figure A-8.** SEP Trajectories to Uranus: SLS Block-1B with 15 kW, 25 kW and 35 kW SEP power.

**Figure A-9** lists all possible launch opportunities between 2025 and 2037. Only the 25 kW case is shown as the trend remains the same across different SEP power levels. The optimal launch period is between 2030 and 2034 but off-nominal dates are also performant. High performance launch opportunities perform similarly across the three launch vehicle options.

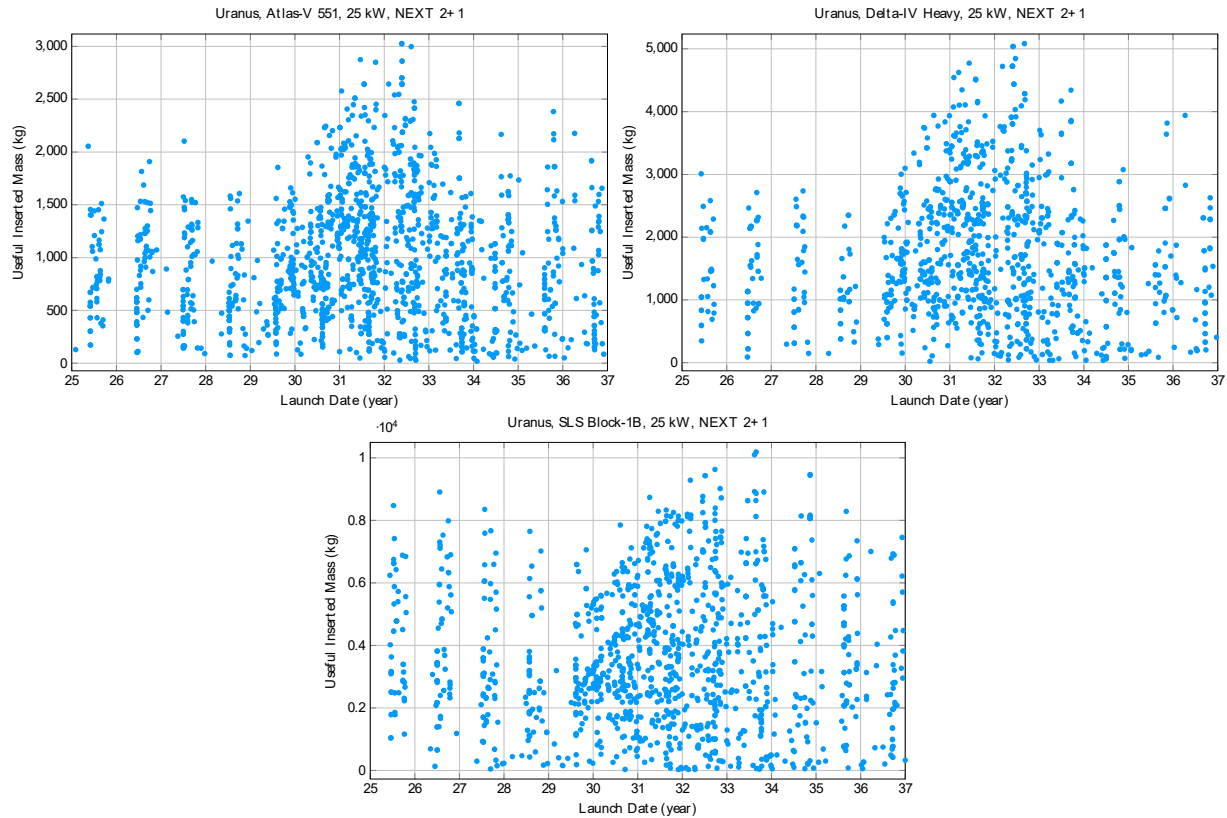

**Figure A-9.** SEP trajectory options to Uranus: launch opportunities for 25 kW SEP power case.

### A.4.4 SEP Trajectories for Neptune

**Figure A-10** highlights the interplanetary trajectory tradespace for SEP missions to Neptune with up to 4 flybys, launching on Atlas V (551) and with three different SEP power levels (15 kW, 25 kW and 35 kW). The three SEP power levels result in different SEP stage dry mass, total propellant throughput and number of engines, as shown in **Table A-4**. Using the 15 kW SEP stage, we get >4,000 kg arrival mass at Neptune in ~12 years from launch. The maximum arrival mass of ~4,700 kg is possible for a 13 years' trajectory. Going to high power levels gives some benefit in terms of arrival mass when launching on the Atlas V 551 as the trajectory performance is limited by launch C3 and number of SEP engines (maximum thrust). A similar trend is noted for the useful inserted mass metric. A 15 kW SEP stage trajectory allows for only ~1,200 kg mass in to Neptune orbit for a ~13-year interplanetary trajectory and going to higher power levels doesn't show significant benefit.

Please note these results are valid for assumptions listed in **Table A-4**. If the SEP stage is considerably heavier than those listed, the sweet spot SEP power will shift from 15 kW towards higher power levels of 25 kW. Furthermore, going to a heavier SEP stage will result in longer interplanetary flight times, as was observed during the Neptune mission Team X session.

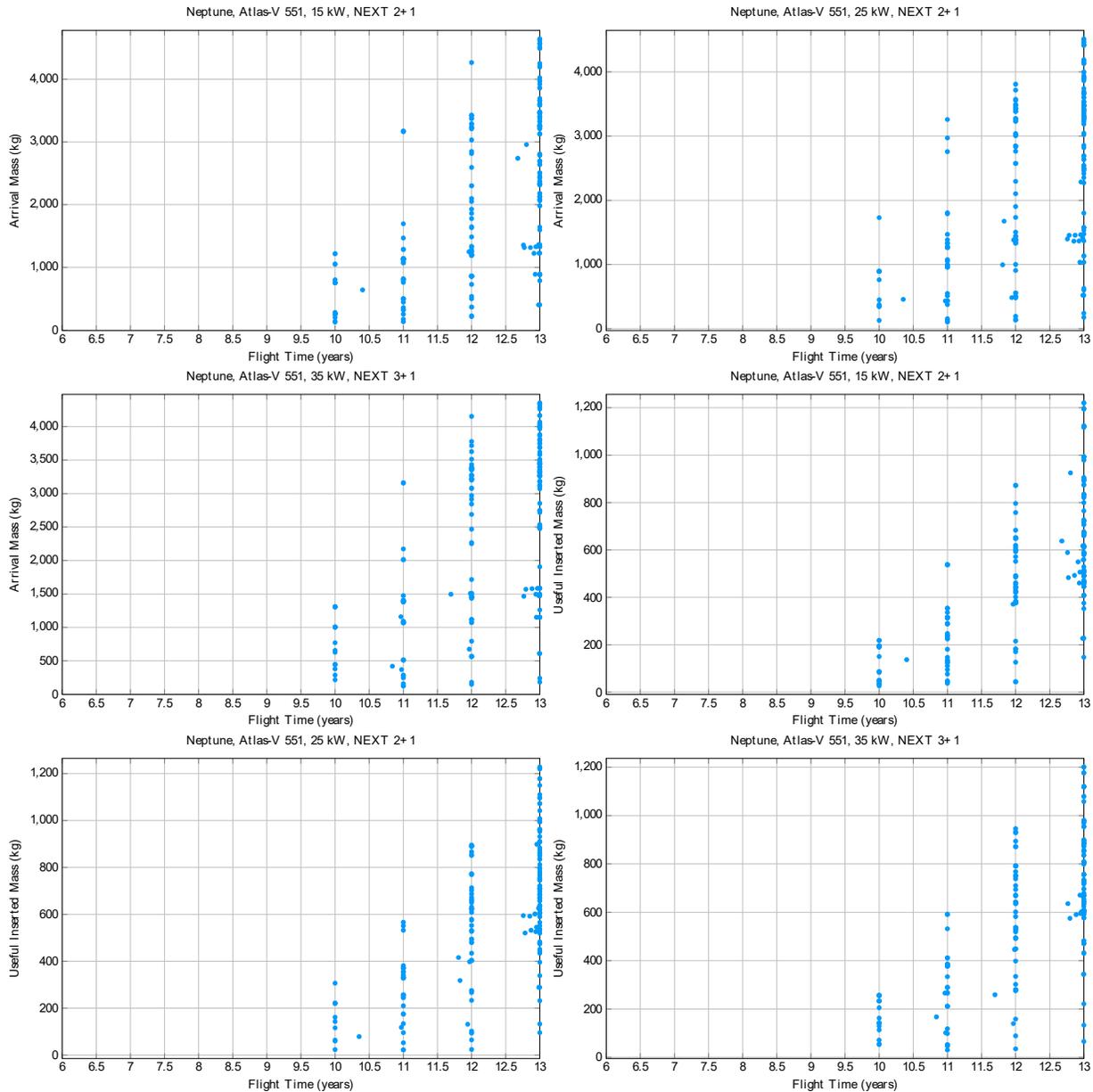

**Figure A-10.** SEP Trajectories to Neptune: Atlas V (551) with 15 kW, 25 kW and 35 kW SEP power.

**Figure A-11** highlights the interplanetary trajectory tradespace for SEP missions to Neptune with up to 4 flybys, launching on the Delta-IV Heavy launch vehicle. The performance trends remain similar and with the exception that going to higher power levels results in some improvement in arrival and useful inserted mass. A 15 kW SEP stage trajectory allows for >2,000 kg mass in Neptune orbit for a ~13 years interplanetary trajectory. For higher power levels, we see a small but steady increase in useful inserted mass for the same 13-year interplanetary flight time.

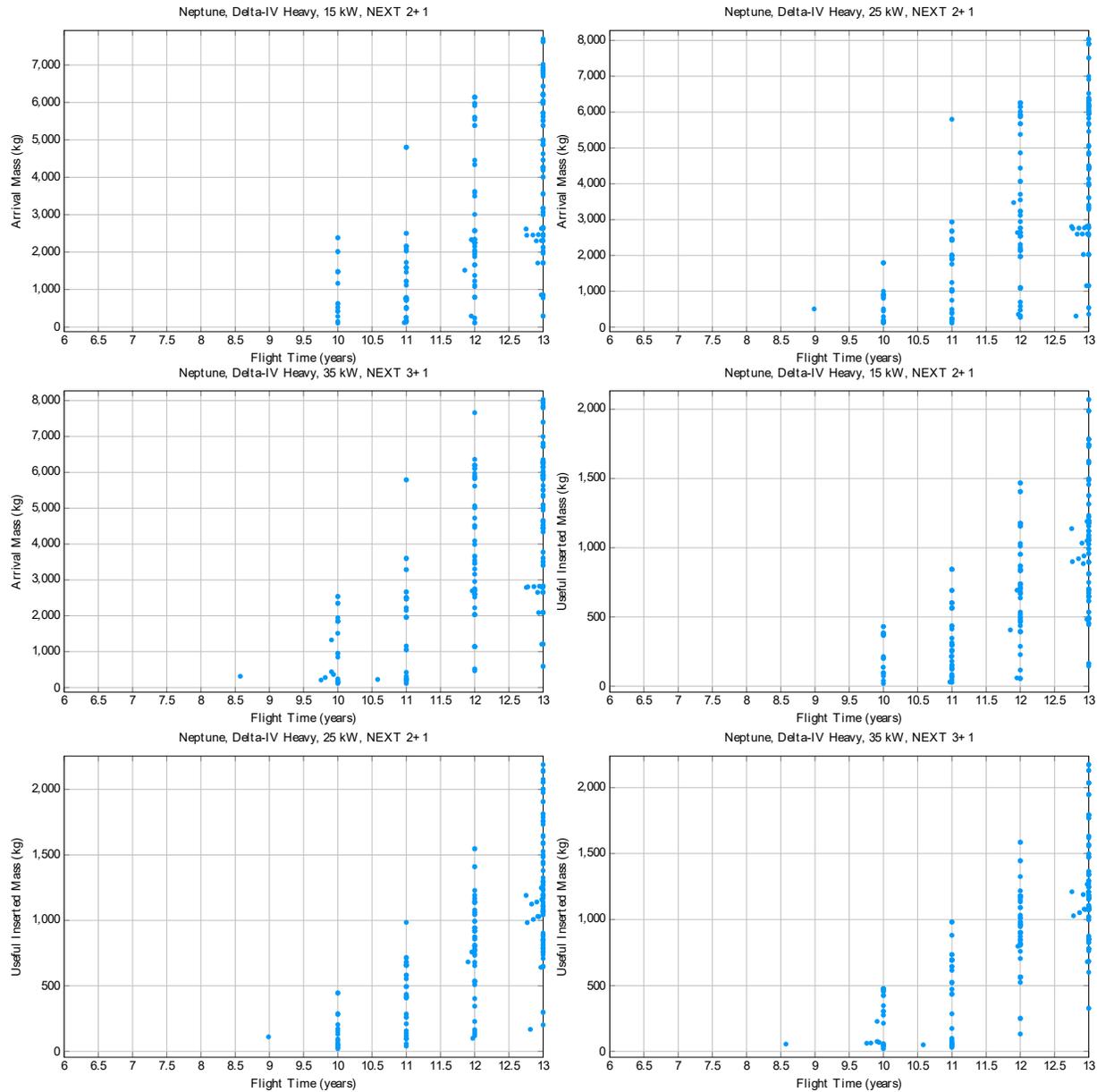

**Figure A-11.** SEP trajectories to Neptune: Delta-IV Heavy with 15 kW, 25 kW and 35 kW SEP power.

**Figure A-12** highlights the interplanetary trajectory tradespace for SEP missions to Neptune with up to 4 flybys, launching on the SLS-Block 1b launch vehicle. As for Uranus, SLS provides a dramatic improvement in inserted mass or reduction on flight time at all power level. Going to higher SEP power levels results in improvement in both arrival mass, useful inserted mass. A 15 kW SEP stage trajectory allows for >2,000 kg mass in to Neptune orbit in ~11.5 years from launch. For a fixed 2,000 kg, useful inserted mass, we see a significant reduction in flight time when we go to high SEP power levels. A 35 kW SEP trajectory, launched on an optimal date, requires only ~10 years to insert ~2,000 kg of mass in to Neptune orbit. This is due to the fact that SLS has better upper stage performance at lower C3s and the bigger SEP stages makes up for the loss in launch energy. One other hand a 15 kW SEP stage can deliver in excess

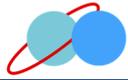

of 4,000 kg in Neptune orbit in 13 years and going to a 35 kW SEP stages increases the inserted mass to > 5000 kg for the same 13 years' cruise time.

As noted before, going to heavier SEP stage results in longer interplanetary flight times and larger SEP propellant usage.

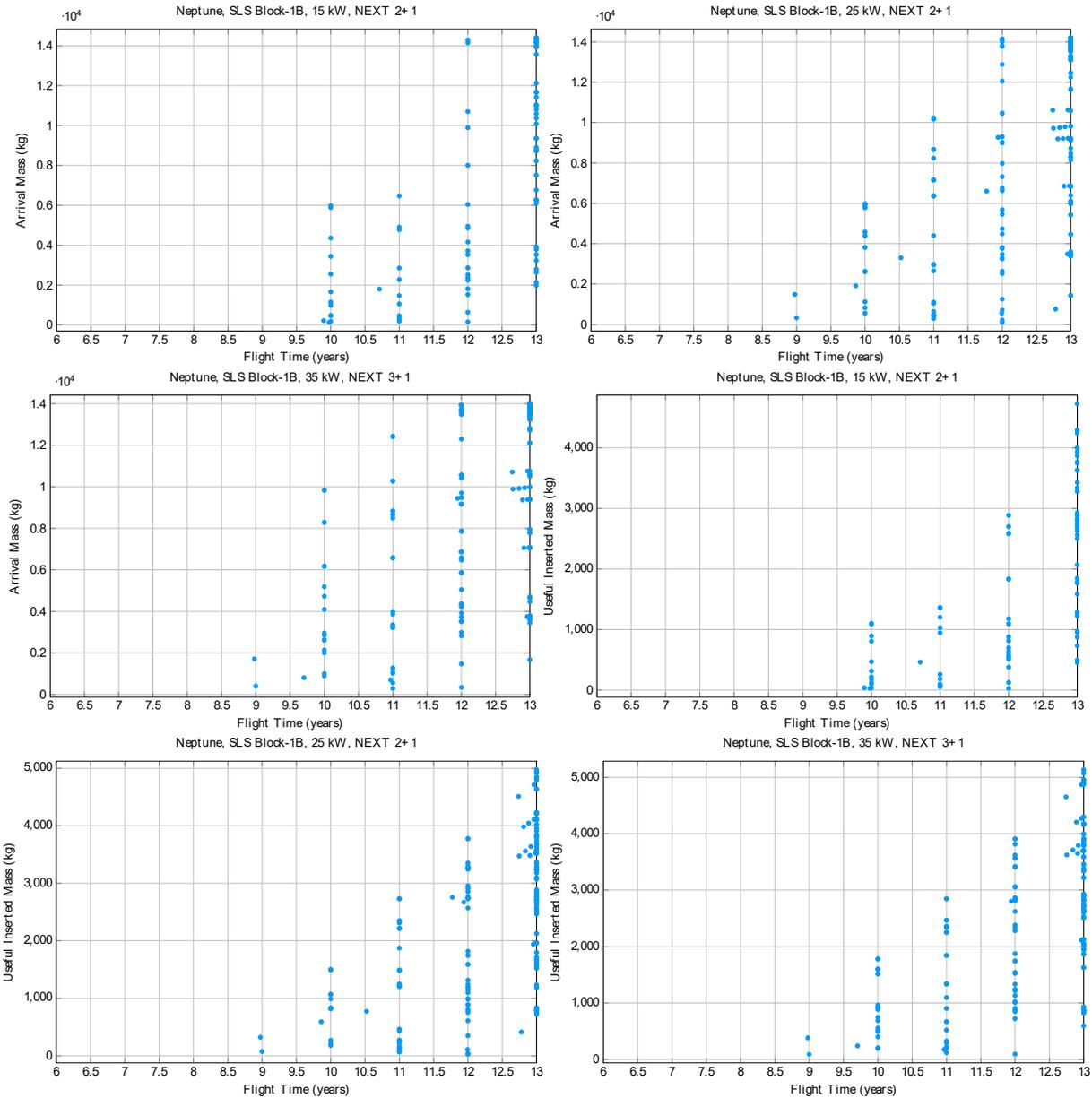

**Figure A-12.** SEP trajectories to Neptune: SLS Block-1B with 15 kW, 25 kW and 35 kW SEP power.

**Figure A-13** lists all possible launch opportunities to Neptune between 2025 and 2037. Only the 25 kW case is shown as the trend remains the same across different SEP power levels. The optimal launch period is between 2029 and 2031 but off nominal dates are also performant. The optimal launch period signifies the availability of Jupiter flyby. High performance launch opportunities perform similarly across the three launch vehicle options.

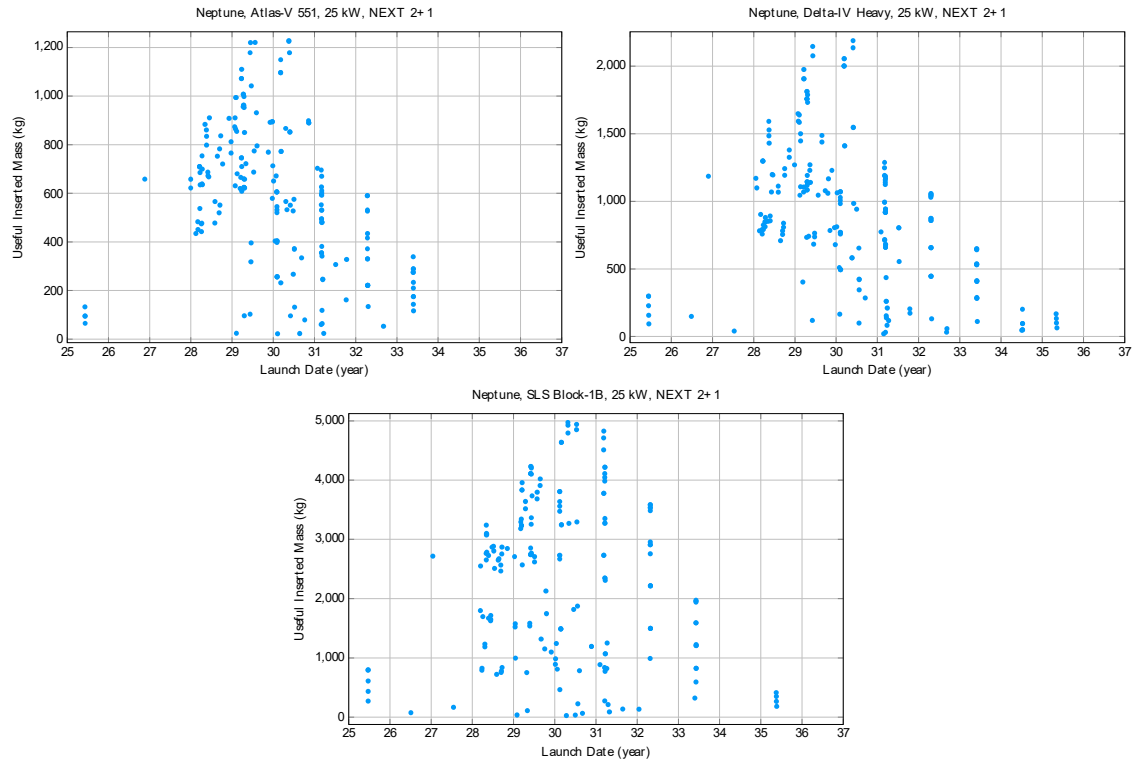

**Figure A-13.** SEP trajectory options to Neptune: launch opportunities for 25 kW SEP power case.

## A.4.5    Uranus and Neptune Orbit Insertion Considerations

Orbit insertion at Uranus and Neptune is challenging and is influenced by the following factors:

a. Arrival velocity: The interplanetary flight time limits results (see **Table A-2**) in high arrival velocities ($V_\infty$) at Uranus and Neptune. The arrival $V_\infty$ are generally higher for Neptune as it is further away from the Sun.

b. Gravity well: Uranus and Neptune have a smaller gravity well than Jupiter or Saturn, this reduces the "Oberth Effect" leading to higher orbit insertion $\Delta V$.

c. Ring avoidance: Spacecraft entering orbit around Uranus and Neptune need to avoid the rings during plane crossing. This is discussed in detail later in this section.

Coupling of these effects results in large orbit insertion $\Delta V$. **Figures A-14** and **A-15** show orbit insertion $\Delta V$ as a function of arrival $V_\infty$ at both planets, respectively. For consistency, an insertion orbit period of 200 days is assumed for these figures. The colored legend represents orbit insertion periapsis in units of planet radii. The Neptune gravity well is stronger than Uranus by ~18%, hence for the same arrival $V_\infty$ and periapsis radius, orbit insertion $\Delta V$ is smaller than that for Uranus.

The science team decided that to ensure a safe orbit insertion (avoiding rings) at either planet, the ring plane crossing should be either at ~1.05 body radii or at ≥1.25 body radii. To reduce orbit insertion $\Delta V$ the mission design team decided on an orbit insertion with a periapsis radius of 1.05 body radii at both planets. For the mission options considered in this study, orbit insertion $\Delta V$ at Uranus ranged from 1.7 km/s to 2.5 km/s and for Neptune it ranged from 2.3 km/s to 3.5 km/s.

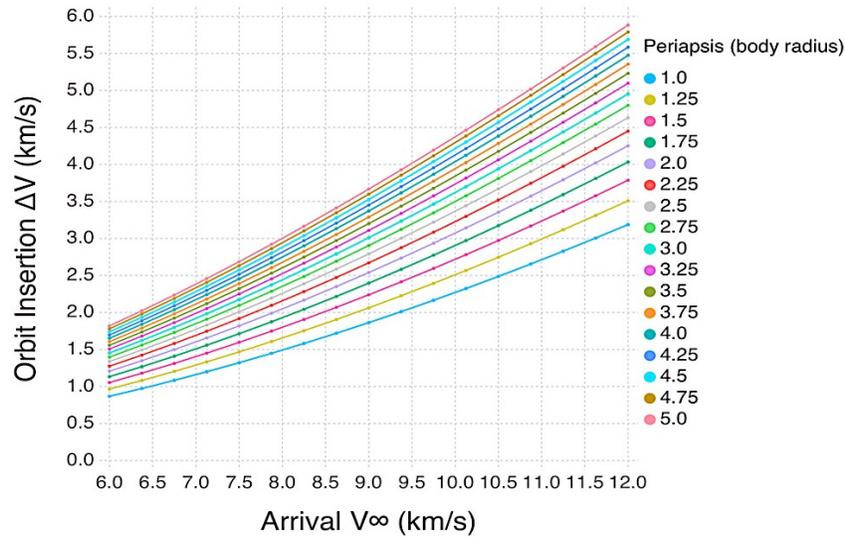

**Figure A-14.** Arrival $V_\infty$ vs. Uranus Orbit Insertion $\Delta V$.

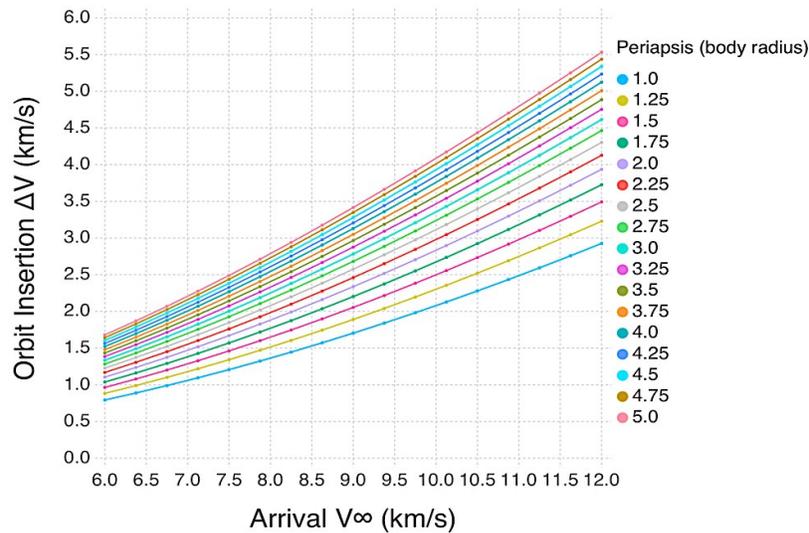

**Figure A-15.** Arrival $V_\infty$ vs. Neptune Orbit Insertion $\Delta V$.

In the remainder of this section, we provide some details of the ring hazard.

In general, an orbit insertion burn should be performed as close to the planet as possible to minimize the amount of fuel (the delta-v) required to be captured into orbit. Passing close to the planet also maximizes the science return from gravity and magnetic field measurements. There is a risk of colliding with ring particles, however, when passing through the ring-plane within the known rings of any giant planet. Several ring particle mitigation strategies were considered, including:

- Do not pass through an area suspected of having high particle densities and allocate additional delta-v for orbit insertion. Using Uranus as an example, performing UOI at 2.0 planetary radii out---in a gap between the epsilon and nu rings---requires about 600 m/s more delta-v than doing orbit insertion at 1.1 radii (2.3 km/s vs. 1.7 km/s). For comparison, the delta-v allocation for a two-year tour of the 5 major satellites is about 60 m/s. Given the large delta-v penalty, we prefer not to baseline this option.

- Before the orbiter arrives, fly a "pathfinder" spacecraft along the desired trajectory to directly measure the particle density. If the region is found to be safe, the orbiter flies the same trajectory. If the region is considered too risky, the orbiter targets a known safe distance (incurring the delta-v penalty discussed above). The pathfinder spacecraft can be launched with the orbiter, and deep space maneuvers used to ensure the pathfinder arrives at the ice giant well in advance of the orbiter. This option is also not our preferred solution, as it requires the spacecraft to be designed for the worst-case delta-v and (if launched together) the orbiter must accommodate the pathfinder spacecraft. Both of these reduce payload mass.

- Attempt observations over the next few years of the ice-giant's upper atmospheres and their rings, along with modeling, to constrain the ring particle risk. (The atmosphere is important in this regard because the hot, extended thermospheres of Uranus and Neptune affect ring-particle orbits.) This could be done with Earth-based measurements (e.g. a stellar occultation campaign or JWST imaging). Close-in measurements at Jupiter and Saturn by the Juno and Cassini spacecraft, respectively, may also teach us something about the ring particle distributions at those planets that help us assess the risk at the ice giants. If these observations do not adequately constrain the ring particle risk, avoid all suspect regions and accept the delta-v penalty as discussed above. This approach should be pursued, but does not help us constrain the ring hazard for the current study.

- Fly through the ring-plane at an altitude where atmospheric drag is high enough to drastically reduce the number of particles, but low enough to present no risk to the spacecraft. This is the approach used for the current studies.

To target our ring-plane crossing periapse altitude, we need to find an atmospheric region dense enough to rapidly remove ring particles, but not so dense as to adversely affect spacecraft three-axis stabilization. It must be emphasized that the calculations we have made are relatively crude. They indicate that it is plausible that there is a safe-zone to fly through, but more detailed calculations are needed to understand the exact size and location of this zone, and to study how well it can be constrained given uncertainties on our knowledge of the upper atmospheres of Uranus and Neptune. (Additional Earth-based observations of the atmospheres may help reduce this uncertainty.)

To determine the altitude at which the spacecraft might become aerodynamically unstable, we relied on calculations performed by the Cassini team for their flybys through Titan's atmosphere. They found that at a Titan density of ~2.0E-9 $kg/m^3$, Cassini would start to tumble (Sarani 2009). Since atmospheric torques on the spacecraft are proportional to the velocity-squared, and our Uranus orbiter has a periapse velocity 3.7 times larger than a typical Titan flyby velocity (22 km/s versus 6 km/s), we estimate our orbiter would start to tumble when flying through an atmospheric density of about 1.5E-10 $kg/m^3$. We note that the atmospheric torque also depends linearly on the cross sectional area of the spacecraft and the effective lever-arm between the center of mass and the center of drag. Our Uranus orbiter is about half the size of Cassini (our largest cross section is about 11 $m^2$, while Cassini's is near 20 $m^2$) so one would expect our orbiter to be more stable than Cassini, but without a detailed calculation we make the conservative choice to ignore this possible correction factor. Extrapolating the recent Uranus atmosphere tabulated in Orton et al. (2014) to lower densities assuming an isothermal, ideal-gas atmosphere, we estimate our spacecraft would become unstable about 2000 km above the 1-bar pressure level. This indicates our spacecraft periapse should be at least 1.08 radii above the equator.

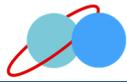

To determine the altitude at which atmospheric drag effectively removes ring particles, we use an expression for the instantaneous decay rate of orbiting particles from Broadfoot et al. (1986). Using the extrapolated Orton et al. atmosphere, we find that a 100 micron particle is unable to complete one orbit within 2500 km above the 1-bar level. Since particles smaller than 100 micron decay at even higher altitudes, and there are thought to be very few particles larger than 100 microns outside the main rings, we use 2500 km (1.1 radii) as the maximum safe altitude to avoid ring particles when crossing the ring-plane. We note that the true lifetime of 100 micron particles at 1.1 radii is actually much less than one orbit, since we neglected the fact that as a particle's orbit starts to decay, the decay rate increases exponentially due to increasing atmospheric density.

These calculations suggest there is a window about 500 km wide, between 1.08 and 1.1 planetary radii, through which our Uranus orbiter could safely fly during a periapse ring-plane crossing. A similar window is expected to exist at Neptune. While Neptune's upper atmosphere is warmer than Uranus', reducing atmospheric density at a given altitude and increasing the pressure scale height, the atmospheric drag that causes a spacecraft to tumble is the same drag that causes ring particle orbits to decay. Thus both the inner and outer radii calculated above should move inward at Neptune, maintaining the safe window. We emphasize again, however, that this promising calculation is only a rough estimate. More detailed calculations are needed to better constrain the location and size of this window, and to study how uncertainties in our knowledge of upper atmospheric properties affects this window. Only then can a final assessment be made of the optimal way to mitigate the ring hazard during orbit insertion.

At the time our study trajectories were finalized, we only had a preliminary estimate of the location of the safe-zone. For this reason, all Uranus and Neptune orbiter trajectories used for the point designs have a periapse altitude of 1.05 radii and cross the ring-plane near 1.1 radii during the orbit insertion burn. We estimate increasing the periapse altitude from 1.05 to 1.1 radii would increase the required orbit insertion delta-v by an insignificant amount from what is found in the mission point designs presented later. This and other changes that would result (e.g. in probe relay) are relatively minor, and will not alter our conclusions or recommendations.

The above discussion focused on the desire to fly close to minimize delta-v requirements for orbit insertion. We also wanted to comment on the science enabled by the low periapse. Doppler tracking of the spacecraft at low altitudes allows the highest-resolution gravity data to be collected, which is most useful for understanding zonal winds. (Traditionally such measurements would also be the primary method of probing interior structure, but as discussed in Section 3.3.1, a seismology instrument such as a Doppler Imager is our preferred way to get that information.) Measurements of the magnetic field from low altitudes will provide valuable constraints on the field generation region. While having many low-altitude periapses would maximize the gravity and magnetic field information, our trajectories do not stay low after orbit insertion. Instead, our missions keep periapse outside the known rings to optimize satellite encounters. Our team felt this approach is the best to achieve our science objectives. Proximal orbits, such as Cassini will be flying, could certainly be used in a scientifically rewarding end-of-mission scenario.

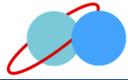

## A.5   Probe Considerations

### A.5.1   Summary

Some of the mission concept options considered in **Table A-2** had an atmospheric probe (for Uranus or Neptune). The probe trajectory was designed to have a coast of 60 days before orbit insertion at either of the planets. The probe entry was in the same direction of motion as the spacecraft trajectory as it flies by or enters into orbit around the target planet. The "standard" approach is similar to that adopted by the Galileo probe and the Huygens probe. This standard hyperbolic probe entry at either of the planets must trade the following design parameters:

a.   Orbit insertion ΔV magnitude

b.   Probe g-load tolerance

c.   Probe-orbiter relay telecommunications requirements (aspect angle and range); related to the probe Entry Flight Path Angle (EFPA)

Orbit insertion ΔV is sensitive to the orbit insertion periapsis altitude (see Section 4.1.5). Higher orbiter periapsis provides better probe telecom relay line-of-sight and longer persistence (lower angular rate relative to probe), but higher orbit insertion ΔV. Shallow EFPA reduces probe g-load, but presents challenging relay geometry and increases cumulative heat load. The entry flight angle for a Uranus probe and Neptune probe were carefully selected to balance probe-orbiter relay geometry (see individual Team X mission options for more details on the relay geometry), probe heat/entry g loads, time between probe entry, and orbit insertion ΔV. Increasing the separation increases the range between orbiter and probe while decreasing this separation puts pressure on mission operations for performing two mission critical events (probe entry and orbit insertion) within a short time. During the Team X design studies, it was found that a 2-hour gap between probe atmospheric entry (~1,000 km altitude) and orbit insertion would suffice for both Uranus and Neptune. An alternative probe entry trajectory, which may alleviate some of the above-mentioned issues is described in Section A.5.5.

The atmospheric entry analysis was conducted for both Ice Giants, for given entry vectors. Several entry trajectories were analyzed for both planets and suitable trajectories were selected for these studies based on current entry systems technology. Analysis was conducted for a mid-density TPS material, HEEET, as well as a high-density material, full density carbon phenolic. These two materials span the range of masses for heatshield design purposes. Current HEEET development is targeted towards Saturn and Venus missions which have lower aerothermal environments, compared to either Uranus or Neptune, however design trajectories were identified that place requirements within the tested capabilities of the HEEET material.

For Uranus, EFPAs of -30° or lower allow for stagnation pressures and heat fluxes that were within the testing range for TPS materials. HEEET was sized for forebody TPS and PICA was sized for aftbody. Carbon phenolic was also sized as a reference for forebody TPS. Mass estimates were provided for all the TPS materials. For Neptune entry, three entry trajectories were considered. The first one at -34° EFPA was not feasible from an entry systems technology perspective due to the resulting extremely high pressures and heat fluxes. The second and third trajectories at -20° and -16° of EFPA were considered for further analysis. While the environments were significantly higher for -20° EFPA, the TPS mass was lower due to the lower heat load. The environments were more reasonable for -16⁰ EFPA, however heat loads for this entry were very high and the TPS mass increased by more than 20%. A follow-on analysis is recommended at an intermediate trajectory at -18° EFPA entry. The study team selected -20 EFPA case for the Neptune Team-X design study. Note that as explained in Section 4.1.6,

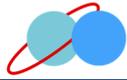

Neptune orbit insertion and the probe entry were retrograde relative to Neptune's rotation direction, this resulted in a minor increase in probe g and peak heat load.

## A.5.2    Entry Probe Analysis – Assumptions and Scope

For both of the probe missions studied (Uranus and Neptune), the entry vehicle was specified as a 1.2-m diameter 45° sphere cone at 325 kg, similar to the Galileo probe. The delivered probe mass (payload) was assigned to be 200 kg, with the remaining 125 kg representing the aeroshell (TPS, structures, etc.).

Atmospheric models for Uranus and Neptune have large uncertainties due to the lack of data. The Voyager-2 single fly-by radio and UV occultation measurements were used for both planetary models. The Uranus atmospheric model was based upon different Voyager-2 publications that were spliced together into a single model. The details for this model are discussed in a 2014 Uranus atmospheric entry study conducted by NASA Ames (Agrawal et al. 2014). The Neptune atmospheric model used NeptuneGRAM as a starting point. The Uranus atmospheric model is particularly challenging due to its motion (the planet rotates sideways and has an obliquity of 98°). Conditions in Uranus' upper atmosphere can vary significantly depending upon the Uranus year. Both Uranus and Neptune have extremely high temperatures in their exospheres as measured by Voyager-2, e.g., 800K for Uranus at 7000 km altitude and 737K for Neptune at 4000 km. The mechanism behind this high temperature is not understood, however it affects the entry interface altitude for atmospheric entry analysis. It is also important to note that atmospheric pressure and temperature distribution affects the entry parameters like peak stagnation pressure, peak heating, peak deceleration load etc. Contrary to the upper atmosphere, the temperatures in the lower atmosphere near the tropo-pause can be extremely cold, e.g., 51K for Neptune at 40 km. The Chemical Equilibrium with Applications (CEA) code (see Section 2.3.3) has an operating temperature range of 200K to 20,000K. The version of CEA used by the TRAJ entry simulation for thermodynamic modeling was modified in Neptune GRAM to account for the extremely low temperature. The low temperature corrections were added in CEA for the ratio-of-specific heats, the isentropic exponent ($\gamma$), as shown in **Figure A-16**. The original CEA version shown by the blue dotted line is valid only above 200K. The modified values based on data from Los Alamos National Laboratory (LANL) are shown with the red line. NeptuneGRAM assumes a constant value of 1.45 for $\gamma$ (shown in the green line), which is correct only for a fixed temperature and gas composition.

The Uranus and Neptune atmospheric models included methane, however methane phase change was not modeled. The high temperature of the Uranus and Neptune exosphere defines a very high entry interface (EI) altitude for atmospheric entry simulation. Unlike for Saturn and Jupiter, which have EIs of 450 km, the EI for Uranus and Neptune was selected to be 1,000 km in order to properly account for the heating that occurs at high altitudes. At 450 km, the stagnation heating for the vehicle at Uranus is over 100 W/cm². Consequently, the EI needed to be considered at a much higher altitude in order to capture the entire heat pulse along the entry trajectory.

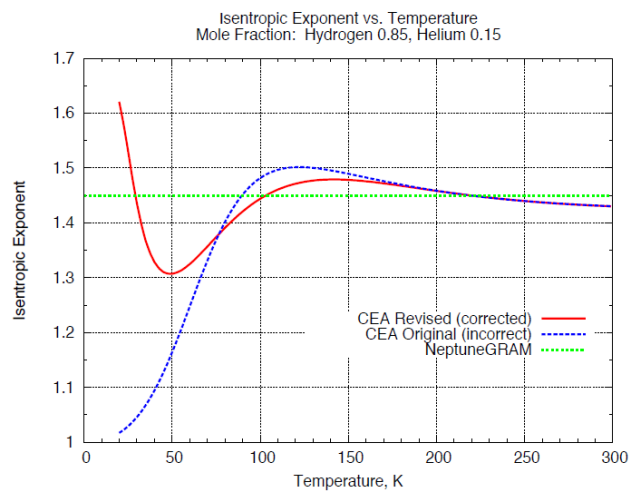

**Figure A-16.** Corrections made to Isentropic Exponent ($\gamma$) in CEA based on data from Los Alamos National Laboratory.

Neptune is more extreme for atmospheric entry than Uranus due to the significantly higher velocities ($V_\infty$). Present point design for Uranus has an entry velocity of 8.4 km/s as opposed to 11.4 km/sec for Neptune. The Jupiter Atmospheric Entry (JAE) aerodynamic and heating models (Milos and Chen 2013) calibrated against the Galileo probe were used to compute the heating environments at the stagnation point of the vehicle. The JAE aerodynamic model is a Newtonian model with corrections added by Michael Tauber at NASA Ames (Tauber et al. 1999). The backshell heating environments were assumed to be 5% of stagnation point heating, which is a conservative estimate based on prior analysis of 45° sphere cone geometries (Wright et al. 2009). Future studies need to be conducted using detailed CFD analysis to obtain a higher fidelity of the aerothermal afterbody environments. The CFD analyses performed during 2014 Uranus atmospheric entry studies show lower heat-flux values at the stagnation point compared to TRAJ predictions (Tauber et al. 1999), which gives confidence that the results presented in this study are conservative estimates.

TPS sizing was performed assuming peak heating at the stagnation point. A detailed CFD analysis will be needed to generate a spatially and temporally varying heat-flux profile and for the exact location of peak heating, especially if there is turbulence. When performing HEEET sizing at a given location on the heat shield, recession layer thickness was selected to be equal to the FIAT-predicted total recession for the specified trajectory. This assumption will ensure that the insulation layer is never exposed to external high-temperature flow. Insulation layer thickness was sized such that the temperature below the HEEET material (known as bond-line temperature) remains below a specified design limit until a time specified by mission designers. Since the amount of TPS recession (and as a result recession layer thickness) depends on the thickness of insulation layer, sizing of the two layers must be done iteratively. Currently, the HEEET project advises 50% margined thickness for recession layer and 10% margined thickness for insulation layer. These margins account for the material uncertainty. The aerothermal environment uncertainties are not included in these margins. For future studies, it is recommended to perform CFD analyses and include margins for aerothermal environment uncertainties. Based on limited recession data from arcjet testing at IHF 3-inch and 6-inch nozzles (Milos et al. 2017), the recession margin is believed to be conservative. Furthermore, the 10% additional thickness used here for the insulation layer results in a bond-line temperature margin higher than what was adopted in MSL and Orion sizing margin policies. For Carbon-Phenolic sizing, a thickness margin of 50% was used.

The aeroshell structure assumes 0.762 mm of HT-424 for adhesive and 0.762 mm of aluminum face sheet for substructure based on the Galileo probe design (Milos et al. 1999). The bond-line temperature limit used in sizing was 260°C. This is the same limit that is being used for HT-424 adhesive with aluminum substructure for Orion. For the CP analysis, two different types of substructure were investigated to determine the effect of the structure on the TPS sizing. One case had a structure of 0.762 mm aluminum face-sheet (same as for the HEEET cases), and the other had 3.175 mm thick aluminum (Pioneer Venus probe design). The TPS initial temperature is assumed to be -10°C. The total forebody heat-shield mass assumes a TPS surface area of 1.5862 m$^2$. For backshell TPS material, PICA was selected to ensure due to the estimated peak heat flux on the afterbody. For backshell geometry, surface area and mass estimates, the Galileo probe was used as a reference. The surface area for sizing of the backshell was calculated as 1.3075 m$^2$. For aftbody substructure 3.175 mm thick aluminum was considered. For PICA sizing, a thickness margin of 50% was used.

### A.5.3 EDL for Uranus – Analysis and Results

Two cases were considered for a Uranus probe atmospheric entry based on orbital mechanics of the mission design. The first case ("Design #1") had a steep entry flight path angle (EFPA) of -35°. The entry parameters and resulting heating and pressure at the stagnation point are shown in **Table A-6**. With inertial velocity of 23.1 km/s this entry would produce 217g of deceleration and 12.0 bar of pressure at the stagnation point. Both the pressure and deceleration values are well above the qualification limits for sensitive science instruments, and represent challenges to ground testing TPS materials at these aerothermal environments. Therefore, a shallower entry at EFPA at -30° ("Design #2") was considered for the second case, which allows for a reduced G-load on sensitive equipment in the payload, and also corresponds to a lower aerothermal environment. The results are summarized in column 2 of **Table A-6**, with both EFPAs highlighted in blue. This shallower entry led to 165 g of deceleration load and reduced pressure and stagnation point heating. The stagnation pressure was calculated to be 9.0 bar with a peak stagnation point heat flux of 2,500 W/cm². This environment is well within the tested range of HEEET, the selected forebody TPS material. The heat load of 41 kJ/cm² also is within the tested capability of HEEET. **Figure A-17** shows the stagnation point heat flux distribution along the entry trajectory for Design #2. The peak heat flux occurs in the lower atmosphere at 140 km. The stagnation pressure profile along the trajectory is shown in **Figure A-18**. The peak stagnation pressure occurs at 100 km altitude (slightly lower than the peak heat flux altitude). Both of these occurrences are in the altitude range where atmospheric uncertainties are highest; it is recommended that a sensitivity analysis to the atmospheric densities be conducted in future studies to determine the potential changes in peak pressure, heat flux, and deceleration and heat load values.

**Table A-6.** Entry parameters and environment for Uranus.

| Entry Parameters | Point Design 1 | Point Design 2 |
|---|---|---|
| Hyperbolic excess velocity (km/s) | 9.91 | 8.41 |
| Entry interface altitude (km) | 1015 | 1000 |
| Radial distance (km) | 26559 | 26553 |
| Inertial velocity (km/s) | 23.10 | 22.52 |
| Entry Flight Path Angle, gamma (deg) | -35 | -30 |
| Inertial heading angle (deg) | -5.82 | -20.02 |
| Latitude (deg) | -9.22 | -5.63 |
| Max deceleration (g loads) | 216.65 | 164.75 |
| Stg pressure (bar) | 12 | 9 |
| Peak convective heat flux (W/cm²) | 3456 | 2498 |
| Peak radiative heat flux (W/cm2) | N/A | N/A |
| Total heat load (J/cm²) | 43572 | 41114 |

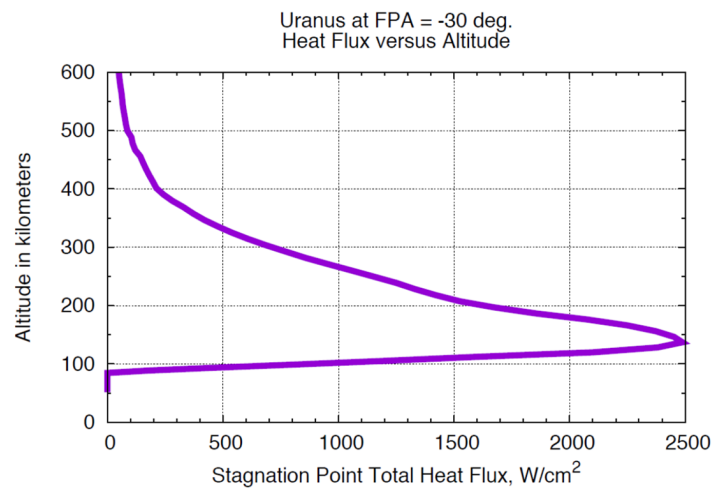

**Figure A-17.** Stagnation point heat-flux distribution for Uranus entry trajectory at -30⁰ EFPA (Design #2).

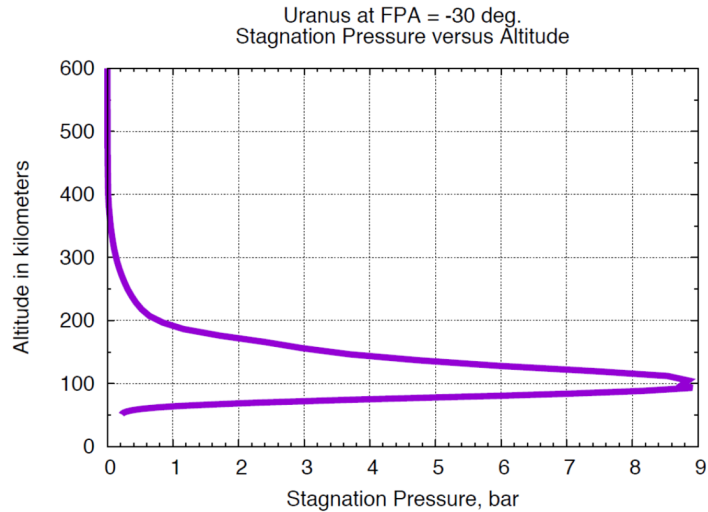

**Figure A-18.** Stagnation point pressure distribution for Uranus entry trajectory at -30⁰ EFPA

The HEEET TPS material sizing for Uranus was performed for Design #2 with EFPA of -30°. The results from TPS sizing are summarized in **Table A-7.** The unmarged mass was calculated to be 24.53 kg. With HEEET material margins policy (not including the aerothermal environment uncertainty) the total forebody TPS mass was calculated as 28.67 kg. This constitutes about 10% of the total vehicle mass. For Design # 2, sizing for Carbon Phenolic (CP) as a reference material is also presented. These calculations also assumed a bond-line temperature of 260°C and 0.762 mm thick aluminum face-sheet, as well as a lower bond-line temperature of 250°C (a value adopted for MSL) and a 3.175 mm thick aluminum substructure that was used for Pioneer Venus. All the computations are summarized in **Table A-7.** For conditions equivalent to HEEET the unmarged CP thickness is 1.75 cm and corresponding mass is 39.8 kg (~40 kg). With 50% margin the total CP TPS mass is 59.75 kg, which is more than a factor of 2 higher than for HEEET and is 20% of the total vehicle mass. The analysis shows that thickness of aluminum substructure does play a role and for a thicker sheet at the same bondline temperature corresponds to 1.6 cm of CP, or 36.5 kg unmarged mass. Lowering the bondline temperature by 10 °C to 250 °C does not significantly impact TPS thickness. The mass increases from 39.8 kg to 40.2 kg.

**Table A-7.** Aeroshell thickness and mass estimates for Uranus Design #2.

| TPS Material | HEET | Carbon-Phenolic | Carbon-Phenolic | Carbon-Phenolic | Carbon-Phenolic | PICA |
|---|---|---|---|---|---|---|
| **Aeroshell** | **Fore-Body** | **Fore-Body** | **Fore-Body** | **Fore-Body** | **Fore-Body** | **Aft-Body** |
| Adhesive Type | HT-424 | RTV-560 | RTV-560 | RTV-560 | RTV-560 | RTV-560 |
| Adhesive Thickness (mm) | 0.76 | 0.38 | 0.38 | 0.38 | 0.38 | 0.38 |
| Adhesive Mass (kg) | 2.42 | 0.84 | 0.84 | 0.84 | 0.84 | 0.69 |
| Substructure Type | Aluminum Face sheet | Aluminum Face sheet | Aluminum Face sheet | Aluminum Structure | Aluminum Structure | Aluminum Structure |
| Substr Thickness *Thin* (mm) | 0.762 | 0.762 | 0.762 | 3.175 | 3.175 | 3.175 |
| Substr Mass (kg) | | | | 14 | 14 | 11.5 |
| Bondline Temperature (°C) | 260 | 260 | 250 | 260 | 250 | 260 |
| Unmargined TPS Thickness (cm) | RL - .244 IL – 1.5621 | 1.75 | 1.77 | 1.6 | 1.62 | 0.68 |
| Unmargined TPS Mass (kg) | 24.53 | 39.8 | 40.23 | 36.48 | 36.88 | 2.372 |
| Margined TPS Thickness (cm) | RL - .366 IL – 1.72 | 2.625 | 2.655 | 2.4 | 2.43 | 1.02 |
| Margined TPS Mass (kg) | 28.67 | 59.7 | 60.345 | 54.72 | 55.32 | 3.558 |

The PICA thickness and mass for aftbody TPS were calculated assuming a peak aftbody heat flux of 5% of the forebody stagnation point heating. The thickness was calculated as 0.68 cm. The corresponding PICA mass for aftbody TPS was calculated to be 2.37 kg. Using a material margin of 50%, the PICA thickness and mass is 1.0 cm and 3.55 kg respectively. The mass values for adhesives and substructures were also estimated and summarized in **Table A-7**.

### A.5.4    EDL for Neptune – Analysis and Results

For the Neptune probe study, three entry cases were considered. The first entry trajectory (Design #3) considered was at -34° EFPA, under the assumption that a probe mission to Neptune would be similar to Uranus. However, Neptune entries have higher velocities and a significantly different atmospheric density profile from Uranus, demonstrating the need to select shallower EFPAs for feasible missions. **Table A-8** summarizes the Neptune entry parameters. For Design #3, for an entry velocity of 26.1 km/s and EFPA of -34°, the aerothermal heating environments and pressures are beyond current testing limits in ground test facilities. At a stagnation pressure of 25 bar and heat flux of ~10,000 W/cm², currently manufacturable TPS materials would not be able to be tested and qualified to these levels. Therefore, shallower entries were investigated. The second Neptune design (Design #4) was at 25.7 km/s entry velocity and EFPA of -20°, leading to a stagnation pressure of 11.5 bar and peak hot-wall heat flux of 5,465 W/cm². **Figure A-19** shows the heat flux variation during the entire entry trajectory. The peak heat flux occurs at 100 km altitude. **Figure A-20** shows the pressure distribution, and peak stagnation pressure occurs at much lower altitude of 70 km. While the HEEET material has been tested for pressures in the range of 14 bar as well as heat fluxes of 7,000 W/cm², the current ground test facilities are not able to test at this combined high heat flux and high pressure range. The engineering correlations used by TRAJ to predict the stagnation peak heat flux and pressure do not estimate the shear environments, which are anticipated to be large at the vehicle shoulder. It is known for other TPS materials that high shear forces can cause material erosion at an accelerated rate and without a suitable test at corresponding conditions, it would not be possible to evaluate TPS performance. A more comprehensive discussion of this topic is provided in Appendix D under the title of "New TPS Technology". Due to these considerations, an even shallower trajectory (Design #5) with EFPA of -16° was considered. The entry parameters associated with Design #5 are shown in **Table A-8**. The heat flux and stagnation pressure profile during the entire trajectory are shown in **Figures A-21** and **A-22**, respectively. For this trajectory, the heat flux and pressures are well within the test range for current arcjet facilities. However, the heat load for this shallower trajectory is very high as the vehicle would spend more time in the atmosphere. In order to protect the payload, thicker TPS material will be required to sustain the higher heat-load while maintaining a recommended bond-line temperature of 260°C.

**Table A-8.** Entry parameters and environment for Neptune.

| Entry Parameters | Design #3 | Design #4 | Design #5 |
|---|---|---|---|
| Hyperbolic excess velocity (km/s) | 1232 | 11.3 | 11.4 |
| Entry interface altitude (km) | 1000 | 1000 | 1000 |
| Radial distance (km) | 25766 | 25690 | 25701.95 |
| Inertial velocity (km/s) | 26.12 | 25.73 | 25.72 |
| Entry Flight Path Angle, gamma (deg) | -34 | -20 | -16 |
| Inertial heading angle (deg) | -99.1 | -84.26 | -86.45 |
| Latitude (deg) | -1.42 | 24.8 | 22.64 |
| Max deceleration (g loads) | 454.91 | 208.71 | 124.51 |
| Stg pressure (bar) | 25 | 11.5 | 6.84 |
| Peak convective heat flux (W/cm²) | 9368.5 | 5362.4 | 4311 |
| Peak radiative heat flux (W/cm2) | 265.68 | 99.12 | 68.2 |
| Total heat load (J/cm²) | 81476 | 109671 | 133874 |

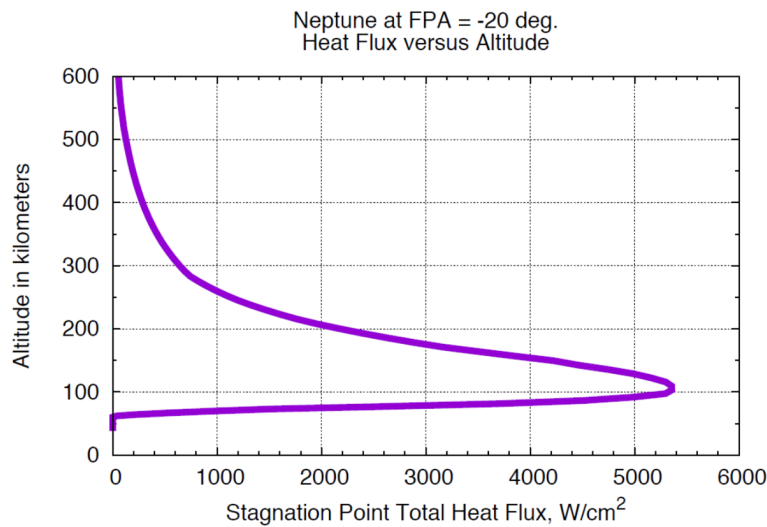

**Figure A-19.** Stagnation point heat-flux distribution for Neptune entry trajectory at -200 EFPA (Design # 4).

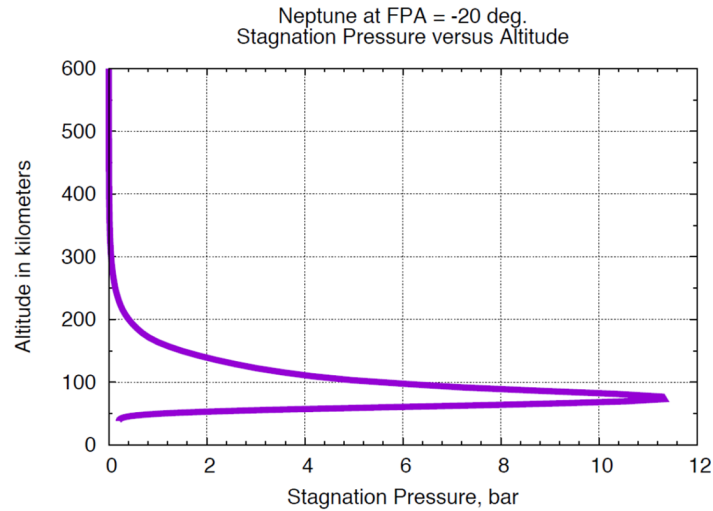

**Figure A-20.** Stagnation point pressure distribution for Neptune entry trajectory at -200 EFPA (Design #4).

The HEEET TPS material sizing was performed for designs #4 and #5 corresponding to EFPAs of -20° and -16°, respectively. Design #3 is not feasible due to the facility testing limitations for TPS materials. The results from TPS sizing for Design #4 and Design #5 are summarized in **Tables A-9** and **A-10**, respectively. The unmargined HEEET TPS mass for Design #4 was calculated to be 31.81 kg. Including material margins, the total HEEET mass was calculated as 39.42 kg. TPS sizing was also computed for CP. Using the same initial temperature, bond-line temperature and 0.762 mm aluminum face-sheet for substructure, the corresponding unmargined CP thickness is 2.135 cm and 48.60 kg. With 50% margin the thickness and mass for CP is 3.20 cm and 72.9 kg. Again, the CP is more than twice the mass of HEEET given the same conditions. For aftbody TPS, PICA was sized assuming a peak afterbody heatflux of 5% of the forebody stagnation point heating. The thickness was calculated as 1.0 cm. The corresponding PICA mass for aftbody TPS was calculated to be 3.85 kg. With 50% margins applied to these computations the thickness will be 1.5 cm and mass will be 5.78 kg. **Table A-9** lists the associated TPS thicknesses and masses for the Neptune design cases.

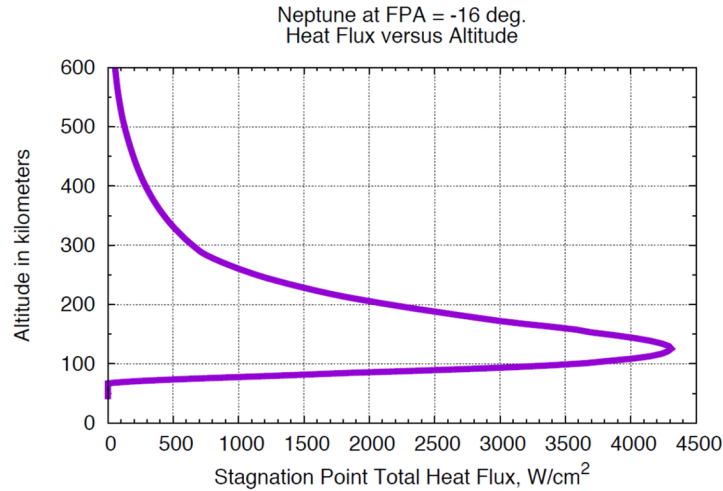

**Figure A-21.** Stagnation point heat-flux distribution for Neptune entry trajectory at -16⁰ EFPA (Design #5).

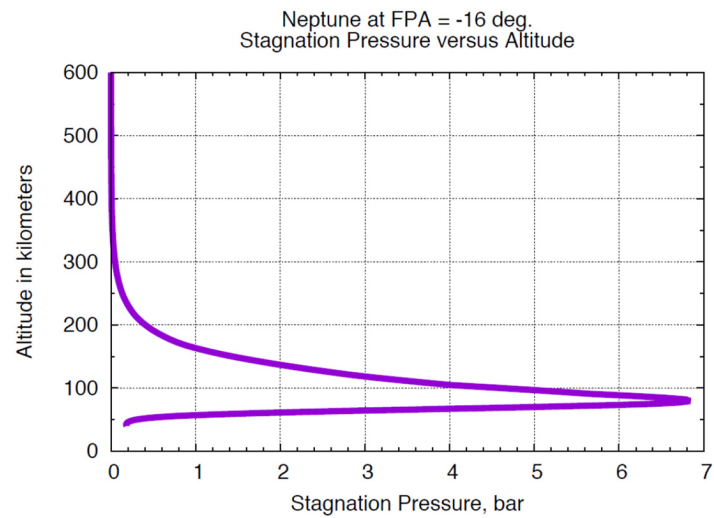

**Figure A-22.** Stagnation point pressure distribution for Neptune entry trajectory at -16⁰ EFPA (Design #5).

For Design #5, due to the higher heat load, the mass of the forebody HEEET TPS increased compared to Design #4. The un-margined HEEET TPS mass was calculated as 38.35 kg and with margins the value went up to 47.25 kg. The unmargined CP thickness was 2.58 cm, resulting in mass of 58.72 kg. With 50% margins the thickness will be 3.87 cm and corresponding mass will be 88.08 kg. This is almost double the mass of HEEET and represents 70% of the total allocation for the aeroshell mass. The aftbody TPS thickness for same conditions was calculated as 1.3 cm, corresponding to mass of 4.53 kg. The margined PICA is 1.95 cm thick with mass at 6.80 kg.

The aluminum substructure and adhesive thicknesses for both heatshield and backshell are listed in **Table A-9** and **A-10** for both of the designs.

Due to the fact that -16° shallow trajectory has certain drawbacks, such as higher heat load leading to higher TPS mass, a potential compromise in communications, as well as a greater risk of skip out, and -20⁰ trajectory leads to extreme pressure and heatflux environments; in the future it will be worthwhile to investigate trajectories at -18° EFPA and attain a more optimum design solution. For this study -20° EFPA probe entry trajectory was selected as the baseline design point for Neptune probe missions.

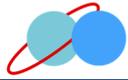

**Table A-9.** Aeroshell thickness and mass estimates for Neptune Design #4.

| TPS Material | HEET | Carbon-Phenolic | Carbon-Phenolic | Carbon-Phenolic | Carbon-Phenolic | PICA |
|---|---|---|---|---|---|---|
| **Aeroshell** | **Fore-Body** | **Fore-Body** | **Fore-Body** | **Fore-Body** | **Fore-Body** | **Aft-Body** |
| Adhesive Type | HT-424 | RTV-560 | RTV-560 | RTV-560 | RTV-560 | RTV-560 |
| Adhesive Thickness (mm) | 0.76 | 0.38 | 0.38 | 0.38 | 0.38 | 0.38 |
| Adhesive Mass (kg) | 2.42 | 0.84 | 0.84 | 0.84 | 0.84 | 0.69 |
| Substructure Type | Aluminum Face sheet | Aluminum Face sheet | Aluminum Face sheet | Aluminum Structure | Aluminum Structure | Aluminum Structure |
| Substr Thickness *Thin* (mm) | 0.762 | 0.762 | 0.762 | 3.175 | 3.175 | 3.175 |
| Substr Mass (kg) | | | | 14 | 14 | 11.5 |
| Bondline Temperature (°C) | 260 | 260 | 250 | 260 | 250 | 260 |
| Unmargined TPS Thickness (cm) | RL - .64 IL – 1.6 | 2.14 | 2.15 | 1.98 | 2.00 | 1.00 |
| Unmargined TPS Mass (kg) | 31.81 | 48.61 | 49.01 | 45.00 | 45.42 | 3.46 |
| Margined TPS Thickness (cm) | RL - .96 IL – 1.76 | 3.20 | 3.23 | 2.97 | 3.00 | 1.50 |
| Margined TPS Mass (kg) | 39.42 | 72.92 | 73.52 | 67.50 | 68.13 | 5.18 |

**Table A-10.** Aeroshell thickness and mass estimates for Neptune Design #5.

| TPS Material | HEET | Carbon-Phenolic | Carbon-Phenolic | Carbon-Phenolic | Carbon-Phenolic | PICA |
|---|---|---|---|---|---|---|
| **Aeroshell** | **Fore-Body** | **Fore-Body** | **Fore-Body** | **Fore-Body** | **Fore-Body** | **Aft-Body** |
| Adhesive Type | HT-424 | RTV-560 | RTV-560 | RTV-560 | RTV-560 | RTV-560 |
| Adhesive Thickness (mm) | 0.76 | 0.38 | 0.38 | 0.38 | 0.38 | 0.38 |
| Adhesive Mass (kg) | 2.42 | 0.84 | 0.84 | 0.84 | 0.84 | 0.69 |
| Substructure Type | Aluminum Face sheet | Aluminum Face sheet | Aluminum Face sheet | Aluminum Structure | Aluminum Structure | Aluminum Structure |
| Substr Thickness *Thin* (mm) | 0.762 | 0.762 | 0.762 | 3.175 | 3.175 | 3.175 |
| Substr Mass (kg) | | | | 14 | 14 | 11.5 |
| Bondline Temperature (°C) | 260 | 260 | 250 | 260 | 250 | 260 |
| Unmargined TPS Thickness (cm) | RL - .731 IL – 2.0 | 2.58 | 2.60 | 2.40 | 2.42 | 1.30 |
| Unmargined TPS Mass (kg) | 38.35 | 58.72 | 59.22 | 54.68 | 55.17 | 4.53 |
| Margined TPS Thickness (cm) | RL – 1.1 IL – 2.17 | 3.87 | 3.90 | 3.60 | 3.63 | 1.95 |
| Margined TPS Mass (kg) | 47.25 | 88.08 | 88.83 | 82.02 | 82.76 | 6.80 |

## A.5.5 "Retrograde" Probe Entry Option

In designing a mission to an ice giant planet that involves both an orbiter spacecraft and an atmospheric entry probe to be delivered upon initial approach, trajectory designers face some aspects of the planet and its position in the solar system that constrain available options. In discussing some of these options, the terms "prograde" and "retrograde" will be used. Usually, these terms refer to an orbiting object's direction of travel with respect to the primary's rotation direction. In this case, however, "prograde" and "retrograde" will refer to a probe's direction of travel around the primary (before entry) with respect to the orbiter's direction of travel. This can be expressed in terms of the directions of the angular momentum vectors of the orbiting objects. The direction of an angular momentum vector (AMV) can be established by the "right hand rule": aligning the fingers of the right hand with the direction of travel of the orbiting object around the primary, the extended thumb points in the direction of the AMV. If the orbiter's AMV and the probe's AMV point in roughly the same direction, the probe's approach is deemed "prograde". If those vectors are roughly anti-aligned, then the probe's approach is deemed "retrograde".

If the specifications and performance parameters of the instruments and thermal protection system (TPS) materials used in this study are valid when the mission is implemented, the prograde probe entry adopted for this study appears feasible. But the modified entry trajectory adopted here requires a steeper-than-usual entry flight path angle (EFPA). This produces higher peak heating rates and significantly higher inertial loads than a shallower EFPA. It is possible that as this mission concept evolves, changes in approach circumstances, less-than-anticipated performance by various components, or other unforeseen drivers could force the prograde approach to an entry that is riskier than the project is willing to accept. The probe mission is not lost in that case. The retrograde approach appears to be a viable alternative that relieves some of the stress on the flight system. That approach is outlined here.

Note that scale drawings in the figures and the occasional numeric specification are based upon the Earth-to-Uranus transfer trajectory adopted by the Ice Giants study for an orbiter plus atmospheric entry probe mission. Changes to that trajectory, primarily changes in the $V_\infty$ of approach, will change the details of the trajectories in the Uranus system but not the qualitative conclusions reached. After the discussion of the Uranus scenario, the applicability of this approach to Neptune will be discussed briefly.

### Prograde Approach

Characteristics of Uranus and the uranian system that impose the constraints mentioned above include the mass of Uranus, Uranus's large heliocentric distance, and its large obliquity, which are 14.5 Earth masses, 19.2 AU, and 97.77°, respectively. The large heliocentric distance requires a relatively large approach $V_\infty$ for reasonable trip durations. This results in large orbit insertion $\Delta V$, a significant driver of the flight system's launch mass. To minimize that $\Delta V$, the orbit insertion maneuver should be performed as close to the planet as possible, so there is strong motivation to keep the orbiter's periapsis as low as possible during the orbit insertion maneuver, and to keep that maneuver as symmetric as possible around periapsis. The design of the entry probe's trajectory must consider several aspects of the mission, including atmosphere-relative entry speed, EFPA and various entry conditions that result (such as peak heating rates, total heat load, and peak inertial loading), and telecommunications to the orbiter spacecraft. These aspects are not independent: there are significant couplings that constrain the flexibility in practical probe entry trajectories.

**Figure A-23** illustrates the problems with the most straight-forward probe entry trajectory. The blue curve is the orbiter's trajectory during its UOI pass. Its periapsis in this diagram is directly to the right of the planet center.

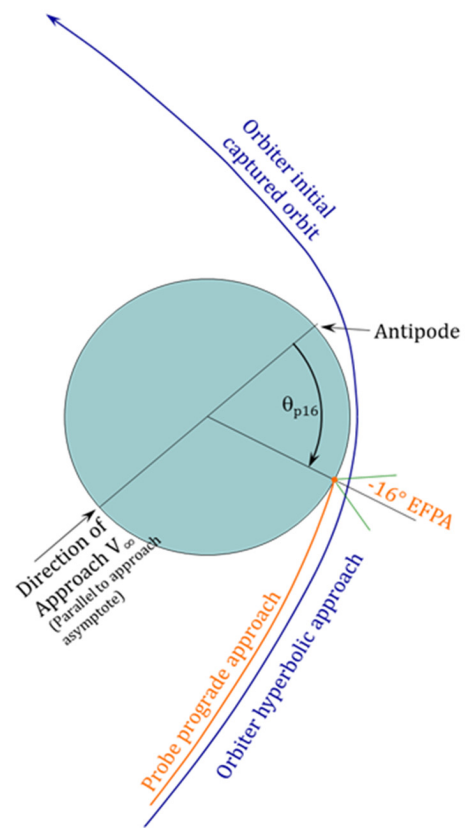

**Figure A-23**. Geometry of orbiter & probe trajectories for a 'straightforward' probe entry.

The offset angle from the antipode and the periapse (not shown here) is a function mostly of the approach $V_\infty$ magnitude, the radius of periapsis, and the mass of the planet. For a given combination of Earth-to-Uranus transfer trajectory and minimum periapse radius, this can be considered fixed. The orange curve is the probe trajectory that yields an EFPA of -16°. That EFPA is in the range that yields what is considered the best combination of low inertial loads, low peak heating rates, and acceptable total heat load. The green lines radiating from the probe entry site represent the usable limits of the probe's radio antenna beam. Knowing that near periapsis the orbiter's speed is quite high, it is obvious that the orbiter is within that usable communications zone for an impractically short period of time, less than 4 minutes. Also, given the orbit insertion ΔV and the thrust level of practical rocket engines for the orbiter, the time required for the UOI maneuver would ideally have that maneuver in progress during this communications period. Project managers are reticent to schedule simultaneous critical events!

One approach to address these problems is to move the probe entry location farther from the orbiter periapsis by increasing the angle θ as shown in **Figure A-24**. This has the orbiter at higher altitude during the relay period, with three advantages: the orbiter is farther from the probe, so its pathlength through the communications region is longer; it is farther from Uranus, so its speed is slower; and the communications period conflicts less with the optimal UOI maneuver window. The first two contribute to a longer communications window. But this also has disadvantages: the EFPA steepens considerably, increasing the peak heating rate and the magnitude of inertial loads. Originally the study team proposed steepening the EFPA to -35°, but the TPS experts advised that the resulting entry conditions exceeded the anticipated margined performance limits for the HEEET TPS material, so the EFPA was reduced to -30°. Though this yields a tractable entry situation, the entry conditions are still challenging, with inertial loads greater than 200 g's, and a communications window duration just barely sufficient for the probe's prime science mission.

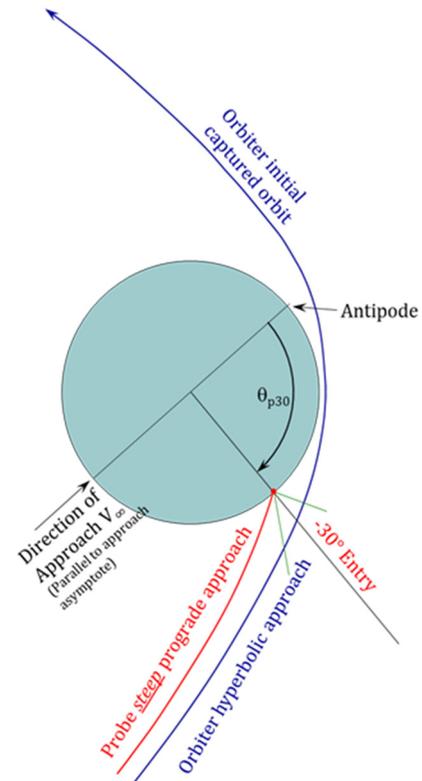

**Figure A-24.** Geometry of orbiter and *steep* (-30° EFPA) probe entry trajectories.

### Retrograde Approach

There is a probe trajectory option that features an EFPA of -16°, a probe data communications window of more than 2 hours duration, and no conflict between the communications window and the UOI maneuver window. This is the "retrograde" probe entry trajectory, shown in **Figure A-25**. For this approach the probe entry occurs at an offset angle θ equal in measure to the θ for the prograde, -16° entry, but whose direction of offset (clock angle around the antipode) is 180° degrees away from that of the prograde entry. The probe entry is timed to occur about one hour after the orbiter's periapsis pass, leaving plenty of time for the orbiter to perform an optimal UOI maneuver. The probe's deceleration and deployment is completed as the orbiter enters the probe's antenna beam, about an hour after the orbiter's periapsis pass, at an orbiter-to-probe range of about 31,000 km. Were there no planetary rotation, the orbiter would then stay within the communications zone for many hours, and its range to the probe would remain less than 100,000 km for about 2.5 hours. The communications distances are larger than for the prograde steep entry, and thus yield lower data rates for a given transmitted power from the probe, but the largest distances are commensurate with those being considered for a Saturn entry probe, and the smallest are considerably smaller. The communications window afforded provides ample time for the probe to reach its "prime mission" depth with significant margin, allowing an "extended mission" to greater depths.

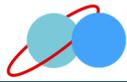

A detailed discussion of the effects of planetary rotation is difficult for this analysis because Uranus's extreme obliquity causes the declination of the approach asymptote to vary seasonally from near-polar to near-equatorial. Likewise, probe entry latitudes can vary from equatorial to polar, while orbiter inclinations vary widely, with huge difference in rotational effects on the data link geometry. One aspect of planetary rotation is its effect on the atmosphere-relative entry speed. Depending upon the latitude of entry and the direction of the entry with respect to the local direction of rotation, the range of possible influence on entry speed goes from subtracting 2.6 km/s from the inertial value, to adding 2.6 km/s. There is a means of dealing with at least some of the rotational effects. The entry offset angle θ can be oriented in any direction from the antipode, not only in the plane of the diagrams of **Figures A-23** to **A-25**. This flexibility in the 3rd dimension allows biasing the initial location (and thus direction) of the probe's antenna beam, at least partially compensating for planetary rotation.

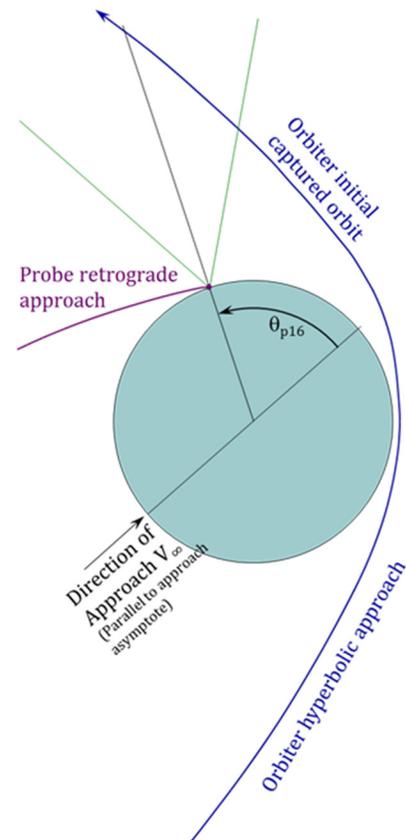

**Figure A-25.** Geometry of the orbiter and retrograde probe entry trajectories.

The situation for using this technique at Neptune is similar in some aspects to the situation at Uranus but there are also some differences. At ~17 Earth masses Neptune is somewhat more massive than Uranus, and its mean radius is a bit smaller. For a given V∞ of approach this decreases θ somewhat, reducing slightly both the probe-to-orbiter distances involved and the time between orbiter periapsis and opening of the data relay window. But since Neptune's heliocentric distance is about 50% larger than Uranus', the approach V∞ for transfer trajectories of reasonable durations is somewhat higher. This increases θ, pressing those distances and the time period to data relay onset toward larger values. This would be true at Uranus also: a larger V∞ of approach would have the same ramifications for the data relay window opening time and the probe-to-orbiter distances during that window. Note that although details of the trajectories, time periods, and probe-to-orbiter distances would change somewhat for a mission to Neptune instead of to Uranus, except in extreme cases the qualitative assessment of the utility of this approach for a Uranus mission also holds for a Neptune mission.

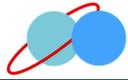

## A.6

### Dual Spacecraft, Single Launch Trajectories

The large lift capability of the SLS-1B, coupled with fortuitous availability of gravity assists in Jupiter's strong gravity well (for both Uranus and Neptune) and the high specific impulse of electric propulsion, allows a dual manifest to be considered, wherein two spacecraft are launched on a single launch vehicle, one destined for Uranus and one for Neptune. The two simplest options to consider are:

a. The two spacecraft are propelled by a single, large SEP stage, until one spacecraft separates, e.g., just before the Jupiter flyby, and continues with only chemical propulsion to one of the Ice Giants.

b. Two spacecraft, each with its own SEP stage, remain joined only until shortly after launch. They separate after Earth escape and use their respective SEP stages to propel them towards the two Ice Giants.

Examples of both of these (for assessing feasibility) are presented in this section. The case of departing on different launch asymptotes was not considered as this would require two different upper stages be used. Many variations on the two simple options are possible, for example, performing a chemical deep-space maneuver directly after separation, using REP instead of SEP, separating before Earth flyby followed by Jupiter flyby etc. A full exploration of the dual-manifest trade space was beyond the scope of this study.

The trifecta of SLS-1B, a Jupiter gravity assist, and EP permit large, flagship-class spacecraft to be inserted into orbit at each of the planets from a dual manifest launch. Removing any one of these three elements would likely disallow a pair of even floor missions. The early 2030s offer the last auspicious phasing of Jupiter for a dual Uranus-Neptune manifest for a number of decades.

The first dual-manifest example presented here uses only a single gravity assist, at Jupiter, after launch. The objective is to insert as much useful mass as possible into orbit at each Ice Giant – the sum total of useful inserted mass at each planet, under previously stated flight time constraints. In the second example, we present a case of minimizing the launch mass for a given useful inserted mass at each planet. These trajectories were computed in a semi-automated way: A grid search determined the general timeframe when opportunities of each type arise, and then further iterations were performed manually.

### A.6.1    Earth Launch – Jupiter – Ice-Giant

This trajectory assumes the spacecraft remain joined until shortly before the Jupiter flyby, at which point they separate, with the Uranus-bound spacecraft keeping the SEP stage for further thrusting and the Neptune-bound spacecraft proceeding ballistically after a small targeting maneuver. In the design, mass can be traded between the two spacecraft. The result presented in **Figures A-26** and **A-27** is simply a representative point design in this option space. An 80-kW, 3500-kg, 9+1 NEXT SEP stage is assumed (a conservative mass estimate). The high power is again put to use after separation by continuing to provide some thrust after the Jupiter flyby to make up for the fact that the flyby occurs at the lower limit on the altitude. The larger SEP stage is also needed to provide the necessary thrust and acceleration.

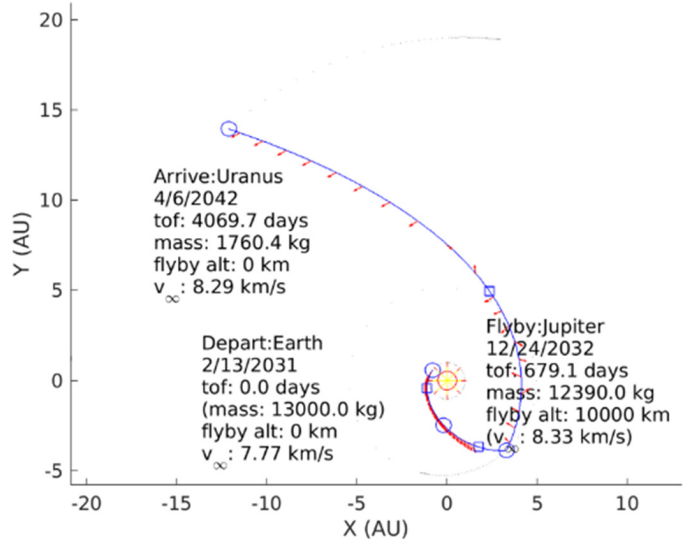

**Figure A-26.** One part of a dual launch manifest.

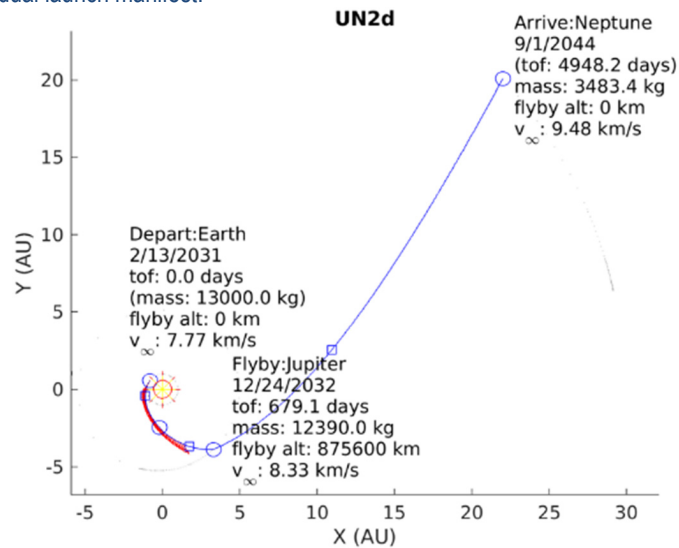

**Figure A-27.** One part of a dual launch manifest.

## A.6.2    Earth Launch – Mars – Jupiter – Ice-Giant

The spacecraft separate just after launch and use their individual SEP stages for thrusting in the inner solar system. Both spacecraft follow similar trajectories with Mars flyby occurring on the same day. After Jupiter flyby, the Neptune spacecraft proceeds ballistically to Neptune, while the Uranus spacecraft continues to thrust for a short time. Note, that unlike the previous case, the Jupiter flybys do not occur on the same day. As with the previous case, mass can be traded between the two spacecraft. Furthermore, post orbit insertion masses can be increased (at the expense of increased SEP propellant load) as SLS-1B is capable of launching twice the combined masses of each spacecraft. The result presented in **Figures A-28** and **A-29** is simply a representative point design in this option space. Both spacecraft use the above described (see **Table A.4**) 35-kW, 1000-kg, 3+1 NEXT SEP stage.

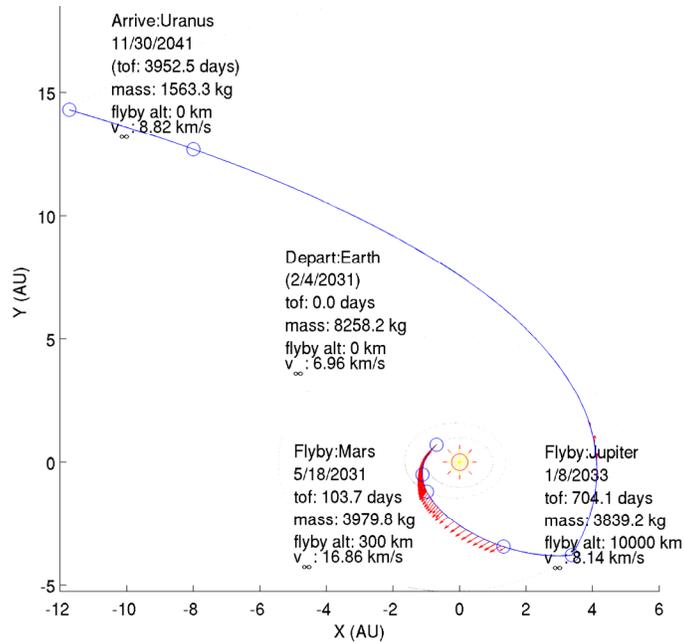

**Figure A-28.** First part of a dual launch manifest: Uranus case.

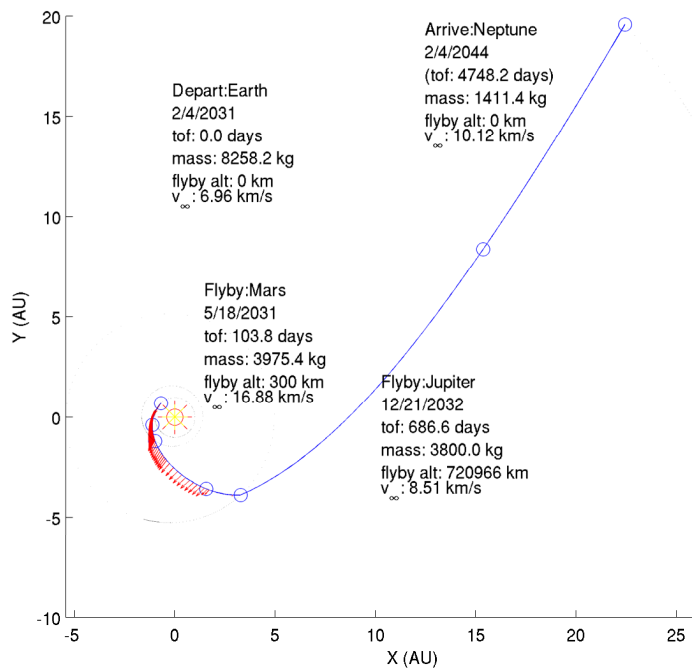

**Figure A-29.** Second part of a dual launch manifest: Neptune case.

## A.7    Radioisotope Electric Propulsion Trajectories

This section briefly summarizes high performing REP mission options to the Ice Giants. Given then the spacecraft can thrust all the way up to few days prior to orbit insertion, these trajectories are expected to have medium to arrival relative velocity with respect to the target planet. This in turn results in substantial reduction in the orbit insertion propellant (Bi-Prop).

**Figure A-30** highlights the interplanetary trajectory tradespace for REP missions to Uranus with up to 4 flybys, launching on Atlas V (551) and Delta-IV Heavy and with two different REP

power levels (1.652 kW, 2.152 kW). The two REP power levels result in different total propellant (xenon) usage. All figures show Useful Inserted Mass as a function of launch date with colors representing interplanetary cruise time in years. Note, that colors are not consistent between figures.

Using a REP propulsion system before orbit insertion significantly increases the delivered useful mass into orbit when compared to pure chemical trajectory (Bi-Prop). It also results in increased launch flexibility allowing launches every year to Uranus for both class of launch vehicles.

Specifically, using 1.65 kW REP system with Altas V (551) LV results in > 1500 kg of useful inserted mass into Uranus with ~11 years of interplanetary cruise time. Going to 25 kW REP stage results in small increase in delivered mass (on average). Note, that the high-power REP option comes with a reduced maximum allowable throughput due to XIPs thruster limitations. See REP assumptions for more details. Going from Atlas v (551) to the Delta-IV Heavy LV results in significant (almost 80%) increase in useful inserted mass at Uranus.

**Figures A-31** highlights the interplanetary trajectory tradespace for REP missions to Neptune with up to 4 flybys, launching on Atlas V (551) and Delta-IV Heavy and with two different REP power levels (1.652 kW, 2.152 kW). As for the Uranus case, the two REP power levels result in different total propellant (xenon) usage.

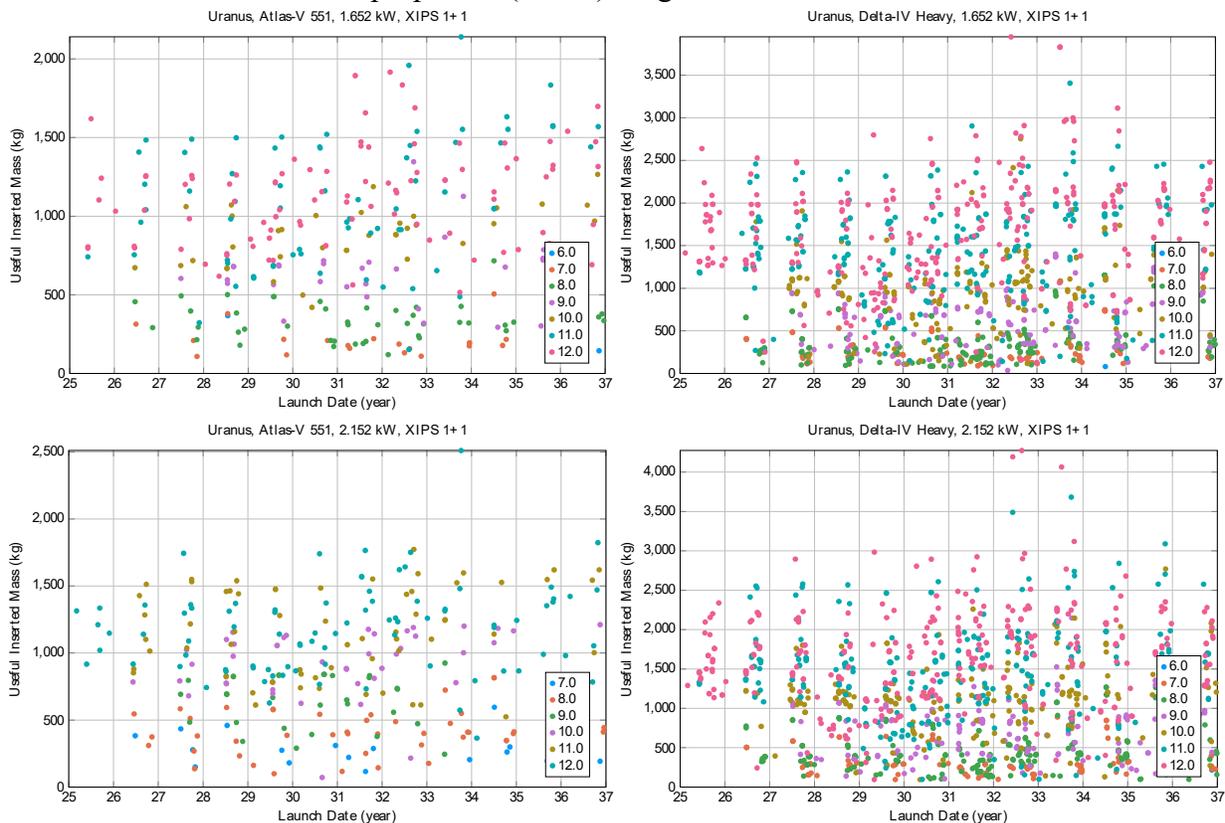

**Figure A-30.** REP trajectory options to Uranus: Atlas V(551) LV left, Delta-IV Heavy (right), 1.652 kW (top), 2.152 kW (top).

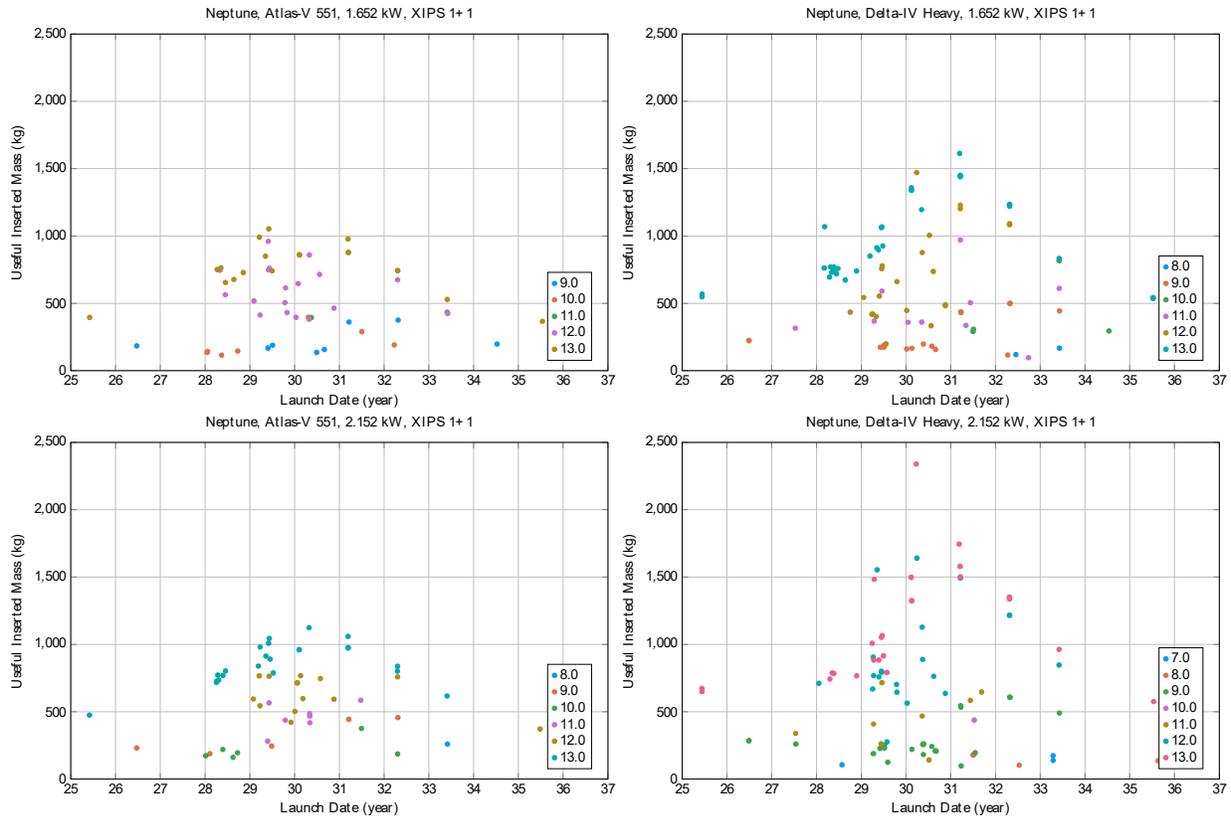

**Figure A-31.** REP trajectory options to Neptune: Atlas V(551) LV left, Delta-IV Heavy (right), 1.652 kW (top), 2.152 kW (top).

Using 1.65 kW REP system with Altas V (551) LV results in ~1000 kg of useful inserted mass into Neptune orbit for a 12-13 years of interplanetary cruise time. Going to 25 kW REP stage results in small increase in delivered mass to ~1100 kg. Note, that the high-power REP option comes with a reduced maximum allowable throughput due to XIPS thruster limitations. Going from Atlas v (551) to the Delta-IV Heavy LV results in significant (50% to 100%) increase in useful inserted mass at Neptune. It is possible using a 2.15 kW powered REP system, to insert >1500 km into Neptune orbit with 12 years of interplanetary flight time.

Another consequence of using a REP trajectory is the significant reduction in orbit insertion ΔV. **Figures A-32** and **A-33** highlight this fact by showing the orbit insertion ΔV as function of the launch year. Only orbit insertion ΔVs for 1.652 kW REP power level trajectories are shown, as these results are representative of other power levels.

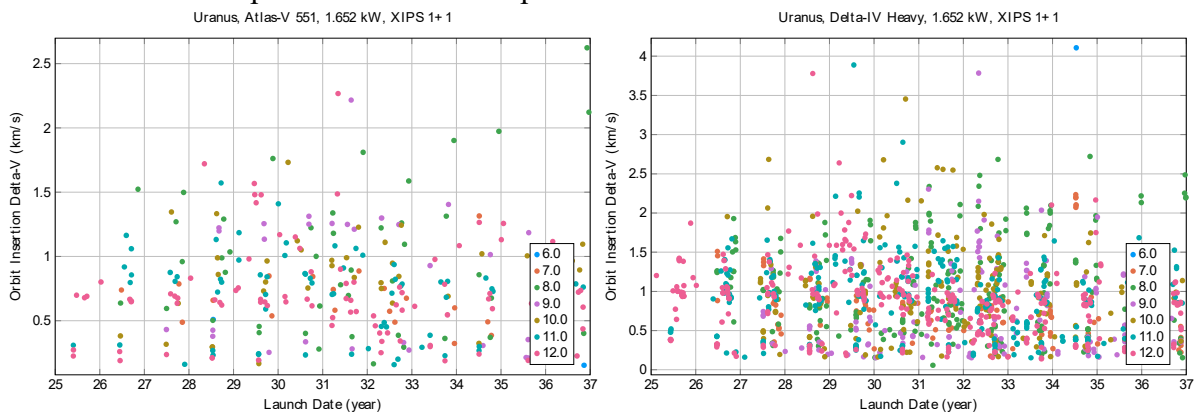

**Figure A-32.** REP trajectory options to Uranus, Orbit Insertion ΔVs : Atlas V(551) LV left, Delta-IV Heavy (right).

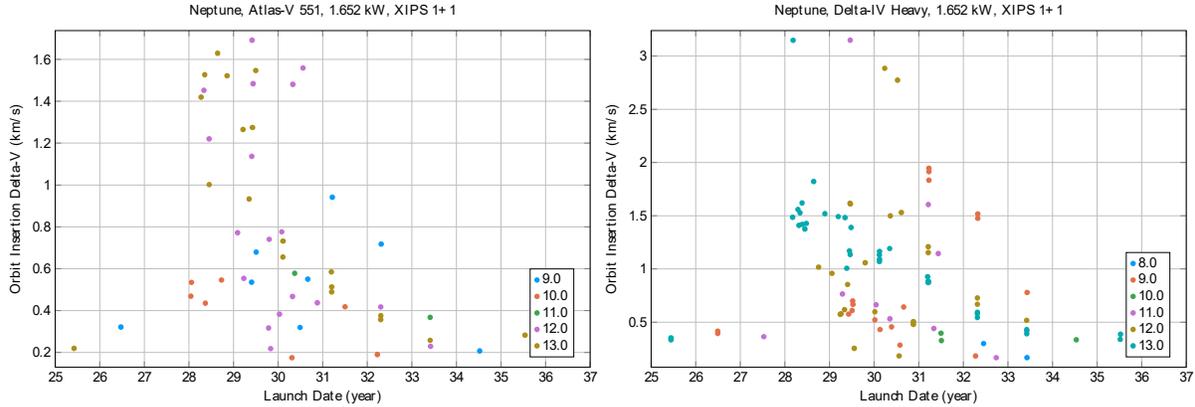

**Figure A-33.** REP trajectory options to Uranus, Orbit Insertion ΔVs: Atlas V(551) LV left, Delta-IV Heavy (right).

Three things immediately stand out:1) orbit insertion ΔVs for longer flight times, hybrid REP and Bi-Prop mission trajectories are significantly less when compared to equivalent Chemical or SEP mission trajectories, 2) orbit insertion ΔVs are smaller for Uranus when compared to Neptune, primarily due to Neptune trajectories having more time for breaking thrust before orbit insertion, and 3) orbit insertion ΔV is higher for launches using the higher performing Delta-IV Heavy LV. This is partially attributed to the fact that a high performing launch vehicle results in a larger launch mass to a given mission launch energy, resulting in less breaking (and higher planet relative velocity) prior to target body orbit insertion.

## A.8 Trajectory Options

This appendix documents some high performing chemical and SEP trajectories for both Uranus and Neptune based on the assumptions outlined above.

### A.8.1 Uranus

#### Chemical Options to Uranus

| Launch_Date | C3 | Path | Arrival_Mass _551 | Arrival_Mass _D4H | Arrival_Mass _SLS | Arrival_Vinf | Arrival_Decli nation | Useful_Mass _551 | Useful_Mass _D4H | Useful_Mass _SLS | TOF | OI_DV | DSM |
|---|---|---|---|---|---|---|---|---|---|---|---|---|---|
| 20250601 | 51.3 | 3367 | 1409.9 | 2754.1 | 10739 | 11.41 | 72.65 | 458.3 | 895.2 | 3490.6 | 10 | 3.03 | 1.21 |
| 20250605 | 52.56 | 3367 | 1499.4 | 2930.2 | 11462.9 | 9.29 | 73.59 | 711.5 | 1390.5 | 5439.7 | 11 | 2.07 | 0.93 |
| 20250930 | 25.3 | 337 | 419.3 | 756.9 | 2745.4 | 7.04 | 77.69 | 271.2 | 489.6 | 1775.9 | 12 | 1.23 | 6.94 |
| 20260524 | 29.68 | 3367 | 1402.9 | 2572.3 | 9449.8 | 11.47 | 72.63 | 450.9 | 826.8 | 3037.5 | 9 | 3.06 | 2.78 |
| 20260808 | 27.33 | 3367 | 1507.7 | 2741.2 | 10001.6 | 7.45 | 74.68 | 927.1 | 1685.6 | 6150 | 11 | 1.37 | 2.72 |
| 20260817 | 26.3 | 337 | 453.4 | 821.2 | 2987.4 | 8.32 | 80.1 | 248.3 | 449.8 | 1636.3 | 10 | 1.69 | 6.62 |
| 20270612 | 30.72 | 3367 | 1360.2 | 2501.5 | 9217.8 | 8.71 | 76.98 | 704.5 | 1295.5 | 4773.9 | 10 | 1.84 | 2.81 |
| 20270715 | 52.56 | 337 | 674.4 | 1318 | 5156.2 | 8.44 | 82.35 | 363.2 | 709.9 | 2777 | 11 | 1.73 | 3.48 |
| 20271017 | 14.87 | 337 | 458.5 | 809.7 | 2847.2 | 8.13 | 79.49 | 257.9 | 455.4 | 1601.2 | 9 | 1.62 | 7.35 |
| 20280808 | 26.62 | 3367 | 512.7 | 929.6 | 3384.9 | 8.45 | 77.13 | 275.7 | 499.9 | 1820.2 | 9 | 1.74 | 6.21 |
| 20290908 | 48.51 | 337 | 511.4 | 993.8 | 3846.9 | 9.69 | 75.21 | 227.3 | 441.7 | 1709.8 | 10 | 2.24 | 4.64 |
| 20291124 | 12.8 | 323357 | 4167.7 | 7335.2 | 25631.3 | 13.1 | 47.99 | 931.2 | 1638.8 | 5726.6 | 11 | 3.9 | 0.44 |
| 20300724 | 17.13 | 332357 | 3377.3 | 5989.5 | 21204.1 | 11.55 | 47.13 | 1067.7 | 1893.6 | 6703.6 | 11 | 3.1 | 0.84 |
| 20300730 | 18.03 | 332357 | 3025.1 | 5374.8 | 19079.8 | 13.99 | 48.46 | 535.9 | 952.1 | 3379.7 | 10 | 4.39 | 1.13 |

| 20310222 | 30.84 | 32237 | 1824.8 | 3357 | 12374 | 7.61 | 71.94 | 1099.3 | 2022.3 | 7454.3 | 12 | 1.43 | 1.86 |
|---|---|---|---|---|---|---|---|---|---|---|---|---|---|
| 20310526 | 11.95 | 323357 | 4003.1 | 7036.7 | 24525.2 | 10.19 | 47.54 | 1634.9 | 2873.8 | 10016.3 | 11 | 2.46 | 0.63 |
| 20310716 | 52.56 | 3357 | 1547.7 | 3024.7 | 11832.6 | 14.05 | 48.6 | 269.8 | 527.2 | 2062.6 | 9 | 4.43 | 0.83 |
| 20310727 | 17.41 | 32357 | 3386.3 | 6008.9 | 21291.1 | 11.54 | 47.12 | 1073.4 | 1904.7 | 6748.8 | 10 | 3.09 | 0.81 |
| 20320503 | 28.5 | 3357 | 2826.6 | 5161.5 | 18896.8 | 8.47 | 47.8 | 1515.6 | 2767.5 | 10132.1 | 11 | 1.75 | 0.63 |
| 20320719 | 27.19 | 3357 | 2590.4 | 4707.2 | 17168 | 9.89 | 47.55 | 1115 | 2026.1 | 7389.6 | 10 | 2.32 | 1.01 |
| 20320726 | 28.44 | 3357 | 2365.4 | 4318.6 | 15808.3 | 12.15 | 47.59 | 659.1 | 1203.4 | 4405 | 9 | 3.4 | 1.21 |
| 20330609 | 29.88 | 3357 | 2118.6 | 3886.9 | 14287.5 | 8.84 | 51.66 | 1076.9 | 1975.7 | 7262.2 | 10 | 1.89 | 1.45 |
| 20330612 | 29.9 | 3357 | 2023.4 | 3712.5 | 13647.1 | 10.84 | 50.76 | 734.8 | 1348.2 | 4956.1 | 9 | 2.76 | 1.6 |
| 20330828 | 27.96 | 3357 | 2100.5 | 3828.1 | 13993.5 | 7.05 | 53.13 | 1357 | 2473 | 9040 | 11 | 1.24 | 1.62 |
| 20330829 | 27.29 | 3357 | 1794.7 | 3262.5 | 11902.4 | 12.91 | 50.16 | 419.9 | 763.3 | 2784.9 | 8 | 3.8 | 2.17 |
| 20340720 | 25.63 | 3357 | 692.2 | 1250.8 | 4541.4 | 10.58 | 50.76 | 263.6 | 476.3 | 1729.4 | 8 | 2.64 | 5.32 |
| 20340725 | 31.09 | 3357 | 722.8 | 1330.5 | 4907.8 | 8.62 | 51.75 | 379.4 | 698.4 | 2576.4 | 9 | 1.8 | 4.8 |
| 20341002 | 48.96 | 3347 | 1529.9 | 2977 | 11536.9 | 7.92 | 56.52 | 885.2 | 1722.5 | 6675.1 | 11 | 1.54 | 1.11 |
| 20341004 | 49.15 | 3347 | 1311.6 | 2553.6 | 9901.1 | 9.45 | 54.36 | 606.5 | 1180.7 | 4578 | 10 | 2.14 | 1.59 |
| 20351007 | 26.37 | 3347 | 1428.6 | 2588.1 | 9416.7 | 9.44 | 54.34 | 662.2 | 1199.7 | 4364.9 | 9 | 2.13 | 2.96 |
| 20351011 | 27.53 | 3347 | 1519.5 | 2764.8 | 10093.9 | 7.89 | 56.55 | 883.1 | 1606.8 | 5866.2 | 10 | 1.53 | 2.68 |
| 20351011 | 27.5 | 3347 | 1610.5 | 2929.9 | 10695.8 | 6.8 | 59.34 | 1071.2 | 1948.8 | 7114.1 | 11 | 1.16 | 2.5 |
| 20360215 | 44.55 | 32337 | 1562.4 | 2996.5 | 11477.7 | 10.72 | 40.69 | 580.6 | 1113.6 | 4265.6 | 11 | 2.7 | 1.37 |
| 20360821 | 49.61 | 3347 | 453.4 | 883.6 | 3430.2 | 8.07 | 56.28 | 257.1 | 501 | 1945 | 9 | 1.59 | 4.94 |
| 20360826 | 29.55 | 3347 | 501.9 | 919.9 | 3378.1 | 9.65 | 54.21 | 224.8 | 412 | 1513.1 | 8 | 2.22 | 6.07 |
| 20360923 | 57.26 | 337 | 476.1 | 932.4 | 3692.3 | 7.96 | 52.67 | 274.1 | 536.8 | 2125.9 | 10 | 1.55 | 4.27 |

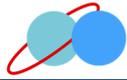

## SEP Trajectories to Uranus with Atlas V (551)

| Launch_Date | Launch_Mass | C3 | Path | Arrival_Mass | Arrival_Vinf | Arrival_Declination | Useful_Mass | TOF | OL_DV | P0 |
|---|---|---|---|---|---|---|---|---|---|---|
| 20250413 | 5607 | 3.95 | 33367 | 4073.6 | 8.99 | 76.81 | 2054.5 | 12 | 1.91 | 25 |
| 20250526 | 5700.8 | 3.09 | 3367 | 3709.2 | 7.75 | 74.55 | 2219.3 | 12 | 1.45 | 35 |
| 20250630 | 5610.7 | 3.91 | 337 | 3374.1 | 8.55 | 80.17 | 1813.7 | 11 | 1.74 | 35 |
| 20260614 | 4256 | 17.9 | 337 | 2875.8 | 8.61 | 80.15 | 1533 | 10 | 1.76 | 25 |
| 20260622 | 5421.5 | 5.68 | 337 | 3385.5 | 8.53 | 82.28 | 1825.4 | 12 | 1.73 | 35 |
| 20260710 | 5108.4 | 8.72 | 3367 | 3212.7 | 8.61 | 77.09 | 1686.8 | 11 | 1.8 | 25 |
| 20260726 | 4287.4 | 17.53 | 3367 | 2785.3 | 7.17 | 78.19 | 1788.4 | 12 | 1.25 | 15 |
| 20260812 | 4776.7 | 12.13 | 337 | 2960.1 | 8.32 | 79.69 | 1641.8 | 10 | 1.65 | 35 |
| 20260827 | 4982.3 | 9.99 | 3367 | 3378.3 | 8.19 | 80.04 | 1908.5 | 12 | 1.6 | 25 |
| 20270601 | 4323.3 | 17.12 | 3367 | 3021 | 8.77 | 76.96 | 1573.4 | 10 | 1.82 | 25 |
| 20270607 | 4372.8 | 16.55 | 337 | 3018.7 | 6.45 | 79.54 | 2102.3 | 12 | 1.03 | 25 |
| 20270701 | 5680.1 | 3.28 | 3367 | 3549.3 | 7.2 | 80.69 | 2271 | 12 | 1.26 | 35 |
| 20270730 | 4815.9 | 11.71 | 3367 | 3120.2 | 7.12 | 78.11 | 2015.5 | 11 | 1.24 | 35 |
| 20270801 | 4782.7 | 12.06 | 3367 | 3074.5 | 8.46 | 77.05 | 1672.9 | 10 | 1.71 | 35 |
| 20280528 | 4774.8 | 12.15 | 3367 | 3031.3 | 7.3 | 80.63 | 1918.2 | 11 | 1.29 | 35 |
| 20280813 | 3928.1 | 21.83 | 337 | 2379.8 | 7.1 | 81.57 | 1542 | 12 | 1.23 | 35 |
| 20280818 | 4757.2 | 12.33 | 3367 | 2996.9 | 8.22 | 80.03 | 1684.5 | 10 | 1.62 | 35 |
| 20280818 | 3511 | 27.23 | 337 | 2367.7 | 7.08 | 81.62 | 1536.7 | 12 | 1.22 | 25 |
| 20280902 | 4229.9 | 18.2 | 3367 | 2835.2 | 8.16 | 80.06 | 1606.4 | 10 | 1.6 | 25 |
| 20290614 | 4728.5 | 12.64 | 337 | 2958 | 8.53 | 80.04 | 1593.4 | 10 | 1.73 | 35 |
| 20290629 | 4244 | 18.04 | 337 | 2874.8 | 8.48 | 80.1 | 1560 | 10 | 1.71 | 25 |
| 20290712 | 4950.2 | 10.32 | 337 | 3364.3 | 8.37 | 73.26 | 1853.3 | 12 | 1.67 | 25 |
| 20290907 | 5420.7 | 5.69 | 327 | 2889.8 | 8.46 | 78.4 | 1573.1 | 11 | 1.7 | 25 |
| 20291106 | 5956.5 | 0.83 | 337 | 3363.1 | 7.95 | 73.67 | 1961.9 | 12 | 1.52 | 35 |
| 20300327 | 5373.2 | 6.14 | 3357 | 4235.7 | 10.13 | 46.63 | 1782.3 | 12 | 2.38 | 15 |
| 20300730 | 5746.1 | 2.69 | 32237 | 4190.8 | 9.94 | 68.42 | 1786 | 11 | 2.35 | 25 |
| 20300827 | 4819.8 | 11.67 | 337 | 3044.7 | 7.03 | 79.15 | 1988.3 | 11 | 1.21 | 35 |
| 20300908 | 4365.8 | 16.63 | 337 | 3009 | 6.18 | 81.05 | 2155.3 | 12 | 0.95 | 25 |
| 20301112 | 5684.7 | 3.24 | 32357 | 4576.3 | 10.86 | 46.88 | 1691.4 | 11 | 2.71 | 15 |
| 20301116 | 5691.2 | 3.18 | 32357 | 4243.2 | 9.01 | 46.48 | 2132.9 | 12 | 1.92 | 35 |
| 20310620 | 5952.3 | 0.86 | 33357 | 4495.1 | 8.26 | 47.81 | 2513.8 | 12 | 1.63 | 35 |
| 20310706 | 4270.3 | 17.73 | 337 | 2898.2 | 8.4 | 73.08 | 1590.6 | 10 | 1.68 | 25 |
| 20310723 | 5682.6 | 3.26 | 3337 | 4081.4 | 8.22 | 65.4 | 2295 | 12 | 1.62 | 25 |
| 20310729 | 5913.2 | 1.2 | 332237 | 4341.9 | 8.32 | 70.37 | 2408.4 | 11 | 1.65 | 15 |
| 20310729 | 5308.9 | 6.75 | 3337 | 3848.9 | 8.2 | 65.39 | 2169.9 | 12 | 1.61 | 15 |
| 20310906 | 4750 | 12.41 | 337 | 2968.9 | 8.11 | 73.57 | 1694 | 10 | 1.58 | 35 |
| 20310923 | 5926 | 1.09 | 332237 | 4263 | 8.12 | 70.69 | 2399.8 | 11 | 1.61 | 25 |
| 20311030 | 6016.7 | 0.3 | 332237 | 4118.4 | 7.99 | 71 | 2388.8 | 11 | 1.53 | 35 |
| 20320421 | 5710.5 | 3.01 | 32237 | 4379.8 | 6.52 | 75.09 | 3026 | 12 | 1.05 | 25 |
| 20320422 | 5851.9 | 1.74 | 32237 | 4204.1 | 6.52 | 75.12 | 2905.1 | 12 | 1.05 | 35 |
| 20320423 | 5696.9 | 3.13 | 32237 | 4337.7 | 7.43 | 72.25 | 2702.1 | 11 | 1.34 | 25 |
| 20320424 | 5848.6 | 1.77 | 32237 | 4177.6 | 7.42 | 72.3 | 2603.1 | 11 | 1.34 | 35 |
| 20320425 | 5664.2 | 3.43 | 32237 | 4349.8 | 7.42 | 72.27 | 2711.1 | 11 | 1.34 | 15 |
| 20320427 | 5683.7 | 3.25 | 32237 | 4211.1 | 8.68 | 69.73 | 2187.5 | 10 | 1.83 | 25 |
| 20320429 | 5621.1 | 3.82 | 32237 | 4176.7 | 8.67 | 69.81 | 2204.9 | 10 | 1.79 | 15 |
| 20320429 | 5826 | 1.97 | 32237 | 4115.7 | 8.68 | 69.81 | 2172 | 10 | 1.79 | 35 |
| 20320706 | 4301.2 | 17.37 | 3357 | 3366.2 | 6.84 | 48.54 | 2246.3 | 12 | 1.15 | 15 |

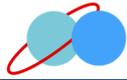

| Launch_Date | Launch_Mass | C3 | Path | Arrival_Mass | Arrival_Vinf | Arrival_Declination | Useful_Mass | TOF | OI_DV | P0 |
|---|---|---|---|---|---|---|---|---|---|---|
| 20321207 | 4912.8 | 10.7 | 32237 | 3684.4 | 7.86 | 71.28 | 2173.9 | 10 | 1.49 | 25 |
| 20321215 | 5916.3 | 1.18 | 3357 | 3845.4 | 7.56 | 48.17 | 2357.3 | 11 | 1.38 | 35 |
| 20321231 | 5292.4 | 6.91 | 32237 | 3401.4 | 7.77 | 71.22 | 2029.2 | 10 | 1.46 | 35 |
| 20330118 | 4652.6 | 13.45 | 32237 | 3206 | 9.11 | 68.92 | 1588.2 | 9 | 1.96 | 25 |
| 20330728 | 4907.3 | 10.76 | 3357 | 3345.7 | 6.07 | 54.83 | 2421.2 | 12 | 0.92 | 35 |
| 20330802 | 4559.1 | 14.47 | 3357 | 3396.1 | 6.06 | 54.86 | 2461.1 | 12 | 0.92 | 25 |
| 20330804 | 4542.5 | 14.65 | 3357 | 3379.9 | 7.13 | 53.04 | 2180.7 | 11 | 1.24 | 25 |
| 20330815 | 4260.1 | 17.85 | 3357 | 3287.9 | 6.03 | 54.94 | 2390.3 | 12 | 0.91 | 15 |
| 20330816 | 4249 | 17.98 | 3357 | 3272.1 | 7.09 | 53.1 | 2121.1 | 11 | 1.23 | 15 |
| 20330817 | 4235.5 | 18.13 | 3357 | 3246.4 | 8.5 | 51.73 | 1756.7 | 10 | 1.72 | 15 |
| 20340125 | 5158.8 | 8.22 | 3357 | 3057.7 | 7.86 | 52.55 | 1802.7 | 10 | 1.49 | 35 |
| 20340713 | 4376 | 16.52 | 337 | 3043.9 | 6.24 | 66.44 | 2166.3 | 12 | 0.97 | 25 |
| 20340714 | 4215.6 | 18.37 | 3357 | 2737.3 | 7.05 | 61.54 | 1765.6 | 11 | 1.24 | 25 |
| 20340717 | 4322.1 | 17.13 | 337 | 2888.6 | 8.27 | 60.7 | 1591.5 | 10 | 1.67 | 25 |
| 20340915 | 3206.6 | 31.48 | 3347 | 2299.2 | 6.86 | 59.16 | 1529.9 | 12 | 1.16 | 15 |
| 20340923 | 5188.5 | 7.93 | 32337 | 3683.9 | 9.5 | 45.88 | 1716.7 | 12 | 2.12 | 35 |
| 20341012 | 5689.4 | 3.2 | 3347 | 3674.4 | 7.88 | 56.08 | 2161.6 | 11 | 1.49 | 35 |
| 20341208 | 3764.2 | 23.9 | 3347 | 2372.3 | 6.64 | 59.51 | 1618 | 12 | 1.09 | 15 |
| 20350905 | 4928.6 | 10.54 | 3347 | 3268.2 | 6.08 | 62.05 | 2364.6 | 12 | 0.92 | 35 |
| 20350906 | 4912.8 | 10.7 | 3347 | 3244.3 | 6.88 | 58.89 | 2154.9 | 11 | 1.16 | 35 |
| 20350908 | 4867.5 | 11.17 | 3347 | 3199.3 | 8 | 56.13 | 1852.3 | 10 | 1.54 | 35 |
| 20350914 | 4552.2 | 14.55 | 3347 | 3285.3 | 6.05 | 61.94 | 2382.4 | 12 | 0.92 | 25 |
| 20350917 | 4517 | 14.93 | 3347 | 3257.4 | 6.85 | 58.86 | 2171.2 | 11 | 1.15 | 25 |
| 20350923 | 4458.2 | 15.59 | 3347 | 3192.4 | 7.95 | 56.16 | 1861.3 | 10 | 1.52 | 25 |
| 20351006 | 4106.1 | 19.66 | 3347 | 2996.1 | 6.8 | 59.04 | 2007.8 | 11 | 1.14 | 15 |
| 20360305 | 5206.8 | 7.75 | 32337 | 4161.7 | 8.74 | 42.53 | 2176 | 12 | 1.81 | 25 |
| 20360714 | 4753.9 | 12.37 | 337 | 3037.4 | 6.22 | 58.06 | 2165.1 | 12 | 0.96 | 35 |
| 20360718 | 4727.4 | 12.65 | 337 | 3010.1 | 7.05 | 54.88 | 1961.4 | 11 | 1.21 | 35 |
| 20360719 | 4707.3 | 12.86 | 337 | 2968.8 | 8.21 | 52.08 | 1670.9 | 10 | 1.61 | 35 |
| 20360722 | 4360.8 | 16.69 | 337 | 2968.8 | 7.03 | 54.75 | 1918.5 | 11 | 1.24 | 25 |
| 20360725 | 4333.4 | 17 | 337 | 2947.8 | 8.2 | 52.01 | 1662.4 | 10 | 1.61 | 25 |
| 20360807 | 3896.1 | 22.23 | 337 | 2769.8 | 7 | 55.1 | 1814.6 | 11 | 1.2 | 15 |
| 20361023 | 4606.2 | 13.96 | 337 | 3124.2 | 7.8 | 44.47 | 1857.4 | 12 | 1.47 | 15 |

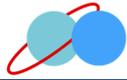

## SEP Trajectories to Uranus with Delta-IV Heavy

| Launch_Date | Launch_Mass | C3 | Path | Arrival_Mass | Arrival_Vinf | Arrival_Declination | Useful_Mass | TOF | OI_DV | P0 |
|---|---|---|---|---|---|---|---|---|---|---|
| 20250425 | 9063.7 | 7.37 | 33367 | 6771.3 | 8.93 | 76.84 | 3444.3 | 12 | 1.89 | 35 |
| 20250504 | 6847.7 | 22.58 | 33367 | 5368.4 | 8.88 | 76.87 | 2748.6 | 12 | 1.87 | 15 |
| 20250505 | 7543.2 | 17.41 | 33367 | 5877.7 | 8.88 | 76.87 | 3010.3 | 12 | 1.87 | 25 |
| 20250813 | 5544.3 | 33.63 | 337 | 4165.5 | 8.3 | 79.57 | 2286.8 | 11 | 1.68 | 25 |
| 20250818 | 4939.3 | 39.54 | 337 | 3863.1 | 8.3 | 79.5 | 2150 | 11 | 1.64 | 15 |
| 20250928 | 7302 | 19.15 | 3367 | 4716.7 | 8.71 | 73.81 | 2478 | 11 | 1.8 | 35 |
| 20260518 | 5128.2 | 37.63 | 3367 | 3960.1 | 9.38 | 73.55 | 1881.1 | 10 | 2.07 | 15 |
| 20260529 | 6604.7 | 24.5 | 3367 | 5147.9 | 9.33 | 73.58 | 2465.1 | 10 | 2.05 | 25 |
| 20260615 | 5960.2 | 29.88 | 3367 | 4506.5 | 11.34 | 72.71 | 1521.1 | 9 | 2.94 | 25 |
| 20260624 | 6971 | 21.63 | 337 | 4891.3 | 8.57 | 80.12 | 2619.1 | 10 | 1.75 | 35 |
| 20260803 | 5377.9 | 35.2 | 3367 | 4192.3 | 7.11 | 78.13 | 2710.9 | 12 | 1.23 | 25 |
| 20260807 | 4999.3 | 38.93 | 3367 | 3988.1 | 7.1 | 78.14 | 2583 | 12 | 1.23 | 15 |
| 20260812 | 5911.9 | 30.31 | 337 | 4230.3 | 8.3 | 81.79 | 2352.8 | 11 | 1.65 | 35 |
| 20260813 | 6955.6 | 21.75 | 3367 | 4955.2 | 10.89 | 72.75 | 1820.9 | 9 | 2.73 | 35 |
| 20260815 | 5599.9 | 33.12 | 337 | 4201.3 | 8.27 | 81.82 | 2316.3 | 11 | 1.67 | 25 |
| 20260817 | 7170 | 20.13 | 337 | 4488.2 | 7.17 | 81.09 | 2882.2 | 12 | 1.25 | 35 |
| 20260821 | 4966.7 | 39.26 | 337 | 3870.2 | 8.27 | 81.77 | 2161.6 | 11 | 1.64 | 15 |
| 20270531 | 4642.8 | 42.66 | 3367 | 3477.4 | 8.78 | 76.96 | 1808.6 | 10 | 1.83 | 15 |
| 20270623 | 6185.3 | 27.95 | 3367 | 4716 | 8.68 | 77.04 | 2486.6 | 10 | 1.79 | 25 |
| 20270711 | 5374.1 | 35.24 | 3367 | 3930.5 | 10.46 | 76.05 | 1561.6 | 9 | 2.53 | 25 |
| 20270809 | 5955.6 | 29.93 | 3367 | 4283.1 | 7.04 | 80.71 | 2793.1 | 12 | 1.21 | 35 |
| 20270810 | 5927.4 | 30.17 | 3367 | 4240.7 | 8.25 | 80.01 | 2374.2 | 11 | 1.63 | 35 |
| 20270816 | 5576.7 | 33.33 | 3367 | 4197.7 | 8.2 | 79.95 | 2335 | 11 | 1.65 | 25 |
| 20270816 | 5395.6 | 35.03 | 3367 | 4185.3 | 7.02 | 80.72 | 2735.7 | 12 | 1.21 | 25 |
| 20270822 | 7162.7 | 20.19 | 3367 | 5133.7 | 8.37 | 77.11 | 2827.4 | 10 | 1.67 | 35 |
| 20270822 | 4990.5 | 39.01 | 3367 | 3957.9 | 7 | 80.73 | 2591.9 | 12 | 1.2 | 15 |
| 20270823 | 4965.2 | 39.27 | 3367 | 3900.1 | 8.21 | 80.04 | 2197.6 | 11 | 1.61 | 15 |
| 20270905 | 6556.2 | 24.89 | 3367 | 4533.8 | 10.06 | 76.1 | 1928.9 | 9 | 2.35 | 35 |
| 20280624 | 7143.9 | 20.33 | 3367 | 5083.9 | 8.5 | 79.98 | 2749.4 | 10 | 1.72 | 35 |
| 20280714 | 8157.1 | 13.17 | 337 | 5787 | 8.39 | 77.04 | 3178.3 | 12 | 1.68 | 35 |
| 20280720 | 6412.2 | 26.06 | 337 | 4315.7 | 10.09 | 81.47 | 1826.9 | 9 | 2.37 | 35 |
| 20280723 | 5410.1 | 34.89 | 3367 | 3961.1 | 8.39 | 80.04 | 2176.8 | 11 | 1.68 | 25 |
| 20280807 | 3972.9 | 50.36 | 3367 | 2858.9 | 8.33 | 80.07 | 1584.7 | 10 | 1.66 | 15 |
| 20280820 | 5928.3 | 30.16 | 337 | 4243.6 | 8.22 | 80.44 | 2385.2 | 11 | 1.62 | 35 |
| 20280823 | 5618.6 | 32.95 | 337 | 4215.1 | 8.19 | 80.42 | 2350 | 11 | 1.64 | 25 |
| 20280829 | 4981.8 | 39.1 | 337 | 3885.2 | 8.19 | 80.42 | 2192.9 | 11 | 1.61 | 15 |
| 20290707 | 7012.8 | 21.32 | 337 | 4945.5 | 8.45 | 80.09 | 2694 | 10 | 1.7 | 35 |
| 20290827 | 5882.4 | 30.57 | 337 | 4218.1 | 8.18 | 77.41 | 2386.3 | 11 | 1.6 | 35 |
| 20290831 | 5574.7 | 33.35 | 337 | 4196.1 | 8.14 | 77.44 | 2356.2 | 11 | 1.62 | 25 |
| 20290905 | 4990.3 | 39.02 | 337 | 3898.1 | 8.14 | 77.41 | 2215 | 11 | 1.59 | 15 |
| 20291012 | 5658 | 32.58 | 337 | 3659.7 | 9.54 | 79.11 | 1694.2 | 9 | 2.14 | 35 |
| 20291027 | 9284.3 | 6.04 | 323357 | 8047.1 | 10.95 | 46.91 | 2923.6 | 12 | 2.76 | 35 |
| 20291028 | 9151.9 | 6.84 | 323357 | 8350.3 | 10.95 | 46.91 | 3036.6 | 12 | 2.75 | 15 |
| 20291129 | 9736.5 | 3.38 | 33357 | 8239.2 | 10.77 | 46.84 | 3098 | 12 | 2.67 | 25 |
| 20300706 | 5790.6 | 31.39 | 337 | 4059.6 | 10.06 | 71.49 | 1727.6 | 10 | 2.35 | 35 |
| 20300721 | 9932.1 | 2.26 | 332357 | 8553.1 | 9.57 | 46.54 | 3940.3 | 12 | 2.15 | 25 |
| 20300728 | 6461.8 | 25.65 | 337 | 4391.5 | 10 | 75.34 | 1888.8 | 9 | 2.33 | 35 |

| Launch_Date | Launch_Mass | C3 | Path | Arrival_Mass | Arrival_Vinf | Arrival_Declination | Useful_Mass | TOF | OI_DV | P0 |
|---|---|---|---|---|---|---|---|---|---|---|
| 20300728 | 9687.8 | 3.66 | 32237 | 7307.4 | 9.95 | 68.42 | 3167.2 | 11 | 2.31 | 35 |
| 20300809 | 10103.3 | 1.3 | 332357 | 8378.8 | 9.49 | 46.55 | 3912.1 | 12 | 2.11 | 35 |
| 20300915 | 9729.7 | 3.42 | 332357 | 8040.3 | 13.62 | 48.18 | 1572.6 | 10 | 4.19 | 25 |
| 20301113 | 9707.4 | 3.55 | 32357 | 8062.8 | 10.83 | 46.83 | 2927.8 | 11 | 2.76 | 25 |
| 20301127 | 8996.1 | 7.79 | 32357 | 7823.6 | 8.96 | 46.48 | 3962.9 | 12 | 1.9 | 15 |
| 20301128 | 8963.1 | 7.99 | 32357 | 7768.3 | 10.8 | 46.84 | 2902.2 | 11 | 2.69 | 15 |
| 20310111 | 9054.5 | 7.43 | 322237 | 7486.4 | 7.74 | 71.54 | 4481.9 | 12 | 1.45 | 15 |
| 20310506 | 9628.8 | 4 | 332237 | 7628.8 | 7.39 | 72.27 | 4773.1 | 12 | 1.33 | 25 |
| 20310612 | 10194.4 | 0.79 | 332237 | 7904.1 | 7.28 | 72.57 | 5008.9 | 12 | 1.29 | 35 |
| 20310619 | 10165.4 | 0.95 | 332237 | 7731.9 | 8.47 | 70.04 | 4198.5 | 11 | 1.71 | 35 |
| 20310713 | 9178.3 | 6.68 | 33357 | 7768.3 | 9.93 | 47.55 | 3379.9 | 11 | 2.3 | 15 |
| 20310714 | 8201.6 | 12.87 | 32357 | 7131.1 | 11.61 | 47.16 | 2283.8 | 10 | 3.07 | 25 |
| 20310714 | 10061.5 | 1.53 | 332237 | 7647 | 8.36 | 70.27 | 4161 | 11 | 1.71 | 25 |
| 20310716 | 7041.1 | 21.1 | 337 | 4984.5 | 8.37 | 73.15 | 2747.9 | 10 | 1.67 | 35 |
| 20310719 | 8107.4 | 13.5 | 32357 | 7134.4 | 11.57 | 47.14 | 2299.7 | 10 | 3.05 | 15 |
| 20310731 | 6492.5 | 25.4 | 337 | 4418.2 | 9.96 | 71.45 | 1913 | 9 | 2.31 | 35 |
| 20320227 | 7753.8 | 15.92 | 32237 | 6216.7 | 7.6 | 71.85 | 3791.5 | 11 | 1.4 | 15 |
| 20320422 | 9656.4 | 3.84 | 32237 | 7629.5 | 6.52 | 75.06 | 5272.8 | 12 | 1.05 | 35 |
| 20320424 | 9644.1 | 3.91 | 32237 | 7550.5 | 7.42 | 72.22 | 4705.1 | 11 | 1.34 | 35 |
| 20320425 | 9441.8 | 5.1 | 32237 | 7319.1 | 8.67 | 69.77 | 3807.7 | 10 | 1.83 | 25 |
| 20320425 | 9620.4 | 4.05 | 32237 | 7651.6 | 7.41 | 72.29 | 4724 | 11 | 1.36 | 25 |
| 20320428 | 9625 | 4.03 | 32237 | 7388.5 | 8.68 | 69.71 | 3897.7 | 10 | 1.79 | 35 |
| 20320511 | 6771.1 | 23.18 | 32237 | 5288.5 | 8.63 | 69.89 | 2810 | 10 | 1.77 | 15 |
| 20320522 | 8962.9 | 7.99 | 33357 | 7432.9 | 7.4 | 52.88 | 4642.8 | 12 | 1.33 | 15 |
| 20320712 | 7178 | 20.07 | 3357 | 6040.3 | 12.25 | 47.6 | 1690.4 | 9 | 3.39 | 25 |
| 20320715 | 6911.3 | 22.09 | 3357 | 5886.4 | 12.23 | 47.6 | 1655.1 | 9 | 3.38 | 15 |
| 20320801 | 9564.9 | 4.38 | 33357 | 7887.5 | 7.14 | 53.03 | 5084.7 | 12 | 1.24 | 25 |
| 20320805 | 6480.4 | 25.5 | 337 | 4441.3 | 9.9 | 67.32 | 1941.5 | 9 | 2.28 | 35 |
| 20321205 | 7775.7 | 15.77 | 32237 | 6278 | 7.86 | 71.16 | 3702.5 | 10 | 1.49 | 25 |
| 20330105 | 7450.1 | 18.08 | 32237 | 5309.2 | 11.14 | 67.22 | 1861.6 | 8 | 2.84 | 35 |
| 20330124 | 7651.3 | 16.64 | 32237 | 5575 | 7.69 | 71.24 | 3360.3 | 10 | 1.43 | 35 |
| 20330126 | 7469.8 | 17.93 | 32237 | 5381.3 | 9.07 | 68.97 | 2680.8 | 9 | 1.94 | 35 |
| 20330810 | 7642.2 | 16.7 | 3357 | 5984.2 | 6.04 | 54.91 | 4344.9 | 12 | 0.91 | 35 |
| 20330814 | 7579.7 | 17.15 | 3357 | 5951.2 | 7.1 | 53.09 | 3854.5 | 11 | 1.23 | 35 |
| 20330817 | 7221.6 | 19.75 | 3357 | 5970.9 | 6.02 | 54.95 | 4343 | 12 | 0.91 | 25 |
| 20330818 | 7192.1 | 19.97 | 3357 | 5947.4 | 7.08 | 53.11 | 3858.7 | 11 | 1.23 | 25 |
| 20330821 | 7109.6 | 20.58 | 3357 | 5802.3 | 10.37 | 50.78 | 2340.8 | 9 | 2.49 | 25 |
| 20330822 | 6842.3 | 22.62 | 3357 | 5745.2 | 6.01 | 55 | 4184.3 | 12 | 0.9 | 15 |
| 20330824 | 6775.1 | 23.15 | 3357 | 5667.2 | 7.07 | 53.15 | 3684.6 | 11 | 1.22 | 15 |
| 20330827 | 6615.7 | 24.41 | 3357 | 5504.8 | 8.46 | 51.77 | 2995.8 | 10 | 1.71 | 15 |
| 20330906 | 6168 | 28.1 | 3357 | 5058.9 | 10.28 | 50.82 | 2074.7 | 9 | 2.45 | 15 |
| 20331206 | 3734 | 53.36 | 3237 | 2564.9 | 7.86 | 70.93 | 1512.2 | 9 | 1.49 | 25 |
| 20340726 | 7096.6 | 20.68 | 337 | 5043 | 8.24 | 60.72 | 2827.6 | 10 | 1.62 | 35 |
| 20340813 | 6587 | 24.64 | 337 | 4521.7 | 9.8 | 58.79 | 2008.6 | 9 | 2.24 | 35 |
| 20340913 | 5971 | 29.79 | 3347 | 4380.1 | 7.99 | 56.36 | 2540 | 11 | 1.53 | 35 |
| 20340916 | 5476.5 | 34.27 | 3347 | 4309.3 | 7.98 | 56.39 | 2503.1 | 11 | 1.53 | 25 |
| 20340917 | 5450.8 | 34.51 | 3347 | 4240.6 | 9.53 | 54.15 | 1966.8 | 10 | 2.13 | 25 |
| 20340920 | 5096.7 | 37.95 | 3347 | 4107.6 | 7.96 | 56.4 | 2390.3 | 11 | 1.52 | 15 |

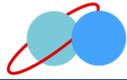

| Launch_Date | Launch_Mass | C3 | Path | Arrival_Mass | Arrival_Vinf | Arrival_Declination | Useful_Mass | TOF | OI_DV | P0 |
|---|---|---|---|---|---|---|---|---|---|---|
| 20340922 | 5073.2 | 38.18 | 3347 | 4011.5 | 9.51 | 54.19 | 1867.7 | 10 | 2.12 | 15 |
| 20340929 | 8501.9 | 10.9 | 32337 | 6681.7 | 9.48 | 45.98 | 3125.8 | 12 | 2.11 | 35 |
| 20341018 | 7776.6 | 15.76 | 32337 | 6482.2 | 9.39 | 46.15 | 3075.6 | 12 | 2.07 | 25 |
| 20341024 | 7594.1 | 17.04 | 32337 | 6414.7 | 9.36 | 46.31 | 3056.2 | 12 | 2.06 | 15 |
| 20350916 | 4546.1 | 43.71 | 3347 | 3416.1 | 6.06 | 62.05 | 2477 | 12 | 0.92 | 15 |
| 20350924 | 7627.8 | 16.81 | 3347 | 5788.8 | 6.03 | 61.99 | 4206.3 | 12 | 0.91 | 35 |
| 20350925 | 7585.7 | 17.1 | 3347 | 5733.5 | 6.83 | 58.84 | 3831.1 | 11 | 1.14 | 35 |
| 20350928 | 7554.8 | 17.32 | 3347 | 5597.9 | 7.93 | 56.06 | 3270.8 | 10 | 1.51 | 35 |
| 20351002 | 5056.5 | 38.35 | 337 | 3957 | 7.9 | 52.84 | 2321.1 | 11 | 1.5 | 15 |
| 20351006 | 4882.1 | 40.13 | 337 | 3773.3 | 9.38 | 50.71 | 1791.2 | 10 | 2.07 | 15 |
| 20351006 | 7126.3 | 20.46 | 3347 | 5485.9 | 6.8 | 58.91 | 3641.7 | 11 | 1.16 | 25 |
| 20351012 | 6671 | 23.97 | 3347 | 5238.3 | 6 | 62.11 | 3818.5 | 12 | 0.9 | 25 |
| 20351014 | 7045.1 | 21.07 | 3347 | 5016.3 | 9.4 | 53.93 | 2374 | 9 | 2.08 | 35 |
| 20351031 | 5825.7 | 31.07 | 3347 | 4416.8 | 7.82 | 56.31 | 2619.9 | 10 | 1.47 | 25 |
| 20351112 | 4892 | 40.02 | 3347 | 3511.2 | 9.26 | 54.16 | 1699.5 | 9 | 2.02 | 25 |
| 20360306 | 8935.4 | 8.16 | 32337 | 7359 | 8.74 | 42.57 | 3848.8 | 12 | 1.81 | 35 |
| 20360307 | 8773.6 | 9.17 | 32337 | 7546.9 | 8.74 | 42.6 | 3948.4 | 12 | 1.81 | 15 |
| 20360307 | 8785.3 | 9.1 | 32337 | 7528 | 8.73 | 42.56 | 3939.6 | 12 | 1.81 | 25 |
| 20360308 | 8913.2 | 8.3 | 32337 | 7224.3 | 10.58 | 40.59 | 2810.1 | 11 | 2.58 | 35 |
| 20360310 | 8756.1 | 9.28 | 32337 | 7264.6 | 10.57 | 40.66 | 2827.6 | 11 | 2.58 | 25 |
| 20360404 | 7345.5 | 18.84 | 32337 | 5896.7 | 10.43 | 40.94 | 2354.6 | 11 | 2.52 | 15 |
| 20360803 | 7140 | 20.36 | 337 | 5075.5 | 8.17 | 52.03 | 2873.9 | 10 | 1.6 | 35 |
| 20360815 | 4849.7 | 40.46 | 337 | 3750.6 | 9.7 | 45.73 | 1694.4 | 10 | 2.2 | 15 |
| 20360927 | 5116.8 | 37.74 | 3347 | 3824.2 | 9.33 | 47.19 | 1830 | 10 | 2.05 | 25 |

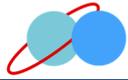

## *Attractive SEP Trajectories to Uranus for SLS*

| Launch_Date | Launch_Mass | C3 | Path | Arrival_Mass | Arrival_Vinf | Arrival_Declination | Useful_Mass | TOF | OI_DV | P0 |
|---|---|---|---|---|---|---|---|---|---|---|
| 20250516 | 6438.8 | 92.52 | 367 | 5094.2 | 10.51 | 67.87 | 2005.6 | 9 | 2.55 | 35 |
| 20250518 | 5705.2 | 97 | 367 | 4635.3 | 10.52 | 67.72 | 1783.2 | 9 | 2.61 | 25 |
| 20250608 | 15315.1 | 51.4 | 3367 | 14063.6 | 7.69 | 74.58 | 8473.4 | 12 | 1.43 | 25 |
| 20250609 | 15336.9 | 51.32 | 3367 | 13602.4 | 11.37 | 72.5 | 4453.6 | 10 | 3.01 | 25 |
| 20250609 | 15298.9 | 51.46 | 3367 | 13945.2 | 9.26 | 73.59 | 6748.5 | 11 | 2.02 | 25 |
| 20250613 | 15554.2 | 50.53 | 3367 | 13763 | 11.34 | 72.68 | 4648.2 | 10 | 2.94 | 35 |
| 20250618 | 12651.9 | 61.76 | 3367 | 11470 | 7.65 | 74.6 | 6945.1 | 12 | 1.42 | 15 |
| 20250622 | 11754.4 | 65.57 | 3367 | 10583.4 | 9.2 | 73.62 | 5171.1 | 11 | 1.99 | 15 |
| 20250705 | 15514.4 | 50.67 | 337 | 13448.2 | 8.52 | 79.94 | 7254.7 | 11 | 1.73 | 35 |
| 20250903 | 15510.6 | 50.68 | 337 | 13562.1 | 7.1 | 77.38 | 8786.4 | 12 | 1.23 | 35 |
| 20260526 | 10705.8 | 70.25 | 3367 | 9502.6 | 8.79 | 76.93 | 4932.6 | 11 | 1.83 | 15 |
| 20260528 | 6375.2 | 92.9 | 367 | 5033.8 | 9.35 | 69.05 | 2403.6 | 9 | 2.05 | 35 |
| 20260528 | 11850.7 | 65.15 | 3367 | 10645 | 7.36 | 77.99 | 6685.4 | 12 | 1.32 | 15 |
| 20260529 | 5911.9 | 95.71 | 367 | 4581.6 | 11.3 | 68.76 | 1558.5 | 8 | 2.92 | 35 |
| 20260530 | 5683.4 | 97.13 | 367 | 4616.1 | 9.35 | 68.85 | 2166.2 | 9 | 2.1 | 25 |
| 20260604 | 4640.4 | 103.91 | 367 | 3899.5 | 7.79 | 69.52 | 2321.9 | 10 | 1.46 | 15 |
| 20260622 | 15660.7 | 50.14 | 3367 | 14000 | 8.65 | 76.85 | 7306.2 | 11 | 1.82 | 25 |
| 20260622 | 15391.2 | 51.12 | 3367 | 14038.3 | 7.27 | 78.06 | 8907 | 12 | 1.29 | 25 |
| 20260628 | 15536.1 | 50.59 | 3367 | 13597.3 | 10.52 | 75.98 | 5345.7 | 10 | 2.56 | 35 |
| 20260703 | 13864.7 | 56.88 | 3367 | 12132.7 | 10.48 | 75.84 | 4700.9 | 10 | 2.59 | 25 |
| 20260824 | 15566.9 | 50.48 | 3367 | 13759.1 | 7.04 | 78.17 | 8976.6 | 12 | 1.21 | 35 |
| 20260825 | 15704.5 | 49.99 | 3367 | 13741.3 | 8.35 | 77.12 | 7587.3 | 11 | 1.67 | 35 |
| 20270608 | 9103.3 | 77.96 | 3367 | 7903.9 | 8.56 | 79.93 | 4241.6 | 11 | 1.74 | 15 |
| 20270609 | 10457.1 | 71.4 | 3367 | 9252.9 | 7.26 | 80.65 | 5880.6 | 12 | 1.28 | 15 |
| 20270611 | 5799.3 | 96.41 | 367 | 4468.9 | 10.12 | 70.68 | 1883.8 | 8 | 2.38 | 35 |
| 20270611 | 6215.6 | 93.86 | 367 | 4873.1 | 8.36 | 71.39 | 2688.6 | 9 | 1.67 | 35 |
| 20270612 | 5558.4 | 97.92 | 367 | 4483.5 | 8.35 | 71.17 | 2443.6 | 9 | 1.7 | 25 |
| 20270613 | 5151.9 | 100.52 | 367 | 4104.4 | 10.11 | 70.49 | 1697.3 | 8 | 2.43 | 25 |
| 20270615 | 4388.3 | 105.63 | 367 | 3645.2 | 8.34 | 71.38 | 2018 | 9 | 1.66 | 15 |
| 20270615 | 4616.4 | 104.07 | 367 | 3872.6 | 6.98 | 72.22 | 2542.3 | 10 | 1.19 | 15 |
| 20270624 | 14053.3 | 56.15 | 3367 | 12275.5 | 8.47 | 79.89 | 6579.4 | 11 | 1.75 | 25 |
| 20270625 | 14574 | 54.15 | 3367 | 13067.1 | 7.21 | 80.68 | 8349.7 | 12 | 1.27 | 25 |
| 20270804 | 10635.2 | 70.58 | 337 | 8941.6 | 9.97 | 81.6 | 3788.2 | 10 | 2.37 | 25 |
| 20270907 | 15602.3 | 50.35 | 3367 | 13506.4 | 8.14 | 80.06 | 7673.4 | 11 | 1.59 | 35 |
| 20270908 | 15692.3 | 50.03 | 3367 | 13795.6 | 6.95 | 80.75 | 9093.8 | 12 | 1.18 | 35 |
| 20270921 | 12609.7 | 61.94 | 3367 | 10583.3 | 9.67 | 79.24 | 4780.5 | 10 | 2.19 | 35 |
| 20280615 | 8393.5 | 81.6 | 337 | 7198 | 8.53 | 80.24 | 3878.1 | 11 | 1.73 | 15 |
| 20280624 | 5920.8 | 95.66 | 367 | 4589.8 | 7.64 | 74.63 | 2785.9 | 9 | 1.41 | 35 |
| 20280625 | 5535.6 | 98.06 | 367 | 4210.1 | 9.19 | 73.64 | 2060.3 | 8 | 1.99 | 35 |
| 20280625 | 5300.6 | 99.56 | 367 | 4245 | 7.61 | 74.42 | 2554.8 | 9 | 1.43 | 25 |
| 20280626 | 4927.7 | 101.99 | 367 | 3876.4 | 9.17 | 73.43 | 1871.2 | 8 | 2.03 | 25 |
| 20280626 | 10602.2 | 70.73 | 337 | 8848.8 | 10.15 | 78.89 | 3635.7 | 10 | 2.44 | 25 |
| 20280627 | 4459.2 | 105.14 | 367 | 3718.8 | 6.43 | 75.7 | 2594.3 | 10 | 1.02 | 15 |
| 20280628 | 3839.9 | 109.47 | 367 | 3101.9 | 9.16 | 73.62 | 1524.4 | 8 | 1.98 | 15 |
| 20280628 | 4204.8 | 106.9 | 367 | 3465.2 | 7.61 | 74.61 | 2109.1 | 9 | 1.4 | 15 |
| 20280628 | 13188.1 | 59.57 | 337 | 11421.1 | 8.45 | 80.22 | 6137.9 | 11 | 1.74 | 25 |
| 20280629 | 13533.3 | 58.18 | 337 | 12027.7 | 7.26 | 81.65 | 7648.3 | 12 | 1.28 | 25 |

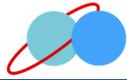

| Launch_Date | Launch_Mass | C3 | Path | Arrival_Mass | Arrival_Vinf | Arrival_Declination | Useful_Mass | TOF | OI_DV | P0 |
|---|---|---|---|---|---|---|---|---|---|---|
| 20280718 | 15730.6 | 49.89 | 337 | 13667.9 | 8.4 | 80.35 | 7496.3 | 11 | 1.68 | 35 |
| 20280728 | 10889.2 | 69.42 | 337 | 9720 | 8.33 | 77.13 | 5386.7 | 12 | 1.66 | 15 |
| 20280917 | 15520.2 | 50.65 | 337 | 13595.9 | 7 | 81.67 | 8909.2 | 12 | 1.2 | 35 |
| 20281015 | 9930.6 | 73.89 | 3367 | 7883.5 | 9.64 | 81.26 | 3593.4 | 10 | 2.18 | 35 |
| 20290709 | 5523 | 98.14 | 367 | 4205.6 | 7.22 | 78.12 | 2685.1 | 9 | 1.27 | 35 |
| 20290710 | 5119 | 100.74 | 367 | 3804.1 | 8.6 | 77.07 | 2030.4 | 8 | 1.76 | 35 |
| 20290710 | 4921.2 | 102.04 | 367 | 3874.8 | 7.19 | 77.99 | 2458.8 | 9 | 1.29 | 25 |
| 20290711 | 4543.8 | 104.56 | 367 | 3496.2 | 8.57 | 76.9 | 1847.9 | 8 | 1.78 | 25 |
| 20290724 | 15513.3 | 50.68 | 337 | 13517.8 | 8.35 | 77.22 | 7463.5 | 11 | 1.67 | 35 |
| 20290801 | 13309.3 | 59.08 | 337 | 11587.8 | 8.3 | 77.22 | 6364.6 | 11 | 1.68 | 25 |
| 20290805 | 13434.4 | 58.58 | 337 | 11288.3 | 9.95 | 75.55 | 4895.1 | 10 | 2.31 | 35 |
| 20290811 | 7381.9 | 87.08 | 337 | 6230.8 | 8.22 | 77.33 | 3502.1 | 11 | 1.62 | 15 |
| 20290822 | 6728.2 | 90.82 | 337 | 5591.4 | 9.85 | 75.71 | 2463.2 | 10 | 2.26 | 15 |
| 20290921 | 15735.5 | 49.87 | 337 | 13767.9 | 6.97 | 79.47 | 9054.9 | 12 | 1.19 | 35 |
| 20291004 | 12126.9 | 63.97 | 337 | 10689.3 | 6.93 | 79.5 | 7056.3 | 12 | 1.18 | 25 |
| 20291013 | 8292.8 | 82.13 | 337 | 7172.7 | 6.91 | 79.59 | 4747.5 | 12 | 1.17 | 15 |
| 20291014 | 8355.2 | 81.8 | 337 | 6946.9 | 9.49 | 76.05 | 3243.5 | 10 | 2.11 | 25 |
| 20300601 | 14698 | 53.68 | 3357 | 13515.9 | 11.84 | 47.31 | 4128.8 | 11 | 3.18 | 15 |
| 20300605 | 15258.1 | 51.61 | 3357 | 14270.8 | 9.78 | 46.55 | 6361.3 | 12 | 2.23 | 15 |
| 20300626 | 6726.8 | 90.83 | 337 | 5544.5 | 10.13 | 71.6 | 2332.3 | 10 | 2.38 | 15 |
| 20300704 | 9403.5 | 76.46 | 337 | 7897.8 | 10.09 | 71.59 | 3345.8 | 10 | 2.37 | 25 |
| 20300709 | 13739.4 | 57.37 | 337 | 12231.4 | 7.18 | 75.62 | 7850.4 | 12 | 1.26 | 25 |
| 20300719 | 13420.9 | 58.63 | 337 | 11247.1 | 10.01 | 71.59 | 4826.1 | 10 | 2.33 | 35 |
| 20300725 | 4606 | 104.14 | 367 | 3304.3 | 8.37 | 80.04 | 1819.8 | 8 | 1.67 | 35 |
| 20300725 | 5008 | 101.46 | 367 | 3693.3 | 7.13 | 80.73 | 2383.9 | 9 | 1.24 | 35 |
| 20300726 | 4050.2 | 107.98 | 367 | 3020 | 8.33 | 79.95 | 1650.7 | 8 | 1.69 | 25 |
| 20300726 | 4436.6 | 105.3 | 367 | 3399.6 | 7.09 | 80.69 | 2183.2 | 9 | 1.25 | 25 |
| 20300729 | 15543.9 | 50.56 | 337 | 13551.9 | 8.31 | 73.52 | 7530.5 | 11 | 1.65 | 35 |
| 20300806 | 13346.8 | 58.93 | 337 | 11615.8 | 8.26 | 73.52 | 6415.4 | 11 | 1.67 | 25 |
| 20300927 | 15734 | 49.88 | 337 | 13778.2 | 6.93 | 76.24 | 9101.7 | 12 | 1.18 | 35 |
| 20310306 | 15110.6 | 52.15 | 32237 | 14094.4 | 8.89 | 69.66 | 7105 | 11 | 1.91 | 25 |
| 20310318 | 9884.8 | 74.11 | 357 | 9114.5 | 11.93 | 48.2 | 2732.6 | 10 | 3.23 | 15 |
| 20310717 | 15294.6 | 51.47 | 3357 | 14096 | 14.04 | 48.59 | 2564.7 | 9 | 4.34 | 25 |
| 20310720 | 15326.2 | 51.36 | 3357 | 14299.6 | 7.95 | 46.49 | 8331.3 | 12 | 1.52 | 25 |
| 20310722 | 12889 | 60.78 | 3357 | 11783.3 | 14.01 | 48.57 | 2162.6 | 9 | 4.32 | 15 |
| 20310723 | 15195.2 | 51.84 | 3357 | 14369.4 | 9.87 | 47.57 | 6308.5 | 11 | 2.27 | 15 |
| 20310723 | 15200.8 | 51.82 | 3357 | 14375.4 | 8.13 | 47.88 | 8187.7 | 12 | 1.58 | 15 |
| 20310802 | 15555.6 | 50.52 | 337 | 13585.4 | 8.26 | 69.53 | 7594.4 | 11 | 1.63 | 35 |
| 20310813 | 13773.6 | 57.24 | 337 | 11622 | 9.83 | 67.53 | 5121.9 | 10 | 2.26 | 35 |
| 20310915 | 15018.5 | 52.49 | 32357 | 13070 | 13.65 | 48.23 | 2634.5 | 9 | 4.12 | 35 |
| 20310916 | 15000 | 52.56 | 32357 | 13705.4 | 11.22 | 46.95 | 4622.7 | 10 | 2.94 | 25 |
| 20311001 | 15734.5 | 49.88 | 337 | 13771.1 | 6.89 | 72.54 | 9132.9 | 12 | 1.17 | 35 |
| 20320515 | 15390 | 51.12 | 32237 | 13353.3 | 7.36 | 72.4 | 8382.5 | 11 | 1.32 | 35 |
| 20320516 | 15355.8 | 51.25 | 32237 | 13260.8 | 7.35 | 72.49 | 8247 | 11 | 1.34 | 25 |
| 20320518 | 14833.3 | 53.18 | 32237 | 12300.7 | 8.6 | 69.86 | 6560.7 | 10 | 1.76 | 35 |
| 20320617 | 15258.3 | 51.6 | 3357 | 14106.4 | 10.07 | 47.55 | 5994.6 | 10 | 2.36 | 15 |
| 20320822 | 15052 | 52.36 | 3357 | 13938.1 | 8 | 47.9 | 8075.8 | 11 | 1.54 | 15 |
| 20320822 | 15219.5 | 51.75 | 3357 | 14117 | 6.69 | 48.67 | 9580.7 | 12 | 1.1 | 15 |

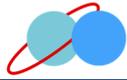

| Launch_Date | Launch_Mass | C3 | Path | Arrival_Mass | Arrival_Vinf | Arrival_Declination | Useful_Mass | TOF | OI_DV | P0 |
|---|---|---|---|---|---|---|---|---|---|---|
| 20320823 | 15358 | 51.24 | 3357 | 14046.1 | 11.93 | 47.52 | 4094.1 | 9 | 3.29 | 25 |
| 20320823 | 15284.9 | 51.51 | 3357 | 14141.8 | 9.7 | 47.56 | 6390.1 | 10 | 2.2 | 25 |
| 20320824 | 15386.3 | 51.14 | 3357 | 13878.7 | 11.94 | 47.57 | 4153.2 | 9 | 3.23 | 35 |
| 20320824 | 15298.6 | 51.46 | 3357 | 14181.4 | 6.68 | 48.67 | 9630.2 | 12 | 1.1 | 25 |
| 20320824 | 15305.8 | 51.43 | 3357 | 13900.4 | 6.68 | 48.67 | 9441.6 | 12 | 1.1 | 35 |
| 20320904 | 13737.4 | 57.38 | 3357 | 12608.2 | 12.87 | 50.2 | 3074.6 | 9 | 3.71 | 15 |
| 20330428 | 9224.1 | 77.35 | 357 | 8435.6 | 8.28 | 46.39 | 4708.6 | 10 | 1.64 | 15 |
| 20330430 | 11014.4 | 68.85 | 357 | 9670.4 | 12.03 | 47.42 | 2836.5 | 8 | 3.28 | 35 |
| 20330501 | 10477.7 | 71.31 | 357 | 9401.9 | 12.01 | 47.36 | 2695.7 | 8 | 3.33 | 25 |
| 20330507 | 13006.4 | 60.3 | 3357 | 11801.1 | 7.46 | 52.82 | 7321.3 | 11 | 1.35 | 15 |
| 20330508 | 13609.9 | 57.88 | 3357 | 12402.2 | 6.3 | 54.55 | 8768.8 | 12 | 0.99 | 15 |
| 20330713 | 15326.5 | 51.35 | 3357 | 13975.2 | 7.22 | 53.1 | 8921.8 | 11 | 1.27 | 25 |
| 20330716 | 14901 | 52.92 | 3357 | 13182.3 | 10.61 | 50.81 | 4989.2 | 9 | 2.65 | 25 |
| 20330726 | 15522.4 | 50.64 | 3357 | 14086.6 | 6.08 | 54.81 | 10189.4 | 12 | 0.92 | 25 |
| 20330930 | 15399.9 | 51.09 | 3357 | 13415 | 10.14 | 50.86 | 5635.1 | 9 | 2.39 | 35 |
| 20330930 | 15377.4 | 51.17 | 3357 | 13605.2 | 8.3 | 51.84 | 7563.9 | 10 | 1.65 | 35 |
| 20330930 | 15369.8 | 51.2 | 3357 | 13692.1 | 6.94 | 53.27 | 9032.2 | 11 | 1.18 | 35 |
| 20330930 | 15623.2 | 50.28 | 3357 | 13903.5 | 5.91 | 55.17 | 10226.6 | 12 | 0.88 | 35 |
| 20331001 | 14404.8 | 54.79 | 3357 | 12754.9 | 8.3 | 51.86 | 7004 | 10 | 1.68 | 25 |
| 20331014 | 8411.2 | 81.51 | 3357 | 7334.8 | 10.05 | 50.86 | 3126.9 | 9 | 2.35 | 15 |
| 20340607 | 9944.1 | 73.83 | 357 | 8609.6 | 10.13 | 47.58 | 3618.9 | 8 | 2.39 | 35 |
| 20340607 | 10235.3 | 72.44 | 357 | 8893.4 | 8.32 | 47.85 | 4930.8 | 9 | 1.66 | 35 |
| 20340608 | 9394.4 | 76.5 | 357 | 8072.8 | 12.53 | 47.65 | 2127.4 | 7 | 3.53 | 35 |
| 20340608 | 9457.2 | 76.19 | 357 | 8399.6 | 10.11 | 47.54 | 3474 | 8 | 2.43 | 25 |
| 20340608 | 9750.9 | 74.76 | 357 | 8684.3 | 8.31 | 47.83 | 4759.5 | 9 | 1.69 | 25 |
| 20340608 | 9916 | 73.96 | 357 | 8841.9 | 6.93 | 48.53 | 5781.7 | 10 | 1.2 | 25 |
| 20340609 | 8913.5 | 78.92 | 357 | 7862.4 | 12.5 | 47.58 | 2020.7 | 7 | 3.59 | 25 |
| 20340609 | 8911.1 | 78.93 | 357 | 8148.9 | 8.31 | 47.83 | 4526.6 | 9 | 1.65 | 15 |
| 20340609 | 9072.9 | 78.11 | 357 | 8312.1 | 6.93 | 48.5 | 5490.7 | 10 | 1.18 | 15 |
| 20340609 | 9185.8 | 77.54 | 357 | 8421.5 | 5.86 | 49.67 | 6224.6 | 11 | 0.86 | 15 |
| 20341012 | 15311.2 | 51.41 | 3347 | 13884.8 | 7.88 | 56.45 | 8164.2 | 11 | 1.5 | 25 |
| 20341012 | 15357.9 | 51.24 | 3347 | 14111.7 | 6.79 | 59.27 | 9466.6 | 12 | 1.13 | 25 |
| 20341012 | 15299.4 | 51.45 | 3347 | 13830.1 | 6.79 | 59.29 | 9278 | 12 | 1.13 | 35 |
| 20341013 | 15577.5 | 50.44 | 3347 | 13850.6 | 7.88 | 56.55 | 8147.8 | 11 | 1.49 | 35 |
| 20341016 | 15170.5 | 51.93 | 3347 | 13053.9 | 9.39 | 54.2 | 6190.3 | 10 | 2.07 | 35 |
| 20341021 | 12576.9 | 62.07 | 3347 | 11441.5 | 6.77 | 59.39 | 7693.5 | 12 | 1.13 | 15 |
| 20350714 | 7901.9 | 84.22 | 357 | 6576.7 | 7.22 | 53.12 | 4198.6 | 9 | 1.27 | 35 |
| 20350714 | 8164.3 | 82.82 | 357 | 6827.8 | 6.13 | 54.95 | 4914.7 | 10 | 0.94 | 35 |
| 20350715 | 7557.9 | 86.11 | 357 | 6239 | 8.67 | 51.78 | 3293.2 | 8 | 1.79 | 35 |
| 20350715 | 7399.4 | 86.99 | 357 | 6347.9 | 7.21 | 53.09 | 4015.3 | 9 | 1.3 | 25 |
| 20350715 | 7659.6 | 85.55 | 357 | 6600.8 | 6.13 | 54.88 | 4716.6 | 10 | 0.96 | 25 |
| 20350716 | 7047.2 | 88.98 | 357 | 5736.4 | 10.63 | 50.84 | 2210.5 | 7 | 2.61 | 35 |
| 20350716 | 7049.8 | 88.96 | 357 | 6004.2 | 8.66 | 51.73 | 3131.3 | 8 | 1.82 | 25 |
| 20350716 | 6751.2 | 90.68 | 357 | 6002.2 | 6.11 | 54.88 | 4326.3 | 10 | 0.93 | 15 |
| 20350716 | 6964 | 89.45 | 357 | 6212.8 | 5.29 | 57.17 | 4838.4 | 11 | 0.72 | 15 |
| 20350717 | 6545.5 | 91.89 | 357 | 5500.3 | 10.61 | 50.78 | 2083.6 | 7 | 2.65 | 25 |
| 20350717 | 6132.6 | 94.36 | 357 | 5386.6 | 8.66 | 51.73 | 2851.1 | 8 | 1.78 | 15 |
| 20350718 | 5606 | 97.62 | 357 | 4862.2 | 10.6 | 50.8 | 1882.2 | 7 | 2.6 | 15 |

| Launch_Date | Launch_Mass | C3 | Path | Arrival_Mass | Arrival_Vinf | Arrival_Declination | Useful_Mass | TOF | OI_DV | P0 |
|---|---|---|---|---|---|---|---|---|---|---|---|
| 20350721 | 10098.8 | 73.09 | 337 | 8893.2 | 7.05 | 55.04 | 5796.7 | 12 | 1.21 | 15 |
| 20350804 | 14170.2 | 55.69 | 337 | 12661.6 | 7.01 | 55.09 | 8287.3 | 12 | 1.2 | 25 |
| 20350822 | 15533.1 | 50.6 | 337 | 13604.2 | 8.1 | 52.54 | 7777.5 | 11 | 1.57 | 35 |
| 20350828 | 13853.1 | 56.92 | 337 | 12126.4 | 8.08 | 52.47 | 6862.7 | 11 | 1.6 | 25 |
| 20351022 | 15478.5 | 50.8 | 337 | 13589.4 | 6.77 | 56.02 | 9133.5 | 12 | 1.13 | 35 |
| 20351205 | 7437.9 | 86.77 | 3347 | 5465.2 | 9.15 | 54.2 | 2687.8 | 9 | 1.98 | 35 |
| 20360826 | 15754.2 | 49.81 | 337 | 13786.4 | 8.07 | 48.18 | 7911.3 | 11 | 1.56 | 35 |
| 20360901 | 14371.7 | 54.92 | 337 | 12213.1 | 9.62 | 45.86 | 5583.7 | 10 | 2.17 | 35 |
| 20360901 | 13863.5 | 56.88 | 337 | 12127.3 | 8.05 | 48.02 | 6890.1 | 11 | 1.59 | 25 |
| 20360901 | 11369.8 | 67.26 | 337 | 10202 | 8.03 | 43.52 | 5887.7 | 12 | 1.55 | 15 |
| 20360911 | 11593.2 | 66.27 | 337 | 9882.5 | 9.57 | 45.92 | 4474.3 | 10 | 2.19 | 25 |
| 20360925 | 8392.2 | 81.61 | 337 | 6735.9 | 11.67 | 44.47 | 2074.6 | 9 | 3.16 | 25 |
| 20361012 | 4119.2 | 107.49 | 337 | 3023.1 | 7.92 | 53 | 1768.3 | 10 | 1.51 | 15 |
| 20361025 | 15603.5 | 50.35 | 337 | 13697.6 | 6.76 | 51.67 | 9222.5 | 12 | 1.12 | 35 |
| 20361105 | 12475.8 | 62.5 | 337 | 11042 | 6.74 | 51.75 | 7454.2 | 12 | 1.12 | 25 |
| 20361116 | 7524.1 | 86.29 | 337 | 6413.9 | 7.73 | 49.02 | 3846.8 | 11 | 1.44 | 15 |

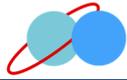

## A.8.2 Neptune

### Chemical Mission Options to Neptune

| Launch Date | C3 | Path | Arrival_Mass_551 | Arrival_Mass_D4H | Arrival_Mass_SLS | Arrival_Vinf | Arrival_Declination | Useful_Mass_551 | Useful_Mass_D4H | Useful_Mass_SLS | TOF | OI_DV | DSM |
|---|---|---|---|---|---|---|---|---|---|---|---|---|---|
| 20261112 | 24.86 | 322238 | 1932 | 3482.9 | 12616.6 | 14.57 | 14.08 | 346.3 | 624.2 | 2261.2 | 13 | 4.37 | 2.11 |
| 20261213 | 52.56 | 322238 | 1340.7 | 2620.2 | 10250.2 | 14.4 | 14.1 | 251.7 | 491.8 | 1924.1 | 13 | 4.27 | 1.29 |
| 20280607 | 22.53 | 323358 | 3274.6 | 5870 | 21118.7 | 14.75 | 8.97 | 559.7 | 1003.3 | 3609.5 | 13 | 4.47 | 0.58 |
| 20290328 | 26.65 | 33458 | 2811.3 | 5098.2 | 18565.1 | 14.43 | 8.48 | 522.8 | 948.1 | 3452.5 | 12 | 4.29 | 0.78 |
| 20290328 | 26.49 | 33458 | 3063.3 | 5552.2 | 20209 | 12.65 | 8.4 | 862.1 | 1562.6 | 5687.5 | 13 | 3.38 | 0.52 |
| 20300607 | 52.56 | 3358 | 1250.9 | 2444.5 | 9563.1 | 12.81 | 8.98 | 340.5 | 665.3 | 2602.8 | 12 | 3.45 | 1.51 |
| 20300611 | 52.56 | 3358 | 1214.1 | 2372.7 | 9282 | 14.7 | 8.94 | 210.3 | 411 | 1607.8 | 11 | 4.44 | 1.6 |
| 20300802 | 11.53 | 33458 | 695.1 | 1221.1 | 4250.7 | 10.72 | 8.36 | 281.1 | 493.8 | 1719 | 13 | 2.48 | 6.23 |

### SEP Trajectories to Neptune for Atlas V (551)

| Launch_Date | Launch_Mass | C3 | Path | Arrival_Mass | Arrival_Vinf | Arrival_Declination | Useful_Mass | TOF | OI_DV | P0 |
|---|---|---|---|---|---|---|---|---|---|---|
| 20290506 | 5812.1 | 2.1 | 33358 | 4494.4 | 12.97 | 8.98 | 1220.4 | 13 | 3.46 | 15 |
| 20290508 | 5861 | 1.66 | 33358 | 4462.1 | 12.95 | 8.98 | 1178.4 | 13 | 3.52 | 25 |
| 20290522 | 5892.2 | 1.39 | 33358 | 4260.9 | 12.89 | 8.98 | 1176.6 | 13 | 3.42 | 35 |
| 20300411 | 4854.6 | 11.31 | 3358 | 3320.1 | 11.49 | 9.05 | 1201.1 | 13 | 2.77 | 35 |
| 20300417 | 4664.3 | 13.33 | 3358 | 3381.3 | 11.47 | 9.05 | 1228.3 | 13 | 2.76 | 25 |
| 20300427 | 4247.8 | 17.99 | 3358 | 3268.3 | 11.42 | 9.05 | 1196.2 | 13 | 2.74 | 15 |

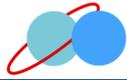

## SEP Trajectories to Neptune for Delta-IV Heavy

| Launch_Date | Launch_Mass | C3 | Path | Arrival_Mass | Arrival_Vinf | Arrival_Declination | Useful_Mass | TOF | OI_DV | P0 |
|---|---|---|---|---|---|---|---|---|---|---|
| 20261025 | 7811 | 15.52 | 322238 | 6522.2 | 14.67 | 14.07 | 1186.3 | 13 | 4.34 | 25 |
| 20261025 | 7837.2 | 15.34 | 322238 | 6352.3 | 14.68 | 14.07 | 1154.2 | 13 | 4.34 | 35 |
| 20261029 | 7500.9 | 17.71 | 322238 | 6439.5 | 14.66 | 14.07 | 1176.4 | 13 | 4.33 | 15 |
| 20271130 | 7483.9 | 17.83 | 3358 | 5855 | 14.25 | 8.85 | 1185.9 | 13 | 4.11 | 35 |
| 20280407 | 8994 | 7.8 | 33358 | 7622.4 | 14.37 | 8.48 | 1497.3 | 13 | 4.18 | 15 |
| 20280407 | 9836.8 | 2.8 | 33358 | 7980.4 | 14.37 | 8.48 | 1568.6 | 13 | 4.18 | 35 |
| 20280412 | 9713.7 | 3.51 | 33358 | 8040.1 | 14.34 | 8.48 | 1591.9 | 13 | 4.16 | 25 |
| 20290319 | 7111.3 | 20.57 | 33458 | 6202.5 | 12.69 | 8.4 | 1782.8 | 13 | 3.33 | 15 |
| 20290320 | 7086 | 20.76 | 33458 | 6143.1 | 14.48 | 8.49 | 1176.2 | 12 | 4.23 | 15 |
| 20290505 | 9844.5 | 2.76 | 33358 | 7905.2 | 12.97 | 8.98 | 2146 | 13 | 3.46 | 25 |
| 20290505 | 9917 | 2.35 | 33358 | 7849.9 | 12.97 | 8.98 | 2130.5 | 13 | 3.46 | 35 |
| 20290513 | 9796.2 | 3.04 | 33358 | 7665.5 | 14.86 | 8.94 | 1325.7 | 12 | 4.44 | 35 |
| 20291023 | 7346.3 | 18.83 | 32358 | 5675.9 | 13.98 | 8.99 | 1229 | 12 | 3.97 | 25 |
| 20300425 | 7592.8 | 17.05 | 3358 | 5954.6 | 11.44 | 9.05 | 2174.9 | 13 | 2.74 | 35 |
| 20300428 | 7365.6 | 18.69 | 3358 | 5978.6 | 11.42 | 9.05 | 2189.5 | 13 | 2.74 | 25 |
| 20300429 | 7511.8 | 17.63 | 3358 | 5885.1 | 13 | 8.98 | 1586.2 | 12 | 3.48 | 35 |
| 20300430 | 7321.8 | 19.01 | 3358 | 5928.4 | 13 | 8.84 | 1548.2 | 12 | 3.55 | 25 |
| 20300505 | 6735.3 | 23.46 | 3358 | 5625.2 | 11.39 | 9.06 | 2069.3 | 13 | 2.72 | 15 |
| 20300510 | 6496.6 | 25.37 | 3358 | 5383.1 | 12.95 | 8.98 | 1467.5 | 12 | 3.45 | 15 |
| 20310203 | 3892.2 | 51.36 | 3458 | 2832.7 | 10.12 | 8.37 | 1287.7 | 13 | 2.18 | 25 |
| 20310204 | 4068.1 | 49.2 | 3458 | 2745.8 | 11.38 | 8.37 | 1012.5 | 12 | 2.72 | 35 |
| 20310204 | 4128.2 | 48.49 | 3458 | 2800 | 10.12 | 8.38 | 1272.5 | 13 | 2.18 | 35 |
| 20310205 | 3384.7 | 58.02 | 3458 | 2629.9 | 10.11 | 8.37 | 1197 | 13 | 2.18 | 15 |
| 20320316 | 3415.4 | 57.59 | 358 | 2091.6 | 9.17 | 9.37 | 1091.5 | 13 | 1.82 | 35 |
| 20320318 | 3071.7 | 62.5 | 358 | 2023.1 | 9.16 | 9.35 | 1057 | 13 | 1.81 | 25 |

## SEP Trajectories to Neptune for SLS

| Launch_Date | Launch_Mass | C3 | Path | Arrival_Mass | Arrival_Vinf | Arrival_Declination | Useful_Mass | TOF | OI_DV | P0 |
|---|---|---|---|---|---|---|---|---|---|---|
| 20261215 | 15073.4 | 52.29 | 322238 | 13671.1 | 14.39 | 14.09 | 2677.3 | 13 | 4.19 | 35 |
| 20280402 | 15244.3 | 51.66 | 3358 | 13880.9 | 13.65 | 8.76 | 3240.2 | 13 | 3.8 | 25 |
| 20280403 | 15306.7 | 51.43 | 3358 | 13798.4 | 13.65 | 8.76 | 3224.8 | 13 | 3.8 | 35 |
| 20280405 | 15221 | 51.74 | 3358 | 14381.9 | 14.39 | 8.48 | 2813.9 | 13 | 4.19 | 15 |
| 20290216 | 15218.8 | 51.75 | 3358 | 14147.6 | 14.66 | 8.49 | 2583.4 | 12 | 4.33 | 15 |
| 20290501 | 15169.7 | 51.93 | 33458 | 13994.2 | 12.5 | 8.39 | 4177.3 | 13 | 3.24 | 35 |
| 20290504 | 15214.2 | 51.77 | 3358 | 14167.5 | 12.49 | 8.39 | 4244.1 | 13 | 3.23 | 15 |
| 20290506 | 15330.6 | 51.34 | 3358 | 14116.1 | 14.23 | 8.36 | 2767.6 | 12 | 4.18 | 25 |
| 20290506 | 15358.5 | 51.24 | 3358 | 13952.3 | 14.22 | 8.47 | 2850 | 12 | 4.1 | 35 |
| 20290506 | 15169.3 | 51.93 | 3358 | 14012.2 | 12.48 | 8.39 | 4205.2 | 13 | 3.22 | 25 |
| 20300325 | 15413.7 | 51.04 | 3358 | 13938.8 | 11.57 | 9.07 | 4973.5 | 13 | 2.8 | 25 |
| 20300331 | 12115.6 | 64.02 | 3358 | 10977.2 | 11.54 | 9.07 | 3933.3 | 13 | 2.79 | 15 |
| 20300404 | 11021.8 | 68.81 | 3358 | 9889.2 | 13.14 | 9 | 2588.9 | 12 | 3.54 | 15 |
| 20300611 | 15425.8 | 50.99 | 3358 | 13563.3 | 12.79 | 8.98 | 3817.3 | 12 | 3.38 | 35 |
| 20300611 | 15346.7 | 51.28 | 3358 | 13612.3 | 11.25 | 9.06 | 5135.5 | 13 | 2.66 | 35 |
| 20300613 | 14463.8 | 54.57 | 3358 | 12431.5 | 14.68 | 8.94 | 2255.2 | 11 | 4.34 | 35 |
| 20300614 | 13697.9 | 57.53 | 3358 | 12052.9 | 12.79 | 8.85 | 3294.2 | 12 | 3.44 | 25 |
| 20300619 | 11854.6 | 65.14 | 3358 | 10234.3 | 14.65 | 8.94 | 1873.7 | 11 | 4.33 | 25 |
| 20300628 | 7537.3 | 86.22 | 3358 | 6472.6 | 14.59 | 8.93 | 1202.1 | 11 | 4.3 | 15 |
| 20310208 | 11239.1 | 67.84 | 3458 | 10186.1 | 12.88 | 8.29 | 2733.1 | 11 | 3.49 | 25 |
| 20310208 | 11518.3 | 66.6 | 3458 | 10459.8 | 11.37 | 8.24 | 3776.3 | 12 | 2.77 | 25 |
| 20310208 | 11893.4 | 64.97 | 3458 | 10565.8 | 11.36 | 8.37 | 3909.5 | 12 | 2.71 | 35 |
| 20310208 | 11639.4 | 66.07 | 3458 | 10590.5 | 10.1 | 8.37 | 4826.8 | 13 | 2.18 | 25 |
| 20310208 | 12094.6 | 64.11 | 3458 | 10758 | 10.1 | 8.37 | 4901.5 | 13 | 2.18 | 35 |
| 20310209 | 11591.2 | 66.28 | 3458 | 10281.2 | 12.87 | 8.41 | 2847.7 | 11 | 3.41 | 35 |
| 20310209 | 11106.4 | 68.43 | 3458 | 10373.8 | 10.1 | 8.36 | 4730.6 | 13 | 2.17 | 15 |
| 20310210 | 11131.5 | 68.32 | 3458 | 9834.5 | 14.69 | 8.5 | 1780.5 | 10 | 4.35 | 35 |
| 20310224 | 6644.1 | 91.29 | 3358 | 5888.8 | 14.61 | 8.49 | 1088.9 | 10 | 4.31 | 15 |
| 20320324 | 8395 | 81.6 | 358 | 7072.9 | 9.14 | 9.35 | 3704.9 | 13 | 1.81 | 35 |
| 20320325 | 7891.4 | 84.28 | 358 | 6578.5 | 11.57 | 9.07 | 2346.9 | 11 | 2.8 | 35 |
| 20320325 | 7680.4 | 85.43 | 358 | 6629.7 | 10.24 | 9.18 | 2956.5 | 12 | 2.23 | 25 |
| 20320325 | 8182.1 | 82.72 | 358 | 6864.2 | 10.24 | 9.19 | 3058.9 | 12 | 2.23 | 35 |
| 20320325 | 7893.1 | 84.27 | 358 | 6838 | 9.14 | 9.34 | 3583.9 | 13 | 1.81 | 25 |
| 20320326 | 7000.7 | 89.24 | 358 | 5960.6 | 13.18 | 8.86 | 1496.9 | 10 | 3.64 | 25 |
| 20320326 | 7478.7 | 86.54 | 358 | 6171.4 | 13.18 | 9 | 1600.9 | 10 | 3.57 | 35 |
| 20320326 | 7414.4 | 86.9 | 358 | 6370.4 | 11.57 | 8.92 | 2216.4 | 11 | 2.86 | 25 |
| 20320327 | 6793.2 | 90.44 | 358 | 6046.8 | 10.23 | 9.17 | 2699.8 | 12 | 2.23 | 15 |
| 20320327 | 7005.9 | 89.21 | 358 | 6258.8 | 9.13 | 9.33 | 3283.6 | 13 | 1.8 | 15 |
| 20320410 | 3597.1 | 111.22 | 3358 | 2855.8 | 11.5 | 9.07 | 1031 | 11 | 2.77 | 15 |
| 20330503 | 5071.6 | 101.05 | 358 | 3760.2 | 8.56 | 11.35 | 2129 | 13 | 1.6 | 35 |
| 20330504 | 4527 | 104.68 | 358 | 3228.4 | 10.69 | 10.76 | 1340 | 11 | 2.42 | 15 |
| 20330504 | 4285.9 | 106.33 | 358 | 3258.2 | 9.52 | 10.78 | 1592.2 | 12 | 1.99 | 25 |
| 20330504 | 4826.5 | 102.66 | 358 | 3515.1 | 9.52 | 11.04 | 1745.1 | 12 | 1.95 | 35 |
| 20330504 | 4521.3 | 104.72 | 358 | 3480.5 | 8.55 | 11.34 | 1972.3 | 13 | 1.6 | 25 |
| 20330505 | 3988.5 | 108.41 | 358 | 2946.8 | 10.68 | 10.74 | 1224.5 | 11 | 2.42 | 25 |
| 20330506 | 3528.4 | 111.73 | 358 | 2796.9 | 8.55 | 11.32 | 1586.6 | 13 | 1.59 | 15 |

# B   A-TEAM STUDY REPORT

This appendix provides the A-Team Study Report from the Ice Giants Workshop (March 29–31, 2016).

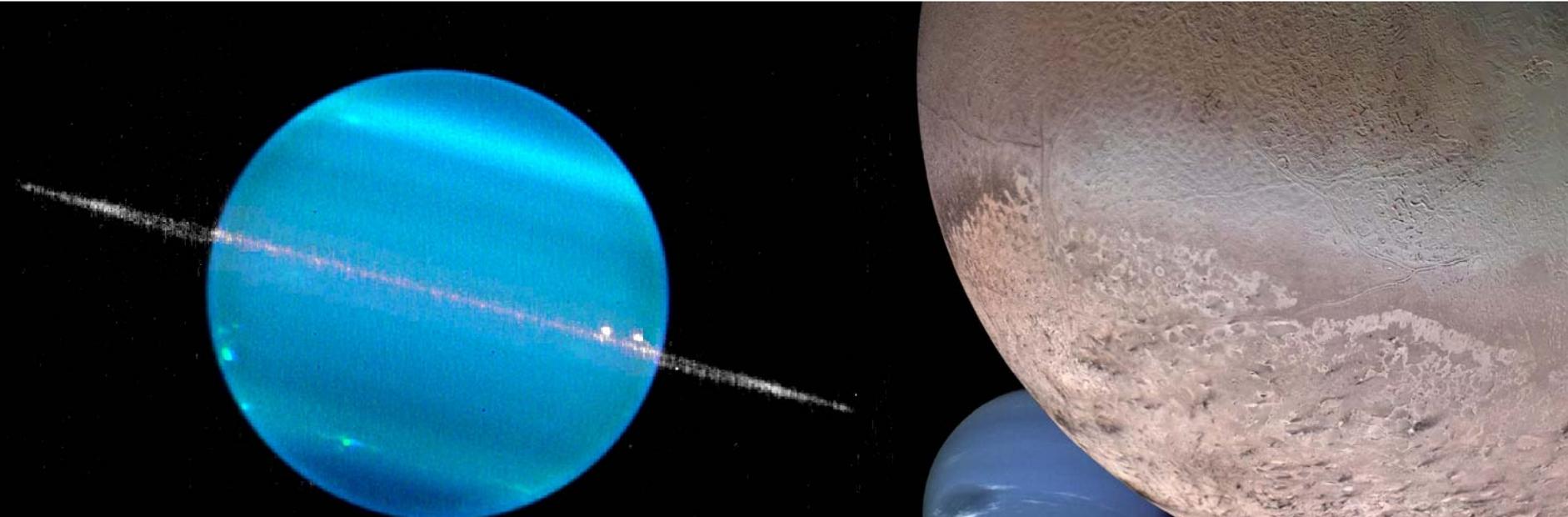

# ICE GIANT WORKSHOP
## A-Team Study
### March 29-31, 2016

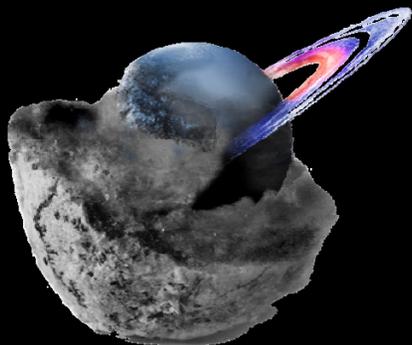

Client:       Jim Cutts
Client Lead:  Mark Hofstadter
Client Lead:  John Elliott
Study Lead:   Paul Johnson
Assist. SL:   Jonathan Murphy
Facilitator:  Randii Wessen

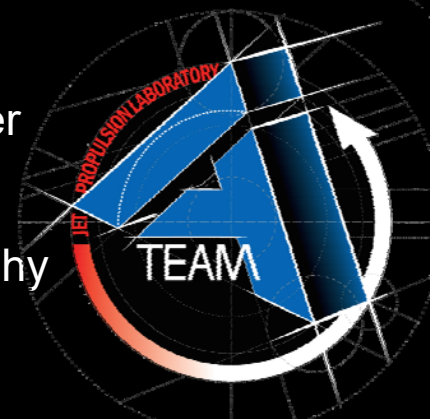

**JPL Innovation Foundry**

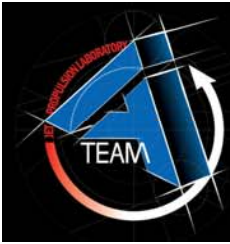

# Study Team

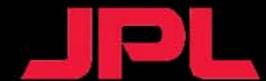

- Jim Cutts — Client
- Mark Hofstadter* — Client Lead-SDT
- John Elliott* — Client Lead
- Paul Johnson* — Study Lead
- Jonathan Murphy* — Assistant Study Lead
- Randii Wessen* — Facilitator
- Melissa Brown* — Logistics
- Terri Anderson* — Costing
- Jessie Kawata* — Visual Strategy
- Barry Colman — Documentarian
- Christopher Guethe — Documentarian
- Samin Asmar — SME -Radio Science
- Joe Lazio — SME -Radio Science
- Chet Borden* — Client Team
- Steve Matousek — Client Team
- Neil Murphy — SME-Doppler Imaging
- Anastassios Petropoulos — SME-Trajectories
- Nitin Arora — SME-Trajectories
- Kim Reh — Client Team
- Tom Spilker — Client Team
- Young Lee — Client Team
- Amy Simon — Client Team-SDT

- Don Banfield — Client Team-SDT
- Jonathan Fortney — Client Team-SDT
- Alexander Hayes — Client Team-SDT
- Matthew Hedman — Client Team-SDT
- George Hospodarsky — Client Team-SDT
- Kathleen Mandt — Client Team-SDT
- Adam Masters — Client Team-SDT
- Mark Showalter — Client Team-SDT
- Krista Soderlund — Client Team-SDT
- Diego Turrini — Client Team-SDT

*A-Team Core Member

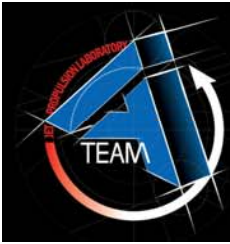
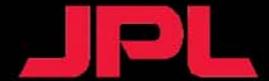

- Study Goal:
  - This was a science focused Workshop to identify the range and scope of science that can be accomplished at Uranus and Neptune. Architectures commensurate with the range of science were identified and evaluated. The goal of the study was to determine which architectures provide the most compelling ice giant science.

- Summary:
  - Findings of the last decadal survey were summarized highlighting changes since.
  - Presentations describing potential ice giant science were given.
  - Results of the Ice Giant Prep study were reviewed.
  - The SDT prioritized science investigations in their working STM.
  - Nominal flyby and orbiter payloads were identified that address the prioritized science.
  - A science value matrix ranking the science value provided by a wide range of considered architectures for each prioritized science investigation.

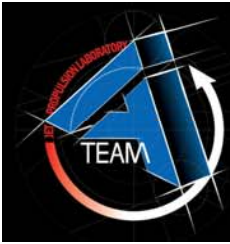
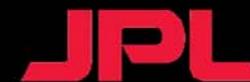

- The ice giants were recognized as an important target for future scientific exploration in the 2013–2022 Planetary Science Decadal Survey, with the number three flagship mission recommendation being a mission to explore Uranus.

- Building on this, NASA began a study of mission options, including science and technology options, for exploring the Ice Giant planets with the intent to provide information for the deliberative process for the next Decadal Survey.

- This A-Team workshop is a part of this broader effort.

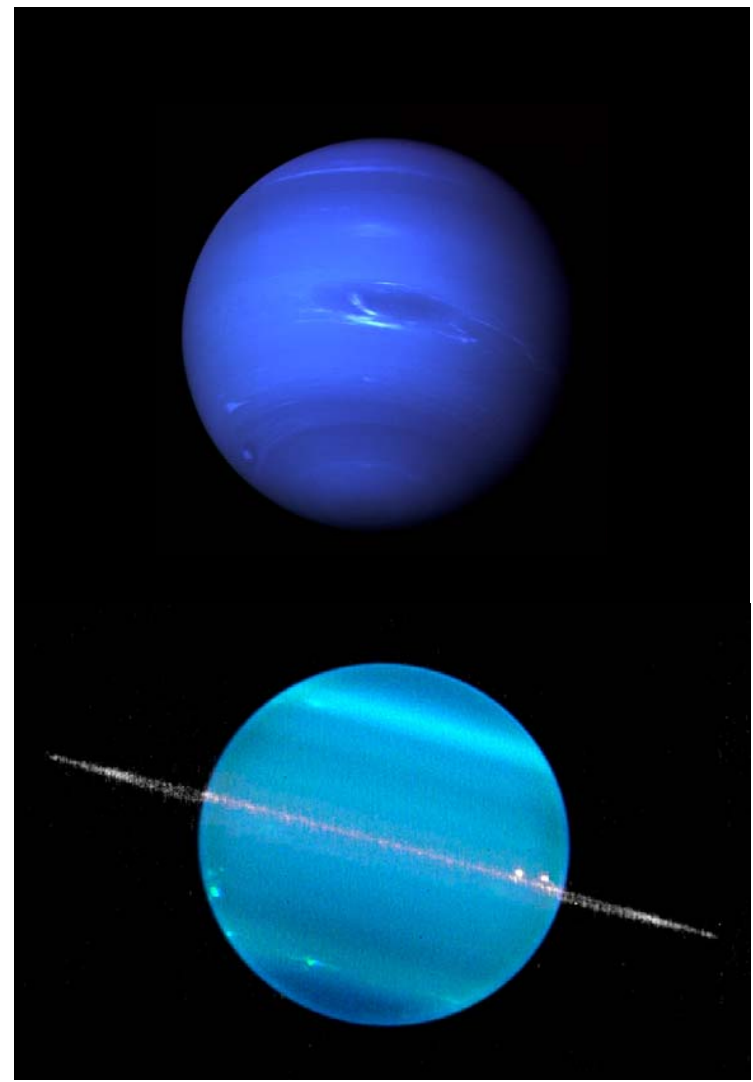

- Identify the top 6-10 architectures under $2B that achieve the most compelling science.

- Flesh out these architectures so that each can be used to generate detailed point designs (i.e., taken to Team X).

- Determine what the 'next steps' are.

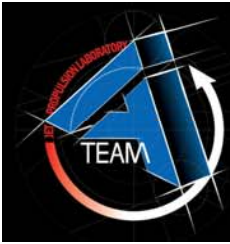

# Methods

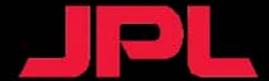

- The study consisted of 3 full day sessions.

- In the first session, findings of the last decadal survey were summarized highlighting changes since. Members of the SDT then gave presentations describing ice giant science that a mission could address. This was was followed by a report on results of the Ice Giant Prep study looking at cost, possible trajectories, power options, possible architectures, etc. Presentations on Radio Science and Doppler Imaging were also given.

- The second session began with the SDT prioritizing the science investigations in their working STM. The full team then discussed and identified nominal small (30kg), medium (100kg), and large (200kg) payloads on flyby or orbiting space craft that could address the prioritized science.

- In the third Session, a Science Value Matrix was developed which assessed how well various architectures were able to address the prioritized science. The top ranking architectures were then assigned rough costs based on the A-Team CML 2 cost tool. This enabled an evaluation of the science value achieved per unit cost ($) for the top architectures. Actions needed going forward were then discussed and identified.

- Throughout the study, preliminary visual prototypes were designed to help convey the science questions discussed during the study. These were intended to initiate the development of visual aids for the larger NASA Ice Giant study.

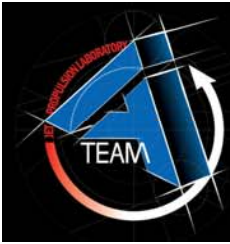

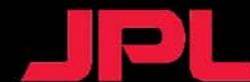

# Results and Discussion

- The following slides present an executive summary of the results.

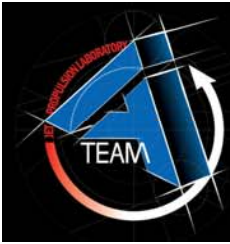
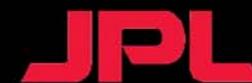

# INPUT FROM PREP STUDY

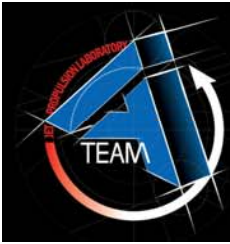

# Ice Giant Prep Study

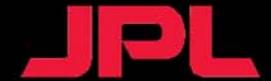

- The weeks prior to the Ice Giant Workshop, a study was held to get the JPL team prepared in terms of general engineering considerations, new technologies and possible architectures.

- Key elements discussed included aerocapture, chemical and SEP trajectory options, cryogenic propulsion, power options, hibernation, optical com, costs, key trades and architecture options.

- Some key results going into the workshop are summarized on the following slides.

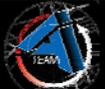

| | | | |  | | |
|---|---|---|---|---|---|
| Primary Flight Elements | Flyby S/C | Orbiter | Dual Orbiter | | | |
| Secondary Flight Elements | Neptune/Uranus Probe | Triton Probe (SYLPH) | Impactor | Free Flying sub-S/C | Lander | Separable KBO flyby S/C |
| Primary Power | Solar Concentrators | eMMRTG | Battery | | | |
| Science Pointing and Ops | Body Fixed | Individual Instrument Scanning | Scan Platform | Gimbaled HGA | Multiple Flybys w/Focused Obj. | |
| Ops and Data Return Technology | Deployable Antenna | Dual Polarization | Arrayed 34m DSN | 70m DSN | Optical Telecom | Autonomous Nav/Ops |
| Launch Vehicles/Options | Falcon 9 Heavy | Delta IV Heavy | Atlas V | Added Upper Stage w/in Fairing | SLS | Dual S/C, 1 Launch |
| Propulsion Systems | Chemical | SEP Stage | REP | REP&SEP | Cryopropulsion | |
| Cruise Approach | Chemical | GA | Low Thrust Inner System | Hibernation | | |
| NOI Approach | Propulsive NOI | Neptune/Uranus Aerocapture | Low Thrust Outer System | Long Term Cryogenics | Flyby | Aerobraking |
| Neptune Trajectory | Flyby | Low Periapse Orbit | Triton Limited Tour | Full Tour | | |
| Satellite Trajectory | Flyby | Multiple Flybys | View from Distance | Lander | Impactor | Orbit/Loosely Captured Orbit |
| Uranus Trajectory | Flyby | Orbit (polar, equatorial, both) | | | | |
| Uranus Rings Trajectory | Fly Nearby | View from Distance | Ring Crossing/Intersection | Become a Ring Particle | Flythrough | |
| "Bonus" Trajectory | KBO | Multiple KBO Flybys (w/luck) | Centaur | | | |
| Time Scales | Time of Flight to Neptune: ≤13 years | Time of Flight to Uranus: ≤11 years | Tour Duration ≤2 years | Primary Neptune Mission: ≤15 years | Primary Uranus Mission: ≤13 years | |

## Uranus Architectures

**Uranus Flyby** 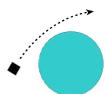

Flyby w/ Probe(s) 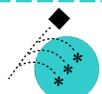

Flyby w/ Sub S/C 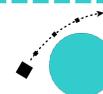

**Uranus Orbiter** 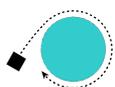

Moon Orbiter 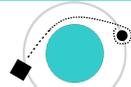

Orbiter w/ Moon Tour 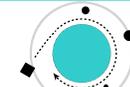

Ring Hover 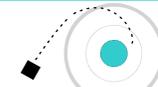

Orbiter w/ Probe(s) 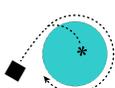

Orbiter w/ Sub S/C 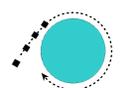

Orbiter w/ Moon Lander 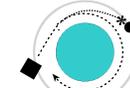

Orbiter w/ Icy Moon Penetrator 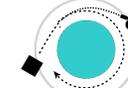

**Uranus Dual Orbiter** 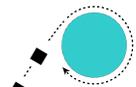

Dual Orbiter w/ Probe 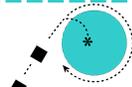

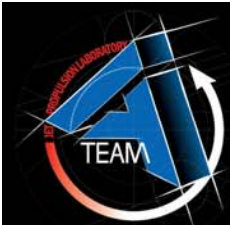

# Architectures Considered

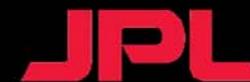

## Neptune Architectures

**Neptune Flyby**

Flyby w/ Probe(s)

Flyby w/ Sub S/C

**Neptune Orbiter**

**Triton** Orbiter

Orbiter w/ Moon Tour

Ring Hover

Orbiter w/ Probe(s)

Orbiter w/ Sub S/C

Orbiter w/ **Triton** Lander

Orbiter w/ **Triton** Penetrator

**Neptune Dual Orbiter**

Dual Orbiter w/ Probe

## Hybrid Architectures

**Multiple Bodies**

Note: Flyby S/C can be instrumented SEP stage

Uranus Orbiter, KBO Flyby

Neptune Orbiter, KBO Flyby

Neptune and Uranus Orbiters

Uranus Orbiter, Neptune Flyby

Neptune Orbiter, Uranus Flyby

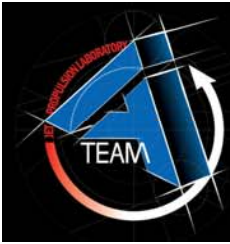

# Preliminary Cost Estimates

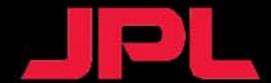

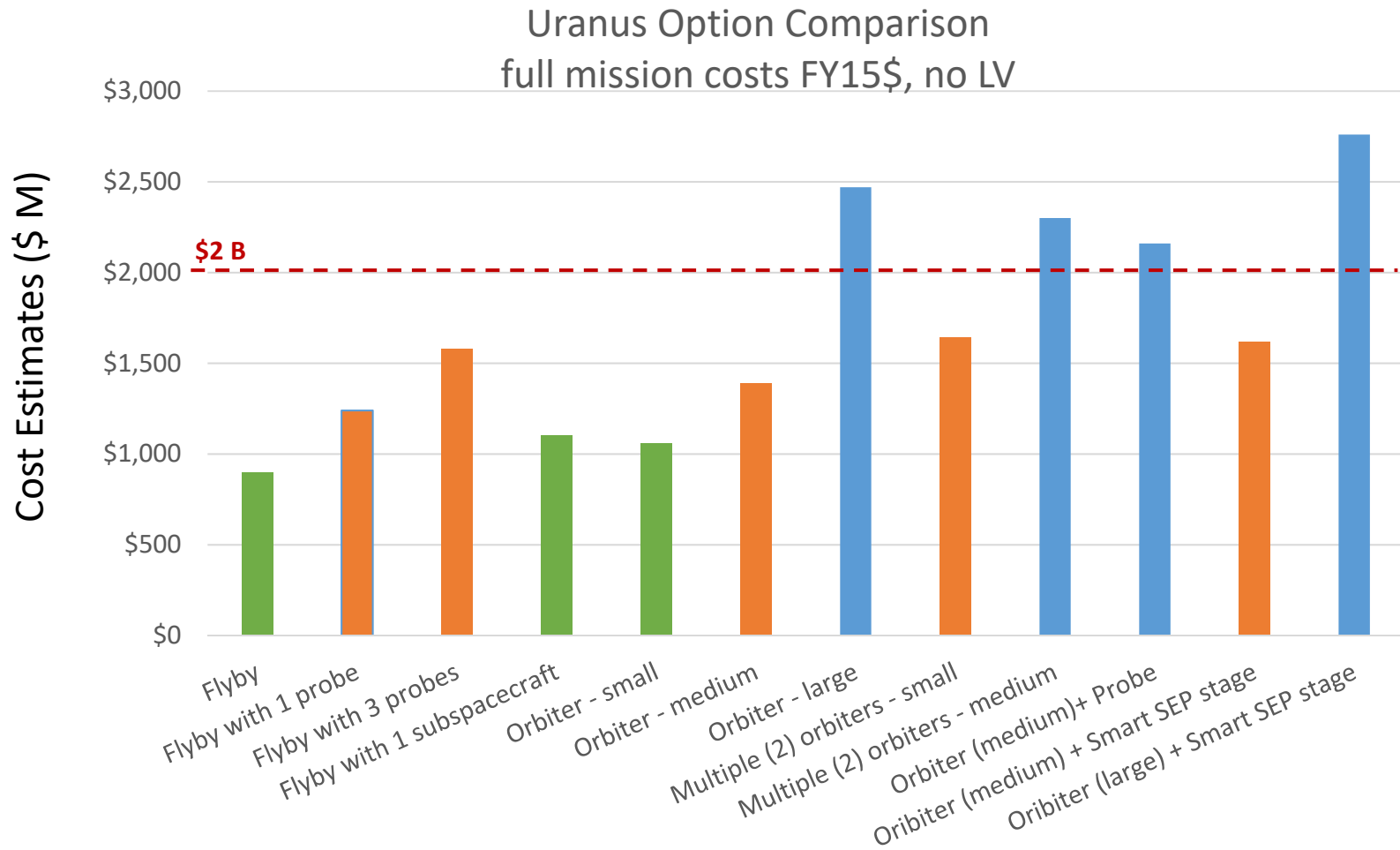

Green = New Frontiers,    Orange = Small Flagship,    Blue = Large Flagship

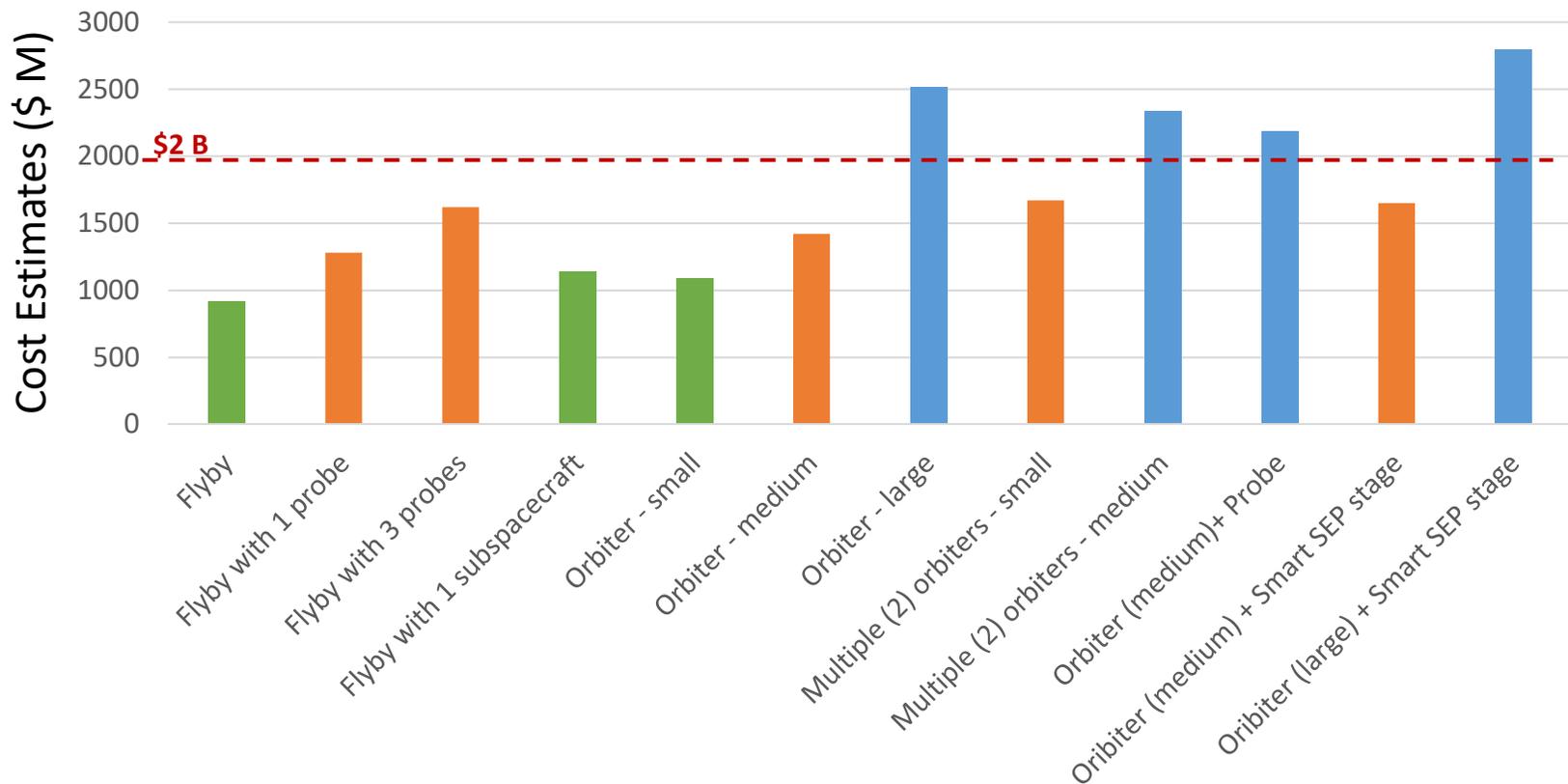

Neptune Option Comparison
full mission costs FY15$, no LV

Green = New Frontiers,   Orange = Small Flagship,     Blue = Large Flagship

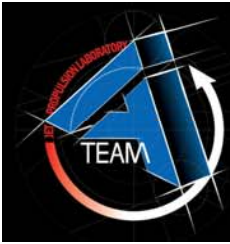
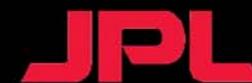

# KEY WORKSHOP FINDINGS

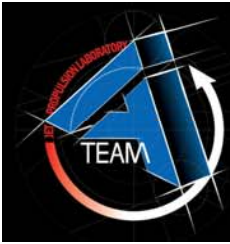

# SDT-Preliminary Science Prioritization

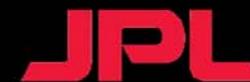

| Science Priority | Votes | Science Investigation | Category |
|---|---|---|---|
| High | 13 | Constrain the structure and characteristics of the interior | Atmosphere |
| High | 12 | Determine bulk composition of planet | Atmosphere |
| Medium | 7 | Determine atmospheric heat balance | Atmosphere |
| Medium | 5 | Determine bulk density, mass distribution, internal structure of the satellites | Satellites |
| Medium | 5 | Surface composition of the rings and moons | Rings/Satellites |
| Medium | 4 | Measure tropospheric 3-D flow | Atmosphere |
| Medium | 4 | Ring temporal changes and fine scale structures | Rings |
| Medium | 3 | Map satellite shape and surface geology | Satellites |
| Medium | 3 | Investigate solar wind - magnetosphere - ionosphere& Constrain plasma transport in magnetosphere | Magnetosphere |

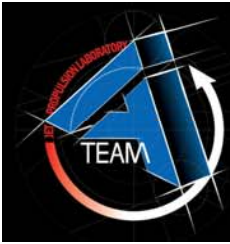

# Top Architectures
## (including *hybrid* options)

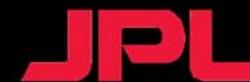

| Science Rank | Description | Science Value Score | Cost | Science/G$ |
|---|---|---|---|---|
| 1 | *Uranus and Neptune orbiters + 2 probes* | 78 | $3,000 M | 26 |
| 2 | *Uranus and Neptune orbiters + 1 probe* | 73 | $2,500 M | 29 |
| 3 | *Uranus and Neptune orbiters* | 67 | $2,000 M | 34 |
| 4 | *Uranus orbiter + Neptune flyby* | 50 | TBD | TBD |
| 5 | Uranus dual orbiter + probe | 45 | $2,450 M | 18 |
| 6 | Neptune dual orbiter + probe | 45 | $2,480 M | 18 |
| 7 | Uranus dual orbiter | 40 | $1,930 M | 21 |
| 8 | Neptune dual orbiter | 40 | $1,960 M | 20 |
| 9 | Uranus orbiter + probe | 39 | $2,010 M | 19 |
| 10 | Neptune orbiter + probe | 39 | $2,040 M | 19 |
| 11 | *Uranus and Neptune flyby + 1 probe* | 39 | TBD | TBD |
| 12 | Triton orbiter | 38 | TBD | TBD |
| 13 | Neptune orbiter | 34 | $1,150 M | 30 |
| 14 | Uranus orbiter | 33 | $1,120 M | 29 |
| 15 | Neptune flyby + probe | 24 | $1,150 M | 21 |
| 16 | Neptune flyby | 17 | $860 M | 20 |

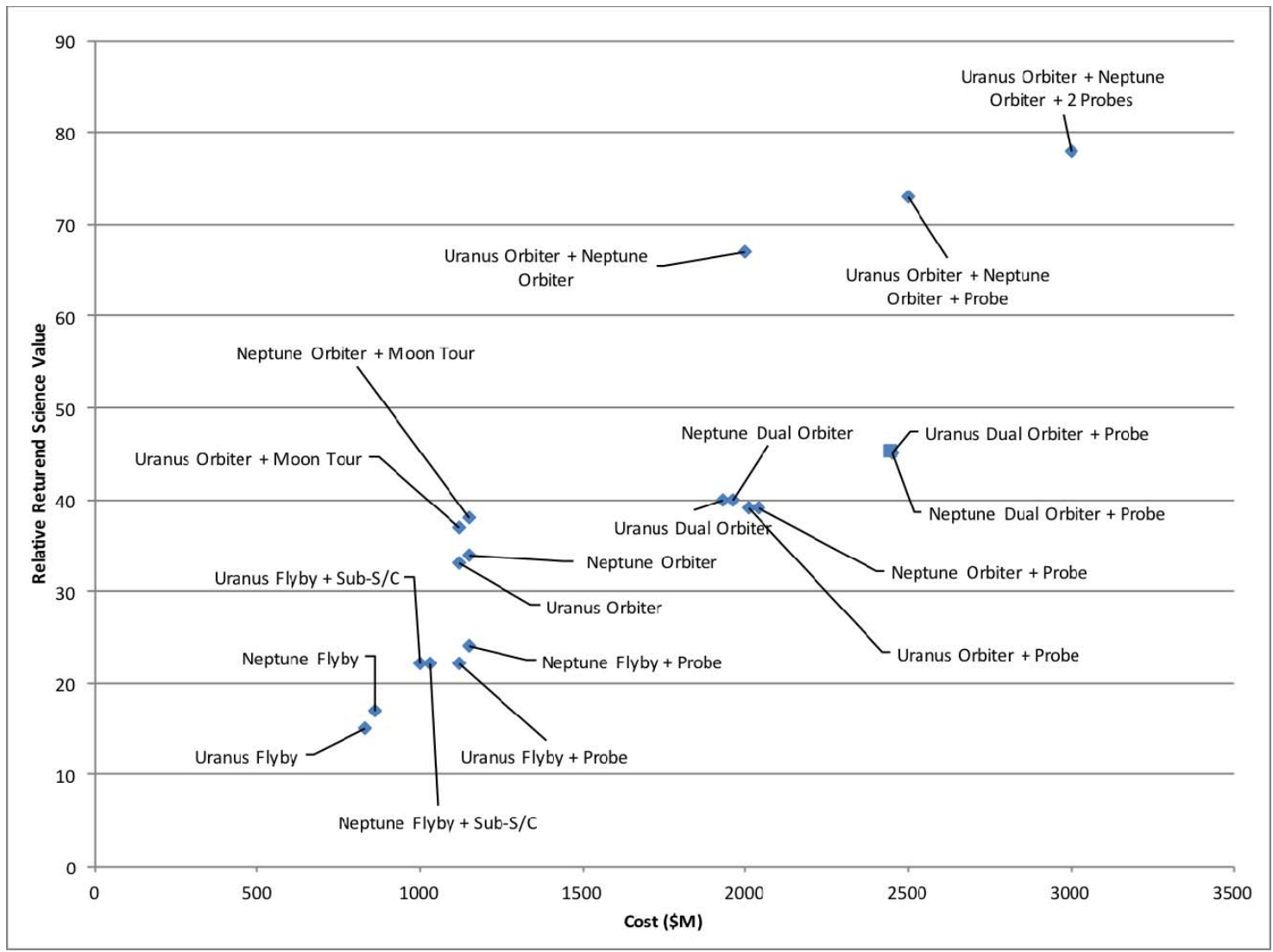

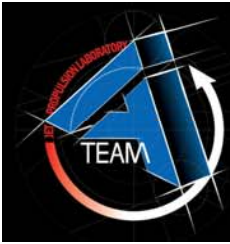

# Visual Prototypes

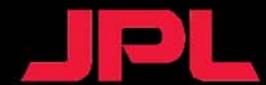

A. CONSTRAIN INTERIOR

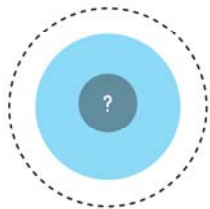

B. BULK COMPOSITION

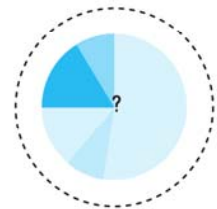

C. HEAT BALANCE

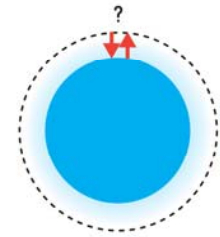

D. SATELLITE BULK DENSITY

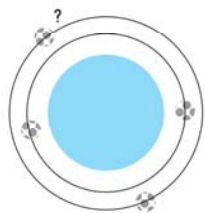

E. SATELLITE SURFACE COMP

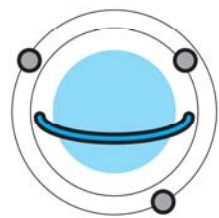

F. TROPOSPHERIC 3D FLOW

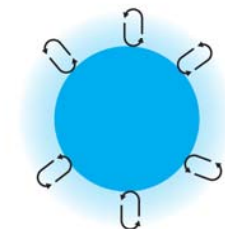

F'. RING STRUCTURE

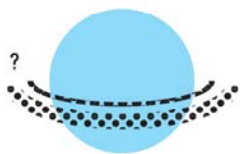

G. SATELLITE SURFACE GEOLOGY

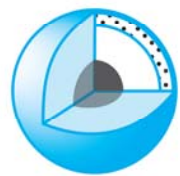

G'. PLANETARY MAGNETOSPHERE

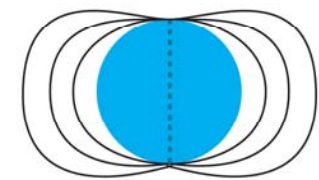

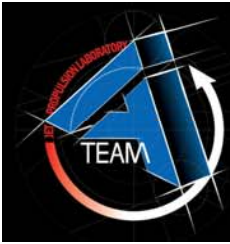
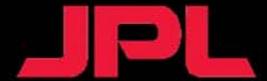

- First cut at top 8 architectures (done)
- Cost for 80 kg payload (+$100M per mission)
- Prioritize science questions

# C    TEAM X STUDY REPORT

This appendix provides the Executive Summary from the Team X Ice Giants Study Report (June and July 2016).

# Ice Giants Study
# Executive Summary

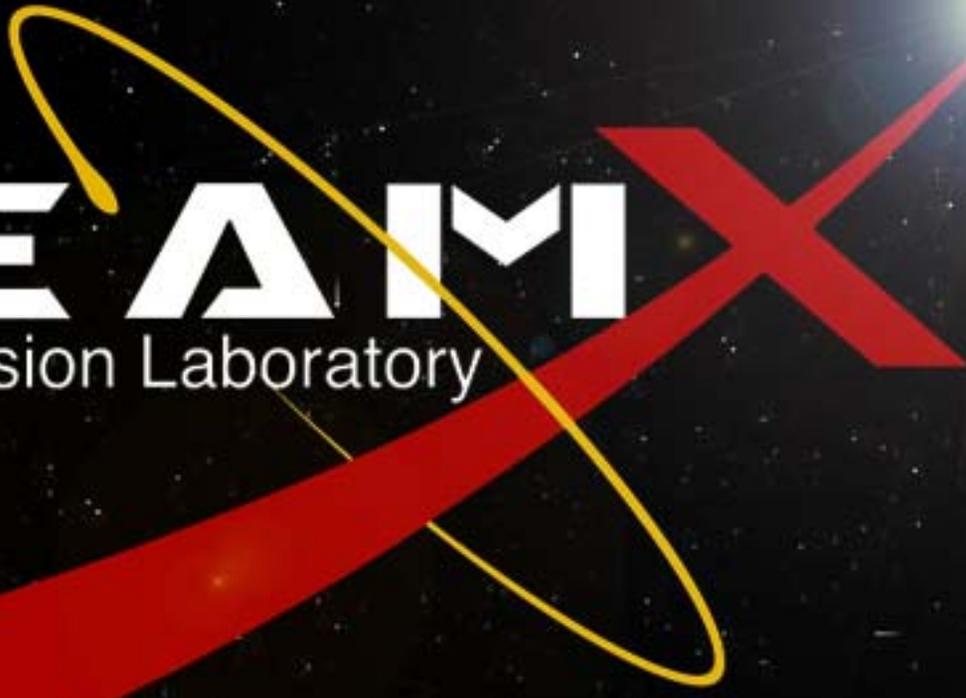

Authors: Adam Nelessen, Bob Kinsey, Alan Didion

Study Dates: June & July 2016

- The data contained in this document may not be modified in any way.
- Distribution of this document is constrained by the terms specified in the footer on each page of the report.

- **This study is part of a range of studies to evaluate mission concepts to the Ice Giants, Uranus and Neptune.**
- **This is a mission study to design and cost 6 options for mission concepts to Uranus and Neptune.**
  - Building on prior mission studies and a recent Ice Giants Probe study
- **Ice Giants Study Goal**
  - Assess science priorities and affordable mission concepts & options in preparation for the next Decadal Survey.
- **Objectives are to identify**
  - Mission concepts that can address science priorities based on what has been learned since the 2013-2022 Decadal
  - Potential concepts across a spectrum of price points
  - Enabling/enhancing technologies
  - Opportunities for international collaboration

✘ **Purpose: The purpose of these studies is to produce subsystem level point designs and cost estimates for affordable missions to the ice giants Uranus and/or Neptune. Affordable is defined as not to exceed FY15 $2B – excluding commercial (non SLS) launch vehicle, but including operations.**

✘ **Objectives: Create a point design and determine the cost for three orbiter configurations:**
- See Options below
- The Ice Giants Atmospheric Probe will have been designed and costed in study 1734 Ice Giants Probe 2016-06
- Develop subsystem level design with MEL & PEL• Provide WBS Level 2 Cost Estimate.

✘ **Option 1:** Uranus Orbiter with 50 kg of payload allocation, SEP stage, & atmospheric probe

✘ **Option 2:** Uranus Orbiter with 150 kg of payload allocation, SEP stage, & no probe

✘ **Option 3:** Neptune Orbiter with 50 kg of payload allocation, SEP stage, & atmospheric probe

✘ **Option 4:** Uranus Flyby with 50 kg of payload allocation, no SEP stage, & atmospheric probe

✘ **Option 5:** Uranus Orbiter with 50 kg of payload allocation, no SEP stage, & atmospheric probe

✘ **Option 6:** Uranus Orbiter with 150 kg of payload allocation, no SEP stage, & no probe

✘ **Common Element Structure:**
- Element 1: Probe
- Element 2: Entry System
- Element 3: Orbiter
- Element 4: SEP Stage

## 1735 Ice Giants Orbiter Participation List

| Subsystem Role | Name |
| --- | --- |
| Facilitator | Kinsey, Robert |
| Systems | Nelessen, Adam |
| Deputy Systems | Didion, Alan |
| ACS | Lee, Austin |
| CDS | Klemm, Roger |
| Configuration | Baez, Enrique |
| Cost | Anderson, Theresa |
| Instruments | Nash, Alfred |
| Mechanical | Landry, Christopher |
| Mechanical | Spaulding, Matthew |
| Mission Design | Schadegg, Maximillian |
| Power | Hall, Ronald |
| Propulsion | Reh, Jonathan |
| Science | Smythe, William |
| Telecom. Hardware | Hansen, David |
| Thermal | Forgette, Daniel |
| Ground Systems | Wenkert, Daniel |
| Software | Hecox, David |
| SVIT | Badaruddin, Kareem |

- **In situ missions to outer planets constrain models of solar system formation and the origin and evolution of atmospheres.**
  - The abundance of noble gases and certain isotopic ratios are diagnostic, and measurements can be made relatively high (pressures < 5 bars)
  - Provides a basis for comparative studies of the gas and ice giants
  - Provides a line to the composition and weather on extrasolar gas giants
- **Uranus and Neptune provide laboratories for study of atmospheric chemistries, dynamics, and interiors of all planets, including Earth.**
  - Provides clues to the local chemical and physical conditions existing at time and location of formation, and to current conditions
  - Tests models of fluid flow and climate evolution under different conditions
  - The deeper we probe, the more knowledge we gain.
- **Entry probe will provide precise vertical profiles of key constituents needed to elucidate chemical processes at work.**
  - Some abundances cannot be measured by remote sensing (no signature or blocked by clouds)
  - Will also collect data on atmospheric structure and winds

| Option | Launch MEV Wet Mass (kg) | Total Cost ($M) | RTGs | Cruise (years) | Downlink Data Rate (kbps) | 34m Ant in GS Array | Description |
|---|---|---|---|---|---|---|---|
| 1 Uranus | 6886 | 1945 | 4 | 11 | 15 | 1 | SEP cruise stage, chemical orbit insertion at Uranus, 3 instruments plus Probe/Entry System |
| 2 Uranus | 7324 | 2259 | 5 | 13 | 30 | 2 | SEP cruise stage, chemical orbit insertion at Uranus, 15 instruments and no Probe/Entry System |
| 3 Neptune | 7365 | 1972 | 4 | 11 | 15 | 3 | SEP cruise stage, chemical orbit insertion at Uranus, 15 instruments and no Probe/Entry System |
| 4 Uranus | 1524 | 1493 | 4 | 10 | 15 | 1 | Chemical cruise, flyby at Uranus, 3 instruments plus Probe/Entry System |
| 5 Uranus | 4345 | 1700 | 4 | 12 | 15 | 1 | Chemical cruise, chemical orbit insertion at Uranus, 3 instruments plus Probe/Entry System |
| 6 Uranus | 4717 | 2005 | 5 | 12 | 30 | 2 | Chemical cruise, chemical orbit insertion at Uranus, 15 instruments and no Probe/Entry System |

# Option 1

- **Option 1: Uranus Orbiter concept w/Probe**
  - 50 kg payload allocation
  - 1 atmospheric probe (previously designed)
  - Includes VVE flybys

- **Class B mission**
- **Dual string redundancy**
- **eMMRTGs could be used for Orbiter power**
  - Carry <u>no mass contingency</u>, because eMMRTG masses provided are "not to exceed" values

- **Xenon residual calculation overridden with Dawn heritage values**
  - ~5 kg residuals on ~1000 kg of propellant
- **Assuming SEP stage based on modified launch vehicle adapter**
  - Affects mechanical/structure masses

**Mission:**
- Launch: 6/2030; Arrival: 2041
- Launch, Venus gravity assist, cruise to Uranus
- SEP stage jettisoned roughly at 5-6 AU
- Probe separation 60 days prior to entry

**Mission Design**
- 11-year cruise to UOI, 4-year science tour
- Orbiter serves as communications relay during probe entry
  - Continuous line of sight between orbiter and probe is critical for telecom
  - Baseline probe EFPA is -35 deg to allow acceptable geometry
- UOI begins ~2 hours after Probe entry.
  - Allows sufficient time for probe data relay prior to the turn-to-burn
  - More than 2 hours results in gain issues for the zenith-pointed Probe UHF antenna (since range and off-pointing angle from zenith increase).
- UOI inserts into 180-day initial orbit, lowered to ~50-day orbit
- Will require optical navigation upon approach to UOI, and during science for targeting moon flybys
  - Doppler imager will be used for OpNav on approach

**Launch Vehicle**
- Delta IV-Heavy (~10,120 kg to C3 of 2.68 $km^2/s^2$)

- **Arrival Vinf / Declination**
    - ~9.9 km/s, 68 degree (spin-axis relative)
- **Orbiter-Probe separation is ~100,000 km at entry time**
    - This range closes as the Orbiter proceeds towards periapse and the Probe decelerates in the Uranus atmosphere.
    - Telecom requires Orbiter within 60 deg of zenith and < 100,000 km range.
- **EFPA of -35 degrees**
    - Imparts 200g's on the probe during entry.
    - Reducing the EFPA results in line-of-sight geometry challenges.

| Event | Rel. Time | Duration | Delta V (m/s) | # Maneuvers | Comments |
|---|---|---|---|---|---|
| Venus Flyby #1 | L+838 days | | | | 1043 km altitude |
| Venus Flyby #2 | L+1483 days | | | | 300 km altitude |
| Earth Flyby | L+1532 days | | | | 300 km altitude |
| SEP Jettison | L+~2400 days | | Total SEP: 5600 | | Timing flexible |
| TCM-1 | E-~400 days | | 10 | 1 | Non-deterministic |
| TCMs 2-5 | | | 13 (Total) | 4 | Non-deterministic |
| Separation/Divert | E-60 days | | 20 | 1 | |
| UOI | L+4017 days | ~1 hr | 2260 | 1 | |
| OTMs 1-5 | | | 290 (Total) | 5 | Includes UOI cleanup, period+incl changes, flyby targeting, and other statistical mnvrs |
| **Total** | | | ~2600 chemical, ~5600 SEP | | |

- **Element 1: Atmospheric Probe**
  - Designed in study 1734 (June 28,30[th])
  - Common for both planets
  - No propulsion, no ACS, no power generation
  - Telecom relay to Orbiter
- **Element 2: Entry System**
  - Heatshield, backshell, structure
- **Element 3: Orbiter**
  - Instrument allocation defined by Option
  - Chemical propulsion
  - eMMRTGs, no solar arrays
- **Element 4: SEP Cruise Stage**
  - "Dumb" cruise stage:
    - No CDS/ACS/telecom
    - No instruments
  - SEP, no chemical propulsion
  - Solar arrays, no RTGs

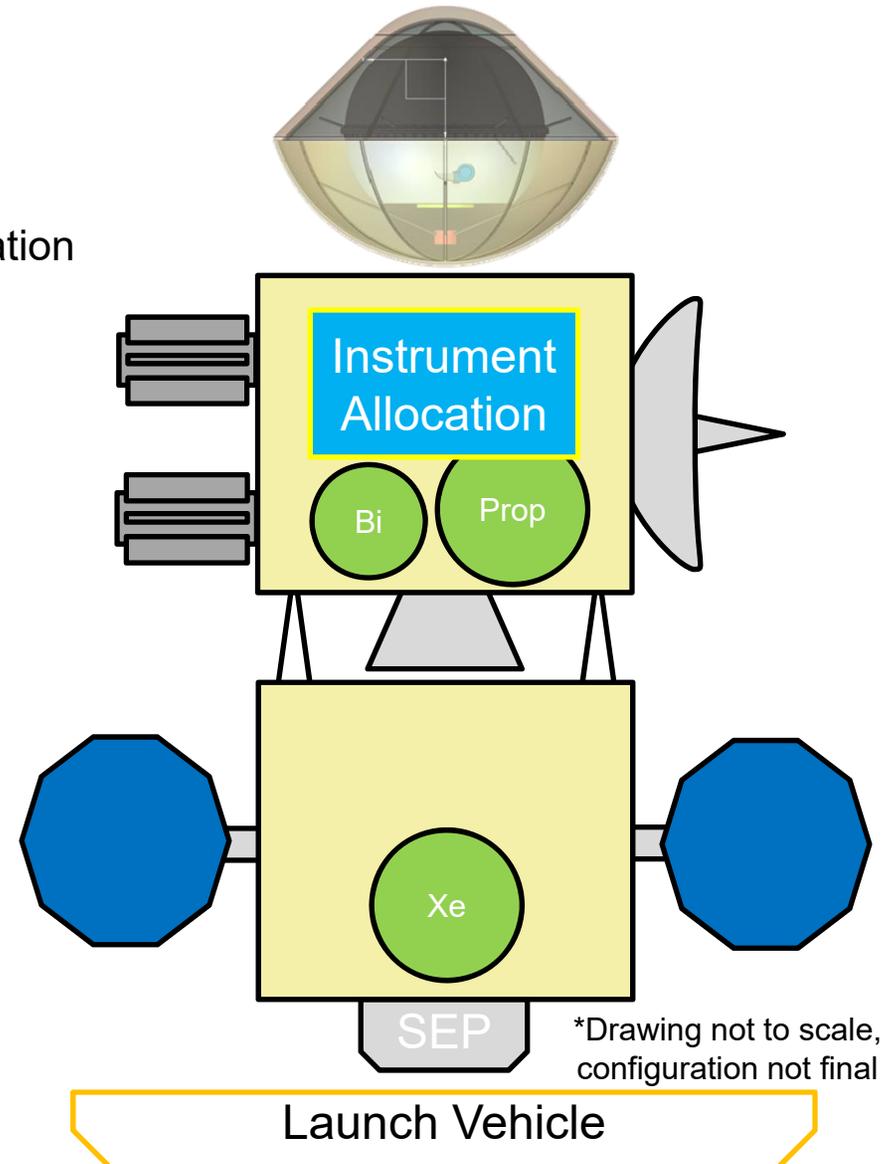

Instrument Allocation

Bi

Prop

Xe

SEP

*Drawing not to scale, configuration not final

Launch Vehicle

**Mission Timeline**

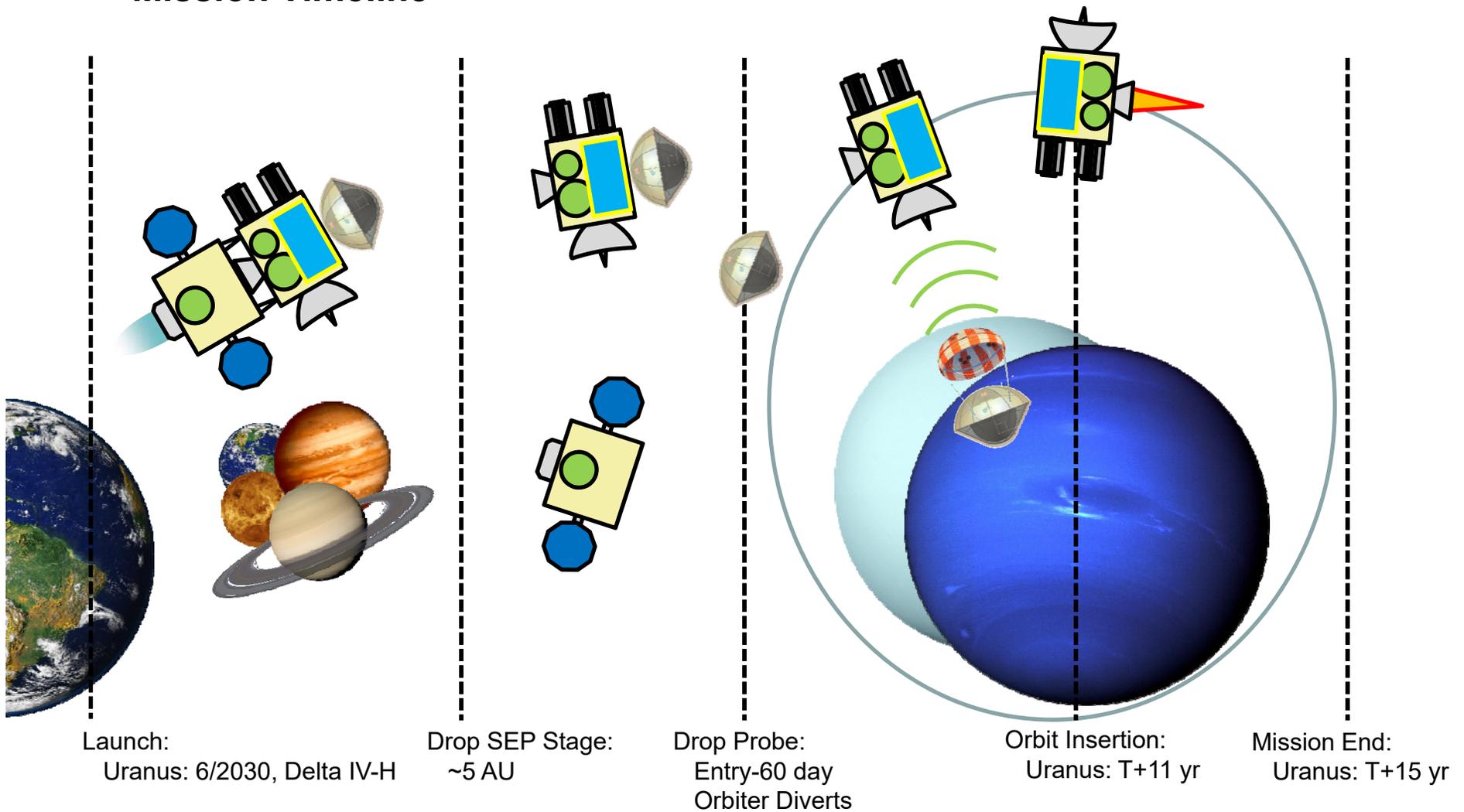

Launch:
Uranus: 6/2030, Delta IV-H

Drop SEP Stage:
~5 AU

Drop Probe:
Entry-60 day
Orbiter Diverts

Orbit Insertion:
Uranus: T+11 yr

Mission End:
Uranus: T+15 yr

**Approach Timeline**

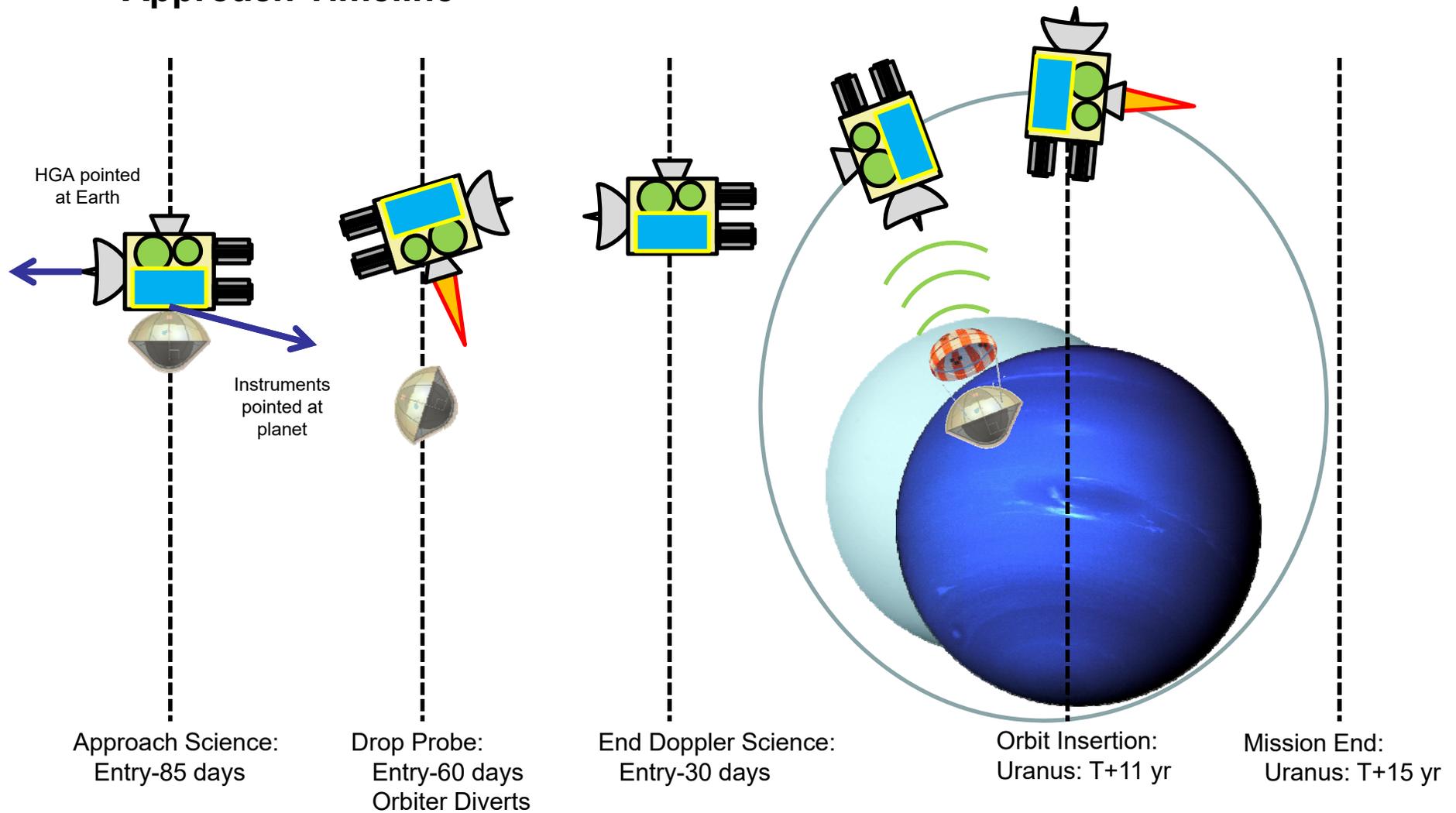

HGA pointed
at Earth

Instruments
pointed at
planet

Approach Science:
Entry-85 days

Drop Probe:
Entry-60 days
Orbiter Diverts

End Doppler Science:
Entry-30 days

Orbit Insertion:
Uranus: T+11 yr

Mission End:
Uranus: T+15 yr

✖ **Entry Timeline**

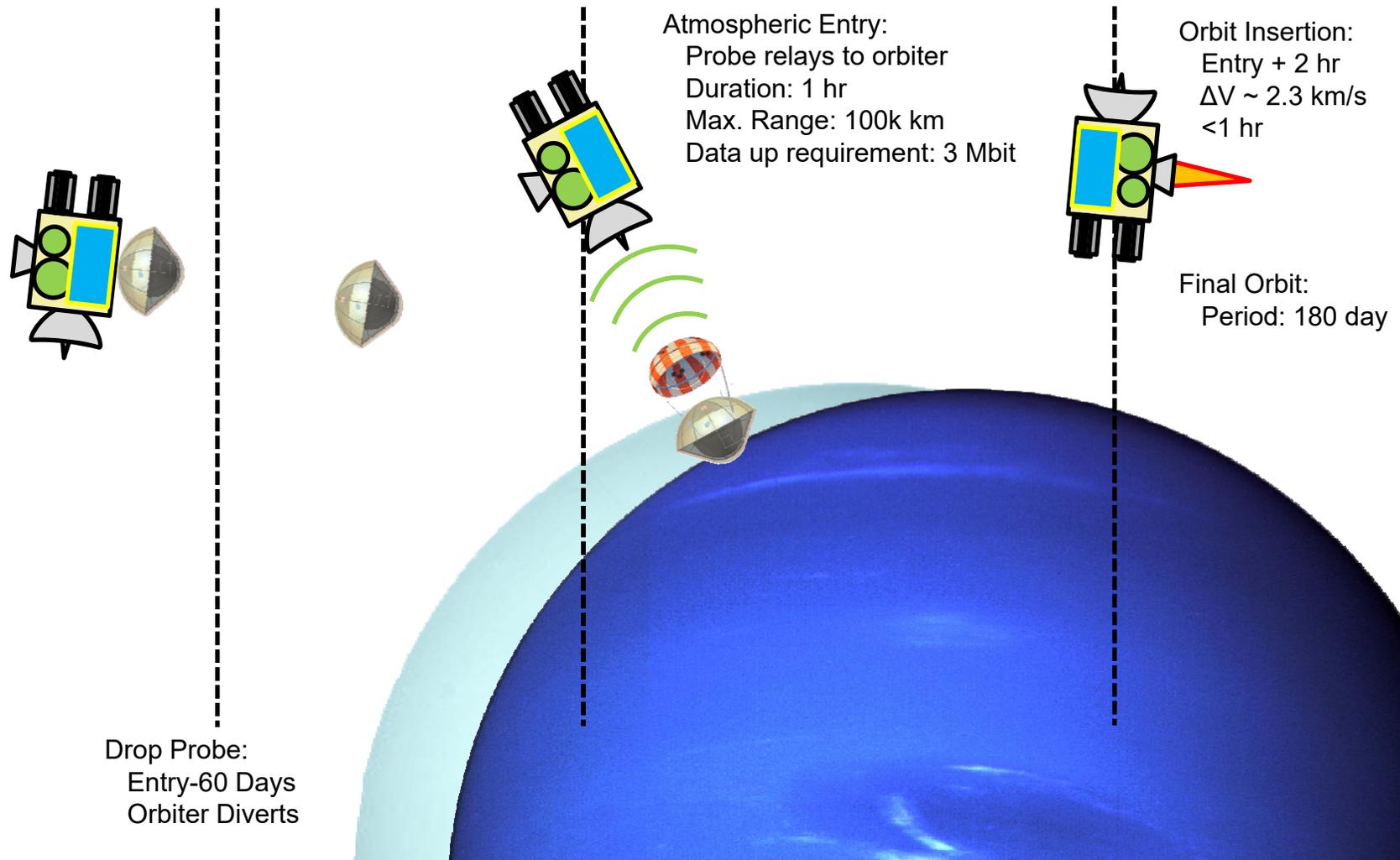

Atmospheric Entry:
Probe relays to orbiter
Duration: 1 hr
Max. Range: 100k km
Data up requirement: 3 Mbit

Orbit Insertion:
Entry + 2 hr
ΔV ~ 2.3 km/s
<1 hr

Final Orbit:
Period: 180 day

Drop Probe:
Entry-60 Days
Orbiter Diverts

Team X Study Guidelines

## Ice Giants Study 2016-07
## SEP Stage

### Project - Study

| | |
|---|---|
| Customer | John Elliott, Kim Reh |
| Study Lead | Bob Kinsey |
| Study Type | Pre-Decadal Study |
| Report Type | Full PPT Report |

### Project - Mission

| | |
|---|---|
| Mission | Ice Giants Study 2016-07 |
| Target Body | Uranus |
| Science | Imaging and Magnetometry |
| Launch Date | 1–Jul–30 |
| Mission Duration | 1 year cruise, 4 years in orbit |
| Mission Risk Class | B |
| Technology Cutoff | 2026 |
| Minimum TRL at End of Phase B | 6 |

### Project - Architecture

| | | |
|---|---|---|
| Probe | on | Entry System |
| Entry System | on | Orbiter |
| Orbiter | on | SEP Stage |
| SEP Stage | on | Launch Vehicle |

| | |
|---|---|
| Launch Vehicle | Delta IV-H |
| Trajectory | VVE Gravity Assists, 180 day orbit, FPA = –35deg |
| L/V Capability, kg | 10120 kg to a C3 of 2 with 0% contingency taken out |
| Tracking Network | DSN |
| Contingency Method | Apply Total System-Level |

| Spacecraft | Orbiter |
|---|---|
| Instruments | None |
| Potential Inst-S/C Commonality | None |
| Redundancy | Dual (Cold) |
| Stabilization | 3-Axis |
| Heritage | TBD |
| Radiation Total Dose | TBD krad behind 100 mil. of Aluminum, with an RDM of 2 added. |
| Type of Propulsion Systems | System 1-SEP, System 2-0, System 3-0 |
| Post-Launch Delta-V, m/s | 0 |
| P/L Mass CBE, kg | 0 kg Payload CBE + 4359 kg Orbiter + Entry System + Probe  (Wet Mass) (alloc) |
| P/L Power CBE, W | 0 |
| P/L Data Rate CBE, kb/s | 0 |
| P/L Pointing, arcsec | TBD |
| EDL Type | None |
| RSDO bus? | NO |
| RSDO Bus Name | NONE |
| Other Guideline | TBD |
| Other Guideline | TBD |
| Hardware Models | Protoflight S/C, EM instrument TBR |

| *Project – Cost and Schedule* | |
|---|---|
| Cost Target | < $2B |
| Mission Cost Category | Flagship – e.g. Cassini |
| FY$ (year) | 2015 |
| Include Phase A cost estimate? | Yes |
| Phase A Start | September 2023 |
| Phase A Duration (months) | 20 |
| Phase B Duration (months) | 16 |
| Phase C/D Duration (months) | 47 |
| Review Dates | PDR – September 2026, CDR – November 2027, ARR – November 2028 |
| Phase E Duration (months) | 179 |
| Phase F Duration (months) | 4 |
| New Development Tests | TBD |
| Project Pays Tech Costs from TRL | 6 TBR |
| Spares Approach | Typical |
| Parts Class | Commercial + Military 883B TBR |
| Launch Site | Cape Canaveral |

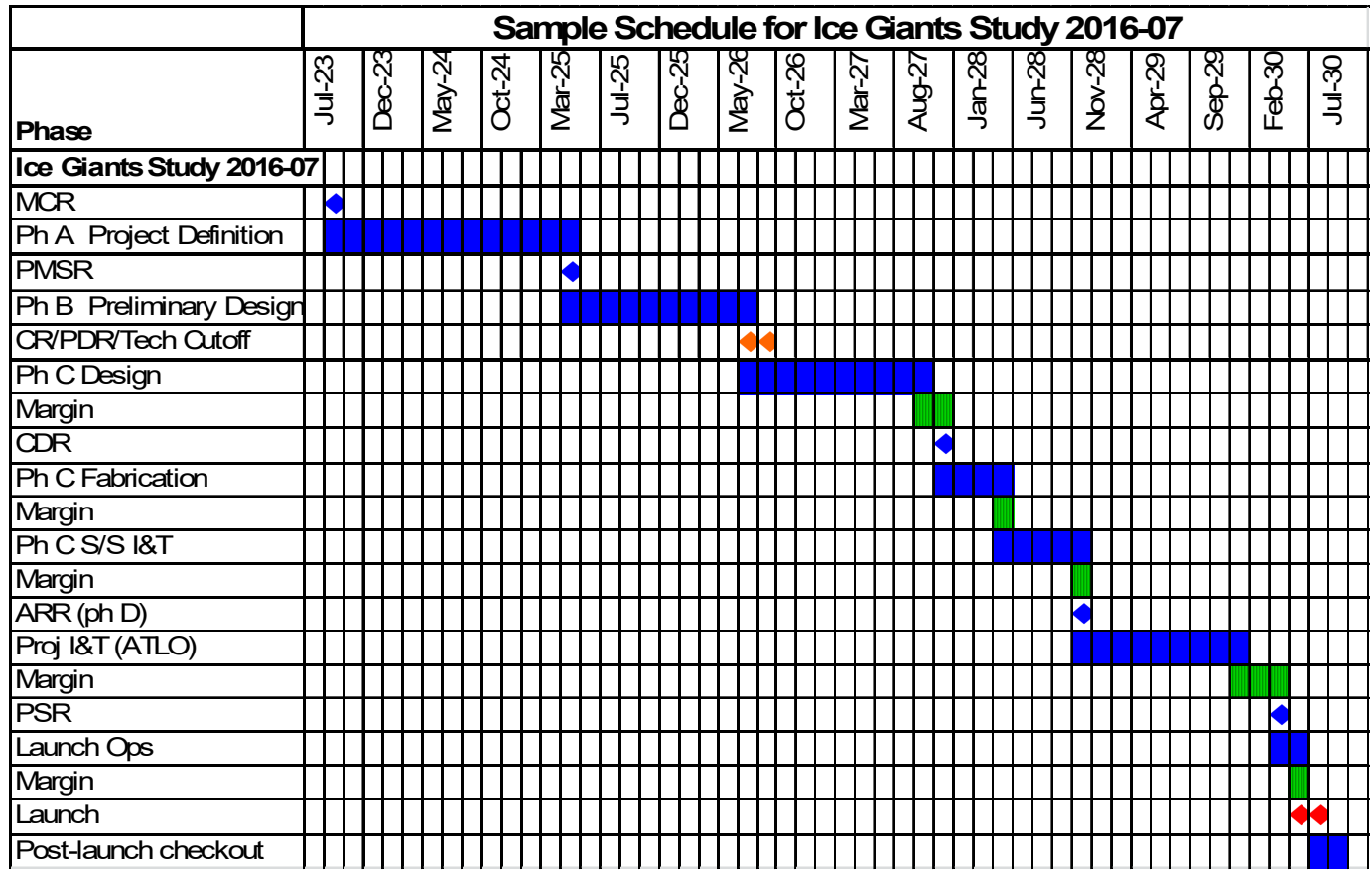

Sample Schedule for Ice Giants Study 2016-07

| Phase | Jul-23 | Dec-23 | May-24 | Oct-24 | Mar-25 | Jul-25 | Dec-25 | May-26 | Oct-26 | Mar-27 | Aug-27 | Jan-28 | Jun-28 | Nov-28 | Apr-29 | Sep-29 | Feb-30 | Jul-30 |
|---|---|---|---|---|---|---|---|---|---|---|---|---|---|---|---|---|---|---|

Proposed development schedule consistent with typical New Frontiers missions and current Europa mission schedule.

Phase A 20 mos., Phase B 16 mos., Phase C/D 47 mos.

Launch July 2030

✗ **Selected instruments come with some extras.**

- NAC has a 2-DOF gimbal.
- Doppler Imager has an internal fast steering mirror (FSM).

| Instrument Name | # units | Heritage | CBE Mass (kg) | Cont. | CBE+Cont. Mass/Unit (kg) | Op. Power CBE per Instrument (W) | Standy Power CBE per Instrument (W) |
|---|---|---|---|---|---|---|---|
| | | | 37 kg | 23% | 45.2 | | |
| Narrow Angle Camera (EIS Europa) | 1 | Inherited design | 12.0 | 15% | 13.8 | 16 W | 2 W |
| Doppler Imager (ECHOES JUICE) | 1 | New design | 20.0 | 30% | 26 | 20 W | 2 W |
| Magnetometer (Gallileo) | 1 | Inherited design | 4.7 | 15% | 5.405 | 8 W | 1 W |

| Instrument Name | Instrument Peak Data Rate | Units |
|---|---|---|
| Narrow Angle Camera (EIS Europa) | 12000 | kbps |
| Doppler Imager (ECHOES JUICE) | 60 | kbps |
| Magnetometer (Gallileo) | 1200 | kbps |

✘ **Instruments**
- Gas Chromatograph Mass Spectrometer (GCMS)
- Atmospheric Structure Instrument (ASI)
- Nephelometer
- Ortho-para Hydrogen Measurement Instrument

✘ **CDS**
- Redundant Sphinx Avionics

✘ **Power**
- Primary batteries
  - In probe:
    - 17.1kg, 1.0 kW-hr EOM
- Redundant Power Electronics

✘ **Thermal**
- RHU heating, passive cooling
- Vented probe design
- Thermally isolating struts

✘ **Telecom**
- Redundant IRIS radio
- UHF SSPA
- UHF Low Gain Antenna (similar to MSL)

✘ **Structures**
- ~50kg Heatshield
  - 1.2m diameter, 45deg sphere cone
- ~15kg Backshell
- ~10kg Parachutes
- ~15kg Probe Aerofairing

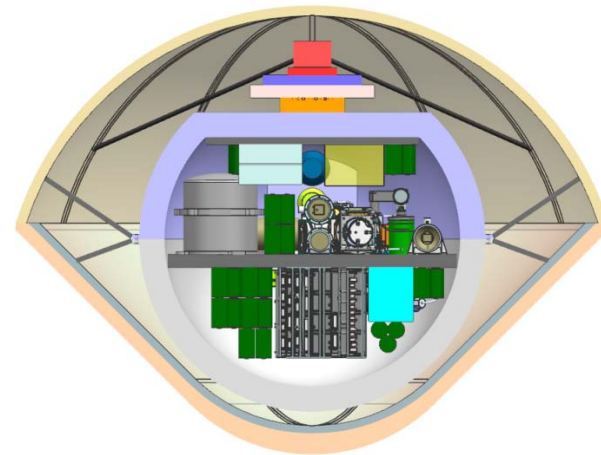

## Instruments
- Narrow Angle Camera
- Doppler Imager
- Magnetometer

## CDS
- JPL reference bus avionics
- Dual string cold redundancy

## Baseline Power System
- 4 eMMRTGs, 45kg each
- 10 A-hr, <5kg, Li Ion Battery

## Telecom
- Radios
  - Two X/X/Ka SDST transponders
  - Two IRIS radio UHF receivers
  - Two 35W Ka-Band TWTAs
  - Two 25W X-Band TWTAs
- Antennas
  - One 3m X/Ka HGA
  - One X-Band MGA
  - Two X-Band LGAs
  - One UHF patch array – 15 dBic gain

## Thermal
- Active and passive thermal control design
- Louvers, heaters, MLI

## ACS
- Four 0.1N Honeywell HR16 reaction wheels
- IMUs, Star Trackers, Sun Sensors

## Propulsion
- Dual-mode bipropellant system provides 2588m/s of delta-V
- Two 200lbf Aerojet main engines
- Four 22N engines
- Eight 1N RCS engines

## Structures
- ~300kg structure
- 40kg ballast
- 86kg harness
- 10m Magnetometer Boom
- Main Engine cover
- SEP Stage and Probe separation mechanisms

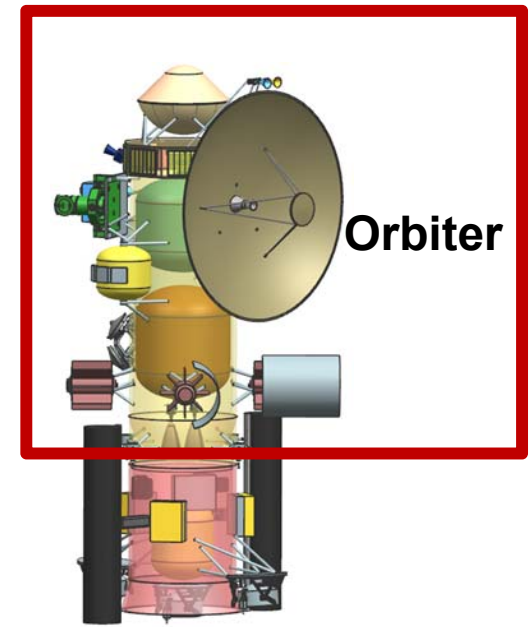

Orbiter

- **Propulsion**
  - Four NEXT main engines
  - Four 35 kg PPUs
  - 1040 kg Xenon Propellant
- **Baseline Power System**
  - Two 54 m$^2$ ROSA Solar Arrays
  - Provide ~29.5 kW at 1AU
  - Redundant JPL Reference Bus Power Electronics
- **CDS**
  - Redundant remote engineering unit
- **ACS**
  - 1DOF solar array gimbal drive electronics
  - 2DOF SEP engine gimbal drive electronics
  - Sun sensors
- **Thermal**
  - Active and passive thermal control design
  - Louvers, heaters, MLI
  - RTG shields
- **Structures**
  - Cylindrical bus shape, made up of stacked CXX adapters

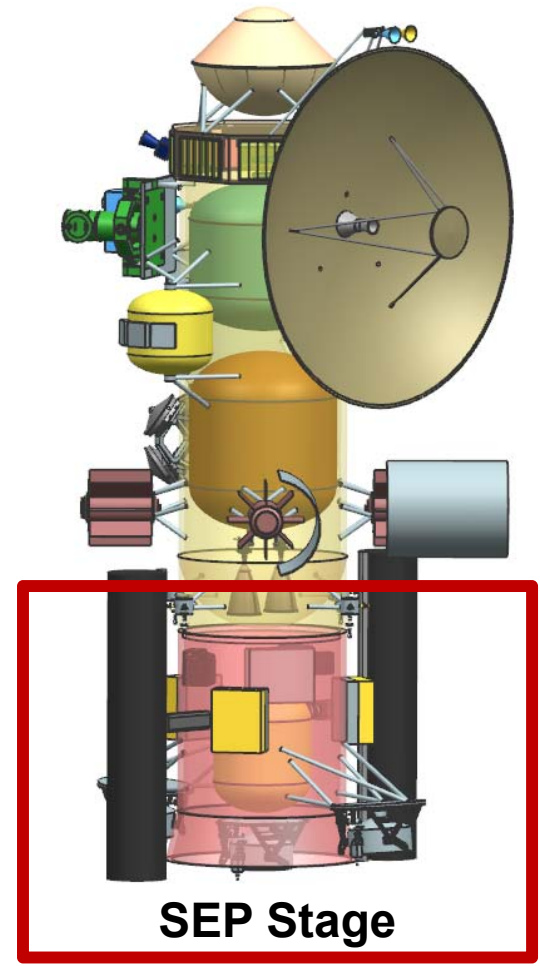

**SEP Stage**

## Probe + Entry System

## Probe

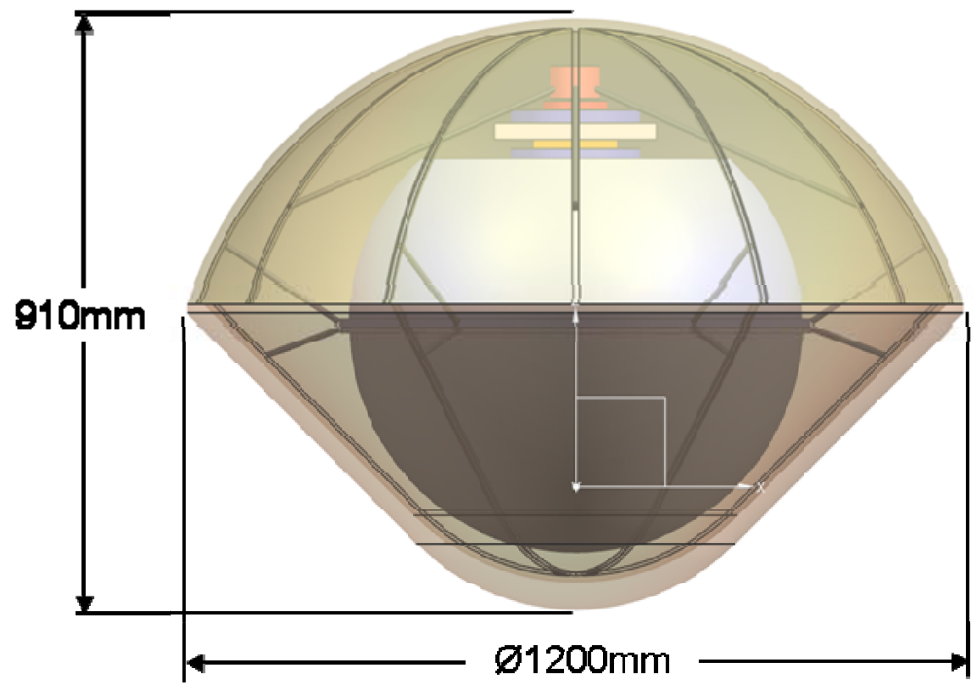

910mm

Ø1200mm

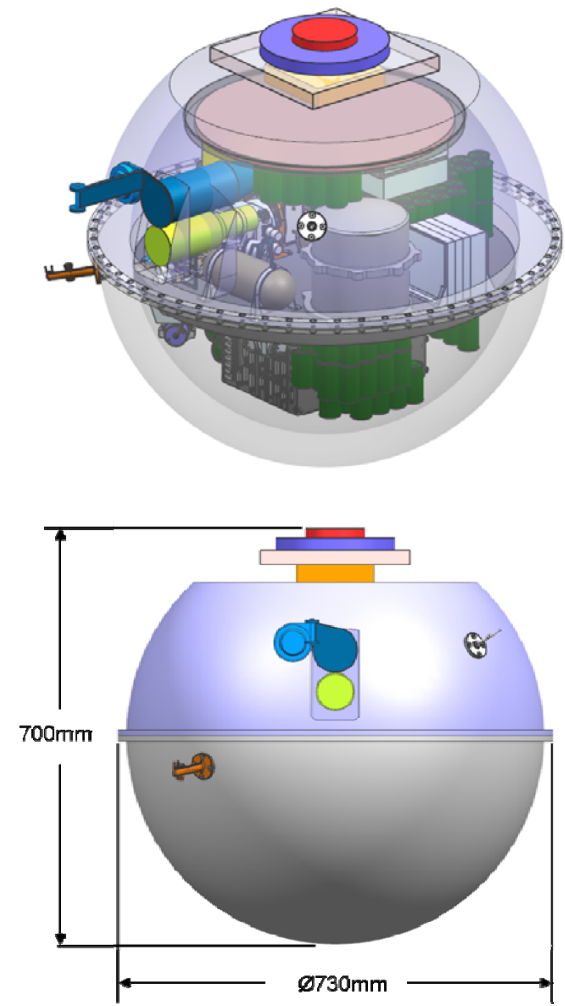

700mm

Ø730mm

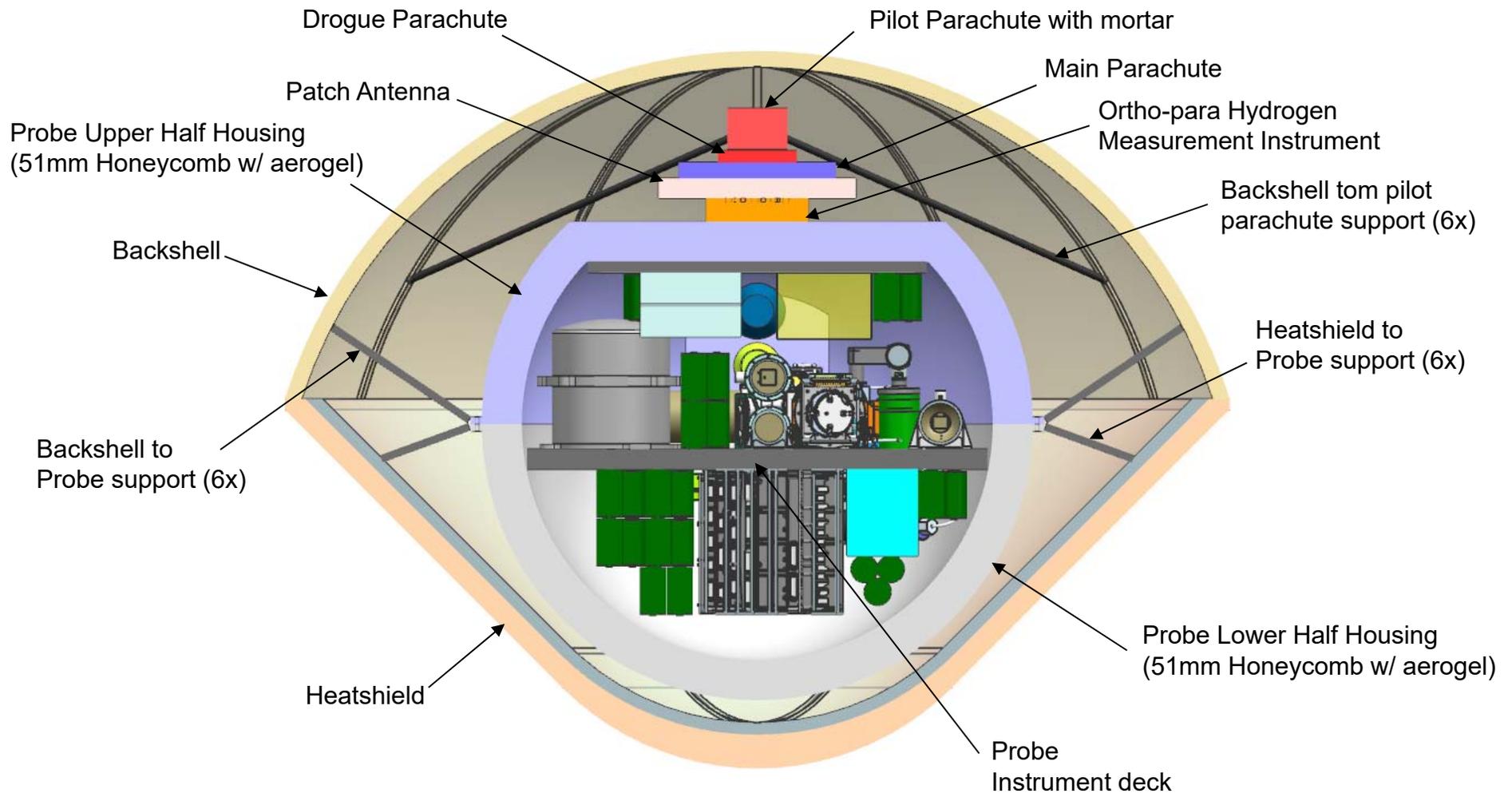

Drogue Parachute

Pilot Parachute with mortar

Patch Antenna

Main Parachute

Probe Upper Half Housing
(51mm Honeycomb w/ aerogel)

Ortho-para Hydrogen
Measurement Instrument

Backshell tom pilot
parachute support (6x)

Backshell

Heatshield to
Probe support (6x)

Backshell to
Probe support (6x)

Heatshield

Probe Lower Half Housing
(51mm Honeycomb w/ aerogel)

Probe
Instrument deck

**✖ Configuration Drawings – Stowed**

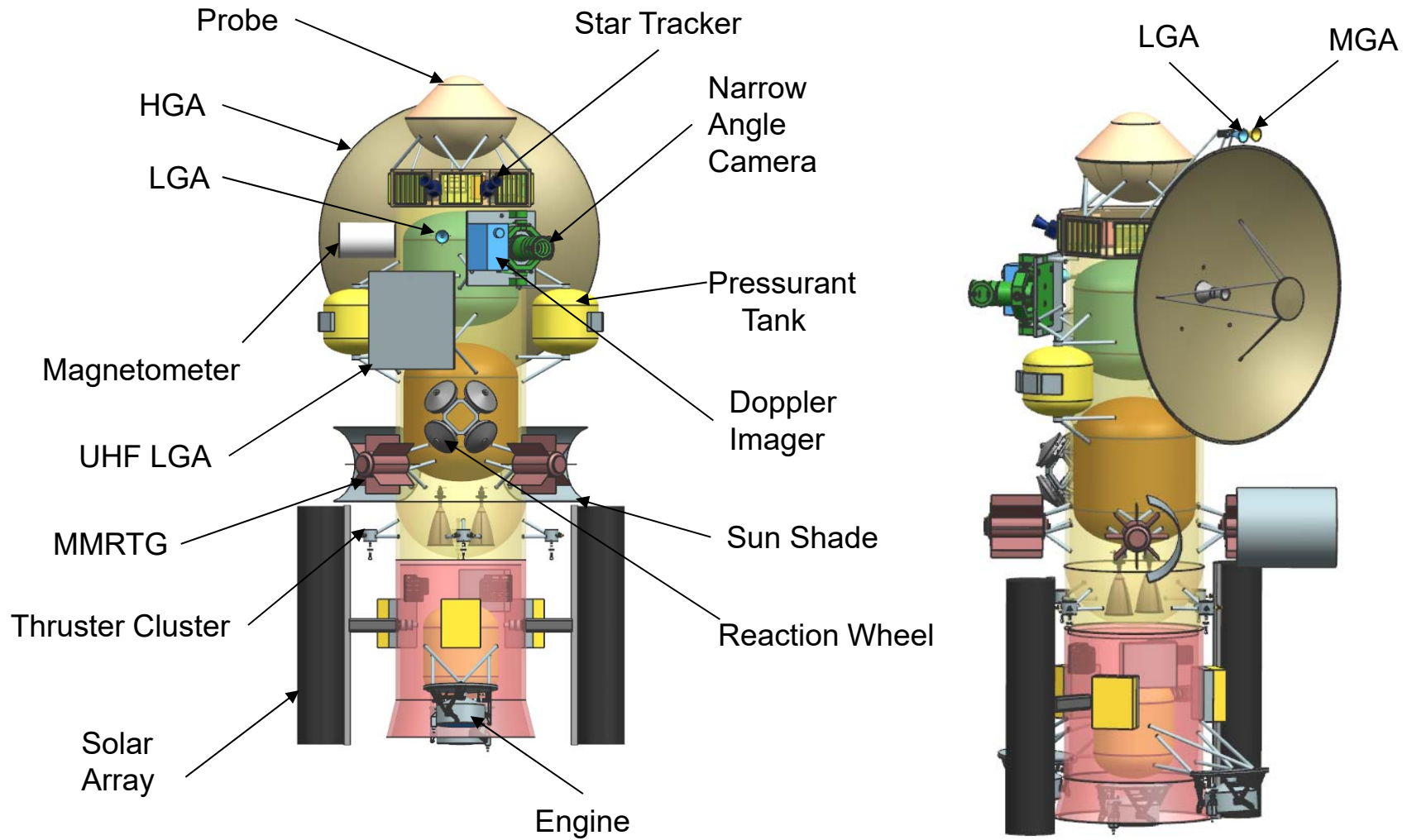

- ★ **Configuration Drawings – Deployed**

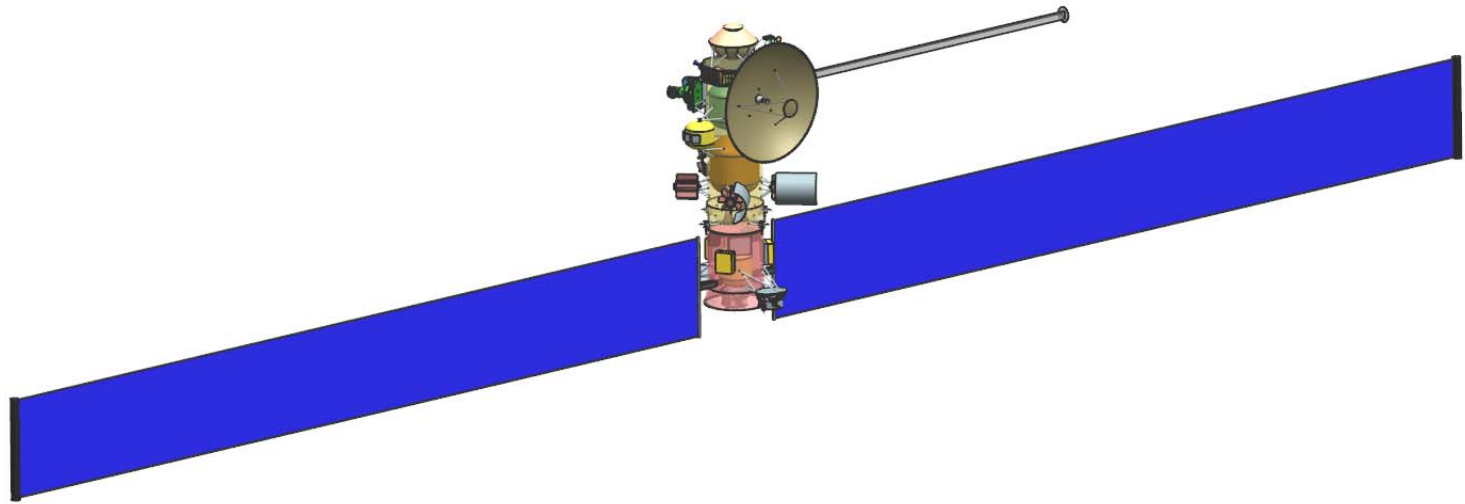

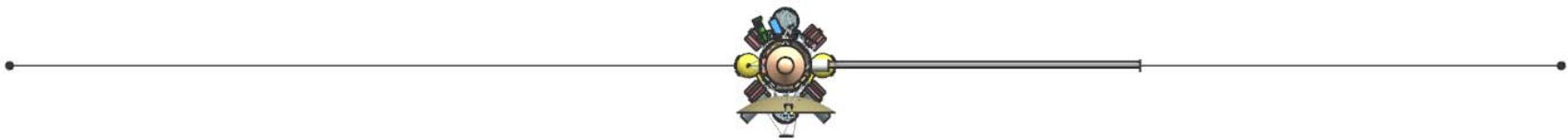

| | | Mass Fraction | Mass (kg) | Subsys Cont. % | CBE+ Cont. (kg) | Mode 1 Power (W) Coast – 60 Days | Mode 2 Power (W) Warmup | Mode 3 Power (W) Science |
|---|---|---|---|---|---|---|---|---|
| *Power Mode Duration (hours)* | | | | | | *24* | *0.5* | *1* |
| **Payload on this Element** | | | | | | | | |
| Instruments | | 21% | 25.3 | 29% | 32.5 | 0 | 91 | 74 |
| Payload Total | | 21% | 25.3 | 29% | 32.5 | 0 | 91 | 74 |
| **Spacecraft Bus** | | | | do not edit formulas below this line, use the calcualtions and overri | | | | |
| Command & Data | | 0% | 0.6 | 17% | 0.7 | 0 | 8 | 8 |
| Power | | 17% | 20.1 | 26% | 25.4 | 0 | 11 | 11 |
| Structures & Mechanisms | | 41% | 49.8 | 30% | 64.7 | 0 | 0 | 0 |
| Cabling | | 9% | 11.5 | 30% | 15.0 | | | |
| Telecom | | 5% | 6.2 | 26% | 7.8 | 0 | 0 | 184 |
| Thermal | | 6% | 7.8 | 3% | 8.1 | 0 | 0 | 0 |
| Bus Total | | | 96.0 | 27% | 121.7 | 0 | 19 | 203 |
| Thermally Controlled Mass | | | | | 121.7 | | | |
| **Spacecraft Total (Dry): CBE & MEV** | | | 121.3 | 27% | 154.2 | 0 | 109 | 277 |
| Subsystem Heritage Contingency | 27% | | 32.9 | SEP Cont | 10% | 0 | 0 | 0 |
| System Contingency | 16% | | 19.3 | | | 0 | 47 | 119 |
| Total Contingency ☐ Include Carried? | 43% | | 52.2 | | | | | |
| **Spacecraft with Contingency:** | | | 173 | of total | w/o addl pld | 0 | 157 | 396 |

| | | Mass Fraction | Mass (kg) | Subsys Cont. % | CBE+ Cont. (kg) |
|---|---|---|---|---|---|
| *Power Mode Duration (hours)* | | | | | |
| **Additional Elements Carried by this Element** | | | | | |
| Probe | | 54% | 121.3 | 43% | 173.5 |
| Carried Elements Total | | 54% | **121.3** | 43% | **173.5** |
| **Spacecraft Bus** | | | | do not edit formulas below | |
| Structures & Mechanisms | | 45% | 101.9 | 30% | 132.4 |
| Cabling | | 0% | 0.9 | 30% | 1.2 |
| Bus Total | | | 102.8 | 30% | 133.6 |
| Thermally Controlled Mass | | | | | 133.6 |
| **Spacecraft Total (Dry): CBE & MEV** | | | **224.1** | 37% | **307.1** |
| Subsystem Heritage Contingency | 37% | | 83.0 | SEP Cont | 10% |
| System Contingency | 6% | | 13.4 | | |
| Total Contingency ☐ Include Carried? | **43%** | | 96.4 | | |
| **Spacecraft with Contingency:** | | | **320** | of total | w/o addl pld |

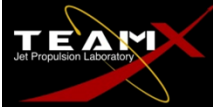
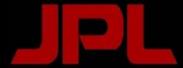

| | Mass Fraction | Mass (kg) | Subsys Cont % | CBE+ Cont (kg) | Mode 1 Power (W) Recharge | Mode 2 Power (W) Approach Science – 85 Days | Mode 3 Power (W) Telecom Downlink | Mode 4 Power (W) UOI Delta-V | Mode 5 Power (W) Orbital Science – Apoapse | Mode 6 Power (W) Orbital Science – Periapse | Mode 7 Power (W) Orbital Science – Moon Tour | Mode 8 Power (W) Safe | Mode 9 Power (W) SEP Thrusting | Mode 10 Power (W) Probe Relay |
|---|---|---|---|---|---|---|---|---|---|---|---|---|---|---|
| *Power Mode Duration (hours)* | | | | | 24 | 24 | 8 | 1.05 | 16 | 16 | 16 | 24 | 24 | 1.5 |
| **Payload on this Element** | | | | | | | | | | | | | | |
| Instruments | 3% | 36.7 | 23% | 45.2 | 4 | 28 | 4 | 4 | 26 | 44 | 26 | 4 | 4 | 4 |
| **Payload Total** | 3% | 36.7 | 23% | 45.2 | 4 | 28 | 4 | 4 | 26 | 44 | 26 | 4 | 4 | 4 |
| **Additional Elements Carried by this Element** | | | | | | | | | | | | | | |
| Entry System + Probe | 15% | 224.1 | 43% | 320.5 | | | | | | | | | | |
| **Carried Elements Total** | 15% | 224.1 | 43% | 320.5 | 0 | 0 | 0 | 0 | 0 | 0 | 0 | 0 | 0 | 0 |
| **Spacecraft Bus** | | do not edit formulas below this line, use the calculations and override tables instead –> | | | | | | | | | | | | |
| Attitude Control | 4% | 63.5 | 10% | 69.8 | 0 | 55 | 55 | 88 | 55 | 55 | 55 | 42 | 55 | 93 |
| Command & Data | 1% | 21.6 | 10% | 23.8 | 57 | 57 | 57 | 57 | 57 | 57 | 57 | 57 | 57 | 57 |
| Power | 15% | 216.6 | 2% | 220.8 | 24 | 40 | 32 | 24 | 24 | 24 | 24 | 40 | 32 | 32 |
| Propulsion1 ☐ SEP1 | 12% | 173.3 | 5% | 182.7 | 31 | 3 | 3 | 151 | 3 | 3 | 3 | 3 | 3 | 3 |
| Structures & Mechanisms | 32% | 462.7 | 30% | 601.5 | 0 | 0 | 0 | 0 | 0 | 0 | 0 | 0 | 0 | 0 |
| Cabling | 6% | 86.3 | 30% | 112.3 | | | | | | | | | | |
| Telecom | 4% | 59.4 | 16% | 68.9 | 12 | 65 | 92 | 71 | 12 | 12 | 12 | 71 | 12 | 32 |
| Thermal | 8% | 112.6 | 23% | 138.8 | 25 | 25 | 25 | 25 | 25 | 25 | 25 | 25 | 25 | 25 |
| **Bus Total** | | 1196.1 | 19% | 1418.6 | 149 | 245 | 264 | 417 | 176 | 176 | 176 | 238 | 184 | 242 |
| Thermally Controlled Mass | | | | 1418.6 | | | | | | | | | | |
| **Spacecraft Total (Dry): CBE & MEV** | | **1456.9** | 22% | **1784.2** | 154 | 272 | 268 | 421 | 202 | 220 | 202 | 242 | 188 | 246 |
| Subsystem Heritage Contingency | 22% | 327.4 | SEP Cont | 10% | 0 | 0 | 0 | 0 | 0 | 0 | 0 | 0 | 0 | 0 |
| System Contingency | 15% | 221.7 | | | 66 | 117 | 115 | 181 | 87 | 95 | 87 | 104 | 81 | 106 |
| Total Contingency ☐ Include Carried? | **38%** | 549.1 | | | | | | | | | | | | |
| **Spacecraft with Contingency:** | | **2006** | of total | w/o addl pld | **220** | **390** | **383** | **602** | **289** | **315** | **289** | **347** | **269** | **352** |
| Propellant & Pressurant with residuals1 | 54% | 2353.7 | For S/C mass = | 4350.8 | Delta-V, Sys 1 | 2588.0 | m/s | residuals = | 61.6 | kg | | | | |
| **Spacecraft Total with Contingency (Wet)** | | **4359.7** | | | | | | | | | | | | |

| | Mass Fraction | | Mass (kg) | Subsys Cont % | CBE+ Cont (kg) | | Mode 8 Power (W) Launch | Mode 9 Power (W) SEP Thrusting | Mode 10 Power (W) Safe |
|---|---|---|---|---|---|---|---|---|---|
| *Power Mode Duration  (hours)* | | | | | | | *2* | *24* | *24* |
| **Additional Elements Carried by this Element** | | | | | | | | | |
| Orbiter + Entry System + Probe  (Wet Mass) | | 79% | 3810.6 | 14% | 4359.7 | | | | |
| **Carried Elements Total** | | 79% | **3810.6** | 14% | **4359.7** | | **0** | **0** | **0** |
| **Spacecraft Bus** | | | do not edit formulas below th | | | s and override tables instead —> | | | |
| Attitude Control | | 0% | 6.0 | 7% | 6.4 | | 0 | 18 | 0 |
| Command & Data | | 0% | 1.6 | 5% | 1.7 | | 4 | 4 | 4 |
| Power | | 5% | 263.9 | 29% | 340.9 | | 0 | 659 | 29 |
| Propulsion1 ☑ SEP1 | | 5% | 245.0 | 22% | 297.8 | | 0 | 25000 | 0 |
| Structures & Mechanisms | | 8% | 367.7 | 30% | 478.0 | | 0 | 0 | 0 |
| Cabling | | 2% | 76.3 | 30% | 99.2 | | | | |
| Thermal | | 2% | 78.8 | 0% | 78.8 | | 0 | 163 | 331 |
| Bus Total | | | 1039.4 | 25% | 1302.8 | | 4 | 25844 | 365 |
| Thermally Controlled Mass | | | | | 1302.8 | | | | |
| **Spacecraft Total (Dry): CBE & MEV** | | | **4850.0** | 17% | **5662.5** | | 4 | 25844 | 365 |
| Subsystem Heritage Contingency | 17% | | 812.5 | SEP Cont | 10% | | 0 | 2500 | 0 |
| System Contingency | 4% | | 183.5 | | | | 2 | 363 | 157 |
| Total Contingency     ☐ Include Carried? | **21%** | | 996.0 | | | | | | |
| **Spacecraft with Contingency:** | | | **5846** | of total | w/o addl pld | | **6** | **28707** | **521** |
| Propellant & Pressurant with residuals1 | | 15% | 1040.4 | For S/C mass = | | 2000.0 | kg | | |
| **Spacecraft Total with Contingency (Wet)** | | | **6886.5** | | | | | | |

| Element Number | Element Name | Dry CBE (kg) | Cont / JPL Margin (kg) | Dry Allocation (kg) | Propellant (kg) | Dry Allocation + Propellant (kg) |
|---|---|---|---|---|---|---|
| 1 | Probe | 121 | 52 | 173 | - | 173 |
| 2 | Entry System | 103 | 44 | 147 | - | 147 |
| 3 | Orbiter minus eMMRTGs | 1,053 | 453 | 1,505 | 2,354 | 3,859 |
| 3.1 | eMMRTGs | 180 | - | 180 | - | 180 |
| 4 | SEP Stage | 1,039 | 447 | 1,486 | 1,040 | 2,526 |
| | **Total Stack** | **2,496** | **996** | **3,492** | **3,394** | **6,886** |
| | | | | Dry Mass Allocation | | 3,492 |
| | | | | JPL Margin (kg / %) | | 996 / 28.5% |
| | | | | JPL Margin without eMMRTG (kg / %) | | 996 / 30% |
| | | | | Delta IV-H Capacity (kg) | | 10,120 |
| | | | | Extra Launch Vehicle Margin (kg) | | 3,234 |

| Element Number | Element Name | Dry CBE (kg) | Cont (%) | Cont. (kg) | MEV (kg) | Dry Allocation (kg) | Propellant (kg) | Dry Allocation + Propellant (kg) |
|---|---|---|---|---|---|---|---|---|
| 1 | Probe | 121 | 27% | 33 | 154 | 173 | - | 173 |
| 2 | Entry System | 103 | 30% | 31 | 134 | 147 | - | 147 |
| 3 | Orbiter minus eMMRTGs | 1,053 | 22% | 231 | 1,284 | 1,505 | 2,354 | 3,859 |
| 3.1 | eMMRTGs | 180 | - | - | 180 | 180 | - | 180 |
| 4 | SEP Stage | 1,039 | 25% | 264 | 1,303 | 1,486 | 1,040 | 2,526 |
| | **Total Stack** | **2,496** | | **559** | **3,054** | **3,492** | **3,394** | **6,886** |

| | |
|---|---|
| Dry Mass Allocation (kg) | 3,492 |
| NASA Margin (kg / %) | 438 / 14% |
| NASA Margin without eMMRTG (kg / %) | 438 / 15% |
| Delta IV-H Capacity (kg) | 10,120 |
| Extra Launch Vehicle Margin (kg) | 3,234 |

- **Probe needs to enter Uranus atmosphere head-on versus shallow**
  - Probe relay antenna is nominally aligned with zenith.
  - Need Orbiter within tens of degrees of zenith to close the link.
  - Head-on: Orbiter is close enough to zenith for one hour.
    - Probe deceleration ~200 g's
  - Shallow: Orbiter is too far from zenith to close the link.
    - On the other hand, Probe deceleration relatively low.
  - Verified that Probe can operate through the higher deceleration.

- **Ka-band transmitter power versus array of 34m ground stations**
  - Downlink 15 kbps using one 35W TWTA to a 34m BWG ground station.
    - Telecom system uses most of one eMMRTG power during downlink.
  - Could increase downlink rate using more power, adding an eMMRTG, or by using an array of two or more 34m ground stations.
  - For this option, make do with 15 kbps downlink rate, per customer request.

✖ **Data downlink strategy for Doppler Imager (DI) on approach**

- DI generates a lot of data continuously for tens of days on approach.
- Configuration with HGA and DI on opposite sides of the cylindrical bus allows pointing DI towards Uranus while pointing HGA towards Earth.
- Can downlink for ~20 hours/day and maintain positive power balance.
  - ◆ Using only 4 eMMRTGs, as opposed to 5
- Data that can't be downlinked before UOI will be downlinked after.

✖ **Configuration that helps to minimize mass and power**

- eMMRTGs outside the cylindrical bus provide heating
  - ◆ Reduces mass and power of thermal subsystem components
- Propellant and oxidizer tanks inside bus, pressurant tanks outside
  - ◆ Pressurant tanks are easier to keep warm than propellant/oxidizer tanks.
- Shorten the stack to minimize structure mass
  - ◆ Single custom propellant tank instead of two tanks
  - ◆ Stow solar arrays perpendicular to bus to reduce height of the SEP stage

## Mechanical

- LV interfaces directly to the SEP Stage; Orbiter interfaces to SEP Stage; Entry System containing Probe is attached to the side of the Orbiter.

- Orbiter: primary structure is the largest mass element.
  - 275 kg CBE out of 463 kg total for Mechanical (1196 kg bus dry mass)
  - Drivers are the large Propulsion and Power masses.

- SEP stage: primary structure is the largest mass element.
  - 217 kg CBE out of 368 kg total for Mechanical (1069 kg SEP Stage dry mass)
  - Due to the Orbiter and other mission elements being carried during launch.
  - Primary structure of the SEP Stage is being utilized as the LVA.

- **Power: Orbiter power bus spans both Orbiter and SEP stage**
  - Dual String Reference Bus electronics heritage
  - New development High Voltage Electronics Assembly for the SEP stage
  - Two Rollout Solar Arrays (ROSAs) support 25kW SEP at 1AU.

- **Propulsion: SEP Stage used to ~5 AU; dual mode bi-prop.**
  - SEP Stage: 2+1 system using NEXT Engines
    - 1033 kg Xe mass allocation + 7 kg residuals for a total of 1040 kg
  - Chemical System: orbit insertion delta V= 2.3km/s desired in < 1hr
    - Probe release after SEP stage separation, before UOI
    - Two 890N main engines used to achieve burn time < 1hr

- **Thermal: Cassini-heritage waste-heat recovery system on Orbiter**
  - RTG end domes each provide 75 W waste heat to propulsion module via conductive and radiative coupling.
  - VRHUs act as primary control mechanism for thruster clusters.
    - Also act as trim heaters for the propulsion module
  - Louvers act as primary control mechanism for avionics module.

- **Telecom: X- and Ka-Band subsystem, plus UHF for Probe data.**
  - Two 35W Ka-Band TWTAs, two 25W X-Band TWTAs
  - Two X/X/Ka SDST transponders, two IRIS radio UHF receivers
  - 3m X/Ka HGA, one X-Band MGA, two X-Band LGAs, UHF patch array.
  - Supports a data rate at Uranus of 15 kbps into 34m BWG ground station.
  - Supports uplink of 3Mbits of probe data

- **CDS: Reference Bus architecture ideally suited for high reliability, long lifetime mission.**
  - Standard JPL spacecraft CDS that is similar to SMAP
    - RAD750 CPU, NVM, MTIF, MSIA, CRC, LEU-A, LEU-D, MREU
    - 128 GBytes storage for science data
    - 1553 and RS-422 ICC/ITC interfaces for subsystems and instruments

- **ACS: 3-axis stabilized with star tracker, sun sensor, gyros, wheels.**
  - All stellar attitude determination to minimize power, conserve gyros.
  - Sun sensor performance may degrade once the Orbiter passes Saturn.
    - May impact safe mode used during star tracker outage.
    - Detailed analysis on Sun sensor performance versus distance is needed.

- **Software: core product line is appropriate since this mission has aspects similar to MSL/M2020/SMAP/Europa.**
  - Complexity rankings range from Medium to High.
    - Medium infrastructure: dual string with warm spare.
    - High fault behaviors: high redundancy, string swapping, critical events.
    - Medium/High ACS: tight pointing requirements, many ACS modes
    - Medium Telecom: dual active UHF, redundant DTE
    - Medium Science data processing, full file system

- **SVIT: Probe testbed, system I&T and V&V costs are included**
  - Cost of assembling and testing RTG's is captured elsewhere
    - Cost of integrating RTG's is included with other ATLO costs
  - Probe with 5 Instruments costed separately; testbed costs included.

- **Ground Systems**
  - Mission specific implementation of standard JPL mission operations and ground data systems
  - Ground network: DSN 34-m BWG; 70-m or equivalent for safe mode
  - Science support: 24x7 tracking on approach; daily contacts on orbit

| COST SUMMARY (FY2015 $M) | Generate ProPricer Input | Team X Estimate | | |
|---|---|---|---|---|
| | | CBE | Res. | PBE |
| **Project Cost** | | **$1594.5 M** | **22%** | **$1945.4 M** |
| **Launch Vehicle** | | **$33.0 M** | **0%** | **$33.0 M** |
| **Project Cost (w/o LV)** | | **$1561.5 M** | **22%** | **$1912.4 M** |
| **Development Cost** | | **$1283.3 M** | **25%** | **$1598.8 M** |
| Phase A | | $12.8 M | 25% | $16.0 M |
| Phase B | | $115.5 M | 25% | $143.9 M |
| Phase C/D | | $1155.0 M | 25% | $1438.9 M |
| **Operations Cost** | | **$278.2 M** | **13%** | **$313.6 M** |

Total mission cost is $1.95B. This is the likely cost within a range that typically can be as much as 10% lower up to 20% higher. The development cost with reserves is $1.60B.

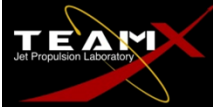
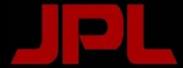

| WBS Elements | NRE | RE | 1st Unit |
|---|---|---|---|
| Project Cost (no Launch Vehicle) | $1351.0 M | $594.4 M | $1945.4 M |
| Development Cost (Phases A - D) | $1004.5 M | $594.3 M | $1598.8 M |
| 01.0 Project Management | $47.3 M | | $47.3 M |
| 1.01 Project Management | $11.4 M | | $11.4 M |
| 1.02 Business Management | $13.6 M | | $13.6 M |
| 1.04 Project Reviews | $2.5 M | | $2.5 M |
| 1.06 Launch Approval | $19.8 M | | $19.8 M |
| 02.0 Project Systems Engineering | $23.7 M | $0.5 M | $24.3 M |
| 2.01 Project Systems Engineering | $8.9 M | | $8.9 M |
| 2.02 Project SW Systems Engineering | $5.2 M | | $5.2 M |
| 2.03 EEIS | $1.5 M | | $1.5 M |
| 2.04 Information System Management | $1.7 M | | $1.7 M |
| 2.05 Configuration Management | $1.5 M | | $1.5 M |
| 2.06 Planetary Protection | $0.2 M | $0.2 M | $0.4 M |
| 2.07 Contamination Control | $1.2 M | $0.3 M | $1.5 M |
| 2.09 Launch System Engineering | $1.0 M | | $1.0 M |
| 2.10 Project V&V | $2.0 M | | $2.0 M |
| 2.11 Risk Management | $0.5 M | | $0.5 M |
| 03.0 Mission Assurance | $52.9 M | $0.0 M | $52.9 M |
| 04.0 Science | $24.8 M | | $24.8 M |
| Orbiter Science | $14.0 M | | $14.0 M |
| Probe Science | $10.8 M | | $10.8 M |
| 05.0 Payload System | $80.2 M | $48.3 M | $128.5 M |
| 5.01 Payload Management | $7.8 M | | $7.8 M |
| 5.02 Payload Engineering | $5.8 M | | $5.8 M |
| Orbiter Instruments | $33.5 M | $24.3 M | $57.8 M |
| Narrow Angle Camera (EIS Europa) | $11.6 M | $8.4 M | $20.0 M |
| Doppler Imager (ECHOES JUICE) | $17.4 M | $12.6 M | $30.0 M |
| Magnetometer (Gallileo) | $4.5 M | $3.3 M | $7.8 M |
| Probe Instruments | $33.1 M | $24.0 M | $57.1 M |
| Mass Spectrometer | $22.9 M | $16.6 M | $39.6 M |
| Atmospheric Structure Investigation (ASI) | $3.4 M | $2.5 M | $5.9 M |
| Nephelometer (Galileo) | $5.3 M | $3.8 M | $9.1 M |
| Ortho-para H2 meas. Expt. | $1.5 M | $1.1 M | $2.6 M |

| WBS Elements | NRE | RE | 1st Unit |
|---|---|---|---|
| 06.0 Flight System | $496.1 M | $386.7 M | $882.8 M |
| 6.01 Flight System Management | $5.0 M | | $5.0 M |
| 6.02 Flight System Systems Engineering | $51.1 M | | $51.1 M |
| 6.03 Product Assurance (included in 3.0) | | | $0.0 M |
| Orbiter | $297.2 M | $236.6 M | $533.8 M |
| 6.04 Power | $94.5 M | $133.1 M | $227.6 M |
| 6.05 C&DH | $31.3 M | $24.3 M | $55.6 M |
| 6.06 Telecom | $28.4 M | $18.1 M | $46.5 M |
| 6.07 Structures (includes Mech. I&T) | $51.9 M | $16.8 M | $68.7 M |
| 6.08 Thermal | $4.2 M | $13.1 M | $17.2 M |
| additional cost for >43 RHUs | $34.0 M | $0.0 M | $34.0 M |
| 6.09 Propulsion | $22.1 M | $16.7 M | $38.8 M |
| 6.10 ACS | $9.4 M | $9.8 M | $19.1 M |
| 6.11 Harness | $4.1 M | $3.8 M | $7.9 M |
| 6.12 S/C Software | $17.1 M | $0.9 M | $18.0 M |
| 6.13 Materials and Processes | $0.4 M | $0.0 M | $0.4 M |
| SEP Stage | $50.6 M | $106.1 M | $156.7 M |
| 6.04 Power | $7.7 M | $50.5 M | $58.1 M |
| 6.05 C&DH | $4.0 M | $2.7 M | $6.7 M |
| 6.06 Telecom | $0.0 M | $0.0 M | $0.0 M |
| 6.07 Structures (includes Mech. I&T) | $10.3 M | $4.5 M | $14.8 M |
| 6.08 Thermal | $3.1 M | $11.2 M | $14.3 M |
| 6.09 Propulsion | $22.2 M | $34.5 M | $56.7 M |
| 6.10 ACS | $0.7 M | $1.2 M | $1.8 M |
| 6.11 Harness | $2.4 M | $1.5 M | $3.9 M |
| 6.12 S/C Software | $0.0 M | $0.0 M | $0.0 M |
| 6.13 Materials and Processes | $0.4 M | $0.0 M | $0.4 M |
| Probe | $26.8 M | $18.1 M | $44.9 M |
| 6.04 Power | $2.9 M | $2.1 M | $5.0 M |
| 6.05 C&DH | $0.3 M | $2.3 M | $2.7 M |
| 6.06 Telecom | $7.9 M | $4.1 M | $12.0 M |
| 6.07 Structures (includes Mech. I&T) | $8.0 M | $3.5 M | $11.5 M |
| 6.08 Thermal | $2.3 M | $4.9 M | $7.2 M |
| 6.11 Harness | $1.8 M | $0.9 M | $2.7 M |
| 6.12 S/C Software | $3.3 M | $0.2 M | $3.5 M |
| 6.13 Materials and Processes | $0.4 M | $0.0 M | $0.4 M |
| Entry System | $57.1 M | $24.4 M | $81.5 M |
| 6.07 Structures (includes Mech. I&T) | $55.3 M | $24.1 M | $79.4 M |
| 6.11 Harness | $1.4 M | $0.3 M | $1.7 M |
| 6.13 Materials and Processes | $0.4 M | $0.0 M | $0.4 M |
| Ames/Langley EDL Engineering/Testing | $3.8 M | $0.0 M | $3.8 M |
| 6.14 Spacecraft Testbeds | $4.5 M | $1.5 M | $6.0 M |

| WBS Elements | NRE | RE | 1st Unit |
|---|---|---|---|
| **07.0 Mission Operations Preparation** | **$27.0 M** | | **$27.0 M** |
| 7.0 MOS Teams | $20.0 M | | $20.0 M |
| 7.03 DSN Tracking (Launch Ops.) | $2.7 M | | $2.7 M |
| 7.06 Navigation Operations Team | $4.2 M | | $4.2 M |
| 7.07.03 Mission Planning Team | $0.0 M | | $0.0 M |
| **09.0 Ground Data Systems** | **$22.1 M** | | **$22.1 M** |
| 9.0A Ground Data System | $20.0 M | | $20.0 M |
| 9.0B Science Data System Development | $1.3 M | | $1.3 M |
| 9A.03.07 Navigation H/W & S/W Development | $0.8 M | | $0.8 M |
| **10.0 ATLO** | **$21.1 M** | **$21.7 M** | **$42.8 M** |
| Orbiter | $15.3 M | $13.3 M | $28.6 M |
| Probe | $5.9 M | $8.4 M | $14.2 M |
| **11.0 Education and Public Outreach** | **$0.0 M** | **$0.0 M** | **$0.0 M** |
| **12.0 Mission and Navigation Design** | **$30.9 M** | | **$30.9 M** |
| 12.01 Mission Design | $2.4 M | | $2.4 M |
| 12.02 Mission Analysis | $11.8 M | | $11.8 M |
| 12.03 Mission Engineering | $1.8 M | | $1.8 M |
| 12.04 Navigation Design | $15.0 M | | $15.0 M |
| **Development Reserves** | **$178.3 M** | **$137.1 M** | **$315.5 M** |

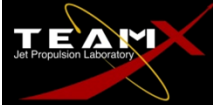

# Executive Summary
## Cost E-F and Launch Nuclear Safety

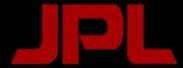

| WBS Elements | NRE | RE | 1st Unit |
|---|---|---|---|
| **Operations Cost (Phases E - F)** | **$313.5 M** | **$0.1 M** | **$313.6 M** |
| **01.0 Project Management** | **$27.1 M** | | **$27.1 M** |
| 1.01 Project Management | $15.3 M | | $15.3 M |
| 1.02 Business Management | $10.7 M | | $10.7 M |
| 1.04 Project Reviews | $1.1 M | | $1.1 M |
| 1.06 Launch Approval | $0.1 M | | $0.1 M |
| **02.0 Project Systems Engineering** | **$0.0 M** | **$0.1 M** | **$0.1 M** |
| **03.0 Mission Assurance** | **$3.6 M** | **$0.0 M** | **$3.6 M** |
| **04.0 Science** | **$69.2 M** | | **$69.2 M** |
| Orbiter Science | $53.4 M | | $53.4 M |
| Probe Science | $15.8 M | | $15.8 M |
| **07.0 Mission Operations** | **$149.5 M** | | **$149.5 M** |
| 7.0 MOS Teams | $79.5 M | | $79.5 M |
| 7.03 DSN Tracking | $42.5 M | | $42.5 M |
| 7.06 Navigation Operations Team | $26.3 M | | $26.3 M |
| 7.07.03 Mission Planning Team | $1.2 M | | $1.2 M |
| **09.0 Ground Data Systems** | **$28.8 M** | | **$28.8 M** |
| 9.0A GDS Teams | $23.1 M | | $23.1 M |
| 9.0B Science Data System Ops | $5.2 M | | $5.2 M |
| 9A.03.07 Navigation HW and SW Dev | $0.6 M | | $0.6 M |
| **11.0 Education and Public Outreach** | **$0.0 M** | **$0.0 M** | **$0.0 M** |
| **12.0 Mission and Navigation Design** | **$0.0 M** | | **$0.0 M** |
| **Operations Reserves** | **$35.3 M** | **$0.0 M** | **$35.4 M** |
| **8.0 Launch Vehicle** | **$33.0 M** | | **$33.0 M** |
| **Launch Vehicle and Processing** | **$0.0 M** | | **$0.0 M** |
| **Nuclear Payload Support** | **$33.0 M** | | **$33.0 M** |

## ✖ Risks related to the Probe

- Only 2 hours between Probe entry and UOI, a critical event
  - ◆ May be operationally challenging to sequence both the Probe relay and UOI for the Orbiter within this time window.
  - ◆ Longer than 2 hours makes the geometry more challenging for telecom.
- May be issues for the relay link margin due to Probe-Orbiter geometry and uncertainties regarding Uranus atmosphere/ potential signal attenuation.
- High g load on the Probe carries some risk.
- Last Probe targeting occurs more than 60 days prior to encounter.
  - ◆ Probe carries no propulsion, so it cannot correct trajectory dispersions.
  - ◆ Need dispersions small enough to ensure safe entry conditions at Uranus.
- Orbit knowledge requirements for science reconstruction need to be determined.
  - ◆ Will drive how the Probe is tracked pre-entry and what telemetry (e.g. IMU) needs to be transmitted with the science data to the Orbiter.
  - ◆ The latter will impact the data budget.

- **Mission duration will push systems to their operating lifetimes.**

- **Science planning risk**
  - Relative velocities between Orbiter and Uranus' satellites will be high.
    - ◆ Flybys occur near periapse

- **Collision avoidance with Uranus' rings needs to be considered.**

- **Uranus stays close to the range of solar conjunction (~4-5 deg)**
  - Doppler measurements may have increased noise levels.

- **Running the Orbiter power bus to the SEP stage makes for a more complex electronics design and adds cabling.**
  - Higher risk than adding a battery on the SEP stage.
  - Chose this to minimize SEP stage mass.

- **eMMRTG still needs some development.**
  - May cause a schedule slip.
  - Performance may degrade at a higher rate than currently predicted.

- **ROSA solar array qualification carries some risk.**

- **Low altitude Venus flybys could pose potential thermal risk.**

- **RTG waste heat recovery design robustness**
  - Approach is highly configuration-dependent and may have high hidden development costs.
  - Less expensive on paper, but the actual implementation could be more expensive than an active system.

- **Component development for both propulsion subsystems**
  - NEXT development for SEP
  - Large bi-prop engines for chemical

- **Sun sensor performance may degrade past Saturn.**
  - May impact safe mode used during star tracker outage.

# Option 2

- **Option 2: <u>Uranus </u>Orbiter Variant Concept**
  - <u>150 kg </u>payload allocation
  - <u>No</u> atmospheric probe
    - No crosslink telecom hardware
  - Includes VVE Flybys

- **Class B mission**
- **Dual string redundancy**
- **eMMRTGs could be used for Orbiter power**
  - Carry <u>no mass contingency</u>, because eMMRTG masses provided are "not to exceed" values

- **Xenon residual calculation overridden with Dawn heritage values**
  - ~7.5 kg residuals on ~1040 kg of propellant
- **Assuming SEP stage based on modified launch vehicle adapter**
  - Affects mechanical/structure masses

- **Mission:**
  - Launch: 9/3/2030; Arrival: 9/3/2041
  - Launch, VVE flybys, cruise to Uranus
  - SEP stage jettisoned roughly at 5 AU

- **Mission Design**
  - 11-year cruise to UOI, 4-year science tour
  - UOI inserts into 180-day initial orbit, lowered to ~50-day orbit
  - Will require optical navigation upon approach to UOI, and during science for targeting moon flybys
    - Doppler imager will be used for OpNav on approach

- **Launch Vehicle**
  - Delta IV-Heavy (~10,120 kg to C3 of 2.68 $km^2/s^2$)

✖ **Arrival Vinf / Declination**

- ~9.74 km/s, 68 degree (spin-axis relative)

| Event | Rel. Time | Duration | Delta V (m/s) | # Maneuvers | Comments |
|---|---|---|---|---|---|
| Venus Flyby #1 | L+838 days | | | | 1043 km altitude |
| Venus Flyby #2 | L+1483 days | | | | 300 km altitude |
| Earth Flyby | L+1532 days | | | | 300 km altitude |
| SEP Jettison | L+~2400 days | | Total SEP: 5600 | | Timing flexible |
| TCM-1 | E-~400 days | | 10 | 1 | Non-deterministic |
| TCMs 2-5 | | | 13 (Total) | 4 | Non-deterministic |
| UOI | L+4017 days | ~1 hr | 2260 | 1 | |
| OTMs 1-5 | | | 290 (Total) | 5 | Includes UOI cleanup, period+incl changes, flyby targeting, and other statistical mnvrs |
| **Total** | | | ~2600 chemical, ~5600 SEP | | |

✖ **Element 1: Atmospheric Probe**

- Designed in study 1~~~~~~~~~(3,30th)
- Common for ~~~~~~
- No pr~~~~~~~~~~~ generation
- ~~~~~~~~~~~

✖ **Eleme~~~~~ ystem**

- Hea~~~~~ backshell, structure
- Specific to each planet

✖ **Element 3: Orbiter**

- Instrument allocation defined by Option
- Chemical propulsion
- eMMRTGs, no solar arrays

✖ **Element 4: SEP Cruise Stage**

- "Dumb" cruise stage:
  - ◆ No CDS/ACS/telecom
  - ◆ No instruments
- SEP, no chemical propulsion
- Solar arrays, no RTGs

*Drawing not to scale

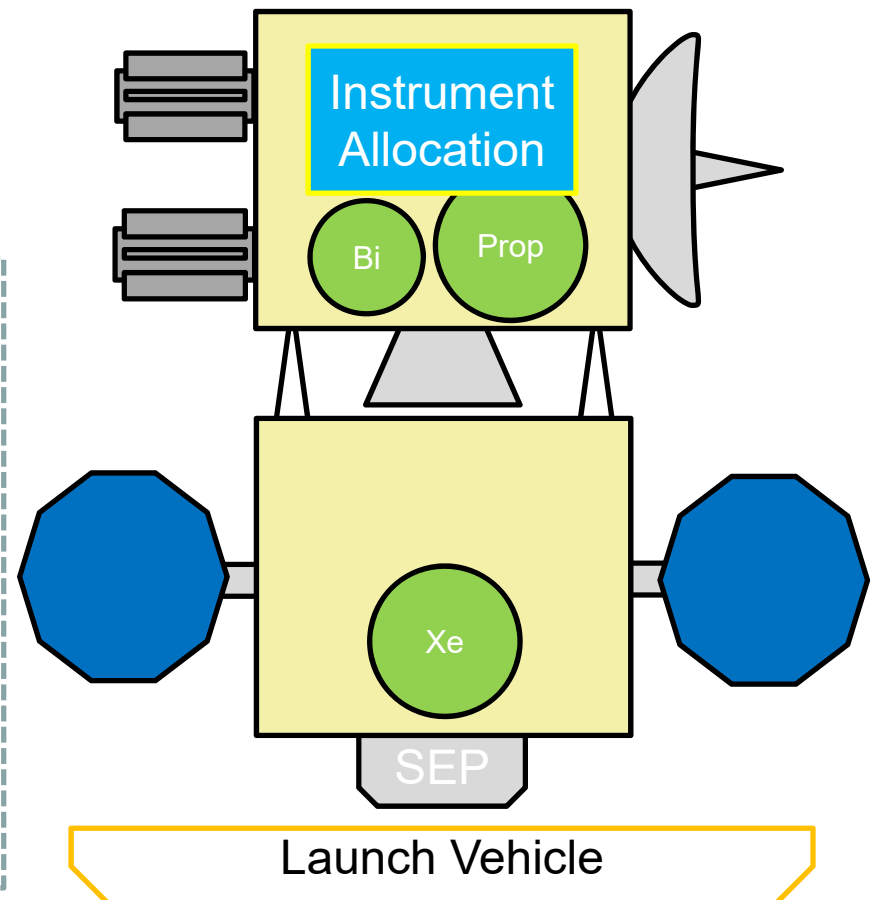

*No probe for this option!*

Instrument Allocation

Bi   Prop

Xe

SEP

Launch Vehicle

**Mission Timeline**

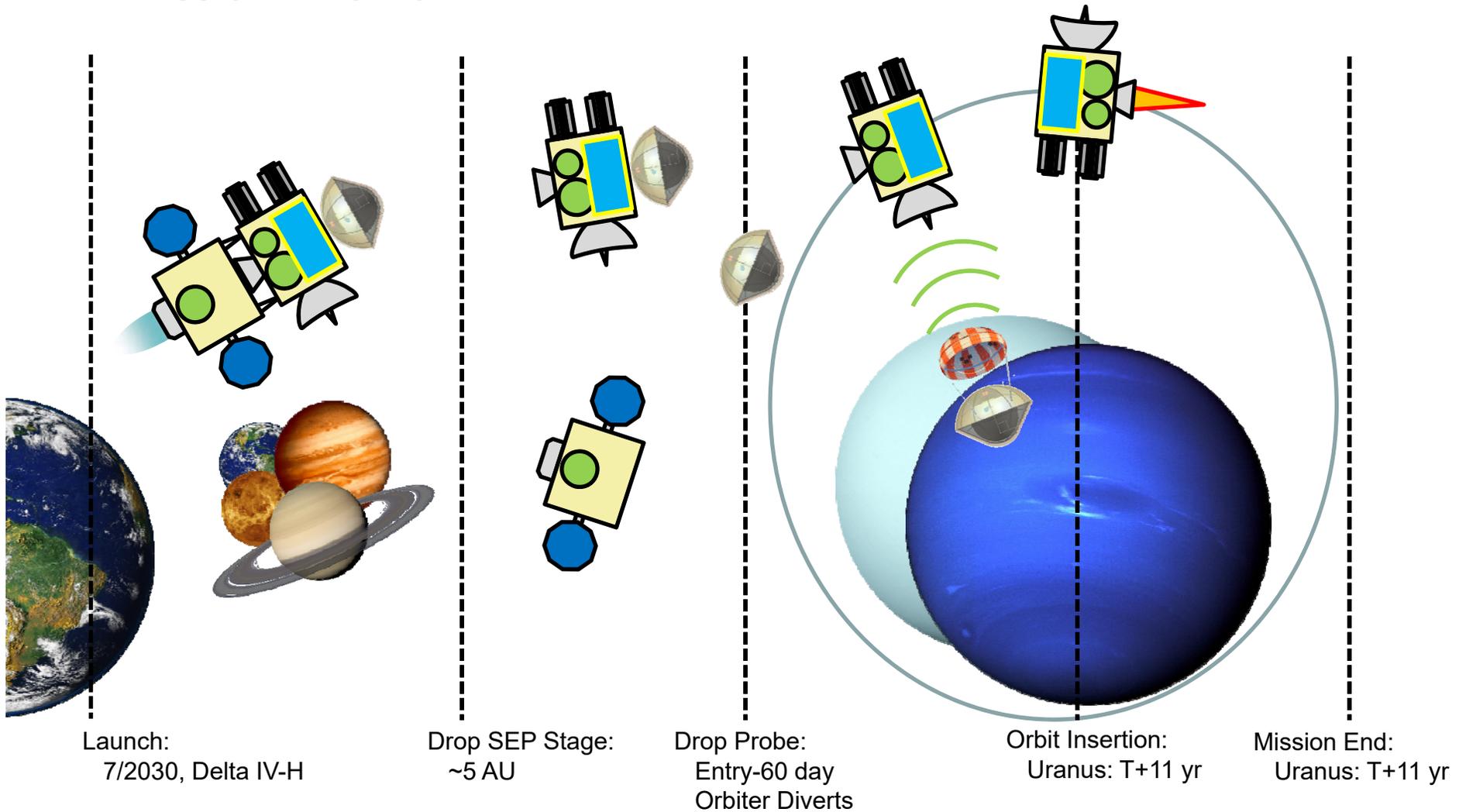

Launch:
7/2030, Delta IV-H

Drop SEP Stage:
~5 AU

Drop Probe:
Entry-60 day
Orbiter Diverts

Orbit Insertion:
Uranus: T+11 yr

Mission End:
Uranus: T+11 yr

**Approach Timeline**

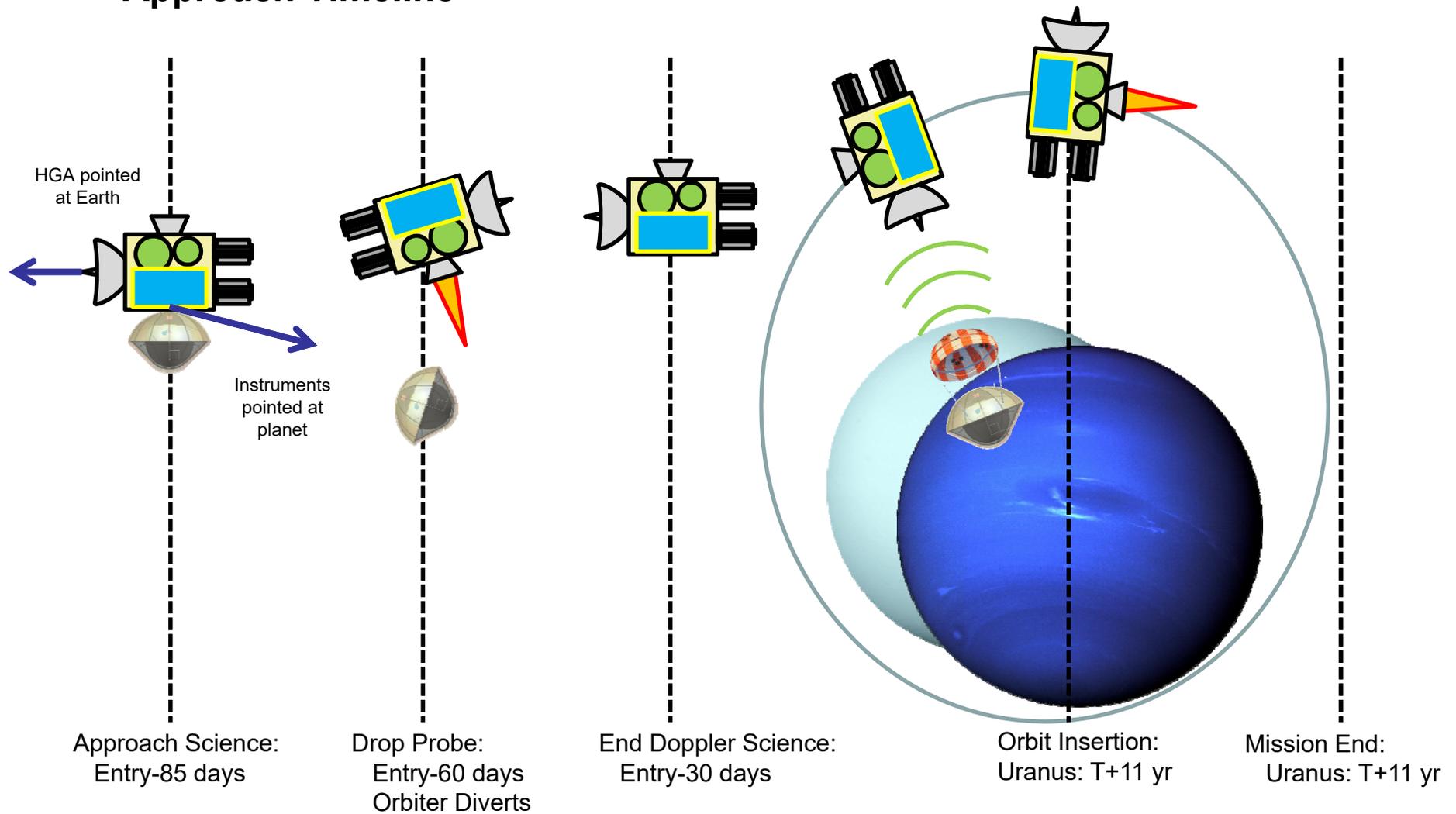

HGA pointed at Earth

Instruments pointed at planet

Approach Science:
Entry-85 days

Drop Probe:
Entry-60 days
Orbiter Diverts

End Doppler Science:
Entry-30 days

Orbit Insertion:
Uranus: T+11 yr

Mission End:
Uranus: T+11 yr

**✸ Entry Timeline**

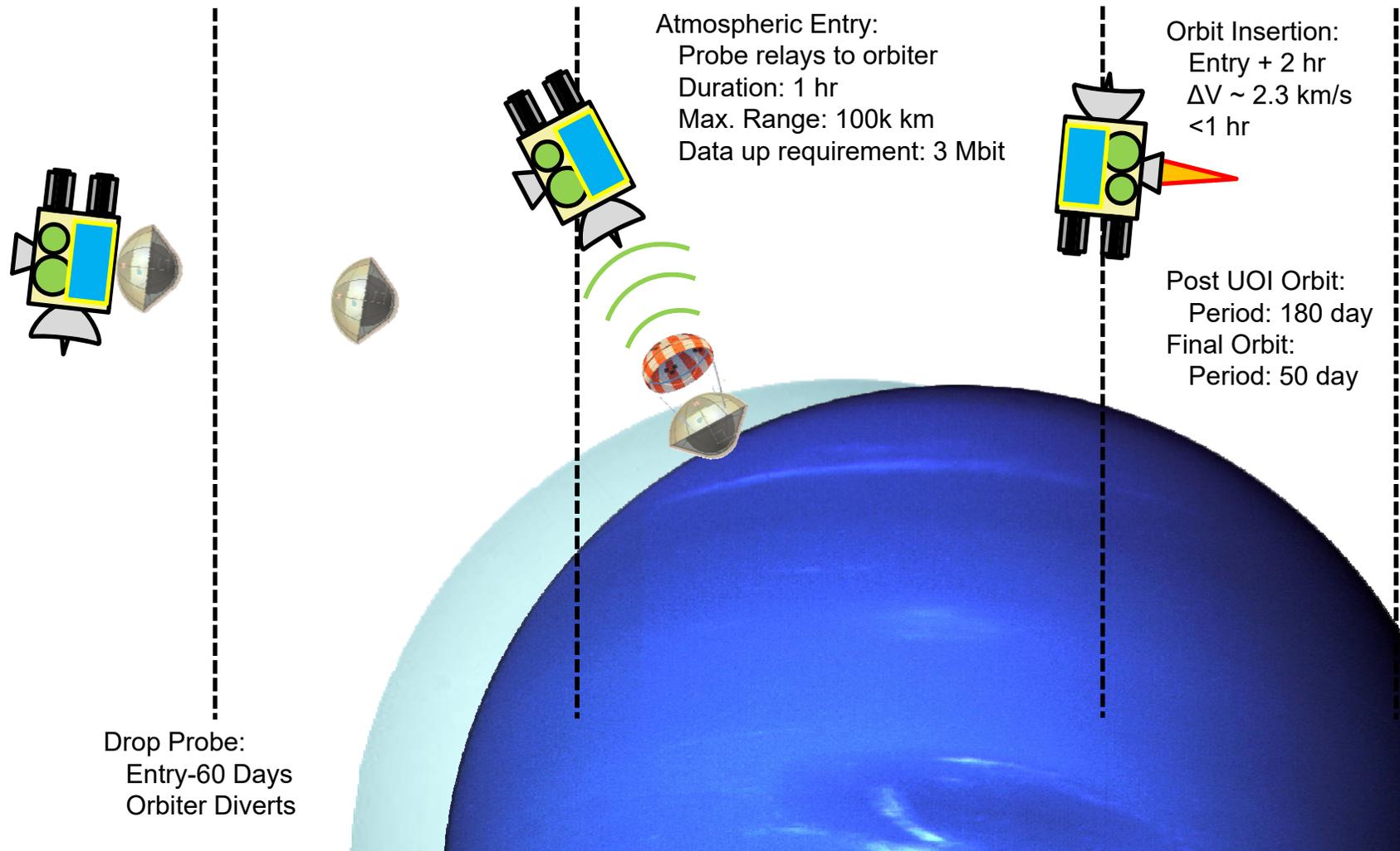

Atmospheric Entry:
  Probe relays to orbiter
  Duration: 1 hr
  Max. Range: 100k km
  Data up requirement: 3 Mbit

Orbit Insertion:
  Entry + 2 hr
  ΔV ~ 2.3 km/s
  <1 hr

Post UOI Orbit:
  Period: 180 day
Final Orbit:
  Period: 50 day

Drop Probe:
  Entry-60 Days
  Orbiter Diverts

Team X Study Guidelines
### Ice Giants Study 2016-07
### Orbiter

**Project - Study**

| | |
|---|---|
| Customer | John Elliott, Kim Reh |
| Study Lead | Bob Kinsey |
| Study Type | Pre-Decadal Study |
| Report Type | Full PPT Report |

**Project - Mission**

| | |
|---|---|
| Mission | Ice Giants Study 2016-07 |
| Target Body | Uranus |
| Science | Imaging and Magnetometry |
| Launch Date | 3-Sep-30 |
| Mission Duration | 11 year cruise, 4 years in orbit |
| Mission Risk Class | B |
| Technology Cutoff | 2026 |
| Minimum TRL at End of Phase B | 6 |

**Project - Architecture**

| | | |
|---|---|---|
| Probe | on | Entry System |
| Entry System | on | Orbiter |
| Orbiter | on | SEP Stage |
| SEP Stage | on | Launch Vehicle |

| | |
|---|---|
| Launch Vehicle | Delta IV-H |
| Trajectory | VVE Gravity Assists, 180 day orbit, FPA = -35deg |
| L/V Capability, kg | 4798 kg to a C3 of 2 with 0% contingency taken out |
| Tracking Network | DSN |
| Contingency Method | Apply Total System-Level |

| Spacecraft | |
|---|---|
| Spacecraft | Orbiter |
| Instruments | Narrow Angle Camera (EIS Europa),Doppler Imager (ECHOES JUICE),Magnetometer (Gallileo),Vis-Near IR Mapping Spectrometer (OVIRS/OSIRIS-Rex),Mid-IR Spectrometer (OTES/OSIRSI-Rex),UV Imaging Spectrometer (Alice/New Horizons),Radio Waves (LPW/Maven),Low Energy Plasma (SWAP/New Horizons),High Energy Plasma (PEPSI/New Horizons),Thermal IR (Diviner/LRO),Energetic Neutral Atoms (INCA/Cassini),Dust Detector (SDC/New Horizons), etc |
| Potential Inst-S/C Commonality | None |
| Redundancy | Dual (Cold) |
| Stabilization | 3-Axis |
| Radiation Total Dose | 29.833 krad behind 100 mil. of Aluminum, with an RDM of 2 added. |
| Type of Propulsion Systems | System 1-Biprop, System 2-0, System 3-0 |
| Post-Launch Delta-V, m/s | 2568 |
| P/L Mass CBE, kg | 144.8 kg Payload CBE |
| P/L Power CBE, W | 84.704 |
| P/L Data Rate CBE, kb/s | 30000 |
| Hardware Models | Protoflight S/C, EM instrument |

| Project - Cost and Schedule | |
|---|---|
| Cost Target | < $2B TBD |
| Mission Cost Category | Flagship - e.g. Cassini |
| FY$ (year) | 2015 |
| Include Phase A cost estimate? | Yes |
| Phase A Start | November 2023 |
| Phase A Duration (months) | 20 |
| Phase B Duration (months) | 16 |
| Phase C/D Duration (months) | 47 |
| Review Dates | PDR - November 2026, CDR - January 2028, ARR - January 2029 |
| Phase E Duration (months) | 179 |
| Phase F Duration (months) | 4 |
| Project Pays Tech Costs from TRL | 6 |
| Spares Approach | Typical |
| Parts Class | Commercial + Military 883B TBR |
| Launch Site | Cape Canaveral |

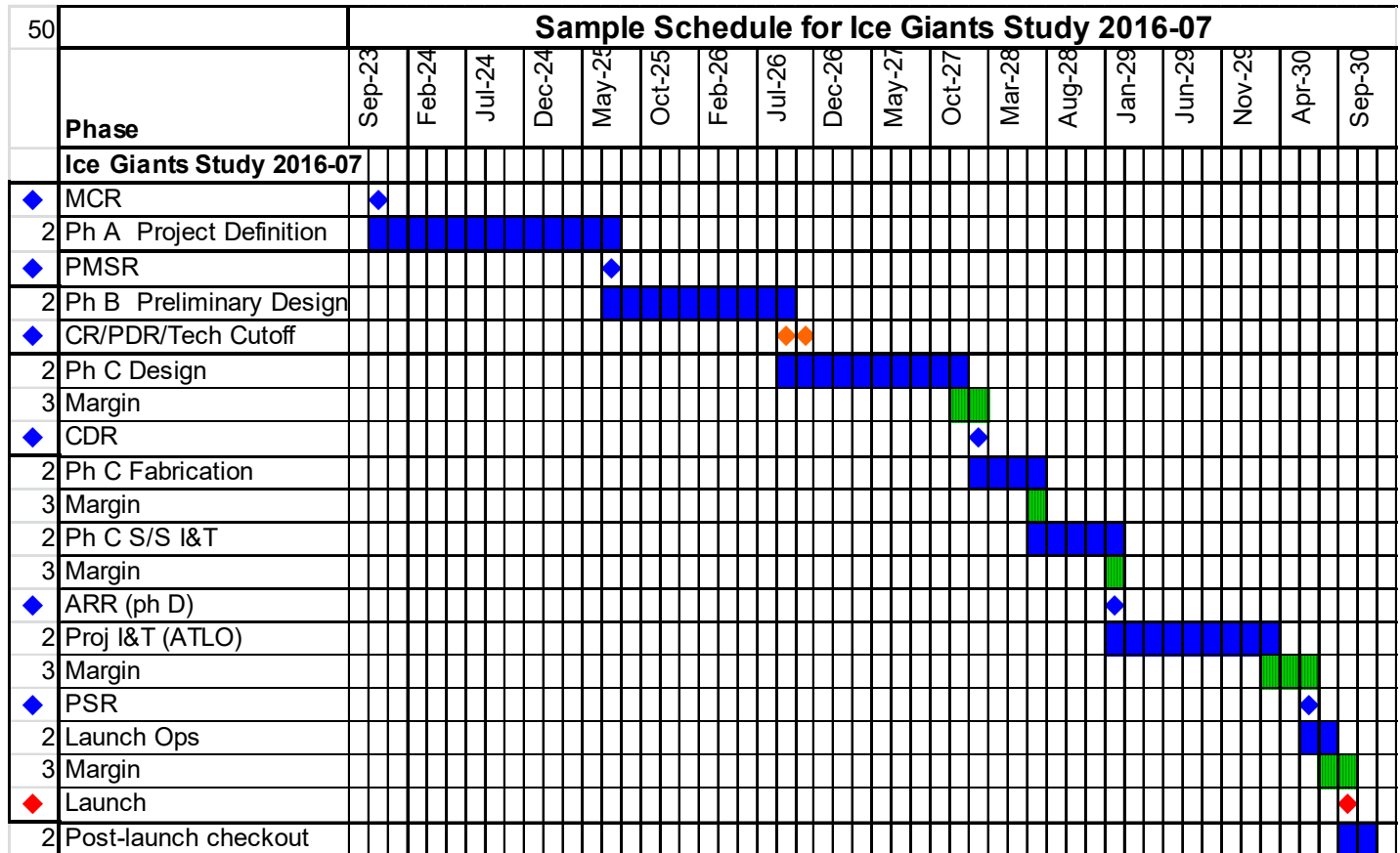

| 50 | Phase | Sample Schedule for Ice Giants Study 2016-07 | | | | | | | | | | | | | | | | | |
|---|---|---|---|---|---|---|---|---|---|---|---|---|---|---|---|---|---|---|---|
| | | Sep-23 | Feb-24 | Jul-24 | Dec-24 | May-25 | Oct-25 | Feb-26 | Jul-26 | Dec-26 | May-27 | Oct-27 | Mar-28 | Aug-28 | Jan-29 | Jun-29 | Nov-29 | Apr-30 | Sep-30 |
| | Ice Giants Study 2016-07 | | | | | | | | | | | | | | | | | | |
| ◆ | MCR | | | | | | | | | | | | | | | | | | |
| 2 | Ph A  Project Definition | | | | | | | | | | | | | | | | | | |
| ◆ | PMSR | | | | | | | | | | | | | | | | | | |
| 2 | Ph B  Preliminary Design | | | | | | | | | | | | | | | | | | |
| ◆ | CR/PDR/Tech Cutoff | | | | | | | | | | | | | | | | | | |
| 2 | Ph C Design | | | | | | | | | | | | | | | | | | |
| 3 | Margin | | | | | | | | | | | | | | | | | | |
| ◆ | CDR | | | | | | | | | | | | | | | | | | |
| 2 | Ph C Fabrication | | | | | | | | | | | | | | | | | | |
| 3 | Margin | | | | | | | | | | | | | | | | | | |
| 2 | Ph C S/S I&T | | | | | | | | | | | | | | | | | | |
| 3 | Margin | | | | | | | | | | | | | | | | | | |
| ◆ | ARR (ph D) | | | | | | | | | | | | | | | | | | |
| 2 | Proj I&T (ATLO) | | | | | | | | | | | | | | | | | | |
| 3 | Margin | | | | | | | | | | | | | | | | | | |
| ◆ | PSR | | | | | | | | | | | | | | | | | | |
| 2 | Launch Ops | | | | | | | | | | | | | | | | | | |
| 3 | Margin | | | | | | | | | | | | | | | | | | |
| ◆ | Launch | | | | | | | | | | | | | | | | | | |
| 2 | Post-launch checkout | | | | | | | | | | | | | | | | | | |

Proposed development schedule consistent with typical New Frontiers missions and current Europa mission schedule.

Phase A 20 mos., Phase B 16 mos., Phase C/D 47 mos.

Launch September 2030

- **First 3 instruments are the same as Option 1.**

- **Total payload mass 145 kg CBE**

- **Largest mass 42 kg CBE Microwave Sounder (MWR/Juno)**

- **Largest power 33 W CBE also MWR/Juno**

- **Total payload cost $219M including**
  - Management
  - Payload Engineering

| Instrument Name | Heritage | CBE Mass (kg) | Cont. | CBE+Cont. Mass (kg) | CBE Op. Power (W) | CBE Standy Power (W) |
|---|---|---|---|---|---|---|
| | | 145 kg | 17% | 169.5 | | |
| Narrow Angle Camera (EIS Europa) | Inherited design | 12.0 | 15% | 13.8 | 16 W | 2 W |
| Doppler Imager (ECHOES JUICE) | New design | 20.0 | 30% | 26 | 20 W | 2 W |
| Magnetometer (Gallileo) | Inherited design | 4.7 | 15% | 5.405 | 8 W | 1 W |
| Vis-Near IR Mapping Spectrometer (OVIRS/OSIRIS-Rex) | Inherited design | 16.5 | 15% | 18.975 | 9 W | 1 W |
| Mid-IR Spectrometer (OTES/OSIRSI-Rex) | Inherited design | 6.3 | 15% | 7.245 | 11 W | 1 W |
| UV Imaging Spec (Alice/New Horizons) | Inherited design | 4.0 | 15% | 4.6 | 10 W | 1 W |
| Radio Waves (LPW/Maven) | Inherited design | 5.6 | 15% | 6.44 | 3 W | 0 W |
| Low Energy Plasma (SWAP/New Horizons) | Inherited design | 3.3 | 15% | 3.795 | 2 W | 0 W |
| High Energy Plasma (PEPSI/New Horizons) | Inherited design | 1.5 | 15% | 1.725 | 3 W | 0 W |
| Thermal IR (Diviner/LRO) | Inherited design | 12.0 | 15% | 13.8 | 25 W | 3 W |
| Energetic Neutral Atoms (INCA/Cassini) | Inherited design | 6.9 | 15% | 7.935 | 3 W | 0 W |
| Dust Detector (SDC/New Horizons) | Inherited design | 5.0 | 15% | 5.75 | 7 W | 1 W |
| Langmuir Probe (RPWS-LP/Cassini) | Inherited design | 1.0 | 15% | 1.15 | 0 W | 0 W |
| Microwave Sounder (MWR/Juno) | Inherited design | 42.0 | 15% | 48.3 | 33 W | 3 W |
| WAC (MDIS-WAC/MESSINGER) | Inherited design | 4.0 | 15% | 4.6 | 10 W | 2 W |

## Instruments
- Narrow Angle Camera
- Doppler Imager
- Magnetometer
- Vis-Near IR Mapping Spectrometer
- Mid-IR Spectrometer
- UV Imaging Spectrometer
- Radio Waves Instrument
- Low Energy Plasma Instrument
- High Energy Plasma Instrument
- Thermal IR
- Energetic Neutral Atoms
- Dust Detector
- Langmuir Probe
- Microwave Sounder

## CDS
- JPL reference bus avionics
- Dual string cold redundancy

## Power
- 5 eMMRTGs, 45kg each
- 10 A-hr, <5kg, Li Ion Battery

## Thermal
- Active and passive thermal control design
- Louvers, heaters, MLI
- eMMRTG shields

## ACS
- Four 0.1N Honeywell HR16 reaction wheels
- IMUs, Star Trackers, Sun Sensors

**Orbiter**

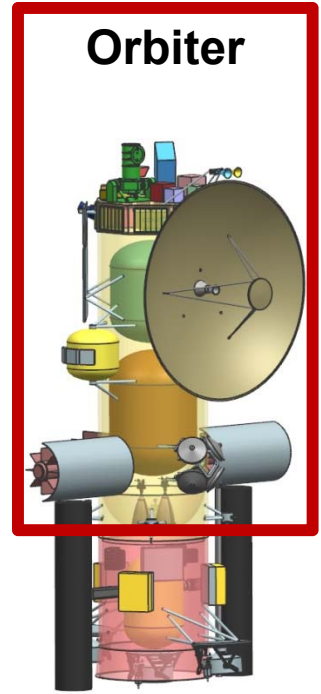

## Telecom
- Radios
  - Two X/X/Ka SDST transponders
  - Two 35W Ka-Band TWTAs
  - Two 25W X-Band TWTAs
- Antennas
  - One 3m X/Ka HGA
  - One X-Band MGA
  - Two X-Band LGAs

## Propulsion
- Dual-mode bipropellant system provides 2568m/s of delta-V
- Two 200lbf Aerojet main engines
- Four 22N engines
- Eight 1N RCS engines

## Structures
- 390kg structure
- 40kg ballast
- 105kg harness
- 10m Magnetometer Boom
- Main Engine cover
- SEP Stage and Probe separation mechanisms
- Main Engine and eMMRTG covers

**Propulsion**
- 2+1 NEXT main engines
- Three 35kg PPUs
- 1040kg Xenon Propellant

**Power**
- Two 54m$^2$ ROSA Solar Arrays
- Provide ~29.5kW at 1AU
- Redundant JPL Reference Bus Power Electronics

**CDS**
- Redundant remote engineering unit

**ACS**
- 1DOF solar array gimbal drive electronics
- 2DOF SEP engine gimbal drive electronics
- Sun sensors

**Thermal**
- Active and passive thermal control design
- Louvers, heaters, MLI

**Structures**
- Cylindrical bus shape, made up of stacked CXX adapters

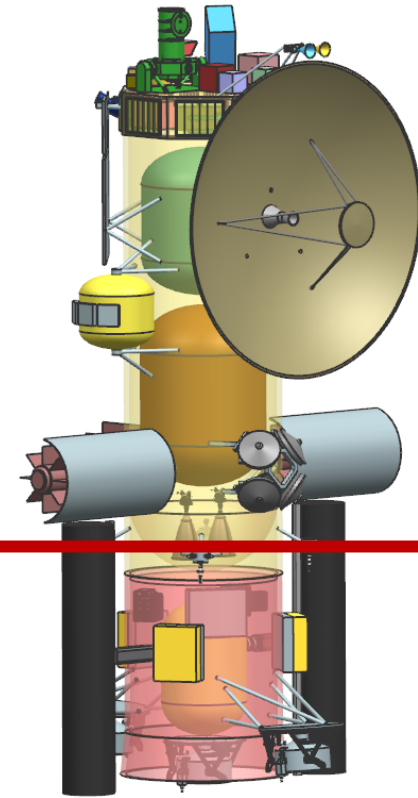

**SEP Stage**

✖ **Configuration Drawings – Stowed**

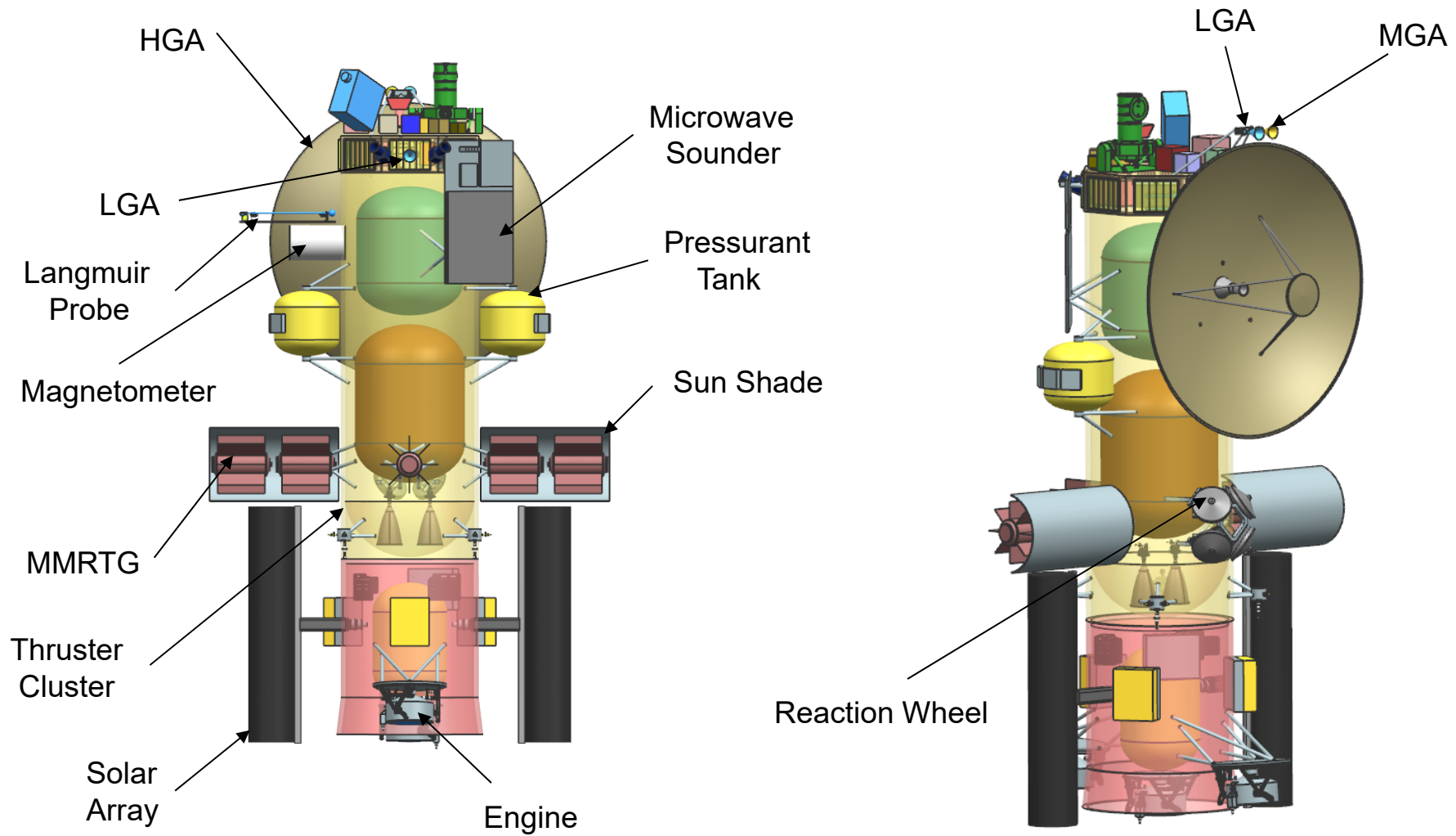

HGA

LGA

MGA

Microwave
Sounder

LGA

Pressurant
Tank

Langmuir
Probe

Magnetometer

Sun Shade

MMRTG

Thruster
Cluster

Reaction Wheel

Solar
Array

Engine

✖ **Configuration Drawings – Deployed**

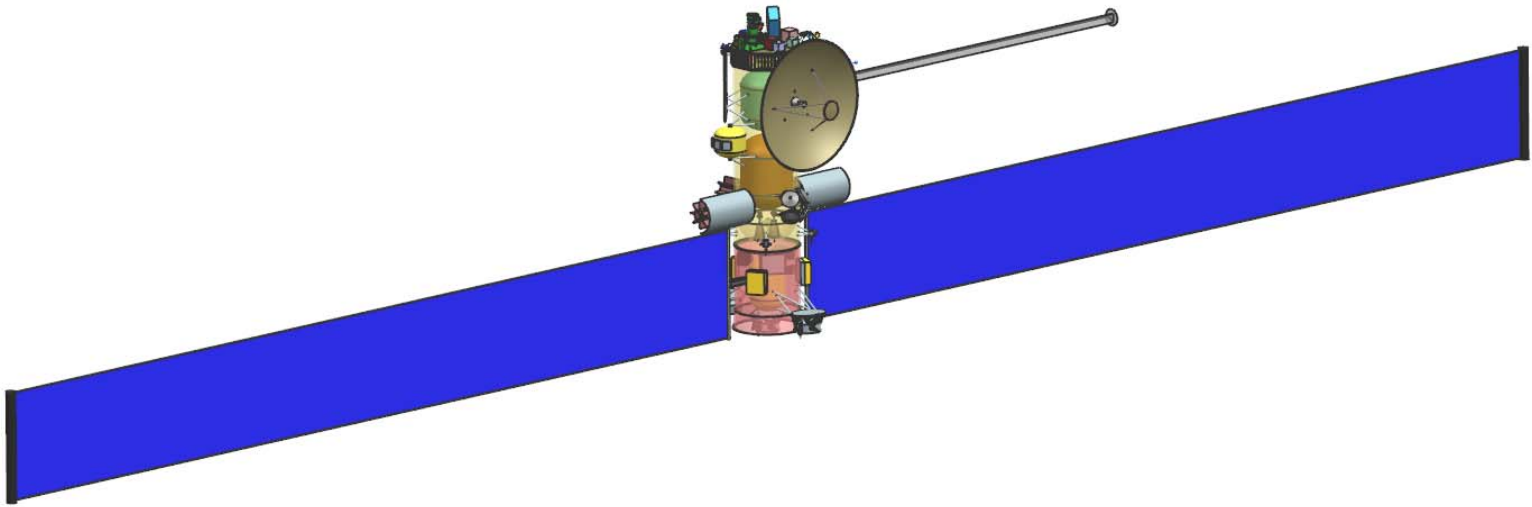

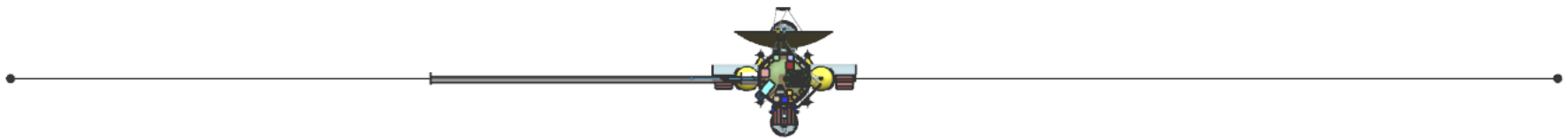

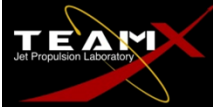

# Executive Summary
## Option 2 – Orbiter Element

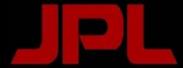

| | Mass Fraction | Mass (kg) | Subsys Cont. % | CBE+ Cont. (kg) | Mode 1 Power (W) Recharge | Mode 2 Power (W) Approach Science - 85 Days | Mode 3 Power (W) Telecom Downlink | Mode 4 Power (W) UOI Delta-V | Mode 5 Power (W) Orbital Science - Apoapse | Mode 6 Power (W) Orbital Science - Periapse | Mode 7 Power (W) Orbital Science - Moon Flyby | Mode 8 Power (W) Safe | Mode 9 Power (W) SEP Thrusting | Mode 10 Power (W) Orbiting Science - Rotation Rate Movie |
|---|---|---|---|---|---|---|---|---|---|---|---|---|---|---|
| *Power Mode Duration  (hours)* | | | | | *24* | *24* | *8* | *1.05* | *16* | *16* | *8* | *24* | *24* | *17* |
| **Payload on this Element** | | | | | | | | | | | | | | |
| Instruments | 10% | 144.8 | 17% | 169.5 | 4 | 28 | 4 | 4 | 16 | 85 | 55 | 4 | 4 | 81 |
| Payload Total | 10% | **144.8** | **17%** | **169.5** | **4** | **28** | **4** | **4** | **16** | **85** | **55** | **4** | **4** | **81** |
| **Spacecraft Bus** | | do not edit formulas below this line, use the calcualtions and override tables instead --> | | | | | | | | | | | | |
| Attitude Control | 4% | 63.5 | 10% | 69.8 | 0 | 55 | 55 | 88 | 55 | 55 | 55 | 42 | 55 | 93 |
| Command & Data | 2% | 27.6 | 18% | 32.4 | 61 | 61 | 61 | 61 | 61 | 61 | 61 | 61 | 61 | 61 |
| Power | 18% | 265.3 | 2% | 269.7 | 28 | 44 | 36 | 28 | 28 | 28 | 28 | 44 | 36 | 36 |
| Propulsion1 ☐ SEP1 | 13% | 188.8 | 5% | 198.9 | 31 | 3 | 3 | 151 | 3 | 3 | 3 | 3 | 3 | 3 |
| Structures & Mechanisms | 35% | 516.4 | 30% | 671.3 | 0 | 0 | 0 | 0 | 0 | 0 | 0 | 0 | 0 | 0 |
| Cabling | 7% | 103.8 | 30% | 135.0 | | | | | | | | | | |
| Telecom | 4% | 55.2 | 15% | 63.4 | 12 | 107 | 155 | 71 | 12 | 42 | 12 | 71 | 12 | 12 |
| Thermal | 8% | 121.1 | 23% | 148.4 | 25 | 25 | 25 | 25 | 25 | 25 | 25 | 25 | 25 | 25 |
| Bus Total | | 1341.6 | 18% | 1589.0 | 157 | 294 | 335 | 424 | 184 | 214 | 184 | 246 | 192 | 230 |
| Thermally Controlled Mass | | | | 1589.0 | | | | | | | | | | |
| **Spacecraft Total (Dry): CBE & MEV** | | **1486.4** | 18% | **1758.6** | 162 | 322 | 339 | 429 | 200 | 298 | 239 | 250 | 196 | 311 |
| Subsystem Heritage Contingency | 18% | 272.1 | SEP Cont | 10% | 0 | 0 | 0 | 0 | 0 | 0 | 0 | 0 | 0 | 0 |
| System Contingency | 18% | 270.3 | | | 69 | 138 | 146 | 184 | 86 | 128 | 103 | 108 | 84 | 134 |
| Total Contingency     ☐ Include Carried? | **36%** | 542.4 | | | | | | | | | | | | |
| **Spacecraft with Contingency:** | | **2029** | of total | w/o addl pld | **231** | **461** | **485** | **613** | **285** | **426** | **342** | **358** | **280** | **444** |
| Propellant & Pressurant with residuals1 | 58% | 2769.8 | For S/C mass = | 4784.0 | | Delta-V, Sys 1 | 2568.0 | m/s | | residuals = | 72.5 | kg | | |
| **Spacecraft Total with Contingency (Wet)** | | **4798.7** | | | | | | | | | | | | |

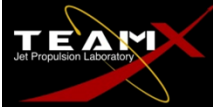
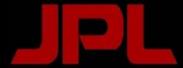

| | | Mass Fraction | Mass (kg) | Subsys Cont. % | CBE+ Cont. (kg) | | Mode 8 Power (W) Launch | Mode 9 Power (W) SEP Thrusting | Mode 10 Power (W) Safe |
|---|---|---|---|---|---|---|---|---|---|
| **Power Mode Duration** *(hours)* | | | | | | | 2 | 24 | 24 |
| **Additional Elements Carried by this Element** | | | | | | | | | |
| Orbiter | | 80% | 4256.3 | 13% | 4798.7 | | | | |
| **Carried Elements Total** | | 80% | 4256.3 | 13% | 4798.7 | | 0 | 0 | 0 |
| **RSDO Option -->** | | | | | | | | | |
| **Spacecraft Bus** | | | do not edit formulas below | | | | ns and override tables instead --> | | |
| Attitude Control | | 0% | 6.0 | 7% | 6.4 | | 0 | 18 | 0 |
| Command & Data | | 0% | 1.6 | 5% | 1.7 | | 4 | 4 | 4 |
| Power | | 5% | 265.5 | 29% | 342.9 | | 0 | 669 | 35 |
| Propulsion1 ☑ SEP1 | | 5% | 245.0 | 22% | 297.8 | | 0 | 25000 | 0 |
| Structures & Mechanisms | | 7% | 364.5 | 30% | 473.8 | | 0 | 0 | 0 |
| Cabling | | 1% | 77.1 | 30% | 100.2 | | | | |
| Thermal | | 1% | 78.8 | 0% | 78.8 | | 0 | 163 | 331 |
| **Bus Total** | | | 1038.5 | 25% | 1301.7 | | 4 | 25855 | 370 |
| Thermally Controlled Mass | | | | | 1301.7 | | | | |
| **Spacecraft Total (Dry): CBE & MEV** | | | **5294.8** | 15% | **6100.4** | | 4 | 25855 | 370 |
| Subsystem Heritage Contingency | 15% | | 805.6 | SEP Cont | 10% | | 0 | 2500 | 0 |
| System Contingency | 3% | | 183.4 | | | | 2 | 367 | 159 |
| Total Contingency ☐ Include Carried? | **19%** | | **989.0** | | | | | | |
| **Spacecraft with Contingency:** | | | **6284** | of total | w/o addl pld | | 6 | 28722 | 529 |
| Propellant & Pressurant with residuals1 | | 14% | 1040.4 | For S/C mass = | 2000.0 | kg | | | |
| **Spacecraft Total with Contingency (Wet)** | | | **7324.3** | | | | | | |
| L/V-Side Adapter | | | 0.0 | Wet Mass for Prop Sizing | 10120 | | | | |
| **Launch Mass** | | | **7324** | Dry Mass for Prop Sizing | 6284 | | | | |
| **Launch Vehicle Capability** | | | **10120** | Delta IV-H | | | | | |
| | | | 5820 | Atlas V 551 | | Launch C3 | 2.68 | | |
| **Launch Vehicle Margin** | | | **2795.7** | Mission Unique LV Contingency | 0% | | | | |

| Element Number | Element Name | Dry CBE (kg) | Cont / JPL Margin (kg) | Dry Allocation (kg) | Propellant (kg) | Dry Allocation + Propellant (kg) |
|---|---|---|---|---|---|---|
| 3 | Orbiter minus eMMRTGs | 1,261 | 542 | 1,804 | 2,770 | 4,574 |
| 3.1 | eMMRTGs | 225 | - | 225 | | 225 |
| 4 | SEP Stage | 1,039 | 447 | 1,485 | 1,040 | 2,525 |
| | **Total Stack** | **2,525** | **989** | **3,514** | **3,810** | **7,324** |
| | | | | Dry Mass Allocation | | 3,514 |
| | | | | JPL Margin (kg / %) | | 989 / 28% |
| | | | | JPL Margin without eMMRTG (kg / %) | | 989 / 30% |
| | | | | Delta IV-H Capacity (kg) | | 10,120 |
| | | | | Extra Launch Vehicle Margin (kg) | | 2,796 |

| Element Number | Element Name | Dry CBE (kg) | Cont (%) | Cont. (kg) | MEV (kg) | Dry Allocation (kg) | Propellant (kg) | Dry Allocation + Propellant (kg) |
|---|---|---|---|---|---|---|---|---|
| 3 | Orbiter minus eMMRTGs | 1,261 | 22% | 272 | 1,534 | 1,804 | 2,770 | 4,574 |
| 3.1 | eMMRTGs | 225 | - | - | 225 | 225 | - | 225 |
| 4 | SEP Stage | 1,039 | 25% | 264 | 1,302 | 1,485 | 1,040 | 2,525 |
| | **Total Stack** | **2,525** | | **536** | **3,061** | **3,514** | **3,810** | **7,324** |

| | |
|---|---|
| Dry Mass Allocation (kg) | 3,514 |
| NASA Margin (kg / %) | 453 / 15% |
| NASA Margin without eMMRTG (kg / %) | 453 / 16% |
| Delta IV-H Capacity (kg) | 10,120 |
| Extra Launch Vehicle Margin (kg) | 2,796 |

✖ **5th eMMRTG enables higher data rate, lower costs on orbit.**

- Expected 5th RTG to be needed for instrument power.
  - ◆ Turns out that only 4 RTGs are needed.
- More instruments while in orbit means higher data volume than Option 1.
  - ◆ Downlink for more passes/day or increase the downlink data rate.
- DSN pricing highly favors one 8-hour pass/day, so avoid more passes.
  - ◆ Cost for 3 passes/day can be 9 times the cost of 1 pass/day.
- 5th RTG enables use of a 70W TWTA instead of 35W, doubles data rate.
  - ◆ 30 kbps compared to 15 kbps for Option 1
- An 8-hour pass to an array of two 34-m ground stations gets the data down.
- Saves a couple of hundred million dollars while on orbit.

- **Data downlink strategy for Doppler Imager (DI) on approach**
  - DI generates a lot of data continuously for tens of days on approach.
  - Configuration with HGA and DI on opposite sides of the cylindrical bus allows pointing DI towards Uranus while pointing HGA towards Earth.
  - Can downlink for 24 hours/day and maintain positive power balance.
  - Any data that can't be downlinked before UOI will be downlinked after.

- **Configuration that helps to minimize mass and power**
  - eMMRTGs outside the cylindrical bus provide heating
    - Reduces mass and power of thermal subsystem components
  - Propellant and oxidizer tanks inside bus, pressurant tanks outside
    - Pressurant tanks are easier to keep warm than propellant/oxidizer tanks.
  - Shorten the stack to minimize structure mass
    - Single custom propellant tank instead of two tanks
    - Stow solar arrays perpendicular to bus to reduce height of the SEP stage

## Mechanical

- LV interfaces directly to the SEP Stage; Orbiter interfaces to SEP Stage; Entry System containing Probe is attached to the side of the Orbiter.

- Orbiter: primary structure is the largest mass element.
  - 321 kg CBE out of 516 kg total for Mechanical (1342 kg bus dry mass)
  - Drivers are the large Propulsion and Power masses.

- SEP stage: primary structure is the largest mass element.
  - 212 kg CBE out of 365 kg total for Mechanical (1039 kg SEP Stage dry mass)
  - Due to the Orbiter and other mission elements being carried during launch.
  - Primary structure of the SEP Stage is being utilized as the LVA.

- **Baseline Power System: Orbiter power bus spans both Orbiter and SEP stage**
  - Dual String Reference Bus electronics heritage
  - New development High Voltage Electronics Assembly for the SEP stage
  - Two Rollout Solar Arrays (ROSAs) support 25kW SEP at 1AU.

- **Propulsion: SEP Stage used to ~5 AU; dual mode bi-prop.**
  - SEP Stage: 2+1 system using NEXT Engines
    - 1033 kg Xe mass allocation + 7 kg residuals for a total of 1040 kg
  - Chemical System: orbit insertion delta V= 2.3km/s desired in < 1hr
    - Two 890N main engines used to achieve burn time < 1hr

- **Thermal: Cassini-heritage waste-heat recovery system on Orbiter**
  - RTG end domes each provide 75 W waste heat to propulsion module via conductive and radiative coupling.
  - VRHUs act as primary control mechanism for thruster clusters.
    - Also act as trim heaters for the propulsion module
  - Louvers act as primary control mechanism for avionics module.

- **Telecom: X- and Ka-Band subsystem.**
  - Two 70W Ka-Band TWTAs, two 25W X-Band TWTAs
  - Two X/X/Ka SDST transponders
  - 3m X/Ka HGA, one X-Band MGA, two X-Band LGAs.
  - Supports a data rate at Uranus of 30 kbps into 34m BWG ground station.

- **CDS: Reference Bus architecture ideally suited for high reliability, long lifetime mission.**
  - Standard JPL spacecraft CDS that is similar to SMAP
    - RAD750 CPU, NVM, MTIF, MSIA, CRC, LEU-A, LEU-D, MREU
    - 128 GBytes storage for science data
    - 1553 and RS-422 ICC/ITC interfaces for subsystems and instruments

- **ACS: 3-axis stabilized with star tracker, sun sensor, gyros, wheels.**
  - All stellar attitude determination to minimize power, conserve gyros.
  - Sun sensor performance may degrade once the Orbiter passes Saturn.
    - May impact safe mode used during star tracker outage.
    - Detailed analysis on Sun sensor performance versus distance is needed.

✖ **Software: core product line is appropriate since this mission has aspects similar to MSL/M2020/SMAP/Europa.**

- Complexity rankings range from Medium to High.
  - ◆ Medium infrastructure: dual string with warm spare.
  - ◆ High fault behaviors: high redundancy, string swapping, critical events.
  - ◆ Medium/High ACS: tight pointing requirements, many ACS modes
  - ◆ High Telecom: 24/7 downlink, redundant DTE
  - ◆ High Science data processing, full file system

✖ **SVIT: Testbed, System I&T and V&V costs are included**

- Cost of assembling and testing RTG's is captured elsewhere
  - ◆ Cost of integrating RTG's is included with other ATLO costs

✖ **Ground Systems**

- Mission specific implementation of standard JPL mission operations and ground data systems
- Ground network: array of DSN 34-m BWG; 70-m or equivalent for safe mode
- Science support: 24x7 tracking on approach; daily 8-hour contacts on orbit

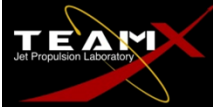
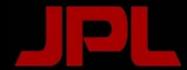

| COST SUMMARY (FY2015 $M) | Generate ProPricer Input | Team X Estimate | | |
|---|---|---|---|---|
| | | CBE | Res. | PBE |
| **Project Cost** | | **$1875.7 M** | **20%** | **$2259.1 M** |
| **Launch Vehicle** | | **$33.0 M** | **0%** | **$33.0 M** |
| **Project Cost (w/o LV)** | | **$1842.7 M** | **21%** | **$2226.1 M** |
| **Development Cost** | | **$1314.8 M** | **24%** | **$1630.7 M** |
| Phase A | | $13.1 M | 24% | $16.3 M |
| Phase B | | $118.3 M | 24% | $146.8 M |
| Phase C/D | | $1183.3 M | 24% | $1467.7 M |
| **Operations Cost** | | **$527.9 M** | **13%** | **$595.4 M** |

Total mission cost is $2.26B. This is the likely cost within a range that typically can be as much as 10% lower up to 20% higher. The development cost with reserves is $1.63B.

| WBS Elements | NRE | RE | 1st Unit |
|---|---|---|---|
| Project Cost (no Launch Vehicle) | $1628.3 M | $630.8 M | $2259.1 M |
| Development Cost (Phases A - D) | $1000.0 M | $630.8 M | $1630.7 M |
| 01.0 Project Management | $47.3 M | | $47.3 M |
| 1.01 Project Management | $11.4 M | | $11.4 M |
| 1.02 Business Management | $13.6 M | | $13.6 M |
| 1.04 Project Reviews | $2.5 M | | $2.5 M |
| 1.06 Launch Approval | $19.8 M | | $19.8 M |
| 02.0 Project Systems Engineering | $24.9 M | $0.8 M | $25.7 M |
| 2.01 Project Systems Engineering | $8.9 M | | $8.9 M |
| 2.02 Project SW Systems Engineering | $5.2 M | | $5.2 M |
| 2.03 EEIS | $1.5 M | | $1.5 M |
| 2.04 Information System Management | $1.7 M | | $1.7 M |
| 2.05 Configuration Management | $1.5 M | | $1.5 M |
| 2.06 Planetary Protection | $0.2 M | $0.2 M | $0.4 M |
| 2.07 Contamination Control | $2.4 M | $0.6 M | $2.9 M |
| 2.09 Launch System Engineering | $1.0 M | | $1.0 M |
| 2.10 Project V&V | $2.0 M | | $2.0 M |
| 2.11 Risk Management | $0.5 M | | $0.5 M |
| 03.0 Mission Assurance | $54.1 M | $0.0 M | $54.1 M |
| 04.0 Science | $66.2 M | | $66.2 M |
| 05.0 Payload System | $147.9 M | $86.3 M | $234.1 M |
| 5.01 Payload Management | $15.8 M | | $15.8 M |
| 5.02 Payload Engineering | $12.9 M | | $12.9 M |
| Orbiter Instruments | $119.1 M | $86.3 M | $205.4 M |
| Narrow Angle Camera (EIS Europa) | $11.6 M | $8.4 M | $20.0 M |
| Doppler Imager (ECHOES JUICE) | $17.4 M | $12.6 M | $30.0 M |
| Magnetometer (Gallileo) | $4.5 M | $3.3 M | $7.8 M |
| Vis-Near IR Mapping Spectrometer (OVIRS/ | $9.7 M | $7.0 M | $16.7 M |
| Mid-IR Spectrometer (OTES/OSIRSI-Rex) | $7.1 M | $5.2 M | $12.3 M |
| UV Imaging Spectrometer (Alice/New Horizo | $5.8 M | $4.2 M | $10.0 M |
| Radio Waves (LPW/Maven) | $3.4 M | $2.4 M | $5.8 M |
| Low Energy Plasma (SWAP/New Horizons) | $2.5 M | $1.8 M | $4.2 M |
| High Energy Plasma (PEPSI/New Horizons) | $2.2 M | $1.6 M | $3.8 M |
| Thermal IR (Diviner/LRO) | $14.7 M | $10.6 M | $25.3 M |
| Energetic Neutral Atoms (INCA/Cassini) | $4.4 M | $3.2 M | $7.7 M |
| Dust Detector (SDC/New Horizons) | $5.7 M | $4.1 M | $9.8 M |
| Langmuir Probe (RPWS-LP/Cassini) | $1.1 M | $0.8 M | $1.9 M |
| Microwave Sounder (MWR/Juno) | $23.3 M | $16.9 M | $40.2 M |
| WAC (MDIS-WAC/MESSINGER) | $5.7 M | $4.1 M | $9.8 M |

| WBS Elements | NRE | RE | 1st Unit |
|---|---|---|---|
| 06.0 Flight System | $391.3 M | $380.5 M | $771.8 M |
| 6.01 Flight System Management | $5.0 M | | $5.0 M |
| 6.02 Flight System Systems Engineering | $35.9 M | | $35.9 M |
| 6.03 Product Assurance (included in 3.0) | | | $0.0 M |
| Orbiter | $295.2 M | $272.6 M | $567.9 M |
| 6.04 Power | $94.5 M | $163.6 M | $258.1 M |
| 6.05 C&DH | $32.0 M | $31.2 M | $63.2 M |
| 6.06 Telecom | $24.6 M | $15.2 M | $39.8 M |
| 6.07 Structures (includes Mech. I&T) | $52.3 M | $17.0 M | $69.3 M |
| 6.08 Thermal | $4.2 M | $13.5 M | $17.6 M |
| additional cost for >43 RHUs | $34.0 M | $0.0 M | $34.0 M |
| 6.09 Propulsion | $22.1 M | $17.0 M | $39.1 M |
| 6.10 ACS | $9.4 M | $9.8 M | $19.1 M |
| 6.11 Harness | $4.5 M | $4.5 M | $8.9 M |
| 6.12 S/C Software | $17.3 M | $0.9 M | $18.2 M |
| 6.13 Materials and Processes | $0.4 M | $0.0 M | $0.4 M |
| SEP Stage | $50.6 M | $106.3 M | $156.9 M |
| 6.04 Power | $7.7 M | $50.7 M | $58.4 M |
| 6.05 C&DH | $4.0 M | $2.7 M | $6.7 M |
| 6.06 Telecom | $0.0 M | $0.0 M | $0.0 M |
| 6.07 Structures (includes Mech. I&T) | $10.2 M | $4.5 M | $14.7 M |
| 6.08 Thermal | $3.1 M | $11.2 M | $14.3 M |
| 6.09 Propulsion | $22.2 M | $34.5 M | $56.7 M |
| 6.10 ACS | $0.7 M | $1.2 M | $1.8 M |
| 6.11 Harness | $2.4 M | $1.5 M | $3.9 M |
| 6.12 S/C Software | $0.0 M | $0.0 M | $0.0 M |
| 6.13 Materials and Processes | $0.4 M | $0.0 M | $0.4 M |
| 6.14 Spacecraft Testbeds | $4.6 M | $1.5 M | $6.1 M |

| WBS Elements | NRE | RE | 1st Unit |
|---|---|---|---|
| **07.0 Mission Operations Preparation** | **$32.2 M** | | **$32.2 M** |
| 7.0 MOS Teams | $26.2 M | | $26.2 M |
| 7.03 DSN Tracking (Launch Ops.) | $2.7 M | | $2.7 M |
| 7.06 Navigation Operations Team | $3.3 M | | $3.3 M |
| 7.07.03 Mission Planning Team | $0.0 M | | $0.0 M |
| **09.0 Ground Data Systems** | **$28.7 M** | | **$28.7 M** |
| 9.0A Ground Data System | $23.3 M | | $23.3 M |
| 9.0B Science Data System Development | $4.6 M | | $4.6 M |
| 9A.03.07 Navigation H/W & S/W Development | $0.8 M | | $0.8 M |
| **10.0 ATLO** | **$16.2 M** | **$17.7 M** | **$33.9 M** |
| **11.0 Education and Public Outreach** | **$0.0 M** | **$0.0 M** | **$0.0 M** |
| **12.0 Mission and Navigation Design** | **$20.8 M** | | **$20.8 M** |
| 12.01 Mission Design | $2.0 M | | $2.0 M |
| 12.02 Mission Analysis | $5.4 M | | $5.4 M |
| 12.03 Mission Engineering | $1.8 M | | $1.8 M |
| 12.04 Navigation Design | $11.6 M | | $11.6 M |
| **Development Reserves** | **$170.4 M** | **$145.6 M** | **$315.9 M** |

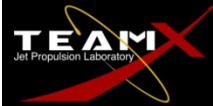

# Executive Summary
## Cost E-F and Launch Nuclear Safety

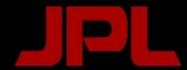

| WBS Elements | NRE | RE | 1st Unit |
|---|---|---|---|
| **Operations Cost (Phases E - F)** | **$595.3 M** | **$0.1 M** | **$595.4 M** |
| **01.0 Project Management** | **$27.1 M** | | **$27.1 M** |
| 1.01 Project Management | $15.3 M | | $15.3 M |
| 1.02 Business Management | $10.7 M | | $10.7 M |
| 1.04 Project Reviews | $1.1 M | | $1.1 M |
| 1.06 Launch Approval | $0.1 M | | $0.1 M |
| **02.0 Project Systems Engineering** | **$0.0 M** | **$0.1 M** | **$0.1 M** |
| **03.0 Mission Assurance** | **$3.6 M** | **$0.0 M** | **$3.6 M** |
| **04.0 Science** | **$243.7 M** | | **$243.7 M** |
| **07.0 Mission Operations** | **$208.3 M** | | **$208.3 M** |
| 7.0 MOS Teams | $107.1 M | | $107.1 M |
| 7.03 DSN Tracking | $78.3 M | | $78.3 M |
| 7.06 Navigation Operations Team | $21.9 M | | $21.9 M |
| 7.07.03 Mission Planning Team | $1.0 M | | $1.0 M |
| **09.0 Ground Data Systems** | **$45.2 M** | | **$45.2 M** |
| 9.0A GDS Teams | $25.7 M | | $25.7 M |
| 9.0B Science Data System Ops | $18.9 M | | $18.9 M |
| 9A.03.07 Navigation HW and SW Dev | $0.6 M | | $0.6 M |
| **11.0 Education and Public Outreach** | **$0.0 M** | **$0.0 M** | **$0.0 M** |
| **12.0 Mission and Navigation Design** | **$0.0 M** | | **$0.0 M** |
| **Operations Reserves** | **$67.4 M** | **$0.0 M** | **$67.5 M** |
| **8.0 Launch Vehicle** | **$33.0 M** | | **$33.0 M** |
| **Launch Vehicle and Processing** | **$0.0 M** | | **$0.0 M** |
| **Nuclear Payload Support** | **$33.0 M** | | **$33.0 M** |

- **Mission duration will push systems to their operating lifetimes.**
- **Science planning risk**
  - Relative velocities between Orbiter and Uranus' satellites will be high.
    - Flybys occur near periapse
- **Collision avoidance with Uranus' rings needs to be considered.**
- **Uranus stays close to the range of solar conjunction (~4-5 deg)**
  - Doppler measurements may have increased noise levels.
- **Running the Orbiter power bus to the SEP stage makes for a more complex electronics design and adds cabling.**
  - Higher risk than adding a battery on the SEP stage.
  - Chose this to minimize SEP stage mass.
- **eMMRTG still needs some development.**
  - May cause a schedule slip.
  - Performance may degrade at a higher rate than currently predicted.
- **ROSA solar array qualification carries some risk.**

- **Low altitude Venus flybys could pose potential thermal risk.**

- **RTG waste heat recovery design robustness**
  - Approach is highly configuration-dependent and may have high hidden development costs.
  - Less expensive on paper, but the actual implementation could be more expensive than an active system.

- **Component development for both propulsion subsystems**
  - NEXT development for SEP
  - Large bi-prop engines for chemical

- **Sun sensor performance may degrade past Saturn.**
  - May impact safe mode used during star tracker outage.

# Option 3

- **Option 3: <u>Neptune</u> Orbiter Concept**
  - 50 kg payload allocation
  - 1 atmospheric probe (previously designed)
  - Does <u>not</u> include a Venus flyby

- **Class B mission**
- **Dual string redundancy**
- **eMMRTGs can be used for Orbiter power**
  - Carry <u>no mass contingency</u>, because eMMRTG masses provided are "not to exceed" values

- **Xenon residual calculation overridden with Dawn heritage values**
  - ~5 kg residuals on ~595kg of propellant
- **Assuming SEP stage based on modified launch vehicle adapter**
  - Affects mechanical/structure masses

- **Mission:**
  - Launch: 4/28/2030; Arrival: 4/28/2043
  - Launch, E-J flybys, cruise to Neptune
  - SEP stage jettisoned roughly at 5 AU
  - Probe separation 60 days prior to entry

- **Mission Design**
  - 13-year cruise to NOI, 2-year science tour
  - Orbiter serves as communications relay during probe entry
    - Continuous line of sight between orbiter and probe is critical for telecom
    - Baseline probe EFPA is -16 deg
  - NOI begins ~2 hours after Probe entry.
    - Allows sufficient time for probe data relay prior to the turn-to-burn
    - More than 2 hours results in gain issues for the zenith-pointed Probe UHF antenna (since range and off-pointing angle from zenith increase).
  - NOI inserts into 180-day initial orbit, lowered to ~50-day orbit
  - Will require optical navigation upon approach to NOI, and during science for targeting moon flybys
    - Doppler imager will be used to OpNav on approach

- **Launch Vehicle**
  - Delta IV-Heavy (~7575 kg to C3 of 18.66 $km^2/s^2$)

- **Arrival Vinf / Declination**
  - ~11.4 km/s, 9.1 degree
- **Orbiter-Probe separation is ~110,000 km at entry time**
  - This range closes as the Orbiter proceeds towards periapse and the Probe decelerates in the Neptune atmosphere.
  - Telecom requires Orbiter within 60 deg of zenith and < 100,000 km range.
- **EFPA of -20 degrees**
  - Imparts 100g's on the probe during entry.
  - Reducing the EFPA results in line-of-sight geometry challenges.

| Event | Rel. Time | Duration | Delta V (m/s) | # Maneuvers | Comments |
|---|---|---|---|---|---|
| Earth Flyby | L+681 days | | | | 300 km altitude |
| Jupiter Flyby | L+1211 days | | | | 1.2e6 km altitude |
| SEP Jettison | L+~2400 days | | Total SEP: 3000 | | Timing flexible |
| TCM-1 | E-~400 days | | 10 | 1 | Non-deterministic |
| TCMs 2-5 | | | 13 (Total) | 4 | Non-deterministic |
| Separation/Divert | E-60 days | | 20 | 1 | |
| NOI | L+4748 days | ~1 hr | 2800 | 1 | |
| OTMs 1-5 | | | 150 (Total) | 5 | Includes NOI cleanup, period changes, flyby targeting, and other statistical mnvrs |
| **Total** | | | ~3000 chemical, ~3000 SEP | | |

- ✖ **Element 1: Atmospheric Probe**
  - Designed in study 1734 (June 28,30[th])
  - Common for both planets
  - No propulsion, no ACS, no power generation
  - Telecom relay to Orbiter
- ✖ **Element 2: Entry System**
  - Heatshield, backshell, structure
- ✖ **Element 3: Orbiter**
  - Instrument allocation defined by Option
  - Chemical propulsion
  - eMMRTGs, no solar arrays
- ✖ **Element 4: SEP Cruise Stage**
  - "Dumb" cruise stage:
    - ◆ No CDS/ACS/telecom
    - ◆ No instruments
  - SEP, no chemical propulsion
  - Solar arrays, no RTGs

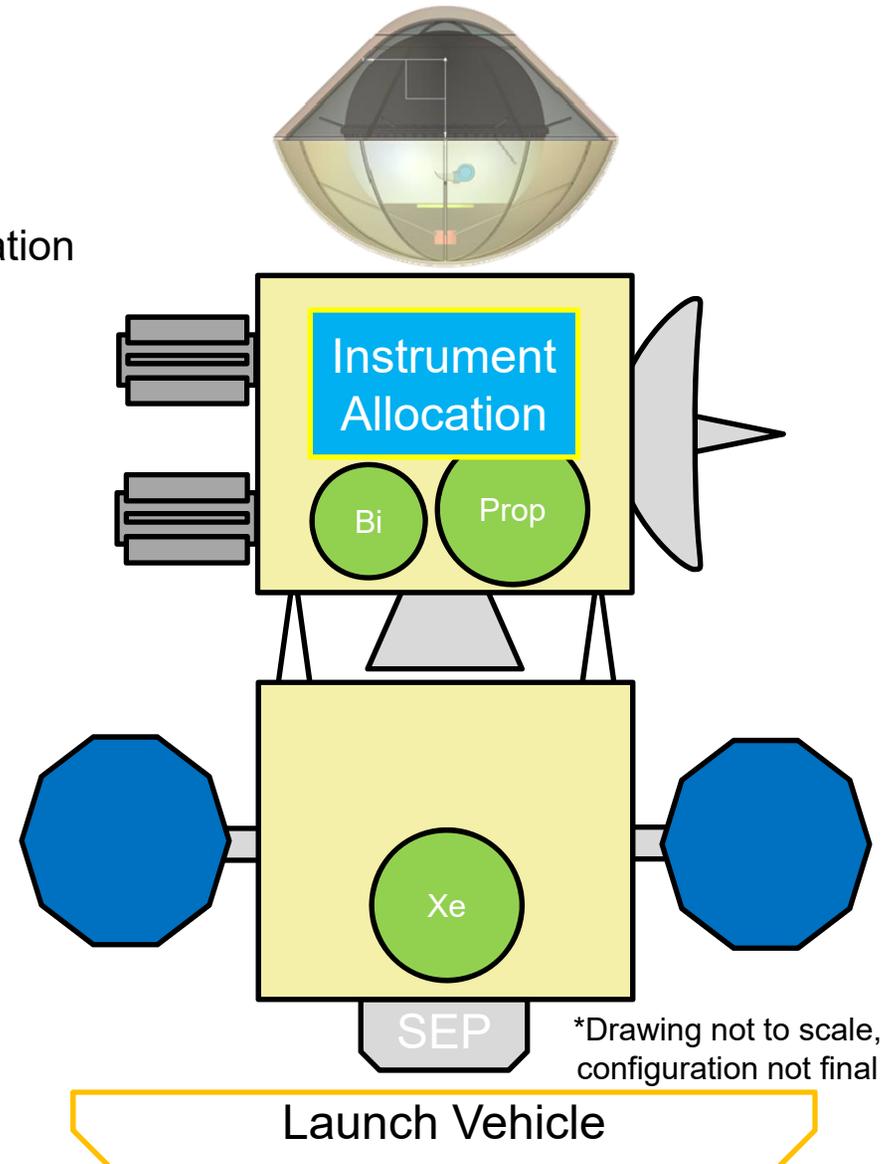

*Drawing not to scale, configuration not final

**Mission Timeline**

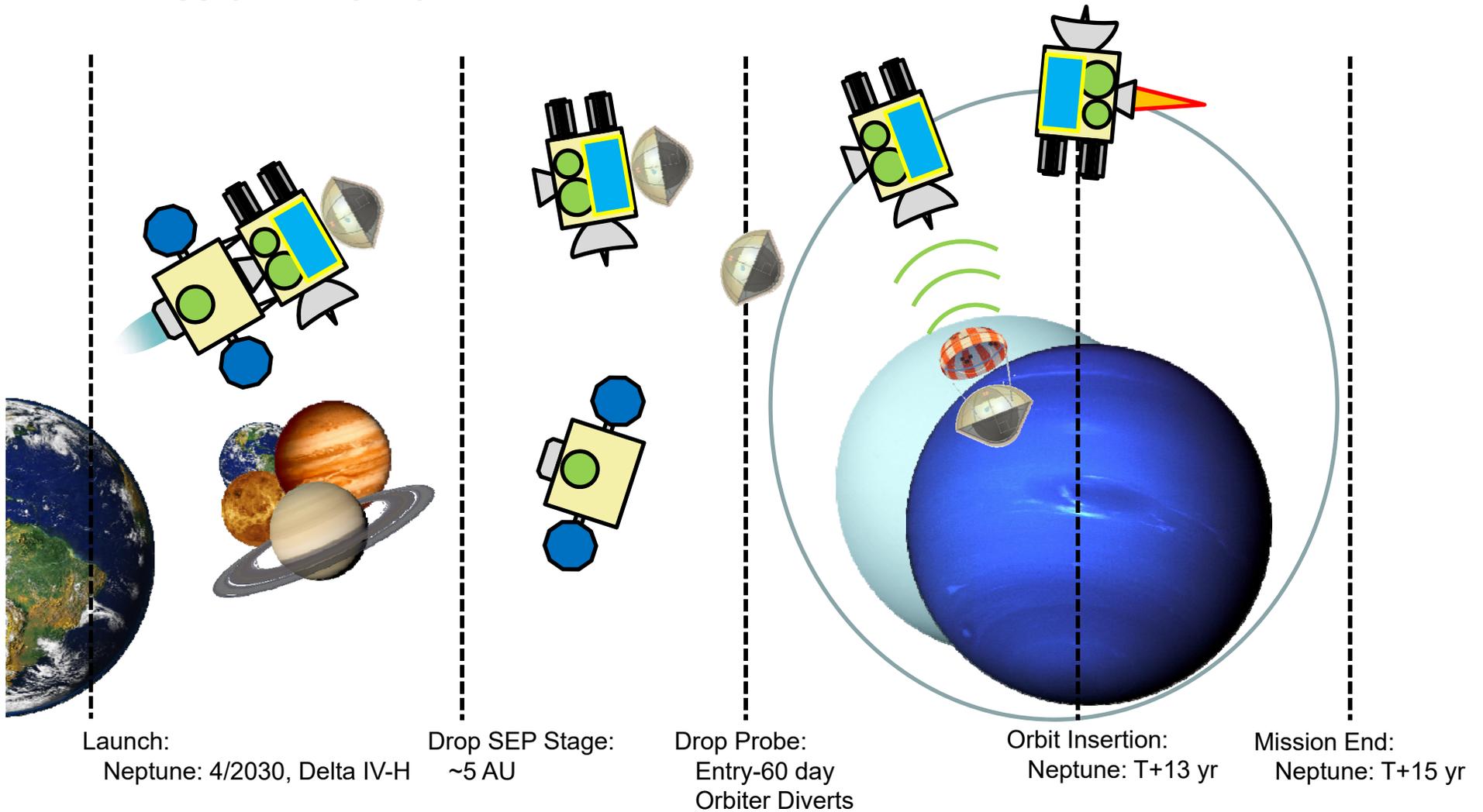

Launch:
  Neptune: 4/2030, Delta IV-H

Drop SEP Stage:
  ~5 AU

Drop Probe:
  Entry-60 day
  Orbiter Diverts

Orbit Insertion:
  Neptune: T+13 yr

Mission End:
  Neptune: T+15 yr

**Approach Timeline**

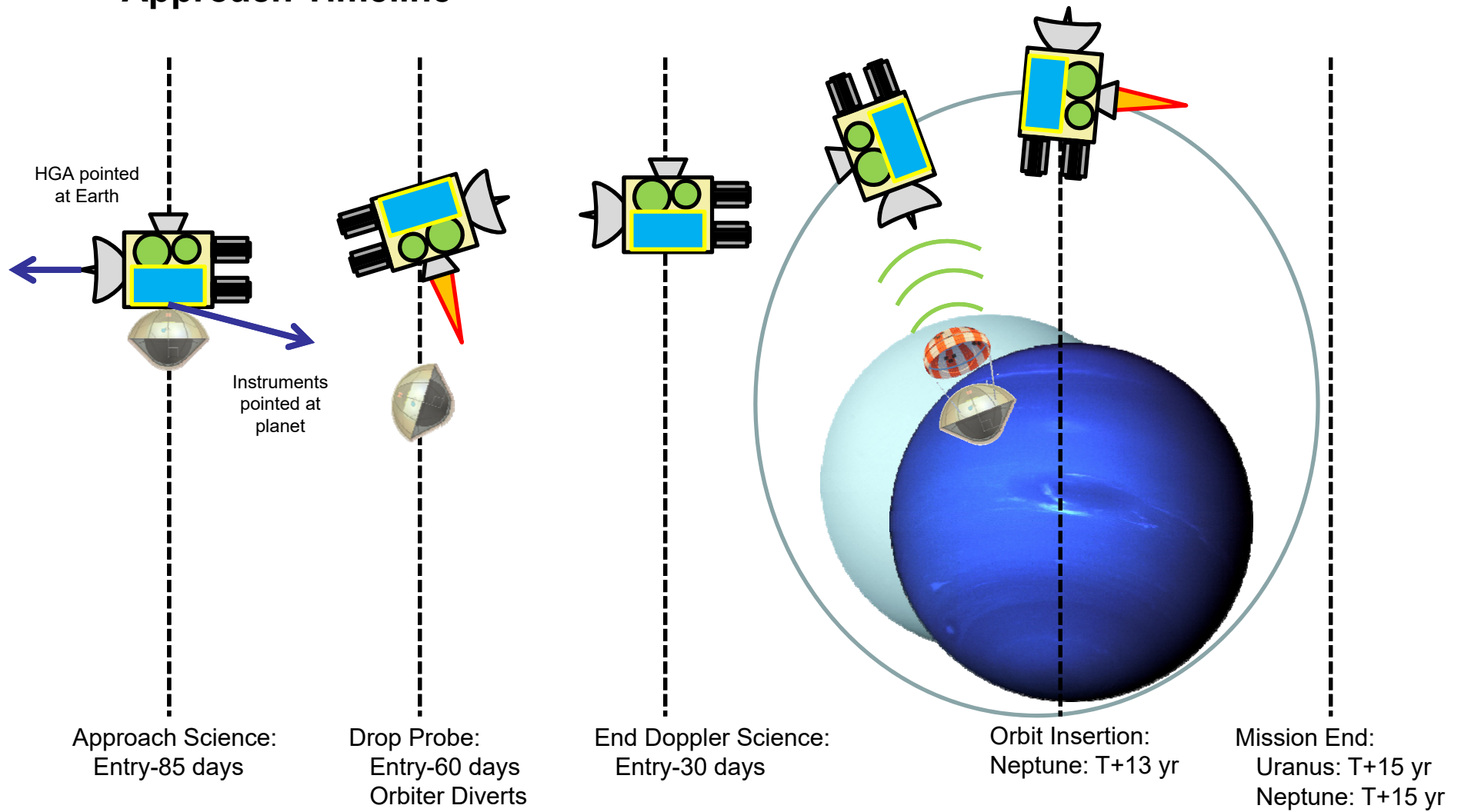

HGA pointed at Earth

Instruments pointed at planet

Approach Science:
Entry-85 days

Drop Probe:
Entry-60 days
Orbiter Diverts

End Doppler Science:
Entry-30 days

Orbit Insertion:
Neptune: T+13 yr

Mission End:
Uranus: T+15 yr
Neptune: T+15 yr

**Entry Timeline**

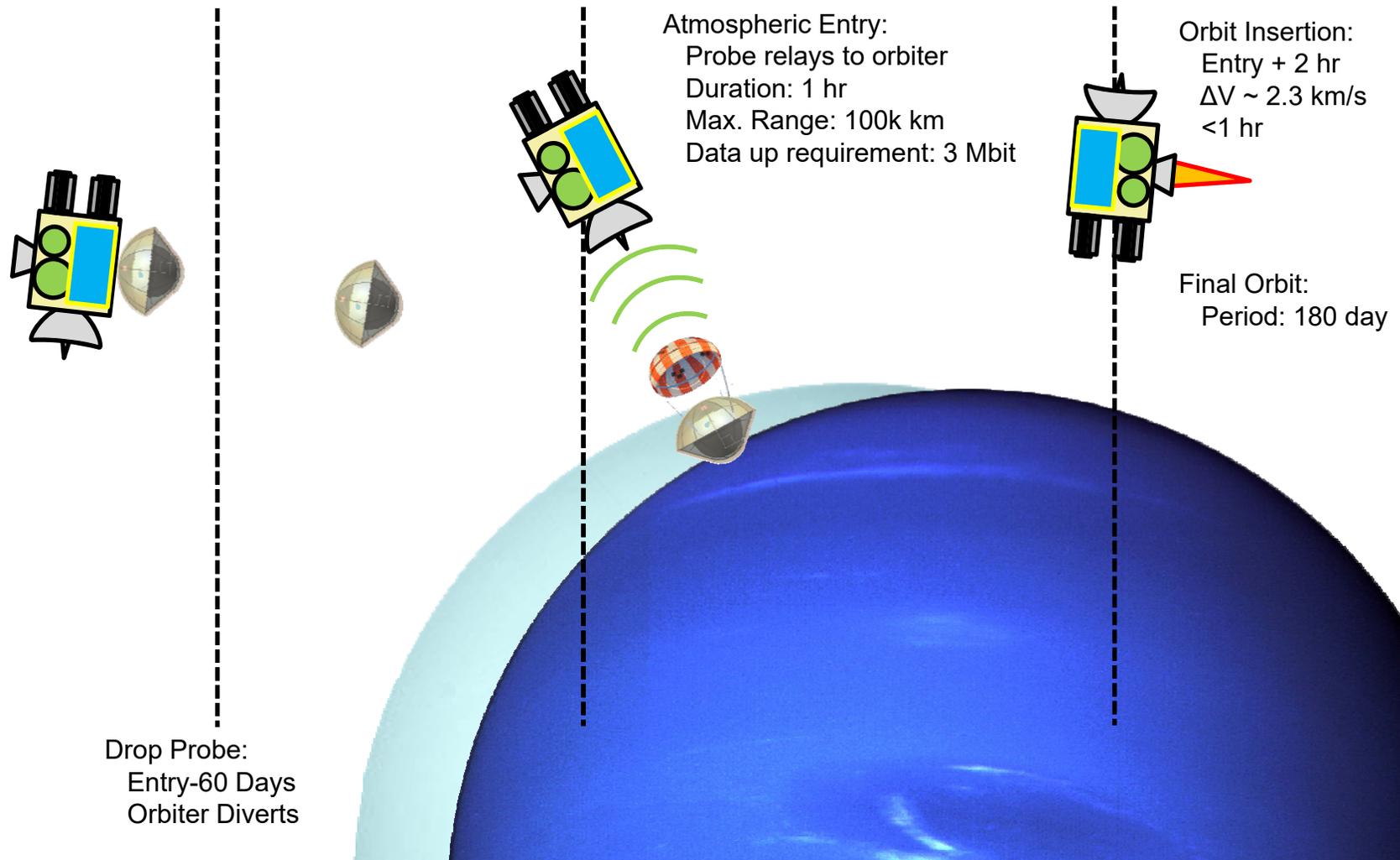

Atmospheric Entry:
  Probe relays to orbiter
  Duration: 1 hr
  Max. Range: 100k km
  Data up requirement: 3 Mbit

Orbit Insertion:
  Entry + 2 hr
  ΔV ~ 2.3 km/s
  <1 hr

Final Orbit:
  Period: 180 day

Drop Probe:
  Entry-60 Days
  Orbiter Diverts

Team X Study Guidelines
### *Ice Giants Study 2016-07*
### *Orbiter*

*Project - Study*

| | |
|---|---|
| Customer | John Elliott, Kim Reh |
| Study Lead | Bob Kinsey |
| Study Type | Pre-Decadal Study |
| Report Type | Full PPT Report |

*Project - Mission*

| | |
|---|---|
| Mission | Ice Giants Study 2016-07 |
| Target Body | Neptune |
| Science | Imaging and Magnetometry |
| Launch Date | 1-Jul-30 |
| Mission Duration | 13 year cruise, 2 years in orbit |
| Mission Risk Class | B |
| Technology Cutoff | 2026 |
| Minimum TRL at End of Phase B | 6 |

*Project - Architecture*

| | | |
|---|---|---|
| Probe | on | Entry System |
| Entry System | on | Orbiter |
| Orbiter | on | SEP Stage |
| SEP Stage | on | Launch Vehicle |

| | |
|---|---|
| Launch Vehicle | Delta IV-H |
| Trajectory | VVE Gravity Assists, 180 day orbit, FPA = -35deg |
| L/V Capability, kg | 5165 kg to a C3 of 18 with 0% contingency taken out |
| Tracking Network | DSN |
| Contingency Method | Apply Total System-Level |

### Spacecraft

| | |
|---|---|
| Spacecraft | Orbiter |
| Instruments | Narrow Angle Camera (EIS Europa),Doppler Imager (ECHOES JUICE),Magnetometer (Galileo) |
| Potential Inst-S/C Commonality | None |
| Redundancy | Dual (Cold) |
| Stabilization | 3-Axis |
| Radiation Total Dose | 29.833 krad behind 100 mil. of Aluminum, with an RDM of 2 added. |
| Type of Propulsion Systems | System 1-Biprop, System 2-0, System 3-0 |
| Post-Launch Delta-V, m/s | 2988 |
| P/L Mass CBE, kg | 36.7 kg Payload CBE + 321 kg Entry System + Probe (alloc) |
| P/L Power CBE, W | 44.4 |
| P/L Data Rate CBE, kb/s | 12000 |
| Hardware Models | Protoflight S/C, EM instrument |

### Project - Cost and Schedule

| | |
|---|---|
| Cost Target | < $2B TBD |
| Mission Cost Category | Flagship - e.g. Cassini |
| FY$ (year) | 2015 |
| Include Phase A cost estimate? | Yes |
| Phase A Start | September 2023 |
| Phase A Duration (months) | 20 |
| Phase B Duration (months) | 16 |
| Phase C/D Duration (months) | 47 |
| Review Dates | PDR - September 2026, CDR - November 2027, ARR - November 2028 |
| Phase E Duration (months) | 179 |
| Phase F Duration (months) | 4 |
| Project Pays Tech Costs from TRL | 6 |
| Spares Approach | Typical |
| Parts Class | Commercial + Military 883B TBR |
| Launch Site | Cape Canaveral |

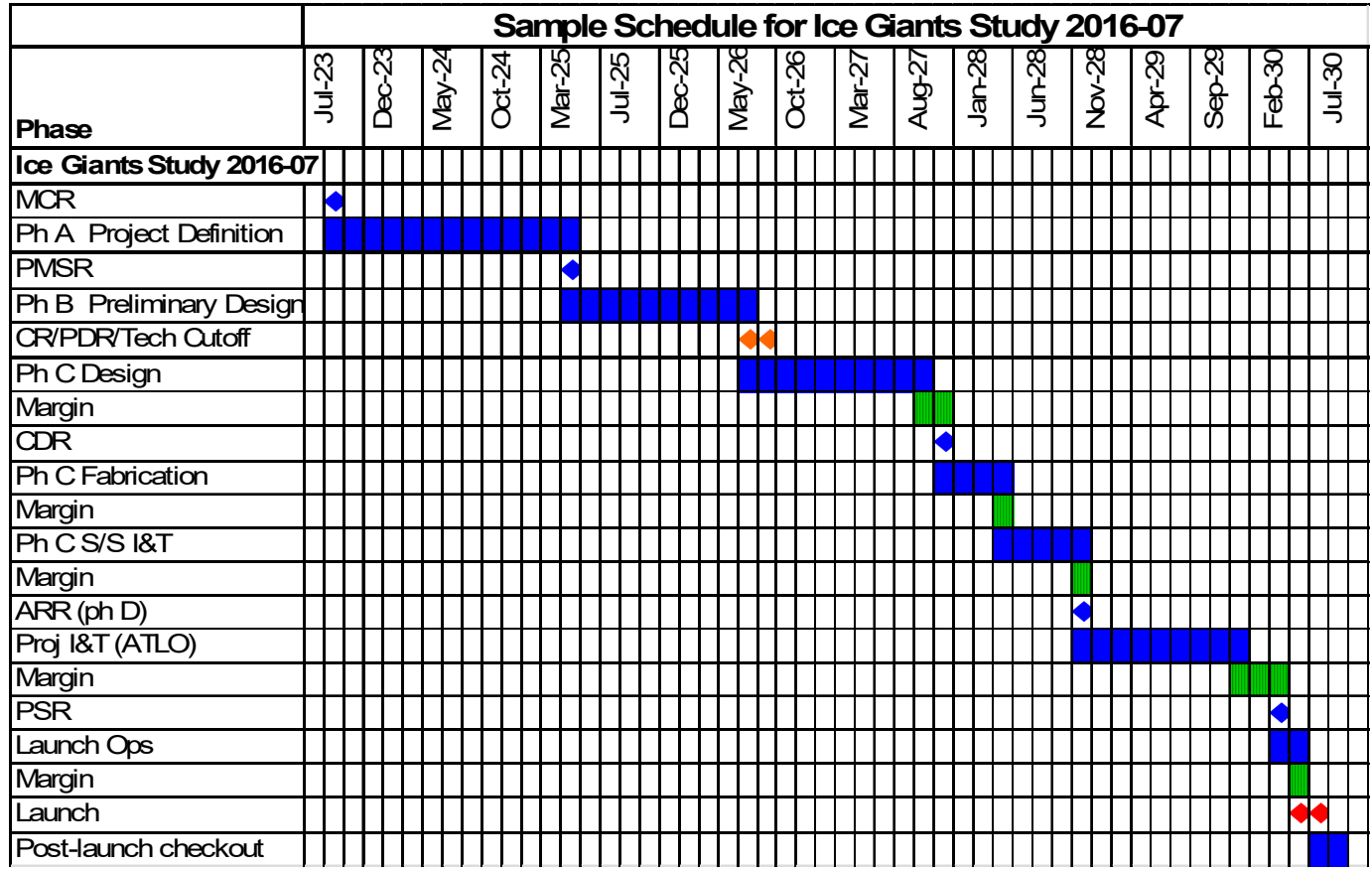

Proposed development schedule consistent with typical New Frontiers missions and current Europa mission schedule.

Phase A 20 mos., Phase B 16 mos., Phase C/D 47 mos.

Launch July 2030

✘ **Selected instruments come with some extras.**

- NAC has a 2-DOF gimbal.
- Doppler Imager has an internal fast steering mirror (FSM).

| Instrument Name | # units | Heritage | CBE Mass (kg) | Cont. | CBE+Cont. Mass/Unit (kg) | Op. Power CBE per Instrument (W) | Standy Power CBE per Instrument (W) |
|---|---|---|---|---|---|---|---|
| | | | 37 kg | 23% | 45.2 | | |
| Narrow Angle Camera (EIS Europa) | 1 | Inherited design | 12.0 | 15% | 13.8 | 16 W | 2 W |
| Doppler Imager (ECHOES JUICE) | 1 | New design | 20.0 | 30% | 26 | 20 W | 2 W |
| Magnetometer (Gallileo) | 1 | Inherited design | 4.7 | 15% | 5.405 | 8 W | 1 W |

| Instrument Name | Instrument Peak Data Rate | Units |
|---|---|---|
| Narrow Angle Camera (EIS Europa) | 12000 | kbps |
| Doppler Imager (ECHOES JUICE) | 60 | kbps |
| Magnetometer (Gallileo) | 1200 | kbps |

- ✳ **Instruments**
  - Gas Chromatograph Mass Spectrometer (GCMS)
  - Atmospheric Structure Instrument (ASI)
  - Nephelometer
  - Ortho-para Hydrogen Measurement Instrument
- ✳ **CDS**
  - Redundant Sphinx Avionics
- ✳ **Power**
  - Primary batteries
    - ◆ In probe:
      - ▪ 17.1kg, 1.0 kW-hr EOM
  - Redundant Power Electronics
- ✳ **Thermal**
  - RHU heating, passive cooling
  - Vented probe design
  - Thermally isolating struts

- ✳ **Telecom**
  - Redundant IRIS radio
  - UHF SSPA
  - UHF Low Gain Antenna (similar to MSL)
- ✳ **Structures**
  - ~50kg Heatshield
    - ◆ 1.2m diameter, 45deg sphere cone
  - ~15kg Backshell
  - ~10kg Parachutes
  - ~15kg Probe Aerofairing

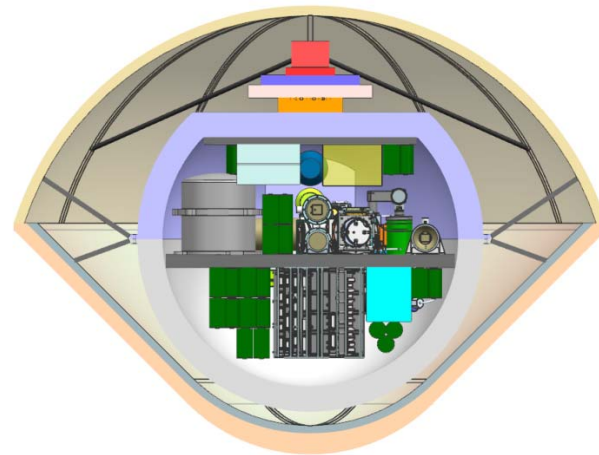

**Instruments**
- Narrow Angle Camera
- Doppler Imager
- Magnetometer

**CDS**
- JPL reference bus avionics
- Dual string cold redundancy

**Baseline Power System**
- 4 eMMRTGs, 45kg each
- 14 A-hr, <5kg, Li Ion Battery

**Telecom**
- Radios
  - Two X/X/Ka SDST transponders
  - Two IRIS radio UHF receivers
  - Two 35W Ka-Band TWTAs
  - Two 25W X-Band TWTAs
- Antennas
  - One 3m X/Ka HGA
  - One X-Band MGA
  - Two X-Band LGAs
  - One UHF patch array – 15 dBic gain

**Thermal**
- Active and passive thermal control design
- Louvers, heaters, MLI

**ACS**
- Four 0.1N Honeywell HR16 reaction wheels
- IMUs, Star Trackers, Sun Sensors

**Propulsion**
- Dual-mode bipropellant system provides 2988m/s of delta-V
- Two 200lbf Aerojet main engines
- Four 22N engines
- Eight 1N RCS engines

**Structures**
- ~375kg structure
- 40kg ballast
- 88kg harness
- 10m Magnetometer Boom
- Main Engine cover
- SEP Stage and Probe separation mechanisms

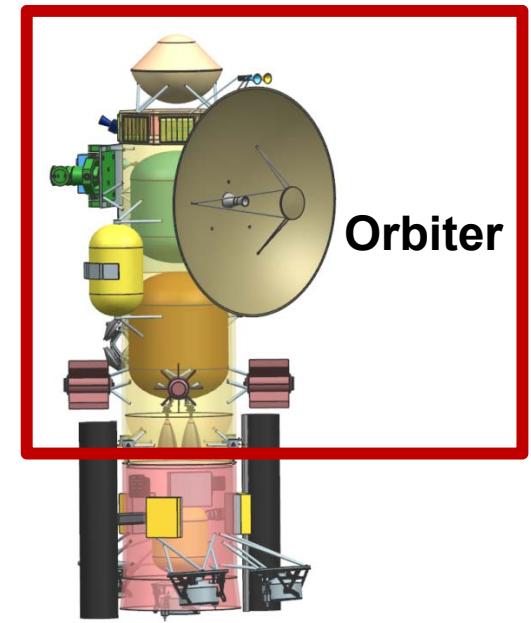

Orbiter

## Propulsion
- Four NEXT main engines
- Four 35kg PPUs
- 595kg Xenon Propellant

## Baseline Power System
- Two 54m$^2$ ROSA Solar Arrays
- Provide ~29.5kW at 1AU
- Redundant JPL Reference Bus Power Electronics

## CDS
- Redundant remote engineering unit

## ACS
- 1DOF solar array gimbal drive electronics
- 2DOF SEP engine gimbal drive electronics
- Sun sensors

## Thermal
- Active and passive thermal control design
- Louvers, heaters, MLI

## Structures
- Cylindrical bus shape, made up of stacked CXX adapters

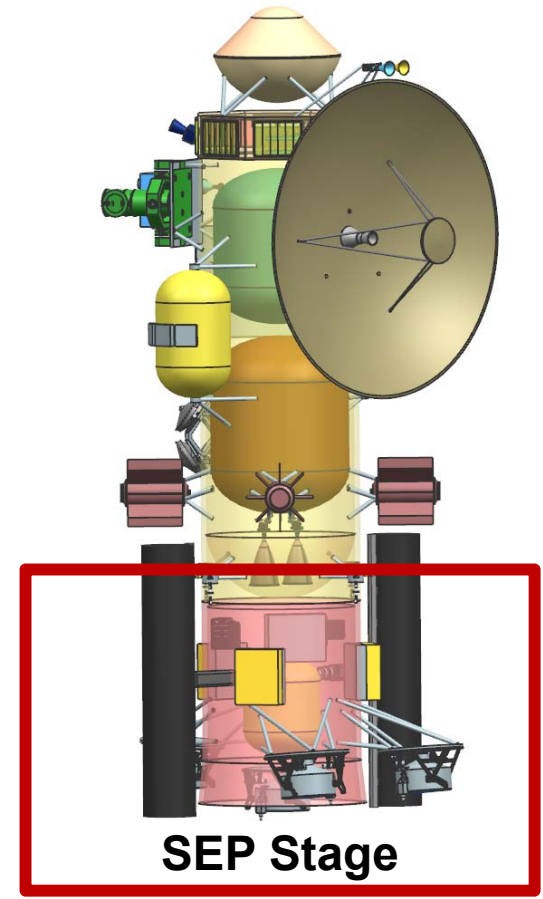

**SEP Stage**

## Probe + Entry System

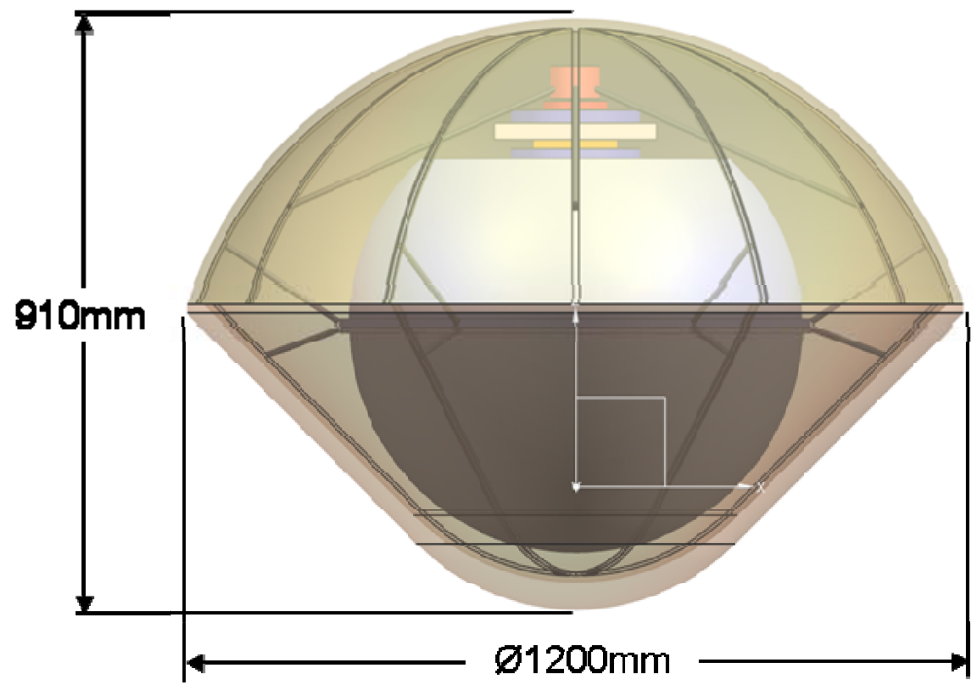

## Probe

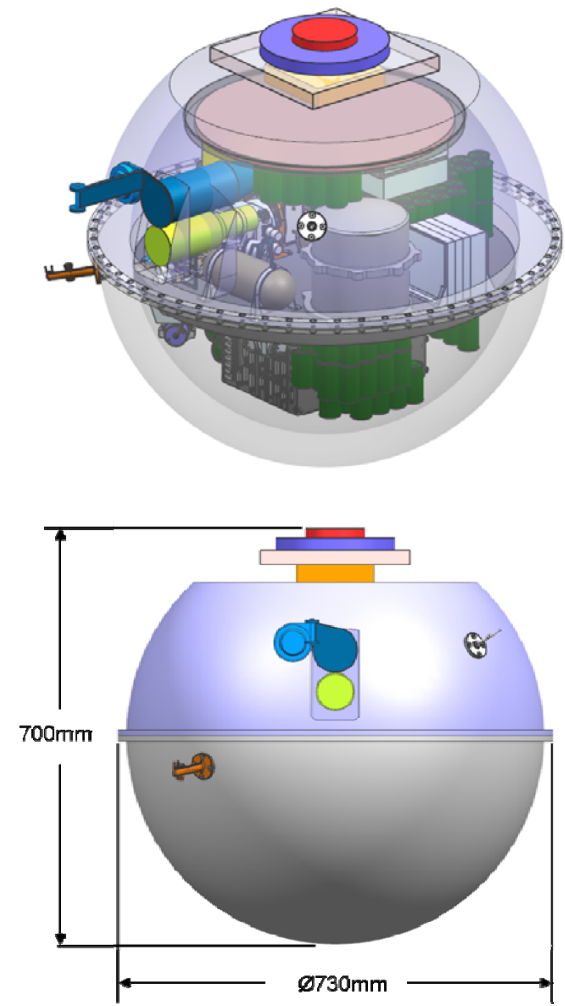

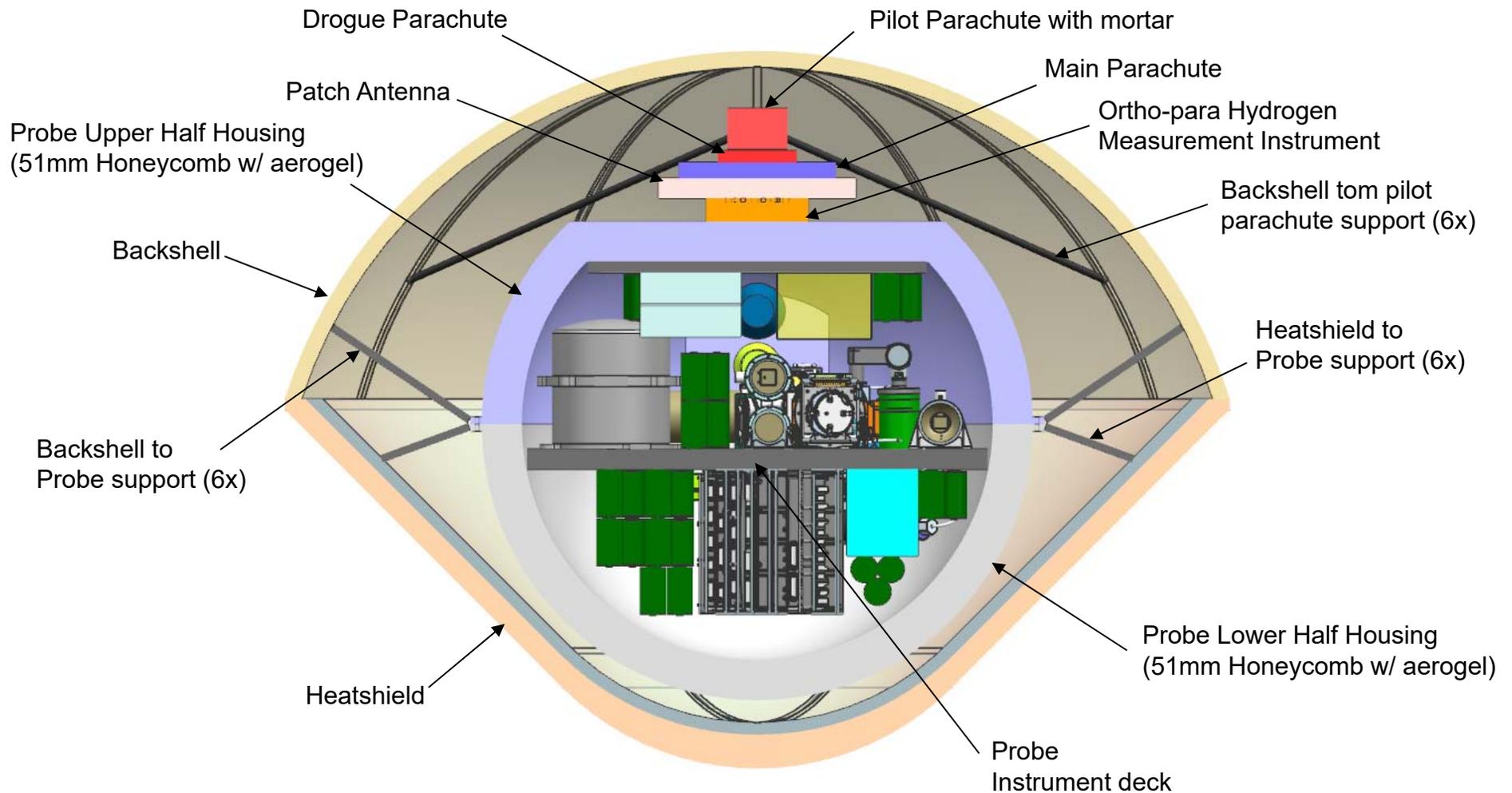

Drogue Parachute

Pilot Parachute with mortar

Patch Antenna

Main Parachute

Probe Upper Half Housing
(51mm Honeycomb w/ aerogel)

Ortho-para Hydrogen
Measurement Instrument

Backshell tom pilot
parachute support (6x)

Backshell

Heatshield to
Probe support (6x)

Backshell to
Probe support (6x)

Heatshield

Probe Lower Half Housing
(51mm Honeycomb w/ aerogel)

Probe
Instrument deck

**✶ Configuration Drawings – Stowed**

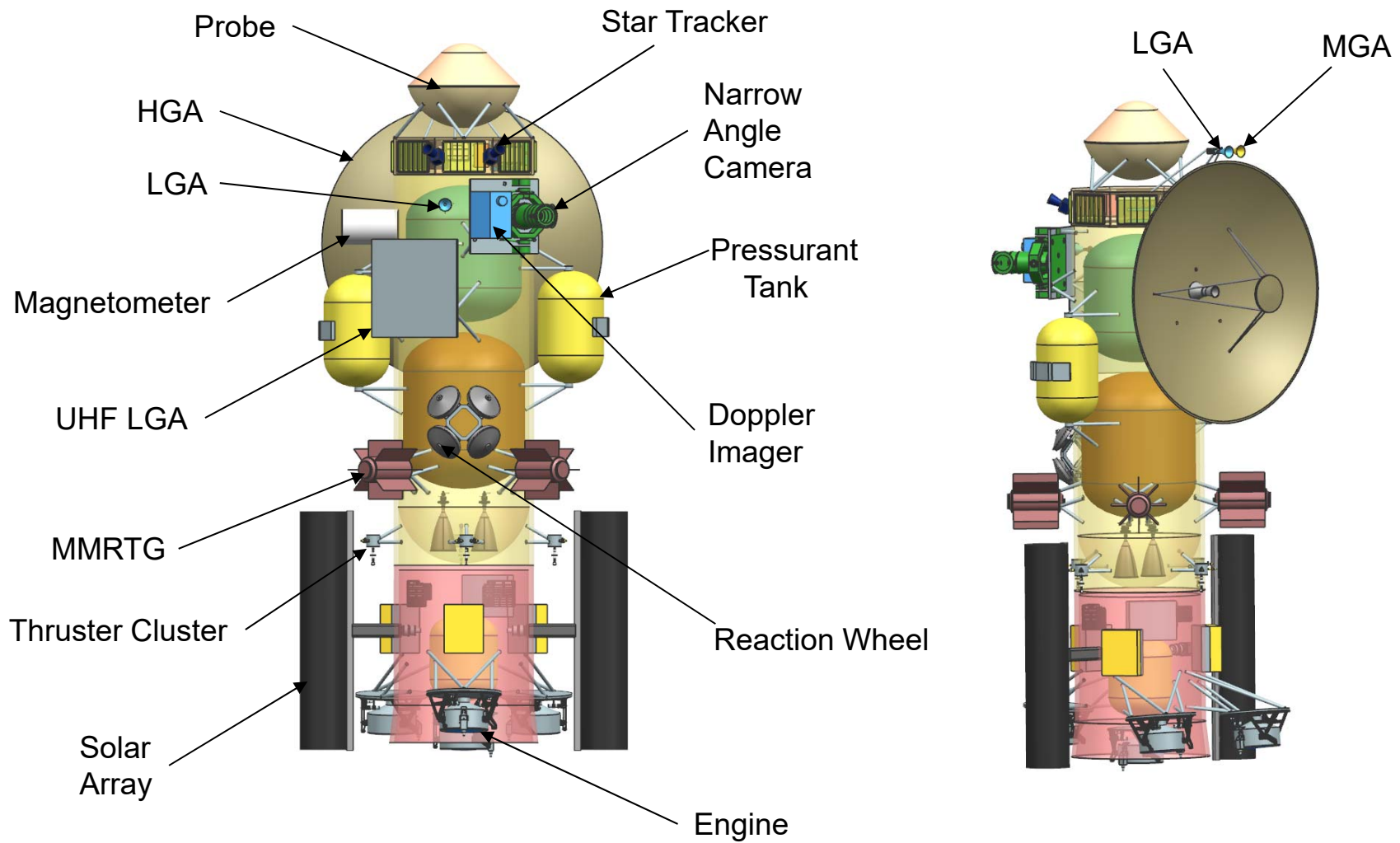

Probe

Star Tracker

HGA

Narrow Angle Camera

LGA

Magnetometer

Pressurant Tank

UHF LGA

Doppler Imager

MMRTG

Thruster Cluster

Reaction Wheel

Solar Array

Engine

LGA    MGA

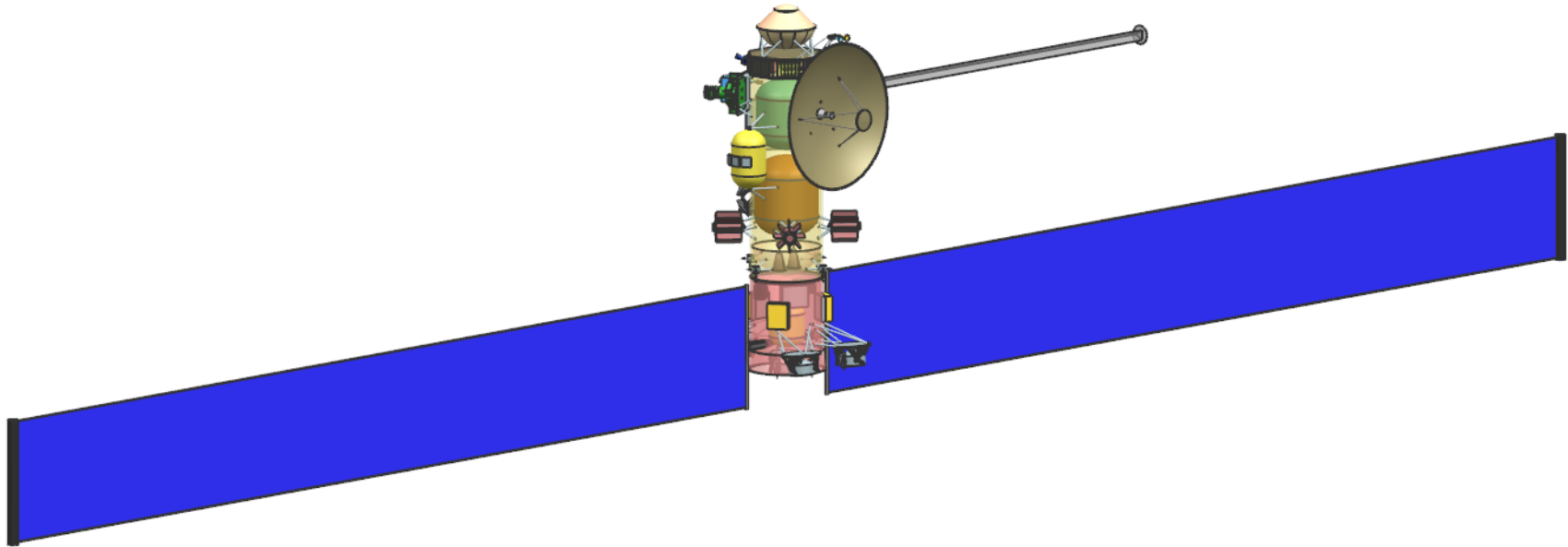

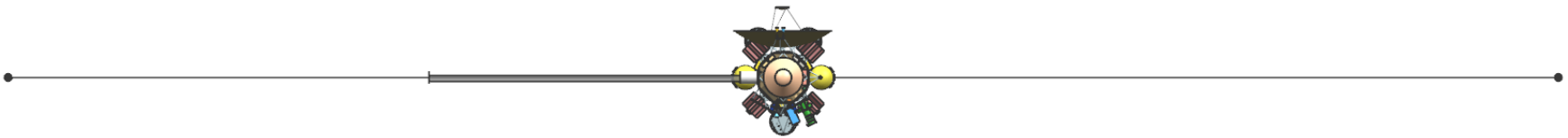

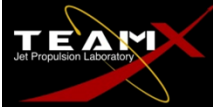

# Executive Summary
## Option 3 – Probe Element

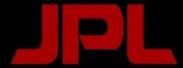

| | Mass Fraction | Mass (kg) | Subsys Cont. % | CBE+ Cont. (kg) | Mode 1 Power (W) Coast - 60 Days | Mode 2 Power (W) Warmup | Mode 3 Power (W) Science |
|---|---|---|---|---|---|---|---|
| **Power Mode Duration   (hours)** | | | | | **24** | *0.5* | *1* |
| **Payload on this Element** | | | | | | | |
| Instruments | 21% | 25.3 | 29% | 32.5 | 0 | 91 | 74 |
| Payload Total | 21% | **25.3** | **29%** | **32.5** | **0** | **91** | **74** |
| **Spacecraft Bus** | | | | do not edit formulas below this line, use the calcualtions and overri | | | |
| Command & Data | 0% | 0.6 | 17% | 0.7 | 0 | 8 | 8 |
| Power | 17% | 20.1 | 26% | 25.4 | 0 | 11 | 11 |
| Structures & Mechanisms | 41% | 49.8 | 30% | 64.7 | 0 | 0 | 0 |
| Cabling | 9% | 11.5 | 30% | 15.0 | | | |
| Telecom | 5% | 6.2 | 26% | 7.8 | 0 | 0 | 184 |
| Thermal | 6% | 7.8 | 3% | 8.1 | 0 | 0 | 0 |
| Bus Total | | **96.0** | **27%** | **121.7** | **0** | **19** | **203** |
| Thermally Controlled Mass | | | | 121.7 | | | |
| **Spacecraft Total (Dry): CBE & MEV** | | **121.3** | **27%** | **154.2** | **0** | **109** | **277** |
| Subsystem Heritage Contingency | 27% | 32.9 | SEP Cont | 10% | 0 | 0 | 0 |
| System Contingency | 16% | 19.3 | | | 0 | 47 | 119 |
| Total Contingency  ☐ Include Carried? | **43%** | 52.2 | | | | | |
| **Spacecraft with Contingency:** | | **173** | of total | w/o addl pld | **0** | **157** | **396** |

| | | Mass Fraction | Mass (kg) | Subsys Cont. % | CBE+ Cont. (kg) |
|---|---|---|---|---|---|
| **Power Mode Duration** *(hours)* | | | | | |
| **Additional Elements Carried by this Element** | | | | | |
| Probe | | 54% | 121.2 | 43% | 173.4 |
| **Carried Elements Total** | | 54% | **121.2** | 43% | **173.4** |
| **Spacecraft Bus** | | | | | do not edit formulas below th |
| Structures & Mechanisms | | 46% | 102.5 | 30% | 133.2 |
| Cabling | | 0% | 0.9 | 30% | 1.2 |
| **Bus Total** | | | 103.4 | 30% | 134.4 |
| Thermally Controlled Mass | | | | | 134.4 |
| **Spacecraft Total (Dry): CBE & MEV** | | | **224.6** | 37% | **307.8** |
| Subsystem Heritage Contingency | 37% | | 83.1 | SEP Cont | 10% |
| System Contingency | 6% | | 13.4 | | |
| Total Contingency | ☐ Include Carried? **43%** | | 96.6 | | |
| **Spacecraft with Contingency:** | | | **321** | of total | w/o addl pld |

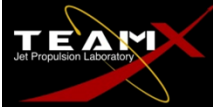

# Executive Summary
## Option 3 – Orbiter Element

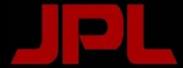

| | Mass Fraction | Mass (kg) | Subsys Cont. % | CBE+ Cont. (kg) | Mode 1 Power (W) Recharge | Mode 2 Power (W) Approach Science - 85 Days | Mode 3 Power (W) Telecom Downlink | Mode 4 Power (W) NOI Delta-V | Mode 5 Power (W) Orbital Science - Apoapse | Mode 6 Power (W) Orbital Science - Periapse | Mode 7 Power (W) Orbital Science - Moon Tour | Mode 8 Power (W) Safe | Mode 9 Power (W) SEP Thrusting | Mode 10 Power (W) Probe Relay |
|---|---|---|---|---|---|---|---|---|---|---|---|---|---|---|
| *Power Mode Duration (hours)* | | | | | *24* | *24* | *8* | *1.22* | *16* | *16* | *16* | *24* | *24* | *1.5* |
| **Payload on this Element** | | | | | | | | | | | | | | |
| Instruments | 2% | 36.7 | 23% | 45.2 | 4 | 28 | 4 | 4 | 26 | 44 | 26 | 4 | 4 | 4 |
| **Payload Total** | 2% | **36.7** | **23%** | **45.2** | **4** | **28** | **4** | **4** | **26** | **44** | **26** | **4** | **4** | **4** |
| **Additional Elements Carried by this Element** | | | | | | | | | | | | | | |
| Entry System + Probe | 15% | 224.6 | 43% | 321.2 | 0 | 0 | 0 | 0 | 0 | 0 | 0 | 0 | 0 | 0 |
| **Carried Elements Total** | 15% | **224.6** | **43%** | **321.2** | **0** | **0** | **0** | **0** | **0** | **0** | **0** | **0** | **0** | **0** |
| **Spacecraft Bus** | | | | do not edit formulas below this line, use the calcuations and override tables instead —> | | | | | | | | | | |
| Attitude Control | 4% | 63.5 | 10% | 69.8 | 0 | 55 | 55 | 88 | 55 | 55 | 55 | 42 | 55 | 93 |
| Command & Data | 1% | 21.6 | 10% | 23.8 | 57 | 57 | 57 | 57 | 57 | 57 | 57 | 57 | 57 | 57 |
| Power | 14% | 217.6 | 2% | 222.2 | 24 | 40 | 32 | 24 | 24 | 24 | 24 | 40 | 32 | 32 |
| Propulsion1 ☐ SEP1 | 12% | 186.8 | 5% | 196.9 | 31 | 3 | 3 | 151 | 3 | 3 | 3 | 3 | 3 | 3 |
| Structures & Mechanisms | 33% | 490.4 | 30% | 637.6 | 0 | 0 | 0 | 0 | 0 | 0 | 0 | 0 | 0 | 0 |
| Cabling | 6% | 87.8 | 30% | 114.2 | | | | | | | | | | |
| Telecom | 4% | 59.4 | 16% | 68.9 | 12 | 65 | 92 | 71 | 12 | 12 | 12 | 71 | 12 | 32 |
| Thermal | 8% | 118.3 | 24% | 146.2 | 25 | 25 | 25 | 25 | 25 | 25 | 25 | 25 | 25 | 25 |
| **Bus Total** | | 1245.6 | 19% | 1479.5 | 149 | 245 | 264 | 417 | 176 | 176 | 176 | 238 | 184 | 242 |
| Thermally Controlled Mass | | | | 1479.5 | | | | | | | | | | |
| **Spacecraft Total (Dry): CBE & MEV** | | **1506.9** | 22% | **1845.9** | 154 | 272 | 268 | 421 | 202 | 220 | 202 | 242 | 188 | 246 |
| Subsystem Heritage Contingency | 22% | 339.0 | SEP Cont | 10% | 0 | 0 | 0 | 0 | 0 | 0 | 0 | 0 | 0 | 0 |
| System Contingency | 15% | 231.6 | | | 66 | 117 | 115 | 181 | 87 | 95 | 87 | 104 | 81 | 106 |
| Total Contingency ☐ Include Carried? | **38%** | 570.6 | | | | | | | | | | | | |
| **Spacecraft with Contingency:** | | **2077** | of total | w/o addl pld | **220** | **390** | **383** | **602** | **289** | **315** | **289** | **347** | **269** | **352** |
| Propellant & Pressurant with residuals1 | 60% | 3088.1 | For S/C mass = | 5170.0 | | Delta-V, Sys 1 | 2988.0 | m/s | | residuals = | 80.9 | kg | | |
| **Spacecraft Total with Contingency (Wet)** | | **5165.6** | | | | | | | | | | | | |

Predecisional information for planning and discussion only
102

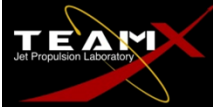
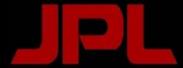

| | Mass Fraction | Mass (kg) | Subsys Cont. % | CBE+ Cont. (kg) | | Mode 8 Power (W) Launch | Mode 9 Power (W) SEP Thrusting | Mode 10 Power (W) Safe |
|---|---|---|---|---|---|---|---|---|
| **Power Mode Duration** *(hours)* | | | | | | 2 | 24 | 24 |
| **Additional Elements Carried by this Element** | | | | | | | | |
| Orbiter + Entry System + Probe (Wet Mass) | 80% | 4595.0 | 12% | 5165.6 | | 0 | 0 | 0 |
| **Carried Elements Total** | 80% | **4595.0** | **12%** | **5165.6** | | **0** | **0** | **0** |
| **RSDO Option -->** | | | | | | | | |
| **Spacecraft Bus** | | | do not edit formulas below t... | ...s and override tables instead ---> | | | | |
| Attitude Control | 0% | 7.0 | 7% | 7.5 | | 0 | 18 | 0 |
| Command & Data | 0% | 1.6 | 5% | 1.7 | | 4 | 4 | 4 |
| Power | 5% | 265.5 | 29% | 342.9 | | 0 | 669 | 35 |
| Propulsion1 ☑ SEP1 | 5% | 289.9 | 24% | 358.1 | | 0 | 25000 | 0 |
| Structures & Mechanisms | 7% | 392.3 | 30% | 510.0 | | 0 | 0 | 0 |
| Cabling | 2% | 86.4 | 30% | 112.4 | | | | |
| Thermal | 1% | 78.9 | 0% | 78.9 | | 0 | 286 | 393 |
| **Bus Total** | | 1121.6 | 26% | 1411.5 | | 4 | 25978 | 432 |
| Thermally Controlled Mass | | | | 1411.5 | | | | |
| **Spacecraft Total (Dry): CBE & MEV** | | **5716.7** | 15% | **6577.1** | | 4 | 25978 | 432 |
| Subsystem Heritage Contingency | 15% | 860.4 | SEP Cont | 10% | | 0 | 2500 | 0 |
| System Contingency | 3% | 192.5 | | | | 2 | 420 | 186 |
| Total Contingency ☐ Include Carried? | **18%** | 1052.9 | | | | | | |
| **Spacecraft with Contingency:** | | **6770** | of total | w/o addl pld | | **6** | **28898** | **618** |
| Propellant & Pressurant with residuals1 | 8% | 595.4 | For S/C mass = | 2000.0 | kg | | | |
| **Spacecraft Total with Contingency (Wet)** | | **7364.9** | | | | | MPV | |

| Element Number | Element Name | Dry CBE (kg) | Cont / JPL Margin (kg) | Dry Allocation (kg) | Propellant (kg) | Dry Allocation + Propellant (kg) |
|---|---|---|---|---|---|---|
| 1 | Probe | 121 | 52 | 173 | - | 173 |
| 2 | Entry System | 103 | 44 | 148 | - | 148 |
| 3 | Orbiter minus eMMRTGs | 1,102 | 474 | 1,576 | 3,088 | 4664 |
| 3.1 | eMMRTGs | 180 | - | 180 | - | 180 |
| 4 | SEP Stage | 1,122 | 482 | 1,604 | 595 | 2,199 |
| | **Total Stack** | **2,628** | **1,053** | **3,681** | **3,683** | **7,365** |
| | | | | Dry Mass Allocation | | 3,681 |
| | | | | JPL Margin (kg / %) | | 1,053 / 28.6% |
| | | | | JPL Margin without eMMRTG (kg / %) | | 1,053 / 30% |
| | | | | Delta IV-H Capacity (kg) | | 7,575 |
| | | | | Extra Launch Vehicle Margin (kg) | | 210 |

| Element Number | Element Name | Dry CBE (kg) | Cont (%) | Cont. (kg) | MEV (kg) | Dry Allocation (kg) | Propellant (kg) | Dry Allocation + Propellant (kg) |
|---|---|---|---|---|---|---|---|---|
| 1 | Probe | 121 | 27% | 33 | 154 | 173 | - | 173 |
| 2 | Entry System | 103 | 30% | 31 | 134 | 148 | - | 148 |
| 3 | Orbiter minus eMMRTGs | 1,102 | 22% | 242 | 1,345 | 1,576 | 3,088 | 4664 |
| 3.1 | eMMRTGs | 180 | - | - | 180 | 180 | - | 180 |
| 4 | SEP Stage | 1,122 | 26% | 289 | 1,411 | 1,604 | 595 | 2,199 |
| | **Total Stack** | **2,628** | | **595** | **3,224** | **3,681** | **3,683** | **7,365** |

| | |
|---|---|
| Dry Mass Allocation (kg) | 3,681 |
| NASA Margin (kg / %) | 457 / 14% |
| NASA Margin without eMMRTG (kg / %) | 457 / 15% |
| Delta IV-H Capacity (kg) | 7,575 |
| Extra Launch Vehicle Margin (kg) | 210 |

- **Probe needs to enter Neptune atmosphere head-on versus shallow**
  - Probe relay antenna is nominally aligned with zenith.
  - Need Orbiter within tens of degrees of zenith to close the link.
  - Head-on: Orbiter is close enough to zenith for one hour.
    - Probe deceleration ~100g's
  - Shallow: Orbiter is far from zenith –more difficult to close the link.
    - On the other hand, Probe deceleration relatively low.
  - Verified that Probe can operate through the higher deceleration.

- **Ka-band transmitter power versus array of 34m ground stations**
  - Downlink 15 kbps using one 35W TWTA to a 34m BWG ground station.
    - Telecom system uses most of one eMMRTG power during downlink.
  - Could increase downlink rate using more power/adding an eMMRTG, or by using an array of two or more 34m ground stations.
  - For this option, don't want to carry additional mass for another eMMRTG.
    - With 4 RTG's, have enough power for 35W TWTA during one 8-hour pass.
    - Use an array of three 34m ground stations during the 8-hour pass.

✖ **Data downlink strategy for Doppler Imager (DI) on approach**

- DI generates a lot of data continuously for tens of days on approach.

- Configuration with HGA and DI on opposite sides of the cylindrical bus allows pointing DI towards Neptune while pointing HGA towards Earth.

- Can downlink for ~20 hours/day and maintain positive power balance.

  ◆ Using only 4 eMMRTGs, as opposed to 5.

- Data that can't be downlinked before NOI will be downlinked after.

✖ **Configuration that helps to minimize mass and power**

- eMMRTGs outside the cylindrical bus provide heating

  ◆ Reduces mass and power of thermal subsystem components

- Propellant and oxidizer tanks inside bus, pressurant tanks outside

  ◆ Pressurant tanks are easier to keep warm than propellant/oxidizer tanks.

- Shorten the stack to minimize structure mass

  ◆ Single custom propellant tank instead of two tanks

  ◆ Stow solar arrays perpendicular to bus to reduce height of the SEP stage

## Mechanical

- LV interfaces directly to the SEP Stage; Orbiter interfaces to SEP Stage; Entry System containing Probe is attached to the side of the Orbiter.

- Orbiter: primary structure is the largest mass element.
  - 314 kg CBE out of 491 kg total for Mechanical (1246 kg bus dry mass)
  - Drivers are the large Propulsion and Power masses.

- SEP stage: primary structure is the largest mass element.
  - 232 kg CBE out of 392 kg total for Mechanical (1122 kg SEP Stage dry mass)
  - Due to the Orbiter and other mission elements being carried during launch.
  - Primary structure of the SEP Stage is being utilized as the LVA.

- **Power: Orbiter power bus spans both Orbiter and SEP stage**
  - Dual String Reference Bus electronics heritage
  - New development High Voltage Electronics Assembly for the SEP stage
  - Two Rollout Solar Arrays (ROSAs) support 25kW SEP at 1AU.

- **Propulsion: SEP Stage used to ~5 AU; dual mode bi-prop.**
  - SEP Stage: 3+1 system using NEXT Engines
    - 591 kg Xe mass allocation + 4 kg residuals for a total of 595 kg
  - Chemical System: orbit insertion delta V= 2.8km/s desired in < 1hr
    - Probe release after SEP stage separation, before NOI
    - Two 890N main engines used to achieve burn time < 1hr

- **Thermal: Cassini-heritage waste-heat recovery system on Orbiter**
  - RTG end domes each provide 75 W waste heat to propulsion module via conductive and radiative coupling.
  - VRHUs act as primary control mechanism for thruster clusters.
    - Also act as trim heaters for the propulsion module
  - Louvers act as primary control mechanism for avionics module.

- **Telecom: X- and Ka-Band subsystem, plus UHF for Probe data.**
  - Two 35W Ka-Band TWTAs, two 25W X-Band TWTAs
  - Two X/X/Ka SDST transponders, two IRIS radio UHF receivers
  - 3m X/Ka HGA, one X-Band MGA, two X-Band LGAs, UHF patch array.
  - Supports a data rate at Neptune of 15 kbps into 34m BWG ground station.
  - Supports uplink of 3Mbits of probe data

- **CDS: Reference Bus architecture ideally suited for high reliability, long lifetime mission.**
  - Standard JPL spacecraft CDS that is similar to SMAP
    - RAD750 CPU, NVM, MTIF, MSIA, CRC, LEU-A, LEU-D, MREU
    - 128 GBytes storage for science data
    - 1553 and RS-422 ICC/ITC interfaces for subsystems and instruments

- **ACS: 3-axis stabilized with star tracker, sun sensor, gyros, wheels.**
  - All stellar attitude determination to minimize power, conserve gyros.
  - Sun sensor performance may degrade once the Orbiter passes Saturn.
    - May impact safe mode used during star tracker outage.
    - Detailed analysis on Sun sensor performance versus distance is needed.

- **Software: core product line is appropriate since this mission has aspects similar to MSL/M2020/SMAP/Europa.**
  - Complexity rankings range from Medium to High.
    - Medium infrastructure: dual string with warm spare.
    - High fault behaviors: high redundancy, string swapping, critical events.
    - Medium/High ACS: tight pointing requirements, many ACS modes
    - Medium Telecom: dual active UHF, redundant DTE
    - Medium Science data processing, full file system

- **SVIT: Probe testbed, system I&T and V&V costs are included**
  - Cost of assembling and testing RTG's is captured elsewhere
    - Cost of integrating RTG's is included with other ATLO costs
  - Probe with 5 Instruments costed separately; testbed costs included.

- **Ground Systems**
  - Mission specific implementation of standard JPL mission operations and ground data systems
  - Ground network: DSN 34-m BWG; 70-m or equivalent for safe mode
  - Science support: 24x7 tracking on approach; daily contacts on orbit

| COST SUMMARY (FY2015 $M) | Generate ProPricer Input | Team X Estimate | | |
|---|---|---|---|---|
| | | CBE | Res. | PBE |
| **Project Cost** | | **$1621.8 M** | **22%** | **$1971.9 M** |
| **Launch Vehicle** | | **$33.0 M** | **0%** | **$33.0 M** |
| **Project Cost (w/o LV)** | | **$1588.8 M** | **22%** | **$1938.9 M** |
| **Development Cost** | | **$1289.3 M** | **25%** | **$1606.6 M** |
| Phase A | | $12.9 M | 25% | $16.1 M |
| Phase B | | $116.0 M | 25% | $144.6 M |
| Phase C/D | | $1160.4 M | 25% | $1445.9 M |
| **Operations Cost** | | **$299.5 M** | **11%** | **$332.3 M** |

Total mission cost is $1.97B. This is the likely cost within a range that typically can be as much as 10% lower up to 20% higher. The development cost with reserves is $1.61B.

| WBS Elements | NRE | RE | 1st Unit |
|---|---|---|---|
| Project Cost (no Launch Vehicle) | $1365.9 M | $606.0 M | $1971.9 M |
| Development Cost (Phases A - D) | $1000.6 M | $605.9 M | $1606.6 M |
| 01.0 Project Management | $47.3 M | | $47.3 M |
| 1.01 Project Management | $11.4 M | | $11.4 M |
| 1.02 Business Management | $13.6 M | | $13.6 M |
| 1.04 Project Reviews | $2.5 M | | $2.5 M |
| 1.06 Launch Approval | $19.8 M | | $19.8 M |
| 02.0 Project Systems Engineering | $23.7 M | $0.5 M | $24.3 M |
| 2.01 Project Systems Engineering | $8.9 M | | $8.9 M |
| 2.02 Project SW Systems Engineering | $5.2 M | | $5.2 M |
| 2.03 EEIS | $1.5 M | | $1.5 M |
| 2.04 Information System Management | $1.7 M | | $1.7 M |
| 2.05 Configuration Management | $1.5 M | | $1.5 M |
| 2.06 Planetary Protection | $0.2 M | $0.2 M | $0.4 M |
| 2.07 Contamination Control | $1.2 M | $0.3 M | $1.5 M |
| 2.09 Launch System Engineering | $1.0 M | | $1.0 M |
| 2.10 Project V&V | $2.0 M | | $2.0 M |
| 2.11 Risk Management | $0.5 M | | $0.5 M |
| 03.0 Mission Assurance | $53.1 M | $0.0 M | $53.1 M |
| 04.0 Science | $24.8 M | | $24.8 M |
| Orbiter Science | $14.0 M | | $14.0 M |
| Probe Science | $10.8 M | | $10.8 M |
| 05.0 Payload System | $80.2 M | $48.3 M | $128.5 M |
| 5.01 Payload Management | $7.8 M | | $7.8 M |
| 5.02 Payload Engineering | $5.8 M | | $5.8 M |
| Orbiter Instruments | $33.5 M | $24.3 M | $57.8 M |
| Narrow Angle Camera (EIS Europa) | $11.6 M | $8.4 M | $20.0 M |
| Doppler Imager (ECHOES JUICE) | $17.4 M | $12.6 M | $30.0 M |
| Magnetometer (Gallileo) | $4.5 M | $3.3 M | $7.8 M |
| Probe Instruments | $33.1 M | $24.0 M | $57.1 M |
| Mass Spectrometer | $22.9 M | $16.6 M | $39.6 M |
| Atmospheric Structure Investigation (ASI) | $3.4 M | $2.5 M | $5.9 M |
| Nephelometer (Galileo) | $5.3 M | $3.8 M | $9.1 M |
| Ortho-para H2 meas. Expt. | $1.5 M | $1.1 M | $2.6 M |

| WBS Elements | NRE | RE | 1st Unit |
|---|---|---|---|
| 06.0 Flight System | $495.4 M | $395.6 M | $891.1 M |
| 6.01 Flight System Management | $5.0 M | | $5.0 M |
| 6.02 Flight System Systems Engineering | $51.1 M | | $51.1 M |
| 6.03 Product Assurance (included in 3.0) | | | $0.0 M |
| Orbiter | $295.4 M | $236.6 M | $532.0 M |
| 6.04 Power | $94.5 M | $133.1 M | $227.6 M |
| 6.05 C&DH | $31.3 M | $24.3 M | $55.6 M |
| 6.06 Telecom | $28.4 M | $18.1 M | $46.5 M |
| 6.07 Structures (includes Mech. I&T) | $50.1 M | $16.2 M | $66.2 M |
| 6.08 Thermal | $4.2 M | $13.4 M | $17.6 M |
| additional cost for >43 RHUs | $34.0 M | $0.0 M | $34.0 M |
| 6.09 Propulsion | $22.1 M | $17.0 M | $39.1 M |
| 6.10 ACS | $9.4 M | $9.8 M | $19.1 M |
| 6.11 Harness | $4.1 M | $3.8 M | $7.9 M |
| 6.12 S/C Software | $17.1 M | $0.9 M | $18.0 M |
| 6.13 Materials and Processes | $0.4 M | $0.0 M | $0.4 M |
| SEP Stage | $51.5 M | $114.9 M | $166.4 M |
| 6.04 Power | $7.7 M | $50.7 M | $58.4 M |
| 6.05 C&DH | $4.0 M | $2.7 M | $6.7 M |
| 6.06 Telecom | $0.0 M | $0.0 M | $0.0 M |
| 6.07 Structures (includes Mech. I&T) | $10.5 M | $4.7 M | $15.2 M |
| 6.08 Thermal | $3.1 M | $11.2 M | $14.3 M |
| 6.09 Propulsion | $22.7 M | $42.8 M | $65.5 M |
| 6.10 ACS | $0.7 M | $1.3 M | $1.9 M |
| 6.11 Harness | $2.4 M | $1.5 M | $3.9 M |
| 6.12 S/C Software | $0.0 M | $0.0 M | $0.0 M |
| 6.13 Materials and Processes | $0.4 M | $0.0 M | $0.4 M |
| Probe | $27.2 M | $18.2 M | $45.4 M |
| 6.04 Power | $2.9 M | $2.1 M | $5.0 M |
| 6.05 C&DH | $0.3 M | $2.3 M | $2.7 M |
| 6.06 Telecom | $7.9 M | $4.1 M | $12.0 M |
| 6.07 Structures (includes Mech. I&T) | $8.0 M | $3.5 M | $11.5 M |
| 6.08 Thermal | $2.3 M | $4.9 M | $7.2 M |
| 6.11 Harness | $2.0 M | $1.0 M | $3.0 M |
| 6.12 S/C Software | $3.3 M | $0.2 M | $3.5 M |
| 6.13 Materials and Processes | $0.5 M | $0.1 M | $0.5 M |
| Entry System | $57.1 M | $24.4 M | $81.5 M |
| 6.07 Structures (includes Mech. I&T) | $55.3 M | $24.1 M | $79.4 M |
| 6.11 Harness | $1.4 M | $0.3 M | $1.7 M |
| 6.13 Materials and Processes | $0.4 M | $0.0 M | $0.4 M |
| Ames/Langley EDL Engineering/Testing | $3.8 M | $0.0 M | $3.8 M |
| 6.14 Spacecraft Testbeds | $4.5 M | $1.5 M | $6.0 M |

| WBS Elements | NRE | RE | 1st Unit |
|---|---|---|---|
| **07.0 Mission Operations Preparation** | **$26.6 M** | | **$26.6 M** |
| 7.0 MOS Teams | $20.0 M | | $20.0 M |
| 7.03 DSN Tracking (Launch Ops.) | $2.7 M | | $2.7 M |
| 7.06 Navigation Operations Team | $3.8 M | | $3.8 M |
| 7.07.03 Mission Planning Team | $0.0 M | | $0.0 M |
| **09.0 Ground Data Systems** | **$22.1 M** | | **$22.1 M** |
| 9.0A Ground Data System | $20.0 M | | $20.0 M |
| 9.0B Science Data System Development | $1.3 M | | $1.3 M |
| 9A.03.07 Navigation H/W & S/W Development | $0.8 M | | $0.8 M |
| **10.0 ATLO** | **$21.1 M** | **$21.7 M** | **$42.8 M** |
| Orbiter | $15.3 M | $13.3 M | $28.6 M |
| Probe | $5.9 M | $8.4 M | $14.2 M |
| **11.0 Education and Public Outreach** | **$0.0 M** | **$0.0 M** | **$0.0 M** |
| **12.0 Mission and Navigation Design** | **$28.8 M** | | **$28.8 M** |
| 12.01 Mission Design | $2.4 M | | $2.4 M |
| 12.02 Mission Analysis | $11.4 M | | $11.4 M |
| 12.03 Mission Engineering | $1.8 M | | $1.8 M |
| 12.04 Navigation Design | $13.3 M | | $13.3 M |
| **Development Reserves** | **$177.4 M** | **$139.8 M** | **$317.3 M** |

| WBS Elements | NRE | RE | 1st Unit |
|---|---|---|---|
| Operations Cost (Phases E - F) | $332.3 M | $0.1 M | $332.3 M |
| 01.0 Project Management | $27.1 M | | $27.1 M |
| 1.01 Project Management | $15.3 M | | $15.3 M |
| 1.02 Business Management | $10.7 M | | $10.7 M |
| 1.04 Project Reviews | $1.1 M | | $1.1 M |
| 1.06 Launch Approval | $0.1 M | | $0.1 M |
| 02.0 Project Systems Engineering | $0.0 M | $0.1 M | $0.1 M |
| 03.0 Mission Assurance | $3.6 M | $0.0 M | $3.6 M |
| 04.0 Science | $69.2 M | | $69.2 M |
| Orbiter Science | $53.4 M | | $53.4 M |
| Probe Science | $15.8 M | | $15.8 M |
| 07.0 Mission Operations | $171.3 M | | $171.3 M |
| 7.0 MOS Teams | $69.8 M | | $69.8 M |
| 7.03 DSN Tracking | $80.5 M | | $80.5 M |
| 7.06 Navigation Operations Team | $20.1 M | | $20.1 M |
| 7.07.03 Mission Planning Team | $1.0 M | | $1.0 M |
| 09.0 Ground Data Systems | $28.2 M | | $28.2 M |
| 9.0A GDS Teams | $22.5 M | | $22.5 M |
| 9.0B Science Data System Ops | $5.2 M | | $5.2 M |
| 9A.03.07 Navigation HW and SW Dev | $0.6 M | | $0.6 M |
| 11.0 Education and Public Outreach | $0.0 M | $0.0 M | $0.0 M |
| 12.0 Mission and Navigation Design | $0.0 M | | $0.0 M |
| Operations Reserves | $32.8 M | $0.0 M | $32.8 M |
| 8.0 Launch Vehicle | $33.0 M | | $33.0 M |
| Launch Vehicle and Processing | $0.0 M | | $0.0 M |
| Nuclear Payload Support | $33.0 M | | $33.0 M |

## Risks related to the Probe

- Only 2 hours between Probe entry and NOI, a critical event
  - May be operationally challenging to sequence both the Probe relay and NOI for the Orbiter within this time window.
  - Longer than 2 hours makes the geometry more challenging for telecom.
- May be issues for the relay link margin due to Probe-Orbiter geometry and uncertainties regarding Neptune atmosphere/ potential signal attenuation.
- High g load on the Probe carries some risk.
- Last Probe targeting occurs more than 60 days prior to encounter.
  - Probe carries no propulsion, so it cannot correct trajectory dispersions.
  - Need dispersions small enough to ensure safe entry conditions at Neptune.
- Orbit knowledge requirements for science reconstruction need to be determined.
  - Will drive how the Probe is tracked pre-entry and what telemetry (e.g. IMU) needs to be transmitted with the science data to the Orbiter.
  - The latter will impact the data budget.

- **Mission duration will push systems to their operating lifetimes.**

- **Science planning risk**
  - Relative velocities between Orbiter and Neptune satellites will be high.
    - Flybys occur near periapse

- **Neptune ring collision avoidance may need to be considered.**

- **Running the Orbiter power bus to the SEP stage makes for a more complex electronics design and adds cabling.**
  - Higher risk than adding a battery on the SEP stage.
  - Chose this to minimize SEP stage mass.

- **eMMRTG still needs some development.**
  - May cause a schedule slip.
  - Performance may degrade at a higher rate than currently predicted.

- **ROSA solar array qualification carries some risk.**

- **RTG waste heat recovery design robustness**
  - Approach is highly configuration-dependent and may have high hidden development costs.
  - Less expensive on paper, but the actual implementation could be more expensive than an active system.
- **Component development for both propulsion subsystems**
  - NEXT development for SEP
  - Large bi-prop engines for chemical
- **Sun sensor performance may degrade past Saturn.**
  - May impact safe mode used during star tracker outage.

# Option 4

✖ **Option 4: Uranus Flyby Variant Concept**

- 50 kg payload allocation
- 1 atmospheric probe (previously designed)
- **Note**: Primary spacecraft named 'orbiter' for consistency between options

✖ **Class B mission**

✖ **Dual string redundancy**

✖ **eMMRTGs could be used for Orbiter power**

- Carry <u>no mass contingency</u>, because eMMRTG masses provided are "not to exceed" values

- **Mission:**
  - Launch: 4/16/2030; Arrival: 4/15/2040
  - Launch, E-J flybys, probe entry at Uranus, 'orbiter' flyby only
  - Probe separation 60 days prior to entry

- **Mission Design**
  - 10-year cruise to UOI, 6-month downlink post-entry
    - Downlink from 'orbiter' post Uranus flyby
  - 'Orbiter' serves as communications relay during probe entry
    - Continuous line of sight between orbiter and probe is critical for telecom
    - Probe EFPA decreased to 30 deg to compensate for increased arrival v_inf
  - Will require optical navigation upon approach to UOI
    - Doppler imager will be used for OpNav on approach

- **Launch Vehicle**
  - Atlas 541 (~1,775 kg to C3 of 52.557 km$^2$/s$^2$)

- **Arrival Vinf / Declination ~14.79 km/s**

- **Orbiter-Probe separation is ~100,000 km at entry time**

  - This range closes as the Flyby S/C proceeds and the Probe decelerates in the Uranus atmosphere.

  - Telecom requires

    - Flyby S/C within 60 deg of zenith and < 100,000 km range.

- **EFPA of -30 degrees**

  - Imparts ~165 g's on the probe during entry

  - Helps to compensate for increased arrival Vinf

  - No line-of-sight geometry issues as there is no insertion

| Event | Rel. Time | Duration | Delta V (m/s) | # Maneuvers | Comments |
|-------|-----------|----------|---------------|-------------|----------|
| DSM | L+409 days | | 240 | 1 | |
| TCMs 1-2 | L+~20 days | | 35 | 2 | Non-deterministic |
| Earth Flyby | L+1119 days | | | | 1650 km altitude |
| Jupiter Flyby | L+1715 days | | | | 13000 km altitude |
| TCMs 3-5 | | | 10 | 3 | Non-deterministic |
| Separation/Divert | E-60 days | | 20 | 1 | |
| **Total** | | | 310 | 7 | |

✖ **Element 1: Atmospheric Probe**
  - Designed in study 1734 (June 28,30[th])
  - Common for both planets
  - No propulsion, no ACS, no power generation
  - Telecom relay to Orbiter

✖ **Element 2: Entry System**
  - Heatshield, backshell, structure

✖ **Element 3: Orbiter**
  - Instrument allocation defined by Option
  - Chemical propulsion
  - eMMRTGs, no solar arrays

✖ **Element 4: SEP Cruise Stage**
  - "Dumb" cru~~ise~~

*Drawing not to scale

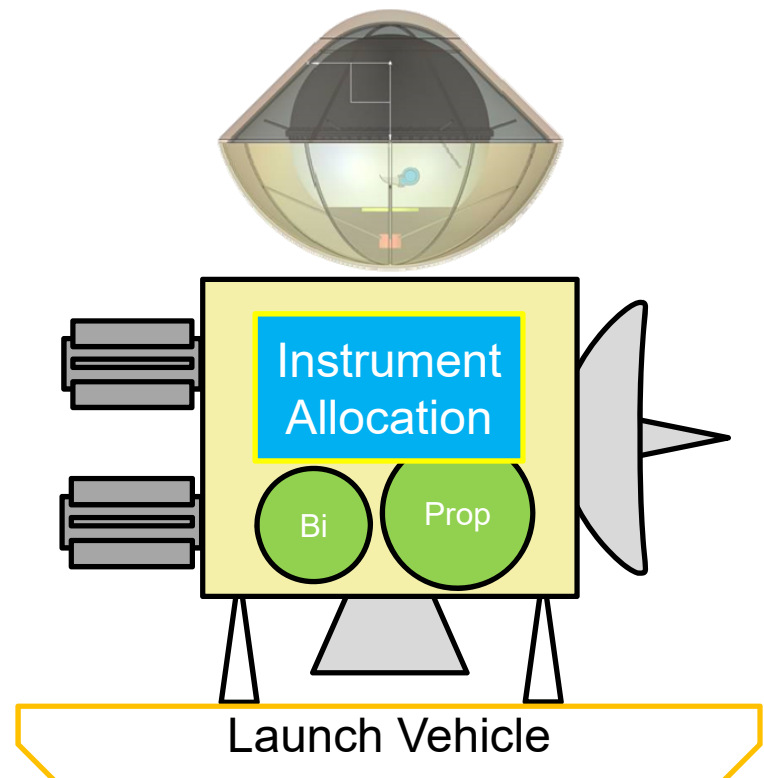

No SEP for this option!

~~...pulsion~~
~~...ys, no RTGs~~

Instrument Allocation

Bi    Prop

Launch Vehicle

**Mission Timeline**

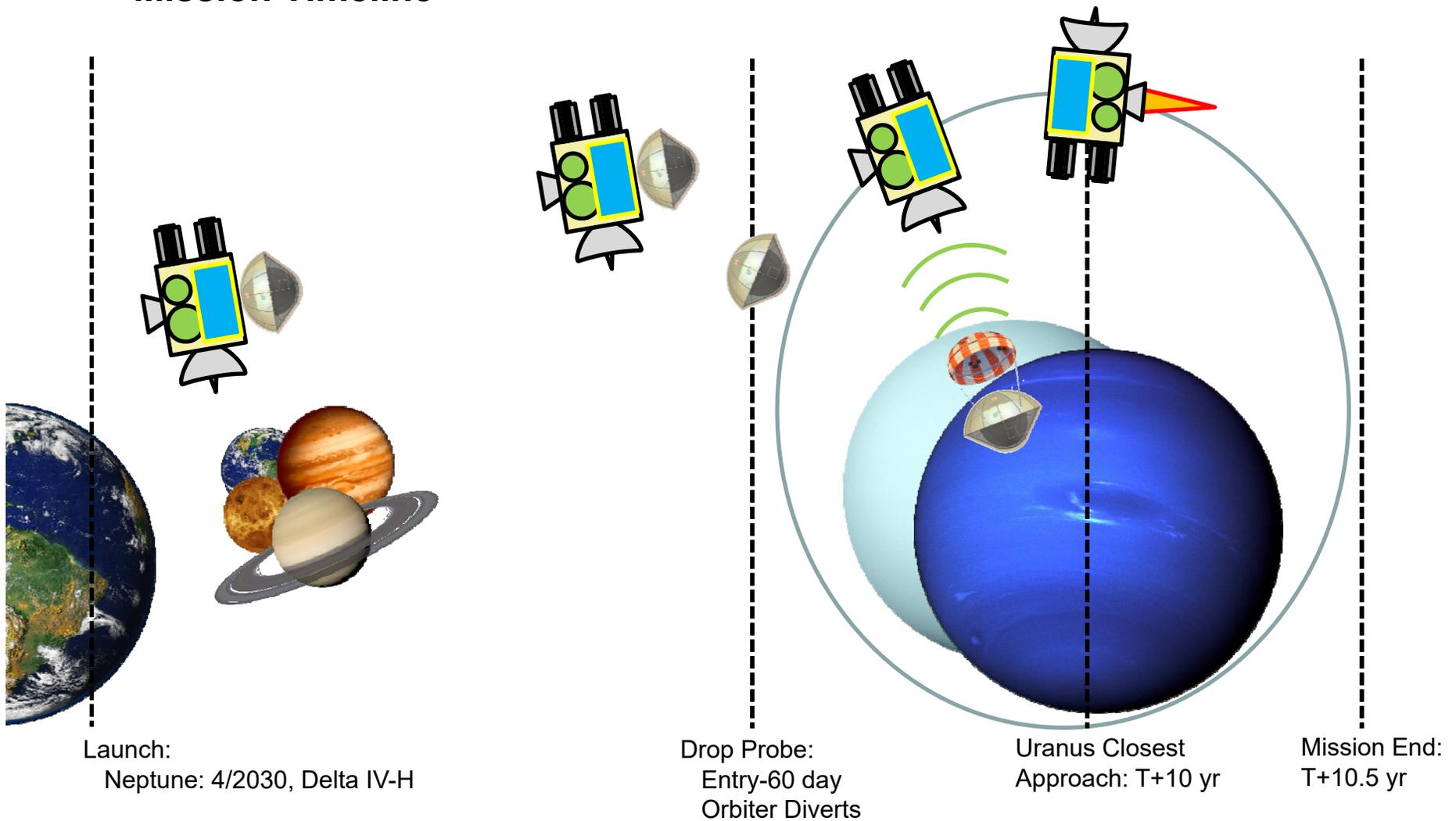

Launch:
  Neptune: 4/2030, Delta IV-H

Drop Probe:
  Entry-60 day
  Orbiter Diverts

Uranus Closest
Approach: T+10 yr

Mission End:
T+10.5 yr

**Approach Timeline**

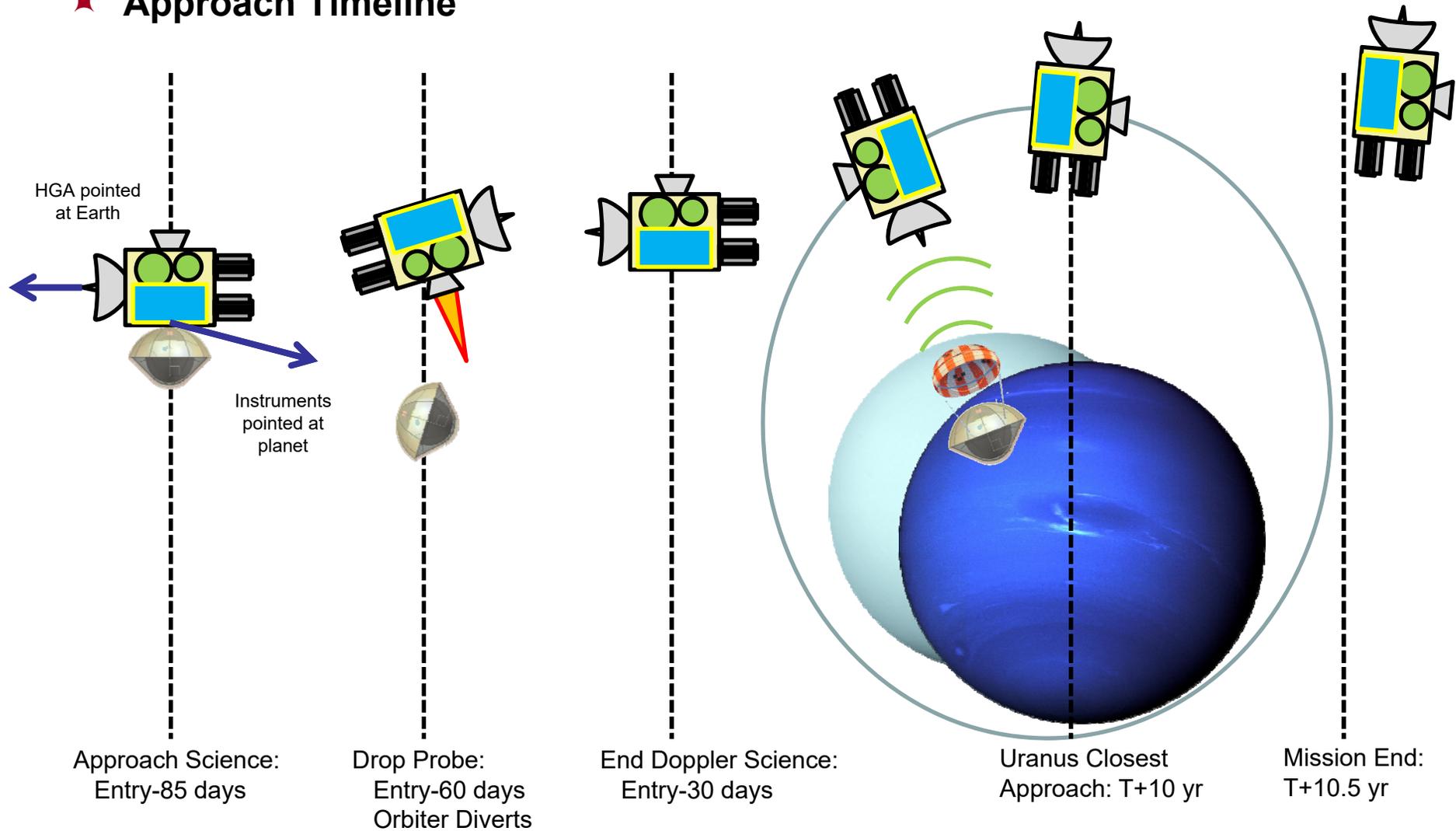

HGA pointed at Earth

Instruments pointed at planet

Approach Science:
Entry-85 days

Drop Probe:
Entry-60 days
Orbiter Diverts

End Doppler Science:
Entry-30 days

Uranus Closest
Approach: T+10 yr

Mission End:
T+10.5 yr

★ **Entry Timeline**

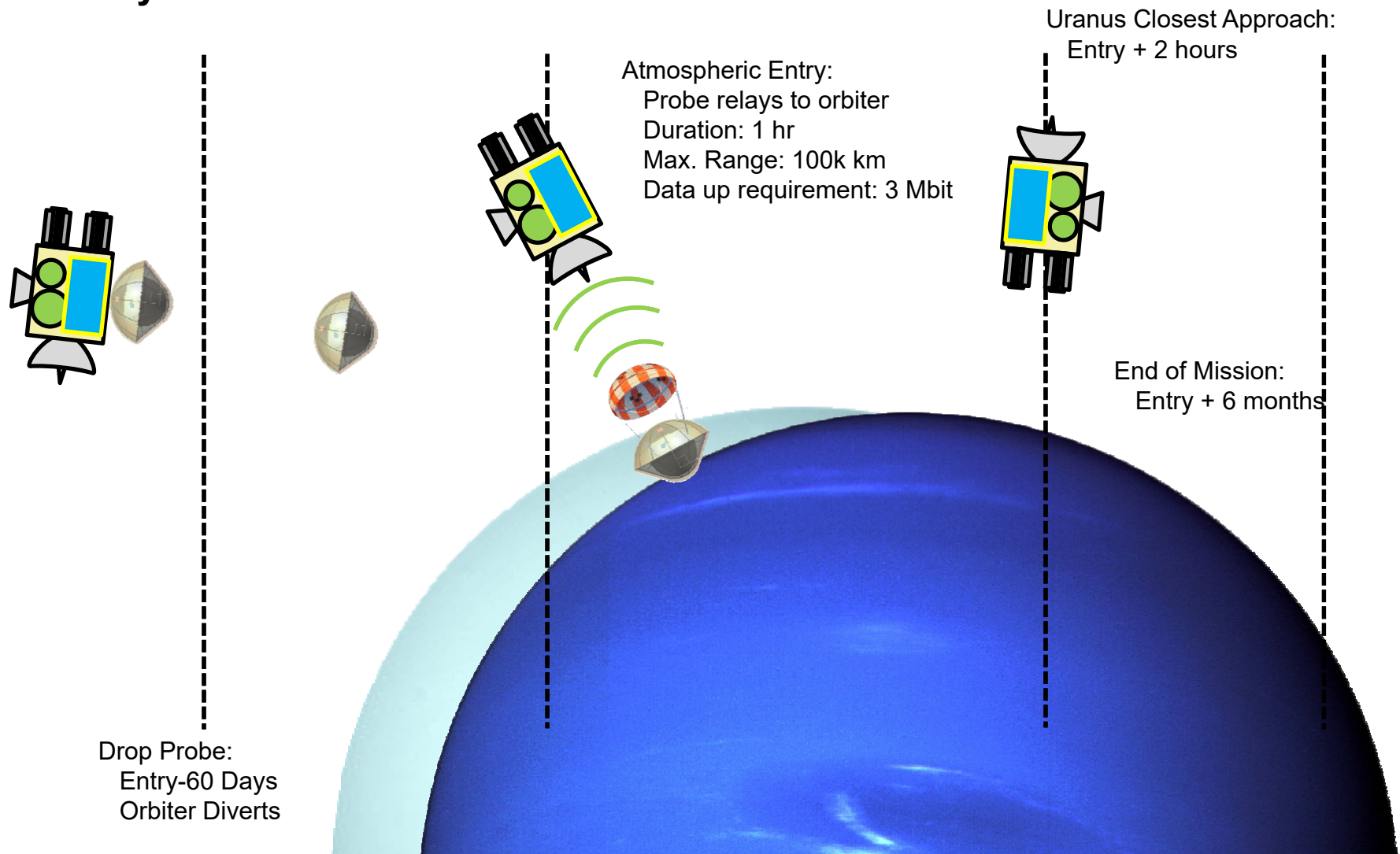

Uranus Closest Approach:
Entry + 2 hours

Atmospheric Entry:
Probe relays to orbiter
Duration: 1 hr
Max. Range: 100k km
Data up requirement: 3 Mbit

End of Mission:
Entry + 6 months

Drop Probe:
Entry-60 Days
Orbiter Diverts

Team X Study Guidelines
### *Ice Giants Study 2016-07*
### *Orbiter*

#### *Project - Study*

| | |
|---|---|
| Customer | John Elliott, Kim Reh |
| Study Lead | Bob Kinsey |
| Study Type | Pre-Decadal Study |
| Report Type | Full PPT Report |

#### *Project - Mission*

| | |
|---|---|
| Mission | Ice Giants Study 2016-07 |
| Target Body | Uranus |
| Science | Imaging and Magnetometry |
| Launch Date | 16-Apr-30 |
| Mission Duration | 10 year cruise, 6 months of downlink after close approach |
| Mission Risk Class | B |
| Technology Cutoff | 2026 |
| Minimum TRL at End of Phase B | 6 |

#### *Project - Architecture*

| | | |
|---|---|---|
| Probe | on | Entry System |
| Entry System | on | Orbiter |
| Orbiter | on | Launch Vehicle |

| | |
|---|---|
| Launch Vehicle | Atlas V 541 |
| Trajectory | E-J Gravity Assists, Probe EFPA = -30 deg |
| L/V Capability, kg | 1775 kg to a C3 of 52 with 0% contingency taken out |
| Tracking Network | DSN |
| Contingency Method | Apply Total System-Level |

| Spacecraft | |
|---|---|
| Spacecraft | Orbiter |
| Instruments | Narrow Angle Camera (EIS Europa), Doppler Imager (ECHOES JUICE), Magnetomiter (Galileo) |
| Potential Inst-S/C Commonality | None |
| Redundancy | Dual (Cold) |
| Stabilization | 3-Axis |
| Heritage | TBD |
| Radiation Total Dose | 29.833 krad behind 100 mil. of Aluminum, with an RDM of 2 added. |
| Type of Propulsion Systems | System 1-Monoprop, System 2-0, System 3-0 |
| Post-Launch Delta-V, m/s | 310 |
| P/L Mass CBE, kg | 36.7 kg Payload CBE + 320 kg Entry System + Probe (alloc) |
| P/L Power CBE, W | 44.4 |
| P/L Data Rate CBE, kb/s | 12000 |

| Project - Cost and Schedule | |
|---|---|
| Cost Target | < $2B TBD |
| Mission Cost Category | Flagship - e.g. Cassini |
| FY$ (year) | 2015 |
| Include Phase A cost estimate? | Yes |
| Phase A Start | September 2023 |
| Phase A Duration (months) | 20 |
| Phase B Duration (months) | 16 |
| Phase C/D Duration (months) | 47 |
| Review Dates | PDR - September 2026, CDR - November 2027, ARR - November 2028 |
| Phase E Duration (months) | 126 |
| Phase F Duration (months) | 4 |
| Spares Approach | Typical |
| Parts Class | Commercial + Military 883B |
| Launch Site | Cape Canaveral |

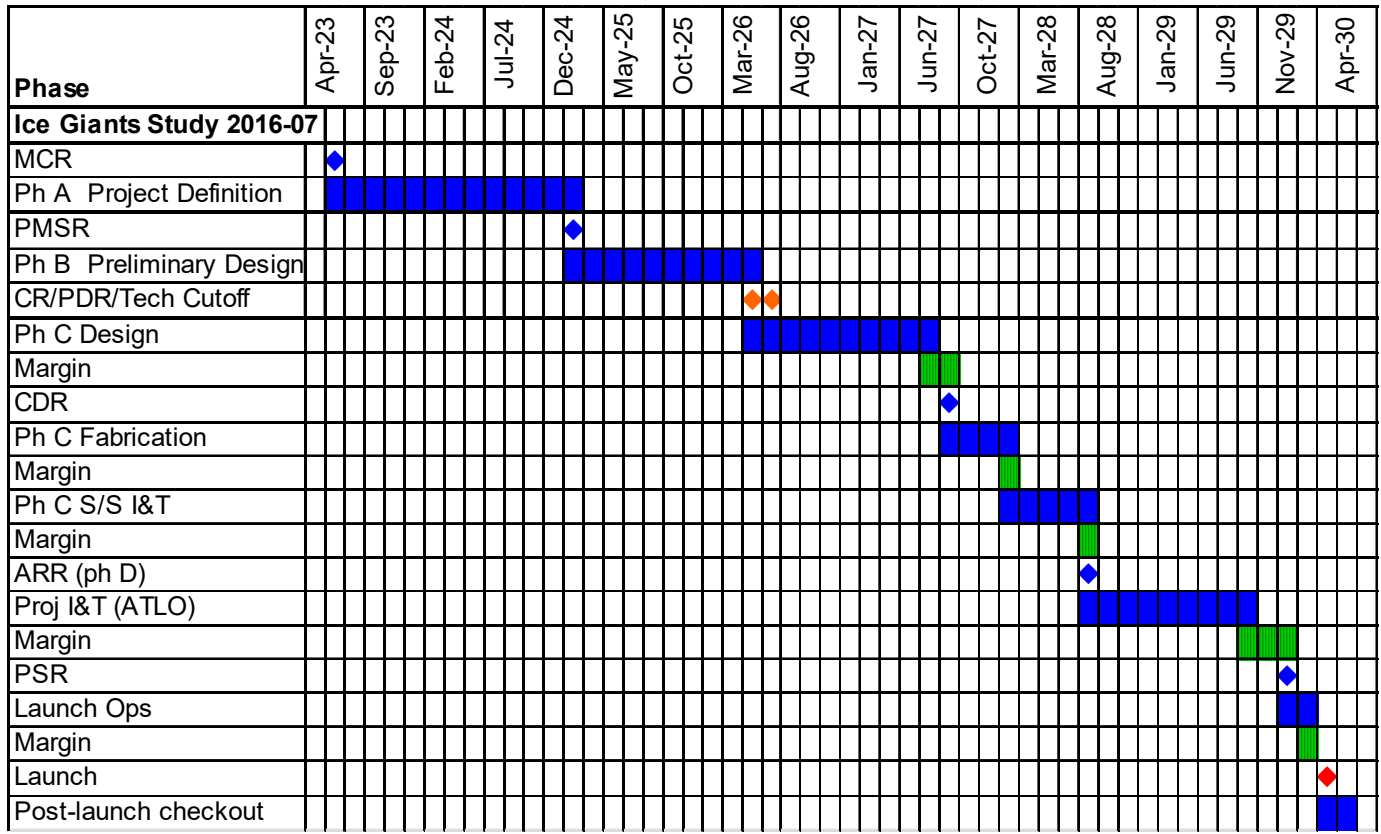

Proposed development schedule consistent with typical New Frontiers missions and current Europa mission schedule.

Phase A 20 mos., Phase B 16 mos., Phase C/D 47 mos.

Launch April 16, 2030

🏹 **Selected instruments come with some extras.**

- NAC has a 2-DOF gimbal.
- Doppler Imager has an internal fast steering mirror (FSM).

| Instrument Name | # units | Heritage | CBE Mass (kg) | Cont. | CBE+Cont. Mass/Unit (kg) | Op. Power CBE per Instrument (W) | Standy Power CBE per Instrument (W) |
|---|---|---|---|---|---|---|---|
| | | | 37 kg | 23% | 45.2 | | |
| Narrow Angle Camera (EIS Europa) | 1 | Inherited design | 12.0 | 15% | 13.8 | 16 W | 2 W |
| Doppler Imager (ECHOES JUICE) | 1 | New design | 20.0 | 30% | 26 | 20 W | 2 W |
| Magnetometer (Gallileo) | 1 | Inherited design | 4.7 | 15% | 5.405 | 8 W | 1 W |

| Instrument Name | Instrument Peak Data Rate | Units |
|---|---|---|
| Narrow Angle Camera (EIS Europa) | 12000 | kbps |
| Doppler Imager (ECHOES JUICE) | 60 | kbps |
| Magnetometer (Gallileo) | 1200 | kbps |

**✶ Instruments**
- Gas Chromatograph Mass Spectrometer (GCMS)
- Atmospheric Structure Instrument (ASI)
- Nephelometer
- Ortho-para Hydrogen Measurement Instrument

**✶ CDS**
- Redundant Sphinx Avionics

**✶ Power**
- Primary batteries
  - In probe:
    - 17.1kg, 1.0 kW-hr EOM
- Redundant Power Electronics

**✶ Thermal**
- RHU heating, passive cooling
- Vented probe design
- Thermally isolating struts

**✶ Telecom**
- Redundant IRIS radio
- UHF SSPA
- UHF Low Gain Antenna (similar to MSL)

**✶ Structures**
- ~50kg Heatshield
  - 1.2m diameter, 45deg sphere cone
- ~15kg Backshell
- ~10kg Parachutes
- ~15kg Probe Aerofairing

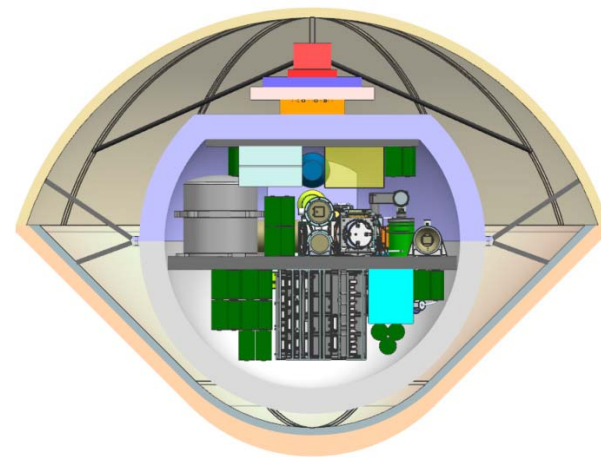

**Instruments**
- Narrow Angle Camera
- Doppler Imager
- Magnetometer

**CDS**
- JPL reference bus avionics
- Dual string cold redundancy

**Baseline Power System**
- 4 eMMRTGs, 45kg each
- 4 A-hr, <2kg, Li Ion Battery

**Telecom**
- Radios
  - Two X/X/Ka SDST transponders
  - Two IRIS radio UHF receivers
  - Two 35W Ka-Band TWTAs
  - Two 25W X-Band TWTAs
- Antennas
  - One 3m X/Ka HGA
  - One X-Band MGA
  - Two X-Band LGAs
  - One UHF patch array – 15 dBic gain

**Thermal**
- Active and passive thermal control design
- Louvers, heaters, MLI

**ACS**
- Four 0.1N Honeywell HR14 reaction wheels
- IMUs, Star Trackers, Sun Sensors

**Propulsion**
- Monopropellant hydrazine system provides 310m/s of delta-V
- One 81lbf Aerojet main engine
- Four 22N engines for main engine control
- Eight 1lbf RCS engines

**Structures**
- 180kg structure
- 24kg ballast
- 70kg harness
- 10m Magnetometer Boom
- Probe separation mechanism

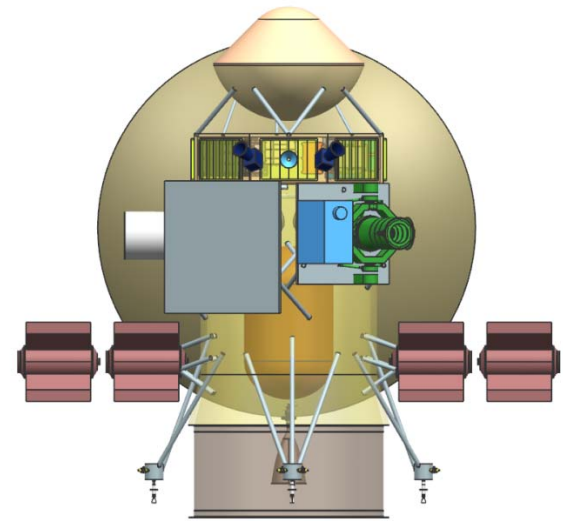

## Probe + Entry System

## Probe

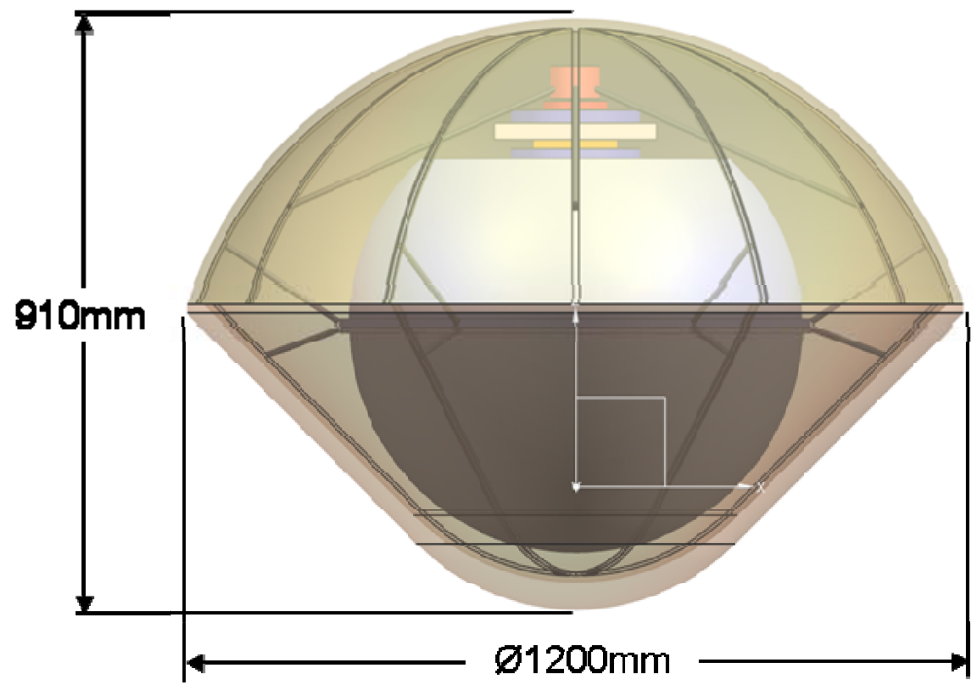

910mm

Ø1200mm

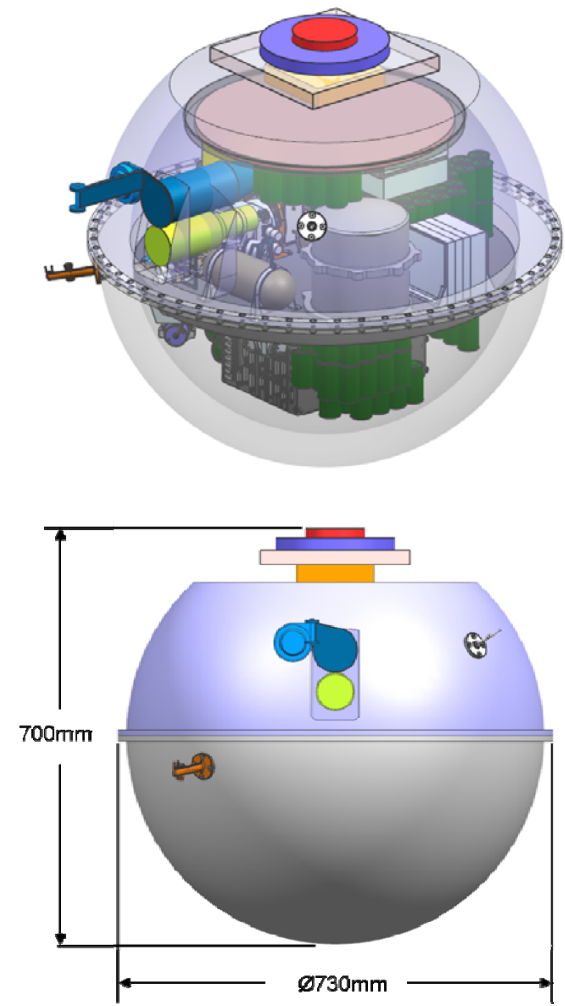

700mm

Ø730mm

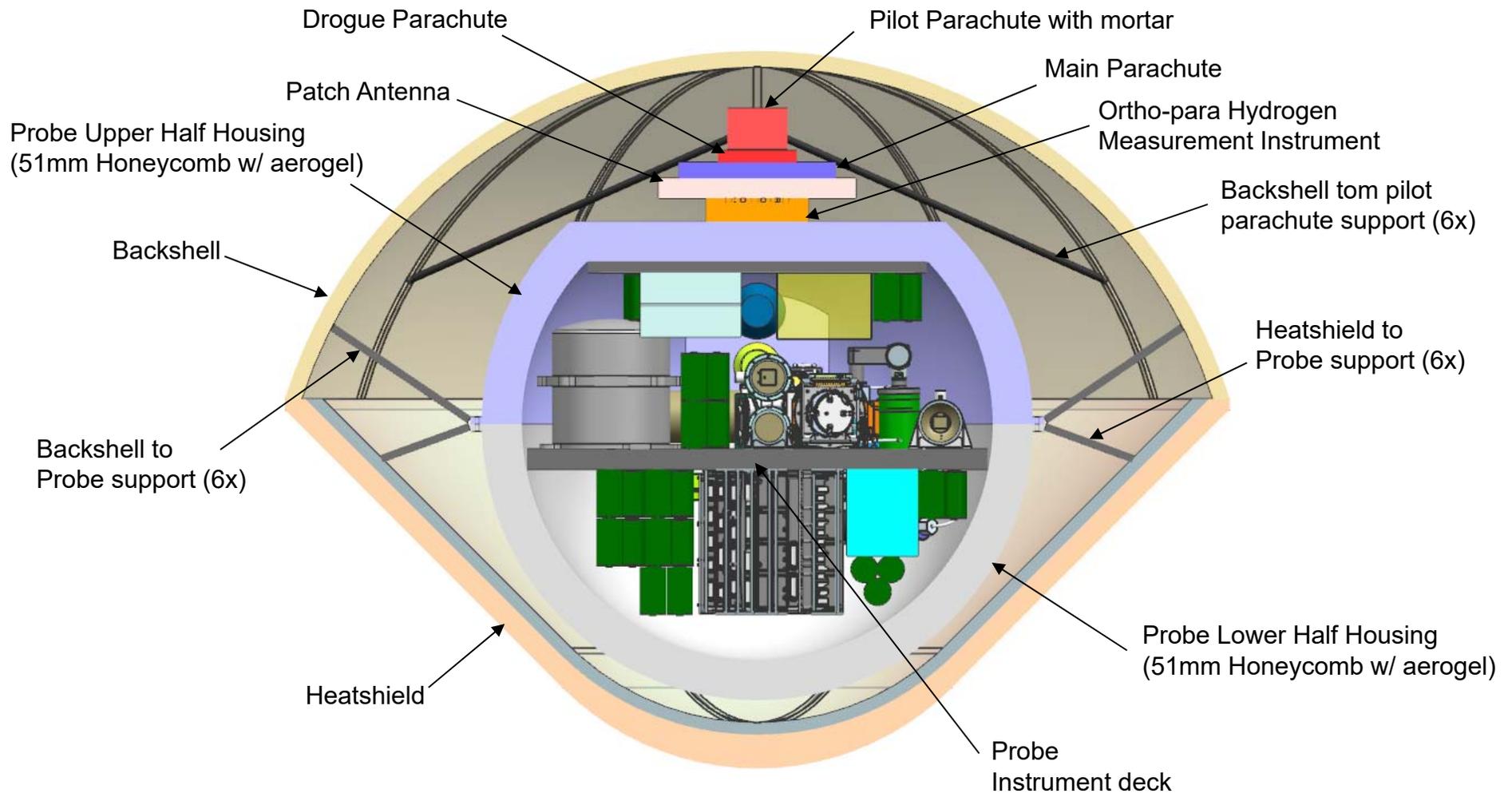

Drogue Parachute

Pilot Parachute with mortar

Patch Antenna

Main Parachute

Probe Upper Half Housing
(51mm Honeycomb w/ aerogel)

Ortho-para Hydrogen
Measurement Instrument

Backshell tom pilot
parachute support (6x)

Backshell

Heatshield to
Probe support (6x)

Backshell to
Probe support (6x)

Heatshield

Probe Lower Half Housing
(51mm Honeycomb w/ aerogel)

Probe
Instrument deck

✖ **Configuration Drawings – Stowed**

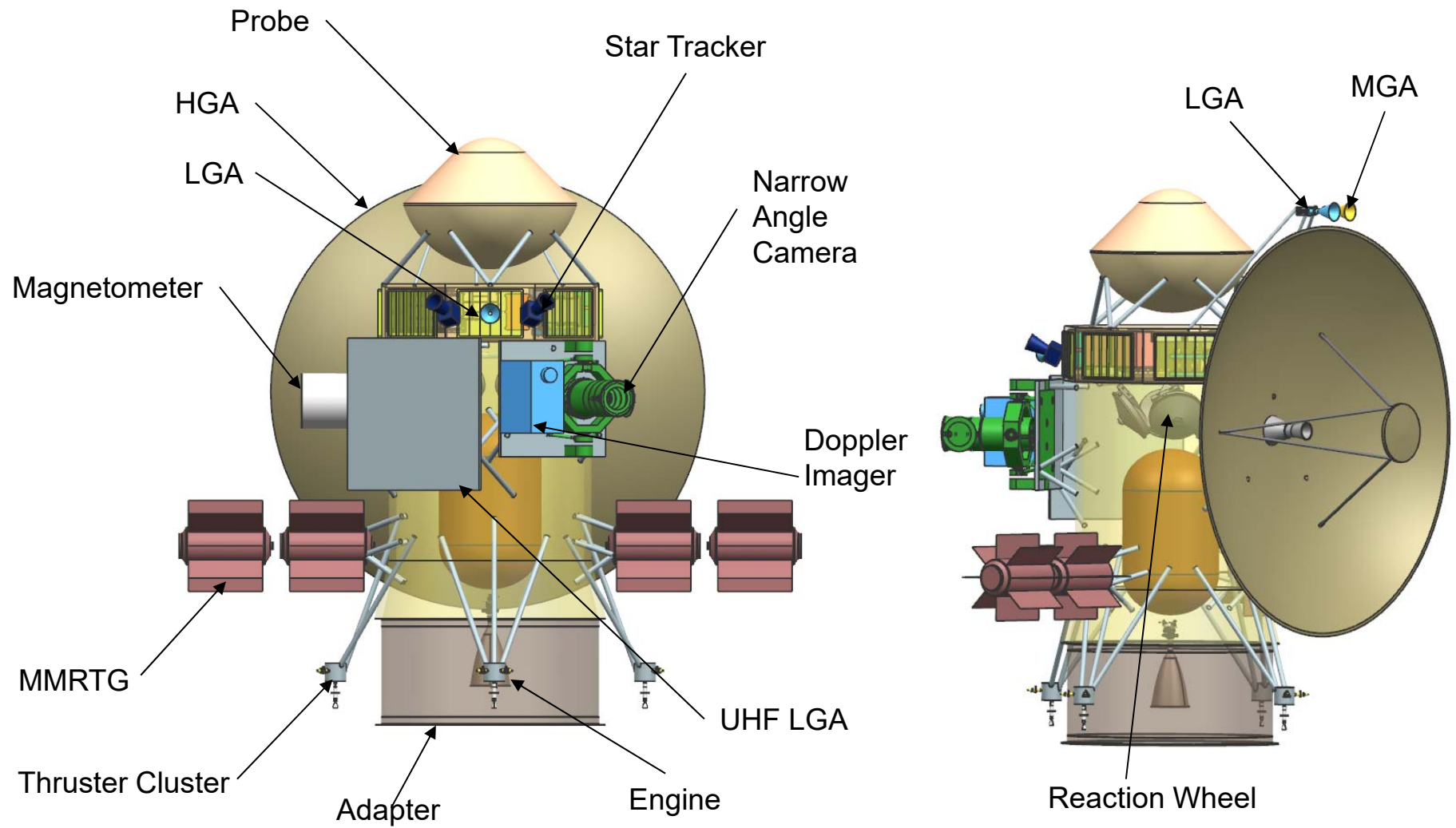

✖ **Configuration Drawings – Deployed**

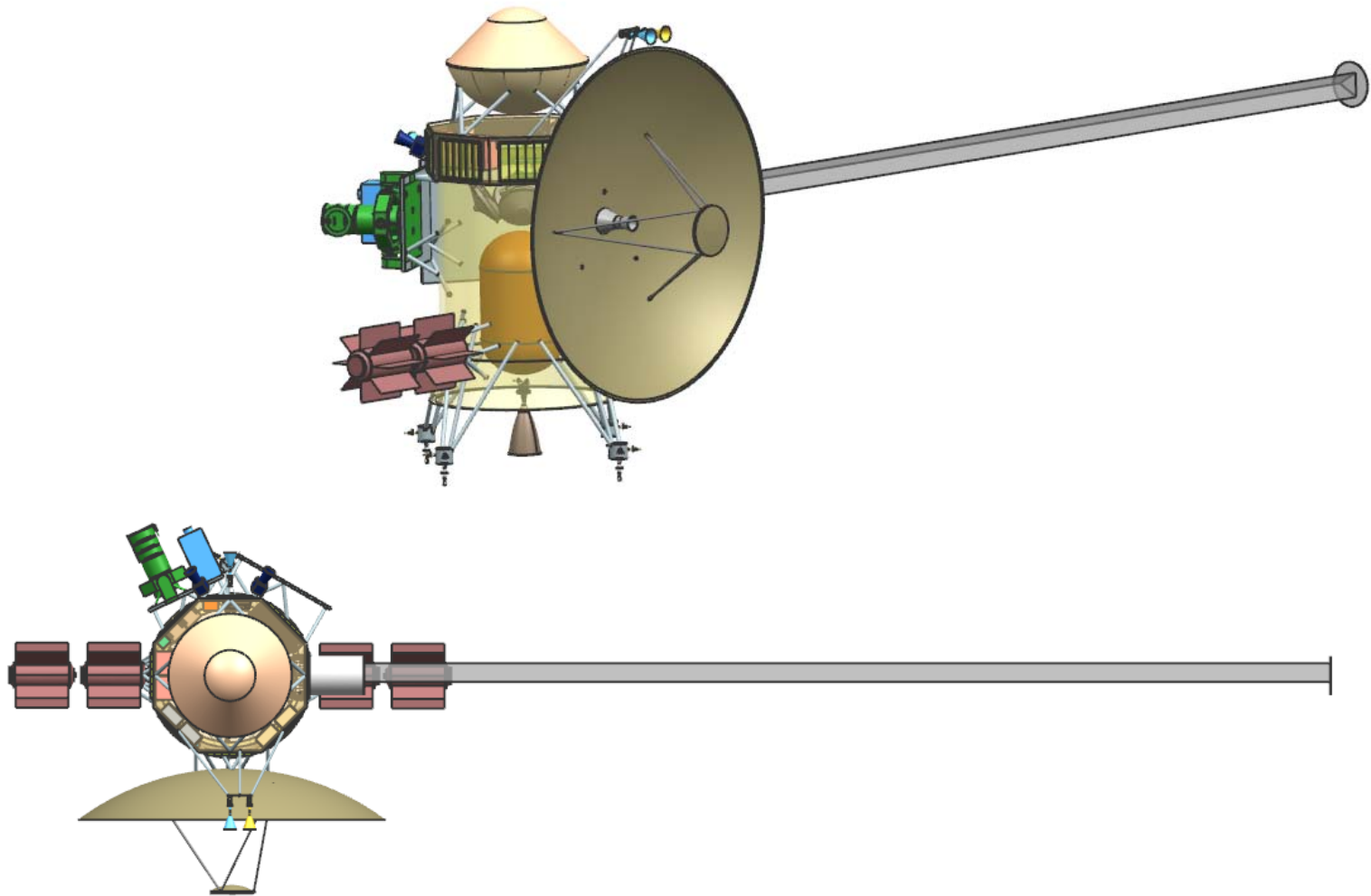

| | Mass Fraction | Mass (kg) | Subsys Cont. % | CBE+ Cont. (kg) | Mode 1 Power (W) Coast - 60 Days | Mode 2 Power (W) Warmup | Mode 3 Power (W) Science |
|---|---|---|---|---|---|---|---|
| **Power Mode Duration (hours)** | | | | | **24** | *0.5* | *1* |
| **Payload on this Element** | | | | | | | |
| Instruments | 21% | 25.3 | 29% | 32.5 | 0 | 91 | 74 |
| Payload Total | 21% | **25.3** | **29%** | **32.5** | **0** | **91** | **74** |
| **Spacecraft Bus** | | | | do not edit formulas below this line, use the calcualtions and overri | | | |
| Command & Data | 0% | 0.6 | 17% | 0.7 | 0 | 8 | 8 |
| Power | 17% | 20.1 | 26% | 25.4 | 0 | 11 | 11 |
| Structures & Mechanisms | 41% | 49.8 | 30% | 64.7 | 0 | 0 | 0 |
| Cabling | 9% | 11.5 | 30% | 15.0 | | | |
| Telecom | 5% | 6.2 | 26% | 7.8 | 0 | 0 | 184 |
| Thermal | 6% | 7.8 | 3% | 8.1 | 0 | 0 | 0 |
| Bus Total | | 96.0 | 27% | 121.7 | 0 | 19 | 203 |
| Thermally Controlled Mass | | | | 121.7 | | | |
| **Spacecraft Total (Dry): CBE & MEV** | | **121.3** | **27%** | **154.2** | 0 | 109 | 277 |
| Subsystem Heritage Contingency | 27% | 32.9 | SEP Cont | 10% | 0 | 0 | 0 |
| System Contingency | 16% | 19.3 | | | 0 | 47 | 119 |
| Total Contingency ☐ Include Carried? | **43%** | 52.2 | | | | | |
| **Spacecraft with Contingency:** | | **173** | of total | w/o addl pld | **0** | **157** | **396** |

| | | Mass Fraction | Mass (kg) | Subsys Cont. % | CBE+ Cont. (kg) |
|---|---|---|---|---|---|
| **Power Mode Duration** _(hours)_ | | | | | |
| **Additional Elements Carried by this Element** | | | | | |
| Probe | | 54% | 121.2 | 43% | 173.4 |
| **Carried Elements Total** | | 54% | **121.2** | 43% | **173.4** |
| **Spacecraft Bus** | | | | do not edit formulas below tl | |
| Structures & Mechanisms | | 46% | 102.5 | 30% | 133.2 |
| Cabling | | 0% | 0.9 | 30% | 1.2 |
| **Bus Total** | | | 103.4 | 30% | 134.4 |
| Thermally Controlled Mass | | | | | 134.4 |
| **Spacecraft Total (Dry): CBE & MEV** | | | **224.6** | 37% | **307.8** |
| Subsystem Heritage Contingency | 37% | | 83.1 | SEP Cont | 10% |
| System Contingency | 6% | | 13.4 | | |
| Total Contingency ☐ Include Carried? | **43%** | | 96.6 | | |
| **Spacecraft with Contingency:** | | | **321** | of total | w/o addl pld |

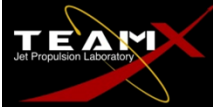
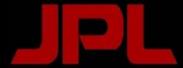

| | Mass Fraction | Mass (kg) | Subsys Cont. % | CBE+ Cont. (kg) | Mode 1 Power (W) Recharge | Mode 2 Power (W) Approach Science - 5x 11 Day Blocks | Mode 3 Power (W) Telecom Downlink | Mode 4 Power (W) TCMs | Mode 5 Power (W) Planet Flyby Science | Mode 6 Power (W) Moon Flyby Science | Mode 7 Power (W) | Mode 8 Power (W) Safe | Mode 9 Power (W) | Mode 10 Power (W) Probe Relay |
|---|---|---|---|---|---|---|---|---|---|---|---|---|---|---|
| *Power Mode Duration (hours)* | | | | | 24 | 24 | 8 | 0.50 | 16 | 8 | | 24 | | 1.5 |
| **Payload on this Element** | | | | | | | | | | | | | | |
| Instruments | 4% | 36.7 | 23% | 45.2 | 4 | 28 | 4 | 4 | 44 | 24 | 0 | 4 | 0 | 4 |
| **Payload Total** | 4% | **36.7** | 23% | **45.2** | 4 | 28 | 4 | 4 | 44 | 24 | 0 | 4 | 0 | 4 |
| **Additional Elements Carried by this Element** | | | | | | | | | | | | | | |
| Entry System+Probe | 23% | 224.1 | 43% | 320.5 | | | | | | | | | | |
| **Carried Elements Total** | 23% | **224.1** | 43% | **320.5** | 0 | 0 | 0 | 0 | 0 | 0 | 0 | 0 | 0 | 0 |
| **Spacecraft Bus** | | do not edit formulas below this line, use the calculations and override tables instead --> | | | | | | | | | | | | |
| Attitude Control | 5% | 49.5 | 10% | 54.4 | 0 | 55 | 55 | 88 | 55 | 55 | 0 | 42 | 0 | 84 |
| Command & Data | 2% | 21.6 | 10% | 23.8 | 57 | 57 | 57 | 57 | 57 | 57 | 57 | 57 | 57 | 57 |
| Power | 22% | 214.9 | 2% | 218.7 | 24 | 40 | 32 | 24 | 24 | 24 | 24 | 40 | 32 | 32 |
| Propulsion1 ☐ SEP1 | 3% | 31.5 | 3% | 32.5 | 27 | 1 | 1 | 71 | 1 | 0 | 0 | 0 | 0 | 0 |
| Structures & Mechanisms | 24% | 236.8 | 30% | 307.8 | 0 | 0 | 0 | 0 | 0 | 0 | 0 | 0 | 0 | 0 |
| Cabling | 7% | 70.3 | 30% | 91.4 | | | | | | | | | | |
| Telecom | 6% | 59.4 | 16% | 68.9 | 12 | 65 | 92 | 71 | 71 | 71 | 12 | 32 | 12 | 32 |
| Thermal | 4% | 39.5 | 19% | 47.0 | 30 | 30 | 30 | 30 | 30 | 30 | 30 | 30 | 30 | 30 |
| **Bus Total** | 16% | **733.4** | 16% | **854.4** | 150 | 247 | 266 | 341 | 237 | 237 | 123 | 201 | 131 | 235 |
| Thermally Controlled Mass | | | | 854.4 | | | | | | | | | | |
| **Spacecraft Total (Dry): CBE & MEV** | | **994.3** | 23% | **1220.1** | 154 | 275 | 271 | 345 | 282 | 261 | 123 | 205 | 131 | 239 |
| Subsystem Heritage Contingency | 23% | 225.8 | SEP Cont | 10% | 0 | 0 | 0 | 0 | 0 | 0 | 0 | 0 | 0 | 0 |
| System Contingency | 13% | 124.3 | | | 66 | 118 | 116 | 149 | 121 | 112 | 53 | 88 | 56 | 103 |
| Total Contingency ☐ Include Carried? | **35%** | 350.1 | | | | | | | | | | | | |
| **Spacecraft with Contingency:** | | **1344** | of total | w/o addl pld | **220** | **393** | **387** | **494** | **403** | **373** | **176** | **293** | **187** | **342** |
| Propellant & Pressurant with residuals1 | 12% | 179.8 | For S/C mass = | 1345.3 | | Delta-V, Sys 1 | 310.0 | m/s | | residuals = | 4.4 | kg | | |
| **Spacecraft Total with Contingency (Wet)** | | **1524.2** | | | | | | | | | | | | |
| L/V-Side Adapter | | 0.0 | Wet Mass for Prop Sizing | 1775 | | BOL Power: | 0.0 | W | | | | | | |
| **Launch Mass** | | **1524** | Dry Mass for Prop Sizing | 1344 | | EOL Power: | 0.0 | W | | | | | | |
| **Allocation** | | **1775** | Atlas V 541 | | | | | | | | | | | |
| | | | | | | Launch C3 | 52.557 | | | | | | | |
| **Launch Vehicle Margin** | | **250.8** | | | | Mission Unique LV Contingency | 0% | | | | | | | |

| Element Number | Element Name | Dry CBE (kg) | Cont / JPL Margin (kg) | Dry Allocation (kg) | Propellant (kg) | Dry Allocation + Propellant (kg) |
|---|---|---|---|---|---|---|
| 1 | Probe | 121 | 52 | 173 | - | 173 |
| 2 | Entry System | 103 | 44 | 148 | - | 148 |
| 3 | Orbiter minus eMMRTGs | 590 | 254 | 844 | 180 | 1024 |
| 3.1 | eMMRTGs | 180 | - | 180 | - | 180 |
| | **Total Stack** | **994** | **350** | **1344** | **180** | **1524** |
| | Dry Mass Allocation | | | | | 1344 |
| | JPL Margin (kg / %) | | | | | 350 / 26% |
| | JPL Margin without eMMRTG (kg / %) | | | | | 350 / 30% |
| | Atlas V 541 Capacity (kg) C3 = 52.557 | | | | | 1775 |
| | Extra Launch Vehicle Margin (kg) | | | | | 251 |

| Element Number | Element Name | Dry CBE (kg) | Cont (%) | Cont. (kg) | MEV (kg) | Dry Allocation (kg) | Propellant (kg) | Dry Allocation + Propellant (kg) |
|---|---|---|---|---|---|---|---|---|
| 1 | Probe | 121 | 27% | 33 | 154 | 173 | - | 173 |
| 2 | Entry System | 103 | 30% | 31 | 134 | 148 | - | 148 |
| 3 | Orbiter minus eMMRTGs | 590 | 22% | 130 | 720 | 844 | 180 | 1024 |
| 3.1 | eMMRTGs | 180 | - | - | 180 | 180 | - | 180 |
| | **Total Stack** | **994** | **20%** | **194** | **1188** | **1344** | **180** | **1524** |

| | |
|---|---|
| Dry Mass Allocation (kg) | 1344 |
| NASA Margin (kg / %) | 156 / 13% |
| NASA Margin without eMMRTG (kg / %) | 156 / 15% |
| Atlas V 541 Capacity (kg) C3 = 52.557 | 1775 |
| Extra Launch Vehicle Margin (kg) | 250.8 |

✘ **Probe needs to enter Uranus atmosphere head-on versus shallow**

- Probe relay antenna is nominally aligned with zenith.
- Need Flyby S/C within tens of degrees of zenith to close the link.
- Head-on: Flyby S/C is close enough to zenith for one hour.
  - ◆ Probe deceleration ~200 g's
- Shallow: Flyby S/C is far from zenith – more difficult to close the link.
  - ◆ On the other hand, Probe deceleration relatively low.
- Verified that Probe can operate through the higher deceleration.

✘ **Ka-band transmitter power versus array of 34m ground stations**

- Downlink 15 kbps using one 35W TWTA to a 34m BWG ground station.
  - ◆ Telecom system uses most of one eMMRTG power during downlink.
- Could increase downlink rate using more power, adding an eMMRTG, or by using an array of two or more 34m ground stations.
- For this option, make do with 15 kbps downlink rate.

- **Data downlink strategy for Doppler Imager (DI) on approach**
  - DI generates a lot of data continuously for tens of days on approach.
  - Configuration with HGA and DI on opposite sides of the cylindrical bus allows pointing DI towards Uranus while pointing HGA towards Earth.
  - Can downlink for ~20 hours/day and maintain positive power balance.
    - Using only 4 eMMRTGs, as opposed to 5
  - Data that can't be downlinked before flyby will be downlinked after.

- **Configuration that helps to minimize mass and power**
  - eMMRTGs outside the cylindrical bus provide heating
  - Reduces mass and power of thermal subsystem components
  - Propellant tank inside bus

- **Mechanical**
  - LV interfaces to the Flyby Spacecraft
  - Entry System containing Probe is attached to the top of the Flyby S/C.
  - Flyby S/C: primary structure is the largest mass element.
    - 121 kg CBE out of 237 kg total for Mechanical (733 kg bus dry mass)
    - Driver is the large Power mass.

- **Baseline Power System**
  - Dual String Reference Bus electronics heritage
  - eMMRTG mass of 45kg/unit, with cooling tubes, is a not to exceed value, so 0% mass uncertainty is applied.
  - 215 kg CBE subsystem dry mass

- **Propulsion: Blow down monoprop; 310 m/s total delta V**
  - One 275N main engine
  - Four 22N thrusters for main engine control; Eight 4.5N thrusters
  - 32.5 kg CBE dry mass; 180 kg propellant and pressurant

- **Thermal: Cassini-heritage waste-heat recovery system on Orbiter**
  - RTG end domes each provide 75 W waste heat to propulsion module via conductive and radiative coupling.
  - VRHUs act as primary control mechanism for thruster clusters.
    - Also act as trim heaters for the propulsion module
  - Louvers act as primary control mechanism for avionics module.

- **Telecom: X- and Ka-Band subsystem, plus UHF for Probe data.**
  - Two 35W Ka-Band TWTAs, two 25W X-Band TWTAs
  - Two X/X/Ka SDST transponders, two IRIS radio UHF receivers
  - 3m X/Ka HGA, one X-Band MGA, two X-Band LGAs, UHF patch array.
  - Supports a data rate at Uranus of 15 kbps into 34m BWG ground station.
  - Supports uplink of 3Mbits of probe data

✘ **CDS: Reference Bus architecture ideally suited for high reliability, long lifetime mission.**

- Standard JPL spacecraft CDS that is similar to SMAP
  - RAD750 CPU, NVM, MTIF, MSIA, CRC, LEU-A, LEU-D, MREU
  - 128 GBytes storage for science data
  - 1553 and RS-422 ICC/ITC interfaces for subsystems and instruments

✘ **ACS: 3-axis stabilized with star tracker, sun sensor, gyros, wheels.**

- All stellar attitude determination to minimize power, conserve gyros.
- Sun sensor performance may degrade once the Flyby S/C passes Saturn.
  - May impact safe mode used during star tracker outage.
  - Detailed analysis on Sun sensor performance versus distance is needed.

- **Software: core product line is appropriate since this mission has aspects similar to MSL/M2020/SMAP/Europa.**
  - Complexity rankings range from Medium to High.
    - Medium infrastructure: dual string with warm spare.
    - High fault behaviors: high redundancy, string swapping, critical events.
    - Medium/High ACS: tight pointing requirements, many ACS modes
    - Medium Telecom: dual active UHF, redundant DTE
    - Medium Science data processing, full file system

- **SVIT: Probe testbed, system I&T and V&V costs are included**
  - Cost of assembling and testing RTG's is captured elsewhere
    - Cost of integrating RTG's is included with other ATLO costs
  - Probe with 5 Instruments costed separately; testbed costs included.

- **Ground Systems**
  - Mission specific implementation of standard JPL mission operations and ground data systems
  - Ground network: DSN 34-m BWG; 70-m or equivalent for safe mode
  - Science support: 24x7 tracking on approach; daily contacts after flyby

| COST SUMMARY (FY2015 $M) | Generate ProPricer Input | Team X Estimate | | |
|---|---|---|---|---|
| | | CBE | Res. | PBE |
| **Project Cost** | | **$1223.1 M** | **22%** | **$1492.8 M** |
| **Launch Vehicle** | | **$33.0 M** | **0%** | **$33.0 M** |
| **Project Cost (w/o LV)** | | **$1190.1 M** | **23%** | **$1459.8 M** |
| | | | | |
| **Development Cost** | | **$1014.8 M** | **24%** | **$1259.9 M** |
| Phase A | | $10.1 M | 24% | $12.6 M |
| Phase B | | $91.3 M | 24% | $113.4 M |
| Phase C/D | | $913.3 M | 24% | $1133.9 M |
| **Operations Cost** | | **$175.4 M** | **14%** | **$199.9 M** |

Total mission cost is $1.49B. This is the likely cost within a range that typically can be as much as 10% lower up to 20% higher. The development cost with reserves is $1.26B.

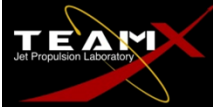

# Executive Summary
# Cost A-D – Option 4

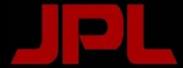

| WBS Elements | NRE | RE | 1st Unit |
|---|---|---|---|
| Project Cost (no Launch Vehicle) | $1064.6 M | $428.2 M | $1492.8 M |
| Development Cost (Phases A - D) | $831.8 M | $428.1 M | $1259.9 M |
| 01.0 Project Management | $47.3 M | | $47.3 M |
| 1.01 Project Management | $11.4 M | | $11.4 M |
| 1.02 Business Management | $13.6 M | | $13.6 M |
| 1.04 Project Reviews | $2.5 M | | $2.5 M |
| 1.06 Launch Approval | $19.8 M | | $19.8 M |
| 02.0 Project Systems Engineering | $23.7 M | $0.5 M | $24.2 M |
| 2.01 Project Systems Engineering | $8.9 M | | $8.9 M |
| 2.02 Project SW Systems Engineering | $5.2 M | | $5.2 M |
| 2.03 EEIS | $1.5 M | | $1.5 M |
| 2.04 Information System Management | $1.7 M | | $1.7 M |
| 2.05 Configuration Management | $1.5 M | | $1.5 M |
| 2.06 Planetary Protection | $0.2 M | $0.2 M | $0.4 M |
| 2.07 Contamination Control | $1.2 M | $0.3 M | $1.5 M |
| 2.09 Launch System Engineering | $1.0 M | | $1.0 M |
| 2.10 Project V&V | $2.0 M | | $2.0 M |
| 2.11 Risk Management | $0.5 M | | $0.5 M |
| 03.0 Mission Assurance | $42.3 M | $0.0 M | $42.3 M |
| 04.0 Science | $24.8 M | | $24.8 M |
| Orbiter Science | $14.0 M | | $14.0 M |
| Probe Science | $10.8 M | | $10.8 M |
| 05.0 Payload System | $80.2 M | $48.3 M | $128.5 M |
| 5.01 Payload Management | $7.8 M | | $7.8 M |
| 5.02 Payload Engineering | $5.8 M | | $5.8 M |
| Orbiter Instruments | $33.5 M | $24.3 M | $57.8 M |
| Narrow Angle Camera (EIS Europa) | $11.6 M | $8.4 M | $20.0 M |
| Doppler Imager (ECHOES JUICE) | $17.4 M | $12.6 M | $30.0 M |
| Magnetometer (Gallileo) | $4.5 M | $3.3 M | $7.8 M |
| Probe Instruments | $33.1 M | $24.0 M | $57.1 M |
| Mass Spectrometer | $22.9 M | $16.6 M | $39.6 M |
| Atmospheric Structure Investigation (ASI) | $3.4 M | $2.5 M | $5.9 M |
| Nephelometer (Galileo) | $5.3 M | $3.8 M | $9.1 M |
| Ortho-para H2 meas. Expt. | $1.5 M | $1.1 M | $2.6 M |

| WBS Elements | NRE | RE | 1st Unit |
|---|---|---|---|
| 06.0 Flight System | $383.2 M | $259.8 M | $643.0 M |
| 6.01 Flight System Management | $5.0 M | | $5.0 M |
| 6.02 Flight System Systems Engineering | $49.1 M | | $49.1 M |
| 6.03 Product Assurance (included in 3.0) | | | $0.0 M |
| Orbiter | $237.0 M | $215.8 M | $452.7 M |
| 6.04 Power | $94.5 M | $133.1 M | $227.6 M |
| 6.05 C&DH | $31.3 M | $24.3 M | $55.6 M |
| 6.06 Telecom | $28.4 M | $18.1 M | $46.5 M |
| 6.07 Structures (includes Mech. I&T) | $38.6 M | $11.8 M | $50.4 M |
| 6.08 Thermal | $4.2 M | $8.3 M | $12.4 M |
| additional cost for >43 RHUs | $0.0 M | $0.0 M | $0.0 M |
| 6.09 Propulsion | $9.1 M | $6.5 M | $15.6 M |
| 6.10 ACS | $9.4 M | $8.9 M | $18.3 M |
| 6.11 Harness | $4.1 M | $3.8 M | $7.9 M |
| 6.12 S/C Software | $17.1 M | $0.9 M | $18.0 M |
| 6.13 Materials and Processes | $0.4 M | $0.0 M | $0.4 M |
| Probe | $26.8 M | $18.1 M | $44.9 M |
| 6.04 Power | $2.9 M | $2.1 M | $5.0 M |
| 6.05 C&DH | $0.3 M | $2.3 M | $2.7 M |
| 6.06 Telecom | $7.9 M | $4.1 M | $12.0 M |
| 6.07 Structures (includes Mech. I&T) | $8.0 M | $3.5 M | $11.5 M |
| 6.08 Thermal | $2.3 M | $4.9 M | $7.2 M |
| 6.11 Harness | $1.8 M | $0.9 M | $2.7 M |
| 6.12 S/C Software | $3.3 M | $0.2 M | $3.5 M |
| 6.13 Materials and Processes | $0.4 M | $0.0 M | $0.4 M |
| Entry System | $57.1 M | $24.4 M | $81.5 M |
| 6.07 Structures (includes Mech. I&T) | $55.3 M | $24.1 M | $79.4 M |
| 6.11 Harness | $1.4 M | $0.3 M | $1.7 M |
| 6.13 Materials and Processes | $0.4 M | $0.0 M | $0.4 M |
| Ames/Langley EDL Engineering/Testing | $3.8 M | $0.0 M | $3.8 M |
| 6.14 Spacecraft Testbeds | $4.5 M | $1.5 M | $6.0 M |

| WBS Elements | NRE | RE | 1st Unit |
|---|---|---|---|
| **07.0 Mission Operations Preparation** | **$26.6 M** | | **$26.6 M** |
| 7.0 MOS Teams | $22.3 M | | $22.3 M |
| 7.03 DSN Tracking (Launch Ops.) | $2.7 M | | $2.7 M |
| 7.06 Navigation Operations Team | $1.6 M | | $1.6 M |
| 7.07.03 Mission Planning Team | $0.0 M | | $0.0 M |
| **09.0 Ground Data Systems** | **$24.0 M** | | **$24.0 M** |
| 9.0A Ground Data System | $22.8 M | | $22.8 M |
| 9.0B Science Data System Development | $0.7 M | | $0.7 M |
| 9A.03.07 Navigation H/W & S/W Development | $0.5 M | | $0.5 M |
| **10.0 ATLO** | **$21.7 M** | **$20.8 M** | **$42.5 M** |
| Orbiter | $15.9 M | $12.4 M | $28.3 M |
| Probe | $5.9 M | $8.4 M | $14.2 M |
| **11.0 Education and Public Outreach** | **$0.0 M** | **$0.0 M** | **$0.0 M** |
| **12.0 Mission and Navigation Design** | **$11.5 M** | | **$11.5 M** |
| 12.01 Mission Design | $1.9 M | | $1.9 M |
| 12.02 Mission Analysis | $3.0 M | | $3.0 M |
| 12.03 Mission Engineering | $1.8 M | | $1.8 M |
| 12.04 Navigation Design | $4.8 M | | $4.8 M |
| **Development Reserves** | **$146.3 M** | **$98.8 M** | **$245.1 M** |

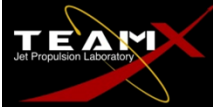
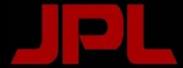

| WBS Elements | NRE | RE | 1st Unit |
|---|---|---|---|
| **Operations Cost (Phases E - F)** | **$199.8 M** | **$0.1 M** | **$199.9 M** |
| 01.0 Project Management | $19.3 M | | $19.3 M |
| 1.01 Project Management | $10.8 M | | $10.8 M |
| 1.02 Business Management | $7.6 M | | $7.6 M |
| 1.04 Project Reviews | $0.8 M | | $0.8 M |
| 1.06 Launch Approval | $0.1 M | | $0.1 M |
| 02.0 Project Systems Engineering | $0.0 M | $0.1 M | $0.1 M |
| 03.0 Mission Assurance | $2.5 M | $0.0 M | $2.5 M |
| 04.0 Science | $46.4 M | | $46.4 M |
| Orbiter Science | $30.6 M | | $30.6 M |
| Probe Science | $15.8 M | | $15.8 M |
| 07.0 Mission Operations | $83.6 M | | $83.6 M |
| 7.0 MOS Teams | $65.5 M | | $65.5 M |
| 7.03 DSN Tracking | $11.7 M | | $11.7 M |
| 7.06 Navigation Operations Team | $5.8 M | | $5.8 M |
| 7.07.03 Mission Planning Team | $0.6 M | | $0.6 M |
| 09.0 Ground Data Systems | $23.5 M | | $23.5 M |
| 9.0A GDS Teams | $17.9 M | | $17.9 M |
| 9.0B Science Data System Ops | $5.1 M | | $5.1 M |
| 9A.03.07 Navigation HW and SW Dev | $0.4 M | | $0.4 M |
| 11.0 Education and Public Outreach | $0.0 M | $0.0 M | $0.0 M |
| 12.0 Mission and Navigation Design | $0.0 M | | $0.0 M |
| Operations Reserves | $24.5 M | $0.0 M | $24.6 M |
| **8.0 Launch Vehicle** | **$33.0 M** | | **$33.0 M** |
| Launch Vehicle and Processing | $0.0 M | | $0.0 M |
| Nuclear Payload Support | $33.0 M | | $33.0 M |

## Risks related to the Probe

- May be issues for the relay link margin due to Probe-Flyby S/C geometry and uncertainties regarding Uranus atmosphere/ potential signal attenuation.

- High g load on the Probe carries some risk.

- Last Probe targeting occurs more than 60 days prior to encounter.
  - Probe carries no propulsion, so it cannot correct trajectory dispersions.
  - Need dispersions small enough to ensure safe entry conditions at Uranus.

- Trajectory knowledge requirements for science reconstruction need to be determined.
  - Will drive how the Probe is tracked pre-entry and what telemetry (e.g. IMU) needs to be transmitted with the science data to the Flyby S/C.
  - The latter will impact the data budget.

- **Mission duration will push systems to their operating lifetimes.**

- **Collision avoidance with Uranus' rings needs to be considered.**

- **Uranus stays close to the range of solar conjunction (~4-5 deg)**
  - Doppler measurements may have increased noise levels.

- **eMMRTG still needs some development.**
  - May cause a schedule slip.
  - Performance may degrade at a higher rate than currently predicted.

- **RTG waste heat recovery design robustness**
  - Approach is highly configuration-dependent and may have high hidden development costs.
  - Less expensive on paper, but the actual implementation could be more expensive than an active system.

- **Sun sensor performance may degrade past Saturn.**
  - May impact safe mode used during star tracker outage.

# Option 5

- **Option 5: Uranus Orbiter Concept with Probe**
  - 50 kg payload allocation
  - 1 atmospheric probe (previously designed)
  - Includes VEEJ flybys

- **Class B mission**
- **Dual string redundancy**
- **eMMRTGs could be used for Orbiter power**
  - Carry <u>no mass contingency</u>, because eMMRTG masses provided are "not to exceed" values

- ✖ **Mission:**
  - Launch: 5/25/2031; Arrival: 5/17/2043
  - Launch, V-E-E-J flybys, cruise to Uranus
  - Probe separation 60 days prior to entry

- ✖ **Mission Design**
  - 12-year cruise to UOI, 3-year science tour
  - Orbiter serves as communications relay during probe entry
    - ◆ Continuous line of sight between orbiter and probe is critical for telecom
    - ◆ Baseline probe EFPA is -30 deg to allow acceptable geometry and probe entry g-loads
  - UOI begins ~2 hours after Probe entry.
    - ◆ Allows sufficient time for probe data relay prior to the turn-to-burn
    - ◆ More than 2 hours results in gain issues for the zenith-pointed Probe UHF antenna (since range and off-pointing angle from zenith increase).
  - UOI inserts into 120-day initial orbit (1.05 Ur periapse), lowered to ~50-day orbit
  - Will require optical navigation upon approach to UOI, and during science for targeting moon flybys
    - ◆ Doppler imager will be used for OpNav on approach

- ✖ **Launch Vehicle**
  - Atlas 541 (~4450 kg to C3 of 11.90 $km^2/s^2$)

- **Arrival Vinf / Declination**
  - ~8.41 km/s,  48 degree (spin-axis relative)
- **Orbiter-Probe separation is ~100,000 km at entry time**
  - This range closes as the Orbiter proceeds towards periapse and the Probe decelerates in the Uranus atmosphere.
  - Telecom requires Orbiter within 60 deg of zenith and < 100,000 km range.
- **EFPA of -30 degrees**
  - Imparts 165 g's on the probe during entry.
  - Reducing the EFPA results in line-of-sight geometry challenges.

| Event | Rel. Time | Duration | Delta V (m/s) | # Maneuvers | Comments |
|-------|-----------|----------|---------------|-------------|----------|
| TCM-1 | L+10 days | | 25 | 1 | Non-deterministic |
| Venus Flyby | L+116 days | | | | 7700 km altitude |
| Earth Flyby #1 | L+383 days | | | | 3400 km altitude |
| TCMs 2-4 | TBD | | 15 | 3 | Non-deterministic |
| DSM-1 | L+787 days | | 327 | 1 | |
| Earth Flyby #2 | L+1098 days | | | | 300 km altitude |
| DSM-2 | L+1305 days | | 238 | 1 | |
| Jupiter Flyby | L+1769 days | | | | 1.3e6 km altitude |
| TCM-5 | E-80 days | | 10 | 1 | Non-deterministic |
| Separation | E-60 days | | Small | 1 | |
| ODDM | E-55 days | | 15 | 1 | Non-deterministic |
| UOI | L+4375 days | ~1 hr | 1687 | 1 | |
| OTMs 1-5 | TBD | | 290 (Total) | 5 | Includes UOI cleanup, period+incl changes, flyby targeting, and other statistical mnvrs |
| **Total** | | | 2607 | 15+ | |

★ **Element 1: Atmospheric Probe**
  - Designed in study 1734 (June 28,30[th])
  - Common for both planets
  - No propulsion, no ACS, no power generation
  - Telecom relay to Orbiter

★ **Element 2: Entry System**
  - Heatshield, backshell, structure

★ **Element 3: Orbiter**
  - Instrument allocation defined by Option
  - Chemical propulsion
  - eMMRTGs, no solar arrays

★ **Element 4: SEP Cruise Stage**
  - "Dumb" cruise

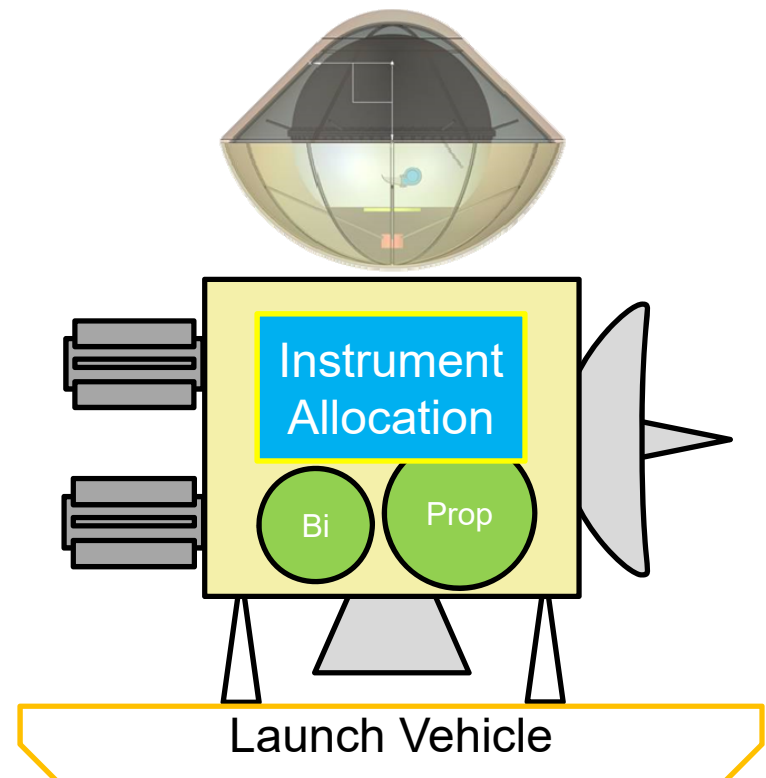

*Drawing not to scale

No SEP for this option!

Instrument Allocation

Bi    Prop

Launch Vehicle

**✖ Mission Timeline**

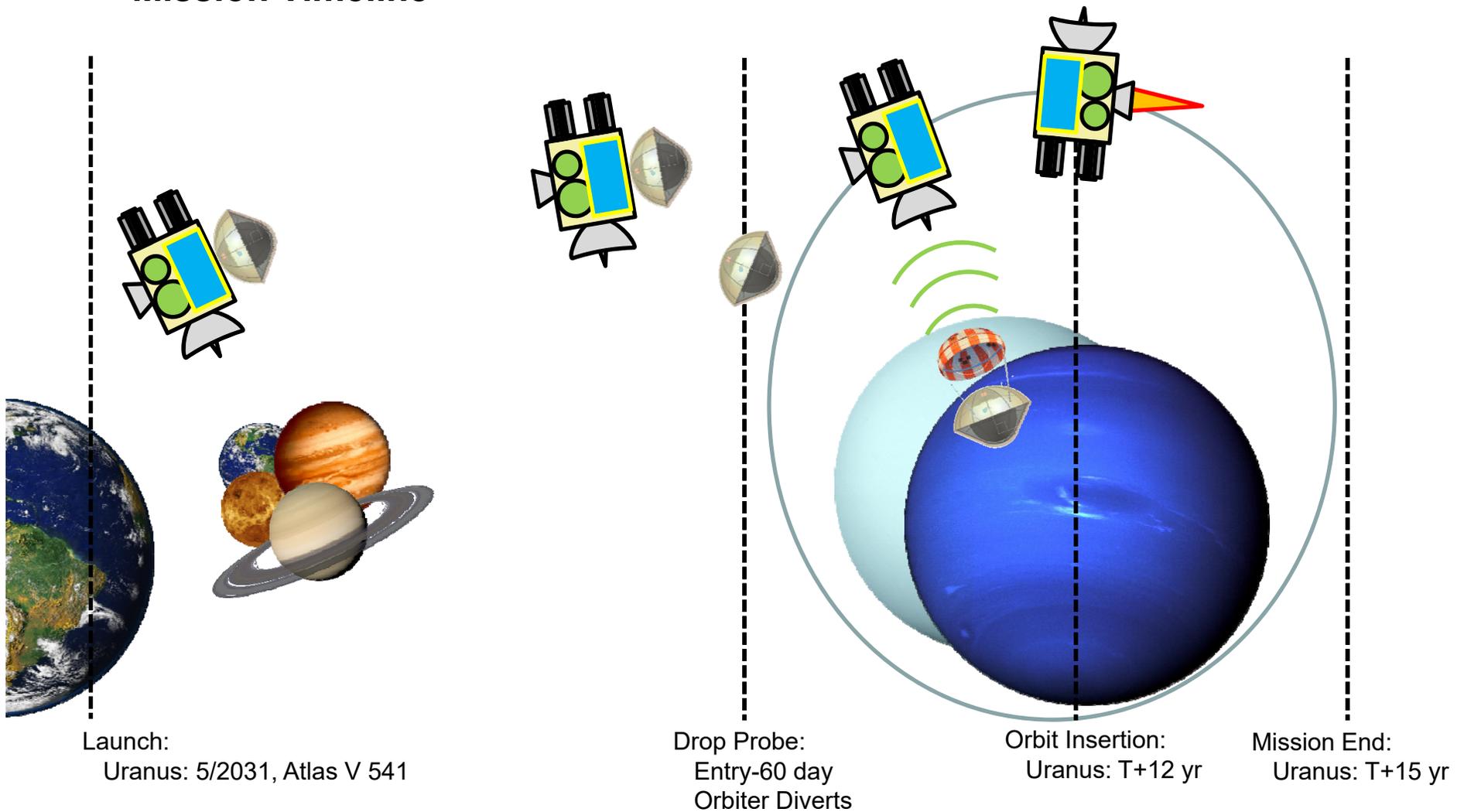

Launch:
   Uranus: 5/2031, Atlas V 541

Drop Probe:
   Entry-60 day
   Orbiter Diverts

Orbit Insertion:
   Uranus: T+12 yr

Mission End:
   Uranus: T+15 yr

**Approach Timeline**

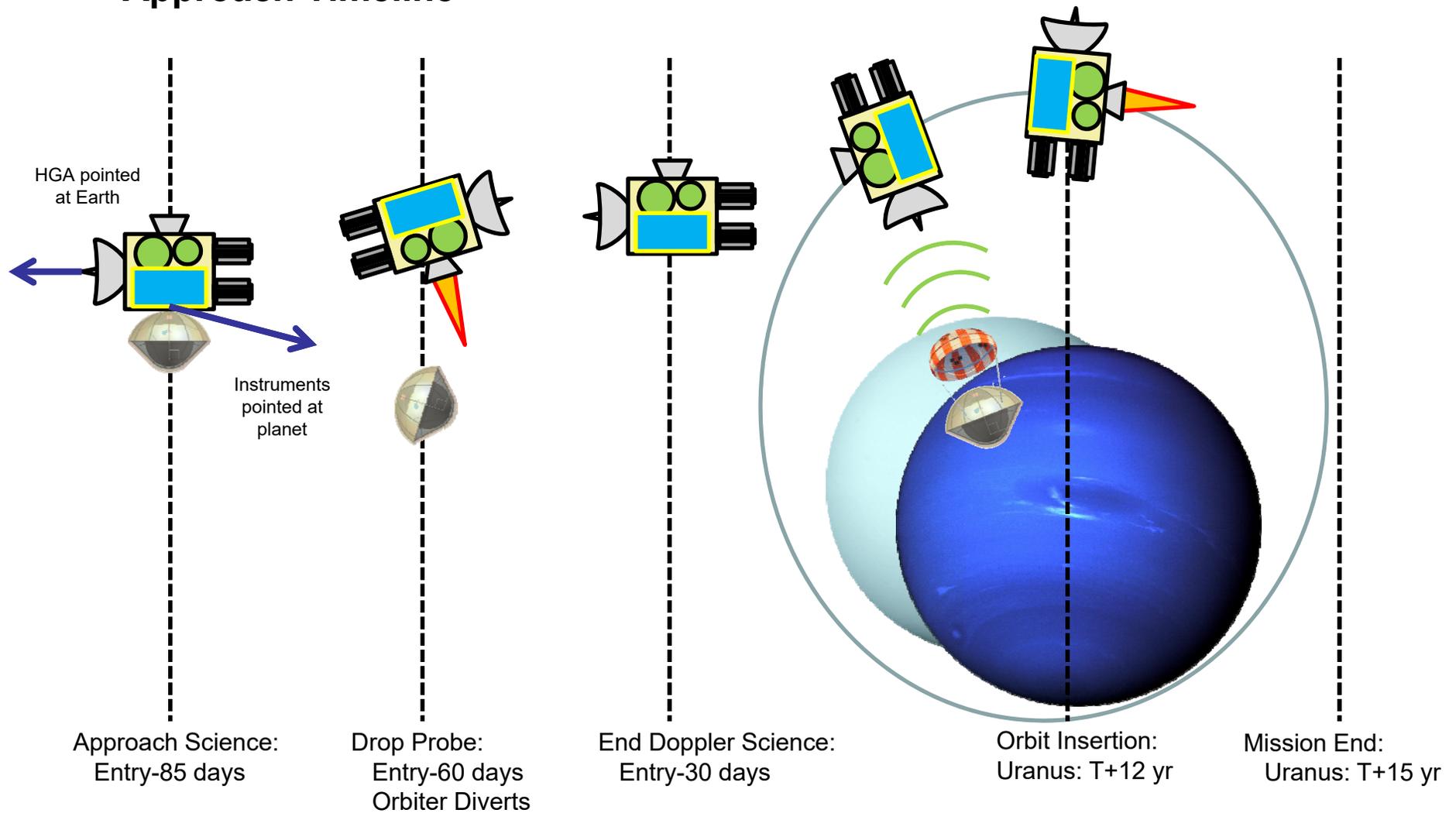

HGA pointed at Earth

Instruments pointed at planet

Approach Science:
Entry-85 days

Drop Probe:
Entry-60 days
Orbiter Diverts

End Doppler Science:
Entry-30 days

Orbit Insertion:
Uranus: T+12 yr

Mission End:
Uranus: T+15 yr

✶ **Entry Timeline**

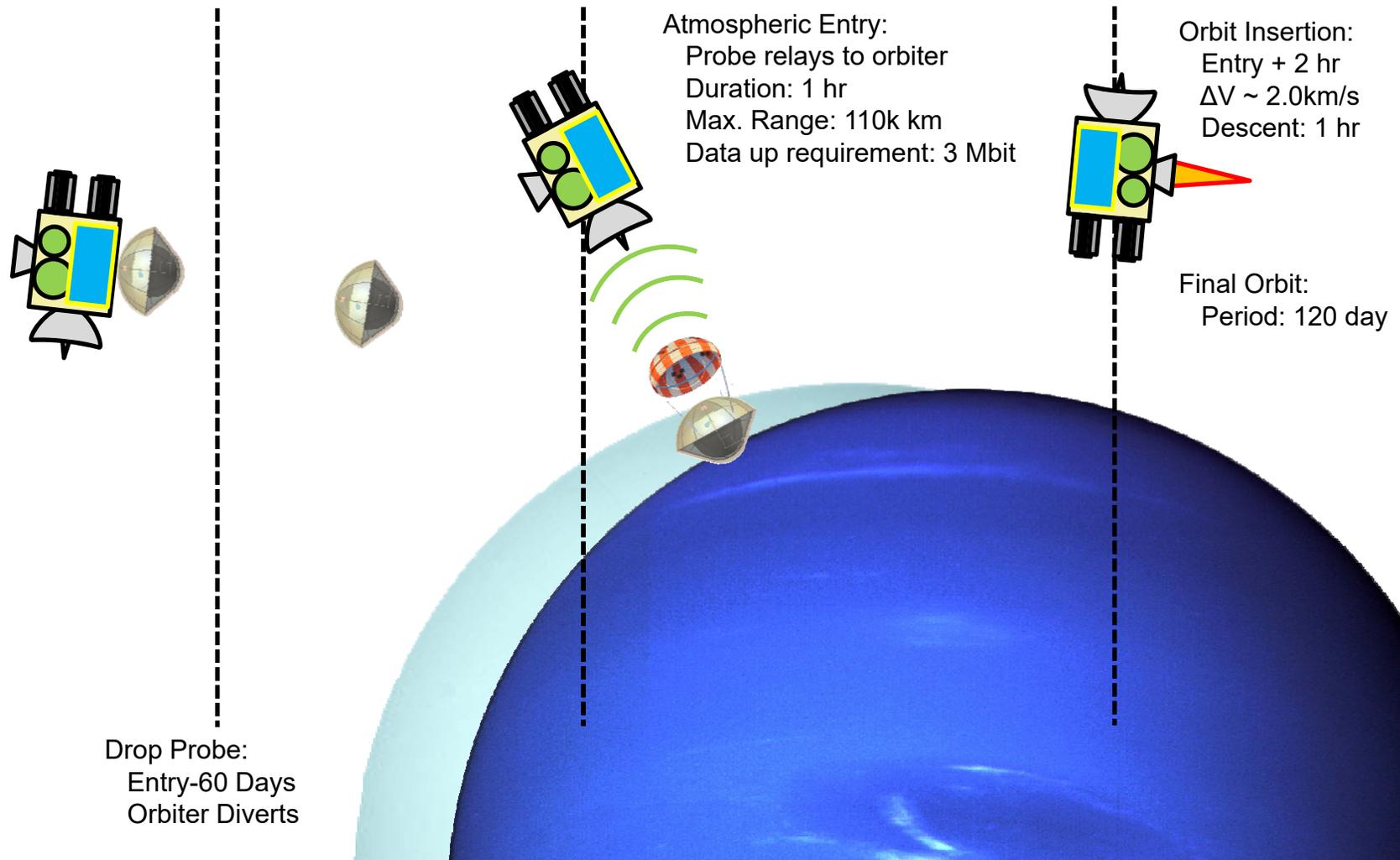

Atmospheric Entry:
  Probe relays to orbiter
  Duration: 1 hr
  Max. Range: 110k km
  Data up requirement: 3 Mbit

Orbit Insertion:
  Entry + 2 hr
  ΔV ~ 2.0km/s
  Descent: 1 hr

Final Orbit:
  Period: 120 day

Drop Probe:
  Entry-60 Days
  Orbiter Diverts

Team X Study Guidelines
## Ice Giants Study 2016-07
## Orbiter

### Project - Study

| | |
|---|---|
| Customer | John Elliott, Kim Reh |
| Study Lead | Bob Kinsey |
| Study Type | Pre-Decadal Study |
| Report Type | Full PPT Report |

### Project - Mission

| | |
|---|---|
| Mission | Ice Giants Study 2016-07 |
| Target Body | Uranus |
| Science | Imaging and Magnetometry |
| Launch Date | 25-May-31 |
| Mission Duration | 12 year cruise, 3 years in orbit |
| Mission Risk Class | B |
| Technology Cutoff | 2027 |
| Minimum TRL at End of Phase B | 6 |

### Project - Architecture

| | | |
|---|---|---|
| Probe | on | Entry System |
| Entry System | on | Orbiter |
| Orbiter | on | SEP Stage |
| SEP Stage | on | Launch Vehicle |

| | |
|---|---|
| Launch Vehicle | Atlas V 541 |
| Trajectory | VEEJ Gravity Assists, 120 day initial orbit, probe FPA = -30deg |
| L/V Capability, kg | 4450 kg to a C3 of 11 with 0% contingency taken out |
| Tracking Network | DSN |
| Contingency Method | Apply Total System-Level |

| Spacecraft | |
|---|---|
| Spacecraft | Orbiter |
| Instruments | Narrow Angle Camera (EIS Europa), Doppler Imager (ECHOES JUICE), Magnetomiter (Galileo) |
| Potential Inst-S/C Commonality | None |
| Redundancy | Dual (Cold) |
| Stabilization | 3-Axis |
| Radiation Total Dose | 29.833 krad behind 100 mil. of Aluminum, with an RDM of 2 added. |
| Type of Propulsion Systems | System 1-Biprop, System 2-0, System 3-0 |
| Post-Launch Delta-V, m/s | 2607 |
| P/L Mass CBE, kg | 36.7 kg Payload CBE + 320 kg Entry System + Probe (alloc) |
| P/L Power CBE, W | 44.4 |
| P/L Data Rate CBE, kb/s | 12000 |

| Project - Cost and Schedule | |
|---|---|
| Cost Target | < $2B TBD |
| Mission Cost Category | Flagship - e.g. Cassini |
| FY$ (year) | 2015 |
| Include Phase A cost estimate? | Yes |
| Phase A Start | July 2024 |
| Phase A Duration (months) | 20 |
| Phase B Duration (months) | 16 |
| Phase C/D Duration (months) | 47 |
| Review Dates | PDR - July 2027, CDR - September 2028, ARR - September 2029 |
| Phase E Duration (months) | 179 |
| Phase F Duration (months) | 4 |
| Spares Approach | Typical |
| Parts Class | Commercial + Military 883B |
| Launch Site | Cape Canaveral |

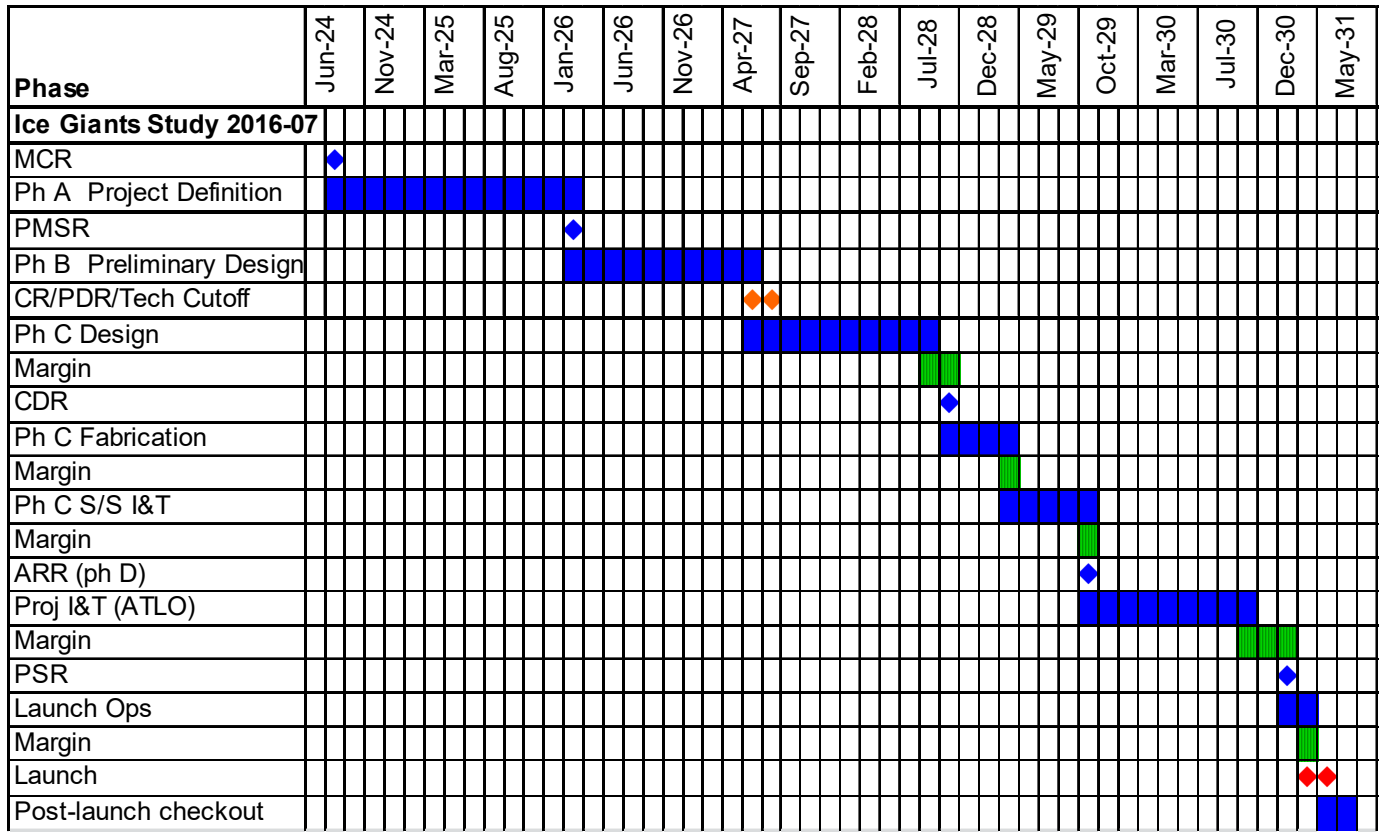

Proposed development schedule consistent with typical New Frontiers missions and current Europa mission schedule.
Phase A 20 mos., Phase B 16 mos., Phase C/D 47 mos.
Launch May 25, 2031

✖ **Selected instruments come with some extras.**

- NAC has a 2-DOF gimbal.
- Doppler Imager has an internal fast steering mirror (FSM).

| Instrument Name | # units | Heritage | CBE Mass (kg) | Cont. | CBE+Cont. Mass/Unit (kg) | Op. Power CBE per Instrument (W) | Standy Power CBE per Instrument (W) |
|---|---|---|---|---|---|---|---|
| | | | 37 kg | 23% | 45.2 | | |
| Narrow Angle Camera (EIS Europa) | 1 | Inherited design | 12.0 | 15% | 13.8 | 16 W | 2 W |
| Doppler Imager (ECHOES JUICE) | 1 | New design | 20.0 | 30% | 26 | 20 W | 2 W |
| Magnetometer (Gallileo) | 1 | Inherited design | 4.7 | 15% | 5.405 | 8 W | 1 W |

| Instrument Name | Instrument Peak Data Rate | Units |
|---|---|---|
| Narrow Angle Camera (EIS Europa) | 12000 | kbps |
| Doppler Imager (ECHOES JUICE) | 60 | kbps |
| Magnetometer (Gallileo) | 1200 | kbps |

- ✖ **Instruments**
  - Gas Chromatograph Mass Spectrometer (GCMS)
  - Atmospheric Structure Instrument (ASI)
  - Nephelometer
  - Ortho-para Hydrogen Measurement Instrument
- ✖ **CDS**
  - Redundant Sphinx Avionics
- ✖ **Baseline Power System**
  - Primary batteries
    - ◆ In probe:
      - 17.1kg, 1.0 kW-hr EOM
  - Redundant Power Electronics
- ✖ **Thermal**
  - RHU heating, passive cooling
  - Vented probe design
  - Thermally isolating struts

- ✖ **Telecom**
  - Redundant IRIS radio
  - UHF SSPA
  - UHF Low Gain Antenna (similar to MSL)
- ✖ **Structures**
  - ~50kg Heatshield
    - ◆ 1.2m diameter, 45deg sphere cone
  - ~15kg Backshell
  - ~10kg Parachutes
  - ~15kg Probe Aerofairing

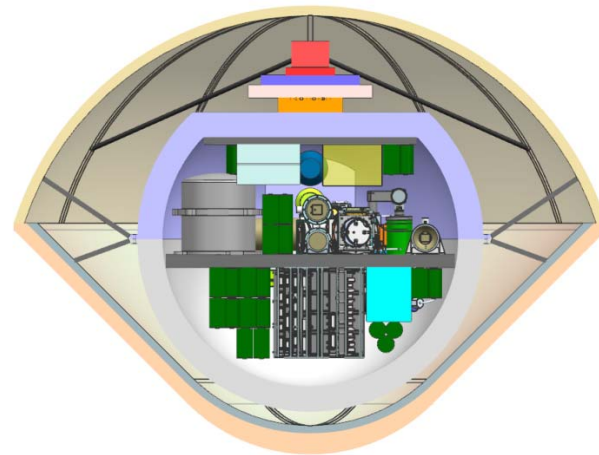

## Instruments
- Narrow Angle Camera
- Doppler Imager
- Magnetometer

## CDS
- JPL reference bus avionics
- Dual string cold redundancy

## Baseline Power System
- 4 eMMRTGs, 45kg each
- 10 A-hr, <5kg, Li Ion Battery

## Telecom
- Radios
  - Two X/X/Ka SDST transponders
  - Two IRIS radio UHF receivers
  - Two 35W Ka-Band TWTAs
  - Two 25W X-Band TWTAs
- Antennas
  - One 3m X/Ka HGA
  - One X-Band MGA
  - Two X-Band LGAs
  - One UHF patch array – 15 dBic gain

## Thermal
- Active and passive thermal control design
- Louvers, heaters, MLI

## ACS
- Four 0.1N Honeywell HR16 reaction wheels
- IMUs, Star Trackers, Sun Sensors

## Propulsion
- Dual-mode bipropellant system provides 2607m/s of delta-V
- Two 200lbf Aerojet main engines
- Four 22N engines
- Eight 1N RCS engines

## Structures
- 315kg structure
- 40kg ballast
- 86kg harness
- 10m Magnetometer Boom
- Main Engine and eMMRTG covers
- Probe separation mechanisms

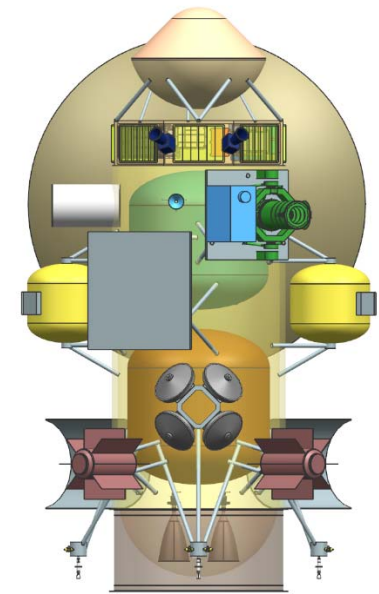

## Probe + Entry System

## Probe

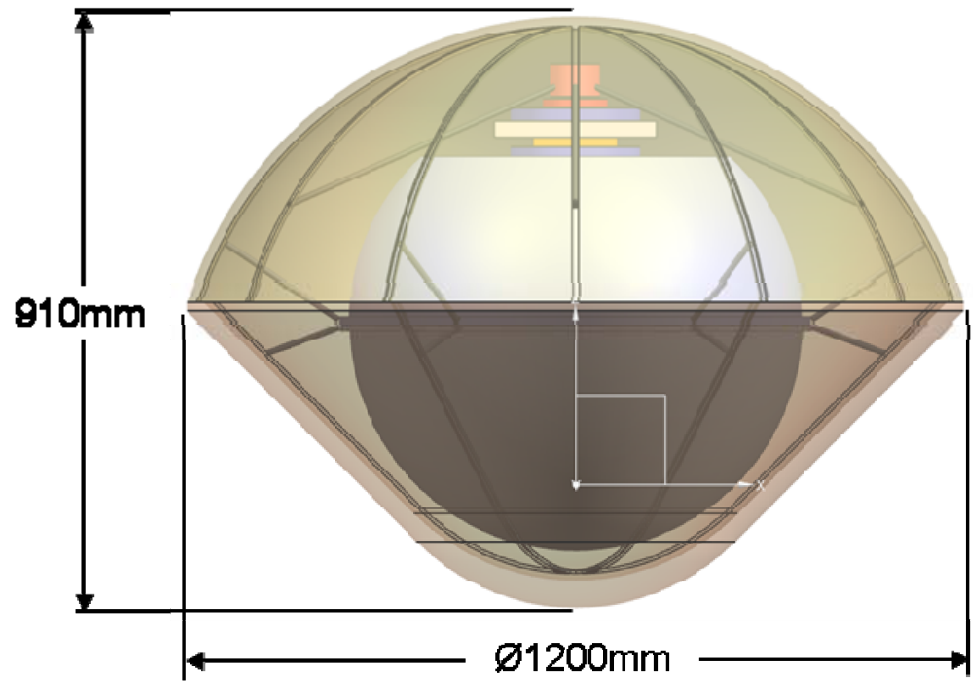

910mm

Ø1200mm

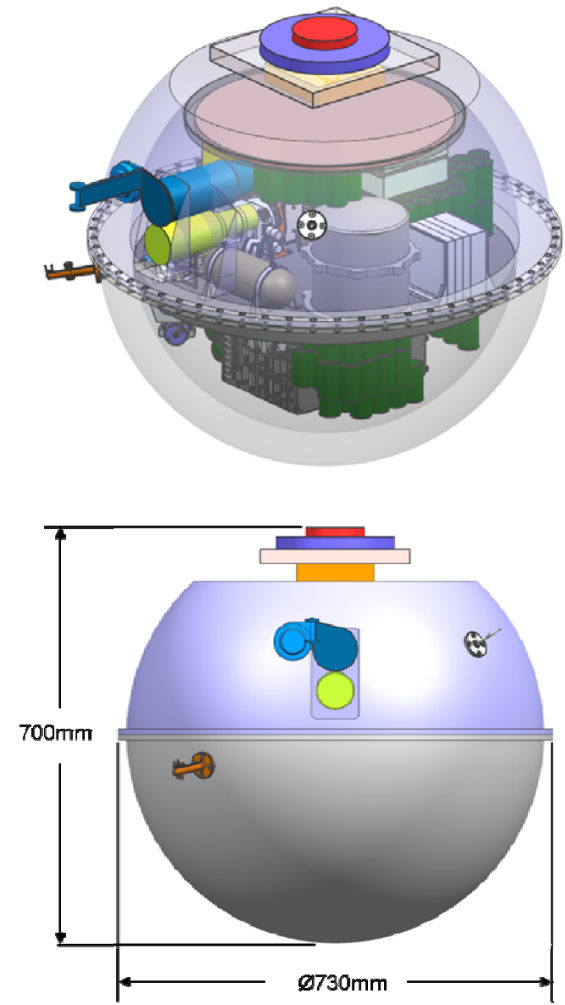

700mm

Ø730mm

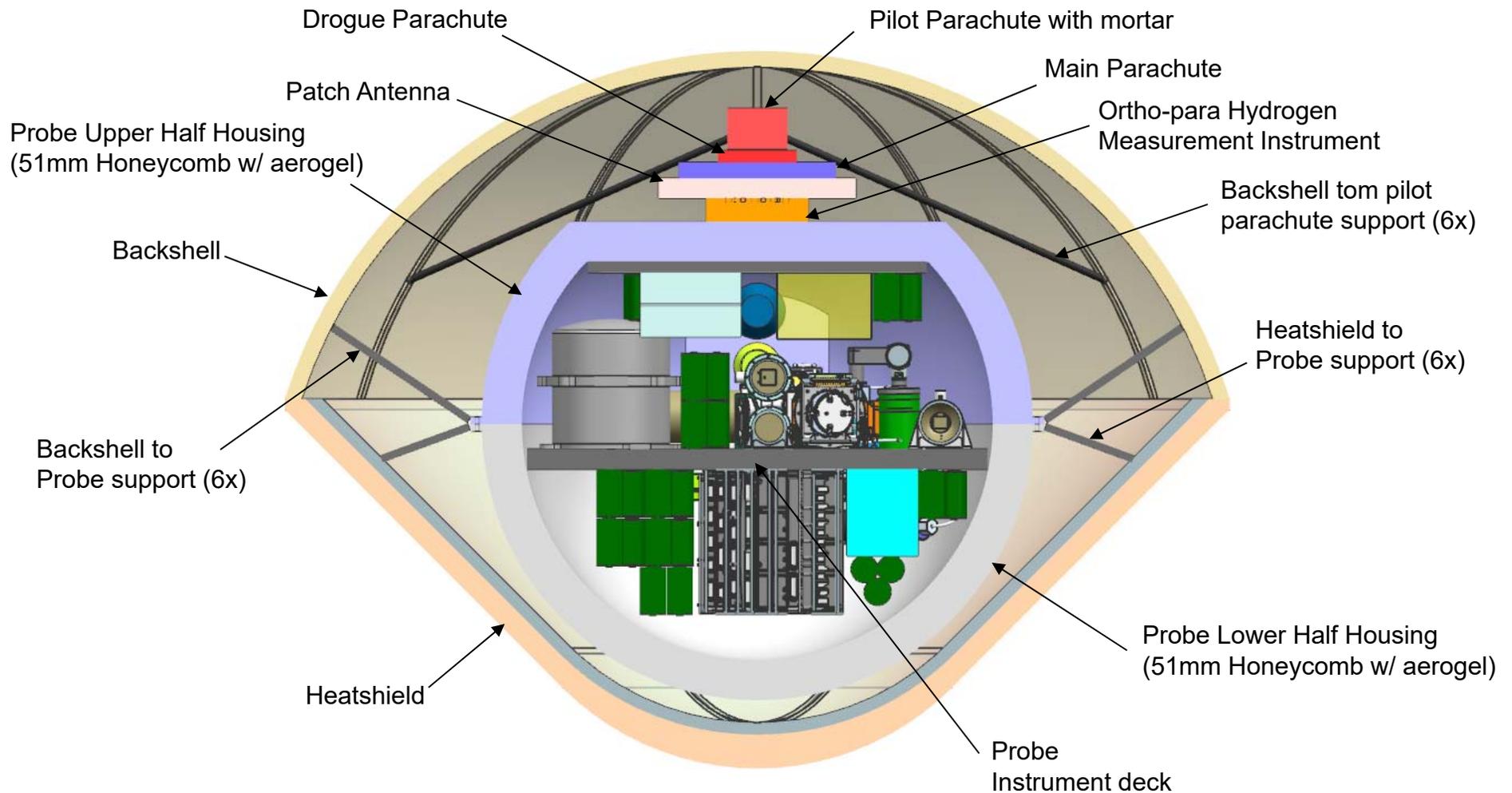

Drogue Parachute

Pilot Parachute with mortar

Main Parachute

Patch Antenna

Ortho-para Hydrogen Measurement Instrument

Probe Upper Half Housing (51mm Honeycomb w/ aerogel)

Backshell tom pilot parachute support (6x)

Backshell

Heatshield to Probe support (6x)

Backshell to Probe support (6x)

Heatshield

Probe Lower Half Housing (51mm Honeycomb w/ aerogel)

Probe Instrument deck

## ✖ Configuration Drawings – Stowed

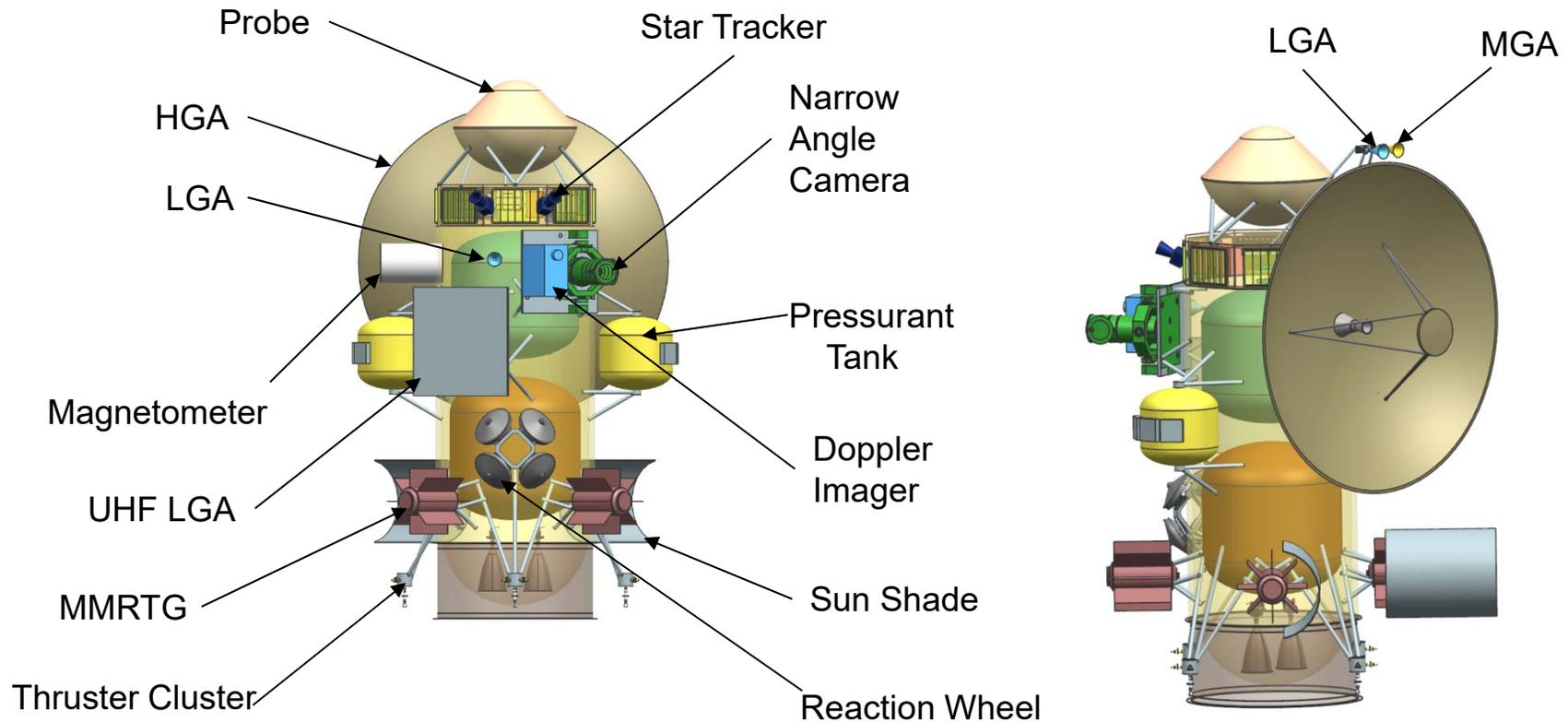

Probe

Star Tracker

HGA

Narrow Angle Camera

LGA

Pressurant Tank

Magnetometer

Doppler Imager

UHF LGA

MMRTG

Sun Shade

Thruster Cluster

Reaction Wheel

LGA

MGA

✖ **Configuration Drawings – Deployed**

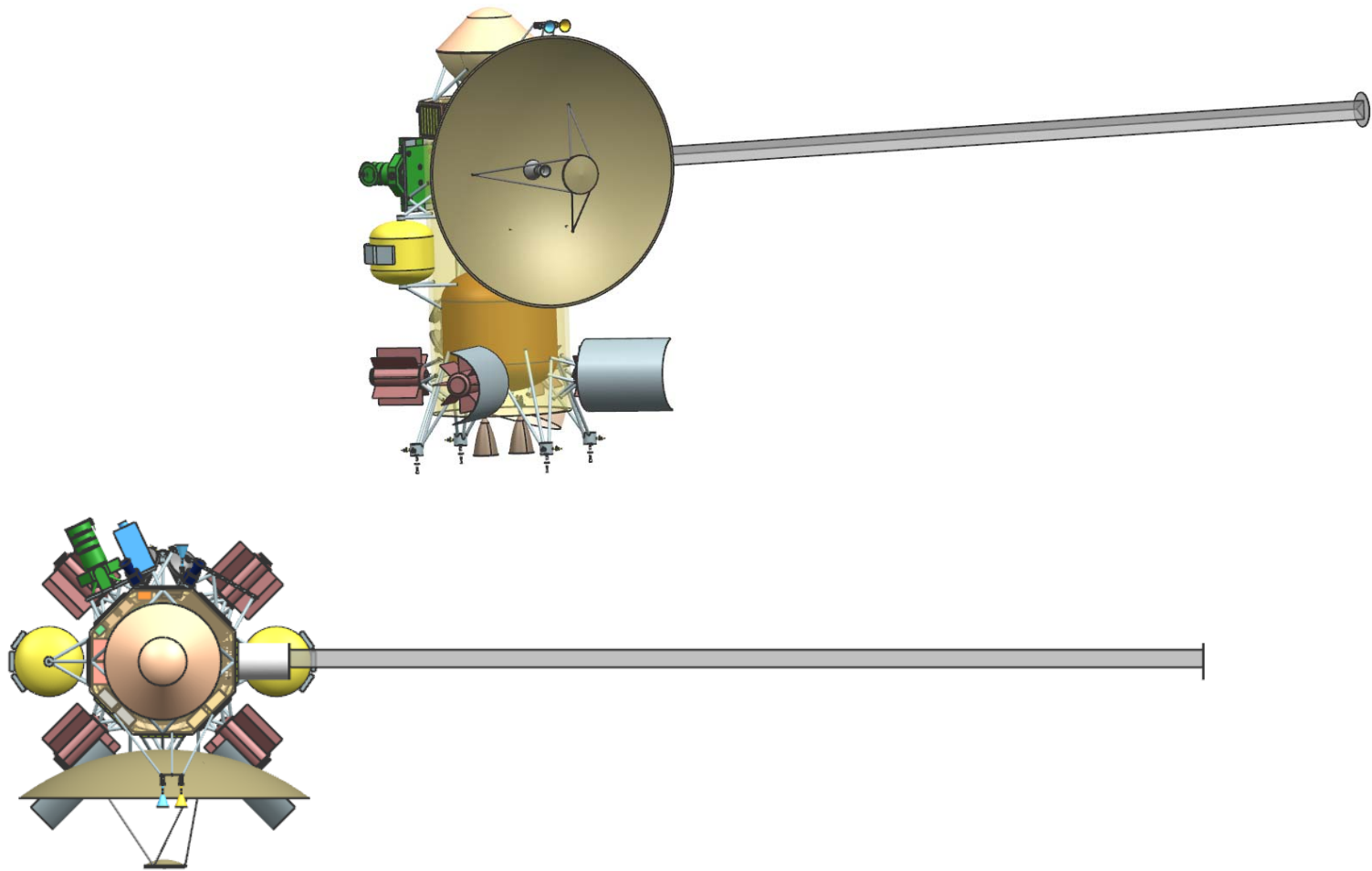

| | Mass Fraction | Mass (kg) | Subsys Cont. % | CBE+ Cont. (kg) | Mode 1 Power (W) Coast – 60 Days | Mode 2 Power (W) Warmup | Mode 3 Power (W) Science |
|---|---|---|---|---|---|---|---|
| *Power Mode Duration  (hours)* | | | | | *24* | *0.5* | *1* |
| **Payload on this Element** | | | | | | | |
| Instruments | 21% | 25.3 | 29% | 32.5 | 0 | 91 | 74 |
| Payload Total | 21% | 25.3 | 29% | 32.5 | 0 | 91 | 74 |
| **Spacecraft Bus** | | | | do not edit formulas below this line, use the calcualtions and overri | | | |
| Command & Data | 0% | 0.6 | 17% | 0.7 | 0 | 8 | 8 |
| Power | 17% | 20.1 | 26% | 25.4 | 0 | 11 | 11 |
| Structures & Mechanisms | 41% | 49.8 | 30% | 64.7 | 0 | 0 | 0 |
| Cabling | 9% | 11.5 | 30% | 15.0 | | | |
| Telecom | 5% | 6.2 | 26% | 7.8 | 0 | 0 | 184 |
| Thermal | 6% | 7.8 | 3% | 8.1 | 0 | 0 | 0 |
| Bus Total | | 96.0 | 27% | 121.7 | 0 | 19 | 203 |
| Thermally Controlled Mass | | | | 121.7 | | | |
| **Spacecraft Total (Dry): CBE & MEV** | | 121.3 | 27% | 154.2 | 0 | 109 | 277 |
| Subsystem Heritage Contingency | 27% | 32.9 | SEP Cont | 10% | 0 | 0 | 0 |
| System Contingency | 16% | 19.3 | | | 0 | 47 | 119 |
| Total Contingency     ☐ Include Carried? | 43% | 52.2 | | | | | |
| **Spacecraft with Contingency:** | | 173 | of total | w/o addl pld | 0 | 157 | 396 |

| | | Mass Fraction | Mass (kg) | Subsys Cont. % | CBE+ Cont. (kg) |
|---|---|---|---|---|---|
| *Power Mode Duration (hours)* | | | | | |
| **Additional Elements Carried by this Element** | | | | | |
| Probe | | 54% | 121.3 | 43% | 173.5 |
| **Carried Elements Total** | | **54%** | **121.3** | **43%** | **173.5** |
| **Spacecraft Bus** | | | | do not edit formulas below | |
| Structures & Mechanisms | | 45% | 101.9 | 30% | 132.4 |
| Cabling | | 0% | 0.9 | 30% | 1.2 |
| **Bus Total** | | | 102.8 | 30% | 133.6 |
| Thermally Controlled Mass | | | | | 133.6 |
| **Spacecraft Total (Dry): CBE & MEV** | | | **224.1** | 37% | **307.1** |
| Subsystem Heritage Contingency | 37% | | 83.0 | SEP Cont | 10% |
| System Contingency | 6% | | 13.4 | | |
| Total Contingency ☐ Include Carried? | **43%** | | 96.4 | | |
| **Spacecraft with Contingency:** | | | **320** | of total | w/o addl pld |

| | Mass Fraction | Mass (kg) | Subsys Cont. % | CBE+ Cont. (kg) | Mode 1 Power (W) Recharge | Mode 2 Power (W) Approach Science - 85 Days | Mode 3 Power (W) Telecom Downlink | Mode 4 Power (W) UCI Delta-V | Mode 5 Power (W) Orbital Science - Apoapse | Mode 6 Power (W) Orbital Science - Periapse | Mode 7 Power (W) Orbital Science - Moon Tour | Mode 8 Power (W) Safe | Mode 9 Power (W) IGNORE THIS MODE | Mode 10 Power (W) Probe Relay |
|---|---|---|---|---|---|---|---|---|---|---|---|---|---|---|
| **Power Mode Duration  (hours)** | | | | | **24** | **24** | **8** | **1.00** | **16** | **16** | **16** | **24** | | **1.5** |
| **Payload on this Element** | | | | | | | | | | | | | | |
| Instruments | 3% | 36.7 | 23% | 45.2 | 4 | 28 | 4 | 4 | 26 | 44 | 26 | 4 | 4 | 4 |
| **Payload Total** | 3% | 36.7 | 23% | 45.2 | 4 | 28 | 4 | 4 | 26 | 44 | 26 | 4 | 4 | 4 |
| **Additional Elements Carried by this Element** | | | | | | | | | | | | | | |
| Entry System + Probe | 16% | 224.1 | 43% | 320.5 | | | | | | | | | | |
| **Carried Elements Total** | 16% | 224.1 | 43% | 320.5 | 0 | 0 | 0 | 0 | 0 | 0 | 0 | 0 | 0 | 0 |
| **Spacecraft Bus** | | | do not edit formulas below this line, use the calculations and override tables instead --> | | | | | | | | | | | |
| Attitude Control | 4% | 63.5 | 10% | 69.8 | 0 | 55 | 55 | 88 | 55 | 55 | 55 | 42 | 55 | 93 |
| Command & Data | 1% | 21.6 | 10% | 23.8 | 57 | 57 | 57 | 57 | 57 | 57 | 57 | 57 | 57 | 57 |
| Power | 15% | 216.6 | 2% | 220.8 | 24 | 40 | 32 | 24 | 24 | 24 | 24 | 40 | 32 | 32 |
| Propulsion1 ☐ SEP1 | 12% | 171.7 | 5% | 181.0 | 31 | 3 | 3 | 151 | 3 | 3 | 3 | 3 | 3 | 3 |
| Structures & Mechanisms | 30% | 433.4 | 30% | 563.4 | 0 | 0 | 0 | 0 | 0 | 0 | 0 | 0 | 0 | 0 |
| S/C-Side Adapter | 1% | 17.7 | 0% | 17.7 | | | | | | | | | | |
| Cabling | 6% | 86.3 | 30% | 112.2 | | | | | | | | | | |
| Telecom | 4% | 59.4 | 16% | 68.9 | 12 | 65 | 92 | 71 | 12 | 12 | 12 | 71 | 12 | 32 |
| Thermal | 8% | 113.2 | 23% | 139.5 | 25 | 25 | 25 | 25 | 25 | 25 | 25 | 25 | 25 | 25 |
| **Bus Total** | | 1183.3 | 18% | 1397.1 | 149 | 245 | 264 | 417 | 176 | 176 | 176 | 238 | 184 | 242 |
| Thermally Controlled Mass | | | | 1397.1 | | | | | | | | | | |
| **Spacecraft Total (Dry): CBE & MEV** | | 1444.2 | 22% | 1762.8 | 154 | 272 | 268 | 421 | 202 | 220 | 202 | 242 | 188 | 246 |
| Subsystem Heritage Contingency | 22% | 318.6 | SEP Cont | 0 | 0 | 0 | 0 | 0 | 0 | 0 | 0 | 0 | 0 | |
| System Contingency | 16% | 224.9 | 10% | | 66 | 117 | 115 | 181 | 87 | 95 | 87 | 104 | 81 | 106 |
| Total Contingency | 38% | ☐ Include Carried? | 543.6 | | | | | | | | | | | |
| **Spacecraft with Contingency:** | | 1988 | of total | wo add pld | 220 | 390 | 383 | 602 | 289 | 315 | 289 | 347 | 269 | 352 |
| Propellant & Pressurant with residuals1 | 54% | 2357.1 | For S/C mass = | 1988.3 | Delta-V, Sys 1 | 2607.0 | m/s | residuals1 | 61.7 | kg | | | | |
| **Spacecraft Total with Contingency (Wet)** | | 4344.8 | | | | | | | | | | | | |
| L/V-Side Adapter | 0.19806094 | 0.0 | Wet Mass for Prop Sizing | 4450 | BCL Power: | 0.0 | W | | | | | | | |
| **Launch Mass** | 19.8% | 4345 | Dry Mass for Prop Sizing | 1988 | ECL Power: | 0.0 | W | | | | | | | |
| **Allocation** | | 4450 | Atlas V 541 | | | | | | | | | | | |
| | 257.7 | | | Launch C3 | 11.9 | | | | | | | | | |
| **Launch Vehicle Margin** | | 105.2 | Mission Unique LV Contingency | 0% | | | | | | | | | | |

| Element Number | Element Name | Dry CBE (kg) | Cont / JPL Margin (kg) | Dry Allocation (kg) | Propellant (kg) | Dry Allocation + Propellant (kg) |
|---|---|---|---|---|---|---|
| 1 | Probe | 121 | 52 | 173 | - | 173 |
| 2 | Entry System | 103 | 44 | 147 | - | 147 |
| 3 | Orbiter minus eMMRTGs | 1040 | 447 | 1487 | 2357 | 3845 |
| 3.1 | eMMRTGs | 180 | - | 180 | - | 180 |
| | **Total Stack** | **1444** | **543** | **1987** | **2357** | **4345** |

| | |
|---|---|
| Dry Mass Allocation | 1987 |
| JPL Margin (kg / %) | 543 / 27% |
| JPL Margin without eMMRTG (kg / %) | 543 / 30% |
| Atlas V 541 Capacity (kg) (C3 = 11.9 km$^2$/s$^2$) | 4450 |
| Extra Launch Vehicle Margin (kg) | 105 |

| Element Number | Element Name | Dry CBE (kg) | Cont (%) | Cont. (kg) | MEV (kg) | Dry Allocation (kg) | Propellant (kg) | Dry Allocation + Propellant (kg) |
|---|---|---|---|---|---|---|---|---|
| 1 | Probe | 121 | 27% | 33 | 154 | 173 | - | 173 |
| 2 | Entry System | 103 | 30% | 31 | 134 | 147 | - | 147 |
| 3 | Orbiter minus eMMRTGs | 1040 | 21% | 222 | 1262 | 1487 | 2357 | 3845 |
| 3.1 | eMMRTGs | 180 | 0% | 0 | 180 | 180 | - | 180 |
| | **Total Stack** | **1444** | **20%** | **286** | **1730** | **1987** | **2357** | **4345** |

| | |
|---|---|
| Dry Mass Allocation | 1987 |
| NASA Margin (kg / %) | 257 / 15% |
| NASA Margin without eMMRTG (kg / %) | 257 / 17% |
| Atlas V 541 Capacity (kg) (C3 = 11.9 km²/s²) | 4450 |
| Extra Launch Vehicle Margin (kg) | 105 |

- **Probe needs to enter Uranus atmosphere head-on versus shallow**
  - Probe relay antenna is nominally aligned with zenith.
  - Need Orbiter within tens of degrees of zenith to close the link.
  - Head-on: Orbiter is close enough to zenith for one hour.
    - Probe deceleration ~165 g's
  - Shallow: Orbiter is far from zenith –more difficult to close the link.
    - On the other hand, Probe deceleration relatively low.
  - Verified that Probe can operate through the higher deceleration.

- **Ka-band transmitter power versus array of 34m ground stations**
  - Downlink 15 kbps using one 35W TWTA to a 34m BWG ground station.
    - Telecom system uses most of one eMMRTG power during downlink.
  - Could increase downlink rate using more power, adding an eMMRTG, or by using an array of two or more 34m ground stations.
  - For this option, make do with 15 kbps downlink rate.

✖ **Data downlink strategy for Doppler Imager (DI) on approach**

- DI generates a lot of data continuously for tens of days on approach.

- Configuration with HGA and DI on opposite sides of the cylindrical bus allows pointing DI towards Uranus while pointing HGA towards Earth.

- Can downlink for ~20 hours/day and maintain positive power balance.
    - Using only 4 eMMRTGs, as opposed to 5

- Data that can't be downlinked before UOI will be downlinked after.

✖ **Configuration that helps to minimize mass and power**

- eMMRTGs outside the cylindrical bus provide heating
    - Reduces mass and power of thermal subsystem components

- Propellant and oxidizer tanks inside bus, pressurant tanks outside
    - Pressurant tanks are easier to keep warm than propellant/oxidizer tanks.

- Shorten the stack to minimize structure mass
    - Single custom propellant tank instead of two tanks

- **Mechanical**
  - LV interfaces to the Orbiter
  - Entry System containing Probe is attached to the side of the Orbiter.
  - Orbiter: primary structure is the largest mass element.
    - 275 kg CBE out of 433 kg total for Mechanical (1183 kg bus dry mass)
    - Drivers are the large Propulsion and Power masses.
- **Baseline Power System**
  - Dual String Reference Bus electronics heritage
  - eMMRTG mass of 45kg/unit, with cooling tubes, is a not to exceed value, so 0% mass uncertainty is applied.
  - 217 kg CBE subsystem dry mass
- **Propulsion: Dual-mode bi-prop; 2.6 km/s total delta V**
  - Orbit insertion delta V = 1.7km/s desired in < 1hr
    - Probe release before UOI
    - Two 890N main engines used to achieve burn time < 1hr
  - 172 kg CBE dry mass; 2357 kg propellant and pressurant

- **Thermal: Cassini-heritage waste-heat recovery system on Orbiter**
  - RTG end domes each provide 75 W waste heat to propulsion module via conductive and radiative coupling.
  - VRHUs act as primary control mechanism for thruster clusters.
    - Also act as trim heaters for the propulsion module
  - Louvers act as primary control mechanism for avionics module.

- **Telecom: X- and Ka-Band subsystem, plus UHF for Probe data.**
  - Two 35W Ka-Band TWTAs, two 25W X-Band TWTAs
  - Two X/X/Ka SDST transponders, two IRIS radio UHF receivers
  - 3m X/Ka HGA, one X-Band MGA, two X-Band LGAs, UHF patch array.
  - Supports a data rate at Uranus of 15 kbps into 34m BWG ground station.
  - Supports uplink of 3Mbits of probe data
  - 

✖ **CDS: Reference Bus architecture ideally suited for high reliability, long lifetime mission.**

- Standard JPL spacecraft CDS that is similar to SMAP
    - ◆ RAD750 CPU, NVM, MTIF, MSIA, CRC, LEU-A, LEU-D, MREU
    - ◆ 128 GBytes storage for science data
    - ◆ 1553 and RS-422 ICC/ITC interfaces for subsystems and instruments

✖ **ACS: 3-axis stabilized with star tracker, sun sensor, gyros, wheels.**

- All stellar attitude determination to minimize power, conserve gyros.
- Sun sensor performance may degrade once the Orbiter passes Saturn.
    - ◆ May impact safe mode used during star tracker outage.
    - ◆ Detailed analysis on Sun sensor performance versus distance is needed.

- **Software: core product line is appropriate since this mission has aspects similar to MSL/M2020/SMAP/Europa.**
  - Complexity rankings range from Medium to High.
    - Medium infrastructure: dual string with warm spare.
    - High fault behaviors: high redundancy, string swapping, critical events.
    - Medium/High ACS: tight pointing requirements, many ACS modes
    - Medium Telecom: dual active UHF, redundant DTE
    - Medium Science data processing, full file system

- **SVIT: Probe testbed, system I&T and V&V costs are included**
  - Cost of assembling and testing RTG's is captured elsewhere
    - Cost of integrating RTG's is included with other ATLO costs
  - Probe with 5 Instruments costed separately; testbed costs included.

- **Ground Systems**
  - Mission specific implementation of standard JPL mission operations and ground data systems
  - Ground network: DSN 34-m BWG; 70-m or equivalent for safe mode
  - Science support: 24x7 tracking on approach; daily contacts on orbit

| COST SUMMARY (FY2015 $M) | Generate ProPricer Input | Team X Estimate | | |
|---|---|---|---|---|
| | | CBE | Res. | PBE |
| Project Cost | | $1402.9 M | 21% | $1700.1 M |
| Launch Vehicle | | $33.0 M | 0% | $33.0 M |
| Project Cost (w/o LV) | | $1369.9 M | 22% | $1667.1 M |
| Development Cost | | $1109.3 M | 24% | $1372.6 M |
| Phase A | | $11.1 M | 24% | $13.7 M |
| Phase B | | $99.8 M | 24% | $123.5 M |
| Phase C/D | | $998.4 M | 24% | $1235.3 M |
| Operations Cost | | $260.5 M | 13% | $294.5 M |

Total mission cost is $1.70B. This is the likely cost within a range that typically can be as much as 10% lower up to 20% higher. The development cost with reserves is $1.37B.

| WBS Elements | NRE | RE | 1st Unit |
|---|---|---|---|
| Project Cost (no Launch Vehicle) | $1243.8 M | $456.3 M | $1700.1 M |
| Development Cost (Phases A - D) | $916.4 M | $456.2 M | $1372.6 M |
| 01.0 Project Management | $47.3 M | | $47.3 M |
| 1.01 Project Management | $11.4 M | | $11.4 M |
| 1.02 Business Management | $13.6 M | | $13.6 M |
| 1.04 Project Reviews | $2.5 M | | $2.5 M |
| 1.06 Launch Approval | $19.8 M | | $19.8 M |
| 02.0 Project Systems Engineering | $23.7 M | $0.5 M | $24.2 M |
| 2.01 Project Systems Engineering | $8.9 M | | $8.9 M |
| 2.02 Project SW Systems Engineering | $5.2 M | | $5.2 M |
| 2.03 EEIS | $1.5 M | | $1.5 M |
| 2.04 Information System Management | $1.7 M | | $1.7 M |
| 2.05 Configuration Management | $1.5 M | | $1.5 M |
| 2.06 Planetary Protection | $0.2 M | $0.2 M | $0.4 M |
| 2.07 Contamination Control | $1.2 M | $0.3 M | $1.5 M |
| 2.09 Launch System Engineering | $1.0 M | | $1.0 M |
| 2.10 Project V&V | $2.0 M | | $2.0 M |
| 2.11 Risk Management | $0.5 M | | $0.5 M |
| 03.0 Mission Assurance | $46.0 M | $0.0 M | $46.0 M |
| 04.0 Science | $24.8 M | | $24.8 M |
| Orbiter Science | $14.0 M | | $14.0 M |
| Probe Science | $10.8 M | | $10.8 M |
| 05.0 Payload System | $80.2 M | $48.3 M | $128.5 M |
| 5.01 Payload Management | $7.8 M | | $7.8 M |
| 5.02 Payload Engineering | $5.8 M | | $5.8 M |
| Orbiter Instruments | $33.5 M | $24.3 M | $57.8 M |
| Narrow Angle Camera (EIS Europa) | $11.6 M | $8.4 M | $20.0 M |
| Doppler Imager (ECHOES JUICE) | $17.4 M | $12.6 M | $30.0 M |
| Magnetometer (Galileo) | $4.5 M | $3.3 M | $7.8 M |
| Probe Instruments | $33.1 M | $24.0 M | $57.1 M |
| Mass Spectrometer | $22.9 M | $16.6 M | $39.6 M |
| Atmospheric Structure Investigation (ASI) | $3.4 M | $2.5 M | $5.9 M |
| Nephelometer (Galileo) | $5.3 M | $3.8 M | $9.1 M |
| Ortho-para H2 meas. Expt. | $1.5 M | $1.1 M | $2.6 M |

| WBS Elements | NRE | RE | 1st Unit |
|---|---|---|---|
| 06.0 Flight System | $445.2 M | $280.5 M | $725.7 M |
| 6.01 Flight System Management | $5.0 M | | $5.0 M |
| 6.02 Flight System Systems Engineering | $51.1 M | | $51.1 M |
| 6.03 Product Assurance (included in 3.0) | | | $0.0 M |
| Orbiter | $297.0 M | $236.5 M | $533.5 M |
| 6.04 Power | $94.5 M | $133.1 M | $227.6 M |
| 6.05 C&DH | $31.3 M | $24.3 M | $55.6 M |
| 6.06 Telecom | $28.4 M | $18.1 M | $46.5 M |
| 6.07 Structures (includes Mech. I&T) | $51.6 M | $16.7 M | $68.3 M |
| 6.08 Thermal | $4.2 M | $13.1 M | $17.2 M |
| additional cost for >43 RHUs | $34.0 M | $0.0 M | $34.0 M |
| 6.09 Propulsion | $22.1 M | $16.7 M | $38.8 M |
| 6.10 ACS | $9.4 M | $9.8 M | $19.1 M |
| 6.11 Harness | $4.1 M | $3.8 M | $7.9 M |
| 6.12 S/C Software | $17.1 M | $0.9 M | $18.0 M |
| 6.13 Materials and Processes | $0.4 M | $0.0 M | $0.4 M |
| Probe | $26.8 M | $18.1 M | $44.9 M |
| 6.04 Power | $2.9 M | $2.1 M | $5.0 M |
| 6.05 C&DH | $0.3 M | $2.3 M | $2.7 M |
| 6.06 Telecom | $7.9 M | $4.1 M | $12.0 M |
| 6.07 Structures (includes Mech. I&T) | $8.0 M | $3.5 M | $11.5 M |
| 6.08 Thermal | $2.3 M | $4.9 M | $7.2 M |
| 6.11 Harness | $1.8 M | $0.9 M | $2.7 M |
| 6.12 S/C Software | $3.3 M | $0.2 M | $3.5 M |
| 6.13 Materials and Processes | $0.4 M | $0.0 M | $0.4 M |
| Entry System | $57.1 M | $24.4 M | $81.5 M |
| 6.07 Structures (includes Mech. I&T) | $55.3 M | $24.1 M | $79.4 M |
| 6.11 Harness | $1.4 M | $0.3 M | $1.7 M |
| 6.13 Materials and Processes | $0.4 M | $0.0 M | $0.4 M |
| Ames/Langley EDL Engineering/Testing | $3.8 M | $0.0 M | $3.8 M |
| 6.14 Spacecraft Testbeds | $4.5 M | $1.5 M | $6.0 M |

| WBS Elements | NRE | RE | 1st Unit |
|---|---|---|---|
| **07.0 Mission Operations Preparation** | **$26.1 M** | | **$26.1 M** |
| 7.0 MOS Teams | $20.0 M | | $20.0 M |
| 7.03 DSN Tracking (Launch Ops.) | $2.7 M | | $2.7 M |
| 7.06 Navigation Operations Team | $3.4 M | | $3.4 M |
| 7.07.03 Mission Planning Team | $0.0 M | | $0.0 M |
| **09.0 Ground Data Systems** | **$22.0 M** | | **$22.0 M** |
| 9.0A Ground Data System | $19.8 M | | $19.8 M |
| 9.0B Science Data System Development | $1.3 M | | $1.3 M |
| 9A.03.07 Navigation H/W & S/W Development | $0.9 M | | $0.9 M |
| **10.0 ATLO** | **$21.1 M** | **$21.7 M** | **$42.8 M** |
| Orbiter | $15.3 M | $13.3 M | $28.6 M |
| Probe | $5.9 M | $8.4 M | $14.2 M |
| **11.0 Education and Public Outreach** | **$0.0 M** | **$0.0 M** | **$0.0 M** |
| **12.0 Mission and Navigation Design** | **$21.9 M** | | **$21.9 M** |
| 12.01 Mission Design | $2.0 M | | $2.0 M |
| 12.02 Mission Analysis | $6.3 M | | $6.3 M |
| 12.03 Mission Engineering | $1.8 M | | $1.8 M |
| 12.04 Navigation Design | $11.8 M | | $11.8 M |
| **Development Reserves** | **$158.0 M** | **$105.3 M** | **$263.3 M** |

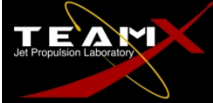

# Executive Summary
## Cost E-F and Launch Nuclear Safety

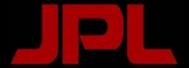

| WBS Elements | NRE | RE | 1st Unit |
|---|---|---|---|
| **Operations Cost (Phases E - F)** | **$294.4 M** | **$0.1 M** | **$294.5 M** |
| **01.0 Project Management** | **$27.1 M** | | **$27.1 M** |
| 1.01 Project Management | $15.3 M | | $15.3 M |
| 1.02 Business Management | $10.7 M | | $10.7 M |
| 1.04 Project Reviews | $1.1 M | | $1.1 M |
| 1.06 Launch Approval | $0.1 M | | $0.1 M |
| **02.0 Project Systems Engineering** | **$0.0 M** | **$0.1 M** | **$0.1 M** |
| **03.0 Mission Assurance** | **$3.6 M** | **$0.0 M** | **$3.6 M** |
| **04.0 Science** | **$69.2 M** | | **$69.2 M** |
| Orbiter Science | $53.4 M | | $53.4 M |
| Probe Science | $15.8 M | | $15.8 M |
| **07.0 Mission Operations** | **$130.6 M** | | **$130.6 M** |
| 7.0 MOS Teams | $80.2 M | | $80.2 M |
| 7.03 DSN Tracking | $34.2 M | | $34.2 M |
| 7.06 Navigation Operations Team | $15.5 M | | $15.5 M |
| 7.07.03 Mission Planning Team | $0.8 M | | $0.8 M |
| **09.0 Ground Data Systems** | **$30.0 M** | | **$30.0 M** |
| 9.0A GDS Teams | $24.2 M | | $24.2 M |
| 9.0B Science Data System Ops | $5.2 M | | $5.2 M |
| 9A.03.07 Navigation HW and SW Dev | $0.6 M | | $0.6 M |
| **11.0 Education and Public Outreach** | **$0.0 M** | **$0.0 M** | **$0.0 M** |
| **12.0 Mission and Navigation Design** | **$0.0 M** | | **$0.0 M** |
| **Operations Reserves** | **$33.9 M** | **$0.0 M** | **$34.0 M** |
| **8.0 Launch Vehicle** | **$33.0 M** | | **$33.0 M** |
| **Launch Vehicle and Processing** | **$0.0 M** | | **$0.0 M** |
| **Nuclear Payload Support** | **$33.0 M** | | **$33.0 M** |

# ✖ **Risks related to the Probe**

- Only 2 hours between Probe entry and UOI, a critical event
    - ◆ May be operationally challenging to sequence both the Probe relay and UOI for the Orbiter within this time window.
    - ◆ Longer than 2 hours makes the geometry more challenging for telecom.
- May be issues for the relay link margin due to Probe-Orbiter geometry and uncertainties regarding Uranus atmosphere/ potential signal attenuation.
- Last Probe targeting occurs more than 60 days prior to encounter.
    - ◆ Probe carries no propulsion, so it cannot correct trajectory dispersions.
    - ◆ Need dispersions small enough to ensure safe entry conditions at Uranus.
- Orbit knowledge requirements for science reconstruction need to be determined.
    - ◆ Will drive how the Probe is tracked pre-entry and what telemetry (e.g. IMU) needs to be transmitted with the science data to the Orbiter.
    - ◆ The latter will impact the data budget.

- **Mission duration will push systems to their operating lifetimes.**
- **Science planning risk**
  - Relative velocities between Orbiter and Uranus' satellites will be high.
    - Flybys occur near periapse
- **Collision avoidance with Uranus' rings needs to be considered.**
- **Uranus stays close to the range of solar conjunction (~4-5 deg)**
  - Doppler measurements may have increased noise levels.
- **eMMRTG still needs some development.**
  - May cause a schedule slip.
  - Performance may degrade at a higher rate than currently predicted.

- **Low altitude Venus flybys could pose potential thermal risk.**

- **RTG waste heat recovery design robustness**
  - Approach is highly configuration-dependent and may have high hidden development costs.
  - Less expensive on paper, but the actual implementation could be more expensive than an active system.

- **Component development for propulsion subsystem**
  - Large bi-prop engines for chemical

- **Sun sensor performance may degrade past Saturn.**
  - May impact safe mode used during star tracker outage.

# Option 6

- **Option 6: Uranus Orbiter Concept <u>without</u> Probe**
  - 150 kg payload allocation
  - 1 atmospheric probe (previously designed)
  - Includes VEEJ flybys

- **Class B mission**
- **Dual string redundancy**
- **eMMRTGs could be used for Orbiter power**
  - Carry <u>no mass contingency</u>, because eMMRTG masses provided are "not to exceed" values

- **Mission:**
  - Launch: 5/25/2031; Arrival: 5/17/2043
  - Launch, VEEJ flybys, cruise to Uranus

- **Mission Design**
  - 12-year cruise to UOI, 3-year science tour
  - UOI inserts into 120-day initial orbit (1.05 Ur periapse), lowered to ~50-day orbit
  - Will require optical navigation upon approach to UOI, and during science for targeting moon flybys
    - Doppler imager will be used for OpNav on approach

- **Launch Vehicle**
  - Atlas 551 (~4880 kg to C3 of 11.90 km$^2$/s$^2$)

✖ **Arrival Vinf / Declination**

- ~8.41 km/s, 48 degree (spin-axis relative)

| Event | Rel. Time | Duration | Delta V (m/s) | # Maneuvers | Comments |
|---|---|---|---|---|---|
| TCM-1 | L+10 days | | 25 | 1 | Non-deterministic |
| Venus Flyby | L+116 days | | | | 7700 km altitude |
| Earth Flyby #1 | L+383 days | | | | 3400 km altitude |
| TCMs 2-4 | TBD | | 15 | 3 | Non-deterministic |
| DSM-1 | L+787 days | | 327 | 1 | |
| Earth Flyby #2 | L+1098 days | | | | 300 km altitude |
| DSM-2 | L+1305 days | | 238 | 1 | |
| Jupiter Flyby | L+1769 days | | | | 1.3e6 km altitude |
| TCM-5 | E-80 days | | 10 | 1 | Non-deterministic |
| UOI | L+4375 days | ~1 hr | 1687 | 1 | |
| OTMs 1-5 | TBD | | 290 (Total) | 5 | Includes UOI cleanup, period+incl changes, flyby targeting, and other statistical mnvrs |
| **Total** | | | 2592 | 13+ | |

✖ **Element 1: Atmospheric Probe**
- Designed in (June 28, 30th)

  ~~No Probe for this option!~~ power generation

  Orbiter

✖ **Element 2: Entry System**
- Heatshield, backshell, structure

✖ **Element 3: Orbiter**
- Instrument allocation defined by Option
- Chemical propulsion
- eMMRTGs, no solar arrays

✖ **Element 4: SEP Cruise Stage**
- "Dumb" cruise

  ~~No SEP for this option!~~ pulsion

  ys, no RTGs

**No Probe for this option!**

**No SEP for this option!**

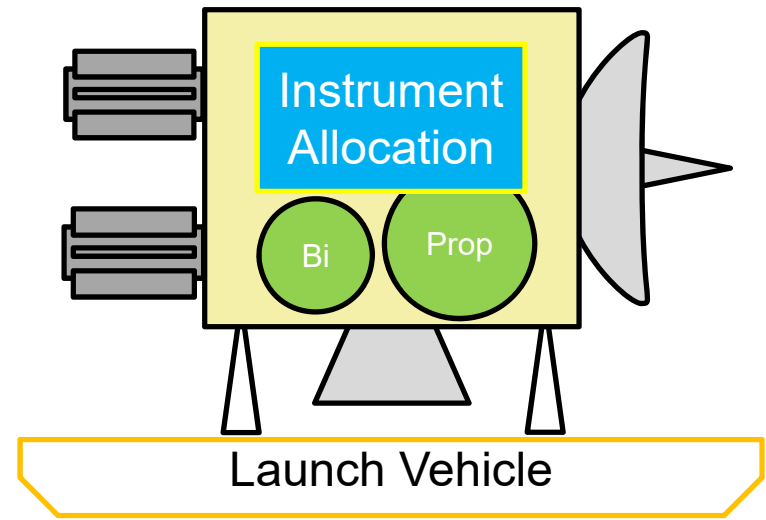

Instrument Allocation

Bi   Prop

Launch Vehicle

*Drawing not to scale

**✷ Mission Timeline**

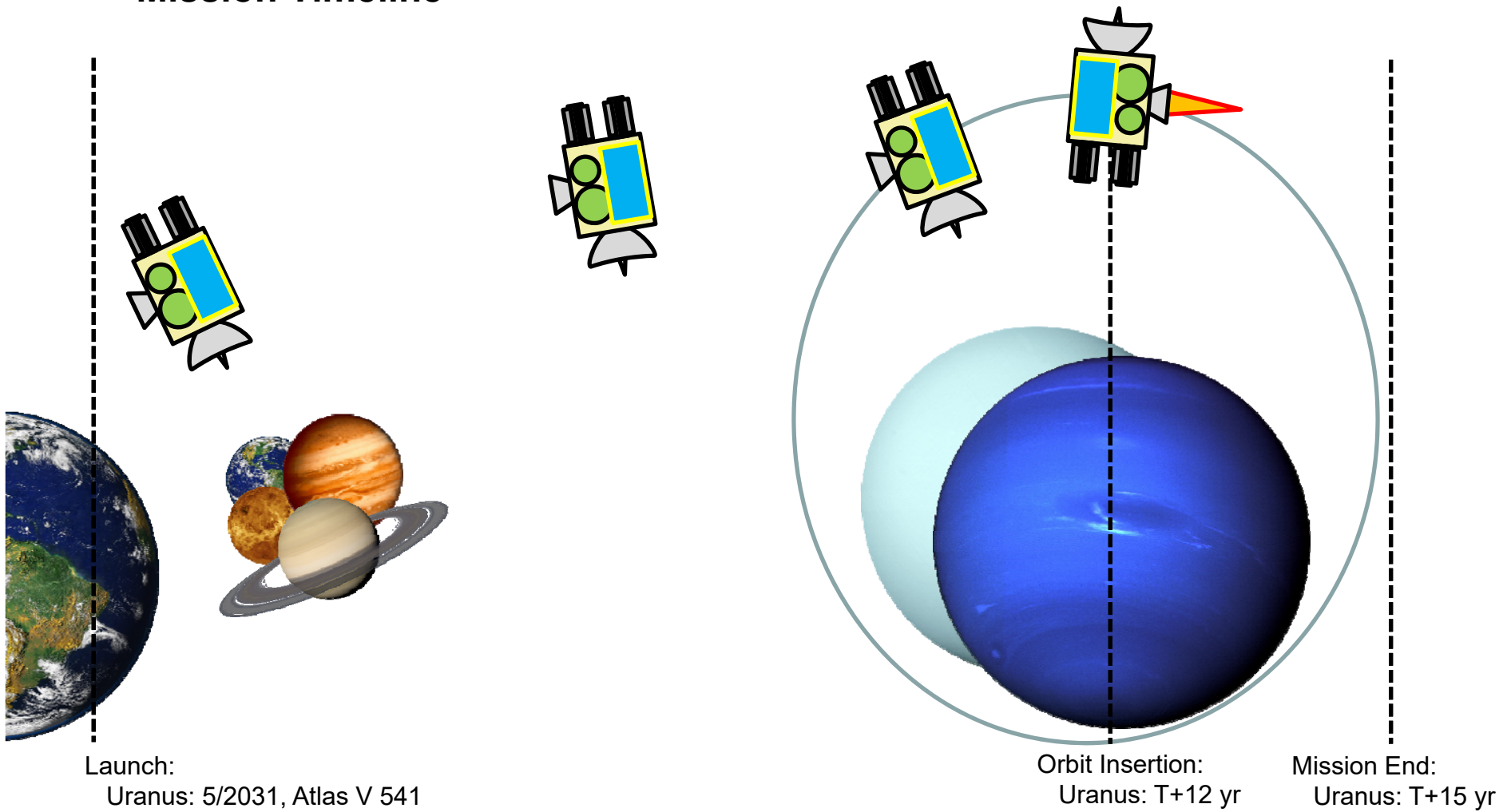

Launch:
   Uranus: 5/2031, Atlas V 541

Orbit Insertion:
   Uranus: T+12 yr

Mission End:
   Uranus: T+15 yr

**Approach Timeline**

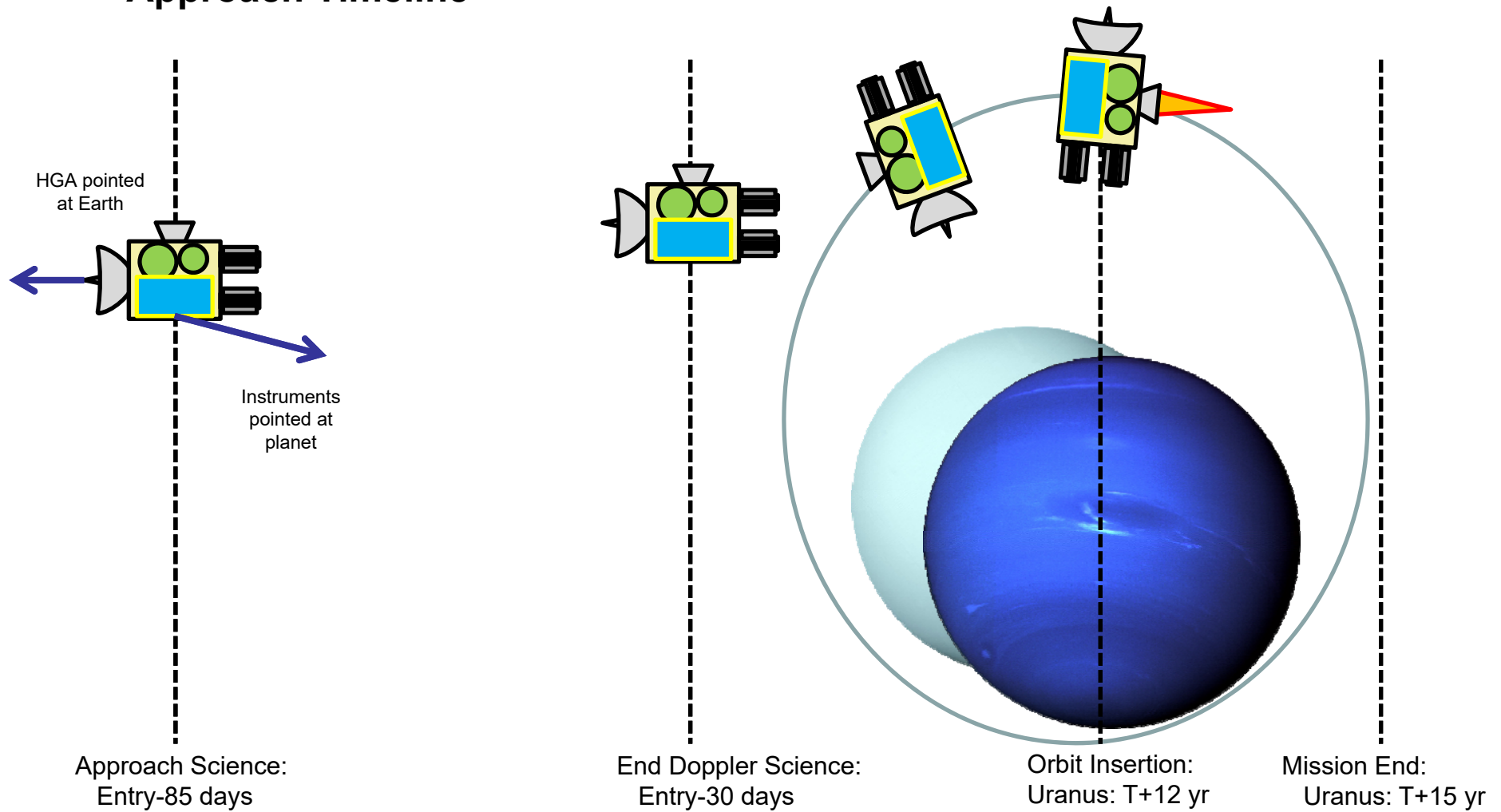

HGA pointed
at Earth

Instruments
pointed at
planet

Approach Science:
Entry-85 days

End Doppler Science:
Entry-30 days

Orbit Insertion:
Uranus: T+12 yr

Mission End:
Uranus: T+15 yr

**Team X Study Guidelines**
## Ice Giants Study 2016-07
## Orbiter

### Project - Study

| | |
|---|---|
| Customer | John Elliott, Kim Reh |
| Study Lead | Bob Kinsey |
| Study Type | Pre-Decadal Study |
| Report Type | Full PPT Report |

### Project - Mission

| | |
|---|---|
| Mission | Ice Giants Study 2016-07 |
| Target Body | Uranus |
| Science | Imaging and Magnetometry |
| Launch Date | 25-May-31 |
| Mission Duration | 12 year cruise, 3 years in orbit |
| Mission Risk Class | B |
| Technology Cutoff | 2027 |
| Minimum TRL at End of Phase B | 6 |

### Project - Architecture

| Orbiter on | Launch Vehicle |
|---|---|
| Launch Vehicle | Atlas V 551 |
| Trajectory | VEEJ Gravity Assists, 120 day initial orbit, Probe EFPA = -30deg |
| L/V Capability, kg | 4880 kg to a C3 of 11 with 0% contingency taken out |
| Tracking Network | DSN |
| Contingency Method | Apply Total System-Level |

| Spacecraft | |
|---|---|
| Spacecraft | Orbiter |
| Instruments | Narrow Angle Camera (EIS Europa),Doppler Imager (ECHOES JUICE),Magnetometer (Gallileo),Vis-Near IR Mapping Spectrometer (OVIRS/OSIRIS-Rex),Mid-IR Spectrometer (OTES/OSIRSI-Rex),UV Imaging Spectrometer (Alice/New Horizons),Radio Waves (LPW/Maven),Low Energy Plasma (SWAP/New Horizons),High Energy Plasma (PEPSI/New Horizons),Thermal IR (Diviner/LRO),Energetic Neutral Atoms (INCA/Cassini),Dust Detector (SDC/New Horizons), etc |
| Potential Inst-S/C Commonality | None |
| Redundancy | Dual (Cold) |
| Stabilization | 3-Axis |
| Heritage | TBD |
| Radiation Total Dose | 29.833 krad behind 100 mil. of Aluminum, with an RDM of 2 added. |
| Type of Propulsion Systems | System 1-Biprop, System 2-0, System 3-0 |
| Post-Launch Delta-V, m/s | 2607 |
| P/L Mass CBE, kg | 144.8 kg Payload CBE |
| P/L Power CBE, W | 84.704 |
| P/L Data Rate CBE, kb/s | 30000 |
| P/L Pointing, arcsec | TBD |
| Hardware Models | Protoflight S/C, EM instrument TBR |

| Project - Cost and Schedule | |
|---|---|
| Cost Target | < $2B TBD |
| Mission Cost Category | Flagship - e.g. Cassini |
| FY$ (year) | 2015 |
| Include Phase A cost estimate? | Yes |
| Phase A Start | July 2024 |
| Phase A Duration (months) | 20 |
| Phase B Duration (months) | 16 |
| Phase C/D Duration (months) | 47 |
| Review Dates | PDR - July 2027, CDR - September 2028, ARR - September 2029 |
| Phase E Duration (months) | 179 |
| Phase F Duration (months) | 4 |
| New Development Tests | TBD |
| Project Pays Tech Costs from TRL | 6 |
| Spares Approach | Typical |
| Parts Class | Commercial + Military 883B TBR |
| Launch Site | Cape Canaveral |

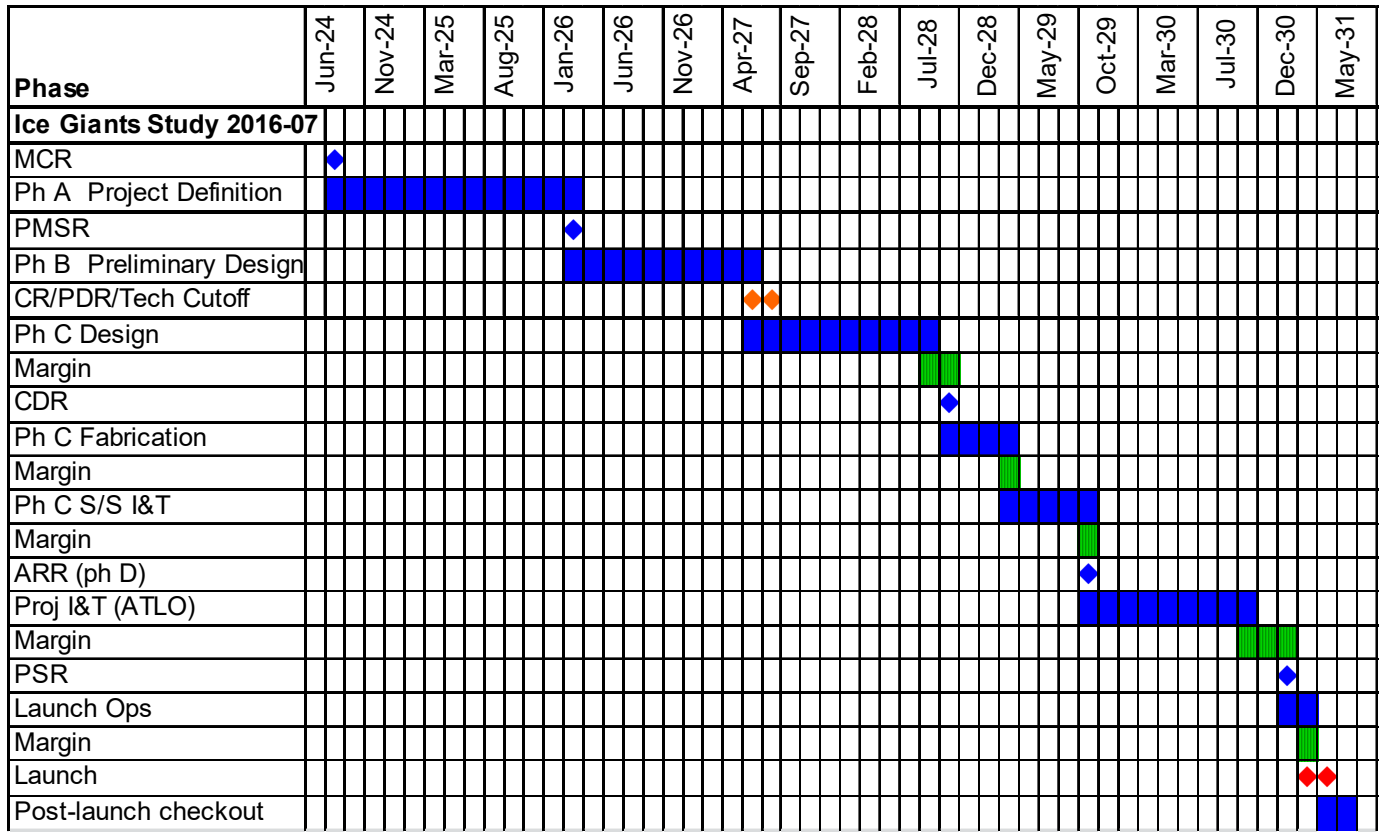

| Phase | Jun-24 | Nov-24 | Mar-25 | Aug-25 | Jan-26 | Jun-26 | Nov-26 | Apr-27 | Sep-27 | Feb-28 | Jul-28 | Dec-28 | May-29 | Oct-29 | Mar-30 | Jul-30 | Dec-30 | May-31 |
|---|---|---|---|---|---|---|---|---|---|---|---|---|---|---|---|---|---|---|
| **Ice Giants Study 2016-07** | | | | | | | | | | | | | | | | | | |
| MCR | | | | | | | | | | | | | | | | | | |
| Ph A  Project Definition | | | | | | | | | | | | | | | | | | |
| PMSR | | | | | | | | | | | | | | | | | | |
| Ph B  Preliminary Design | | | | | | | | | | | | | | | | | | |
| CR/PDR/Tech Cutoff | | | | | | | | | | | | | | | | | | |
| Ph C Design | | | | | | | | | | | | | | | | | | |
| Margin | | | | | | | | | | | | | | | | | | |
| CDR | | | | | | | | | | | | | | | | | | |
| Ph C Fabrication | | | | | | | | | | | | | | | | | | |
| Margin | | | | | | | | | | | | | | | | | | |
| Ph C S/S I&T | | | | | | | | | | | | | | | | | | |
| Margin | | | | | | | | | | | | | | | | | | |
| ARR (ph D) | | | | | | | | | | | | | | | | | | |
| Proj I&T (ATLO) | | | | | | | | | | | | | | | | | | |
| Margin | | | | | | | | | | | | | | | | | | |
| PSR | | | | | | | | | | | | | | | | | | |
| Launch Ops | | | | | | | | | | | | | | | | | | |
| Margin | | | | | | | | | | | | | | | | | | |
| Launch | | | | | | | | | | | | | | | | | | |
| Post-launch checkout | | | | | | | | | | | | | | | | | | |

Proposed development schedule consistent with typical New Frontiers missions and current Europa mission schedule.
Phase A 20 mos., Phase B 16 mos., Phase C/D 47 mos.
Launch May 25, 2031

- **First 3 instruments are the same as Option 5.**

- **Total payload mass 145 kg CBE**

- **Largest mass 42 kg CBE Microwave Sounder (MWR/Juno)**

- **Largest power 33 W CBE also MWR/Juno**

- **Total payload cost $219M including**
  - Management
  - Payload Engineering

| Instrument Name | Heritage | CBE Mass (kg) | Cont. | CBE+Cont. Mass (kg) | CBE Op. Power (W) | CBE Standy Power (W) |
|---|---|---|---|---|---|---|
| | | 145 kg | 17% | 169.5 | | |
| Narrow Angle Camera (EIS Europa) | Inherited design | 12.0 | 15% | 13.8 | 16 W | 2 W |
| Doppler Imager (ECHOES JUICE) | New design | 20.0 | 30% | 26 | 20 W | 2 W |
| Magnetometer (Galileo) | Inherited design | 4.7 | 15% | 5.405 | 8 W | 1 W |
| Vis-Near IR Mapping Spectrometer (OVIRS/OSIRIS-Rex) | Inherited design | 16.5 | 15% | 18.975 | 9 W | 1 W |
| Mid-IR Spectrometer (OTES/OSIRSI-Rex) | Inherited design | 6.3 | 15% | 7.245 | 11 W | 1 W |
| UV Imaging Spec (Alice/New Horizons) | Inherited design | 4.0 | 15% | 4.6 | 10 W | 1 W |
| Radio Waves (LPW/Maven) | Inherited design | 5.6 | 15% | 6.44 | 3 W | 0 W |
| Low Energy Plasma (SWAP/New Horizons) | Inherited design | 3.3 | 15% | 3.795 | 2 W | 0 W |
| High Energy Plasma (PEPSI/New Horizons) | Inherited design | 1.5 | 15% | 1.725 | 3 W | 0 W |
| Thermal IR (Diviner/LRO) | Inherited design | 12.0 | 15% | 13.8 | 25 W | 3 W |
| Energetic Neutral Atoms (INCA/Cassini) | Inherited design | 6.9 | 15% | 7.935 | 3 W | 0 W |
| Dust Detector (SDC/New Horizons) | Inherited design | 5.0 | 15% | 5.75 | 7 W | 1 W |
| Langmuir Probe (RPWS-LP/Cassini) | Inherited design | 1.0 | 15% | 1.15 | 0 W | 0 W |
| Microwave Sounder (MWR/Juno) | Inherited design | 42.0 | 15% | 48.3 | 33 W | 3 W |
| WAC (MDIS-WAC/MESSINGER) | Inherited design | 4.0 | 15% | 4.6 | 10 W | 2 W |

## Instruments
- Narrow Angle Camera
- Doppler Imager
- Magnetometer
- Vis-Near IR Mapping Spectrometer
- Mid-IR Spectrometer
- UV Imaging Spectrometer
- Radio Waves Instrument
- Low Energy Plasma Instrument
- High Energy Plasma Instrument
- Thermal IR
- Energetic Neutral Atoms
- Dust Detector
- Langmuir Probe
- Microwave Sounder

## CDS
- JPL reference bus avionics
- Dual string cold redundancy

## Baseline Power System
- 5 eMMRTGs, 45kg each
- 10 A-hr, <5kg, Li Ion Battery

## Thermal
- Active and passive thermal control design
- Louvers, heaters, MLI
- eMMRTG shields

## ACS
- Four 0.1N Honeywell HR16 reaction wheels
- IMUs, Star Trackers, Sun Sensors

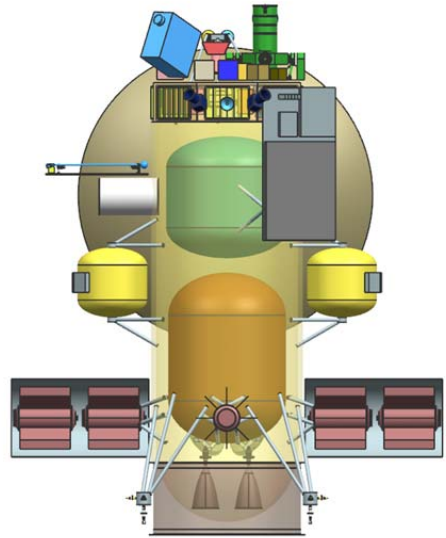

## Telecom
- Radios
  - Two X/X/Ka SDST transponders
  - Two 35W Ka-Band TWTAs
  - Two 25W X-Band TWTAs
- Antennas
  - One 3m X/Ka HGA
  - One X-Band MGA
  - Two X-Band LGAs

## Propulsion
- Dual-mode bipropellant system provides 2607m/s of delta-V
- Two 200lbf Aerojet main engines
- Four 22N engines
- Eight 1N RCS engines

## Structures
- 365kg structure
- 40kg ballast
- 105kg harness
- 10m Magnetometer Boom
- Main Engine cover
- SEP Stage and Probe separation mechanisms
- Main Engine and eMMRTG covers

✖ **Configuration Drawings – Stowed**

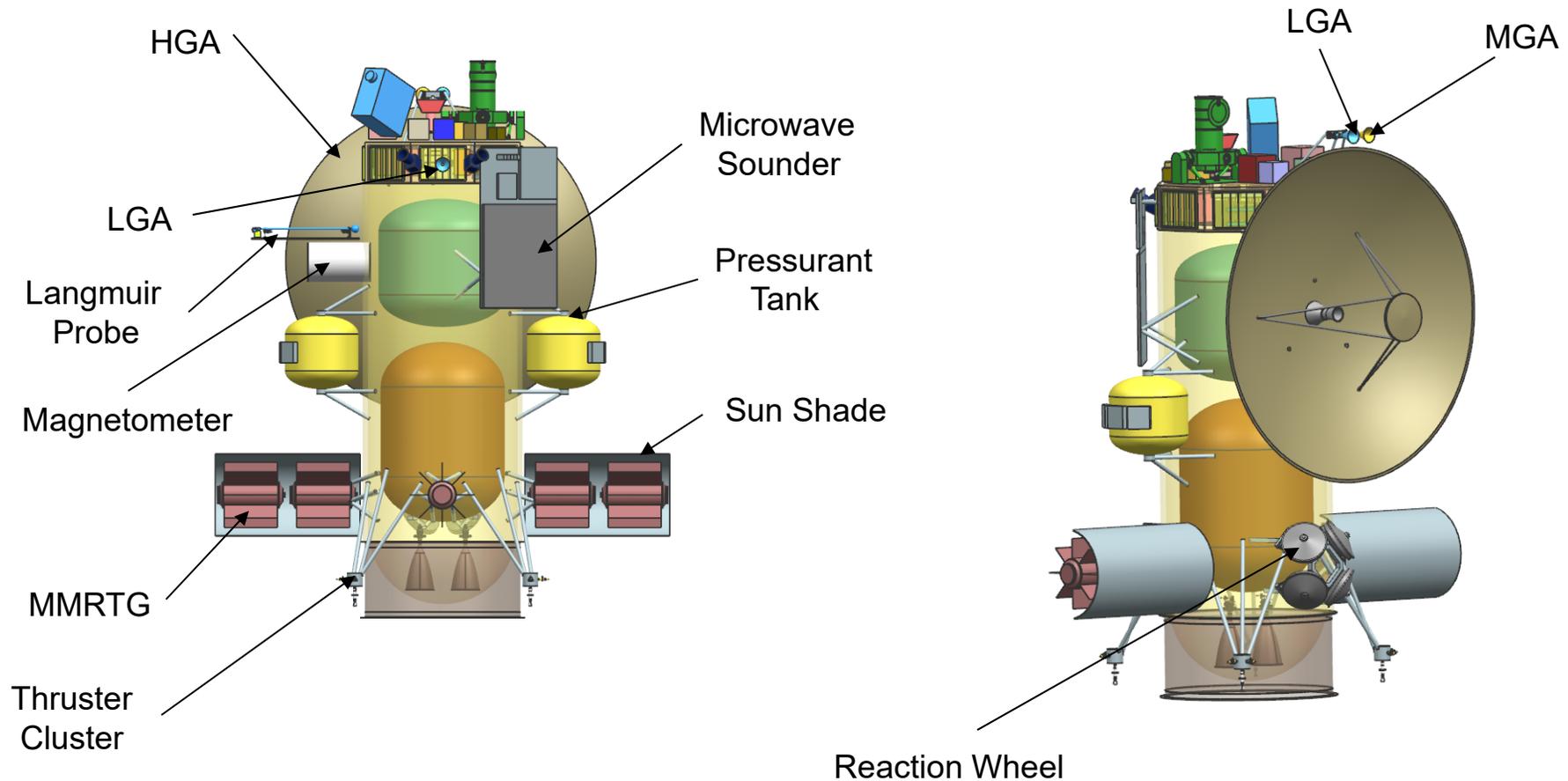

HGA

LGA

Microwave Sounder

Langmuir Probe

Pressurant Tank

Magnetometer

Sun Shade

MMRTG

Thruster Cluster

LGA

MGA

Reaction Wheel

✘ **Configuration Drawings – Deployed**

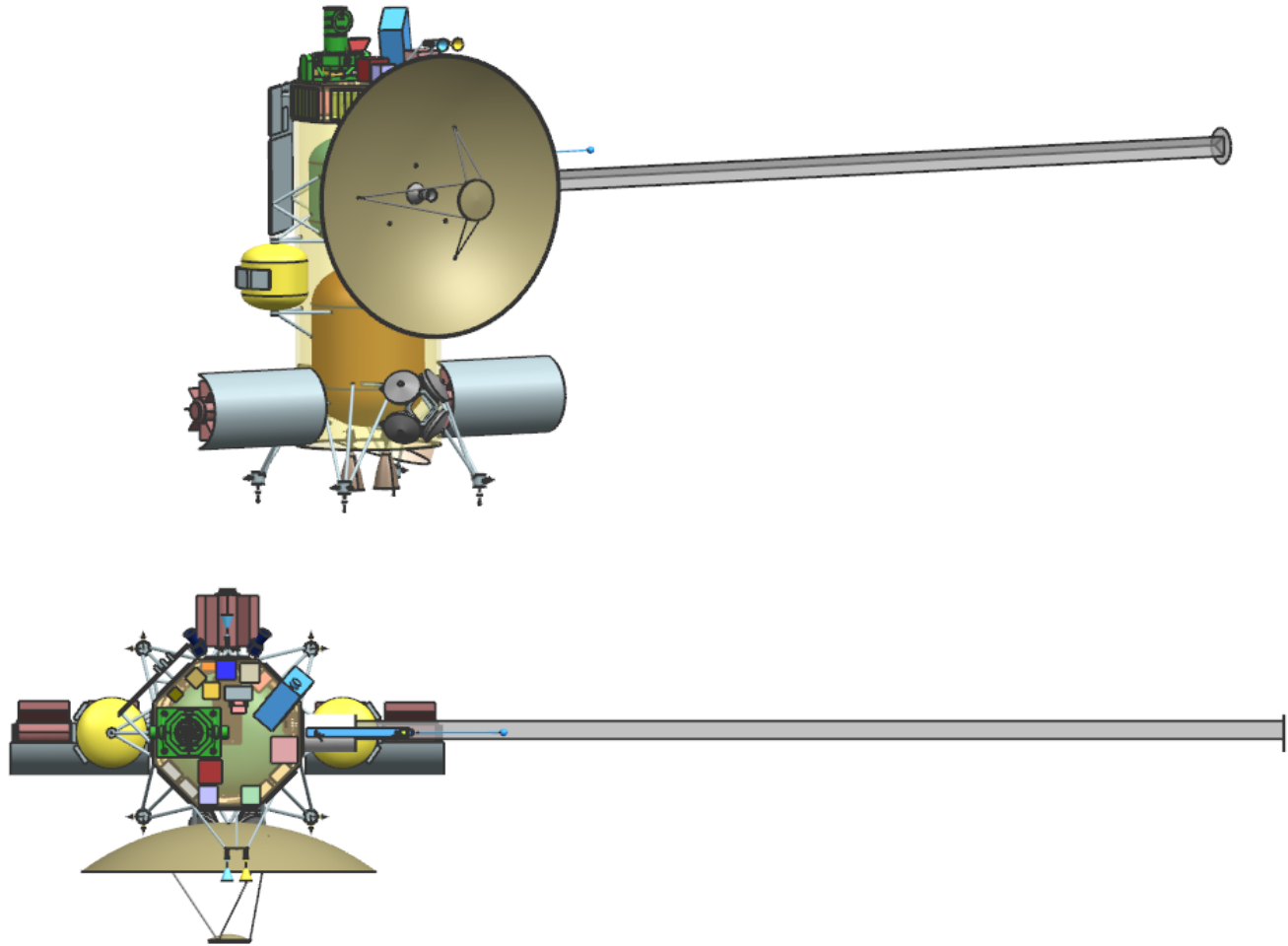

| | Mass Fraction | Mass (kg) | Subsys Cont. % | CBE+ Cont. (kg) | Mode 1 Power (W) Recharge | Mode 2 Power (W) Approach Science - 85 Days | Mode 3 Power (W) Telecom Downlink | Mode 4 Power (W) UOI Delta-V | Mode 5 Power (W) Orbital Science - Apoapse | Mode 6 Power (W) Orbital Science - Periapse | Mode 7 Power (W) Orbital Science - Moon Flyby | Mode 8 Power (W) Safe | Mode 9 Power (W) IGNORE THIS MODE | Mode 10 Power (W) Orbiting Science - Rotation Rate Movie |
|---|---|---|---|---|---|---|---|---|---|---|---|---|---|---|
| **Power Mode Duration (hours)** | | | | | 24 | 24 | 8 | 1.05 | 16 | 16 | 8 | 24 | | 17 |
| **Payload on this Element** | | | | | | | | | | | | | | |
| Instruments | 10% | 144.8 | 17% | 169.5 | 4 | 28 | 4 | 4 | 16 | 85 | 55 | 4 | 4 | 81 |
| **Payload Total** | 10% | **144.8** | 17% | **169.5** | 4 | 28 | 4 | 4 | 16 | 85 | 55 | 4 | 4 | 81 |
| **Spacecraft Bus** | | | do not edit formulas below this line, use the calculations and override tables instead --> | | | | | | | | | | | |
| Attitude Control | 4% | 63.5 | 10% | 69.8 | 0 | 55 | 55 | 88 | 55 | 55 | 55 | 42 | 55 | 93 |
| Command & Data | 2% | 27.6 | 18% | 32.4 | 61 | 61 | 61 | 61 | 61 | 61 | 61 | 61 | 61 | 61 |
| Power | 18% | 265.3 | 2% | 269.7 | 28 | 44 | 36 | 28 | 28 | 28 | 28 | 44 | 36 | 36 |
| Propulsion1 ☐ SEP1 | 12% | 179.9 | 5% | 189.7 | 31 | 3 | 3 | 151 | 3 | 3 | 3 | 3 | 3 | 3 |
| Structures & Mechanisms | 33% | 482.2 | 30% | 626.8 | 0 | 0 | 0 | 0 | 0 | 0 | 0 | 0 | 0 | 0 |
| S/C-Side Adapter | 1% | 19.5 | 0% | 19.5 | | | | | | | | | | |
| Cabling | 7% | 103.4 | 30% | 134.4 | | | | | | | | | | |
| Telecom | 4% | 114.6 | 15% | 148.4 | 12 | 107 | 155 | 71 | 12 | 42 | 12 | 71 | 12 | 12 |
| Thermal | 8% | 121.1 | 23% | 148.4 | 25 | 25 | 25 | 25 | 25 | 25 | 25 | 25 | 25 | 25 |
| **Bus Total** | | **1317.6** | 18% | **1554.1** | 157 | 294 | 335 | 424 | 184 | 214 | 184 | 246 | 192 | 230 |
| Thermally Controlled Mass | | | | 1554.1 | | | | | | | | | | |
| **Spacecraft Total (Dry): CBE & MEV** | | **1462.4** | 18% | **1723.7** | 162 | 322 | 339 | 429 | 200 | 298 | 239 | 250 | 196 | 311 |
| Subsystem Heritage Contingency | 18% | 261.3 | SEP Cont | 10% | 0 | 0 | 0 | 0 | 0 | 0 | 0 | 0 | 0 | 0 |
| System Contingency | 19% | 270.8 | | | 69 | 138 | 146 | 184 | 86 | 128 | 103 | 108 | 84 | 134 |
| Total Contingency | **36%** | 532.1 | | | | | | | | | | | | |
| **Spacecraft with Contingency:** | | **1994** | of total | w/o add'l pld | 231 | 461 | 485 | 613 | 285 | 426 | 342 | 358 | 280 | 444 |
| Propellant & Pressurant with residuals1 | 58% | 2722.6 | For S/C mass = | 1994.7 | | Delta-V, Sys 1 | 2607.0 | m/s | | residuals = | 71.3 | kg | | |
| **Spacecraft Total with Contingency (Wet)** | | **4717.1** | | | | | | | | | | | | |
| L/V-Side Adapter | | 0.0 | Wet Mass for Prop Sizing | 4880 | | BOL Power: | 0.0 | W | | | | | | |
| **Launch Mass** | | **4717** | Dry Mass for Prop Sizing | 1994 | | EOL Power: | 0.0 | W | | | | | | |
| **Allocation** | | **4880** | Atlas V 551 | | | | | | | | | | | |
| | | | | Launch C3 | 11.9 | | | | | | | | | |
| **Launch Vehicle Margin** | | **162.9** | Mission Unique LV Contingency | 0% | | | | | | | | | | |

| Element Number | Element Name | Dry CBE (kg) | Cont / JPL Margin (kg) | Dry Allocation (kg) | Propellant (kg) | Dry Allocation + Propellant (kg) |
|---|---|---|---|---|---|---|
| 3 | Orbiter minus eMMRTGs | 1237 | 532 | 1770 | 2723 | 4493 |
| 3.1 | eMMRTGs | 225 | - | 225 | - | 225 |
| | **Total Stack** | **1462** | **532** | **1995** | **2723** | **4717** |
| | | | | Dry Mass Allocation | | 1995 |
| | | | | JPL Margin (kg / %) | | 533 / 27% |
| | | | | JPL Margin without eMMRTG (kg / %) | | 533 / 30% |
| | | | | Atlas V 551 Capacity (kg) (C3 = 11.9 km$^2$/s$^2$) | | 4880 |
| | | | | Extra Launch Vehicle Margin (kg) | | 163 |

| Element Number | Element Name | Dry CBE (kg) | Cont (%) | Cont. (kg) | MEV (kg) | Dry Allocation (kg) | Propellant (kg) | Dry Allocation + Propellant (kg) |
|---|---|---|---|---|---|---|---|---|
| 3 | Orbiter minus eMMRTGs | 1237 | 21% | 261 | 1498 | 1770 | 2723 | 4493 |
| 3.1 | eMMRTGs | 225 | 0% | 0 | 225 | 225 | - | 225 |
| | **Total Stack** | **1462** | **18%** | **261** | **1723** | **1995** | **2723** | **4717** |

| | |
|---|---|
| Dry Mass Allocation | 1995 |
| NASA Margin (kg / %) | 272 / 16% |
| NASA Margin without eMMRTG (kg / %) | 272 / 18% |
| Atlas V 541 Capacity (kg) (C3 = 11.9 $km^2/s^2$) | 4880 |
| Extra Launch Vehicle Margin (kg) | 163 |

- **5th eMMRTG would enable higher data rate, lower costs on orbit.**
  - Expected 5th RTG to be needed for instrument power.
    - Turns out that only 4 RTGs are needed.
  - More instruments while in orbit means higher data volume than Option 5.
    - Downlink for more passes/day or increase the downlink data rate.
  - DSN pricing highly favors one 8-hour pass/day, so avoid more passes.
    - Cost for 3 passes/day can be 9 times the cost of 1 pass/day.
  - 5th RTG enables use of a 70W TWTA instead of 35W, doubles data rate.
    - 30 kbps compared to 15 kbps for Option 1
  - An 8-hour pass to an array of two 34-m ground stations gets the data down.
  - Saves a couple of hundred million dollars while on orbit.

- **Data downlink strategy for Doppler Imager (DI) on approach**
  - DI generates a lot of data continuously for tens of days on approach.
  - Configuration with HGA and DI on opposite sides of the cylindrical bus allows pointing DI towards Uranus while pointing HGA towards Earth.
  - Can downlink for 24 hours/day and maintain positive power balance.
  - Any data that can't be downlinked before UOI will be downlinked after.

- **Configuration that helps to minimize mass and power**
  - eMMRTGs outside the cylindrical bus provide heating
    - Reduces mass and power of thermal subsystem components
  - Propellant and oxidizer tanks inside bus, pressurant tanks outside
    - Pressurant tanks are easier to keep warm than propellant/oxidizer tanks.
  - Shorten the stack to minimize structure mass
    - Single custom propellant tank instead of two tanks

- **Mechanical**
  - LV interfaces to the Orbiter.
  - Entry System containing Probe is attached to the side of the Orbiter.
  - Orbiter: primary structure is the largest mass element.
    - 317 kg CBE out of 482 kg total for Mechanical (1318 kg bus dry mass)
    - Drivers are the large Propulsion and Power masses.

- **Baseline Power System**
  - Dual String Reference Bus electronics heritage
  - eMMRTG mass of 45kg/unit, with cooling tubes, is a not to exceed value, so 0% mass uncertainty is applied.
  - 265 kg CBE subsystem dry mass

- **Propulsion: Dual-mode bi-prop for 2.6 km/s total delta V**
  - Orbit insertion delta V=  1.7 km/s desired in < 1hr
    - Two 890N main engines used to achieve burn time < 1hr
  - 180 kg CBE dry mass; 2723 kg propellant and pressurant

- **Thermal: Cassini-heritage waste-heat recovery system on Orbiter**
  - RTG end domes each provide 75 W waste heat to propulsion module via conductive and radiative coupling.
  - VRHUs act as primary control mechanism for thruster clusters.
    - Also act as trim heaters for the propulsion module
  - Louvers act as primary control mechanism for avionics module.

- **Telecom: X- and Ka-Band subsystem.**
  - Two 70W Ka-Band TWTAs, two 25W X-Band TWTAs
  - Two X/X/Ka SDST transponders
  - 3m X/Ka HGA, one X-Band MGA, two X-Band LGAs.
  - Supports a data rate at Uranus of 30 kbps into 34m BWG ground station.

- **CDS: Reference Bus architecture ideally suited for high reliability, long lifetime mission.**
  - Standard JPL spacecraft CDS that is similar to SMAP
    - RAD750 CPU, NVM, MTIF, MSIA, CRC, LEU-A, LEU-D, MREU
    - 128 GBytes storage for science data
    - 1553 and RS-422 ICC/ITC interfaces for subsystems and instruments

- **ACS: 3-axis stabilized with star tracker, sun sensor, gyros, wheels.**
  - All stellar attitude determination to minimize power, conserve gyros.
  - Sun sensor performance may degrade once the Orbiter passes Saturn.
    - May impact safe mode used during star tracker outage.
    - Detailed analysis on Sun sensor performance versus distance is needed.

✖ **Software: core product line is appropriate since this mission has aspects similar to MSL/M2020/SMAP/Europa.**

- Complexity rankings range from Medium to High.
  - ◆ Medium infrastructure: dual string with warm spare.
  - ◆ High fault behaviors: high redundancy, string swapping, critical events.
  - ◆ Medium/High ACS: tight pointing requirements, many ACS modes
  - ◆ High Telecom: 24/7 downlink, redundant DTE
  - ◆ High Science data processing, full file system

✖ **SVIT: Testbed, System I&T and V&V costs are included**

- Cost of assembling and testing RTG's is captured elsewhere
  - ◆ Cost of integrating RTG's is included with other ATLO costs

✖ **Ground Systems**

- Mission specific implementation of standard JPL mission operations and ground data systems
- Ground network: array of DSN 34-m BWG; 70-m or equivalent for safe mode
- Science support: 24x7 tracking on approach; daily 8-hour contacts on orbit

| COST SUMMARY (FY2015 $M) | Generate ProPricer Input | Team X Estimate | | |
|---|---|---|---|---|
| | | CBE | Res. | PBE |
| Project Cost | | $1673.6 M | 20% | $2005.1 M |
| Launch Vehicle | | $33.0 M | 0% | $33.0 M |
| Project Cost (w/o LV) | | $1640.6 M | 20% | $1972.1 M |
| Development Cost | | $1140.9 M | 23% | $1404.6 M |
| Phase A | | $11.4 M | 23% | $14.0 M |
| Phase B | | $102.7 M | 23% | $126.4 M |
| Phase C/D | | $1026.8 M | 23% | $1264.2 M |
| Operations Cost | | $499.7 M | 14% | $567.5 M |

Total mission cost is $2.01B. This is the likely cost within a range that typically can be as much as 10% lower up to 20% higher. The development cost with reserves is $1.40B.

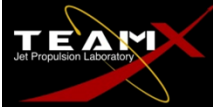

# Executive Summary
## Cost A-D

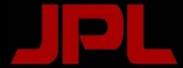

| WBS Elements | NRE | RE | 1st Unit |
|---|---|---|---|
| Project Cost (no Launch Vehicle) | $1514.0 M | $491.1 M | $2005.1 M |
| Development Cost (Phases A - D) | $913.6 M | $491.0 M | $1404.6 M |
| 01.0 Project Management | $47.3 M | | $47.3 M |
| 1.01 Project Management | $11.4 M | | $11.4 M |
| 1.02 Business Management | $13.6 M | | $13.6 M |
| 1.04 Project Reviews | $2.5 M | | $2.5 M |
| 1.06 Launch Approval | $19.8 M | | $19.8 M |
| 02.0 Project Systems Engineering | $24.8 M | $0.8 M | $25.6 M |
| 2.01 Project Systems Engineering | $8.9 M | | $8.9 M |
| 2.02 Project SW Systems Engineering | $5.2 M | | $5.2 M |
| 2.03 EEIS | $1.5 M | | $1.5 M |
| 2.04 Information System Management | $1.7 M | | $1.7 M |
| 2.05 Configuration Management | $1.5 M | | $1.5 M |
| 2.06 Planetary Protection | $0.2 M | $0.2 M | $0.4 M |
| 2.07 Contamination Control | $2.3 M | $0.6 M | $2.9 M |
| 2.09 Launch System Engineering | $1.0 M | | $1.0 M |
| 2.10 Project V&V | $2.0 M | | $2.0 M |
| 2.11 Risk Management | $0.5 M | | $0.5 M |
| 03.0 Mission Assurance | $47.3 M | $0.0 M | $47.3 M |
| 04.0 Science | $66.2 M | | $66.2 M |
| 05.0 Payload System | $147.9 M | $86.3 M | $234.1 M |
| 5.01 Payload Management | $15.8 M | | $15.8 M |
| 5.02 Payload Engineering | $12.9 M | | $12.9 M |
| Orbiter Instruments | $119.1 M | $86.3 M | $205.4 M |
| Narrow Angle Camera (EIS Europa) | $11.6 M | $8.4 M | $20.0 M |
| Doppler Imager (ECHOES JUICE) | $17.4 M | $12.6 M | $30.0 M |
| Magnetometer (Gallileo) | $4.5 M | $3.3 M | $7.8 M |
| Vis-Near IR Mapping Spectrometer (OVIRS/ | $9.7 M | $7.0 M | $16.7 M |
| Mid-IR Spectrometer (OTES/OSIRSI-Rex) | $7.1 M | $5.2 M | $12.3 M |
| UV Imaging Spectrometer (Alice/New Horizo | $5.8 M | $4.2 M | $10.0 M |
| Radio Waves (LPW/Maven) | $3.4 M | $2.4 M | $5.8 M |
| Low Energy Plasma (SWAP/New Horizons) | $2.5 M | $1.8 M | $4.2 M |
| High Energy Plasma (PEPSI/New Horizons) | $2.2 M | $1.6 M | $3.8 M |
| Thermal IR (Diviner/LRO) | $14.7 M | $10.6 M | $25.3 M |
| Energetic Neutral Atoms (INCA/Cassini) | $4.4 M | $3.2 M | $7.7 M |
| Dust Detector (SDC/New Horizons) | $5.7 M | $4.1 M | $9.8 M |
| Langmuir Probe (RPWS-LP/Cassini) | $1.1 M | $0.8 M | $1.9 M |
| Microwave Sounder (MWR/Juno) | $23.3 M | $16.9 M | $40.2 M |
| WAC (MDIS-WAC/MESSINGER) | $5.7 M | $4.1 M | $9.8 M |

| WBS Elements | NRE | RE | 1st Unit |
|---|---|---|---|
| 06.0 Flight System | $333.9 M | $273.0 M | $606.9 M |
| 6.01 Flight System Management | $5.0 M | | $5.0 M |
| 6.02 Flight System Systems Engineering | $35.9 M | | $35.9 M |
| 6.03 Product Assurance (included in 3.0) | | | $0.0 M |
| Orbiter | $288.4 M | $271.5 M | $559.9 M |
| 6.04 Power | $94.5 M | $163.6 M | $258.1 M |
| 6.05 C&DH | $32.0 M | $31.2 M | $63.2 M |
| 6.06 Telecom | $24.6 M | $15.2 M | $39.8 M |
| 6.07 Structures (includes Mech. I&T) | $44.1 M | $14.9 M | $59.0 M |
| 6.08 Thermal | $4.2 M | $13.5 M | $17.6 M |
| additional cost for >43 RHUs | $34.0 M | $0.0 M | $34.0 M |
| 6.09 Propulsion | $22.1 M | $16.7 M | $38.8 M |
| 6.10 ACS | $9.4 M | $9.8 M | $19.1 M |
| 6.11 Harness | $5.6 M | $5.6 M | $11.2 M |
| 6.12 S/C Software | $17.3 M | $0.9 M | $18.2 M |
| 6.13 Materials and Processes | $0.7 M | $0.1 M | $0.7 M |
| SEP Stage | $0.0 M | $0.0 M | $0.0 M |
| 6.14 Spacecraft Testbeds | $4.6 M | $1.5 M | $6.1 M |
| 07.0 Mission Operations Preparation | $32.1 M | | $32.1 M |
| 7.0 MOS Teams | $26.2 M | | $26.2 M |
| 7.03 DSN Tracking (Launch Ops.) | $2.7 M | | $2.7 M |
| 7.06 Navigation Operations Team | $3.1 M | | $3.1 M |
| 7.07.03 Mission Planning Team | $0.0 M | | $0.0 M |
| 09.0 Ground Data Systems | $28.7 M | | $28.7 M |
| 9.0A Ground Data System | $23.3 M | | $23.3 M |
| 9.0B Science Data System Development | $4.6 M | | $4.6 M |
| 9A.03.07 Navigation H/W & S/W Development | $0.8 M | | $0.8 M |
| 10.0 ATLO | $16.2 M | $17.7 M | $33.9 M |
| 11.0 Education and Public Outreach | $0.0 M | $0.0 M | $0.0 M |
| 12.0 Mission and Navigation Design | $18.8 M | | $18.8 M |
| 12.01 Mission Design | $2.0 M | | $2.0 M |
| 12.02 Mission Analysis | $4.1 M | | $4.1 M |
| 12.03 Mission Engineering | $1.8 M | | $1.8 M |
| 12.04 Navigation Design | $10.9 M | | $10.9 M |
| Development Reserves | $150.4 M | $113.3 M | $263.7 M |

| WBS Elements | NRE | RE | 1st Unit |
|---|---|---|---|
| Operations Cost (Phases E - F) | $567.4 M | $0.1 M | $567.5 M |
| 01.0 Project Management | $27.1 M | | $27.1 M |
| 1.01 Project Management | $15.3 M | | $15.3 M |
| 1.02 Business Management | $10.7 M | | $10.7 M |
| 1.04 Project Reviews | $1.1 M | | $1.1 M |
| 1.06 Launch Approval | $0.1 M | | $0.1 M |
| 02.0 Project Systems Engineering | $0.0 M | $0.1 M | $0.1 M |
| 03.0 Mission Assurance | $3.6 M | $0.0 M | $3.6 M |
| 04.0 Science | $243.7 M | | $243.7 M |
| 07.0 Mission Operations | $177.6 M | | $177.6 M |
| 7.0 MOS Teams | $114.7 M | | $114.7 M |
| 7.03 DSN Tracking | $47.9 M | | $47.9 M |
| 7.06 Navigation Operations Team | $14.4 M | | $14.4 M |
| 7.07.03 Mission Planning Team | $0.6 M | | $0.6 M |
| 09.0 Ground Data Systems | $47.7 M | | $47.7 M |
| 9.0A GDS Teams | $28.2 M | | $28.2 M |
| 9.0B Science Data System Ops | $18.9 M | | $18.9 M |
| 9A.03.07 Navigation HW and SW Dev | $0.6 M | | $0.6 M |
| 11.0 Education and Public Outreach | $0.0 M | $0.0 M | $0.0 M |
| 12.0 Mission and Navigation Design | $0.0 M | | $0.0 M |
| Operations Reserves | $67.8 M | $0.0 M | $67.8 M |
| 8.0 Launch Vehicle | $33.0 M | | $33.0 M |
| Launch Vehicle and Processing | $0.0 M | | $0.0 M |
| Nuclear Payload Support | $33.0 M | | $33.0 M |

- **Mission duration will push systems to their operating lifetimes.**
- **Science planning risk**
  - Relative velocities between Orbiter and Uranus' satellites will be high.
    - Flybys occur near periapse
- **Collision avoidance with Uranus' rings needs to be considered.**
- **Uranus stays close to the range of solar conjunction (~4-5 deg)**
  - Doppler measurements may have increased noise levels.
- **eMMRTG still needs some development.**
  - May cause a schedule slip.
  - Performance may degrade at a higher rate than currently predicted.

- **Low altitude Venus flybys could pose potential thermal risk.**

- **RTG waste heat recovery design robustness**
  - Approach is highly configuration-dependent and may have high hidden development costs.
  - Less expensive on paper, but the actual implementation could be more expensive than an active system.

- **Component development for propulsion subsystem**
  - Large bi-prop engines for chemical

- **Sun sensor performance may degrade past Saturn.**
  - May impact safe mode used during star tracker outage.

# D   NEW TECHNOLOGY REPORTS

This appendix describes a set of new technologies that are highly relevant to the Ice Giants mission. Two of them—the proposed enhanced Multi Mission Radioisotope Thermoelectric Generator (eMMRTG) and the High Energy Environment (HEEET) thermal protection technology were deemed to be ready for the mission and are incorporated in the baseline. Other technologies were determined to be either not ready in the time frame of this report or lacking in the information needed to adequately judge their readiness. This appendix describes these new technologies and outlines the impact on the capabilities of missions to the Ice Giants.

## D.1   Thermal Protection System Technology Validation

A high performance Thermal Protection System (TPS) is required for the entry probes that are part of the baseline Ice Giants missions as well as for aerocapture, which is one of the advanced technologies for achieving orbital capture at Uranus or Neptune that was considered but not adopted. This section considers some of the challenges of developing and testing thermal protection systems and some of the recent developments that are critical to both probes and aerocapture.

### D.1.1   TPS Materials for Extreme Entry Environments

Uranus and Neptune are among a family of planetary targets which also includes Venus for which the entry environments are more severe than those experienced at Mars or even the Earth. For many years, heat shields designed for such targets (for the Pioneer Venus and Galileo Jupiter probes) used high density carbon phenolic as a TPS material. However, because of both availability, reproducibility and adaptability challenges in the intervening years, NASA's Ames Research Center (ARC) has investigated other options and conducted a feasibility study of a Woven Thermal Protection System technology (WTPS) intended to address these problems.

The successful demonstration of the Woven Thermal Protection System (WTPS) architecture during this feasibility study led NASA's Game Changing Development (GCD) program led ARC to define a 3-year technology maturation project, "Heat Shield for Extreme Entry Environment Technology (HEEET)." HEEET started in FY14 with the goal of maturing the WTPS technology to Technology Readiness Level (TRL) 6 in 3 years with a primary objective of enabling upcoming Discovery and New Frontier Announcement of Opportunity missions that need robust, mass-efficient and cost-effective ablative TPS.

The near-term mission targets include, but are not limited to, Venus Lander, Venus and Saturn probe missions, and high speed sample return missions, where entry environments exceed the capability of currently available TPS materials such as PICA and Avcoat, and for which heritage like carbon phenolic (tape wrapped and chop molded) is either mass inefficient and/or mass infeasible and hence science limiting. HEEET is more than just a carbon phenolic replacement; it provides targeted capability development to enable mission designs that cannot be conceived of today. HEEET has been baselined for the Uranus and Neptune probes described in this study report and also for the aerocapture concepts discussed in Section D.2.

### D.1.2   Ground Testing Challenges

Every entry mission faces the challenge that limitations in ground-based test facilities do not allow for a test-as-you-fly approach for TPS. This challenge manifests itself in multiple ways. First, there is the inability to test full size subsystems, e.g., a full size heat shield, even at the relatively small size of the probes under consideration within this study. Additionally, existing ground-based test facilities cannot achieve all the entry environment parameters at flight relevant conditions, in a single test configuration, even for missions with relatively benign entry environments such as MSL. This results in what has been termed "piecewise certification", wherein testing attempts to envelop each of the key entry environment parameters, such as heat flux, pressure and shear, individually (or in some cases maybe two of the parameters), but it's never been feasible to bound all the environments in a single test. For example, stagnation testing maybe utilized to bound heat flux and pressure while a wedge test is used to bound shear but at heat fluxes and pressures that maybe lower than in flight. Validating the material response models and exploration for any failure modes, through this piecewise testing set of bounding environments, is then considered to sufficiently reduce risks for flight (when coupled with the TPS sizing margin policy for the specific mission). These challenges are then exacerbated at the very high entry conditions anticipated for missions to the Ice Giants. **Figure D-1** provides a comparison of HEEET arcjet test points compared to Neptune and Uranus missions under consideration in this study. As one can observe, in stagnation test configurations, current ground-based capabilities can get close to bounding the peak heat flux and pressure in the IHF 3″ nozzle and bound maximum pressure in AEDC H3. In order to achieve relevant shears the wedge test article configurations are utilized. However, in wedge tests, the heat fluxes and pressures values can be lower than the environments experienced during flight at maximum shear. There are limited test options to get at pressures and heat fluxes intermediate between IHF 3″ and AEDC H3 2″ flat face (higher pressures than IHF 3″).

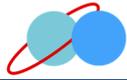

In addition, to achieve very high heat fluxes requires use of small test article sizes, on the order of 1–2 inches diameter. This raises questions about the test article size relative to unit cell sizes in the TPS materials, and particularly about any features such as gaps and seams that are often inherent to TPS designs due to manufacturing limitations. Also, the small size of the test articles can lead to challenges with test article design and performance in these aggressive environments. Questions can then be raised about whether the test results are influenced by the small test article size, or might there be edge effects, etc., such that applied environments that are not representative of flight.

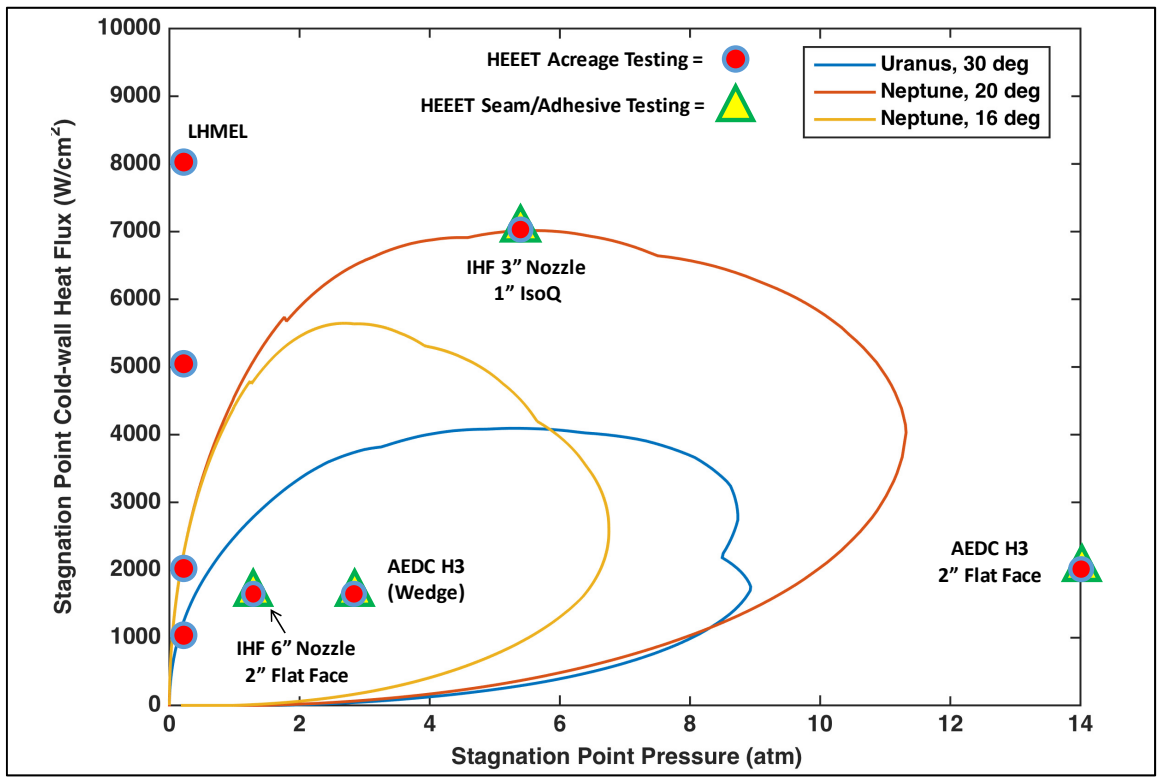

**Figure D-1.** Comparison of HEEET arcjet test points to entry environments for Uranus and Neptune missions under consideration in this study. Figure illustrates the limitations of ground-based test facilities and need for piecewise certification.

### D.1.3     Ice Giant Specific Challenges

Some of the proposed mission designs for this Ice Giants Study have regions in the entry environment that, although bounded in a piecewise fashion as shown in **Figure D-1**, require extrapolations with differences between ground and flight that are greater than typical in recent missions. The Neptune 20 deg entry in particular has high heat fluxes, greater than 4000 W/cm$^2$ and pressures greater than 8 atm where there are no current ground based test points. This requires extrapolation of the material's performance to higher conditions with greater differences between entry environments and test points than has been the case in recent missions.

It should be reiterated that the test points shown in **Figure D-1** represent the current limitations for ground based testing that HEEET is pursuing within its project, and do not represent a threshold or cliff in HEEET material performance. Extrapolation beyond these test points is possible and the magnitude of that extrapolation becomes somewhat dependent on a particular mission's risk posture. It should also be pointed out that the entry environments represented in Figure 1 are from TRAJ, are not full CFD calculations, and are not margined for aerothermal uncertainties, so these environments are likely to go up once these factors are taken into account.

In terms of specific features for a HEEET heatshield, current limitations in weaving width result in a system consisting of a series of tiles with associated seams between them and these seams then become the weak points within the system, both structurally and aerothermally. These seams consist of a gap filler which is bonded to the surrounding acreage TPS. The width of the gap filler is on the scale of that for the small test articles that are utilized to achieve the high heat fluxes and pressures under consideration for these missions, this can pose challenges in designing arcjet test articles and interpreting their results, but these challenges have been overcome in the past.

Of greater concern is extrapolating the performance of the seams to conditions beyond those which can be achieved in ground based testing. We don't know where the limit is in the aerothermal performance of the seams, but it is reasonable to assume that the seam will have a lower limit than that for the acreage material, given the nature of the seam. The acreage HEEET is very robust and testing to date has not revealed any failure modes. As mission entry environments push to the limits, or beyond, of ground-based test capability, there is more confidence in pushing the acreage material than the seams, as a system without seams will be inherently more robust, both structurally and aerothermally.

### D.1.4     Recommendations on Ground-Based Testing

The following recommendations would help to address the concerns with the challenges/limitations associated with ground based aerothermal testing:

1. Single Piece Heatshield: A single piece heatshield would eliminate the weakest link in the HEEET system both aerothermally as well as structurally. The current weaving vendor believes it is feasible to expand the weaving infrastructure to allow weaving of material that would support a single piece heatshield for the size range of these missions. An investment in scaling up the weaving infrastructure, likely the highest risk item in the implementation of a single piece HEEET heatshield is another avenue to reduce risk for these missions. Such investments have the potential to benefit multiple missions such as potential Mars Sample Return.

2. Flight Heatshield Instrumentation: It is likely that HEEET will be implemented in missions prior to the Uranus/Neptune missions under consideration in this study. Gaining actual flight data on the performance of HEEET would be very valuable in assessing the systems performance and robustness to actual flight environments. Instrumentation of the seams in particular would allow evaluation of their performance in flight to be compared back to ground-based test observations. Advocacy from this group for the integration of flight instrumentation into HEEET systems utilized in missions prior to these, would clearly help reduce risk for implementation of HEEET for these missions.

3. Expand Ground-Based Test Environments: Experience gained during the HEEET project has identified some opportunities to expand the test envelope in the existing test facilities, although exploration of those opportunities is outside the scope of the HEEET project. Examples include:

   a. Different nozzle and test article configurations at AEDC to achieve higher shears.

   b. Orion is currently funding the installation of laser heating facilities in the IHF arcjet test facility to allow augmented heating on panel and wedge test articles. It may be feasible to re-route the lasers to allow augmented heating on stagnation test articles, such as in the IHF 3″ nozzle, to increase the applied heat flux.

   c. A thorough study of the opportunities and implementation of the most beneficial environments would be one way to address the risks associated with ground based testing for Ice Giant missions.

   d. Expand Ground–Based Test Atmospheres: Current testing at these extreme environments is limited to tests conducted in air, while these planets have a $H_2$/He atmosphere. Primarily mechanisms for ablation in air are oxidation and sublimation while in $H_2$/He there is no oxidation. Separating these two mechanisms apart in ground-based testing in air at these extreme conditions is difficult as there are substantial contributions from both. There are opportunities to test in $N_2$ and $CO_2$ in existing facilities, however, at much lower conditions. Currently, assumptions are made that the materials performance in a $H_2$/He atmosphere can be accounted for through our models. It would be highly beneficial to be able to test in non-oxidizing environments at higher conditions and to be able to test in $H_2$/He to verify the ability to predict performance in these environments and to understand whether there is any unpredicted behavior when not testing in air. A couple of opportunities (and with additional study more may be feasible) include:

      i. Configure the NASA ARC Interactive Heating Facility (IHF) to run on pure $N_2$. This is feasible and a modest cost and would provide the opportunity to test at extreme conditions in a non-oxidizing environment.

      ii. Bring on line the Developmental Arcjet Facility (DAF) at NASA ARC that would enable testing in $H_2$/He environments. It would not allow the extreme conditions to be achieved that can be done in the IHF, but DAF will allow exploration of the materials response in $H_2$/He at lower conditions to build confidence in our ability to model the materials behavior. Bringing DAF on-line is a much more modest proposition compared to attempting to reconfigure the IHF to run on $H_2$/He.

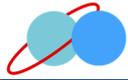

## D.2    Aerocapture Options for Ice Giants Missions

Aerocapture is a potentially useful technology for ice giant planet missions. Two recent studies have addressed that potential utility, one by NASA's Planetary Science Division (PSD), a part of the Science Mission Directorate, and one by the Ice Giants Mission Study (IGMS) team, spearheaded by researchers at Purdue University. The PSD study was a global assessment of the state of readiness of aerocapture in general, and technologies needed to implement aerocapture at various solar system destinations, including the ice giants. The aerocapture study conducted as part of the IGMS went into more technical depth, examining in detail how aerocapture could be implemented for ice giant missions, performance requirements on the hardware technologies involved, and the magnitude of potential benefits to be derived. The PSD study results are available in a NASA/JPL document, D-97058, submitted to NASA in February of 2016.

The PSD-led study found two issues that limited the depth and fidelity of the study: the lack of architectural-level hardware sizing relations for fundamental aerocapture hardware, such as aeroshell masses; and the uncertainties in aerothermodynamics calculations. The lack of hardware sizing relations meant that it is not possible to derive sufficiently precise values for benefits such as mass delivered to orbit. The study identified this as a gap in aerocapture-associated technologies that makes high-level assessments of the potential benefits of aerocapture at various destinations difficult, if not impossible. Previous studies attempting to estimate those benefits have had to do subsystem-level designs and considerable detailed analyses, such as CFD and radiative transfer calculations, detail well beyond that required for the great majority of architectural-level assessments. This is in contrast to subsystems such as chemical rocket propulsion systems, where fairly accurate sizing relations are well known.

For aerocapture, not only are the sizing relations very uncertain, but some of the fundamental physics regarding the conditions that the vehicle will encounter are quite uncertain. Calculations of the levels of convective heating and radiative heating at high hypersonic speeds, using different relations derived by different research groups, give very different answers. There is ample need and motivation for laboratory investigations of such phenomena to better describe the environments produced by hypersonic entries.

Nonetheless, the IGMS-led study has produced some very useful results. After a discussion of general characteristics of the aerocapture technique, this appendix will summarize the high-level conclusions of that study.

### D.2.1    Aerocapture Fundamentals

#### D.2.1.1    Description of the Technique

Aerocapture is one form of *aeroassist maneuver*, a general category that also includes such techniques as *aerobraking* and *aerogravity assist* maneuvers. In the past, instances of references to a maneuver that would be properly called an *aerocapture* maneuver as an *aerobraking* maneuver have led to confusion. The two are related but distinctly different concepts.

At the simplest level, aerocapture is the judicious use of aerodynamic forces (e.g., lift and drag) generated during a vehicle's controlled flight through a planetary-sized body's atmosphere to change an unbound (hyperbolic) approach orbit into a desired bound (captured) orbit. Thus, it is a means of achieving orbit insertion at the body without reliance on a propulsive maneuver, usually performed with rocket engines, for the majority of the ΔV required. The concept of aerocapture is not new (Cruz 1979) but has yet to be implemented on a space flight mission.

**Figure D-2** illustrates the profile of a typical aerocapture maneuver. It begins with a spacecraft's hyperbolic approach to its destination. During this period, the operations team navigates the spacecraft to a trajectory providing an atmospheric entry within the acceptable *entry corridor*, the range of entry conditions (such as flight path angle and speed) over which the flight system can guide to an acceptable exit state. Several aspects influence establishing the entry corridor, including vehicle constraints such as maximum deceleration, navigation and approach trajectory control accuracies, and uncertainties in the destination's atmospheric structure and the aerothermodynamics of its gas mixture, and the vehicle's mass properties, aerodynamics, and TPS response. This navigation task makes use of knowledge of the planet's gravitational field and its *ephemeris*, its location in space as a function of time. As with a lander mission this likely involves late navigation measurements and trajectory correction maneuvers (TCMs), possibly done autonomously by the spacecraft. Beginning a few hours or days before entry the spacecraft performs and reconfigurations needed for entry and comes to the proper entry attitude. This might involve ejections of now-unneeded hardware, such as a solar electric propulsion (SEP) stage, deployments to provide aerodynamic force modulation, or stowage of hardware that is needed after aerocapture but that must be protected during the aerocapture maneuver.

Once sufficiently dense atmosphere is encountered the vehicle begins its atmospheric flight phase. Using knowledge of the planet's gravity field, the atmosphere's composition and density profile and their uncertainties, and inertial data from onboard sensors (e.g., acceleration and attitude), the spacecraft autonomously controls its atmospheric flight to dissipate the desired amount of energy, emerging from the atmosphere at the desired atmospheric exit point state conditions. There might be a programmatic requirement that the vehicle must report its progress and performance to Earth during this phase. In case of a catastrophic failure, critical event telecom provides the project team with data that could be key in diagnosing the failure's cause. If the atmospheric flight phase includes periods where communication to Earth is not possible because the planet occults the communication path, it might be necessary to provide a relay asset that remains outside the atmosphere, receiving the flight vehicle's data for relay to Earth, similar to the MarCO (Mars Cube One) CubeSats being used in combination with *InSight* (Interior Exploration using Seismic Investigations, Geodesy and Heat Transport) at Mars. This data link has no ground-in-the-loop control duties at all. Even if communication interruptions could be reliably prevented, the time delay inherent in communications to Earth from distant destinations makes such control irrelevant. Relay of these data can be done after the fact, possibly well after atmospheric exit.

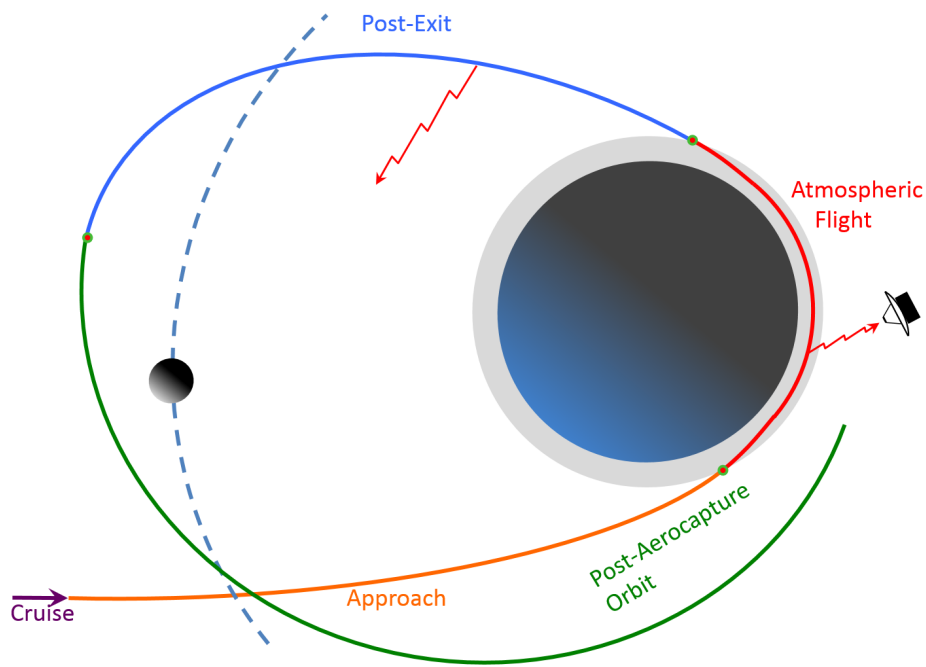

**Figure D-2.** Profile of a typical aerocapture maneuver.

Upon atmospheric exit certain actions must be accomplished quickly. For large heat loads, the heat-soaked aeroshell must be ejected to prevent damage to the orbiter spacecraft; other ejections might be necessary as well. Navigation measurements must be made, probably autonomously, to verify the accuracy of the vehicle's exit state, and to design and execute a post-exit TCM. The closer to the planet the TCM is executed, the smaller is the ΔV required, so prompt action saves propellant mass. This TCM is particularly important if the desired exit speed is very near escape speed, as is the case in past studies of aerocaptured missions to the Neptune system (Lockwood 2006; Hall, Noca, & Bailey, March–April 2005) In the relatively unlikely event that errors in the aerocapture maneuver are large enough that the actual exit speed is greater than escape speed, a TCM should be executed to reduce the orbit energy to a captured state. A post-exit TCM also can adjust the apoapsis altitude for the most efficient subsequent maneuvers to the desired science orbit, including the periapse raise maneuver (PRM), and adjust the "wedge angle" that is related to the argument of periapsis.

Other post-exit activities can occur on a somewhat less pressing time scale than the initial post-exit TCM. Any hardware stowed for the aerocapture maneuver must be redeployed, and any deployments of previously unused hardware, such as a deployable high gain antenna (HGA), might be done. During the flight from atmospheric exit to apoapsis, the spacecraft could relay to Earth more detailed data about the aerocapture maneuver's performance. Upon atmospheric exit the departure orbit has a periapsis radius that is within the planet's atmosphere. The PRM at apoapsis raises the periapsis to prevent reentry into the atmosphere, and typically would raise it to the desired periapsis for the initial science mission orbits. For short-period post-aerocapture orbits this could be a canned maneuver. For long-period orbits, ground control might be involved. Typically, one or more subsequent propulsive maneuvers would fine-tune the initial science orbit.

Uranus and Neptune have large satellites so the propulsive PRM could possibly be replaced or assisted by a gravity-assist flyby of a large satellite, designed to raise periapsis as needed, saving the propellant mass required for up to hundreds of m/s of ΔV. This requires very tight control of the atmospheric flight phase and very accurate post-exit navigation and TCMs to ensure an accurate satellite flyby. Missions to bodies with large satellites would certainly target close flybys of at least one of them, and would probably use one or more of the moons as "tour engines", using multiple planned gravity assists to effect a comprehensive "tour" of the entire system, much as the *Cassini* spacecraft is using Titan to explore the Saturn system. However, there is no fundamental requirement for the first outbound orbit leg to encounter a large moon. The PRM can be performed propulsively, in a way that allows subsequent orbital evolution and TCMs to provide a later initial satellite flyby that begins the tour. A first outbound leg encounter with a large moon could save much propellant, but that must be weighed against the increased risk.

A hybrid aerocapture/propulsive approach is a relatively new concept under consideration. This would have aerocapture provide the majority, but not all, of the ΔV needed for orbit insertion, and have a rocket propulsion system provide the remainder, including any orbit fine-tuning in that propulsive maneuver. Conversely, if the target apoapsis would require an exit velocity very near escape velocity, the aerocapture maneuver could aim for an exit state with a lower exit velocity (and thus a lower apoapsis), then use a propulsive maneuver to boost the apoapsis to that required for the science mission. Although currently not scheduled, future studies might determine if this technique offers potential risk or performance advantages.

### D.2.1.2 Potential Benefits

There are three categories of potential benefits from using aerocapture instead of propulsive orbit insertion. The first is that given a particular launch vehicle, aerocapture can often deliver more payload mass to orbit at the destination. In those cases, the mass of the hardware needed for the aerocapture maneuver and ancillary propulsive maneuvers is less than the mass of propulsion hardware and propellant needed to perform the insertion entirely propulsively. This difference is available for a wide range of potential benefits, such as increased science payload, increased propellant for a more comprehensive orbital tour, or more capable spacecraft subsystems to support the science. An example of the latter would be increasing the mass of the electric power source so it yields more power, and using that power (and a bit more mass) to increase the downlink data rate, thus increasing the mission data volume.

The second category of benefits is that given a particular launch vehicle, aerocapture can reduce the trip time from launch at Earth to the destination. This arises from the fact that as a consequence of the higher $V_\infty$ of approach that results from shortening a mission's trip time, the $\Delta V$ for orbit insertion increases. For a purely propulsive insertion the propellant mass needed for that $\Delta V$ increases quasi-exponentially with $\Delta V$, while the mass of the hardware needed for aerocapture increases approximately linearly with $\Delta V$. Thus, for distant destinations such as the ice giants, the transfer orbit from Earth can arrive with a higher $V_\infty$ of approach that would drive an all-propulsive insertion to an impractical propellant mass, while an aerocaptured insertion could accommodate the higher $\Delta V$ with relatively modest increases in aerocapture hardware mass. It turns out that for a given aerocapture flight system, the accuracy of the exit state achieved actually *improves* as the $V_\infty$ of approach increases. The reduction in trip time has several potential direct benefits. It reduces total operations costs because the total mission duration is shorter. It can reduce the cost of reliability engineering, a significant issue for missions with durations of more than ten years. And it would reduce the amount of degradation of radioisotope power source (RPS) output, so more power is available for science mission operations, or possibly the number of RPS units required for the mission could be decreased.

The third category of benefits is that given a fixed science payload and trajectory, aerocapture could allow launching on a less costly launch vehicle.

In some cases, a mission could benefit from both of the first two categories, delivering more mass to orbit with a shorter trip time.

### D.2.1.3 Applicability to Ice Giant Orbital Missions

Various characteristics of the ice giant planets make aerocapture an attractive option for orbital missions there. They orbit the sun at large heliocentric distances, so aerocapture's ability to handle a high $V_\infty$ of approach efficiently is significant. Uranus and Neptune have masses that are about 14.5 and 17.1 Earth masses, respectively, much smaller than Jupiter or Saturn, so the entry speeds at Uranus and Neptune can be smaller than at the gas giant planets, and the entries are less challenging for thermal protection system (TPS) technologies. Finally, their hydrogen-dominated atmospheres have large scale heights, which helps to reduce the aerodynamic control authority required to successfully execute an aerocapture maneuver, and can reduce the peak heating intensity experienced by the TPS, possibly reducing its mass.

#### D.2.1.4  Challenges for Ice Giant Orbital Missions

Some characteristics of the ice giant planets present challenges to the design of aerocapture systems and maneuvers there. Uranus and Neptune have been visited only by the *Voyager 2* spacecraft which did only flybys of both, so our knowledge of many aspects of those systems still has large uncertainties. Notably, uncertainties in atmospheric structure (pressure, temperature, and density as a function of depth) are considerably larger than for the atmospheres of Jupiter, Saturn, or Titan. This drives hardware designs toward higher required aerodynamic control authority and narrower entry corridor width. If relatively high control authority is indeed necessary, this could require an R&D effort for higher-L/D aeroshell shapes.

Those aeroshells, though entering an ice giant atmosphere at speeds slower than would be required at Jupiter, nonetheless experience harsh entry conditions. Typical entry speeds at the ice giants are in the 20–30 km/s range, so peak heating rates and heat loads are substantial. Many commonly-used TPS materials cannot handle such conditions, so the TPS must be carefully designed and the TPS material carefully chosen. In some cases of entry speeds on the high end of that range, qualification of candidate materials for conditions beyond currently qualified limits might be necessary, or possibly even development and qualification of more advanced materials.

Both ice giant planets have ring systems. While they are much less massive than Saturn's spectacular ring system they still pose a potential collision hazard to a spacecraft crossing the ring planes between some minimum and maximum planetocentric radii. Currently there is no theory unanimously accepted among ring scientists of what those maximum and minimum radii might be. Some argue that on the minimum side, a significant collision hazard exists all the way down to the upper reaches of the planet's atmosphere, where atmospheric drag quickly precipitates the particles. Others argue that between the inner edges of the observable rings and the atmosphere, orbital period resonances with moons can drive the eccentricities of the ring particle orbits, precipitating them into the atmosphere, thus making the ring particle densities inward of the resonances very sparse. Except for a relatively narrow range of interplanetary approach declinations, in most cases an aerocapture maneuver can be designed to have the vehicle within the planet's atmosphere while crossing the ring plane, avoiding this hazard. But if the mission is also delivering an atmospheric entry probe upon approach, this places additional constraints on the approach trajectory, and it might not be possible to have the orbiter and its aerocapture system in position to provide both data relay for the entry probe and an aerocapture trajectory that is within the planet's atmosphere at ring plane crossing.

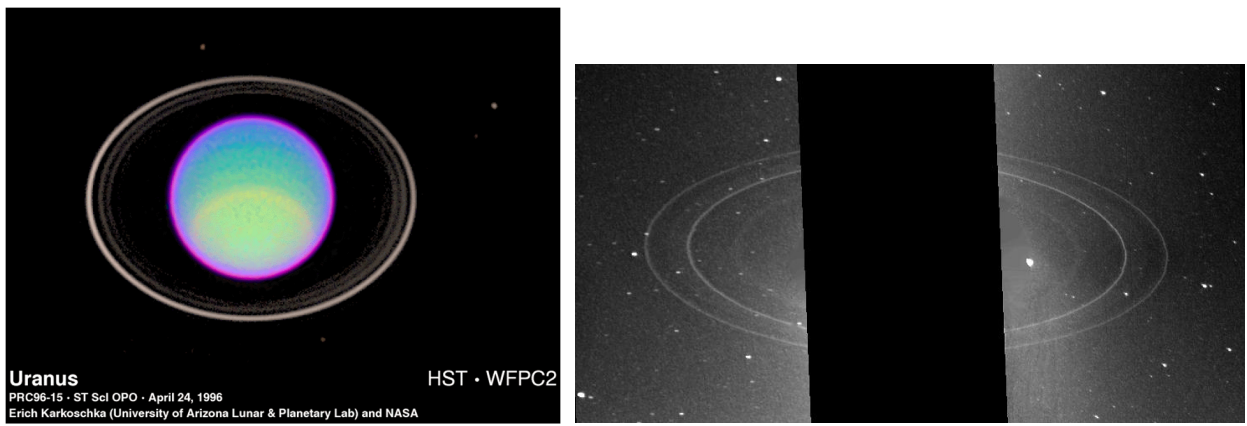

**Figure D-3.**  Uranus ring system from Hubble image (left); Neptune ring system viewed by Voyager 2 (right).

## D.2.2    Specific Aerocapture Options and Requirements at the Ice Giant Planets

### D.2.2.1  Introduction

The Ice Giants Mission study team conducted a collaborative study of the applicability and available options of aerocapture for ice giant planet missions.  The study was led by Purdue University, with participation and review by personnel from JPL, NASA's Ames Research Center and NASA's Langley Research Center.  Here we describe and summarize the high-level results of that study.  More detail is available in the final report generated by the Purdue team (Saikia et al. 2016).  This study made significant use of the results of a previous in-depth study by NASA's Aerocapture Systems Analysis Team (ASAT) of aerocapture at Neptune (Lockwood 2006).

### D.2.2.2  Critical Parameters and Characteristics of Aerocapture at the Ice Giants

There are a variety of characteristics of the ice giant planets and the aerocapture technique, and parameters of associated technologies, that influence feasibility assessments and system designs for aerocaptured missions to those destinations.  First are the fundamental characteristics of the destinations, covered above.  Also important are science goals and objectives that govern the target orbits for the science mission.  Programmatic aspects such as the maximum allowed duration of the Earth-to-destination transfer, or specified launch date windows that affect seasonally-varying approach circumstances, can affect assessments, and also can change with a changing programmatic environment, even after the flight project has been approved and funded.  Finally, there are many technological aspects that directly influence assessments and designs.  Various components, such as science instruments, avionics, etc., have environmental constraints, such as limits to inertial loading.  Entry and atmospheric flight system technologies have performance limits: TPS materials have limits to the temperatures, heating rates, pressure loads, and shear loads they can withstand; various aeroshell shapes have different ranges of lift-to-drag ratios (L/D) they can provide, notably with maximum performance limits; spacecraft trajectory control accuracy depends on a variety of guidance, navigation, and control (GNC) technologies with varying performance limits.

All the above combine to define an *entry corridor*, the maximum range of entry circumstances the aerocapture vehicle can tolerate and still yield a successful aerocapture maneuver, usually defined in terms of the entry flight path angle (EFPA), $\gamma$, the angle between the vehicle's flight path and the local horizontal plane.  If the aerocapture maneuver uses lift modulation for flight path control instead of drag modulation or some other means of control, the theoretical corridor width (TCW) is defined as the angular difference between $\gamma_{MAX}$, the steepest EFPA that allows the vehicle to reach the target apoapsis while flying lift-up over the entire maneuver duration, and $\gamma_{MIN}$, the shallowest EFPA that allows the vehicle to reach the target apoapsis while flying lift-down over the entire maneuver duration.  The TCW for a given flight system varies with a large number of conditions and parameters, but notably it varies with the approach circumstances, especially the atmosphere-relative entry speed and the chosen nominal EFPA.  For aerocapture purposes, those two are not entirely independent.  For instance, an entry speed that is acceptable for a given choice of nominal EFPA can become unacceptable for a steeper or shallower choice of EFPA.

Given a value for the TCW, the GNC performance for a flight project must be sufficient to deliver the aerocapture vehicle within the entry corridor.  Errors in spacecraft navigation and planetary ephemerides principally contribute to variation in the EFPA; errors in entry speed are relatively smaller and usually do not pose serious issues.  Errors in spacecraft navigation are primarily errors in the determination of the spacecraft's absolute location with respect to the destination.  For a spacecraft 20 AU (~3 billion km) from the Sun, knowledge of the spacecraft's heliocentric position to within 100 km represents exquisite accuracy, ~3 parts in $10^8$. But if the spacecraft is 50,000 km from its destination, and the destination's position is also known only to within 100 km, the position of the spacecraft with respect to the destination is known to only ~3 parts in $10^3$, roughly 5 orders of magnitude worse. This position error leads to errors in the b-plane aimpoint, and that leads to errors in the EFPA: the smaller the b-plane aimpoint is from the destination's center, the steeper the EFPA. A primary goal of aerocapture mission design is to design a system, using available or anticipated technologies, that provides an entry corridor at the destination sufficiently wide that the flight system can deliver the aerocapture vehicle within that entry corridor, despite the errors inherent in the spacecraft's GNC and the destination's ephemeris.

### D.2.2.3  Collaborative Study

#### Study Objectives

The Purdue study examined the application of aerocapture for missions to the ice giant planets, with a list of objectives to pursue and questions to answer:

4.  Given the mass of the spacecraft and entry speed as parameters, and making rough assumptions about the vehicle configuration, develop rough design rules for:
    e.  Mass sizing of aerocapture systems
    f.  Plane change that can be attained

5.  Can aerocapture bump up the "mass class" of a mission? E.g., from medium class with chemical propulsion to flagship class with aerocapture, where the criterion is the useful mass inserted into orbit. For example, assuming a Delta-IVH-class launch vehicle, could aerocapture give a flagship mission to Uranus for ~11-year time of flight (TOF) when Jupiter and Saturn are unavailable for gravity assists, or to Neptune for an ~13-year TOF?

6.  Can aerocapture reduce flight time for orbital missions by three or more years, while maintaining the useful inserted mass and not switching to a larger launch vehicle?

7.  Based on the analyses, can we:
    a.  Develop design relationships that can be used in flight system sizing for future studies involving aerocapture?
    b.  Determine which technologies and related investments would be required?
    c.  Identify knowledge gaps?
    d.  Identify tasks and pathways for further investigations?

#### Challenges

The study identified three characteristics of the problem that lead to significant challenges. First, both the limitations on TOF and need for an acceptable TCW steer toward very high entry speeds and thus high aerothermodynamic loads. Second, the desire for capture into orbits with periods of 20 days or more yields an aerocapture maneuver exit speed greater than 99% of the destination's escape speed, so there must be reliable control of the post-aerocapture state. Finally, it appears that high values of the lift ballistic coefficient are needed, presenting challenges for spacecraft packaging and flight path control authority.

#### Results of Analyses

This study conducted a series of analyses addressing many aspects of aerocapture at the ice giant planets, drawing upon the results of the ASAT study (Lockwood 2006) when possible. Some of the most important analyses and their results are summarized in this section. More detail will be available in a publication by the Purdue team (Saikia et al. 2016).

One of the issues the study addressed is the level of L/D needed for aerocapture at an ice giant, given the large uncertainties in those atmospheres' structures and the TCW needed to accommodate anticipated GNC accuracies. It turns out that the *slower* the $V_\infty$ of approach, the higher the L/D must be, as shown in **Figures D-4** and **D-5**, and this sets the lower limit for the approach $V_\infty$. The option of adopting increased approach $V_\infty$ to minimize L/D requirements and reduce exit state uncertainties has its limits, because above some approach $V_\infty$ value the TPS materials available can no longer handle the harsh entry conditions arising from high-speed entries. The conclusion of the ASAT team (Lockwood 2006) is that for both Uranus and Neptune, an aeroshell with an L/D in the range of 0.6–0.8 is needed, probably in the high end of that range. The Purdue team concurs with that result.

The harsh entry conditions mentioned above include peak deceleration rates, peak heating rates, and absorbed heat loads. These values increase quickly with increasing approach $V_\infty$, especially the heating rates and loads. **Figures D-6** and **D-7** show peak decelerations vs. approach $V_\infty$ at Uranus and Neptune, respectively, parametric in the ballistic coefficient $\beta$ and L/D. **Figures D-8** and **D-9**

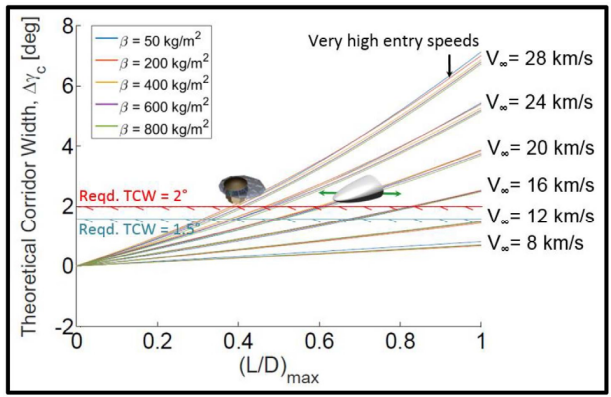

Uranus

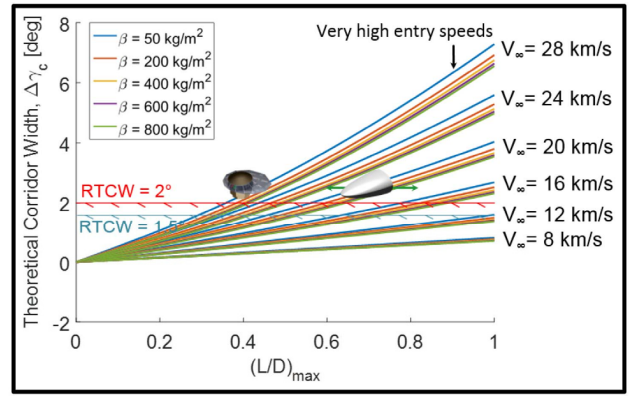

Neptune

**Figure D-4.** Key Parameters determining the achievable theoretical corridor width for Uranus. Higher L/D and entry speeds are advantageous. Ballistic coefficient has only a small effect.

**Figure D-5.** Key Parameters determining the achievable theoretical corridor width for Neptune. Relationships are similar to those for Uranus (**Figure D-4**) but ballistic coefficient has a larger effect for Neptune.

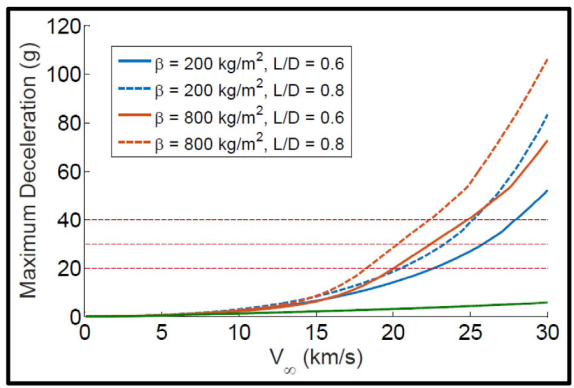

Uranus

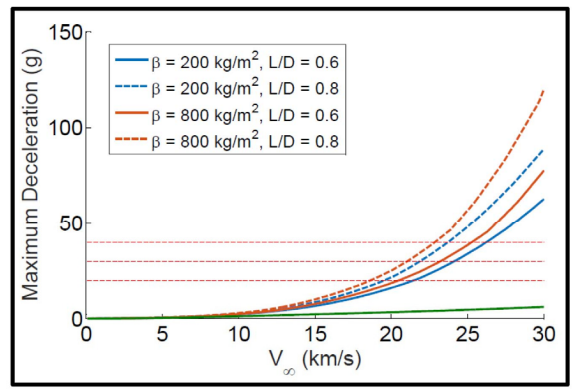

Neptune

**Figure D-6.** Dependence of maximum deceleration on entry velocity, L/D and ballistic coefficient for Uranus.

**Figure D-7.** Dependence of maximum deceleration on entry velocity, L/D and ballistic coefficient for Neptune.

show similar plots of peak heating rates. There is a horizontal dotted line in those figures at 7 kW/cm², which is 1 kW/cm² less than the current tested limit of the HEEET TPS material being qualified by NASA's ARC (Ellerby 2014). **Figure D-10** is another similar plot for Uranus, showing heat load vs. approach $V_\infty$.

Another interesting result is that the choice of initial post-aerocapture orbit period, and thus the apoapsis radius of that orbit, influences several aspects of the aerocapture maneuver and conditions experienced during the maneuver. For instance, for orbit periods less than ~10 days, the shorter the orbit period, the wider the TCW as constrained solely by the control aspect; other aspects, such as heating rates and loads as discussed above, can further constrain the TCW. But this TCW constraint is nearly constant for orbit periods of ~10 days or longer, as shown in **Figures D-12** and **D-13**. Another aspect is the maximum deceleration rate (and thus inertial loads) needed to reach a given initial orbit period, and the closely associated maximum dynamic pressure, shown for the Uranus case in **Figures D-14** and **D-15** for two values of the ballistic coefficient. Similarly, for orbits with periods less than ~10 days the TCW increases with decreasing orbit period, but is roughly constant for orbit periods greater than ~10 days.

From the constraints above, and characteristics of structural materials, TPS materials, and aerocapture-associated components, it was possible to derive rough scaling laws for the mass of the aerocapture system needed for the maneuvers at ice giants. There is a caveat: there is no empirical relation giving the TPS mass needed to handle a given heat load. A relatively accurate estimate requires CFD analyses, beyond the scope of this investigation. The previous ASAT study performed such CFD analyses (Lockwood 2006) and can be used for rough estimates, but those analyses were based on different TPS materials such as heritage carbon phenolic, so there is added uncertainty in applying those results to this study. **Figure D-16** is a plot of the total aeroshell mass required given the 'Total Aerocapture Mass', i.e., the vehicle's total mass at the entry interface, parametric in the 'payload mass' and the cleanup ΔV propellant. 'Payload mass' is the Total Aerocapture Mass less the mass of the aerocapture system jettisoned after exit. For a given launch vehicle and transfer trajectory the Total Aerocapture Mass has an upper limit given by the launch capacity of the launch vehicle, decremented by any propellant used and any equipment jettisoned during the transfer. Jettisoned equipment might include items such as a SEP stage or a transfer-phase-only telecom antenna that would not be stowed within the aeroshell for the aerocapture maneuver.

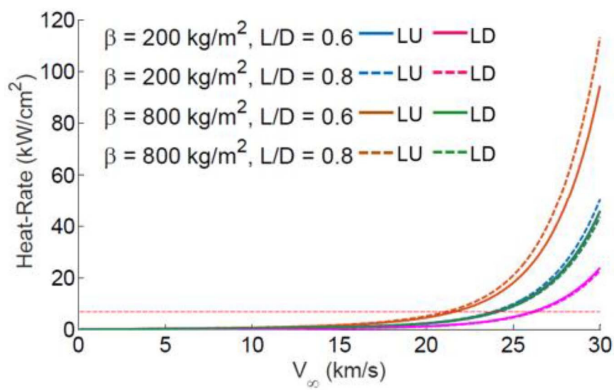

**Figure D-8.** Dependence of Heating Rate on key parameters for Uranus. LU indicates Lift Up and LD indicates Lift Down.

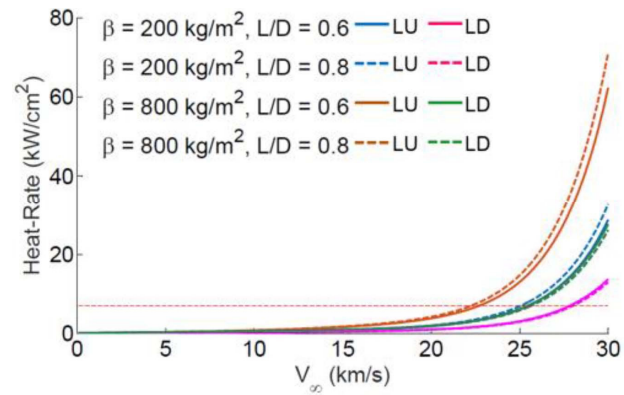

**Figure D-9.** Dependence on Heating Rate on key parameters for Neptune.

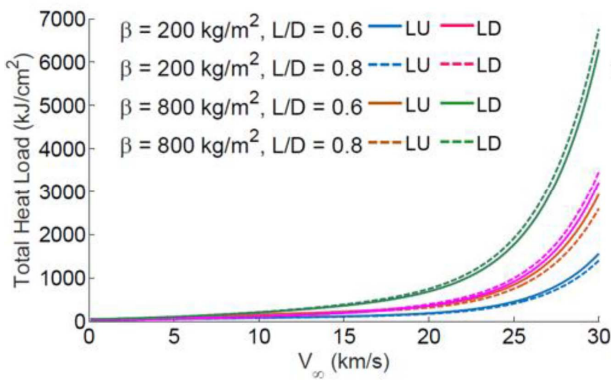

**Figure D-10.** Dependence of Total Heat Load on key driving parameters for Uranus.

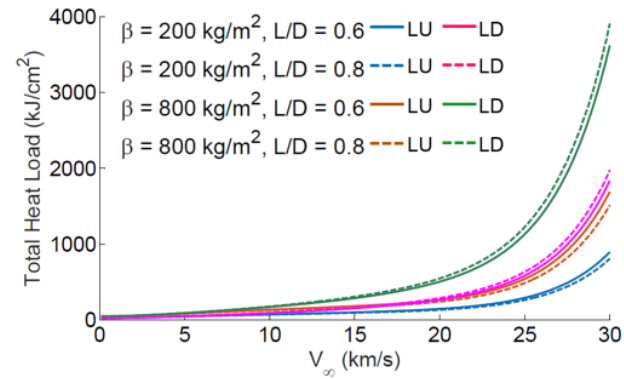

**Figure D-11.** Dependence of Total Heat Load on key driving parameters for Neptune.

Parameters of the interplanetary approach influences significant aspects of the initial post-aerocapture orbit. As seen above, the approach $V_\infty$ is one of these parameters. Another is the 'declination of the approach asymptote' (DAP), the angle between the approach $V_\infty$ vector and the destination's equatorial plane. This is particularly important for Uranus due to its extreme obliquity and thus wide seasonal variation of the available DAPs, and its retinue of several large natural satellites, all orbiting very near the equatorial plane. A Uranus system tour including those satellites is best done from a prograde orbit within or very near the equatorial plane because that inclination minimizes the flyby velocities at the satellites. The difficulty of maneuvering to the equatorial orbit from the initial orbit is a strong function of the DAP. At Neptune there is not such a strong preference for an equatorial orbit, but there is still science motivation for an orbit that allows multiple flybys of Triton, and the difficulty of attaining that orbit also is a function of the DAP. **Figures D-17** and **D-18** show the DAPs for Uranus and Neptune, respectively, for various families of transfer trajectories and arrival dates. It is obvious from those plots that the range of variation is far greater for Uranus than for Neptune. Elements of the plot are color-coded to show values of the approach $V_\infty$.

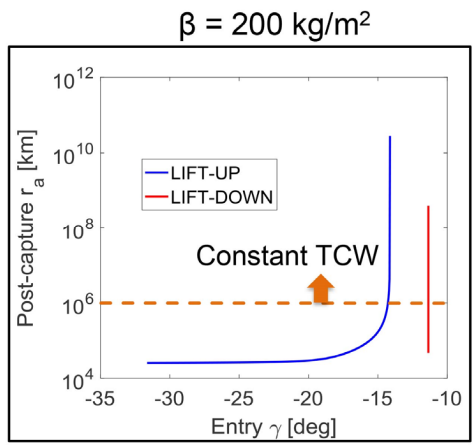

**Figure D-12.** Post Capture apoapsis radius as a function of entry angle for a Beta = 200kg/m² for Uranus.

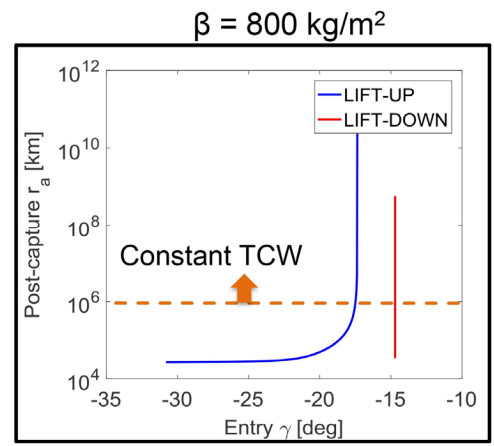

**Figure D-13.** Post Capture apoapsis radius as a function of entry angle for a Beta = 800kg/m² for Uranus.

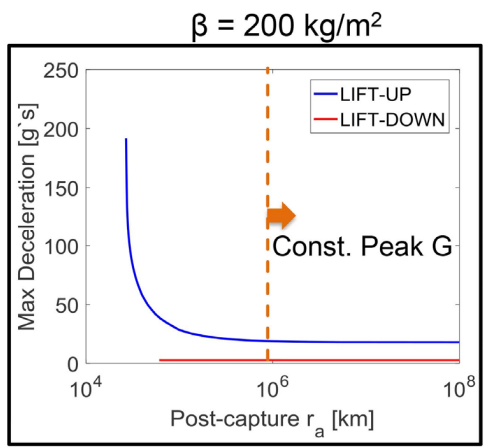

**Figure D-14.** Maximum deceleration as function of postcapture apoapsis radius for Uranus.

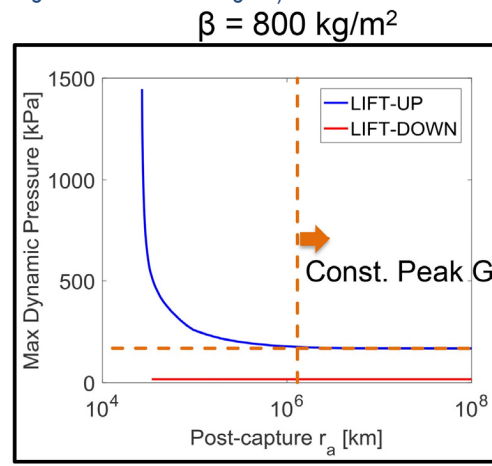

**Figure D-15.** Maximum dynamic Pressure as a function of posture apoapsis radius for Uranus.

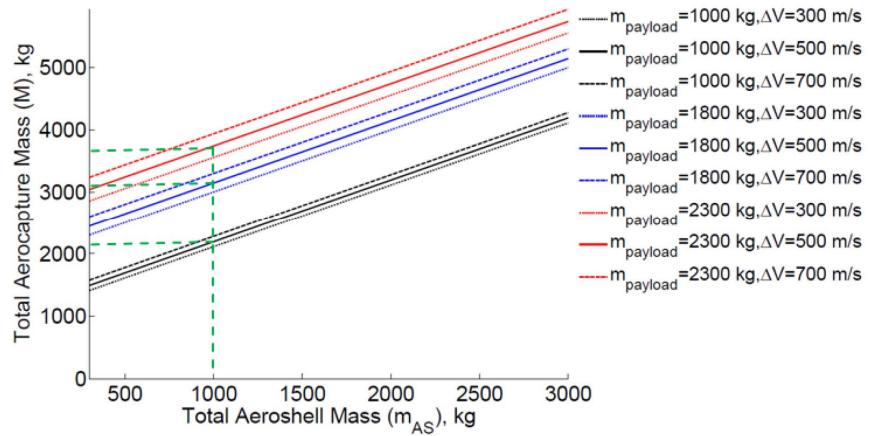

**Figure D-16.** Total Aerocapture Mass as a function of Total Aeroshell Mass and other key parameters.

Two of the orbit adjustments that can be needed after capture into an initial orbit are to rotate the line of apsides, the line between the orbit's periapsis and apoapsis, and to change the orbit's plane (rotating the line of apsides does not change the orbit plane). These might be done to achieve a particular desired orbit plane, such as the equatorial plane for Uranus. **Figure D-19** illustrates the high $\Delta V$ required for such rotations of the line of apsides if done propulsively. For orbits with periods greater than a few days, this $\Delta V$ cost is relatively insensitive to orbit period. But an aerocapture maneuver can be designed to perform some intentional rotation of the line of apsides so the initial orbit has its line of apsides at or nearer to the desired orientation, as shown in **Figure D-20**. This could save considerable propellant mass. At the same time, the aerocapture maneuver can be designed to provide some plane change ('crank' or 'twist' maneuvers), again at a savings of

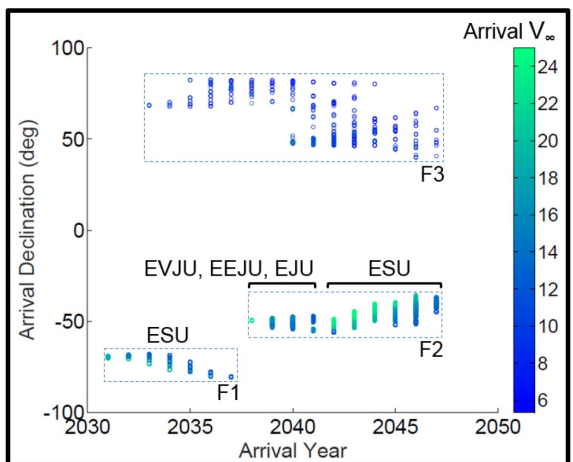

Uranus

**Figure D-17.** Arrival declination for different trajectories for Uranus. Color coding indicates arrival $V_\infty$.

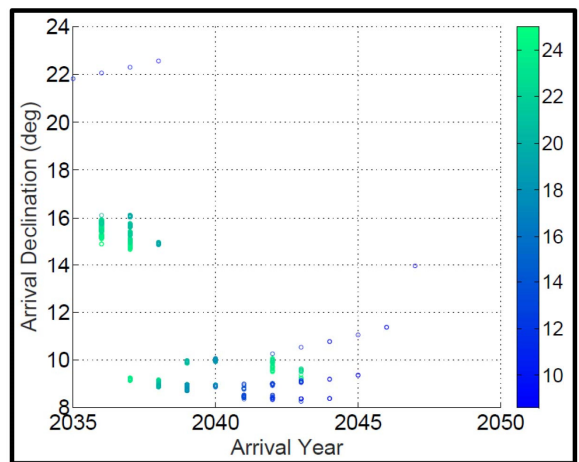

Neptune

**Figure D-18.** Arrival declination for different trajectories for Neptune. Color coding indicates arrival $V_\infty$.

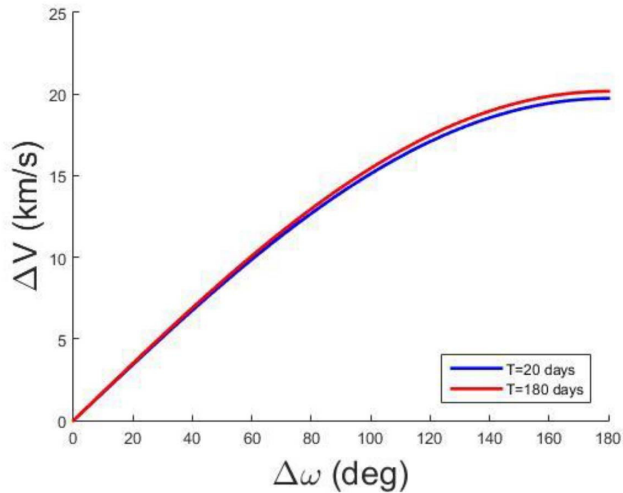

**Figure D-19.** Required ΔV for propulsive rotation of the line of apsides.

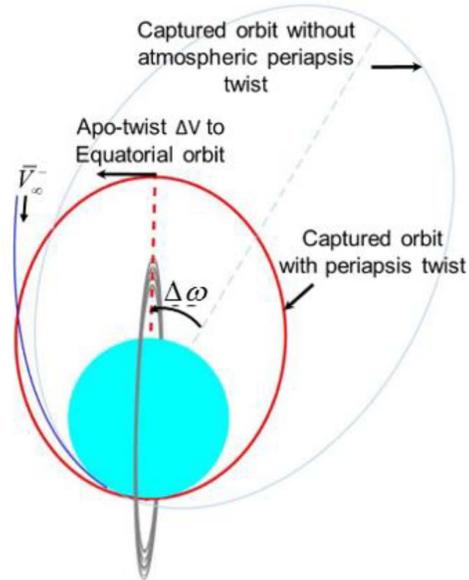

**Figure D-20.** Use of aerocapture maneuver to accomplish initial rotation of the line of apsides.

propellant mass. **Figure D-21** shows the ΔV cost of plane changes (done at apoapsis) as a function of orbit period. Those costs can be rather high for short-period orbits. A technique often considered for such adjustments is to perform the crank at the apoapsis of a long-period orbit, then reduce the orbit period via a periapsis maneuver. While this reduces the ΔV cost, that cost can still be substantial. Having the aerocapture maneuver include a plane-change component can save the propellant mass associated with a propulsive plane-change maneuver. The maximum plane-change capability is a function of several parameters, but primarily depends upon the approach $V_\infty$ and the vehicle's L/D, of course limited by the ability of the vehicle to withstand the entry conditions encountered. **Figure D-22** shows the maximum possible crank angle as a function of approach $V_\infty$, parametric in L/D (and $\beta$, which has only a minor influence) for both Uranus (left) and Neptune (right). For the approach circumstances considered in this study the maximum $\Delta i$ appears to be roughly 10° for high approach $V_\infty$'s, and is considerably lower for lower approach $V_\infty$'s.

### High-Level Study Results

The Purdue team designed a single chart, an Aerocapture Applicability and Feasibility (AAF) chart, to capture the results of many different analyses and show how they establish an envelope of feasibility for use of the aerocapture technique (Saikia et al. 2016). **Figure D-23** shows an example of such a chart. The chart axes are the approach $V_\infty$ of the interplanetary transfer trajectory for the abscissa (x axis), and the resulting orbit insertion ΔV for the ordinate (y axis). For the same approach $V_\infty$ the effective insertion ΔV is slightly different between chemical and aerocapture insertion because the effective altitudes of the maneuvers are different, so there are two closely spaced curves representing those $V_\infty$–ΔV relations. There are several colored and stippled vertical lines at various values of the approach $V_\infty$. These represent constraints on the approach $V_\infty$ arising from aspects of the aerocapture hardware and navigation capabilities. The stipples are on the disallowed side of the line, so stipples on the left side of the line indicate a lower limit on the approach $V_\infty$, and stipples on the right side of the line indicate an upper limit. Two red lines represent lower $V_\infty$ limits for two values of minimum TCW. If the example mission's navigation systems could deliver to a 1.5° TCW, then an approach $V_\infty$ as low as ~12 km/s could be tolerated. But if a 2° TCW is needed, then the approach $V_\infty$ would have to be ~16 km/s or more. Two blue lines represent upper limits to the inertial load-handling capability of the vehicle. If that limit is 20 g, then the maximum tolerable $V_\infty$ is ~18.5 km/s. If that limit is 30 g, then the maximum tolerable $V_\infty$ is slightly more than 20 km/s. The highest upper limit line, in violet, represents the ~21 km/s maximum capability of the HEEET TPS material to withstand the entry conditions, given a vehicle with an L/D of 0.8.

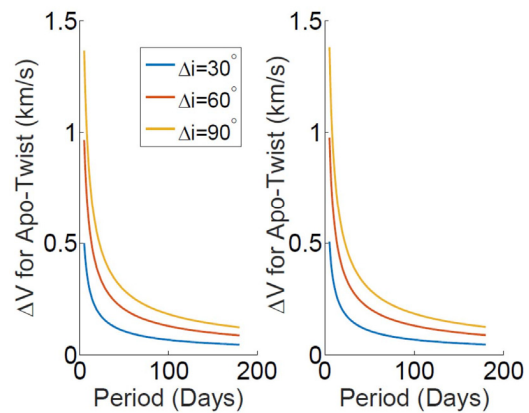

**Figure D-21.** ΔV cost of plane changes as a function of orbital period.

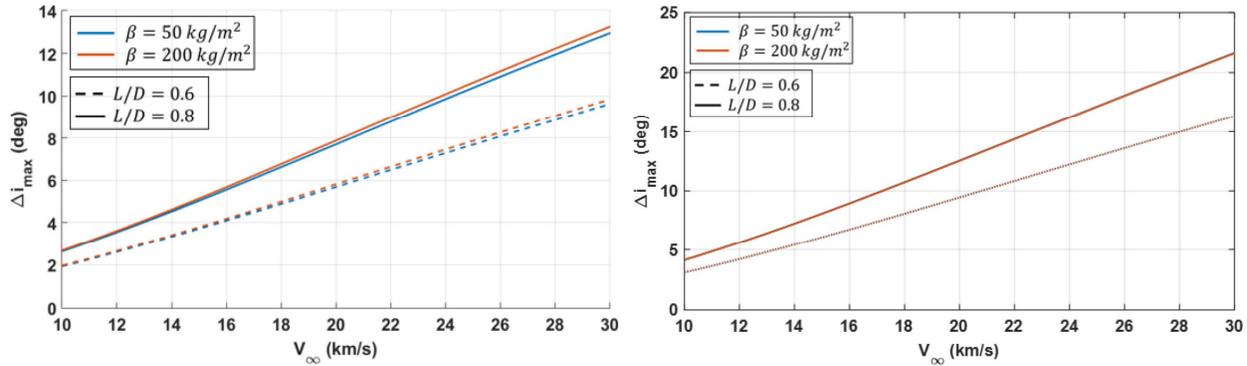

**Figure D-22.** Crank angle as a function of approach $V_\infty$ for Uranus (Left) and Neptune (right).

The single horizontal stippled line represents the maximum insertion $\Delta V$ capability given a limit to the aerocapture mass fraction (ACMF), the fraction of the vehicle's entry mass that is hardware or propellant dedicated to the aerocapture maneuver. In this example chart it is shown at ~9.5 km/s.

The range of feasible approach $V_\infty$'s for aerocapture is that range of $V_\infty$'s that are higher than the greatest lower limit and lower than the least upper limit. In this example, for a 2° TCW minimum, that window is from ~16 km/s to ~18.5 km/s; if the minimum TCW is 1.5°, the window is much wider due to extension of the low side down to ~12 km/s.

The chart has two more stippled lines associated with chemical insertion instead of aerocapture. The black horizontal line represents the ~3.5 km/s $\Delta V$ limit for space-storable, bipropellant chemical systems. This yields an upper limit to the approach $V_\infty$ of just less than 12.5 km/s. The Purdue team notes that if a 2° TCW is required, the aerocapture approach $V_\infty$ window and the chemical $V_\infty$ window do not overlap, and that this is by far the most common situation.

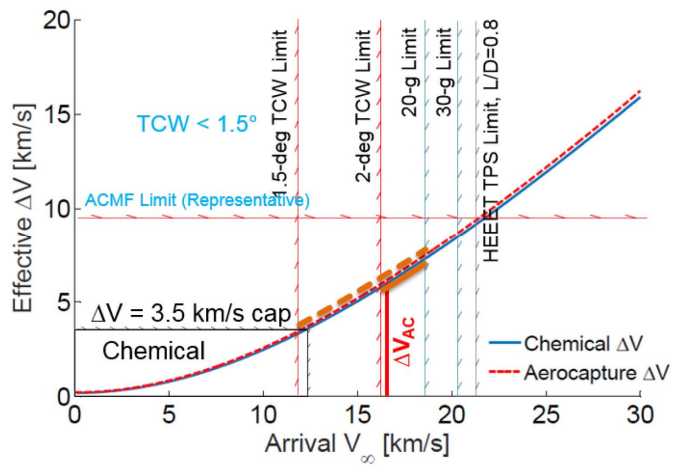

**Figure D-23.** Aerocapture Applicability and Feasibility (AAF) chart showing performance envelope for aerocapture technique.

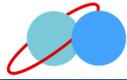

**Figures D-24** and **D-25** give the Purdue team's assessment of the AAF charts for Uranus, for two values of the vehicle's ballistic coefficient. **Figures D-26** and **D-27** give similar charts for Neptune. Note in **Figures D-24** and **D-25** that for a TCW of 1.5° there is some overlap between the approach $V_\infty$ windows for aerocapture and chemical propulsive insertion. For a TCW of 2° there is a relatively narrow $V_\infty$ window, somewhat wider for a $\beta$ of 200 kg/m² than for a $\beta$ of 800 kg/m². Both have a lower limit of ~16 km/s established by the 2° TCW, but with (for instance) a 20-g limit on inertial loading, the upper limit for $\beta$=800 is ~18.5 km/s, where for $\beta$=200 it moves up to 20 km/s. The AAFs for Neptune are similar, with the primary difference being that the approach $V_\infty$ lower limit for a 1.5° TCW increases by more than 2 km/s, so there is no overlap between the aerocapture and chemical propulsive windows.

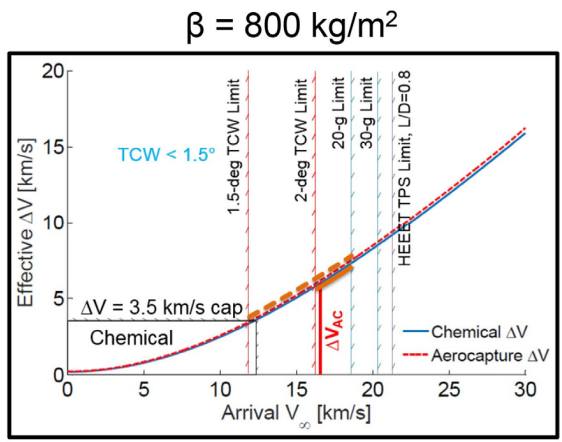

**Figure D-24.** AAF chart for Uranus for ballistic coefficient of 800kg/m².

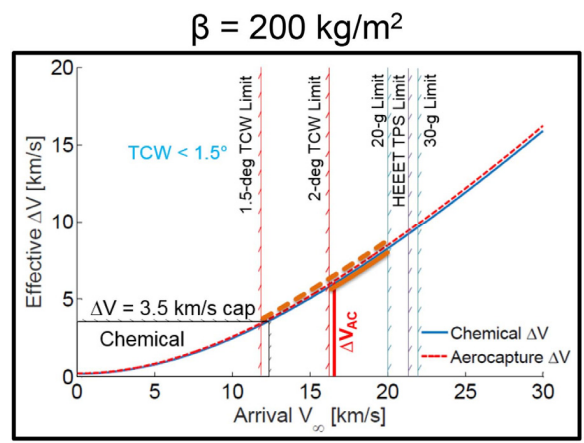

**Figure D-25.** AAF chart for Uranus for ballistic coefficient of 200kg/m².

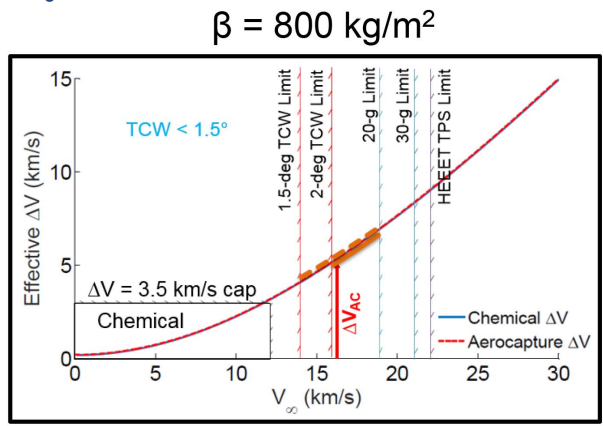

**Figure D-26.** AAF chart for Neptune for ballistic coefficient of 800kg/m².

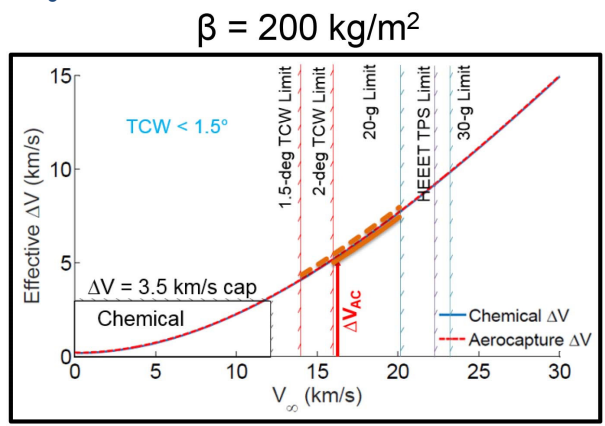

**Figure D-27.** AAF chart for Neptune for ballistic coefficient of 200kg/m².

Several high-level results are presented as responses to the objectives and questions in C.1 above. These are treated in order below.

1. *a. Given the mass of the spacecraft and entry speed as parameters, and making rough assumptions about the vehicle configuration, develop rough design rules for mass sizing of aerocapture systems.*

    Estimating the mass of an aeroshell for aerocaptured missions in general was found to be a non-trivial task. Estimates established for the Ice Giants Mission study relied heavily upon the results of the ASAT study of aerocapture at Neptune that did detailed

computational analyses such as CFD based upon advanced aerothermodynamical models. There were no indications that simple relations independent of those detailed study results were forthcoming without significant new research efforts.

1. b. *Given the mass of the spacecraft and entry speed as parameters, and making rough assumptions about the vehicle configuration, develop rough design rules for the amount of plane change that can be attained*

   For the range of arrival $V_\infty$ for which aerocapture is feasible, the maximum theoretical plane change possible during aerocapture is ≤10°. Practical plane change limits will be smaller than that.

2. *Can aerocapture bump up the "mass class" of a mission? E.g., from medium class with chemical propulsion to flagship class with aerocapture, where the criterion is the useful mass inserted into orbit. For example, assuming a Delta-IVH-class launch vehicle, could aerocapture give a flagship mission to Uranus for ~11-year time of flight (TOF) when Jupiter and Saturn are unavailable for gravity assists, or to Neptune for an ~13-year TOF?*

   There are promising trajectories involving aerocapture with adequate arrival $V_\infty$, short time-of-flight (TOF) of between 4–8 years, and low launch $C_3$, which could be within Delta IV-H capability. In addition, for Neptune, there are trajectories with promising low TOF between 4–8 years and low launch $C_3$. While the delivered mass at both the planets are not known exactly at this time, if the promising trajectories and the Delta IV-H launch capacity can deliver about 3.5–4 tons to Uranus or Neptune, then a flagship class mission might be possible using aerocapture. To verify, the delivered mass needs to be investigated in more detail.

3. *Can aerocapture reduce flight time for orbital missions by three or more years, while maintaining the useful inserted mass and not switching to a larger launch vehicle?*

   Given the explanation in #2 above, there are short TOF (4–8 years) trajectories that could deliver sufficient mass to Uranus or Neptune. If such short TOF trajectories are possible for a reasonably high delivered mass, then aerocapture can reduce flight time for orbital missions by three or more years without switching to a larger launch vehicle.

4. a. *Based on the analyses, can we develop design relationships that can be used in flight system sizing for future studies involving aerocapture?*

   (See 1.a above)

4. *b, c, & d. Based on the analyses, can we a) determine which technologies and related investments would be required? b) identify knowledge gaps? c) identify tasks and pathways for further investigations?*

   Given the current state of knowledge about the atmospheres, and current technological and programmatic constraints for a flagship-class mission to an ice giant planet in the time frame considered, use of aerocapture for such a mission would require developing, testing, and qualifying a mid-L/D (0.7–0.8) aeroshell for the vehicle. This would require a commitment of substantial funding over an effort lasting about 10 years. Currently no other exploration endeavor calls for such a technology so the development burden would rest entirely with the ice giant mission project. However, it is anticipated that once the technology became available, several other useful applications would be evident, such as single aerogravity-assist trajectories at Venus or Mars to allow ~2.5-year transfers to Jupiter with launch $C_3$s in the 12–25 km²/s² range.

Alternate technological pathways might allow less costly approaches. For example, reducing the current uncertainties in atmosphere models, navigation, and vehicle aerodynamics might reduce the required TCW to the point that heritage low-L/D (~0.4) vehicles would be sufficient. Other options include use of a gravity assist maneuver at a large satellite to reduce the aerocapture ΔV requirement, or use of a hybrid of aerocapture and chemical propulsion, where aerocapture removes part of the excess energy and a chemical propulsion system removes the rest.

### D.2.3    Summary

Although aerocapture has not yet been demonstrated with a space flight mission, decades of study and development of useful hardware components have it poised for implementation on flight missions. The ice giant planets' remote locations in the solar system, the relatively large uncertainties in models of their atmospheres, and current performance limits of applicable technologies (such as GNC accuracies) combine to make aerocapture at the ice giants challenging but feasible, if certain technologies are developed. If no advances are made in the areas of atmosphere models, GNC, or vehicle aerodynamics that reduce their associated uncertainties, the primary development needed would be a mid-L/D (0.7–0.8) vehicle. That development would involve significant commitments of time and funds, but if started soon could be ready for a launch in the mid- to late-2020's. Funding of R&D tasks that yield significant advances in atmosphere models, GNC, and vehicle aerodynamics might allow aerocapture at the ice giants with a heritage low-L/D (~0.4) vehicle.

Aerocapture offers three primary classes of advantages for ice giant missions. It can increase the fraction of the approach mass that is delivered to orbit, it can decrease significantly the TOF from Earth to the destination, or it can allow launch on a smaller launch vehicle. All of these allow reduction of costs for a given science return or increased science return for a fixed project budget.

In addition to capturing into orbit at the destination, adjusting such orbit parameters as apoapsis radius and orbit period, there are other orbit adjustments that can be made during the aerocapture maneuver using the lift generated by the vehicle, though the range of these adjustments are limited. One is to change the plane of the post-aerocapture orbit away from the plane of the approach orbit. Given the values of the aerocapture vehicle's L/D and the approach $V_\infty$, the magnitude of possible plane change during an aerocapture maneuver is limited to ~10° or less. Another adjustment is to rotate the orbit's line of apsides (the line between the orbit's periapsis and apoapsis) into a target plane, such as the destination's equatorial plane.

### D.3    Cryogenic Propulsion

Typically, orbital missions with shorter trip times to Ice Giants will require larger delta V to achieve orbit. One approach to overcoming the limitations of conventional propellants is to use aerocapture as discussed in Section A.4.1. Another option is the use of higher performance propellants. The principal challenge with higher performance propellants is that they require cryogenic storage and to be kept at cryogenic temperatures up until the point of use. For many applications, this requires powerful refrigeration systems consuming large amounts of power. However, for an Ice Giants mission where most of the mission is spent in the outer solar system this situation is eased considerably and cryogenic propellant may be an attractive alternative for fast missions to Uranus and Neptune.

### D.3.1    Benefits of Cryogenics Propulsion

An informative metric for assessing the relative merits of various propulsion systems is that of useable or useful mass inserted into orbit. This is the mass of the spacecraft post-orbit-insertion at the destination planet minus the propulsion system mass, namely the mass of the propellant tank, the engine itself, the plumbing, any cryogenic systems, and other propulsion-specific hardware. For a conventional bi-propellant propulsion system, the specific impulse may be expected to be around 325 s, and the propulsion system mass may be up to about 15% of the propellant mass. For a Near-Zero-Boil-off cryogenic LOX/LH2 propulsion system for an Outer Planets mission, the propulsion system mass can be expected to be around 25% of the propellant mass, and a typical specific impulse is about 450 s.

The useable mass fraction under each of these assumptions can then be easily computed as a function of the orbit insertion $\Delta V$. This is shown graphically in **Figure D-28**, where the useable mass is plotted assuming an initial, pre-burn, spacecraft wet mass of 4000 kg. For example, for a $\Delta V$ of 4 km/s, the conventional system and the cryogenic system (bold red and dashed blue curves) have useable masses of about 690 kg and 1000 kg, respectively. That is, the cryogenic system has a useable mass that is almost 45% higher than the conventional system. The cryogenic system offers clear mass advantages over the conventional system over a wide range of $\Delta V$s. Even for values as low as 2 km/s, the useable mass increases by about 300 kg (~15% improvement), and for values as high as 5 km/s, the increase is still about 250 kg (~90% improvement).

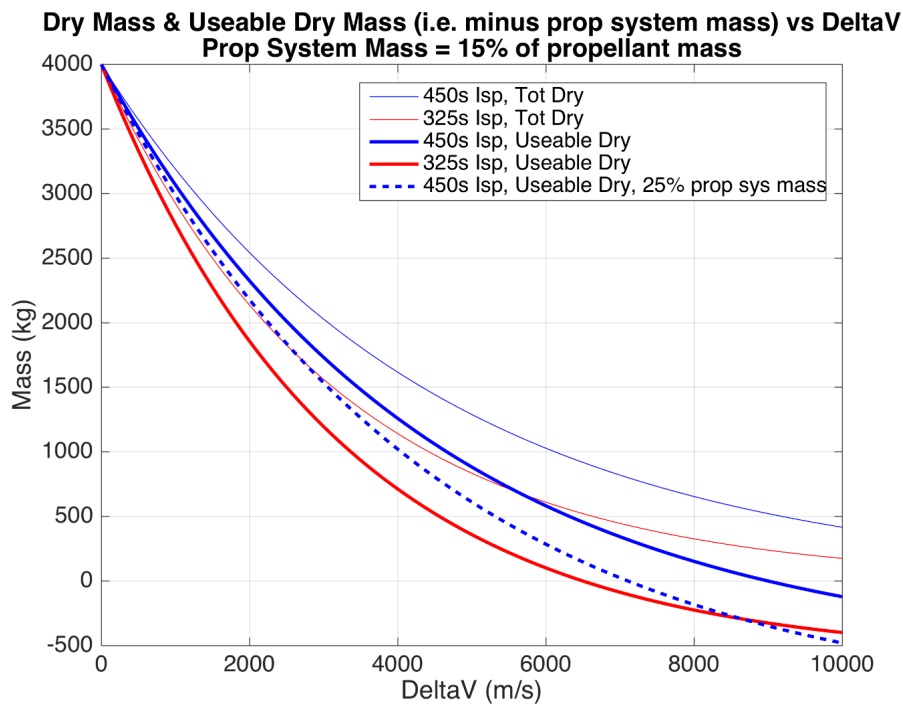

**Figure D-28.** Dry mass and useable dry mass for conventional and cryogenic propulsion systems.

A further potential benefit is that if the cryogenic system were jettisoned after orbit insertion, the $\Delta V$s for flying a science tour would be considerably less costly in terms of propellant mass (at 25%, the cryogenic system mass for a 4 km/s burn on a 4000 kg wet-mass spacecraft is about 600 kg).

The sweet-spot ΔV for cryogenic systems may be about 4 km/s. At these values, flight times to the Ice Giants can be reduced by about 2–4 years. These faster missions will typically not perform any large ΔVs in the inner solar system, and will often recede monotonically from the Sun after launch, which eliminates the need for cryogenic coolers, relying instead on a sun-shade system for thermal control.

### D.3.2    Zero Boil Off Cryogenic Storage

Outer planet missions are very favorable for implementing a Near Zero Boil off cryogenic storage system because they can be shielded from direct sunlight during the early part of the cruise phase near the earth and by the time that they need to use propulsion they are so far from the Sun that deviations from the sunline are tolerable. A new analysis was not required for this study; a previous analysis of a Titan Explorer mission carried out in 2014 is quite adequate.

In contrast, the use of cryogenic coolers is likely to be impractical. They require a lot of power and power is expensive for an outer planet spacecraft which relies on low specific power radioisotopic power sources.

The baseline configuration is shown in **Figure D-29**. It is launched as shown in the inverted position and uses a single set of tanks and a side mounted entry vehicle. This design minimizes the structural requirements on the LH2 tank by placing it and the engine system at the top of the stack. By using a single LOX tank in line with the LH2 tank, the LH2 tank's view to space is maximized. This design allows for relatively straight forward shielding and isolation of the tanks from spacecraft thermal loads. It also provides access to the spacecraft bus and RTGs from the standard payload fairing doors. This configuration effectively decouples the propulsion stage design from the spacecraft bus design (for the most part), allowing this configuration to be used as a starting point for other studies.

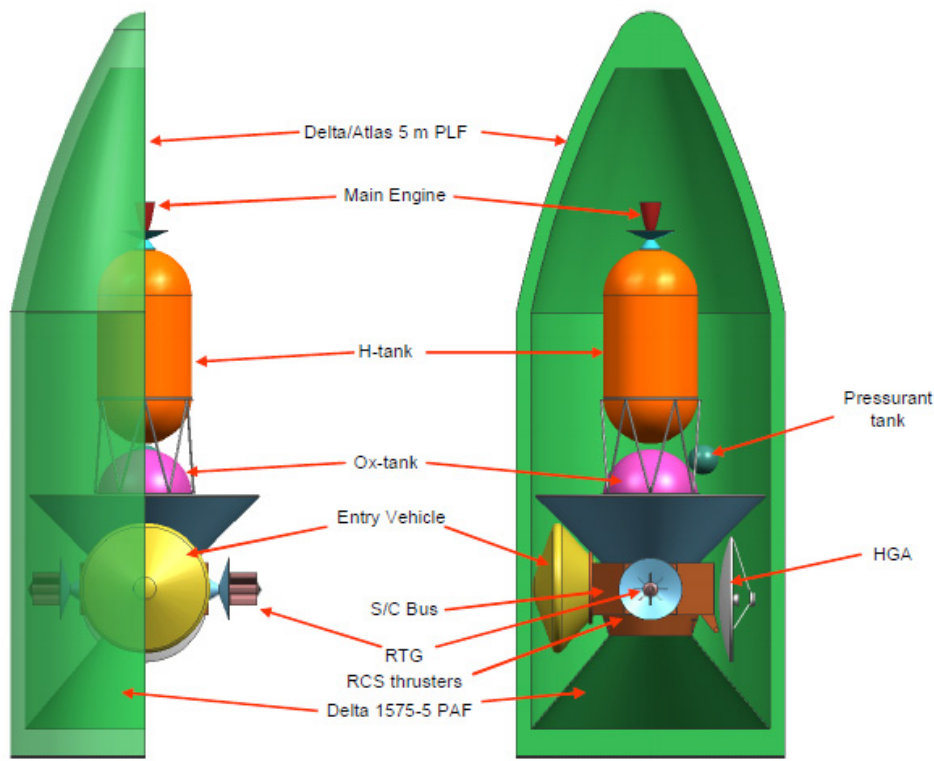

**Figure D-29.** Baseline spacecraft configuration - Titan Explorer shown.

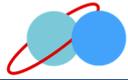

A thermal shield is placed between the tanks and the bus to shield the tanks from the Sun and the spacecraft bus under normal operation. The structure supporting the tanks consists of gamma-alumina bipod struts using the passive orbit disconnect strut (PODS) end fittings to minimize the thermal loads after launch). Gamma alumina was assumed to maintain heritage with the Gravity Probe B helium dewar supports. However, other materials may be more optimal for LH2 supports. The spacecraft normal pointing mode is aligned with the Sun vector such that the tanks are shaded. This design can tolerate short durations of time with the Sun directly in the field of view of the tanks for maneuvers, communications, and safe-mode events. A thermal model of the concept is shown in **Figure D-30**.

A Zero Boil-Off Cryogenic Propellant Systems analysis was performed for the Titan study used several different analysis tools. Thermal loads from environmental conditions were calculated using Thermal Analysis System (TAS), which uses a finite difference solution to solve complex, nonlinear models with temperature dependent properties. The thermal loads generated from TAS were used as inputs to the Cryogenic Analysis Tool (CAT). In turn CAT determined the time dependent fluid conditions of the propellants over the entire mission life. CAT was used to ensure the specific ZBO design did not overpressure the propellant tanks.

When using the SOA design on the Titan Explorer, it was found that the 12 struts between the LOX and LH2 tanks would leak about 0.2W. Although the magnitude of this heat leak is low, this is significant because 20K cryocoolers of this size have very low Carnot and mechanical efficiency, 10% and 2%, respectively. Thus, 0.2W of heat leak would require 100W of power. Therefore it was decided to incorporate PODS. A schematic of a PODS is sketched in **Figure D-31**. When in space the heat path is through the small diameter composite tube, a longer path which also has a much smaller cross section area than the load bearing end housing. During launch, the load path is through a shorter, stronger path with a much larger cross-sectional area. By using the PODS the heat leak can easily be reduced to 1/10th of that of the SOA struts.

PODS were used on the Gravity Probe-B Program, which launch in April 2004, and may be used on many other missions. Note that PODS greatly reduce the orbital natural frequency of the structure. Therefore before choosing PODS as a support method, structural and flight control analyses need to be performed to verify that they can be safely used with the spacecraft system in question. Reduced on-orbit stiffness can impact attitude slew rate and settling times following attitude changes.

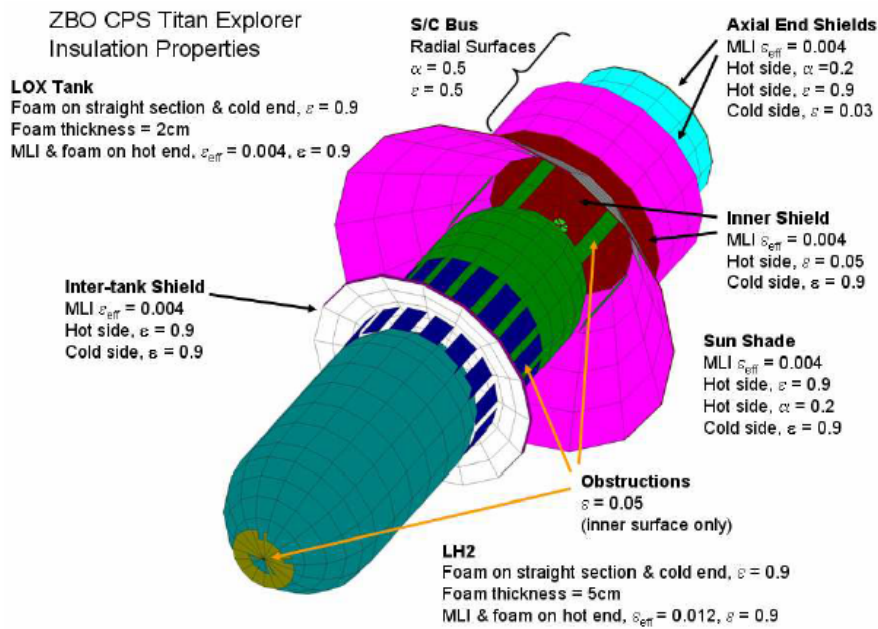

**Figure D-30.** Thermal model of the LOX H2 propulsion stage for Titan Explorer concept.

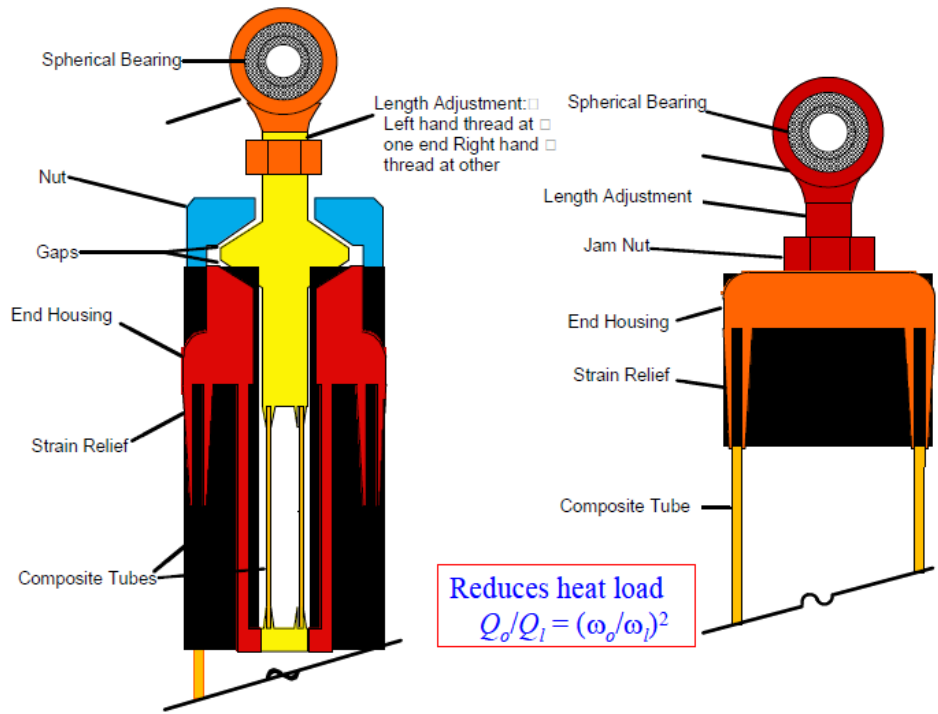

**Figure D-31.** Schematic of passive on-orbit disconnect struts (PODS) to improve thermal isolation.

### D.3.2.1  Discussion

The most significant findings was that for missions that are not required to perform major propulsive maneuvers after long periods in low orbit around planetary bodies it is possible to store the LOX/ LH2 passively, without active cryocoolers. This would greatly reduce the cost, complexity, and risk associated with using cryogenic propulsion on such missions. This result was obtained during the Titan Explorer missions study. That mission study evolved a configuration architecture that maximized the propellant tanks' view to deep space while minimizing heat loads from the spacecraft bus. Transient studies showed that this design could tolerate 24 hours off of the desired orientation (with the propellant tanks shaded from the sun) with negligible tank pressure increase. In addition to configuration and spacecraft pointing constraints, passive storage was enabled by a unique combination of sun shades, inter-tank radiation shields, low conductivity Passive On-orbit Disconnect Struts (PODS), optimized insulation arrangements, and optimal surface properties.

A number of issues still remain before this approach could be baselined.

1. The volume of the propellant tanks for H2-LOX systems is much larger than for an equivalent biprop system despite the greater efficiency because of the very low density of liquid hydrogen. The implications of this for the fixed mass of the system remain to be thoroughly understood.

2. There are currently no flight qualified thrusters in the thrust range needed for an Ice Giants mission available in the U.S. There may be thrusters available in Europe but this has not been confirmed. This issue would need to be resolved before proceeding.

3. To address the development of both the zero pressure systems and thrusters a development cost of order $100M was estimated in the original study in 2005. This needs to be updated to reflect subsequent developments.

## D.4   Optical Communications

Optical communications with laser transceivers deployed on space platforms offer higher downlink capacity that can benefit science data return from future NASA missions. By controlling the pointing of narrow laser beams from space, higher power densities can be delivered to Earth receivers, resulting in increased communications capacity. Additionally, lasers operating in the near infrared spectral region spell relief for the inevitable bandwidth congestion that will limit state of the art radio frequency telecommunication from expanding to meet the ever increasing information capacity demands needed for space exploration and science.

### D.4.1    Near Earth to Lunar Ranges

Successful technology demonstrations of optical communications from near earth ranges extending to the moon occurred in the past two decades. Near Earth (LEO and GEO ranges) spacecraft are transitioning to the use of optical communications for operational services. These efforts set the stage for the development of key technologies for demonstrating optical communications from deep space ranges extending to approximately 3 AU.

### D.4.2    Deep Space Optical Communications (DSOC) Project

Specifically, the Deep Space Optical Communications (DSOC) Project at JPL is currently funded to develop a deep space optical communication system to conduct a risk retiring demonstration in the early part of the next decade. The goals of the project is a capability for using this technology out to 3 AU.

### D.4.3    Extending the Range of Optical Communications to the Ice Giants

The current deep space laser communications architecture consists of three essential elements:

- Flight Laser Transceiver (FLT) with photon counting sensitivity pointing detector
- Ground Laser Transmitter (GLT) for delivering a beacon pointing reference to the FLT
- Ground Laser Receiver (GLR) with a photon counting receiver that can detect faint signals transmitted by the FLT
- A simple operational view of the architecture currently being developed for a deep space optical communications technology demonstration to a range of 3 AU is shown in **Figure D-32**. For cost effectiveness use of existing ground assets is planned.

Based on link analysis for optical communications from the Icy Giants, upscaling the FLT aperture and laser power to increase the emitted isotropic radiated power (EIRP) and the GLR to provide increased receiver gain are both compelling choices. This upscaling will be constrained by technology. For example, a 50 cm FLT aperture diameter and 20 W average power laser with 11.8 m diameter ground-based earth receiver are reasonable choices for the next decade.

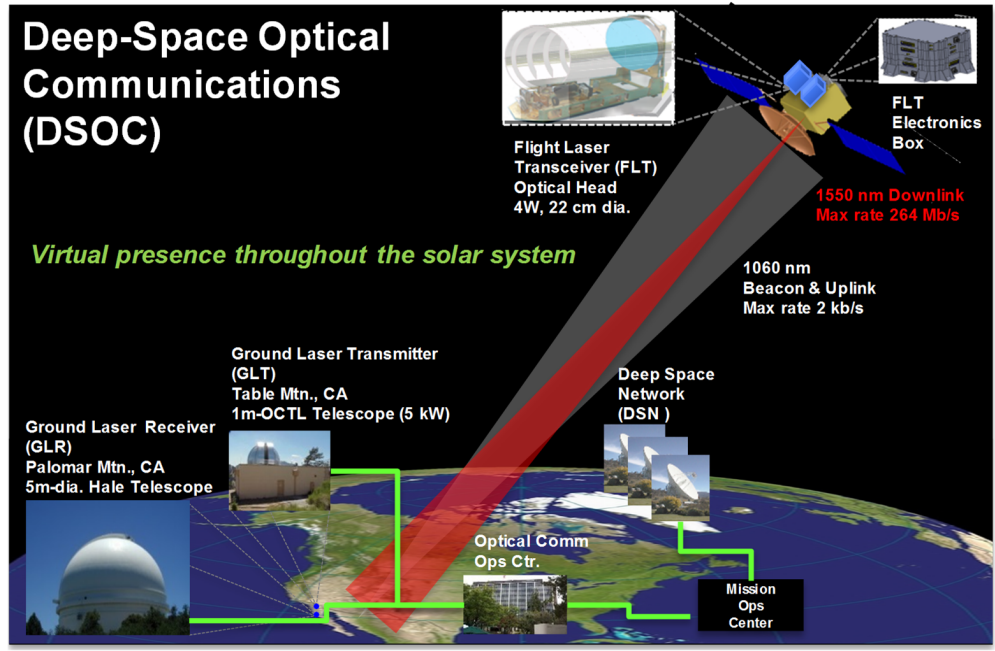

**Figure D-32.** Operational view for a near term technology demonstration of optical communication from deep space.

### D.4.3.1 Link Analysis for Uranus and Neptune Distances

Using reasonable link allocations analysis using these parameters were performed for Neptune and Uranus distances. **Table D-1** summarizes the achievable data rate with 3-dB link margin.

**Table D-1.** Link summary using optical communication based on a 50-cm diameter FLT with a 20 W average power laser modulated with 128-ary or 256-ary pulse position modulation; ground based aperture diameters of 5 m and 11.8 m are compared.

| URANUS | | | 128 PPM | 256 PPM | 128 PPM | 256 PPM |
|---|---|---|---|---|---|---|
| Range | Link Condition | SEP | 11.8 m (Mb/s) | | 5 m (Mb/s) | |
| 21.68 | Day | 5 | 0.113 | 0.192 | 0.033 | 0.054 |
| 17.6 | Night | 150 | 11.35 | 11.7 | 1.557 | 1.647 |

| NEPTUNE | | | 128 PPM | 256 PPM | 128 PPM | 256 PPM |
|---|---|---|---|---|---|---|
| Range | Link Condition | SEP | 11.8 m (Mb/s) | | 5 m (Mb/s) | |
| 31.84 | Day | 5 | 0.025 | 0.043 | 0.0122 | 0.0204 |
| 28.7 | Night | 150 | 0.946 | 1.24 | 0.181 | 0.233 |

The links presume 70% efficient photon-counting detector arrays suitably sized to receive and detect the faint deep space signals after propagation through the Earth's atmosphere.

**Table D-2** provides an initial notional estimate of the resources required by an FLT with the characteristics required to support the links in **Table D-1**.

Implementing the links described above will involve overcoming several challenges that will not be addressed by the planned near term technology demonstrations and these are discussed next.

### D.4.4 Required Capabilities not Addressed by DSOC Program

The increased aperture diameter motivated for the higher antenna gain will result in beam-widths of approximately 3 μrad compared to approximately 7 μrad that is being implemented in the near term. The pointing control will have to be tighter for the Icy Giant links. The improvement in pointing can be achieved without major technology development. The most formidable challenge arises is the implementation of a beacon that can serve as a pointing reference at the distances under consideration. Options to overcome this challenge are discussed next.

**Table D-2.** Initial notional estimate of mass and power required for a 50-cm 20 W FLT.

|  | CBE Mass (kg) | CBE Power (W) |
|---|---|---|
| Flight Transceiver Telescope | 45.6 | 0 |
| Small Optics & Actuators | 0.7 | 1 |
| Laser | 10.8 | 155 |
| Pointing Detector | 2 | 11 |
| Electronics | 6.3 | 20 |
| Thermal/Structure | 5.3 | 6 |
| TOTAL | 70.7 | 193 |

### D.4.4.1 A Beacon that Will Be Useful at Ice Giants Distances

The atmosphere limits transmission of diffraction limited lasers through the turbulence induced refractive index fluctuations. This problem becomes more challenging as more power has to be pumped through the atmosphere to provide sufficient signal at the distance of the ice planets. Three general options have been considered to deal with this problem

*Deploy a beacon laser above the atmosphere*

The ability to deploy beacon laser systems above the denser parts of the atmosphere will overcome this difficulty. The options that come to mind are balloon, space-borne or even lunar-based beacon laser systems. All of these concepts need to be matured through study. This could also mitigate the regulatory restrictions on transmitting high power laser beams through free-space.

*Utilizing the Earth disc imaged in the thermal infrared as a beacon*

This was studied under a JPL funded strategic initiative R&TD and is documented in an IPN Progress report (http://ipnpr.jpl.nasa.gov/progress_report/42-167/167D.pdf).

*Star field tracking and other approaches*

Conceptually the use of a star field or an optimal combination of all these concepts could also be considered.

### D.4.4.2 Technical Challenges of Long-Lived Missions

Another noteworthy challenges that will need to be carefully studied is laser lifetime and reliability. Fiber-based lasers in particular are prone to irrecoverable radiation induced darkening if exposed to cold dark space for extended periods of time, whereas, frequent use results in annealing out of radiation induced darkening. This and other factors related to laser components will have to be studied for the anticipated long lifetimes required for missions to Icy Giants.

### D.4.4.3 Need for Ground Infrastructure

The availability of ground infrastructure in the form of > 8 m ground based collectors is presumed in the links presented above. These are currently being studied and it is reasonable to assume the availability of such assets in ample time to support missions to Icy Giants.

### D.4.5 Conclusions

In summary the prospect of implementing laser links from Icy Giants will result in a significant boost to returned information capacity from the outer solar system. The links presented in this report are conservative and based on reasonable extrapolation of existing technologies. The most formidable challenge is the formulation of an affordable architecture to support pointing back of the laser beam from the outer reaches of the solar system. If this problem is solved the data-rates reported here can be further improved by scaling the laser power either through technology development or wavelength multiplexing of existing laser technology.

## D.5    Radioisotope Power Systems

The NASA Radioisotope Power Systems Program (RPSP) delivered a set of data to the Ice Giants study team so that a set of verifiable and consistent information would be available for this first mission study in preparation for the next Decadal Survey. In addition to the data, the RPSP supported the mission study team by answering questions and by supplying input to the final report. The delivered information covered the Multi-Mission Radioisotope Thermoelectric Generator (MMRTG), the proposed enhanced Multi-Mission Radioisotope Thermoelectric Generator (eMMRTG) and Lightweight Radioisotope Heater Unit (RHU). The Ice Giants Study decided to baseline the eMMRTG and RHUs but in this appendix we also include information on more advanced power sources which may be important for alternate implementations of Ice Gian missions. The data provided included the assumptions, the RPS performance characteristics such as power, mass, fuel thermal energy, degradation rate, electrical load condition, the RPS operational characteristics such as conversion efficiency and heat rejection requirements, and the RPS cost. The next section describes RPS assumptions and constraints the study used. The following section describe relevant information about the eMMRTG, RHU and Mission RPS Cost, respectively. Section D.4.2 provides the description of future RPS concepts that could provide about 1 kWe power to those mission concepts with Radioisotope Electric Propulsion (REP). More details are provided in the references (McNutt 2015; Report 2013).

### D.5.1    Proposed Enhanced Multi-Mission Thermoelectric Generator (eMMRTG)

The eMMRTG was used as the baseline power system in this study. The following definitions will be used:

- Beginning of Life (BOL) is defined as time of fueling
- Beginning of Mission (BOM) is defined as the point at which the MMRTG reaches equilibrium after launch, in space vacuum, with no solar flux, albedo, or Earth IR flux. BOM can be as long as 3 years after BOL
- End of Design Life (EODL) is 17 years after BOL

The eMMRTG is proposed to use Step-2 General Purpose Heat Source (GPHS) modules as heat sources. GPHS modules were estimated at 244–256 Wt at BOL, assuming an average output of 250 Wt. The design life for these systems would be 17 years: up to 3 years of storage, followed by 14 years of operation. The eMMRTG functions both in atmosphere and in vacuum. For deep space environment, a 4 K sink is assumed, leading to a fin root temperature of 420 K. The worst design case would be the Venus Gravity Assist scenario in which the RPS would reach a 450–470 K average fin root temperature, just under the maximum allowable temperature of 473 K. The allowable operating voltage range would be 22–34 volts of direct current (VDC); for the Ice Giants study 32 volts VDC was assumed for charging Li-ion batteries. The waste heat produced by the GPHS modules would need to be rejected by radiation or convection through the outer housing and fins, or removed conductively through cooling loops that can be attached at the base of the fins. This waste heat could be routed to other parts of the spacecraft if needed to warm spacecraft components.

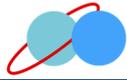

For the assembly, test, launch and operations (ATLO) phase, the integration of no more than 4 RPS into a spacecraft is recommended as a best practice based on previous RPS studies addressing ATLO phase. Storage capacity at the Department of Energy's Idaho National Laboratory is insufficient for 5 RPS and would require an extension to the facility or other imaginative solution to stage and store the RPS prior to shipment to KSC. Current storage and staging facility at KSC is insufficient for 5 RPS so another facility must be identified and prepared for use two years prior to launch. These factors would increase the costs of the mission. In addition, the RPS units must be mounted so they can be installed through doors in existing launch fairing. Four systems will provide a challenge to integrate and five will exacerbate the problem adding additional complexity and costs to the launch vehicle. Any costs associated with 5 units do not include these additional and likely substantial costs. In addition, as of 2016 DOE can fuel 4 generators including the one for the Mars 2020. In order to provide fuel for an additional 2 generators for a total 5 generators, it would require approximately 6 additional years for fuel processing.

It is important to consider RPS-induced radiation and thermal impacts on spacecraft and on instruments. Radiation could have noise effects on instrument measurements, and long-term effects of instrument damage. Gamma dose on order of 1 krad over 10 years with 1-meter separation. Neutron fluence on order of 6E10 1 MeV n/cm$^2$ over 10 years with 1-meter separation (see **Figure D-33**). RPS could produce between 244 to 256 Wth/GPHS that must be considered during spacecraft and instrument design and integration since instrument pointing/field-of-view may be constrained.

### D.5.1.1 eMMRTG Configuration and Characteristics

eMMRTG is a design concept based upon replacing the thermocouples in the MMRTG with new advanced thermoelectric materials, while maintaining the same structural frame and design of the MMRTG and preserving or improving its functional requirements. As with the MMRTG, the eMMRTG is designed to operate in planetary atmospheres as well as in vacuum. Produced by the DOE, each GPHS contains four pellets of plutonium dioxide fuel, each clad in iridium metal and several layers of graphite and carbon-fiber material for protection during potential accident conditions. As with the MMRTG, the eMMRTG would use eight GPHS modules. See **Figure D-34** for eMMRTG configuration.

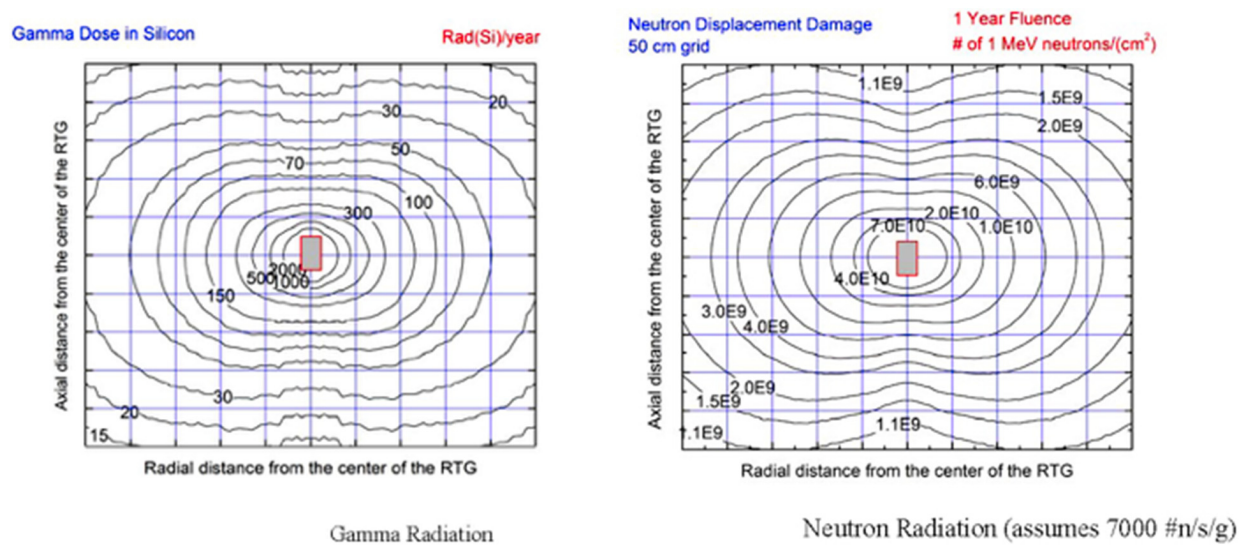

**Figure D-33.** Gamma dose and neutron fluence contour lines with a 50-cm grid.

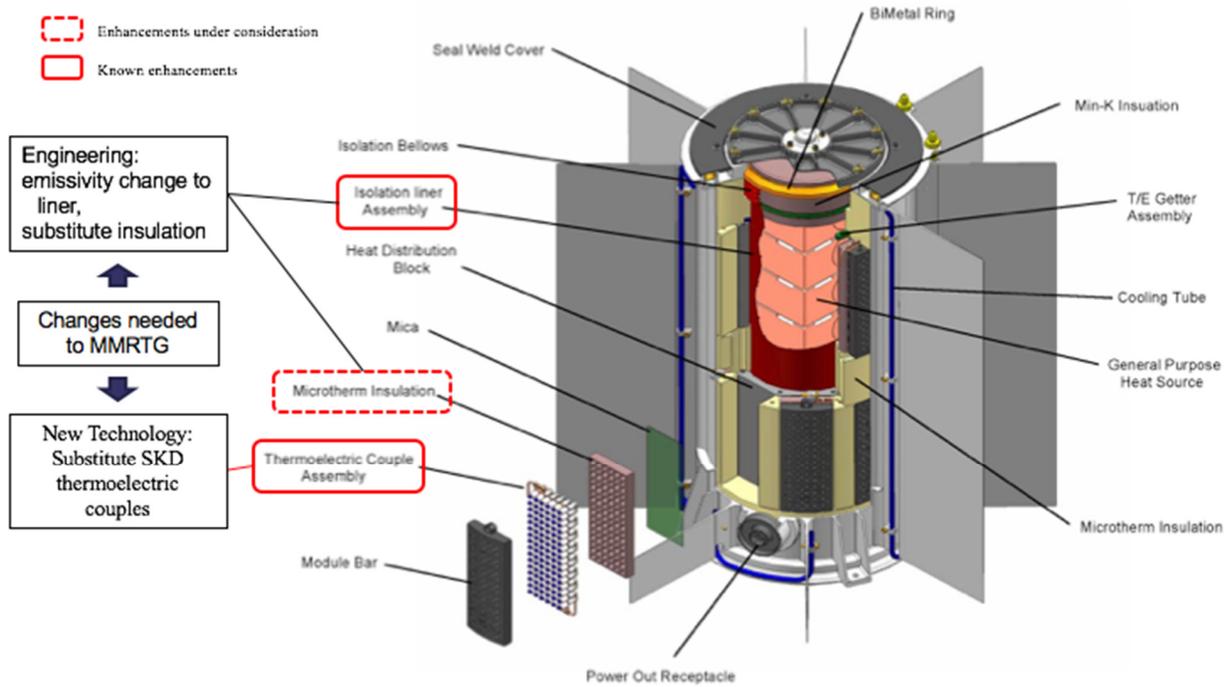

**Figure D-34.** eMMRTG proposed configuration.

There are 8 GPHS modules, assuming 250 Wth per GPHS at BOL. Expected power degradation rate (not yet validated) is ~2.5% per year (including fuel decay). **Table D-3** provides detailed information about eMMRTG's performance characteristics. **Table D-4** describes eMMRTG's nominal operating characteristics. **Table D-5** and **Table D-6** show eMMRTG's random vibration requirements and pyroshock requirements, respectively.

**Table D-3.** Projected eMMRTG performance characteristics.

| Parameter | eMMRTG Value |
|---|---|
| Power (Vacuum, 4 K (–269°C) sink, 250 Wt) | 144 We BOM (At time of fueling) at 32V |
| | 134 We BOM (After a max of 3 years of storage) at 32V |
| | 94 We EODL (After 17 years) at 32V |
| System mass | ~45.0 kg (with cooling tubes) |
| System mass | ~43.6 kg (without cooling tubes) |
| Dimensions | 0.69 m in length, 0.65 m from fin tip to fin tip |
| Operating Environments | Vacuum, planetary atmospheres, launch, landing, shock, etc. |
| System Lifetime | 17 years (3 years storage plus 14 years mission life) |

**Table D-4.** eMMRTG Nominal Operating Characteristics.

| Parameter | eMMRTG Value | | Comments |
|---|---|---|---|
| Heat Rejection Requirement (vacuum, thermal output less electrical output) | 1819 Wt [BOM] | 1655 Wt [EODL] | Assumes 250 Wt per GPHS module (BOL); actual waste heat will vary |
| eMMRTG-induced vibration | none | | Static system |
| Magnetic field | <25 nT | | At 1 m from eMMRTG |
| G-loading limit | 25g | | At launch + 1 year |
| Random-vibe loading limit | <0.2 g$^2$/Hz peak | | During Launch |
| Max allowable average fin-root temp | 200°C | | Not to be exceeded |

**Table D-5.** eMMRTG random vibration requirements.

| Frequency, Hz | EELV | |
|---|---|---|
| | Qual Test | FA Test |
| 20–50 | + 3dB/oct. | +3 dB/oct. |
| 50–250 | 0.20 g²/Hz | 0.10 g²/Hz |
| 250–350 | -6.0 dB/oct. | -6.0 dB/oct. |
| 350–1000 | 0.10 g²/Hz | 0.05 g²/Hz |
| 1000–2000 | -12 dB/oct. | -12 dB/oct. |
| Overall | 12.4 gRMS | 8.7 gRMS |

**Table D-6.** eMMRTG pyroshock requirements.

| Frequency, Hz | Peak SRS Response (Q=10) |
|---|---|
| 100 | 40 g |
| 100–2000 | +10.0 dB/oct. |
| 2000–10000 | 6000 g |

The change in nominal eMMRTG output power over a 14-year mission lifetime assuming a BOM power output of 134 We (after a maximum of 3 years in storage) and a total degradation rate of 2.5% per year (including fuel decay) is shown in **Figure D-35**.

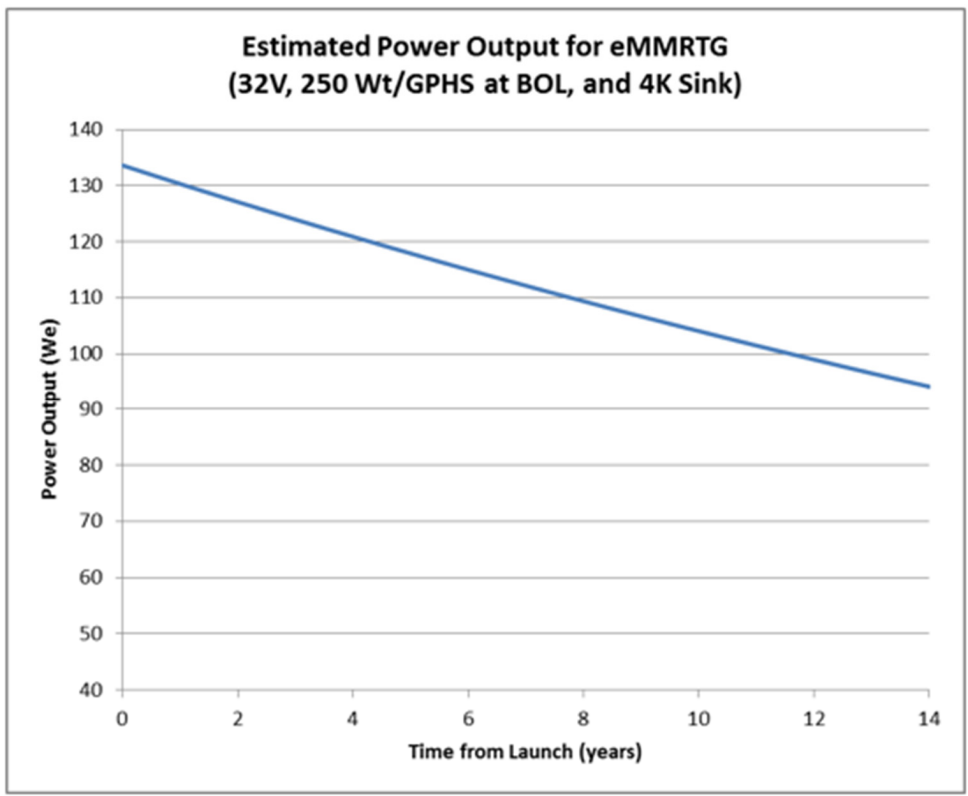

**Figure D-35.** MMRTG estimated power output.

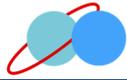

### D.5.1.2 LWRHU Configuration and Characteristics

Light Weight Radioisotope Heating Units (LWRHUs consist of three main components: the radioisotope fuel pellet encased in a platinum-rhodium alloy "clad," graphite insulation, and a graphite aeroshell to provide a thermal shield to protect the fuel in case of a spacecraft failure leading to fire or reentry. See **Figure D-35** for RHU configuration.

See **Table D-7** for RHU performance characteristics. The expected heat degradation rate is ~0.79% per year. The values in are for new RHUs, which would need a production campaign. Currently available RHUs are at 0.89 Wth.

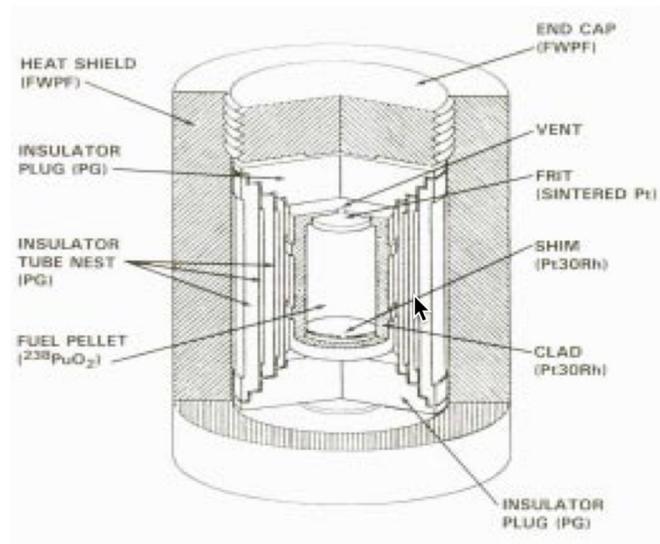

**Figure D-36.** LWRHU configuration.

**Table D-7.** RHU performance characteristics.

| Parameter | RHU Value |
|---|---|
| System mass | 40 g |
| Dimensions | 0.032 m length, 0.026 m diameter |
| Heat generated | 1 Wt |
| Pu-238 mass | 1.9 g |
| Operating environments | Vacuum, atmosphere |
| System Lifetime | >14 years |

### D.5.1.3 Mission RPS Cost

The following costs were supplied to the mission study, these costs are based on using a Delta IV Heavy and are provided in FY15 dollars. For 4 eMMRTGs, the cost estimates for the hardware (including all costs to DOE and support to launching) is $172M, Launch Nuclear Safety is $33M, and the LWRHU additional cost due to needing more than 43 LWRHUs is $34M. For 5 eMMRTGs, the cost estimates for the hardware (including all costs to DOE and support to launching) is $184M, the launch nuclear safety-unique mission costs are $33M, and the LW RHU additional cost due to needing more than 43 LWRHUs is $34M.

### D.5.2 Development of High Power RPS Technology

The RPS Program is pursuing technology development for both the thermoelectric (TE) and Stirling power conversion options. As a possible follow-on to the eMMRTG effort, JPL is leading the development of an advanced thermoelectrics based on the segmented skutterudite, La3-xTe4, and Zintl couple technology. One potential generator design based on this technology is a Segmented Modular Radioisotope Thermoelectric Generator (SMRTG). The SMRTG would be composed of sections of GPHS modules, as shown in **Figure D-37**. Each section could be one, two, or four GPHS modules high, depending on the final design. The minimum size of the SMRTG would be a single section, while the maximum size of the SMRTG would have 16 GPHS modules.

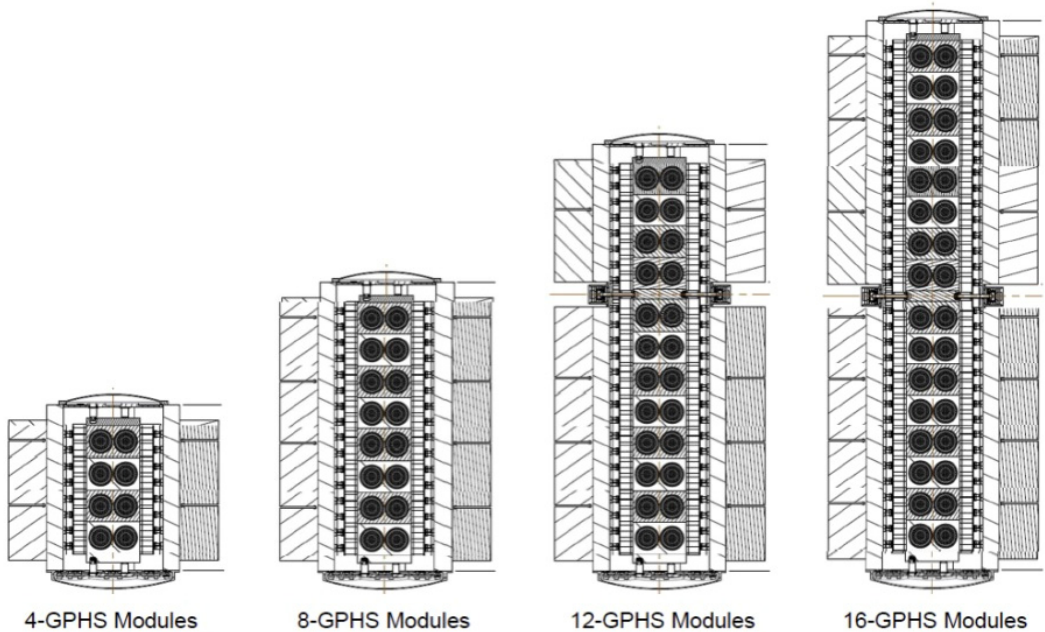

4-GPHS Modules  8-GPHS Modules  12-GPHS Modules  16-GPHS Modules

**Figure D-37.** Conceptual SMRTG configurations based on 4 GPHS module stackable segment design.

Like the GPHS-RTG, the SMRTG would operate only in vacuum. An SMRTG would use segmented TEs in a two GPHS module assembly that produces about 43 We BOM. The design is scalable up to 16 GPHS modules in increments of two, producing just under 500 We BOM. The SMRTG is designed to operate in vacuum at a hot-shoe temperature of 1273K and a cold-shoe of 498K with an overall projected system efficiency of about 8–11%. The operating temperatures and system efficiency would represent major improvements over the eMMRTG, which is projected to achieve about 7% system efficiency at 873K/473K. Some of the key component technologies needed to realize the SMRTG concept include the TE multi-couple module development, lightweight high-temperature MLI, compliant cold-shoes, aerogel encapsulation, sublimation control, and life verification.

GRC is leading the RPS Stirling technology development that includes research and technology development of: hot-end components, cold-end components, and systems/testbeds. Under hot-end components, work has focused on improved thermal insulation materials, Stirling-specific MLI packaging, and heat source backup cooling using variable conductance heat pipes (VCHP). The cold-end components effort has developed advanced NdFeB magnets for higher temperature alternators, high temperature organic adhesives and wire insulation, and titanium-water heat pipes for heat rejection. The systems/testbeds task has demonstrated new fault-tolerant electrical controller architectures for both single- and dual-Stirling convertor systems and pursued mechanical balancers that permit single, unopposed convertors to operate with low vibration.

The High Power Stirling Radioisotope Generator (HPSRG) is a notional group of designs for a higher power generators in the SRG family. The SRG family uses Stirling dynamic power conversion technology to convert thermal energy into electrical with high conversion efficiencies (>25%) and high specific powers. Both distributed heat source systems as well as centralized heat source generators are under consideration. Sharing the heat of multiple GPHS modules may provide the ability to continue operation in the event of a Stirling convertor failure. **Figure D-38** shows an example of a 4 GPHS Stirling generator with 4 Stirling convertors sharing a centralized stack of GPHS modules. One conceptual design from the HPSRG design family modifies the basic ASRG layout to increase output power; it can be thought of as a larger ASRG consisting of two dual-opposed Stirling converters, using the thermal output of four (4-GPHS SRG), six (6-GPHS SRG), or eight (8-GPHS SRG) GPHS modules. **Figure D-39** shows ½ of a 6 GPHS generator. Though this family of design concept uses technologies developed for the ASRG, the HPSRG remains at a conceptual phase and none of the concepts have been built or tested as a system. This advanced concept has undergone some conceptual study to determine feasibility.

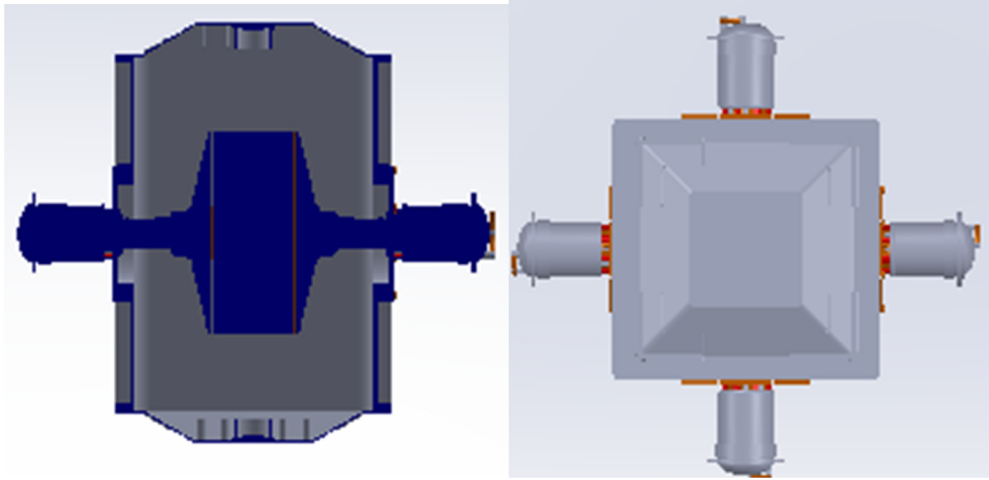

**Figure D-38.** Example of a 4 GPHS Stirling generator with 4 Stirling convertors sharing a centralized stack of GPHS modules.

Greater details on these high power RPS concepts can be found in the Nuclear Power Assessment Study Final Report1 and Radioisotope Power Systems Reference Book for Mission Designers and Planners (Version 1.1)2.

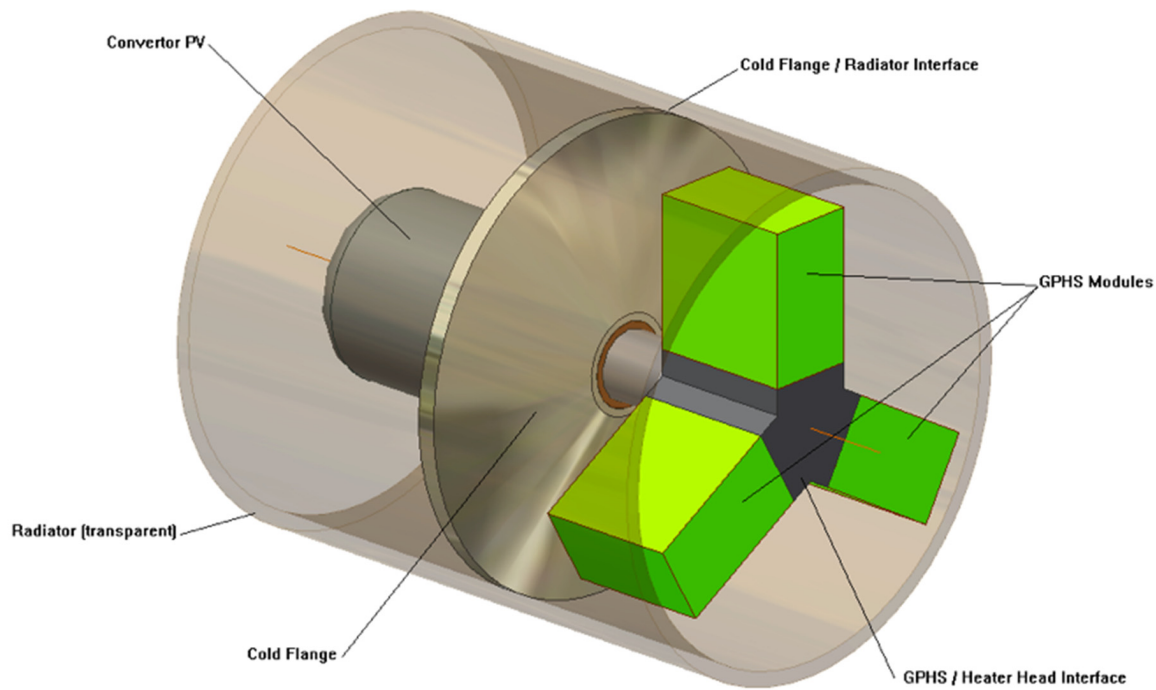

**Figure D-39.** One half of a potential 6 GPHS HPSRG.

## D.6    Small Spacecraft Technology

The ability to provide most of the capabilities of a current spacecraft in a vehicle much smaller in size could make it possible to reduce the trip time for certain ice giant missions and/or reduce mission cost. However, the pathway to achieving this is not well defined. The primary reason for that is that for many subsystems size is important for a deep space spacecraft. For a conventional spacecraft, size dictates the amount of solar energy that can be captured from the sun and the size of antenna that can be deployed to beam data back to Earth. Communicating data from the vast distances of the outer solar system, a very small spacecraft will have low data rates. For example, the New Horizons spacecraft required an entire year to send back the data from a single flyby of Pluto and Charon.

To be able to achieve useful science on an orbital mission to the gas giants using a small spacecraft then we are going to need to use advanced technologies. Optical communications described in section D.4 can achieve high data rates with small aperture and mass instruments and can be made very power efficient. Advances in sensor technology will continue to permit reductions in the size of remote sensing instruments. The development of low power, high-capability computing and data handling systems (C&DH), which has recently been initiated under NASA's Game Changing Development (GCD) program will be a vital part of this development.

Projected advances in solar power technology will be important in missions out to Saturn but are much less likely to impact the Ice Giants where solar fluxes are between a factor of 4 and a factor of nine lower. Accordingly progress in radioisotope power systems will be important but we cannot expect dramatic progress beyond the state of the art. As a result, the power system may represent a larger fraction of the spacecraft mass than for current outer planets spacecraft. Use of a large RPS system as part of a Radioisotope Electric Propulsion System combined with a low mass spacecraft could enable fast flight times Uranus and Neptune and leave vast amounts of power for data communications.

## D.7    Advanced Mission Operations

With deep space missions ever more ambitious and DSN support becoming more contentious, future spacecraft missions would be greatly enhanced with advanced Mission Operations Systems and onboard autonomy. Modern spacecraft missions are expected to meet these challenging objectives with fewer resources and at lower cost. Additionally, spacecraft missions are being flown more frequently than ever before and the increased number of interplanetary missions is straining the DSN's ability to support each individual mission timeline. By incorporating advanced mission planning techniques and onboard spacecraft autonomy, complex mission operations can be implemented at lower cost, with fewer resources of the ground systems, and streamlining onboard the spacecraft operations.

### D.7.1    Advanced Mission Planning Tools

Advanced mission planning tools have begun to take a larger role in recent missions, such as the APGEN (Activity Plan Generator) adapted for the Europa mission (Maldague 2014) and ASPEN (Automated Scheduling and Planning Environment) used on the Rosetta mission (Chien 2015), as well as similar tools on other missions. Examples of mission tools with varying levels of automation can be found in **Table D-8**. These tools are used primarily to generate the plethora of mission related geometries, constraints, etc. and find science opportunities. Some facets of the mission and science planning activities can be performed with separate tools, including PDT (Pointing Design Tool) for Cassini, JPLAN for Juno, and GeoVis/NHAngles/SciPlan for New

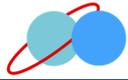

Horizons mission (Birath 2012). Increased level of automation is delivered by incorporating a search algorithm to optimize a limited resource (time, data, pointing, etc.) for particular science opportunities, as found with APGEN, SciBox for MESSENGER (Choo 2014), and SOA (Science Observation Analyzer) for the Dawn mission (Rabideau 2014). Where these tools capabilities differ is by ASPEN's implementation of advanced search and optimization algorithms for an optimal mission plan.

**Table D-8.** Mission and science planning tool examples for recent spacecraft missions.

| Tool | Mission | Adaptability | Planning Capability | Automation Capability |
|------|---------|--------------|---------------------|-----------------------|
| CIMS | Cassini | Mission Specific | Full sequencing database for managing data, telemetry, opmodes, pointing, etc. | **None** - Manual implementation of spacecraft activities |
| PDT/KPT | Cassini | Mission Specific | Detailed science activity design, modeling, and simulation with constraint checks | **None** - Manual implementation of science activity designs |
| GeoVis/NHAngles | New Horizons | Mission Specific | Mission planning event geometries and science activity constraint checks | **None** - Manual implementation of spacecraft activities |
| SciPlan | New Horizons | Mission Specific | Sequencing database for managing data, opmodes, pointing, etc | **None** - Manual implementation of spacecraft activities |
| JPLAN | Juno | Mission Specific | Data and power management | **None** - Manual implementation of spacecraft activities |
| SOAP | Juno | Multi-mission | Mission planning event geometries and constraint checks | **Minimal** - Some limited search/optimization by manual implementation |
| APGEN | Europa | Multi-mission | Detailed activity planning, modeling, and science opportunity search with constraint checks | **Modest** - Non-iterative science opportunity search/optimization for science requirements |
| SciBox | Messenger | Mission Specific | Detailed activity planning, modeling, and science opportunity search with constraint checks | **Modest** - Iterative science opportunity search/optimization for limited science requirements |
| SOA | Dawn | Multi-mission | Detailed science planning, visualization, modeling, and constraint checks. | **Modest** - A Percy search engine incorporated to identify science opportunities. |
| ASPEN | Dawn | Multi-mission | Used as data manangement utiliy, related science activity modeling, and constraint checks | **Advanced** - Data management and science planning optimization tools |
| ASPEN | Rosetta | Multi-mission | Full Mission and Science planning and optimization with constraint checks | **Advanced** - Adaptable full mission events and science opportunity search/optimization for complex science requirements |

The next generation of advanced mission planning tools will likely employ a similar methodology to that demonstrated by ASPEN, where mission constraints and science observation requirements are achieved through an iterative process of numerical search and optimization algorithms. The fidelity of the defined mission constraints and science objectives inputs can be progressively incorporated at different levels of the planning timeline. A typical planning scheme would have 3 levels: long-term planning, medium term planning, and short term planning. The higher-level architecture of the mission design can be executed in the long term planning stage, where the trajectory is designed to target major science goals. The mission planners and science teams only need to provide high level requirements that aid in solidifying the trajectory, such as any close encounter specifications, occultation geometries, etc. (Chien 2015).

## D.7.2 Medium Term Planning Stage

The medium term planning stage incorporates a more complete science targeting plan and pointing schedule. The science teams must provide well-defined science observation requirements, including targeting geometries, lighting conditions, observation cadence, attitude/pointing control, etc. for a sequence. The automated scheduling search rapidly optimizes all of the science observations, engineering activities, and DSN support within the sequence, while adhering to all flight rules and other restrictions. A visual example of this process is shown in **Figure D-40** for the Rosetta mission using ASPEN, where all of the individual spacecraft activities and science observations are optimized for a 4-week sequence. A single run of ASPENS's search and optimization routine took only 20 minutes for this 4 week sequence, handling 2027 observations and attitude targeting and 10,000's of constraints. This sequence planning and optimization process can be iterated with additional mission planner and/or science team inputs to produce an optimal spacecraft operations plan. The automated numerical search and optimization process provides a larger pool of potential science opportunities to be considered. This is especially true when the automated search is able to combine multiple instrument activities to efficiently accomplish separate scientific goals at once, where more traditional and manual planning methods (that of Galileo and Cassini missions) may over look such occasions. Another advantage of the automated planning is the ability to appropriately allocate high and low rate science activities into areas of the sequence that can better accommodate the data storage requirements and/or DSN support. This prevents periods of underutilized spacecraft time, as well as other occasions of over subscribing the onboard data management system.

**Figure D-40.** Example mission plan for a 4-week sequence, 2027 observations, 2160 pointings/slews, 63 science campaigns, 10,000's of constraints checked and over 1400 downlink activities (Chien 2015).

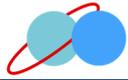

### D.7.3 Short-Term Planning

The short-term planning stage consists of completing the detailed science sequencing and instrument timelines. Any remaining gaps in the spacecraft activity plan are negotiated to be filled by additional science activities. The short-term planning stage is also where late changes to the spacecraft activity plan are accommodated. Unexpected events that can impact the mission such as DSN station downtimes, spacecraft anomalies, or even discovery of natural transient phenomena can be rapidly incorporated into the spacecraft plan by using strategic adjustment of the automated scheduling priorities. With constant improvements of processing speed and optimization algorithms, rapid readjustment and analysis of a spacecraft sequence could be performed to react to an event within a day if not hours of the next uplink opportunity.

There are also many events that can occur in a spacecraft mission that would benefit from a more timely response, rather than waiting for the next uplink opportunity and two-way light time for commanding (5.5 hours two-way light time for Uranus and 8 hours for Neptune). A spacecraft with onboard autonomy capability would be able to quickly recover from all but the most serious of instrument and flight systems faults. An intelligent recovery system of an autonomous system mitigates long instrument and/or spacecraft downtimes, saving multiple days of valuable science observing time, as well as any induced operations costs. An autonomous spacecraft system would also enable the dynamic retasking and rescheduling of science observations in reaction to recent onboard observation analysis. An example would be reacting to a plume discovery during a close flyby, in order to target more observations in that region. Such a capability is especially beneficial for unique encounters, where certain favorable observation conditions may pass too quickly for ground systems cannot react in time. This level of onboard observation data processing power would also be able to post-process the science data for efficient data reduction and improved overall science return of the mission.

Another feature of an autonomous spacecraft system, is executing guidance and navigation processes onboard the spacecraft, known as AutoNav. AutoNav systems use the same observation analysis algorithms for feature detection of celestial bodies in order to determine the sensitive local position and velocity states of the spacecraft. The collected state information can then be processed onboard to autonomously plan and execute any necessary trajectory maneuvers. In certain celestial environments the typical navigation timeline between collecting actionable spacecraft state data back to the ground systems and commanding a maneuver is prohibitive if not impossible, therefore requiring AutoNav. More details of the AutoNav (Onboard GNC) systems can be found in the recent survey of future GNC technology (Quadrelli 2015).

### D.7.4 Summary

The key to successful automated spacecraft planning systems is providing well-defined science observation requirements alongside balanced science priorities and maintaining them to evolve with the mission. Mission and science planning automation has been progressively incorporated more and more into the traditional negotiations-style planning methodology. Yet, there continues to remain hesitation on the part of science and operations teams of complete adoption. By further embracing automation methods, the full operational and science benefits can be drawn. Though some potential benefits of automated spacecraft planning systems is reduced operational personnel and cost, there will always be a need for a core set of mission and science planners to run these tools and manage the process, especially developing and maintaining the science definition inputs.

## D.8    Bibliography

Birath, E., Rose, D., & Harch, A. (2012). *Science Operations Tools for the New Horizons Encounter with Pluto*, AIAA 12th International Conference on Space Operations, Stockholm, Sweden.

Cruz, M. I. (1979). The Aerocapture Vehicle Mission Design Concept," May 1979. *Proceedings of the Conference on Advanced Technology for Future Space Systems*, AIAA paper #790893.

Chien, S. et al. (July 2015). *Activity-based Scheduling of Science Campaigns for the Rosetta Orbiter*, International Joint Conference on Artificial Intelligence (IJCAI 2015), Buenos Aires, Argentina.

Choo, T. H. et al. (2014). *SciBox: An Automated End-to-End Science Planning and Commanding System*, Acta Astronautica, 93, pp. 490-496.

Ellerby, D. (2014.). "Heatshield for Extreme Entry Environment Technology (HEEET) for Missions to Saturn and Beyond,. *NASA Outer Planets Assessment Group Meeting Bethesda MD, August.*

Hall, J. L., Noca, M., & Bailey, R. W. (March–April 2005). Cost-Benefit Analysis of the Aerocapture Mission Set," *Journal of Spacecraft and Rockets*, Vol. 42, No. 2.

Lockwood, M. K. (2006). *Aerocapture Systems Analysis for a Neptune Mission,*" NASA Technical Memorandum TM-2006-214300, pp. 4.

Maldague, P. F. et al. (May 2014). *APGEN scheduling: 15 years of Experience in Planning Automation*, SpaceOps 2014 Conference.

McNutt, R. (2015). NPAS Report Nuclear Power Assessment Study Final Report, TSSD-23122, Johns Hopkins University Applied Physics Laboratory, Laurel, MD, Feb. 2015. http://solarsystem.nasa.gov/rps/docs/NPAS.pdf.

Quadrelli, M. B. et al. (2015). *Guidance, Navigation, and Control Technology Assessment for Future Planetary Science Missions*, Journal of Guidance, Control, and Dynamics, Vol. 38, No. 7, pp. 1165-1186.

Rabideau, G. et al. (June 2014). *A Constraint-Based Data Management Tool for Dawn Science Planning*, Proceedings of International Symposium on Artificial Intelligence, Robotics, and Automation for Space, European Space Agency/ESTEC, Montreal, Canada.

Report, J. (2013). *2Radioisotope Power Systems Reference Book for Mission Designers and Planners (version 1.1), JPL Publication 15-6, Jet Propulsion Laboratory, California Institute of Technology, Pasadena, CA, Aug 8, 2016.*

Saikia, S., J. et al., "Aerocapture Assessment at Uranus and Neptune for NASA's Ice Giant Studies," Technical Report, PU-AAC-2016-MC-0002, Sept. 2016, West Lafayette, US.

# E   AEROSPACE CORPORATION COSTING REPORT

This appendix provides the Independent Cost Estimate prepared by Aerospace Corporation.

## H.3.5 Cost Validation: Independent Cost Estimate (TBR)

An Independent Cost Estimate (ICE) for the JPL pre-decadal study concepts 3-6 was performed by The Aerospace Corporation using a combination of analogy-based estimates and parametric cost models. The ICE results show a total life cycle cost estimate of $2,279M, $1,643M, $1,992M, and $2,321M for Options 3, 4, 5, and 6 in FY15$ at a 70th percentile confidence level, respectively. Table 1 summarizes the Aerospace estimates for each option by WBS.

*Table 1 Summary of Ice Giants Study Option Estimate*

| Aerospace ICE (FY15 $M) | Option 3 | Option 4 | Option 5 | Option 6 |
|---|---|---|---|---|
| Phase A | $24.7 | $19.5 | $21.8 | $21.9 |
| Phase B/C/D | $1,854.5 | $1,376.7 | $1,537.4 | $1,687.0 |
| Mission PM/SE/MA | $158.0 | $121.8 | $133.5 | $143.0 |
| Payload[1] | $124.6 | $124.6 | $124.8 | $272.5 |
| Flight System[2] | $1,012.4 | $714.1 | $810.0 | $740.4 |
| Pre-Launch GDS/MOS | $120.7 | $89.0 | $99.3 | $107.5 |
| *Launch Vehicle Nuclear Support* | *$33.0* | *$33.0* | *$33.0* | *$33.0* |
| Reserves | $405.7 | $294.3 | $336.9 | $390.7 |
| Phase E/F | $400.7 | $247.0 | $433.3 | $612.2 |
| MO&DA - Science | $347.8 | $204.4 | $361.2 | $509.4 |
| Reserves | $52.9 | $42.6 | $72.1 | $102.8 |
| **Total** | **$2,279.9** | **$1,643.2** | **$1,992.6** | **$2,321.1** |

For each of the study options, project Work Breakdown Structure (WBS) elements were categorized and analyzed using appropriate methods for cost estimation. Table 2 shows a detailed break-down of the WBS elements and the methods used to develop the cost estimate.

*Table 2 Summary of estimation methodology used for various WBS elements*

| WBS Element | Parametric Models | Adjusted Analogy Cost | Analogous % Wrap on Hardware Cost | Pass Through | Aerospace Cost Risk Analysis |
|---|---|---|---|---|---|
| PM/SE/MA | | | ● | | |
| Payloads | ● | ● | ● | | |
| Atmospheric Probe | ● | | | | |
| Entry System | ● | ● | | | |
| Orbiter | ● | ● | | | |
| SEP Stage | ● | ● | | | |
| RTG | | | | ● | |
| Pre-Launch GDS/MOS | | | ● | | |
| MO&DA | | ● | | | |
| Reserves | | | | | ● |
| LV Nuclear Support | | | | ● | |
| DSN | | | | ● | |
| Phase A | | | | ● | |

Instruments were estimated with a combination of analogies, NASA Instrument Cost Model (NICM), the Multivariate Instrument Cost Model (MICM), and Aerospace's Space Based Optical

Sensor Cost Model (SOSCM), where applicable based on the instrument type. Instrument estimates for each option are provided in Table 3, with comparisons between estimate, modeled, and analogies are depicted in Figure 1, Figure 2, Figure 3, Figure 4, Figure 5, Figure 6 for each class of instruments. Analogy-based estimates were performed by adjusting the actual costs of analogous hardware elements by the ratio of a relevant Cost Estimating Relationship (CER) calculated for the proposed element and the analogous element. For comparison purposes, Options 3-5 and Option 6 carry 50kg and 150kg payload allocations respectively.

*Table 3 Summary of Instrument Estimates based on Option*

| Aerospace ICE (FY15 $M) | JPL Estimate | Option 3 | Option 4 | Option 5 | Option 6 |
|---|---|---|---|---|---|
| Payload PM/SE/MA | $ 13.6 | $ 16.1 | $ 16.1 | $ 16.2 | $ 35.3 |
| OPH | $ 2.6 | $ 4.8 | $ 4.8 | $ 4.8 | N/A |
| ASI | $ 5.9 | $ 6.9 | $ 6.8 | $ 6.9 | N/A |
| Nephelometer | $ 9.1 | $ 9.5 | $ 9.5 | $ 9.5 | N/A |
| GCMS | $ 39.5 | $ 31.0 | $ 31.0 | $ 31.0 | N/A |
| NAC | $ 20.0 | $ 23.5 | $ 23.5 | $ 23.5 | $ 23.5 |
| Doppler Imager | $ 30.0 | $ 24.9 | $ 24.9 | $ 25.0 | $ 25.0 |
| Magnetometer | $ 7.8 | $ 7.9 | $ 7.9 | $ 7.9 | $ 7.9 |
| VNIRMS | $ 16.7 | N/A | N/A | N/A | $ 30.2 |
| MIRS | $ 12.3 | N/A | N/A | N/A | $ 20.7 |
| UVIS | $ 10.0 | N/A | N/A | N/A | $ 10.0 |
| Radio Waves | $ 5.8 | N/A | N/A | N/A | $ 7.5 |
| Low Energy Plasma | $ 4.1 | N/A | N/A | N/A | $ 4.3 |
| High Energy Plasma | $ 3.8 | N/A | N/A | N/A | $ 6.4 |
| Thermal IR | $ 25.3 | N/A | N/A | N/A | $ 27.8 |
| Energetic Neural Atoms | $ 7.6 | N/A | N/A | N/A | $ 13.2 |
| Dust Detector | $ 9.8 | N/A | N/A | N/A | $ 7.6 |
| Langmuir Probe | $ 1.9 | N/A | N/A | N/A | $ 1.5 |
| Microwave Sounder | $ 40.2 | N/A | N/A | N/A | $ 42.2 |
| WAC | $ 9.8 | N/A | N/A | N/A | $ 9.3 |
| Total | $ 275.8 | $ 124.6 | $ 124.6 | $ 124.8 | $ 272.5 |

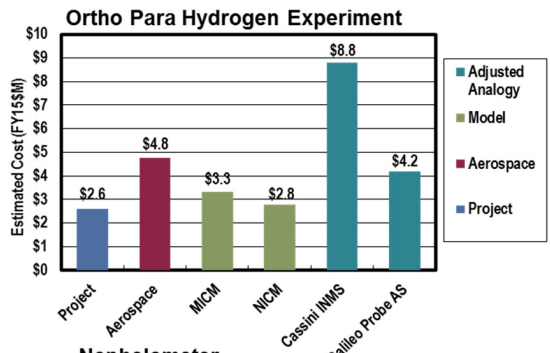
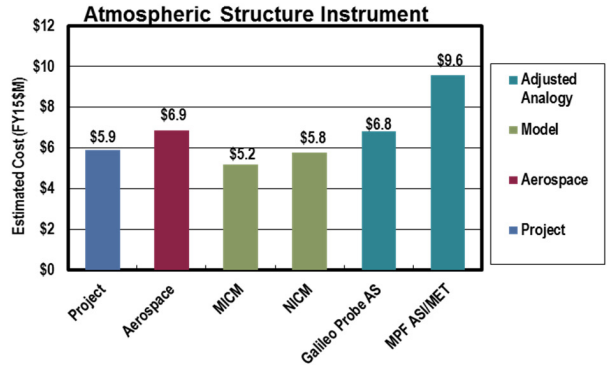

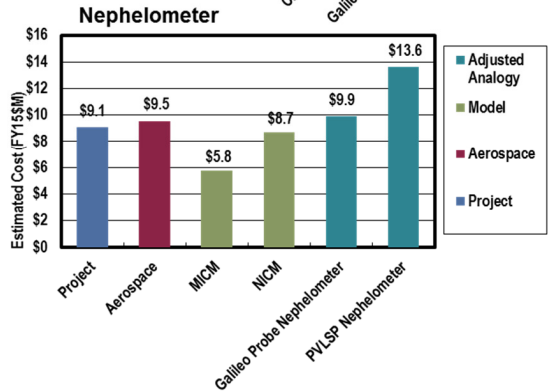
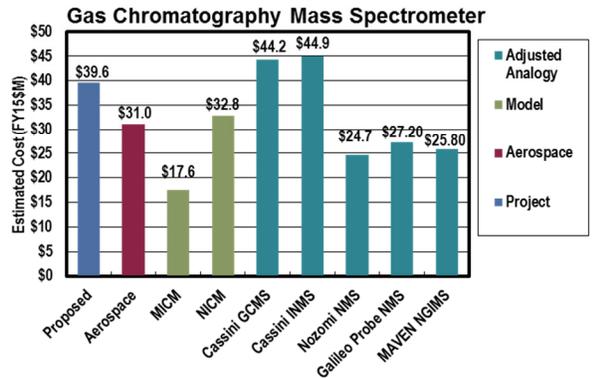

*Figure 1 Probe instrument Estimates based on Analogies and Modeled costs*

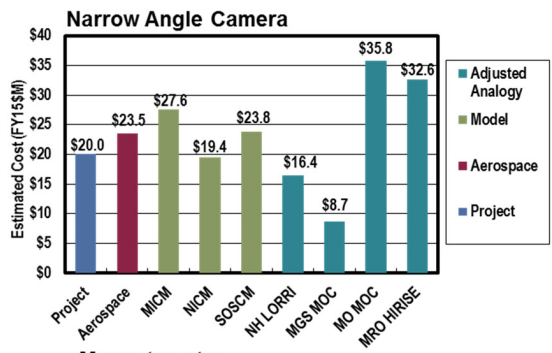
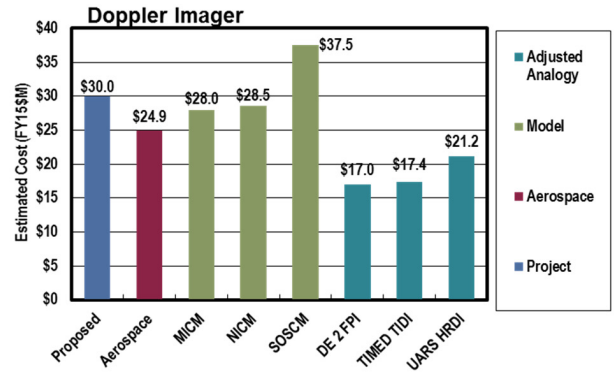

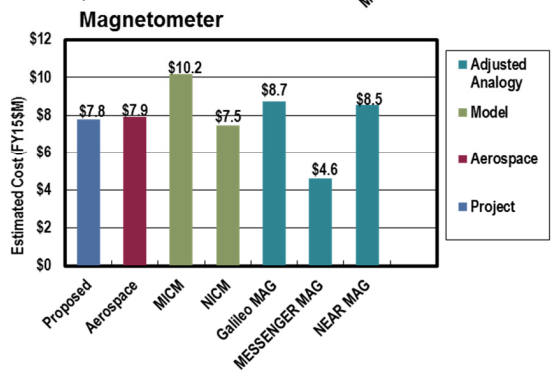

*Figure 2 Option 3-5 Orbiter/Flyby instrument Estimates based on Analogies and Modeled costs*

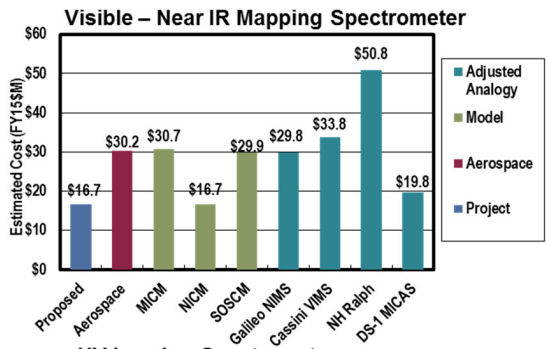

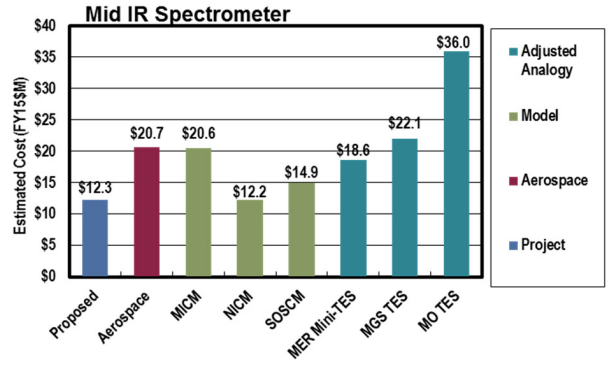

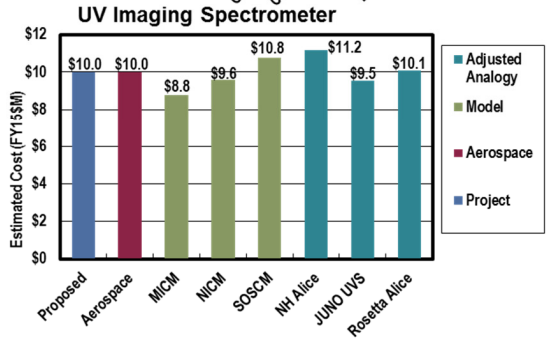

*Figure 3 Option 6 Spectrometer Estimates based on Analogies and Modeled costs*

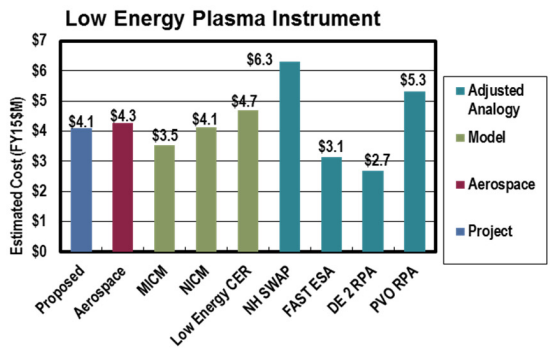

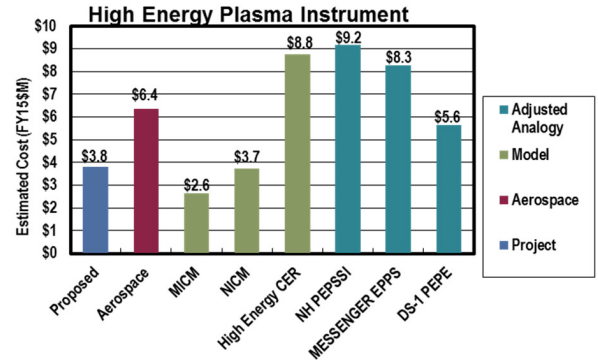

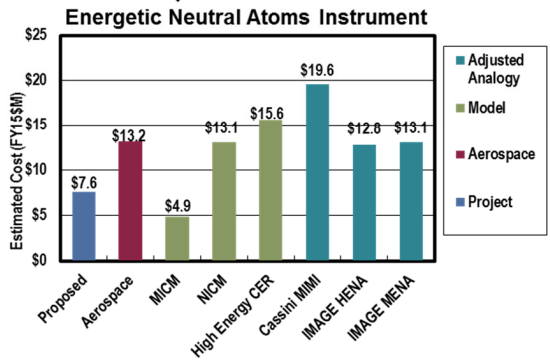

*Figure 4 Option 6 Energy Instrument Estimates based on Analogies and Modeled costs*

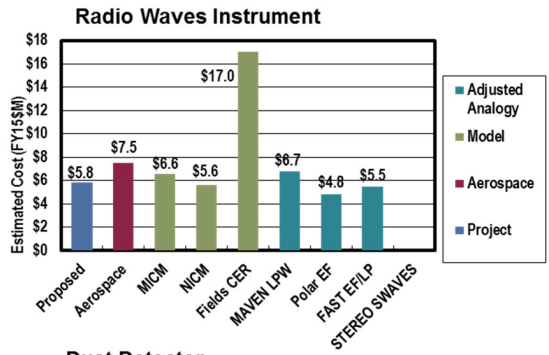

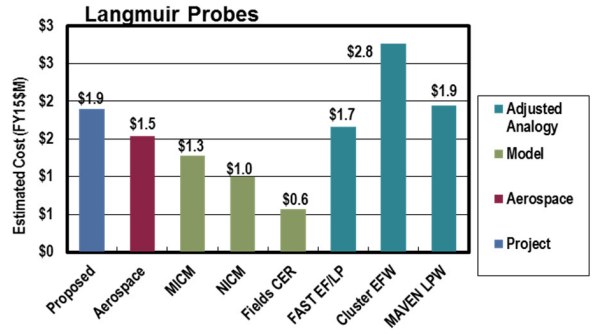

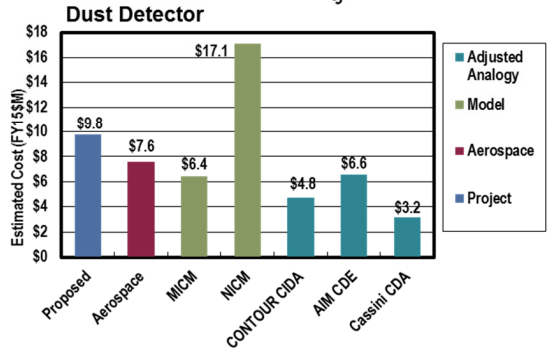

*Figure 5 Option 6 Particle and Fields Instrument Estimates based on Analogies and modeled costs*

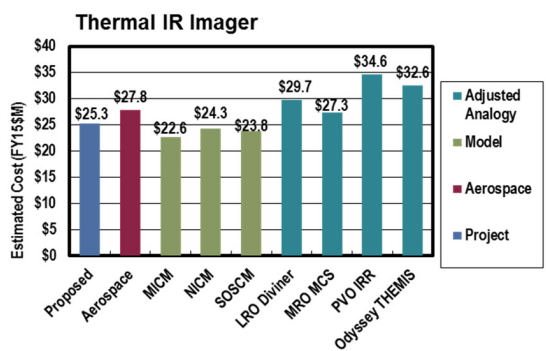

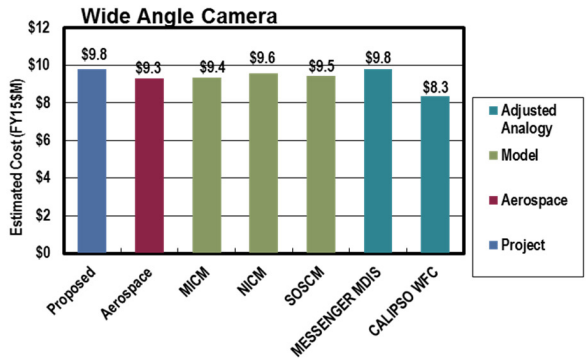

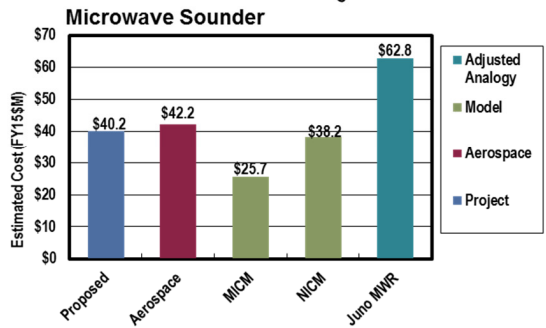

*Figure 6 Option 6 Imager Estimates based on Analogies and Modeled costs*

Orbiter, flyby, and probe elements were estimated using multiple analogies including Aerospace Small Spacecraft Cost Model (SSCM) and Project Cost Estimating Capability (PCEC) model. Other flight system elements, such as the Solar Electric Propulsion (SEP) stage and entry vehicle, in which these cost models were not applicable, used SEER-H and Aerospace developed CERs based on historical mission data. Table 4 identifies the analogies used for WBS element wrap rates and various flight system elements.  Table 5 provides a summary of flight system elements for each option.  Comparison of each estimate to the set of analogies used is shown in in Figure 7, Figure 8, and Figure 9.

*Table 4 Summary of Program Wraps and Flight System Element Analogies*

| Analogy | Wraps | | Adjusted Analogies | | | |
| | PM/SE/MA | Pre-Launch GDS/MOS & Science | Entry System | Orbiter | SEP Bus | MO&DA |
|---|---|---|---|---|---|---|
| Cassini | | | | • | | |
| Dawn | • | • | | | • | • |
| Genesis | • | • | • | | | |
| Juno | • | • | | | • | • |
| MER | • | • | • | | | • |
| MPL | | | • | | | |
| MRO | • | • | | | | • |
| MSL | • | • | • | | | • |
| New Horizons | | | | • | | • |
| Pathfinder | | | • | | | |
| Phoenix | • | • | • | | | • |
| Stardust | • | • | • | | | |
| Deep Impact | • | • | | | | |

*Table 5 Flight System Elements Estimate Summary*

| Aerospace ICE (FY15 $M) | Option 3 | Option 4 | Option 5 | Option 6 |
|---|---|---|---|---|
| Flight System PM/SE/MA | $ 92.5 | $ 65.2 | $ 74.0 | N/A |
| Atmospheric Probe | $ 42.2 | $ 47.0 | $ 47.2 | N/A |
| Entry System | $ 27.5 | $ 27.5 | $ 28.5 | N/A |
| Orbiter/Flyby Bus | $ 464.8 | $ 375.5 | $ 461.6 | $ 515.4 |
| *RTG* | *$ 195.0* | *$ 195.0* | *$ 195.0* | *$ 225.0* |
| SEP Bus | $ 63.3 | N/A | N/A | N/A |
| SEP IPS | $ 77.1 | N/A | N/A | N/A |
| SEP HVPS | $ 46.2 | N/A | N/A | N/A |
| *Ames/Langley EDL Testing* | *$ 3.8* | *$ 3.8* | *$ 3.8* | *N/A* |
| **Total** | **$ 1,012.4** | **$ 714.1** | **$ 810.0** | **$ 740.4** |

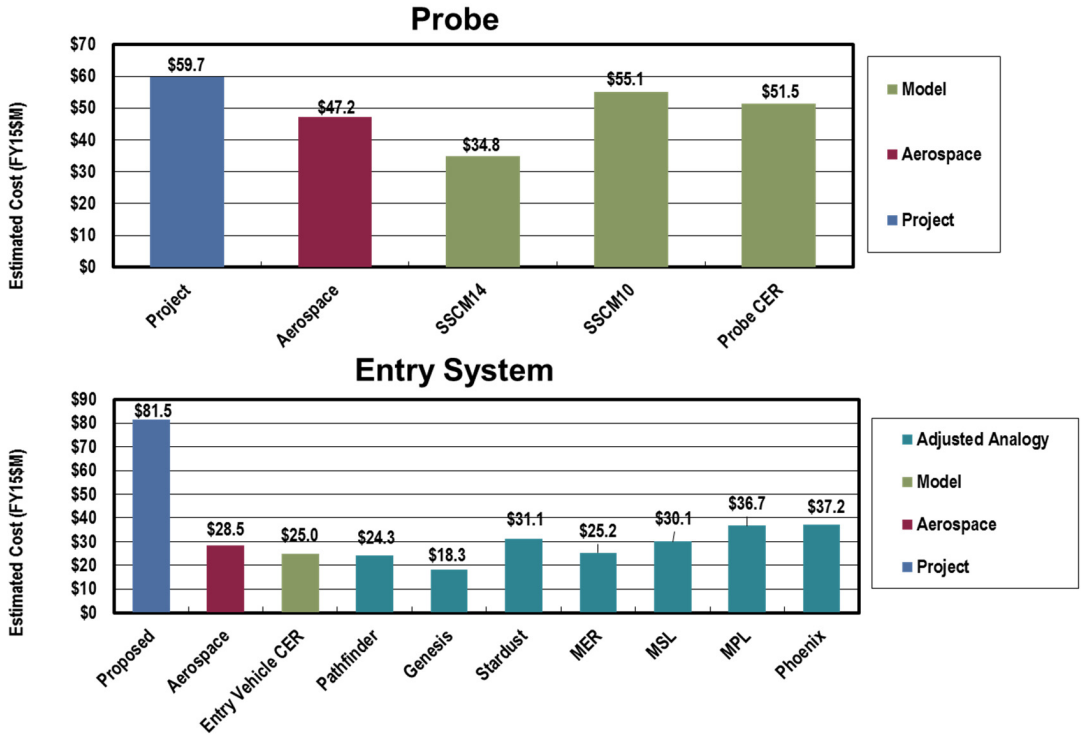

Figure 7 Probe and Entry System Estimates based on Analogies and modeled cots

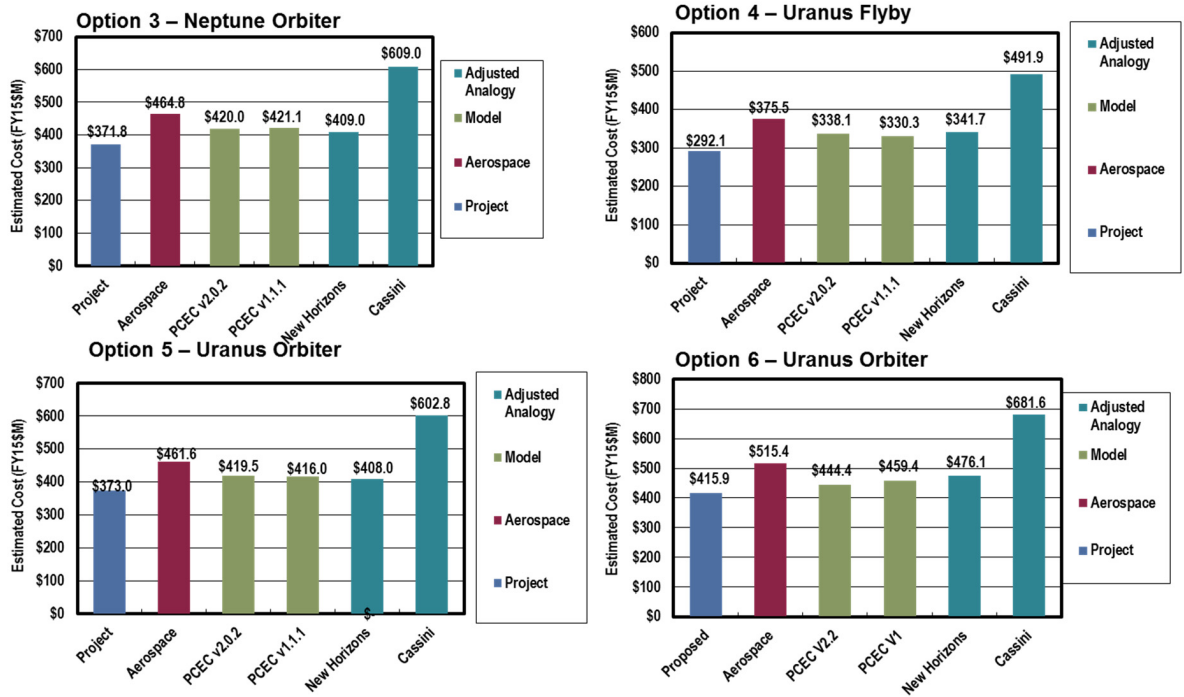

Figure 8 Orbiter and Flyby Estimates based on Analogies and modeled cots

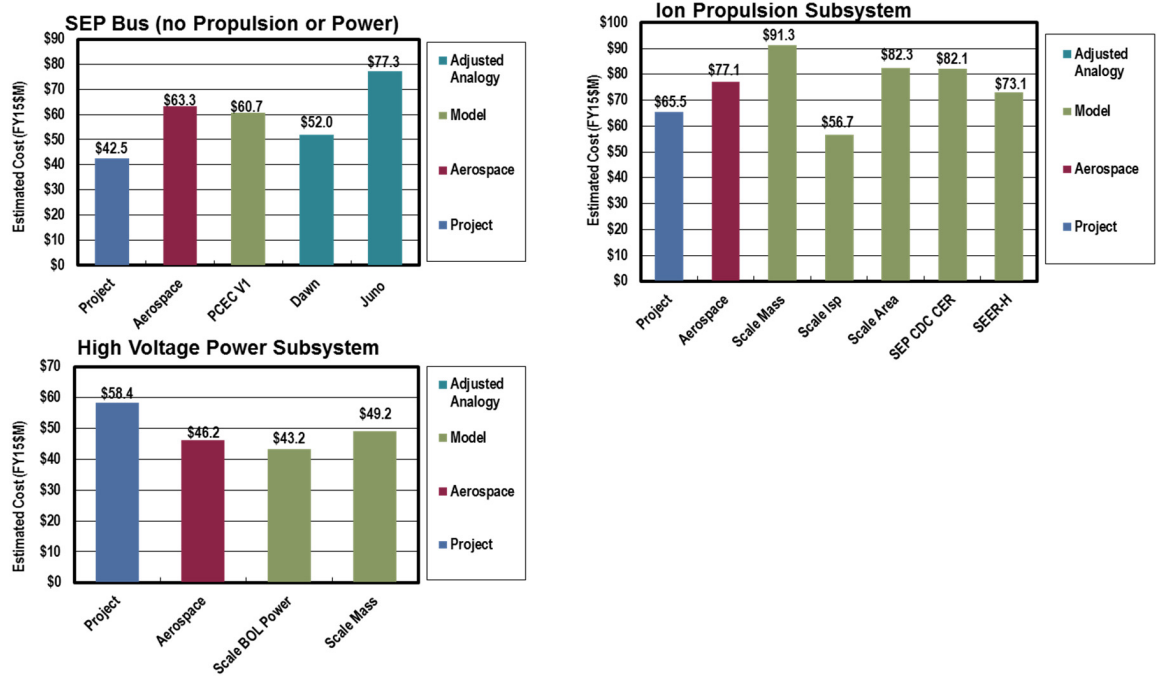

*Figure 9 Option 3 SEP Stage Estimates based on Analogies and modeled cots*

Project Management, System Engineering, and Mission Assurance (PM/SE/MA) and pre-launch Ground Data Systems, Mission Operations Systems, and Science (GDS/MOS/Science) were estimated using wrap factors derived from analogous missions. Phase E Mission Operations and Data Analysis (MO/DA) costs were estimated based on annual spend rates from analogous missions. Analogies used for these elements are included in Table 4 with percentage ranges for PM/SE/MA and Pre-launch ground of 7.2% to 19.8% and 4.6% to 18.9% respectively. Figure 10, Figure 11, Figure 12, and Figure 13 provide a comparison of Phase E MO/DA estimates and analogies for various mission phases and total MO/DA as it compares to the JPL project estimates respectively. Phase A costs were estimated as 1.5% of the total Phase B through Phase D.

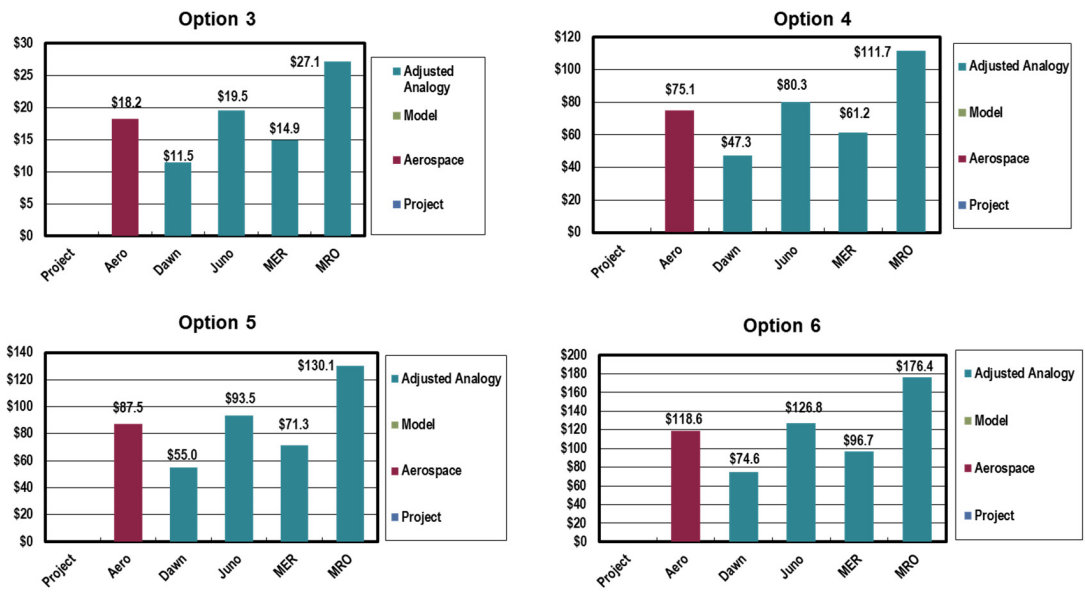

*Figure 10 Inner Cruise and Check Out Estimates based on Analogies*

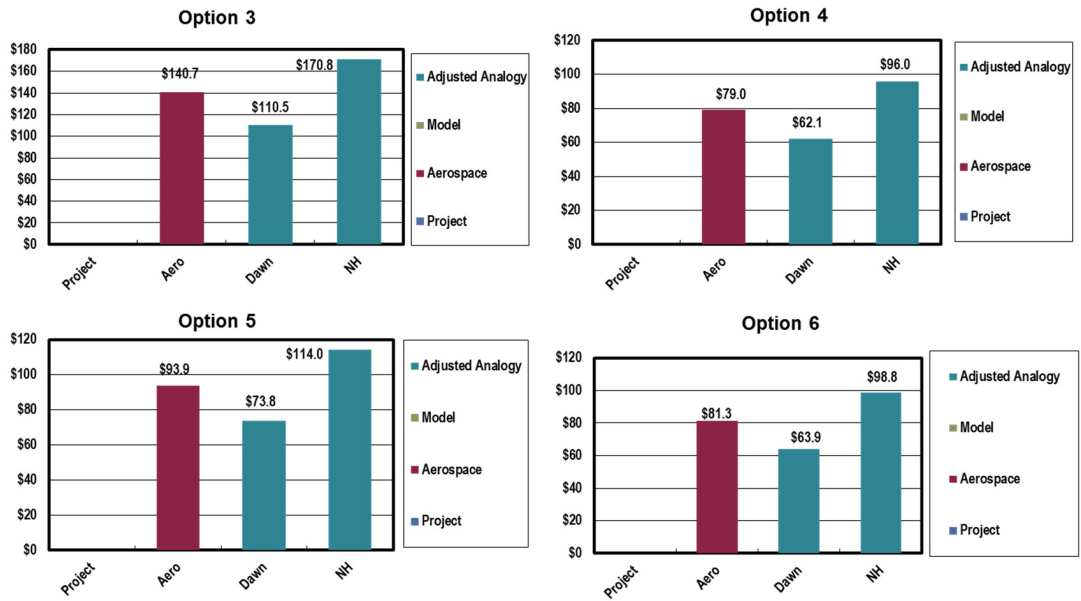

*Figure 11 Quiescent Cruise Estimates based on Analogies*

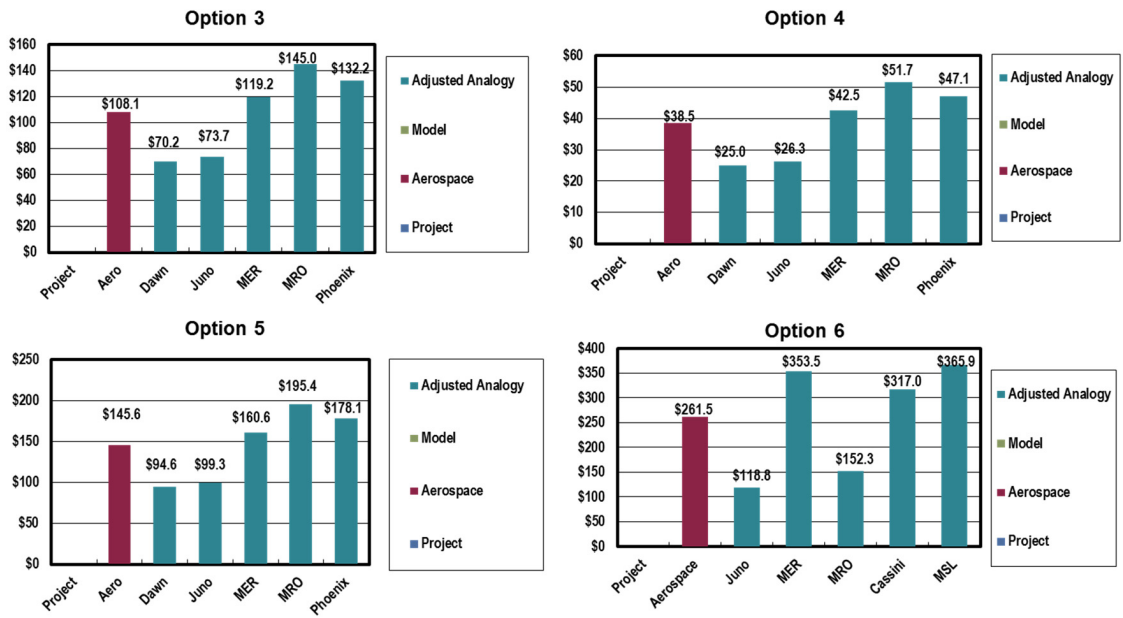

*Figure 12 Science Orbit Operations Estimates based on Analogies*

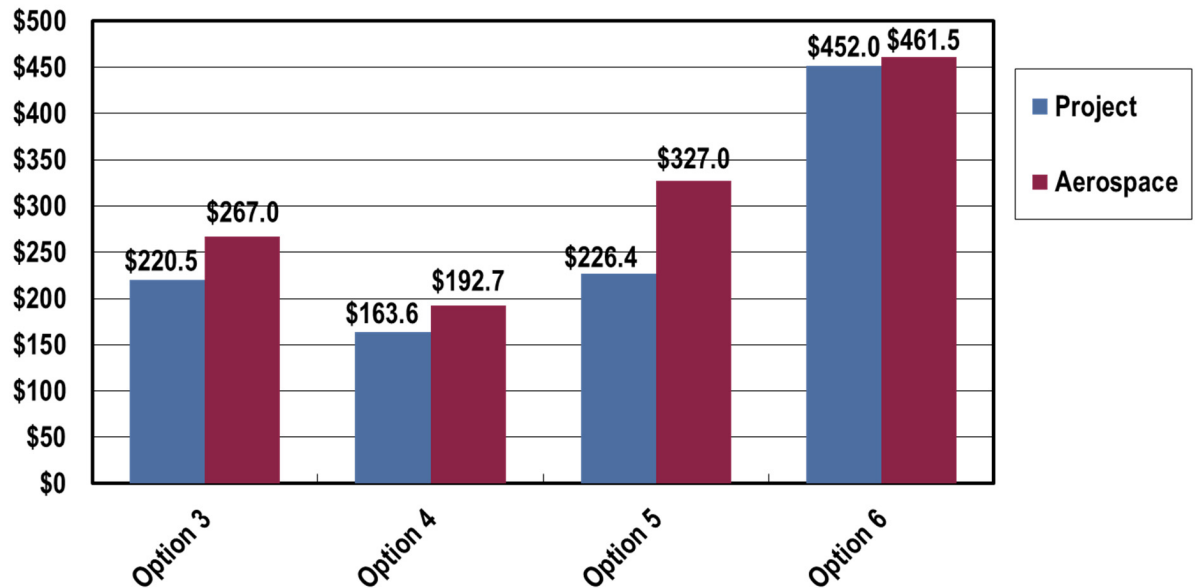

*Figure 13 Overall Comparison of Total MO/DA estimates for all options*

Launch Vehicle Nuclear Payload Support, Deep Space Network (DSN), Ames/Langley EDL testing, and Radioisotope Thermal Electric Generators (RTGs) costs are passed through from the JPL provided estimate at $33M, $80.8M, $195M/$225M, and $3.8M (FY15), respectively.

Probabilistic (Monte Carlo) methods were then used to quantify the cumulative distribution function (Figure 14) for the cost estimates generated from the different methods. For each WBS element a triangular distribution of possible costs are developed from the various estimates, modeled or analogous. The low value of the triangle is set by the lowest of the estimates, the peak or mode of the triangle is set by the average of estimates, and the high value of the triangle is set by the highest of estimates multiplied by a design maturity factor, which ensures that the high value represents the true worst case scenario. The use of multiple models and analogies ensures that the results are not biased by a single approach. Cost reserves are derived from the difference between the sum of the individual element mode values and the 70[th] percentile value from the project-level probability distribution. Reserves for each options are provided in the summary, Table 1.

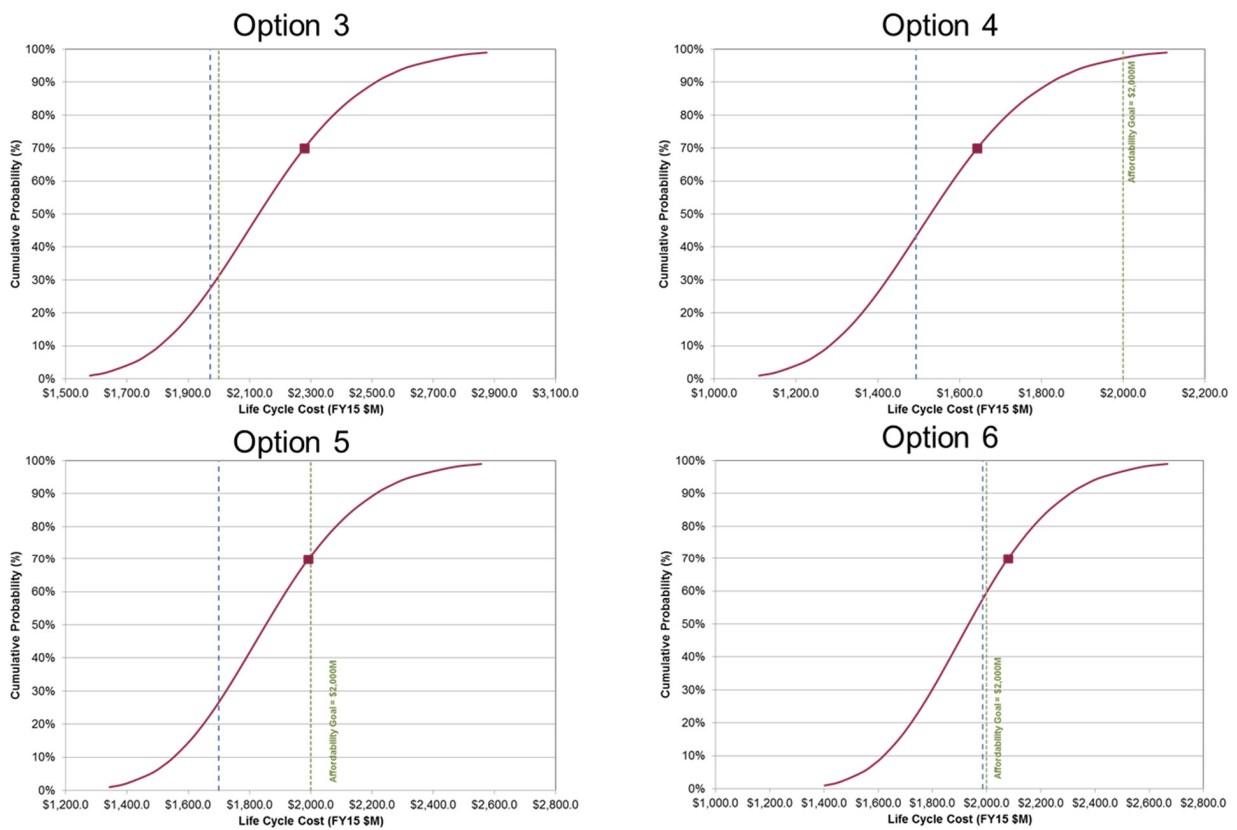

*Figure 14 Cumulative Distribution Function "S-Curve" of The Aerospace Corporation's Monte Carlo simulation of probability life cycle costs for Options 3-6*

# F   SCIENCE QUESTIONS FROM A-TEAM STUDY

As part of the A-Team study, the Science Definition Team (SDT) prepared an extensive list of science questions (~90) it was interested in answering. The questions were traced to the Visions and Voyages Planetary Science Decadal Survey's goals and objectives, and a preliminary assessment was made of the measurements and instruments that could address them.  This list was prioritized and reformatted through several iterations to become the twelve-item Science Traceability Matrix (STM) presented in **Table 3-2**.  The STM contains the high-priority science objectives for an ice giant Flagship mission.

This appendix contains the larger, un-prioritized list of questions the SDT started with.  It is presented for completeness, and as a potential resource for future activities.   The SDT emphasizes, however, that this list represents an early stage in development of our mission science objectives.  The STM in **Table 3-2** should be used for most purposes.

| Decadal Survey Science Goals | Decadal Objective | Discipline (Int, Atm, Ring/Moon, Sat, Mag) | Investigation | Measurements | Instrument | Functional Requirement | Color code: |
|---|---|---|---|---|---|---|---|
| | | | | | | | Must Do / Nice to Have |
| | 1A. Understand heat flow and radiation balance in giant planets. | Int, Atm | 1A1. Determine the atmospheric heat balance | Spatial and temporal variation of net thermal emission | IR spec or radiometer | Need view of at least 90 deg phase, if can't get all probe | |
| | | | | temperature and pressure profile | ASI | | |
| | | | | Measure Bond albedo | IR /vis imaging | | |
| | | Atm | 1A2. Measure tropospheric 3-D flow (zonal, meridional, vertical) including winds, waves, storms and their lifecycles | repeat maps of cloud tracers at multiple wavelengths, including methane absorption bands | UV, Vis, Near-IR imager | global mapping and feature tracks | |
| | | | | Remote Doppler measurements | Doppler imager or spectrometer | | |
| | | | | Doppler vertical wind profile | USO | probe | |
| | | | | Distribution of disequilibrium species | Near-IR/IR spectrometer | | |
| | | Atm | 1A3. Characterize cloud properties, resolved spatially and temporally | repeat maps of cloud tracers at multiple wavelengths, including methane absorption bands, and multiple viewing angles | UV, Vis, Near-IR imager | feature tracking | |
| | | | | In situ probe sampling | nephelometer/net flux radiometer | probe | |
| | | Atm,MAG | 1A4. Search for evidence of moist convection/ lightning | high phase imaging of lightning | Narrow-band imaging | High phase imaging | |
| | | | | repeat maps of cloud tracers at multiple wavelengths, including methane absorption bands, and multiple viewing angles | UV, Vis, Near-IR imager | feature tracking | |
| | | | | In situ observations of RF interefence/whistlers | Probe RF/acoustics | | |
| | | | | Remote observations of RF interefence/whistlers | (1)Radio and Plasma Wave Receiver | 10's Hz to MHz, waveforms for whistlers | |
| | | Int, Mag | Repeat "1C4. Constrain the structure and characteristics of the interior, including layering, locations of convective and stable regions, internal dynamics, and asymmetries in gravity field" from "Probe the interiors" Decadal Objective here, which covers the deep interior heat flow contribution as intended | | | | |
| | | Atm, Int, Mag | Repeat "Quantify behavior of deep zonal winds" from "Atmos dynamics" Decadal Objective | | | | |
| | | Atm | Repeat "Charac. bottom of clouds" from "Atmos dynamics" Decadal Objective | | | | |
| | | Atm | Do all energy balance measurements at different seasons | Compare ground-based and Voyager to new mission results | | | |
| | 1B. Investigate the chemistry of giant planet atmospheres. | Atm, Int | 1B1. Determine bulk composition, including abundances and isotopes of heavy elements, He and heavier noble gases | In situ measurements by entry probe | mass spec/tls | probe | |
| | | | | Remote IR spectroscopic and radio occultation measurements to supplement certain probe data | IR, radio | | |
| | | Atm | 1B2. Global distribution and vertical profiles of trace gases in the stratosphere and upper troposphere including hydrocarbons, nitriles, CO, etc. | Limb observations of IR emission features (stratosphere) | IR spec | Limb viewing | |
| | | Atm | | IR spectral nadir observations (troposphere) | IR spec | | |
| | | Atm | 1B3. Map the distribution of hydrogen ortho-para ratio | Observations of H2 spectral lines | | | |
| | | Atm, Int | 1B4. Constrain deep convection | Map distribution of disequilibrium species, PH3, GeH4, SiH4, CO with IR spectroscopy | | | |
| | | | | Measure disequilibrium species with entry probe | | | |
| | | Atm | 5. Determine the energy sources, temperature, density of upper atmosphere. | Near-IR and FUV spectroscopy. Ionospheric outflow, charged particle precipitation. | | | |
| | | Atm | 6. Determine the cloud structure and characteristics | IR remote sensing, supplemented by entry probe | | | |
| | | Atm, Int, Mag | 1C1. Confirm or improve upon the Voyager-derived rotation rates | Magnetic field observations, Radio emission observations | Magnetometer, Radio Receiver | Radio 10s to 100s kHz | |
| | | | | Gravity passes at multiple inclinations | | | |

| Decadal Survey Science Goals | Decadal Objective | Discipline (Int, Atm, Ring/Moon, Sat, Mag) | Investigation | Measurements | Instrument | Functional Requirement | Color code: |
|---|---|---|---|---|---|---|---|
| | | | | | | | Must Do / Nice to Have |
| Giant Planets as Ground Truth for Exoplanets | 1C. Probe the interiors of giant planets. | Int, Ring/Moon | 1C2 Improve precision of low and high order gravitational moments and time variations | Track ring structures with sub-km precision. Measure precise apsidal precession rates and nodal regression rates of the main rings and inner moons. Look for changes over time scales of years to decades. | | | |
| | | Int, Atm, Mag | 1C3 Measure planetary shape (oblateness) | Occultations at multiple latitudes in order to measure the polar and equatorial radii and determine shape | | | |
| | | Int, Mag, Ring/moons | 1C4. Constrain the structure and characteristics of the interior, including layering, locations of convective and stable regions, internal dynamics, and asymmetries in gravity field | B-field (power spectrum, and secular variations). | | | |
| | | | | Gravity passes at multiple inclinations | | | |
| | | | | Auroral observations and satellite footprints. | | | |
| | | | | Energetic particle measurements of microsignatures due to moon 'bite-outs' | | | |
| | | | | Search for resonant features in the inner rings possibly driven by planetary oscillation modes. | imager | | |
| | | | | Search for planetary oscillations | Doppler imager or spectrometer | | |
| | | Int, Atm, Mag | 1C5. Determine the atmospheric heat balance, same as 1A1, above | | | | |
| | | Int, Atm, Mag | 1C6. Determine the atmospheric bulk composition, same as 1B1. | | | | |
| | | Int, Atm | 1C7. Extend knowledge of the temperature structure into the fully convective region | IR observations of thermal emission and temperature sensitive molecules | | | |
| | | | | Radio and IR Occultations | | | |
| | | | | In situ measurement of vertical temperature vs pressure profile | | | |
| | | Int, Atm, Mag | 1C8. Improve knowledge of the structure of the intrinsic planetary magnetic fields | B-field (power spectrum, and secular variations). | | | |
| | | | | Auroral observations and satellite footprints. | | | |
| | | | | Energetic particle measurements of microsignatures due to moon 'bite-outs' | | | |
| | | Int | Measure the moment of inertia. (May not be possible for Uranus due to the absence of large moons.) | Very precise ring astrometry with a long time baseline, archived. | | | |
| | | Int, Atm | Vigor of deep atmosphere convection: distribution of disequilibrium species, PH3, GeH4, SiH4, CO (see also 1B4) | Remote IR spectroscopy; ground truth with entry probe | | | |
| | | Int, Atm | Quantify behavior of deep zonal winds | Probe sampling | | | |
| | | | | Shape/gravity? | | | |
| | | Int, Atm | Characterize the bottom of cloud/precipitation structures and any alternation of P/T profile | Cloud particle abundance | probe sampling/nephelometer | | |
| | | | | water or methane abundance? | microwave radiometer? | | |
| | | Sat, Ring/Moon | Measure the moment of inertia. (May not be possible for Uranus due to the absence of large moons.) | Astrometry of the rings and innermost moons to solve for the gravity field. | Imager | | |
| | | Int, Ring/Moon | Track planetary precession | Astrometry of the rings and innermost moons to solve for the pole orientation. | Imager | | |
| | | Mag | 1D2. Constrain plasma transport, energy exchange, reconnection, and loss in the magnetosphere | Measure neutrals, negative ions, thermal plasma energy and composition, magnetic fields, charged dust, radio and plasma waves, energetic particles, auroral imaging and spectroscopy. | (1)Magnetometer (1)Particle Instruments (1)Radio and Plasma Waves (2)ENA Instrument | Multiple Orbits, Apogee out to at least 25 planetary radii on dayside, 50 R on nightside, Good Local Time and Latitude Coverage | |

| Decadal Survey Science Goals | Decadal Objective | Discipline (Int, Atm, Ring/Moon, Sat, Mag) | Investigation | Measurements | Instrument | Functional Requirement | Color code: | Must Do | Nice to Have |
|---|---|---|---|---|---|---|---|---|---|

| Decadal Survey Science Goals | Decadal Objective | Discipline (Int, Atm, Ring/Moon, Sat, Mag) | Investigation | Measurements | Instrument | Functional Requirement |
|---|---|---|---|---|---|---|
| | 1D. Analyze the properties and processes in planetary magnetospheres. | Mag | 1D3. investigate solar wind-magnetosphere-ionosphere interaction variation with season, planetary rotation and solar wind conditions, and the effect on the system's energy and momentum budget | Measure neutrals, negative ions, thermal plasma energy and composition, magnetic fields, charged dust, radio and plasma waves, energetic particles, auroral imaging and spectroscopy (auroral morphology), ionospheric electron density. | (1)Magnetometer (1)Particle Instruments (1)Radio and Plasma Waves (2)ENA | Multiple Orbits, Apogee out to at least 25 planetary radii on dayside, 50 radii on nightside, Good Local Time and Latitude Coverage |
| | | Mag, Ring/Moon, Sat, Atm | 1D4. Study satellite/ring/magnetospheric plasma interaction, coupling to the planetary atmosphere, and variation with planetary rotation and season Could probably combine 1D4 and 1D5 | Measure neutrals, negative ions, thermal plasma energy and composition, magnetic fields, charged dust, radio and plasma waves, energetic particles, auroral imaging and spectroscopy. | (1)Magnetometer (1)Particle Instruments (1)Radio and Plasma Waves (2)ENA | Multiple Orbits, Apogee out to at least 25 planetary radii on dayside, 50 R on nightside, Good Local Time and Latitude Coverage |
| | | Mag | 1D5. Determine if any moon is a source of plasma or has a dynamic atmosphere | Measure neutrals, negative ions, thermal plasma energy and composition, magnetic fields, charged dust, radio and plasma waves, energetic particles, auroral imaging and spectroscopy. | (1)Magnetometer (1)Particle Instruments (1)Radio and Plasma Waves (2)ENA | Multiple Close Flybys |
| | | Ring/Moon | Ring dust as a way to study the inner magnetosphere remotely. | Search for spokes, periodic structures in faint rings (G and D rings at Saturn are examples.) Measure drag terms via numerical modeling. | | |
| | | Ring/Moon/MAG | Use the rings as a laboratory for studies of dusty plasmas | In situ measurements far into the inner parts of the ring system. | (1)Dust Instrument (2) Plasma wave Instrument | Waveform Measurments |
| | 1E. Use ring systems *and inner moons* as laboratories for planetary formation processes. | Ring/Moon | 1E1. Obtain a complete inventory of small moons, including embedded source bodies in dusty rings and moons that could sculpt and shepherd dense rings. | Identify all bodies >0.5 km in radius within 250,000 km of the planet's center | | |
| | | Ring/Moon | 1E2.Characterize the full 3-D distribution of dust within the system, identify the sources and sinks for dust particles and the processes that influence their orbital evolution. | High-phase deep imaging for rings, low-phase deep imaging for source bodies. | | |
| | | | | in situ dust | | |
| | | Ring/Moon | 1E3. Determine the precise orbits of inner moons and study their changes over time scales of years to decades. Identify evidence for orbital migration, chaos and/or mutual perturbations | Long-term, regular, precise astrometry for orbit determinations. | | |
| | | Ring/Moon | 1E4. Determine the rotation states of the inner moons and search for unusual spin states that might be evidence of recent impacts. | Repeated imaging photometry of the small moons at small phase angles for spin rates; high-resolution imaging for spin poles. | | |
| | | Ring/Moon | 1E5. Study temporal changes in the rings, including spreading rates and arc formation/evolution/confinement. | High-resolution imaging, stellar and radio occultations, comparisons to Voyager and Earth-based data. | | |
| | | Ring/Moon | 1E6. Determine surface composition of rings and moons; search for variations among moons, and evidence of long-term mass exchange between the inner moons and irregular satellites. | Spatially resolved spectral imaging. | | |
| | | | | Spatial and temporal variation of net thermal emission | IR /vis spec, IR radiometer | Need view of at least 90 deg phase, if can't get all probe |
| | | | | temperature and pressure profile | probe ASI | |
| | | | | Measure Bond albedo | Visible? | |

| Decadal Survey Science Goals | Decadal Objective | Discipline (Int, Atm, Ring/Moon, Sat, Mag) | Investigation | Measurements | Instrument | Functional Requirement | Color code: Must Do / Nice to Have |
|---|---|---|---|---|---|---|---|
| | 1F. Explore planetary extrema in the solar system's giant planets. | Atm | Assess the effects of internal vs external heat flow and minimal solar influence | repeat maps of cloud tracers at multiple wavelengths, including methane absorption bands | UV, Vis, Near-IR imager | global mapping and feature tracks | |
| | | | | Remote Doppler measurements | Doppler imager or spectrometer | | |
| | | | | Doppler vertical wind profile | USO | probe | |
| | | | | Vis/Near-IR spectral mapping with repeat views | | | |
| | | | | feature tracking at multiple wavelengths, including methane absorption bands, and multiple viewing angles | | | |
| | | | | In situ probe sampling | | | |
| | | | | high phase angle imaging of lightning | Narrow-band imaging | | |
| | | | | Vis/Near-IR spectral imaging for cloud structure | | | |
| | | | | In situ observations of RF interefence/whistlers | Probe RF/acoustics | | |
| | | | | Probe sampling | | | |
| | | | | Shape/gravity? | | | |
| | | Atm | Search for characteristic features and properties of an extreme axial tilt or seasonal cycle | Zonal and 2D wind fields, at multiple wavelengths, including IR for thermal winds | | | |
| | | Ring/Moon | Determine the dust influence at the Icy Giants | Study the ring/planet boundary for evidence of time variations in the ring locations and properties. | | | |
| | | | | Measure the meteoritic influx. | Dust Instrument | | |
| 2. Giant Planets Role in Promoting a Habitable Planetary System | 2A. Search for evidence of planetary migration. | Atm | 2A1. Determine bulk composition, same as 1B1 | | | | |
| | | Sat | Study the composition and evolution of satellite surfaces (See 4A2) | Spectral mapping, Isotopic ratios (esp. D/H, 14N/15N), noble gases | | | |
| | | | | gravity | tracking during flybys | | |
| | | Sat | Satellite bulk compositions (See 4A1) | magnetic field (induced, intrinsic?) | magnetometer, plasma instrument | Close Flybys | |
| | | | | shape from limb profiles ~100-m resolution | imaging | | |
| | | | | occultations | imaging (radio?) | | |
| | | | | Satellite dynamics? | imaging | | |
| | | Ring/Moon | Measure orbits and physical properties of irregular satellites. This population strongly constrains the migration models. | Photometry and astrometry of known objects during approach and while in orbit; disk-resolved imaging and spectrometry | | | |
| | 2B. Explore the giant planets' role in creating our habitable Earth through large impacts. | Sat, Ring/Moon | Determine crater populations on satellites, including the small inner moons. (See 4A4) | Near-global imaging with <1km resolution | Imaging | | |
| | 2C. Determine the role of surface modification through smaller impacts. | Sat, Ring/Moon | Search for evidence of micrometeorite bombardment and surface gardening on satellites, including the small inner moons. (See 4B1 and others) | High resolution spectral imaging | Imaging | | |
| | | | | Spectral and phase properties of surface | Imaging, spectral capability | | |
| | | Atm | 3A1. Determine the atmospheric heat balance, same as 1A1 | | | | |
| | | Atm | 3A2. Measure tropospheric 3-D flow (zonal, meridional, vertical) , same as 1A2. | | | | |
| | | Atm | 3A3. Study the spatial and temporal variation of stratospheric tracers, same as 1B2 | | | | |
| | | Atm | 3A4. Characterize cloud properties, resolved spatially and temporally, same as 1A3 | | | | |

| Decadal Survey Science Goals | Decadal Objective | Discipline (Int, Atm, Ring/Moon, Sat, Mag) | Investigation | Measurements | Instrument | Functional Requirement | Color code: | Must Do | Nice to Have |
|---|---|---|---|---|---|---|---|---|---|

| Decadal Survey Science Goals | Decadal Objective | Discipline (Int, Atm, Ring/Moon, Sat, Mag) | Investigation | Measurements | Instrument | Functional Requirement |
|---|---|---|---|---|---|---|
| 3. Giant Planets as Laboratories for Properties and Processes on Earth | 3A. Investigate atmospheric dynamical processes in the giant-planet laboratory. | Atm | 3A5. Search for evidence of moist convection/ lightning, same as 1A4 | | | |
| | | Atm | 6. Quantify behavior of deep zonal winds | Probe sampling | | |
| | | Atm | | Shape/gravity? | | |
| | | Atm | 7. Characterize atmospheric waves and their effects | UV/Vis/Near-IR mapping, repeat views of cloud tracers | | |
| | | Atm | 8. Characterize the bottom of cloud/precipitation structures | Cloud particle abundance | probe sampling/nephelometer | |
| | | | | water or methane abundance? | microwave radiometer? | |
| | | | | maps of what? | gravity | |
| | 3B. Elucidate seasonal change on giant planets. | Atm | 1. Do all "Invest. atmos. dyn. processes" at different seasons | Compare ground-based and Voyager to new mission results | | |
| | | Ring/Moon | Study the role of radiation pressure in the dusty rings. | Faint ring imaging to obtain 3-D structure. | | |
| | 3C. Assess tidal evolution within giant-planet systems. | Sat, Ring/Moon | Study orbital changes in the inner and classical moons. (Tidal evolution may not be measurable because it depends on the moon mass, and these moons are small.) (See 4C2, 4C3) | Precision astrometry over a long time baseline, including Earth-based and orbiter-based measurements. | Imaging | |
| | 3D. Evaluate solar wind and magnetic-field interactions with planets. | MAG | Same ad 1D3 | | | |
| | 4A. What were the conditions during satellite and ring formation? | Sat, Ring/Moon | Determine surface compositions of rings and small inner satellites. Seek compositional variations across the system. | Spectral imaging at multiple phase angles; albedo determinations | | |
| | | Ring/Moon | Detect trace elements from the ring atmosphere and the dust clouds. | In situ sampling. | | |
| | | Ring/Moon | Determine the masses and densities of the rings and inner moons. | Long-term study of orbital interactions; close flybys of a few inner moons; fly over rings at end of mission? | | |
| | | Ring/Moon | Study evolution of the rings and satellites (small and classical) over time scales of billions of years. | Numerical simulations. | | |
| | | Sat, Ring/Moon | 4A1. Determine the density, mass distribution, bulk composition of major satellites and, where possible, small inner satellites and irregular satellites | gravity | tracking during flybs | Multiple satellite flybys s/c tracking during close flybys, ~50 km altitude. |
| | | | | magnetic field (induced, intrinsic?) | magnetometer, plasma instrument | |
| | | | | shape from limb profiles ~100-m resolution | imaging | |
| | | | | occultations | imaging (radio?) | |
| | | | | Satellite dynamics? | imaging | |
| | | Sat | 4A2. Map satellite surface composition, including organic material | Spectral mapping, Isotopic ratios (esp. D/H, 14N/15N), noble gases | spectral imaging | |
| | | Sat | 4A3.Determine major satellite exosphere/atmospheric composition and their variation with season and planetary rotation | composition | mass spectrometer | |
| | | Sat | 4A4. Constrain crater distributions, relative surface ages, including the small inner moons | Near-global surface mapping at better than km resolution | imaging | |
| | | Ring/Moon | Detect trace elements from the ring atmosphere and the dust clouds. | In situ sampling by dust detector or INMS. | | |
| | | Ring/Moon | Determine the masses of the rings. | Gravity passes over rings at beginning and end of mission if this is feasible. | | |
| | | | | surface mapping for crater formation, population and morphologies | Imaging | |

| Decadal Survey Science Goals | Decadal Objective | Discipline (Int, Atm, Ring/Moon, Sat, Mag) | Investigation | Measurements | Instrument | Functional Requirement | Color code: | Must Do | Nice to Have |
|---|---|---|---|---|---|---|---|---|---|
| 4. How did the satellites and rings of the outer solar system form and evolve? | 4B. What determines the abundance and composition of satellite and ring volatiles? What exchange processes are at work? | Sat, Ring/Moon | 4B1. Search for evidence of past and current surface modification and volatile transport (major satellites and small inner moons). | in situ & spectral mapping of sputtering products; spectral and phase properties of the surface | Mass spectrometer, (spectral) imaging | Satellite flybys with specific illumination and resolution requirements. | | | |
| | | | | high phase angle and direct imaging and spectral plume searches, temporal variability, dynamics | Imaging, spectral capability (short wavelength) | | | | |
| | | | | thermal mapping, thermal inertia, spectral mapping, albedo mapping, glacial processes (Triton; rheology; structural mapping) | Thermal IR | | | | |
| | | | | Glacial processes (rheology; structural/topographic mapping) | Imaging | | | | |
| | 4C. How are satellite and ring thermal and orbital evolution and internal structure related? What energy sources are available? Did/do satellites harbor internal oceans and habitable environments? | Sat | 4C1. Correlate gravity & shape/topography, geologic mapping and subsurface structure | surface mapping and topography | imaging (stereo drives flyby geometry) | Satellite flybys with specific illumination and resolution requirements. | | | |
| | | | | subsurface sounding | sounding radar | | | | |
| | | | | magnetic sounding | magnetometer, plasma instrument | | | | |
| | | Sat | 4C2. Search for sources of endogenic heating | Thermal mapping | thermal IR | | | | |
| | | | | High phase angle and direct imaging and spectral plume searches, temporal variability, dynamics | visible and uv imaging for plume searches and monitoring (preferred wavelength for Triton plumes?) | | | | |
| | | Sat | 4C3. Study internal structure using satellite libration, Cassini states, pole position, orbital motion, tides, induced/intrinsic magnetic fields, etc. | mapping at different true anomalies | imaging | | | | |
| | | | | magnetic field (induced, intrinsic?) | (accompanying analysis) | | | | |
| | | | | interior modeling | | | | | |
| | | Ring/Moon | Study small? satellite libration, Cassini states, pole position, etc. | High-resolution imaging at frequent intervals over a time scale of years. | | | | | |
| | 4D. What is the diversity of geologic activity and how has it changed over time? Are active endogenic processes at work? To what extent has material been exchanged between satellite surfaces and their interiors? | Sat | 4D1. Global-scale (80%) surface and structural mapping of major satellites <0.5-km-scale. | High spatial resolution imaging at multiple phase angles and wavelengths, topography and slopes | imaging, spectral capability (filters or spectrometer) | Requires satellite flybys s/c tracking during close flybys, ~50 km altitude. | | | |
| | | | | Topography, slopes, subsurface structure to identify dynamically supported topography | imaging (stereo drives flyby geometry) | | | | |
| | | | | | altimetry (also useful for sounding of cloud altitudes?) potential if in orbit around satellite (e.g., Triton) | | | | |
| | | | | spectral mapping | sounding radar | | | | |
| | | | | | imaging spectrometer | | | | |
| | | | | thermal mapping | thermal IR | | | | |
| | | Sat | 4D2. Determine the interior mass distributions of the classical satellites | gravity mapping | tracking during close flybys | | | | |

| Decadal Survey Science Goals | Decadal Objective | Discipline (Int, Atm, Ring/Moon, Sat, Mag) | Investigation | Measurements | Instrument | Functional Requirement | Color code: | Must Do | Nice to Have |
|---|---|---|---|---|---|---|---|---|---|
| | | Sat | satellites.<br>(Same as 4A1) | Precise measurements of polar orientation and librations. | imaging | | | | |
| | | Ring/Moon | Search for dust rings in the orbits of satellites as an indicator of geologic activity | Deep, high-phase imaging | | | | | |
| | 4E. What is the range of ring-moon interactions, and how have these interactions changed over time? | Ring/Moon | 4E1. Determine the rotation states of the inner moon, same as 1E4. | orbiter photometry for rotation rates; spatially resolved images for poles; repeated observations for librations. | imager | | | | |
| | | Ring/Moon | 4E2. Determine the precise orbits of inner moons and study their changes over time scales of decades. Identify evidence for chaos and/or mutual perturbations., same as 1E3 | Precision astrometry spanning multiple years; precision timing of stellar occultations and mutal events; numerical modeling. | | | | | |
| | | Ring/Moon | 4E3. Study evidence of long-term mass exchange between the inner moons and irregular satellites, same as 1E6 | Resolved spectral imaging of moons and dust rings; high-phase searches for additional dust rings. | | | | | |
| | | Ring/Moon | 4E4. Study evidence for cataclysmic change in the history of the satellite system. | Covered by other investigations, e.g., properties of Nereid and of Neptune's inner ring-moon system. | | | | | |
| | | Ring/Moon | 4E5. Search for additional unseen moons, same as 1E1 | Fine-resolution imaging at low phase angles. | | | | | |
| | | Ring/Moon | Seek evidence for past mutual collisions between inner moons and where possible irregular satellites | Shape determinations and cratering history of moons via high-resolution imaging, searches for moonlet belts via deep, low-phase imaging. | | | | | |
| | | Ring/Moon | Determine the lifetimes of ring-moons by measuring the torques from the nearby rings. | Precision astrometry spanning multiple years; precision timing of stellar occultations and mutal events; archive of precision of photometry from Earth; numerical modeling. | | | | | |
| | 5A. How do active endogenic processes shape the satellites' and rings' surfaces and influence their interiors? | Sat | (See 4D1) | | | | | | |
| | 5B. What processes control the chemistry and dynamics of satellite atmospheres? | Mag, Sat | 5B1. Search for exospheres/ionospheres at any moons and how they vary with planetary rotation and season, same as 4A3 | In situ and remote composition measurements. Occultations. Radio and Plasma Wave measurements (to determine UHR). Magnetic Field, particles, radiation environment, auroral footprints. Modeling | | | | | |
| | | Mag | 5B2. Determine if any moon is a source of plasma or has a dynamic atmosphere, same as 1D5 | In situ and remote composition measurements. Occultations. Radio and Plasma Wave measurements (to determine UHR). Magnetic Field, particles, radiation environment, auroral footprints. Modeling | | | | | |
| | | Sat | 5B3. Determine the composition, density, structure and variability of Triton's atmosphere. | In situ and remote composition measurements, including isotopes (D:H, N14:15), noble gases | | | | | |
| | | Sat | 5B4. Search for evidence of Triton's atmospheric winds and dynamics. | occultations<br>Map topography, dunes, aeolian features Triton Plumes | | | | | |
| | | Sat | 5B5. Determine the source of Triton's atmosphere. | Observe volatile transport, sputtering, plumes as above; composition & albedo of surface | | | | | |
| | | Sat | 5B6. Determine Triton's atmospheric loss rate. | Measurements of mag field, particles, radiation environment<br>In situ measurements<br>Auroral footprints?, occultations? | | | | | |
| | | Sat | Charged particle precipitation into the atmosphere, scavenging, plasma-neutral chemistry (including tholin production), sputtering. | Neturals, negative ions, thermal plasma energy and composition, magnetic fields, charged dust, plasma waves, energetic particles, auroral imaging. | INMS, plasma, particle number density/compositio n, magnetometer, dust detector | | | | |

| Decadal Survey Science Goals | Decadal Objective | Discipline (Int, Atm, Ring/Moon, Sat, Mag) | Investigation | Measurements | Instrument | Functional Requirement | Color code: Must Do / Nice to Have |
|---|---|---|---|---|---|---|---|
| 5. What processes control the present-day behavior of the satellites and rings? | | Atm | 1. Determine the composition, density, structure and variability of satellite atmospheres | In situ and remote composition measurements, including isotopes (D:H, N14:15), noble gases | | | |
| | | Atm | 2. Search for evidence of atmospheric winds and dynamics | Occultations | | | |
| | | | | Map topography, dunes, aeolian features Triton Plumes | | | |
| | 5C. How do exogenic processes modify these bodies? | Ring/Moon | 5C1. Determine the role of solar radiation pressure on dust in the system, same as 1E2 | Imaging of rings for changes in 3-D structure; time variation in the inner dust rings. | | | |
| | | Ring/Moon | Search for impact-derived ripple patterns in the rings. | Imaging of rings at multiple phase angles | | | |
| | 5D. How do satellites (and rings) influence their own magnetospheres and those of their parent planets? | Mag, Sat, Ring/Moon, Atm | 5D1 Study satellite/ring/magnetospheric plasma interaction, coupling to the planetary atmosphere, and variation with planetary rotation and season, same as 1D4 | Measure neutrals, negative ions, thermal plasma energy and composition, magnetic fields, charged dust, radio and plasma waves, energetic particles, auroral imaging and spectroscopy. Modeling. | (1)Magnetometer (1)Particle Instruments (1)Radio and Plasma Waves (2)ENA | Multiple Orbits, Apogee out to at least 25 planetary radii on dayside, 50 R on nightside, Good Local Time and Latitude Coverage | |
| | | Mag, Sat | 5D2 Determine if any moon is a source of plasma or has a dynamic atmosphere, same as 1D5 | Measure neutrals, negative ions, thermal plasma energy and composition, magnetic fields, charged dust, radio and plasma waves, energetic particles, auroral imagingand spectroscopy. Modeling. | (1)Magnetometer (1)Particle Instruments (1)Radio and Plasma Waves (2)ENA | Multiple Close Flybys | |
| | | Mag | Do any of the moons have an intrinsic magnetic field? | Magnetic field, plasma | | Must-have for Triton? | |
| | | Mag | Are any of the moons associated with an induced magnetic field potentially caused by a subsurface ocean? | Magnetic field | | Must-have for Triton? | |
| | | Ring/Moon | Study sources and sinks of magnetospheric materials throughout the system. | In situ measurements far into the inner parts of the ring system. | | | |
| | | Ring/Moon | Assess the role and rate of loss of ring dust into the planet. | High-phase imaging of the inner rings, repeated to observe variations over time scales of years. | | | |
| | 5E. What are the roles of ongoing interactions between rings, moons and the planet? | Ring/Moon | 5E1. Characterize the particle size distribution and packing density in dusty and dense rings, same as 1E2 | In situ dust detections on near-polar orbits; very high phase imaging from within the planet's shadow. | Dust detector, imaging | | |
| | | | | Multiple occultations from UV to radio wavelengths. Imaging down to the scale of the largest ring particles? | | | |
| | | | | Sensitive imaging at many phase angles and wavelengths, sensitive stellar occultation studies for optical depth. | | | |
| | | Ring/Moon | 5E2. Determine the numbers and locations of embedded source bodies within the dusty rings, same as 1E1 | High-resolution imaging, low phase, complete coverage in orbital longitude, repeated multiple times. | | | |
| | | Ring/Moon | 5E3. Study the evolution of arcs and seek confinement mechanisms, same as 1E5 | Fine-resolution imaging, variations with time, dynamical modeling. | | | |
| | | Ring/Moon | 5E4.Characterize the fine-scale structures in the rings, including edge waves, density and bending waves, normal modes, propellors and other localized ring features | Fine-resolution imaging, uplink radio occultation, high-sensitivity stellar occultations. | | | |
| | | Ring/ Moon | 5E5. Characterize the particle size distribution and packing density in dusty and dense rings | (Dust) Imaging and spectroscopy at high phase angles, in-situ measurements (Dense Rings) Extremely high-resolution imaging,photometry and occultations at multiple wavelengths | | | |
| | | Ring/Moon | Study the inner dust rings to assess the role and rate of loss into the planet. | High-phase imaging of the inner rings, repeated to observe variations over time scales of years. | | | |
| | | Ring/Moon | Determine the electric charges on the system dust | In situ dust detector; dust cloud imaging and 3-D modeling, dynamical modeling. | | | |

| Decadal Survey Science Goals | Decadal Objective | Discipline (Int, Atm, Ring/Moon, Sat, Mag) | Investigation | Measurements | Instrument | Functional Requirement | Color code: Must Do / Nice to Have |
|---|---|---|---|---|---|---|---|
| | | Ring/Moon | Determine the rate of thermal transport through the ring as a measure of thermal inertia, rotation rates, dynamical transport. | Thermal IR measurements on both sides of ring, map out dependence with longitude and season | | | |
| | | Mag | What are the implications of the moon-magnetosphere interaction for the surfaces of each satellite? | In situ & spectral mapping of sputtering. Spectral mapping. Measure neutrals, negative ions, thermal plasma energy and composition, magnetic fields, charged dust, radio and plasma waves, energetic particles, auroral imaging and spectroscopy. Modeling | | | |
| 6. What are the processes that result in habitable environments? | 6A. Where are subsurface bodies of liquid water located, and what are their characteristics, ~~and~~ histories, and exchange processes? | Sat, Mag | 6A1. Search for subsurface oceans, same as 4C1, 4C3 | Magnetic field, libration | | | |
| | 6B. What are the sources, sinks, exchange processes, and evolution ~~of~~ affecting organic material? | Sat, Ring/Moon | Search for organics on the surfaces of satellites, inner moons and rings; exchange throughout the system (See 4A2, 4B1, 4C2 -- multiple investigations captured in original STM) | Spectral imaging of inner moons and rings | | | |
| | 6C. What energy sources are available to sustain or constrain life? | Sat | Study orbital motion and tides on the classical satellites (4C3) | | | | |
| | | | Correlate gravity & shape/topography, thermal and geologic mapping, and subsurface structure (See 4C1) | | | | |
| | | | Determine the composition, density, structure and variability of satellite atmospheres (See 4A3, 5B3) | | | | |
| | | | Charged particle interactions with the atmosphere and surface. Energy input into conductive layers. (See 1D4) | | | | |
| | 6D. Is there evidence for ~~life~~ habitable environments on the satellites? | Sat | Map surface composition (See 4A2) | | | | |
| | | | Search for evidence of biomarkers | | | | |
| | | | Radiation destruction of biomarkers. | | | | |